\documentclass{cernyrep}
\usepackage{epsfig}

\def\lsim{\raise0.3ex\hbox{$\;<$\kern-0.75em\raise-1.1ex\hbox{$\sim\;$}}}
\def\gsim{\raise0.3ex\hbox{$\;>$\kern-0.75em\raise-1.1ex\hbox{$\sim\;$}}}

\newcommand{\iddrei}  {{{\mathchoice {\rm 1\mskip-4mu l} {\rm 1\mskip-4mu l}
{\rm 1\mskip-4.5mu l} {\rm 1\mskip-5mu l}}}}

\newcommand {\eq} [1] {Eq.~(\ref{#1})}
\newcommand {\eqs} [2] {Eqs.~(\ref{#1}) and (\ref{#2})}
\newcommand {\Bfig} [1] {Figure~\ref{#1}}
\newcommand {\fig} [1] {Fig.~\ref{#1}}
\newcommand {\sect} [1] {Sect.~\ref{#1}}
\newcommand {\tab} [1] {Table~\ref{#1}}

\newcommand{\ETmi}{\ensuremath{E\!\!\!/}_T\xspace}
\newcommand{\ptmiss}{p_T\!\!\!\!\!\!\!\!\!\!\! \not \,\,\,\,\,\,\,\,\,}

\newcommand{\chapt}[2]{chapter~\ref{#1}.\ref{#2}}

\newcommand{\mrm}[1]{\ensuremath{\mathrm{#1}}}

\newcommand{\numentry}[2]{
\begin{minipage}[t]{1.3cm}\flushright\texttt{#1}\end{minipage}
\hspace{5mm}: 
\begin{minipage}[t]{13cm}\noindent
#2\end{minipage}\\[2mm]
}

\newcommand{\snumentry}[2]{
\begin{minipage}[t]{1.2cm}\flushright\texttt{#1}\end{minipage}
\hspace{2mm}:
\begin{minipage}[t]{11cm}\noindent
#2\end{minipage}\\[1mm]
}

\newcommand{\entry}[1]{
\begin{minipage}[t]{1cm}\ \end{minipage} 
\begin{minipage}[t]{13.5cm}\noindent
#1\end{minipage}\\[2mm]
}

\newcommand{\DRbar}{{\ensuremath{\overline{\mathrm{DR}}}}}
\newcommand{\MSbar}{{\ensuremath{\overline{\mathrm{MS}}}}}
\newcommand{\GeV}{\mathrm{GeV}}
\newcommand{\mgut}{\ensuremath{M_{\mathrm{input}}}}

\newcommand{\arrdes}[1]{\begin{center}\framebox{\parbox{\textwidth}{#1}}%
\end{center}}

\begin{document}

\title{Collider aspects of flavour physics at high Q
\footnote{Report of Working Group 1 of the CERN Workshop ``Flavour in
  the era of the LHC'', Geneva, Switzerland, November 2005 -- March 2007. }}

\author{
F.~del Aguila$^{1}$,
J.~A.~Aguilar--Saavedra$^{1,*}$,
B.~C.~Allanach$^{2,*}$,
J.~Alwall$^{3}$,
Yu.~Andreev$^{4}$,
D.~Aristizabal Sierra$^{5}$,
A.~Bartl$^{6}$, 
M.~Beccaria$^{7,8}$,
S.~B{\'e}jar$^{9,10}$,
L.~Benucci$^{11}$
% ...
}
\institute{
$^{1}$~Departamento de F\'{\i}sica Te{\'o}rica y del Cosmos and CAFPE, Universidad de Granada, E-18071 Granada, Spain; \\
$^{2}$ DAMTP, CMS, University of Cambridge, Cambridge, CB3 0WA, United Kingdom,\\ 
$^{3}$~Centre for Particle Physics and Phenomenology (CP3), Universit{\'e} Catholique de Louvain, 
B-1348 Louvain-la-Neuve, Belgium; \\
$^{4}$ Institute for Nuclear Research RAS,
                       Moscow, 117312, Russia \\
$^{5}$ AHEP Group, Instituto de F\'{\i}sica Corpuscular
 (CSIC, Universitat de Valencia),
E--46071 Val{\`e}ncia, Spain.\\
$^{6}$ Institut f\"ur Theoretische Physik, Universit\"at Wien,  
A-1090 Vienna, Austria \\
$^{7}$~Dipartimento di Fisica, Universit\`a di Salento, 
73100 Lecce, Italy \\
$^{8}$~INFN, Sezione di Lecce, Italy \\
$^{9}$~Grup de F{\'\i}sica Te{\`o}rica, Universitat Aut{\`o}noma de Barcelona, 
E-08193
Barcelona, Spain \\
$^{10}$~Institut de F{\'\i}sica d'Altes Energies, Universitat Aut{\`o}noma de Barcelona, E-08193 
Barcelona, 
Spain \\
$^{11}$~INFN and Universit\`a di Pisa, Pisa, Italy \\
$^{12}$ Institut f\"ur Theoretische Physik, Universit\"at Karlsruhe, 
D-76128 Karlsruhe, Germany \\
$^{13}$~Instituto de F{\'\i}sica, Universidade de S{\~a}o Paulo, S{\~a}o Paulo SP 05508-900, Brazil \\
$^{14}$~LIP - Departamento de F{\'\i}sica, Universidade de Coimbra, 3004-516 Coimbra, Portugal \\
$^{15}$ Universit\'e Libre de Bruxelles, Bruxelles, Belgique\\
$^{16}$ Departamento de F\'{\i}sica e Qu\'{\i}mica, Universidade Estadual
  Paulista, Guaratinguet\'a -- SP, Brazil \\
$^{17}$ Dept.\ of Physics \& Astron., Northwestern University, 
Evanston, IL 60208, USA. \\
$^{18}$ University of Oxford, Denys Wilkinson Building, Keble Road, Oxford, OX1 3RH, UK\\
$^{19}$ Laboratoire de Physique Th\'eorique, Univ. Paris-Sud,  F-91405 Orsay,
 France\\
$^{20}$ Uni. Athens, Physics Dept., Panepistimiopolis, Zografou
157 84 Athens, GREECE \\
$^{21}$~Centro de F{\'\i}sica Te\'orica e Computacional, Faculdade de Ci{\^e}ncias, Universidade de Lisboa, 
1649-003, Lisboa, Portugal \\
$^{22}$ LPSC, Universit\'e Grenoble I/CNRS-IN2P3, 
F-38026 Grenoble, France \\
$^{23}$~Theory group, IPNS, KEK, Tsukuba, 305-0801, Japan \\
$^{24}$~Institute of Theoretical Physics, Warsaw University, 
PL-00681 Warsaw, Poland \\
$^{25}$~Departament de F{\'\i}sica Fonamental, Universitat de Barcelona, 
E-08028 Barcelona, 
Spain \\
$^{26}$ Max-Planck-Institut f\"ur Physik, 
D-80805 Munich, Germany \\
$^{27}$~Instituto de Fisica de Cantabria IFCA (CSIC--UC), E-39005 Santander, Spain \\
$^{28}$~National Institute of Chemical Physics and Biophysics, Ravala 10,
Tallinn 10143, Estonia\\
$^{29}$ Department of Physics, Tokyo Gakugei University, Koganei, 
Tokyo 184-8501, Japan\\
$^{30}$ Department of Physics, National Taiwan University, Taipei, Taiwan 10617\\
$^{31}$ Theory Division, Physics Department, CERN, CH-1211 Geneva, Switzerland. \\
$^{32}$ SLAC, Stanford University, Stanford, CA 94309, USA. \\
$^{33}$ Deutsches Elektronen-Synchrotron DESY, D-22603 Hamburg, Germany \\
$^{34}$~Institute of Nuclear Physics, NCSR ``Demokritos'', Athens, Greece \\
$^{35}$ INR Moscow, Russia\\
$^{36}$ IPPP Durham, Department of Physics, University of Durham, Durham DH1 3LE, United Kingdom\\
$^{37}$ Universit\'a degli Studi di Milano and INFN, 
I-20133 Milano, Italy. \\
$^{38}$~Dipartimento di Fisica Teorica, Universit\`a di Trieste, 
Miramare, Trieste, Italy \\
$^{39}$~INFN, Sezione di Trieste,  I-34014 Trieste, Italy \\
$^{40}$ Faculdade de Engenharia,
Centro Universit\'ario Funda\c{c}\~ao Santo Andr\'e,
Santo Andr\'e -- SP, Brazil \\
$^{41}$ Institut f\"ur Hochenergiephysik der \"Osterreichischen Akademie 
der Wissenschaften, A-1050 Wien, Austria\\
$^{42}$ Universit{\'e} de Montr{\'e}al, D{\'e}partement de Physique, Montr{\'e}al, Canada.\\
$^{43}$ Institute of Physics, Academy of Sciences, Baku, Azerbaijan.\\
$^{44}$ ETH Zurich, CH-8093 Zurich, Switzerland\\
$^{45}$ Univ. College London, Physics and Astronomy Dept., London, UK\\
$^{46}$ Dipartimento di Fisica Teorica, Universit\`a di Torino and INFN,Sezione di Torino, Italy, \\
$^{47}$ INFN, Sezione di Pavia, 
I-27100 Pavia, Italy. \\
$^{48}$ Institut f\"ur Theoretische Physik und Astrophysik,
Universit\"at W\"urzburg, 97074 W\"urzburg, Germany \\
 $^{49}$ SINP MSU, Russia,\\
$^{50}$~Laboratoire de Physique Th\'{e}orique et Astroparticules, UMR 5207, Universit\'{e} Montpellier II, F-34095 Montpellier Cedex 5., France \\
$^{51}$ Instituto de F\'{\i}sica, Universidad de Antioquia - Colombia \\
$^{52}$ Institute for Theoretical Physics, TU Dresden, Dresden, 01062, Germany,\\
$^{53}$ CERN, Physics Dept., CH-1211 Geneva 23, Switzerland\\
$^{54}$ Service de Physique Th\'eorique, CEA Saclay, F91191 Gif--sur--Yvette, France \\
$^{55}$ Theoretical Physics, Fermi National Accelerator Laboratory, Batavia, IL 60510, USA,\\
$^{56}$ LAPTH, CNRS, UMR 5108, Chemin de Bellevue BP110, F-74941, Annecy-le-Vieuy, France \\
$^{57}$~HEP Group, Dept. Estructura i Constituents de la Mat\`eria, Universitat de Barcelona, 
E-08028 Barcelona, Spain. \\
$^{58}$ Paul Scherrer Institute, CH-5232 Villigen PSI, Switzerland,\\
$^{59}$ TOBB University of Economics and Technology, Physics Department,
 Ankara, Turkey,\\
 $^{60}$ Diparimento di Fisica e Astronomia, Universita di Catania, 
I-95123 Catania, Italy. \\
$^{61}$ Univ. California at Irvine, Physics and Astronomy Dept., Irvine, USA,\\
$^{62}$ Gazi University, Physics Department, Ankara, Turkey.\\
$^*$ editor
}

\maketitle

\begin{abstract}
This chapter of the report of the ``Flavour in ther era of LHC'' workshop
discusses flavour related issues in the production and decays
of heavy states at LHC,
both from the experimental side and from the theoretical side. We review
top quark physics and discuss flavour aspects of several extensions
of the Standard Model, such as supersymmetry, little Higgs model or
models with extra dimensions. This includes discovery aspects as well
as measurement of several properties of these heavy states. 
We also present public available
computational tools related to this topic.
\end{abstract}

\tableofcontents


\chapter{Introduction}
\setcounter{page}{1}

\section{Tasks of WG1}

The origin of flavour structures and CP violation remains as one of
the big question in particle physics. Within the Standard Model (SM)
the related phenomena are successfully parametrised with the help
of the CKM matrix in the quark sector and the PMNS matrix in the lepton
sector. In both sectors intensive studies of flavour aspects have been
carried out and are still going on as discussed in the reports by WG2 and WG3.
Following the unification idea originally proposed by Einstein 
it is strongly believed that eventually both sectors can be explained
by a common underlying theory of flavour. 
Although current SM extensions rarely include a theory
of flavour, many of them tackle the flavour question with the help of
some special ansatz leading to interesting predictions for future collider
experiments as the LHC.

This chapter of the ''Flavour in the era of LHC''
report gives a comprehensive overview of the theoretical and
experimental status on: (i) How flavour physics can be explored in the
production of heavy particles like the top quark or new states predicted
in extensions of the SM. (ii) How flavour aspects impact the discovery
and the study of the properties of these new states. We discuss in detail
the physics of the top quark, supersymmetric models, Little Higgs models,
extra dimensions, grand unified models and models explaining neutrino data.

Section \ref{chap:top} discusses flavour aspects related to the top
quark which is expected to play an important role due to its heavy mass.
The LHC will be a top quark factory allowing to study several of
its properties in great detail. The $Wtb$ coupling is an important
quantity which in the SM is directly related to the CKM element
$V_{tb}$. In SM extensions new couplings can be presented which can
be studied with the help of the angular distribution of the top decay 
products and/or in single top production.
In extensions of the SM  also sizable flavour changing neutral
currents decays can be induced, such as $t \to q Z$,  $t \to q \gamma$ or 
 $t \to q g$.
The SM expectations for the corresponding branching ratios 
are of the order $10^{-14}$ for the electroweak decays
and order $10^{-12}$ for the strong one. In extensions like two-Higgs
doublet models, supersymmetry or additional exotic quarks they can be 
 up to order $10^{-4}$. The anticipated sensitivity
of ATLAS and CMS for these branching ratio is of
order $10^{-5}$. New physics contribution will also affect single and
pair production of top quarks at LHC either via loop effects or due
to resonances which is discussed in the third part of this section.

In section~\ref{chap:susy} we consider flavour aspects of supersymmetric
models. This class of models predict partners for the SM particles which
differ in spin by $1/2$. In a supersymmetric world flavour would be
described by the usual Yukawa couplings. However, we know that supersymmetry
(SUSY) must be broken which is most commonly parameterized in terms
of soft SUSY breaking terms.
 After a brief overview of the additional flavour structures
in the soft SUSY breaking sector
we first discuss the effect of lepton flavour violation in models
with conserved R-parity. They can significantly modify  di-lepton spectra,
which play an important role in the determination of the SUSY parameters,
despite the stringent constraints from low energy data such as $\mu\to e\gamma$.
We also discuss the possibilities to discover supersymmetry using the
$e^\pm, \mu^\mp$ + missing energy signature. Lepton flavour violation
plays also an important role in long lived stau scenarios with the
gravitino as lightest supersymmetric particle (LSP).
In models with broken R-parity neutrino physics predicts certain
ratios of branching ratios of the LSP 
in terms of neutrino mixing angle (in case of a gravitino LSP the
prediction will be for the next to lightest SUSY particle).
 Here LHC will be important to establish
several consistency checks of the model.
Flavour aspects affect the squark sector in several ways. Firstly one
expects that the lightest squark will be the lightest stop due to
effects of the large top Yukawa coupling. Various aspects of its
properties are studied here in different scenarios. Secondly it leads
to flavour violating squark production and flavour violating decays of squarks
and gluinos despite the stringent constraints from low energy data
such as $b\to s\gamma$.

Also other non-supersymmetric extensions of the SM, such as grand unification
and little Higgs or extra dimensional models, predict new flavour phenomena
which are presented
in section~\ref{chap:NS}. Such SM extensions introduce new fermions (quarks
and leptons), gauge bosons (charged and/or neutral) and scalars. We study the
LHC capabilities to discover these new mass states, paying a special
attention on how to distinguish among different theoretical models.
We start with the phenomenology of additional quarks and leptons, studying in
detail their production at LHC and decay channels available. It turns out
that particles up to a mass of 1-2 TeV can be discovered and studied. Besides
the discovery reach we discuss the possibilities to measure their mixing with
SM fermions. They are also sources of Higgs bosons (produced in their decay)
and hence they can significantly enhance the Higgs discovery potential of
LHC. Extended gauge structures predict additional heavy gauge bosons and,
depending on the mass hierarchy, they can either decay to new fermions or be
produced int their decay. In particular, the production of heavy neutrinos
can be enhanced when the SM gauge group is extended with an extra
$\text{SU}(2)_R$, which predicts additional $W_R$ bosons. E also discuss
flavour aspects for the discovery of the new gauge bosons. This is specially
important for the case of an extra $Z'$, which appears in any extension of
the SM gauge group, and for which model discrimination is crucial. The
presence or not of new $W'$ bosons also helps identify additional $\text{SU}
(2)$ gauge structures. Finally,
several SM extensions predict anew scalar particles. In some cases the new
scalars are involved in the neutrino mass generation mechanism, e.g.~in
some Little Higgs models and in the Babu-Zee model, which are realisations of the
type II seesaw mechanism (involving a scalar triplet). In these two cases,
high energy observables, such as decay branching ratios of doubly charged
scalars, can be related to the neutrino mixing parameters measured in
neutrino oscillations.

Last but not least computational tools play an important role in the study
of flavour aspects at LHC. In section~\ref{chap:tools} we give 
an overview of the public available tools ranging from spectrum
calculators over decay packages to Monte Carlo programs. In addition
we briefly discuss
the latest version of SUSY Les Houches Accord which serves as an
interface between various programs and now includes flavour aspects.

\section{The ATLAS and CMS experiments}

The CERN Large Hadron Collider (LHC) is currently being installed in the 
27-km ring  previously used for the LEP $e^+ e^-$ collider. This machine 
will push back the high energy frontier by one order of magnitude, 
providing $pp$ collisions at a center-of-mass energy of $\sqrt{s} = 14$~TeV.

\begin{figure}[t]
\begin{center}
\includegraphics[width=12cm,angle=270]{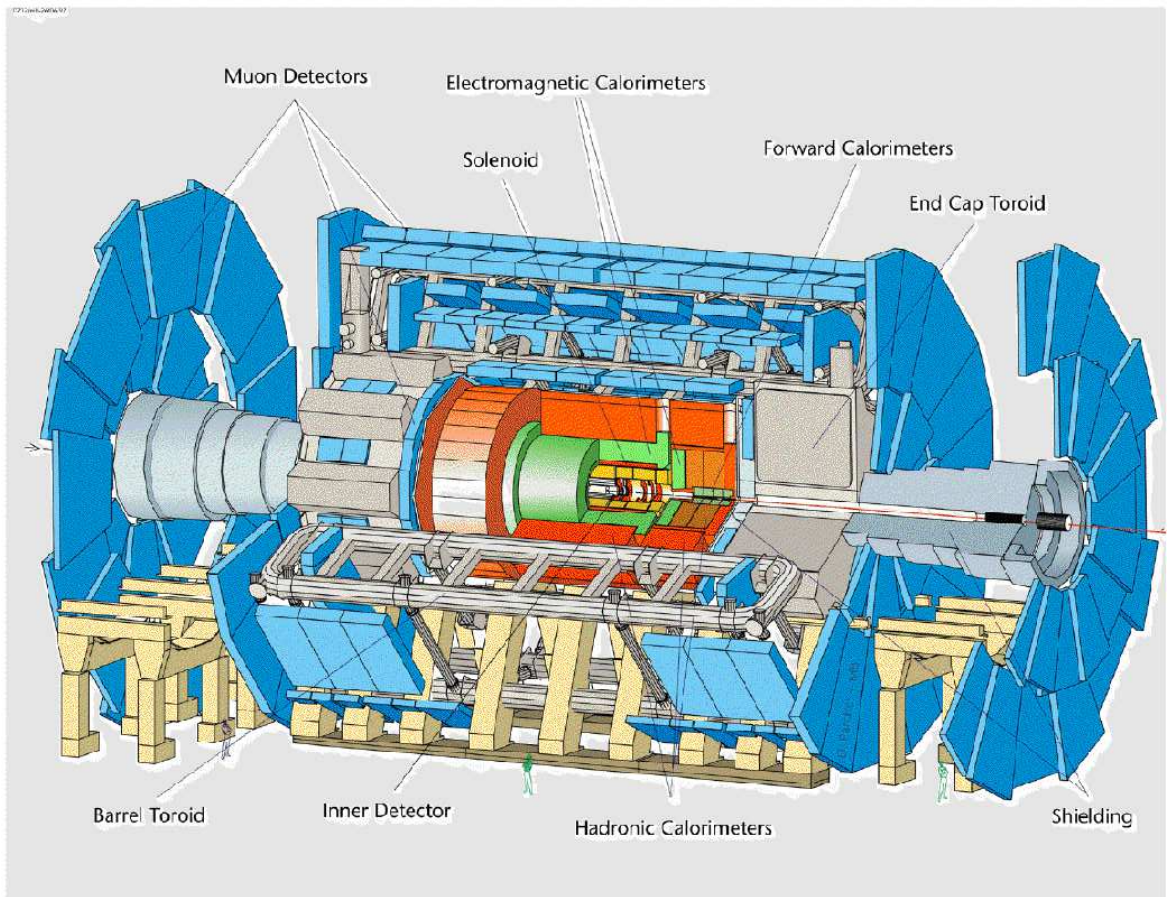}
\end{center} 
\caption{\label{WG1:fig:atlas} An exploded view of the ATLAS detector. }
\end{figure}

\begin{figure}[t]
\begin{center}
\includegraphics[width=12cm]{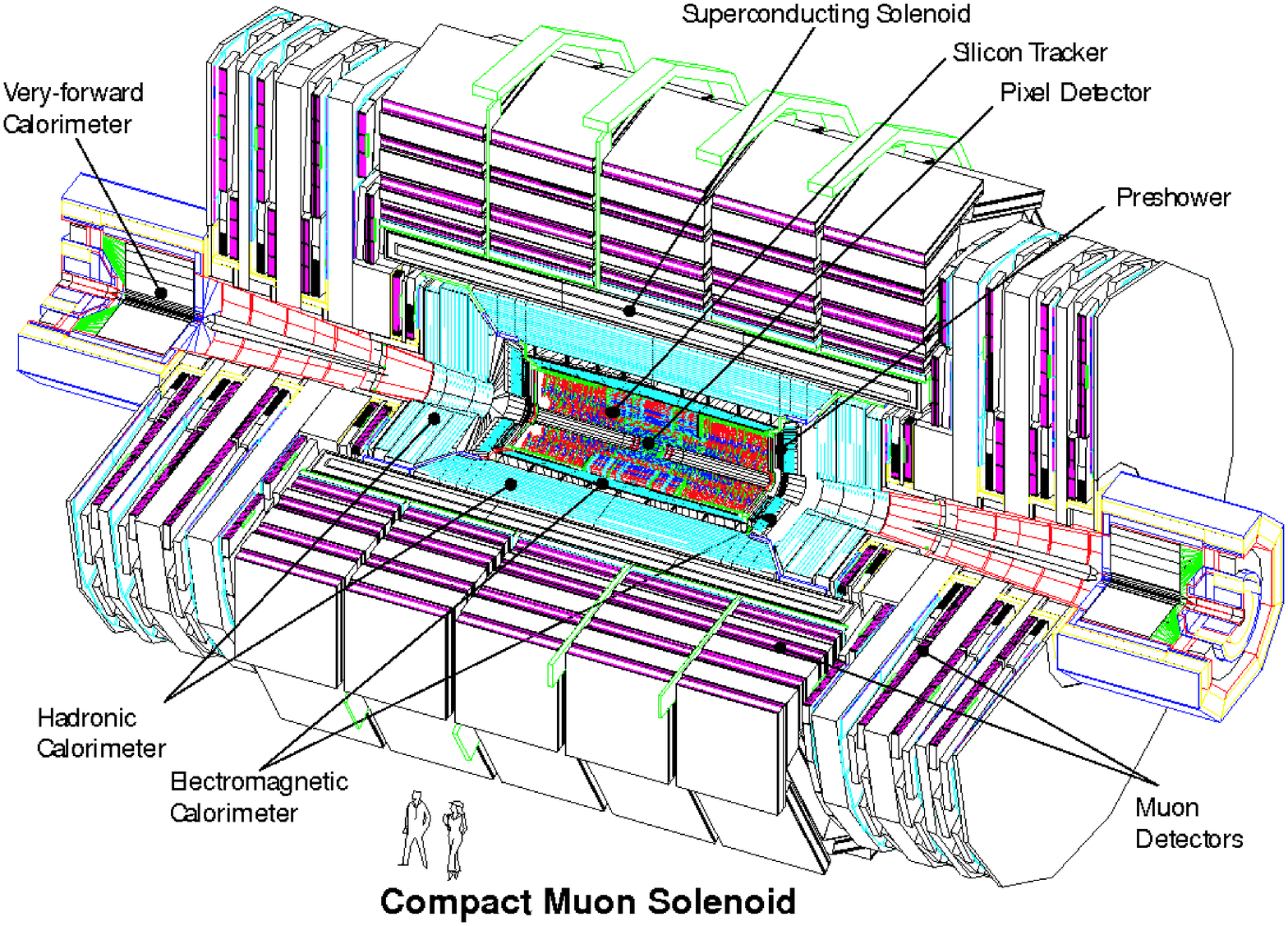}
\end{center} 
\caption{\label{WG1:fig:cms} An exploded view of the CMS detector. }
\end{figure}

Four main experiments will benefit from this accelerator: two 
general-purpose detectors, ATLAS (\fig{WG1:fig:atlas}) and 
CMS (\fig{WG1:fig:cms}), designed to explore the 
physics at the TeV scale; one experiment, LHCb, dedicated to the 
study of $B$-hadrons and CP violation; and one experiment, 
ALICE, which will study heavy ion collisions. Here only the 
ATLAS and CMS experiments and their physics programs are discussed 
in some detail. 

The main goal of these experiments is the verification of the Higgs 
mechanism for the electroweak symmetry breaking and the study of 
the ``new'' (i.e. non-Standard Model) physics which is expected to 
manifest itself at the TeV scale to solve the hierarchy problem. 
The design luminosity of $10^{34} \; \mbox{cm}^{-2} \mbox{s}^{-1}$
of the new accelerator will also allow to collect very large samples of 
B hadrons, W and Z gauge bosons and top quarks, allowing stringent 
tests of the Standard Model predictions. 

Since this programme implies the sensitivity to a very broad range of 
signatures and since it is not known how new physics may manifest itself, 
the detectors have been designed to be able to detect as many particles 
and signatures as possible, with the best possible precision. 

In both experiments the instrumentation is placed around the interaction 
point over the whole solid angle, except for the LHC beam pipe. 
As the particles leave the interaction point, they traverse the Inner Tracker, 
which reconstructs the trajectories of charged particles, the Electromagnetic
and Hadronic calorimeters which absorbe and measure the total energy of 
all particles except neutrinos and muons, and the Muon Spectrometer 
which is used to identify and measure the momentum of muons.  
The presence of neutrinos (and other hypothetic weakly interacting 
particles) is revealed as a non-zero vector sum of the particle 
momenta in the plane transverse to the beam axis.

Both the Inner Tracker and the Muon spectrometer need to be placed inside 
a magnetic field in order to measure the momenta of charged particles 
using the radius of curvature of their trajectories. The two experiments 
are very different in the layout they have chosen for the magnet system.
In ATLAS, a solenoid provide the magnetic field for the Inner Tracker, 
while a system of air-core toroids outside the calorimeters provide the 
field for the Muon Spectrometer. In CMS, the magnetic field is
provided by a single very large solenoid which contains both the Inner 
Tracker and the calorimeters; the muon chambers are embedded in the 
iron of the solenoid return yoke. The magnet layout determines the size, 
the weight (ATLAS is larger but lighter) and even the name of the two 
experiments. 

The CMS Inner Detector consists of Silicon Pixel and Strip detectors, 
placed in a 4~T magnetic field. The ATLAS Inner Tracker is composed by a 
smaller number of Silicon Pixel and Strip detectors and a Transition Radiation 
detector (TRT) at larger radii, inside a 2~T magnetic field. Thanks mainly to 
the larger magnetic field, the CMS tracker has a better momentum resolution, 
but the ATLAS TRT contributes to the electron/pion identification 
capabilities of the detector.

The CMS electromagnetic calorimeter is composed by $\mbox{PbWO}_4$ with 
excellent intrinsic energy resolution 
($\sigma(E)/E \sim 2-5\%/\sqrt{E(GeV)}$). The ATLAS electromagnetic 
calorimeter is a lead/liquid argon sampling calorimeter. While the 
energy resolution is worse ($\sigma(E)/E \sim 10\%/\sqrt{E(GeV)}$), 
thanks to a very fine lateral and longitudinal segmentation the ATLAS 
calorimeter provides more robust particle identification capabilities 
than the CMS calorimeter.

In both detectors the hadronic calorimetry is provided by sampling 
detectors with scintillator or liquid argon as the active medium.  
The ATLAS calorimeter has a better energy resolution for jets  
($\sigma(E)/E \sim 50\%/\sqrt{E(GeV)} \oplus 0.03$) than CMS 
($\sigma(E)/E \sim 100\%/\sqrt{E(GeV)} \oplus 0.05$) because it is 
thicker and has a finer sampling frequency.

The chamber stations of the CMS muon spectrometer are embedded into the
iron of the solenoid return yoke, while those of ATLAS are in air.
Because of multiple scattering in the spectrometer, and the larger field 
in the Inner Tracker the CMS muon reconstruction relies on the 
combination of the informations from the two systems; the ATLAS muon 
spectrometer can instead reconstruct the muons in standalone mode, though 
combination with the Inner detector improves the momentum resolution 
at low momenta. The momentum resolution for 1~TeV muons is about 7\% 
for ATLAS and 5\% for CMS.

\begin{figure}[t]
\begin{center}
\includegraphics[width=12cm,height=9cm]{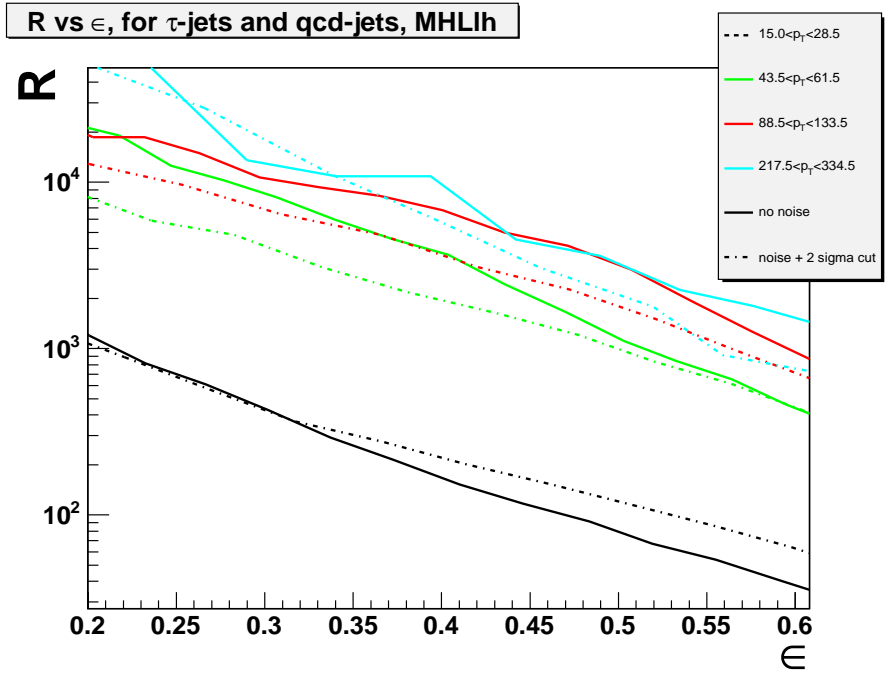}
\end{center} 
\caption{\label{WG1:fig:tau} The QCD jet rejection (inverse of mistagging 
efficiency) as a function of $\tau$ tagging efficiency is reported for 
the ATLAS detector. The four full curves correspond to simulation without 
electronic noise in the calorimeters and different transverse momentum 
ranges, increasing from the lowest to the highest curve. The dashed curves 
correspond to simulation with electronic noise~\cite{Heldmann:923980}. }
\end{figure}

Muons can be unambiguously identified as they are the only particles which 
are capable to reach the detectors outside the calorimeters. Both detectors 
have also an excellent capability to identify electrons that are 
isolated (that is, they are outside hadronic jets). For example, 
ATLAS expects an electron identification efficiency of about 70\% with 
a probability to misidentify a jet as an electron of the order of 
$10^{-5}$~\cite{ATLASTDR}.
The tau identification relies on the hadronic decay modes, since leptonically 
decaying taus cannot be separated from electrons and muons. The 
jets produced by hadronically decaying taus are separated from those 
produced by quark and gluons since they produce narrower jets with a 
smaller number of tracks. The capability of the ATLAS detector to 
separate $\tau$-jets from QCD jets is shown in \fig{WG1:fig:tau}.  

\begin{figure}[t]
\begin{center}
\includegraphics[width=12cm,height=9cm]{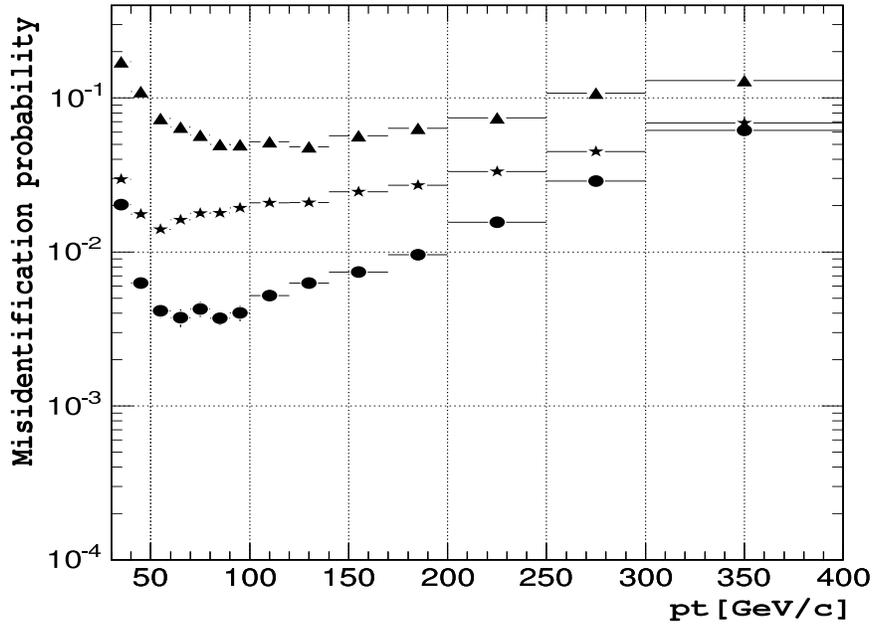}
\end{center} 
\caption{\label{WG1:fig:btag} The non-$b$ jet mistagging efficiency for a fixed 
$b$-tagging efficiency of 0.5 as a function of jet transverse momentum 
for $c$-jets (triangles), $uds$ jets (circles) and gluon jets 
(stars) obtained for the CMS detector with an event sample of QCD jets 
and the secondary vertex tagging algorithm~\cite{DellaNegra:942733a}. }
\end{figure}

The identification of the flavour of a jet produced by a quark is more 
difficult and it is practically limited to the identification of 
$b$ jets, which are tagged by the vertex detectors using the relatively 
long lifetime of $B$ mesons; the presence of soft electron and muon inside 
a jet is also used to improve the $b$-tagging performances. 
In \fig{WG1:fig:btag} the probability of mis-tagging a light jet as a 
$b$ jet is plotted as a function of the $b$-tagging efficiency for the 
CMS detector~\cite{ATLASTDR}; comparable performances are expected for 
ATLAS.

\chapter{Flavour phenomena in top quark physics}
{\small G.~Burdman and N.~Castro}
\label{chap:top}

\section{Introduction}

The top quark is the heaviest and least studied quark of the Standard
Model (SM). Although its properties have already been investigated at
colliders, the available centre of mass energy and the collected
luminosity have not yet allowed for precise measurements, with
exception of its mass. The determination of other fundamental
properties such as its couplings requires larger top samples, which will
be available at the LHC. Additionally, due to its large mass, close to
the electroweak scale, the top quark is believed to offer a unique
window to flavour phenomena beyond the SM. 

Within the SM, the $Wtb$ vertex is purely left-handed, and its size is
given by the Cabibbo-Kobayashi-Maskawa (CKM) matrix element $V_{tb}$,
related to the top-bottom charged current. In a more general way,
additional anomalous couplings such as right-handed vectorial couplings and
left and right-handed tensorial couplings can also be considered. The study
of the angular distribution of the top decay products at the LHC will allow
precision measurements of the structure of the $Wtb$ vertex, providing an
important probe for flavour physics beyond the SM.

In the SM there are no flavour changing neutral current (FCNC)
processes at the tree level 
and at one-loop they can be induced by charged-current
interactions, but they are suppressed by the Glashow-Iliopoulos-Maiani (GIM)
mechanism~\cite{Glashow:1970gm}.
These contributions limit the FCNC decay branching ratios to extremely small
values in the SM. However, there are extensions of the SM which predict the
presence of FCNC contributions already at the tree level and significantly
enhance the top FCNC decay branching ratios\cite{Liu:2004qw, Li:1993mt,
Eilam:2001dh,
Huang:1998wh,Lopez:1997xv,deDivitiis:1997sh,Aguilar-Saavedra:2004wm,
Grzadkowski:1990sm,Bejar:2000ub, Atwood:1995ud, Aguilar-Saavedra:2002ns, 
delAguila:1998tp}.  
Also loop-induced FCNCs could be greatly enhanced in some scenarios beyond
the SM. In all these cases, such processes might be observed at the LHC.

In its first low luminosity phase (10 fb$^{-1}$ per year and per
experiment), the LHC will produce several million top quarks, mainly in
pairs through gluon fusion $g g \to t \bar t$ and quark-antiquark
annihilation $q \bar q \to t \bar t$, with a total cross section of $\sim
833$~pb~\cite{Beneke:2000hk}. Single top
production~\cite{Stelzer:1998ni,Belyaev:1998dn,Tait:2000sh,Sullivan:2004ie,
Campbell:2005bb} will also occur, dominated by the $t$-channel process, $b
q \to t q'$, with a total expected cross section of $\sim
320$~pb~\cite{Sullivan:2004ie,Campbell:2005bb}.  SM extensions, such as
SUSY, may contribute with additional top quark production processes. The
theoretical and experimental knowledge of single top and $t \bar t$
production processes will result in important tests for physics beyond the SM.
Moreover, besides the direct detection of new states (such as SUSY particles and Higgs bosons),
new physics can also be probed via the virtual effects of the additional
particles in precision observables. Finally, in addition to the potential
deviations of the top couplings, it is possible that the top quark couples
strongly to some sector of the new physics at the TeV scale, in such a way
that the production of such states might result in new top quark signals.
This possibility typically involves modifications of the top production
cross sections, either for $t\bar t$ or single top, through the appearance
of resonances or just excesses in the number of observed events. In some of
these cases, the signal is directly associated with a theory of flavour, or
at least of the origin of the top mass.

In this chapter different flavour phenomena associated to top quark physics
are presented, starting with anomalous charged and neutral top couplings.  
In section~\ref{top:wtb} the $Wtb$ vertex structure and the measurement of
$V_{tb}$ are discussed.  Studies related to top quark FCNC processes are
presented in section~\ref{top:fcnc}. Finally, possible contributions of new
physics to top production are discussed in section~\ref{top:corr_xsec},
including the effects of anomalous couplings in $t\bar t$ and single top
production, as well as the possible observation of resonances which
strongly couple to the top quark.


\section{$Wtb$ vertex}
 \label{top:wtb}

In extensions of the SM, departures from the SM expectation $V_{tb} \simeq
0.999$ are possible~\cite{Aguilar-Saavedra:2002kr,DelAguila:2001pu}, as
well as new radiative contributions to the $Wtb$
vertex~\cite{Cao:2003yk,Wang:2005ra}. These deviations might be
observed in top decay processes at the LHC and can be parametrized with the
effective operator formalism by considering the most general $Wtb$
vertex (which contains terms up to dimension five) according to
\begin{eqnarray}
{\mathcal L} & = & - \frac{g}{\sqrt 2} \bar b \, \gamma^{\mu} \left( V_L
P_L + V_R P_R
\right) t\; W_\mu^-
- \frac{g}{\sqrt 2} \bar b \, \frac{i \sigma^{\mu \nu} q_\nu}{M_W}
\left( g_L P_L + g_R P_R \right) t\; W_\mu^- + \mathrm{h.c.} \,,
\label{top:ec:1}
\end{eqnarray}
with $q=p_t-p_b$ (the conventions of Ref.~\cite{delAguila:2002nf} are
followed with slight simplifications in the notation). If $CP$ is
conserved in the decay, the couplings can be taken to be 
real.\footnote{A
general $Wtb$ vertex also contains terms proportional to
$(p_t+p_b)^\mu$, $q^\mu$ and $\sigma^{\mu \nu} (p_t+p_b)_\nu$ Since $b$
quarks are on shell, the $W$ bosons decay to light particles (whose
masses can be neglected) and the top quarks can be approximately assumed
on-shell, these extra operators can be rewritten in terms of the ones in
Eq.~(\ref{top:ec:1}) using Gordon identities.}

\subsection{$Wtb$ anomalous couplings}
 \label{top:anomalouscoup}

Within the SM, $V_L \equiv V_{tb} \simeq 1$ and $V_R$, $g_L$, $g_R$
vanish at the tree level, while nonzero values are generated at one loop
level~\cite{Do:2002ky}. Additional contributions to $V_R$, $g_L$, $g_R$
are possible in SM extensions, without spoiling the agreement with
low-energy measurements. The measurement of $BR(b \to s \gamma)$ is
an important constraint to the allowed values of the $Wtb$ 
anomalous couplings.

At the LHC, the top production and decay processes will allow to probe
in detail the $Wtb$ vertex. Top pair production takes place through the QCD
interactions without involving a $Wtb$ coupling. Additionally, it is
likely that the top quark almost exclusively decays in the channel $t
\to W^+ b$. Therefore, its cross section for production and decay
$gg,q\bar q \to t \bar t \to W^+ b W^- \bar b$ is largely insensitive to the
size and structure of the $Wtb$ vertex. However, the angular
distributions of (anti)top decay products give information about its
structure, and can then be used to trace non-standard couplings.
Angular distributions relating top and antitop decay products probe not
only the $Wtb$ interactions but also the spin correlations among the
two quarks produced, and thus may be influenced by new production
mechanisms as well.

\subsubsection{Constraints from $B$ physics}
 \label{top:const_bfac}

Rare decays of the $B$-mesons as well as the $B\bar B$ mixing provide
important constraints on the anomalous $Wtb$ couplings because they
receive large contributions from loops involving the top quark and the
$W$ boson. In fact, it is the large mass of the top quark that
protects the corresponding FCNC amplitudes against GIM
cancellation. Thus, order-unity values of $V_L-V_{tb}$, $V_R$, $g_L$
and $g_R$ generically cause ${\cal O}(100\%)$ effects in the FCNC
observables.  For $V_R$ and $g_L$, an additional
enhancement~\cite{Cho:1993zb,Fujikawa:1993zu} by $m_t/m_b$ occurs in
the case of $\bar{B}\to X_s\gamma$, because the SM chiral suppression
factor $m_b/M_W$ gets replaced by the order-unity factor $m_t/M_W$.

Deriving specific bounds on the anomalous $Wtb$ couplings from loop
processes requires treating them as parts of certain gauge-invariant
interactions. Here, we shall consider the following dimension-six
operators~\cite{Buchmuller:1985jz}
\begin{eqnarray} 
O^{V_R} &=& \bar{t}_R \gamma^\mu b_R \left( \widetilde{\phi}^\dagger i D_{\mu} 
\phi\right) 
             + {\rm h.c.},\nonumber\\
O^{V_L} &=& \bar{q}_L \tau^a \gamma^\mu q_L \left( \phi^\dagger \tau^a i 
D_{\mu} \phi\right) 
           -\bar{q}_L \gamma^\mu q_L \left( \phi^\dagger i D_{\mu} \phi\right) 
             + {\rm h.c.},\nonumber\\
O^{g_R} &=& \bar{q}_L \sigma^{\mu\nu} \tau^a t_R \widetilde{\phi} W^a_{\mu\nu}
             + {\rm h.c.},\nonumber\\
O^{g_L} &=& \bar{q\,}_{\!\!L}' \sigma^{\mu\nu} \tau^a b_R \phi W^a_{\mu\nu} + 
{\rm h.c.},
\label{top:BWops}
\end{eqnarray} 
where~~ 
$q_L =  \left( t_L,~ V_{tb} b_L + V_{ts} s_L + V_{td} d_L \right)$,~~
${q\,}_{\!\!L}' = \left( V^*_{tb} t_L + V^*_{cb} c_L + V^*_{ub} u_L,~ b_L 
\right)$,~
and $\phi$ denotes the Higgs doublet.
Working in terms of gauge-invariant operators renders the loop results
meaningful, at the expense of taking into account {\em all} the interactions
that originate from Eq.~(\ref{top:BWops}), not only the $Wtb$ ones.

As an example, let us consider the $b\to s\gamma$ transition. Since
it involves low momenta only, one usually treats it in the framework
of an effective theory that arises from the full electroweak model (SM
or its extension) after decoupling the top quark and the heavy
bosons. The leading contribution to the considered decay originates
from the operator
\begin{equation}
O_7 = \frac{e}{16\pi^2} m_b \bar{s}_L \sigma^{\mu\nu} b_R F_{\mu\nu}.
\end{equation}
The SM value of its Wilson coefficient $C_7$ gets modified when the anomalous
$Wtb$ couplings are introduced. Moreover, the presence of $O_7$ also above the
decoupling scale $\mu_0$ becomes a necessity, because counter-terms
involving $O_7$ renormalize the UV-divergent $b\to s\gamma$ diagrams
with $O^{g_L}$ and $O^{g_R}$ vertices. Thus, we are led to consider the ${\bar 
B}\to
X_s\gamma$ branching ratio as a function of not only $V_L$, $V_R$, $g_L$ and
$g_R$ but also $C_7^{(p)}$, i.e. the ``primordial'' value of $C_7$ before
decoupling. Following the approach of Ref.~\cite{Misiak:2006ab}, one finds
\begin{eqnarray} 
{BR}(B\to X_s\gamma)\times 10^4 &=& (3.15 \pm 0.23)
- 8.18\, (V_L - V_{tb}) + 427\, V_R \nonumber\\  
&-&  712\, g_L + 1.91\, g_R - 8.03\, C_7^{(p)}(\mu_0)\nonumber\\ 
&+& {\cal O}\left[ \left(V_L-V_{tb},V_R,g_L,g_R,C_7^{(p)}\right)^2\right],
\label{top:bsgWtbpolyn}
\end{eqnarray}
for $E_\gamma > 1.6\,$GeV and $\mu_0=160\,$GeV in the
$\overline{\rm MS}$ scheme.\footnote{
  The negative coefficient at $V_L$ differs from the one in Fig.~1 of
  Ref.~\cite{Burdman:1999fw} where an anomalous $Wcb$ coupling was
  effectively included, too.}
As anticipated, the coefficients at $V_L$ and $g_R$ are of the same order as
the first (SM) term, while the coefficients at $V_R$ and $g_L$ get
additionally enhanced. The coefficients at $g_L$ and $g_R$ depend on $\mu_0$
already at the leading order, and are well-approximated by 
$-379 - 485 \ln \mu_0/M_W$ and $-0.87 + 4.04 \ln \mu_0/M_W$, respectively. 
This $\mu_0$-dependence and the one of $C_7^{(p)}(\mu_0)$ compensate each 
other in Eq.~(\ref{top:bsgWtbpolyn}).

Taking into account the current world average~\cite{Barberio:2007cr}:
\begin{equation} \label{top:bsg.hfag}
{BR}(\bar{B} \to X_s \gamma)~ =~ 
\left(3.55\pm 0.24{\;}^{+0.09}_{-0.10}\pm0.03\right)\times 10^{-4},
\end{equation}
a thin layer in the five-dimensional
space $(V_L-V_{tb}, V_R, g_L, g_R, C_7^{(p)})$ is found to
be allowed by $b\to s\gamma$.  When one parameter at a time is varied
around the origin (with the other ones turned off), quite narrow
$95\%\,$C.L. bounds are obtained. They are listed in
Table~\ref{top:tab:bsgWtb.bounds}.

\begin{table}[h]
\begin{center}
\caption{ The current $95\%\,$C.L. bounds from
  Eq.~(\ref{top:bsgWtbpolyn}) along the parameter
  axes.}
\begin{tabular}{|r|r|r|r|r|r|} \hline
       & $V_L - V_{tb}$ & $V_R$~~ & $g_L$~~ & $g_R$~ & $C_7^{(p)}(\mu_0)$\\\hline
upper bound &$  0.03~~~ $&$  0.0025 $&$  0.0004 $&$  0.57 $&$  0.04$ \\
lower bound &$ -0.13~~~ $&$ -0.0007 $&$ -0.0015 $&$ -0.15 $&$ -0.14$ \\\hline
\end{tabular}
\label{top:tab:bsgWtb.bounds}
\end{center}
\end{table}

If several parameters are simultaneously turned on in a correlated manner,
their magnitudes are, in principle, not bounded by $b\to s\gamma$ 
alone. However, the larger they are, the tighter the necessary correlation is,
becoming questionable at some point. 

The bounds in Table~\ref{top:tab:bsgWtb.bounds} have been obtained under the assumption that the
non-linear terms in Eq.~(\ref{top:bsgWtbpolyn}) are negligible with respect to the linear ones.
If this assumption is relaxed, additional solutions to that equation arise.
Such solutions are usually considered to be fine-tuned. In any case, they are
expected to get excluded by a direct measurement of the $Wtb$ anomalous
couplings at the LHC (see section~\ref{top:wtb_anom_coup}).

Considering other processes increases the number of constraints but
also brings new operators with their Wilson coefficients into the
game, so long as the amplitudes undergo ultraviolet
renormalization. Consequently, the analysis becomes more and more
involved. Effects of $V_L$ and $V_R$ on $b\to s l^+ l^-$ have been
discussed, e.g., in Refs.~\cite{Burdman:1999fw,Lee:2002bm}. These
analyses need to be updated in view of the recent measurements, and
extended to the case of $g_L$ and $g_R$. The same refers to the $B\bar
B$ mixing, for which (to our knowledge) no dedicated calculation has
been performed to date. Exclusive rare decay modes in the presence of
non-vanishing $V_R$ have been discussed in 
Refs.~\cite{Lee:2002iz,Lee:2006qv}.

\subsubsection{ATLAS sensitivity to Wtb anomalous couplings}
  \label{top:wtb_anom_coup}

The polarisation of the $W$ bosons produced in the top decay is
sensitive to non-standard couplings~\cite{Kane:1991bg}. $W$ bosons can
be produced with positive, negative or zero helicity, with
corresponding partial widths $\Gamma_R$, $\Gamma_L$, $\Gamma_0$ which
depend on $V_L$, $V_R$, $g_L$ and $g_R$. General expressions for
$\Gamma_R$, $\Gamma_L$, $\Gamma_0$ in terms of these couplings can be
found in Ref.~\cite{Aguilar-Saavedra:2006fy} and were included in the 
program {\tt TopFit}. Their absolute
measurement is rather difficult, so it is convenient to consider
instead the helicity fractions $F_i \equiv \Gamma_i/\Gamma$, with
$\Gamma = \Gamma_R + \Gamma_L + \Gamma_0$ the total width for $t \to
Wb$. Within the SM, $F_0 = 0.703$, $F_L = 0.297$, $F_R = 3.6 \times
10^{-4}$ at the tree level, for $m_t = 175$~GeV, $M_W = 80.39$~GeV,
$m_b = 4.8$~GeV. We note that $F_R$ vanishes in the $m_b=0$ limit
because the $b$ quarks produced in top decays have left-handed
chirality, and for vanishing $m_b$ the helicity and the chirality
states coincide. These helicity fractions can be measured in leptonic
decays $W \to \ell \nu$. Let us denote by $\theta_{\ell}^*$ the angle
between the charged lepton three-momentum in the $W$ rest frame and the
$W$ momentum in the $t$ rest frame. The normalised angular distribution
of the charged lepton can be written as
\begin{equation}
\frac{1}{\Gamma} \frac{d \Gamma}{d \cos \theta_{\ell}^*} = \frac{3}{8}
(1 + \cos \theta_{\ell}^*)^2 \, F_R + \frac{3}{8} (1-\cos \theta_{\ell}^*)^2 \, F_L
+ \frac{3}{4} \sin^2 \theta_{\ell}^* \, F_0 \,,
\label{top:ec:dist}
\end{equation}
with the three terms corresponding to the three helicity states and
vanishing interference~\cite{Dalitz:1991wa}.  A fit to the $\cos
\theta_{\ell}^*$ distribution allows to extract, from experiment, the
values of $F_i$, which are not independent but satisfy $F_R + F_L + F_0
= 1$. From these measurements one can constrain the anomalous couplings
in Eq.~(\ref{top:ec:1}). Alternatively, from this distribution one can
measure the helicity ratios~\cite{Aguilar-Saavedra:2006fy}
\begin{equation}
\rho_{R,L} \equiv \frac{\Gamma_{R,L}}{\Gamma_0} = \frac{F_{R,L}}{F_0} 
\,,
\label{top:ec:rho}
\end{equation}
which are independent quantities and take the values $\rho_R = 5.1
\times 10^{-4}$, $\rho_L = 0.423$ in the SM. As for the helicity
fractions, the measurement of helicity ratios sets bounds on $V_R$,
$g_L$ and $g_R$. A third and simpler method to extract information
about the $Wtb$ vertex is through angular asymmetries involving the
angle $\theta_{\ell}^*$. For any fixed $z$ in the interval $[-1,1]$,
one can define an asymmetry
\begin{equation}
A_z = \frac{N(\cos \theta_{\ell}^* > z) - N(\cos \theta_{\ell}^* < 
z)}{N(\cos \theta_{\ell}^* > z) +
N(\cos \theta_{\ell}^* < z)} \,.
\end{equation}
The most obvious choice is $z=0$, giving the forward-backward (FB) 
asymmetry
$A_\mathrm{FB}$~\cite{Lampe:1995xb,delAguila:2002nf}.\footnote{Notice 
the 
difference in sign with respect to
the definitions in Refs.~\cite{Lampe:1995xb,delAguila:2002nf}, where the angle $\theta_{\ell b} = \pi -
\theta_{\ell}^*$ between the charged lepton and $b$ quark is used.} The 
FB asymmetry is
related to the $W$ helicity fractions by
\begin{equation}
A_\mathrm{FB} = \frac{3}{4} [F_R - F_L] \,.
\label{top:ec:afb}
\end{equation}
Other convenient choices are $z = \mp (2^{2/3}-1)$. Defining
$\beta = 2^{1/3}-1$, we have
\begin{eqnarray}
z = -(2^{2/3}-1) & \rightarrow & A_z = A_+ = 3 \beta [F_0+(1+\beta) 
F_R] \,,
\notag \\
z = (2^{2/3}-1) & \rightarrow & A_z = A_- = -3 \beta [F_0+(1+\beta) 
F_L] \,.
\label{top:ec:apm}
\end{eqnarray}
Thus, $A_+$ ($A_-$) only depend on $F_0$ and $F_R$ ($F_L$). The SM
values of these asymmetries are $A_\mathrm{FB} = -0.2225$, $A_+ =
0.5482$, $A_- = -0.8397$. They are very sensitive to anomalous $Wtb$
interactions, and their measurement allows us to probe this vertex
without the need of a fit to the $\cos \theta_{\ell}^*$ distribution.
It should also be pointed out that with a measurement of two of these
asymmetries the helicity fractions and ratios can be reconstructed.

In this section, the ATLAS sensitivity to $Wtb$ anomalous couplings is
reviewed. The $t \bar t \to W^+ b W^- \bar b$ events in which one of the
$W$ bosons decays hadronically and the other one in the leptonic channel $W
\to \ell \nu_\ell$ (with $\ell=e^\pm,\mu^\pm$), are considered as signal
events.\footnote{From now on, the $W$ boson decaying hadronically and its
parent top quark will be named as ``hadronic'', and the $W$ decaying
leptonically and its parent top quark will be called ``leptonic''.} Any
other decay channel of the $t \bar t$ pair constitutes a background to this
signal. Signal events have a final state topology characterised by one
energetic lepton, at least four jets (including two $b$-jets) and large
transverse missing energy from the undetected neutrino.  Top pair
production, as well as the background from single top production, is
generated with {\tt TopReX}~\cite{Slabospitsky:2002ag}.  Further
backgrounds without top quarks in the final state, i.e. $b\bar{b}$,
$W+$jets, $Z/\gamma^*+$jets, $WW$, $ZZ$ and $ZW$ production processes, are
generated using {\tt PYTHIA}~\cite{Sjostrand:2000wi}. In all cases CTEQ5L
parton distribution functions (PDFs)~\cite{Lai:1999wy} were used. Events are
hadronised using {\tt PYTHIA}, taking also into account both initial and
final state radiation. Signal and background events are passed through the
ATLAS fast simulation~\cite{Richter-Was:683751} for particle reconstruction
and momentum smearing. The $b$-jet tagging efficiency is set to 60\%, that
corresponds to a rejection factor of 10 (100) for $c$ jets (light quark and
gluon jets).

A two-level probabilistic analysis, based on the construction of a
discriminant variable which uses the full information of some
kinematical properties of the event was developed and is described
elsewhere~\cite{Aguilar-Saavedra:952732,Aguilar-Saavedra:2007rs}.  
After this analysis, 220024 signal events (corresponding to an
efficiency of 9\%{}) and 36271 background events (mainly from $t \bar t
\to \tau \nu b \bar b q \bar q'$) were selected, for a luminosity of
$10$ fb$^{-1}$. The hadronic $W$ reconstruction is done from the two
non-$b$ jets with highest transverse momentum. The mass of the hadronic
top, is reconstructed as the invariant mass of the hadronic $W$ and the
$b$-jet (among the two with highest $p_T$) closer to the $W$. The
leptonic $W$ momentum cannot be directly reconstructed due to the
presence of an undetected neutrino in the final state. Nevertheless, the
neutrino four-momentum can be estimated by assuming the transverse
missing energy to be the transverse neutrino momentum. Its longitudinal
component can then be determined, with a quadratic ambiguity, by
constraining the leptonic $W$ mass (calculated as the invariant mass of
the neutrino and the charged lepton) to its known on-shell value $M_W= 
80.4$~GeV. In order to solve the twofold quadratic ambiguity in
the longitudinal component it is required that the hadronic and the
leptonic top quarks have the minimum mass difference.

The experimentally observed $\cos \theta_{\ell}^*$ distribution, which
includes the $t \bar t$ signal as well as the SM backgrounds, is
affected by detector resolution, $t \bar t$ reconstruction and selection
criteria. In order to recover the theoretical distribution, it is
necessary to: (i) subtract the background; (ii) correct for the effects
of the detector, reconstruction, etc.  The asymmetries are measured with
a simple counting of the number of events below and above a specific
value of $\cos \theta_{\ell}^*$.  This has the advantage that the
asymmetry measurements are not biased by the extreme values of the
angular distributions, where correction functions largely deviate from
unity and special care is required.

Due to the excellent statistics achievable at the LHC, systematic errors
play a crucial role in the measurement of angular distributions and
asymmetries already for a luminosity of 10 fb$^{-1}$. A thorough
discussion of the different systematic uncertainties in the determination
of the correction functions is therefore compulsory. The systematic
errors in the observables studied (asymmetries, helicity fractions and
ratios) are estimated by simulating various reference samples and
observing the differences obtained. Uncertainties originating from Monte
Carlo generators, PDFs, top mass dependence, initial and final state
radiation, $b$-jet tag efficiency, jet energy scale, background cross
sections, pile-up and $b$ quark fragmentation were considered. The
results of the simulation, including statistical and systematic
uncertainties, are summarized in Table~\ref{top:tab:results}.

\begin{table}[tb]
\begin{center}
\caption{Summary of the results obtained from the simulation for the
observables studied, including statistical and systematic uncertainties.}
\begin{tabular}{|c|rll|}
\hline
Observable & \multicolumn{3}{c|}{Result} \\
\hline
$F_0$   & $0.700$  & $\pm 0.003 \,\mathrm{(stat)}$  & $ \pm 0.019 \,\mathrm{(sys)}$ \\
$F_L$   & $0.299$  & $\pm 0.003 \,\mathrm{(stat)}$  & $ \pm 0.018 \,\mathrm{(sys)}$ \\
$F_R$   & $0.0006$ & $\pm 0.0012 \,\mathrm{(stat)}$ & $ \pm 0.0018 \,\mathrm{(sys)}$ \\
$\rho_L$  & $0.4274$ & $ \pm 0.0080\,\mathrm{(stat)}$ & $ \pm 
0.0356\,\mathrm{(sys)}$ \\
$\rho_R$  & $0.0004$ & $ \pm 0.0021\,\mathrm{(stat)}$ & $ \pm 
0.0016\,\mathrm{(sys)}$ \\
$A_\mathrm{FB}$  & $-0.2231$ & $ \pm 0.0035\,\mathrm{(stat)}$ & $ \pm 
0.0130\,\mathrm{(sys)}$ \\
$A_+$   & $0.5472$ & $ \pm 0.0032\,\mathrm{(stat)}$ & $ \pm 
0.0099\,\mathrm{(sys)}$ \\
$A_-$   & $-0.8387$ & $ \pm 0.0018\,\mathrm{(stat)}$ & $ \pm 
0.0028\,\mathrm{(sys)}$ \\
\hline
\end{tabular}
\label{top:tab:results}
\end{center}
\end{table}

\begin{table}[tb]
\begin{center}
\caption{The $1\sigma $ limits on anomalous couplings obtained from the 
combined measurement
of $A_\pm$, $\rho_{R,L}$ are shown.
In each case, the couplings which are fixed to be zero are denoted by a cross.}
\begin{tabular}{|c|ccc|}
\hline
& $V_R$ & $g_L$ & $g_R$ \\
\hline
$A_\pm$, $\rho_{R,L}$
        & $[-0.0195,0.0906]$ & $\times$            & $\times$ \\
$A_\pm$, $\rho_{R,L}$
        & $\times$           & $[-0.0409,0.00926]$ & $\times$ \\
$A_\pm$, $\rho_{R,L}$
        & $\times$           & $\times$            & $[-0.0112,0.0174]$ \\
$A_\pm$, $\rho_{R,L}$
        & $\times$           & $[-0.0412,0.00944]$ & $[-0.0108,0.0175]$ \\
$A_\pm$, $\rho_{R,L}$
        & $[-0.0199,0.0903]$ & $\times$            & $[-0.0126,0.0164]$ \\
\hline
\end{tabular}
\label{top:tab:lim2}
\end{center}
\end{table}

With this results, and considering the parametric dependence of the
observables on $V_R$, $g_L$ and $g_R$ (see
Ref.~\cite{Aguilar-Saavedra:2006fy}), constraints on the anomalous
couplings were set using {\tt TopFit}. Assuming only one nonzero
coupling at a time, $1\sigma$ limits from the measurement of each
observable can be 
derived~\cite{Aguilar-Saavedra:952732,Aguilar-Saavedra:2007rs}.  These
limits can be further improved by combining the measurements of the
four observables $\rho_{R,L}$ and $A_\pm$, using the correlation
matrix~\cite{Aguilar-Saavedra:2007rs}, obtained from
simulation.\footnote{We point out that the correlations among $A_\pm$,
$\rho_{R,L}$ do depend (as they must) on the method followed to extract
these observables from experiment. In our case, the correlations have
been derived with the same procedure used to extract $A_\pm$,
$\rho_{R,L}$ from simulated experimental data.} Moreover, the
assumption that only one coupling is nonzero can be relaxed. However,
if $V_R$ and $g_L$ are simultaneously allowed to be arbitrary,
essentially no limits can be set on them, since for fine-tuned values of
these couplings their effects on helicity fractions cancel to a large
extent. In this way, values $O(1)$ of $V_R$ and $g_L$ are possible
yielding minimal deviations on the observables studied. Therefore, in
the combined limits, which are presented in Table~\ref{top:tab:lim2},
it is required that either $V_R$ or $g_L$ vanishes. 

\begin{figure}[bt]
\begin{center}
\epsfig{file=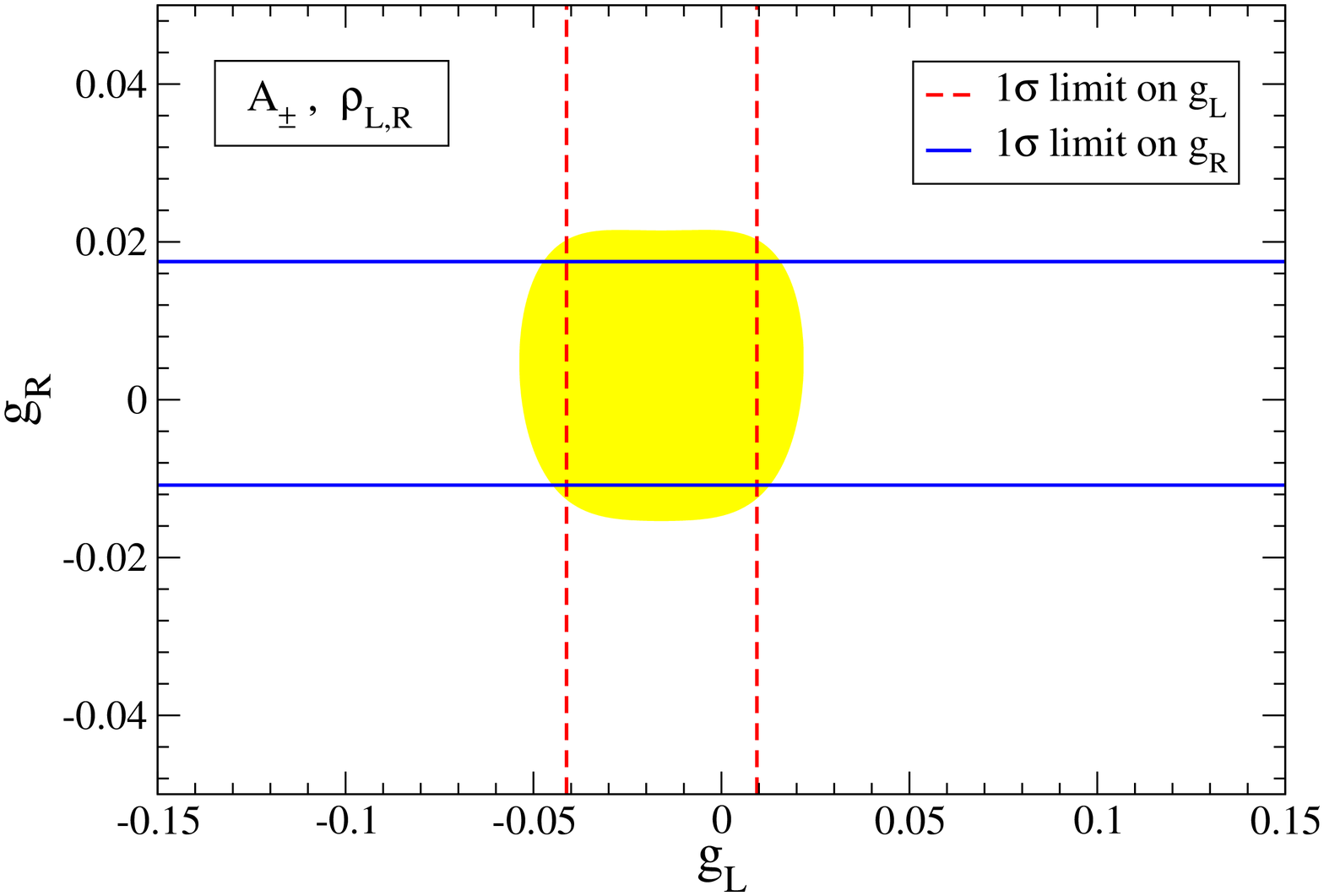,height=5.48cm,clip=}
\epsfig{file=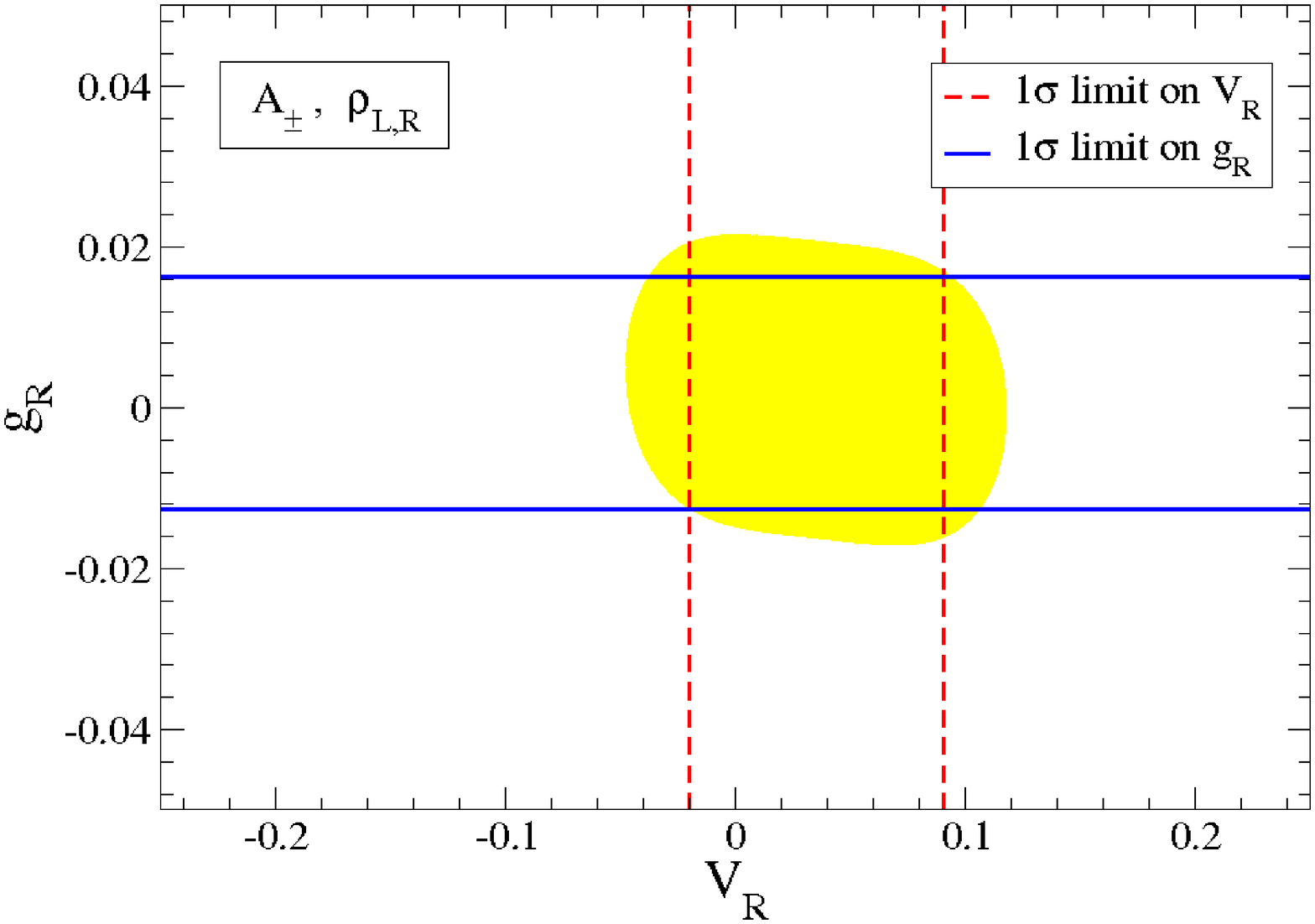,height=5.48cm,clip=}
\caption{68.3\% CL confidence regions on anomalous couplings: $g_L$ and $g_R$,
for $V_R=0$ (left); $V_R$ and $g_R$, for $g_L = 0$ (right).
The $1\sigma$ combined limits in Table~\ref{top:tab:lim2} are also displayed.}
\label{top:fig:reg}
\end{center}
\end{figure}

Finally, with the same procedure, the 68.3\% confidence level (CL) regions
on the anomalous couplings are obtained (Fig.~\ref{top:fig:reg}). The
boundary of the regions has been chosen as a contour of constant $\chi^2$.
In case that the probability density functions (p.d.f.) of $V_R$ and $g_L$
were Gaussian, the boundaries would be ellipses corresponding to $\chi^2 =
2.30$ (see for instance Ref.~\cite{cowan}). In our non-Gaussian case the
$\chi^2$ for which the confidence regions have 68.3\% probability is
determined numerically, and it is approximately 1.83 for the $(g_L,g_R)$
plot and 1.85 for $(V_R,g_R)$.

\subsection{Measurement of $V_{tb}$ in single top production}

The value of the CKM matrix element $V_{tb}$, is often considered to be
known to a very satisfactory precision ($0.9990<|V_{tb}|<0.9992$ at
90\% CL~\cite{Eidelman:2004wy}). However, this range
is determined by assuming the unitarity of the $3\times3$ CKM matrix which
can be violated by new physics effects.
The Tevatron measurements of $R\equiv\frac{|V_{tb}|^2}{|V_{td}|^2+|V_{ts}|^2+|V_{tb}|^2}$ are based on the relative
number of $t\bar t$-like events with zero, one and two
tagged $b$-jets. The resulting values for $R$ are
$1.12^{+0.27}_{-0.23}~\rm(stat.+syst.)$~\cite{Acosta:2005hr}
and $1.03^{+0.19}_{-0.17}~\rm(stat.+syst.)$
\cite{Abazov:2006bh}
for CDF and D\O\ respectively.
Note that $V_{tb}$ determination from $R$, giving
$|V_{tb}| > 0.78$ at 95\% CL, is obtained assuming $|V_{td}|^2+|V_{ts}|^2+|V_{tb}|^2=1$.
In fact, $R\simeq 1$ only implies $|V_{tb}| >> |V_{ts}|, |V_{td}|$.
Therefore the single top production whose cross section is directly proportional to $|V_{tb}|$ is crucial in order to reveal the complete picture of
the CKM matrix.

\begin{figure}[t]
\hspace*{-2cm}
\begin{minipage}{17cm}
\includegraphics[width=17cm]{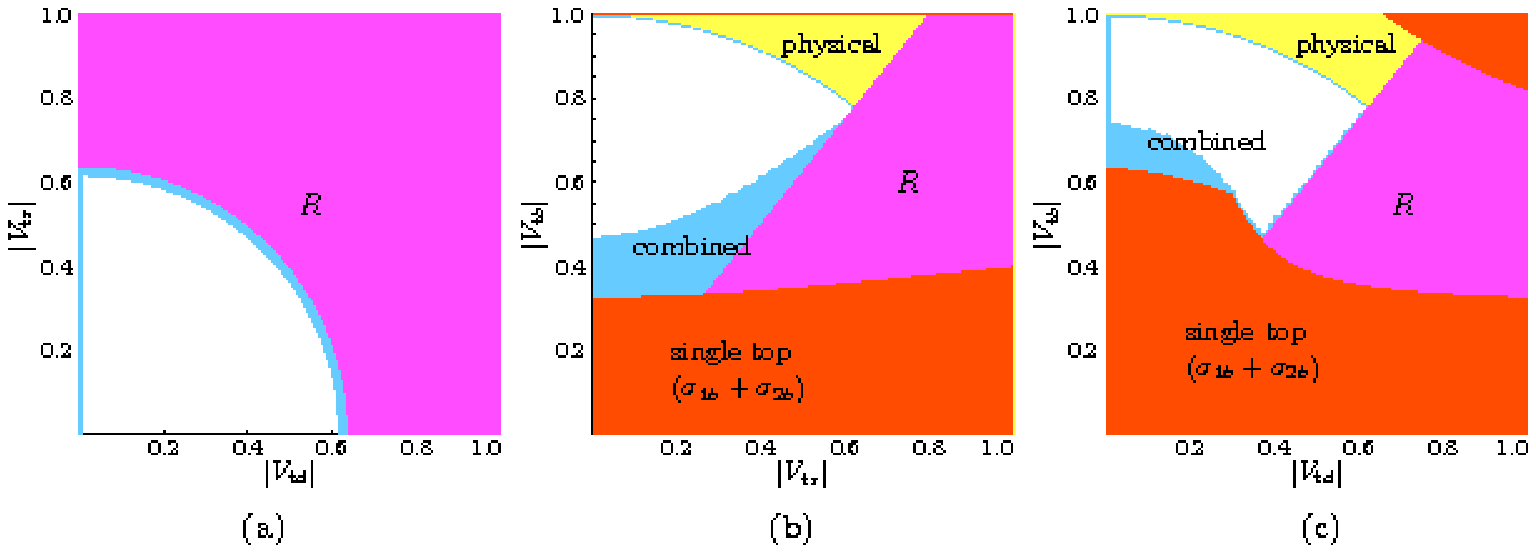}\\
\includegraphics[width=17cm]{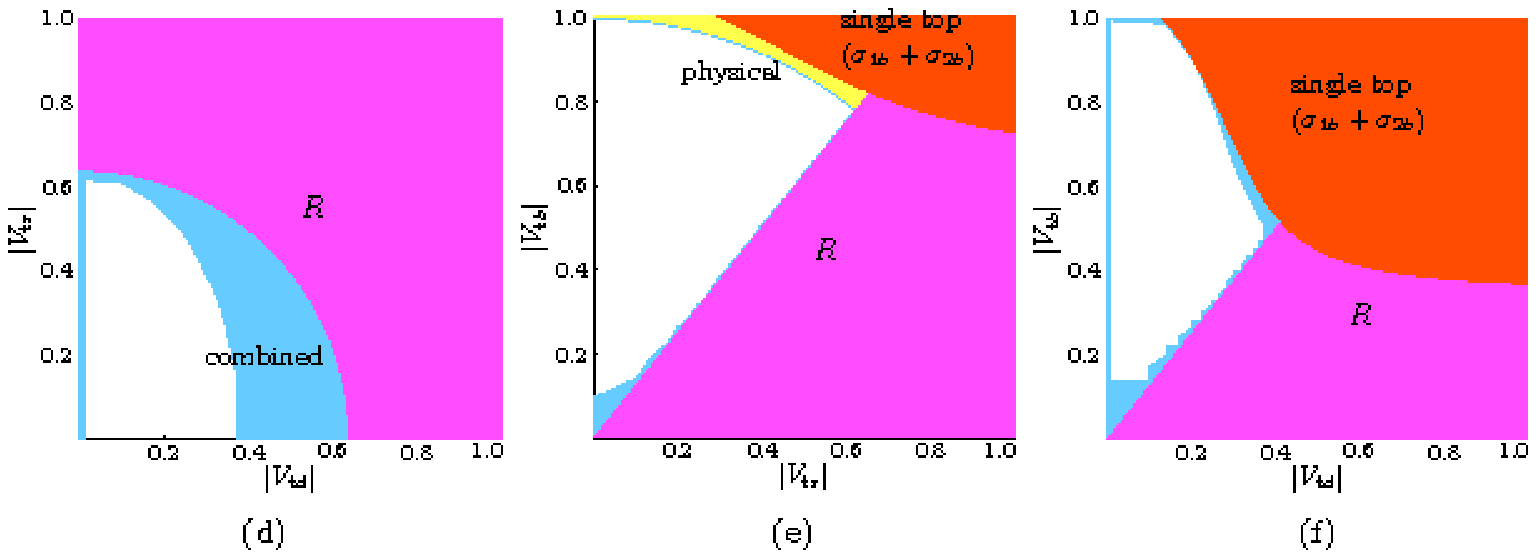}
\end{minipage}
\caption{
Excluded regions for $|V_{td}|$, $|V_{ts}|$, and $|V_{tb}|$, obtained 
from 
the measurement of $R$ and from the single top production,
$\sigma_{1b}+\sigma_{2b}$, at 95 \% C.L. The figures (a)-(c) and 
(d)-(f) are obtained by using, respectively, the latest D\O\ (see 
Eq.~(\ref{top:eq:D0single}))
and CDF (see Eq.~(\ref{top:eq:CDFsingle})) data on the single top 
production.  The combination of both bounds 
provides an additional excluded region.
The physical bound $|V_{td}|^2+|V_{ts}|^2+|V_{tb}|^2<1$ is also 
considered.} 
\label{top:fig:D0con}
\end{figure}

Recently, the D\O\ collaboration announced the 
first observation of the single top production. The 
corresponding results for the $t$ and $s$-channels 
are~\cite{Abazov:2006gd}:
\begin{eqnarray}
\sigma^{\rm s-channel}+\sigma^{\rm t-channel} &=& 4.9\pm 1.4 \ {\rm pb} 
\nonumber \\
\sigma^{\rm s-channel} &=& 1.0\pm 0.9 \ {\rm pb } \nonumber\\
\sigma^{\rm t-channel} &=& 4.2^{+1.8}_{-1.4}\ {\rm pb }  
\label{top:eq:D0single}
\end{eqnarray}
This result can be compared to the SM prediction with $|V_{tb}|=1$~\cite{Sullivan:2004ie}: 
$\sigma^{\rm s-channel}_{\rm SM} = 0.88\pm 0.11 \ {\rm pb }$, 
$\sigma^{\rm t-channel}_{\rm SM} = 1.98\pm 0.25 \ {\rm pb }$.
Taking these results into account and considering the limit $R>0.61$ at 
95\% C.L., excluded regions for $|V_{ti}|$ were obtained and are shown in
Fig.~\ref{top:fig:D0con} (a)-(c) (see~\cite{Alwall:2006bx} for the 
detailed 
computation). From this figure, the allowed values for $|V_{ti}|$ are 
found to be $0\lesssim
|V_{td}|\lesssim 0.62$, $0\lesssim |V_{ts}|\lesssim 0.62$ and 
$0.47\lesssim 
|V_{tb}|\lesssim 1$. The new data on the single top production 
provides, for the first time, the
lower bound of $V_{tb}$.
However, we have to keep in mind that the latest 95\% CL upper 
limits 
on the single top production by the CDF collaboration~\cite{CDF8585}
are lower than those by D\O:
\begin{eqnarray}
\sigma^{\rm s-channel}+\sigma^{\rm t-channel} &<& 2.7 \ {\rm pb} 
\nonumber 
\\
\sigma^{\rm s-channel} &<& 2.5 \ {\rm pb } \nonumber\\
\sigma^{\rm t-channel} &<& 2.3\ {\rm pb }  \label{top:eq:CDFsingle}
\end{eqnarray}
Using this bound, different constraints on $|V_{ti}|$ can be found, as 
shown in Fig.~\ref{top:fig:D0con} (d)-(f).

Going from Tevatron to LHC, the higher energy and luminosity will 
provide
better possibilities for a precise determination of $V_{tb}$.  Among all 
three
possible production mechanisms, the $t$-channel ($q^2_{\mathrm{W}}<0$) 
is the
most promising process due to its large cross section, $\sigma \simeq 
245$
pb~\cite{Smith:1996ij, Stelzer:1997ns, Campbell:2005bb} and $V_{tb}$ 
could be
determined at the 5\% precision level already with 10 fb$^{-1}$ of 
integrated
luminosity, 
assuming a total error of 10\% for the
$t$-channel cross section measurement~\cite{CMSTDR}. The
precision of this result is limited by the systematic uncertainty and 
might be
well improved with better understanding of the detector and background.  
The
other channels, $W-$associated ($q^2_{\mathrm{W}}$=$M^2_{\mathrm{W}}$) 
and
$s$-channel ($q^2_{\mathrm{W}}>0$), are more challenging due to a much 
larger
systematic uncertainty. However, a measurement of these production 
mechanisms
will also be important to further understand the nature of the top quark
coupling to the weak current, 
especially because new physics could affect differently the different single top production channels (see e.g.~\cite{Tait:2000sh}). 

Since $V_{tb}$ is not known, the $V_{tb} \neq 1$ alternative should be
still acceptable. 
If $V_{tb}$ is considerably smaller than one, that would mean that $t 
(b)$ couples not
only to $b (t)$ but also to the extra quarks. Thus, a measurement of
$V_{tb}\neq 1$ would be an evidence for new heavy quarks. 
Their existence is in fact predicted by many extensions of the
SM~\cite{Candelas:1985en, Hewett:1988xc, DelAguila:2001pu,
Arkani-Hamed:2002qx} and furthermore, the current electroweak precision data 
allows such possibility~\cite{Choudhury:2001hs, Carena:2007ua}. 
In this class of models, the familiar $3\times3$
CKM matrix is a sub-matrix of a $3\times4$, $4\times3$,
$4\times4$ or even larger matrix. Those matrices could also be 
constrained,
e.g. by the $4\times 4$ unitarity condition. 
Although the $3\times 4/4\times 3$ matrix, which is often induced by 
the vector-like quark models, breaks the GIM mechanism, 
the current tree level FCNC measurement do not lead to strong constraints. However, the vector-like 
models with down-type quark (models with $3\times 4$ matrix) modifies the tree-level $Zb\bar{b}$ coupling 
by
a factor of $\cos^2\theta_{34}$, where the 3rd-4th generation mixing
$\theta_{34}$ parameterizes the $3\times 4$ matrix together with the 
usual
CKM parameters ($\theta_{12}, \theta_{23}, \theta_{13}$). Since  $V_{tb}$ is written as $V_{tb}\simeq\cos\theta_{34}$ in the same
parameterization,  the measurement of $R_b$ ratio, $R_b=\Gamma (Z\to
b\bar{b})/\Gamma(Z\to \mbox{hadron})$, forbids $V_{tb}$ significantly
different from one  in this type of models.

In  the models with a singlet up-type quark ($4\times 3$ matrix case) or one
complete generation ($4\times 4$ matrix case), the constraint from 
$R_b$ measurement on $V_{tb}$ can be milder.  
In the SM, $R_b$ comes from the tree diagram mentioned above and 
the $t$ quark loop contribution which is
proportional to $|V_{tb}|$ is sub-dominant.  If there is an extra fermion $t^{\prime}$,
$V_{tb}$ can be reduced. On the other hand, we obtain
an extra loop contribution from $t^{\prime}$, which is proportional to
$|V_{t^{\prime}b}|$.  In general, $V_{t^{\prime}b}$ increases when 
$V_{tb}$ decreases.  Thus, the constraint on $V_{tb}$ depends on
the $t^{\prime}$ mass. Using the current CDF upper limit,
$m_{t^\prime}>258$~GeV~\cite{CDF8495}, it can be shown that
$|V_{tb}|>0.95$ (see \chapt{chap:NS}{subsec_q_2_3}).  This result relies on the 
assumption that the corrections to $R_b$ and to $S$, $T$, $U$
parameters~\cite{Peskin:1990zt,Peskin:1991sw}
induced by loop effects are only coming from
the $t$ and $t^{\prime}$.  Therefore more sophisticated models with an
extended particle content may be less constrained. For a more precise
argument in any given model, all the well measured experimental data
from loop processes, such as the $B\to X_s\gamma$ branching ratio and 
the electroweak precision data must be comprehensively
analysed. Nevertheless it should be emphasized that the usual claim that 
the $S$ parameter excludes the fourth generation is based 
on the assumption that ${\rm T}\simeq 0$. The fourth generation model 
increases $S$
and $T$ simultaneously, and thus leaves a larger parameter space for 
this
model than the $R_b$
measurement 
alone~\cite{Alwall:2006bx,Novikov:2002tk,He:2001tp,Holdom:2006mr}.
Further discussion on the search for extra quarks at
the LHC can be found in chapter~\ref{chap:NS} and
in Refs.~\cite{Aguilar-Saavedra:2005pv, Alwall:2006bx}.


\section{FCNC interactions of the top quark 
\label{top:fcnc}}

If the top quark has FCNC anomalous couplings to the gauge bosons, its
production and decay properties will be affected. FCNC processes associated
with the production~\cite{LEP-Exotica-WG-2001-01,Chekanov:2003yt,Aktas:2003yd}
and decay~\cite{Abe:1997fz} of top quarks have been studied at colliders and
the present direct limits on the branching ratios are: $BR({t\to
qZ})<7.8\%{}$~\cite{LEP-Exotica-WG-2001-01}, $BR({t\to
q\gamma})<0.8\%{}$~\cite{Chekanov:2003yt} and $BR({t\to
qg})<13\%{}$~\cite{Ashimova:2006zc}.  Nevertheless, the amount of data
collected up to now is not comparable with the statistics expected at the LHC
and thus either a discovery or an important improvement in the current limits
is expected~\cite{delAguila:1999ac,delAguila:1999ec,Fox:2007in,AguilarSaavedra:2000aj}.

In the top quark sector of the SM, the small FCNC contributions limit the
corresponding decay branching ratios to the gauge bosons ($Z$, $\gamma$ and
$g$) to below $10^{-12}$~\cite{Diaz-Cruz:1989ub, Eilam:1991zc, Mele:1998ag,
Huang:1999bt, Aguilar-Saavedra:2002ns}.  There are however extensions of the
SM, like supersymmetric models including R-parity violation~\cite{Liu:2004qw,
Li:1993mt, Eilam:2001dh,
Huang:1998wh,Lopez:1997xv,deDivitiis:1997sh,Aguilar-Saavedra:2004wm},
multi-Higgs doublet models~\cite{Grzadkowski:1990sm,Bejar:2000ub, 
Atwood:1995ud} and extensions
with exotic (vector-like) quarks~\cite{Aguilar-Saavedra:2002ns,
delAguila:1998tp}, which predict the presence of FCNC contributions already at
the tree level and significantly enhance the FCNC decay branching ratios. The
theoretical predictions for the branching ratios of top FCNC decays within the
SM and some of its extensions are summarized in
Table~\ref{top:fcnc:tab:table_1}.

In addition, theories with additional sources of FCNCs may result in flavour violation in the interactions 
of the  scalar sector with the top quark. For example, this is the case in Topcolor-assisted 
Technicolor~\cite{Hill:1994hp,Buchalla:1995dp}, where tree-level FCNCs are present. 
In the theories the scalar sector responsible for the 
top quark mass can be discovered through its FCNC decay~\cite{Burdman:1999sr} 
$h_t\to t j$, where $j$ is a jet mainly 
of a charm quark. Also, and as we will see in detail in Section~\ref{top:2hdm}, models with 
multi-Higgs doublets contain additional sources of flavour violation at one loop 
that may lead to FCNC decays of the Higgs.

\begin{table}[tb]
  \begin{center} 
    \caption{Branching ratios for FCNC top quark decays predicted by different models.}
    \begin{tabular}{|l||c|c|c|c|} 
      \hline
      Decay             &  SM       &  two-Higgs &SUSY with & Exotic Quarks \\
                        &           &            &R-parity violation& \\
      \hline
      $t \to qZ$      & $ \sim 10^{-14}$& $\sim 10^{-7}$ & $\sim 10^{-5}$& $ \sim 10^{-4}$ \\
      $t \to q\gamma$ & $ \sim 10^{-14}$& $\sim 10^{-6}$ & $\sim 10^{-6}$& $ \sim 10^{-9}$ \\
      $t \to qg$      & $ \sim 10^{-12}$& $\sim 10^{-4}$ & $\sim 10^{-4}$& $ \sim 10^{-7}$ \\ 
      \hline
    \end{tabular}
    \label{top:fcnc:tab:table_1}
  \end{center}
\end{table}

\subsection{Top quark production in the effective lagrangian approach 
\label{top:effectivelagrangianFCNC}}

If strong FCNC exists associated to the top quark sector, it is
expected that it influences the production of single top events through the
process $p p \to t + q,g$. This single top production channel is thus an
excellent probe for flavour phenomena beyond the SM. In this section, the
phenomenology of strong flavour changing single top production in the
effective lagrangian approach is considered. The approach is model independent
and makes use of a subset of all dimension five and six operators that
preserve the gauge symmetries of the SM as written in
ref.~\cite{Buchmuller:1985jz}. The subset chosen contains all operators that
contribute to strong FCNC including the four fermion interactions. This
methodology has been used by many authors to study single top quark production
using the SM as its low energy limit but also in other models like
Supersymmetry, two-Higgs doublet models and
others~\cite{Malkawi:1995dm,Han:1996ce,Han:1996ep,Whisnant:1997qu,
Hosch:1997gz,Han:1998tp,Tait:2000sh,Rizzo:1995uv,Tait:1996dv,
Boos:1999dd,Espriu:2001vj,Guasch:2006hf,Arhrib:2006sg}.

The effective lagrangian is a series in powers of $1/\Lambda$,
$\Lambda$ being the scale of new physics. Therefore, the terms that
originate from mixing with SM charged currents, that is, with diagrams
with a charged boson, either as virtual particle or in the final state
will be first considered. These are processes of the type $ p~p
\rightarrow \left( \bar{q}~q \right) \rightarrow \bar{q}~t + X $ and $
p~p \rightarrow \left( g~q \right) \rightarrow W~t + X$ and the charge
conjugate processes. Due to CKM suppression and small parton density
functions contributions from the incoming quarks, these $\Lambda^{-2}$
terms are much smaller than the $\Lambda^{-4}$ terms. There are several
contributions of order $\Lambda^{-4}$ to the cross section of single
top production. These are summarized in Table~\ref{top_fcnc_lambda4}. A
more detailed discussion can be found in~\cite{Ferreira:2005dr}. Cross
sections for these processes were calculated
in~\cite{Ferreira:2006xe,Ferreira:2006in}.

\begin{table}[tb]
\begin{center}
\caption{Contributions of order $\Lambda^{-4}$ to the cross section of top production.}
\begin{tabular}{|l|l|}
\hline
direct production & $ p~p \rightarrow \left( g~q \right) \rightarrow  t + X $ \\
\hline
top + jet production & $ p~p \rightarrow \left( g~g \right) \rightarrow \bar{q}~t + X $ \\
                     & $ p~p \rightarrow \left( g~q \right) \rightarrow g~t + X $ \\
                     & $ p~p \rightarrow \left(  \bar{q}~q \right) \rightarrow \bar{q}~t + X $ \\
                     & \hspace*{1cm}                            (including 4-fermion interactions)\\
\hline
top + anti-top production & $ p~p \rightarrow \left( g~g \right) \rightarrow \bar{t}~ t + X $ \\
                          & $p~p \rightarrow \left(  \bar{q}~q \right) \rightarrow \bar{t}~t+ X $ \\
\hline
top + gauge boson production & $ p~p \rightarrow \left( g~q \right) \rightarrow \gamma~t + X $ \\
                             & $ p~p \rightarrow \left( g~q \right) \rightarrow Z~t + X $ \\
                             & $ p~p \rightarrow \left( g~q \right) \rightarrow Wt + X $ \\
\hline
top + Higgs production & $ p~p \rightarrow \left( g~q \right) \rightarrow h~t + X $ \\
\hline
\end{tabular}
\label{top_fcnc_lambda4}
\end{center}
\end{table}

The main goal of this work was to produce all cross sections and decay
widths related to strong FCNC with a single top quark in a form appropriate
for implementation in the {\tt TopReX}
generator~\cite{Slabospitsky:2002ag}.  This implies that all cross sections
had to be given in differential form with the top spin taken into account.
Most of the processes were already inserted in the generator (see release
4.20 of {\tt TopReX}) and the remaining ones will be inserted in the near
future.

In this section, a joint analysis of the results obtained
in~\cite{Ferreira:2005dr,Ferreira:2006xe,Ferreira:2006in} is performed. To
investigate the dependence of the cross sections on the values of the
anomalous couplings, which are denoted by constants $\alpha_{ij}$ and
$\beta_{ij}$, random values for $\alpha_{ij}$ and $\beta_{ij}$ were
generated and the resulting cross sections were plotted against the
branching ratio of the top quark for the decay $t \rightarrow g\,u$. The
motivation for doing this is simple: the top quark branching ratios for
these decays may vary by as much as eight orders of magnitude, from
$\sim\,10^{-12}$ in the SM to $\sim\,10^{-4}$ for some supersymmetric
models. This quantity is therefore a good measure of whether any physics
beyond that of the standard model exists.

\begin{figure}[tb]
\epsfysize=8.5cm \centerline{\epsfbox{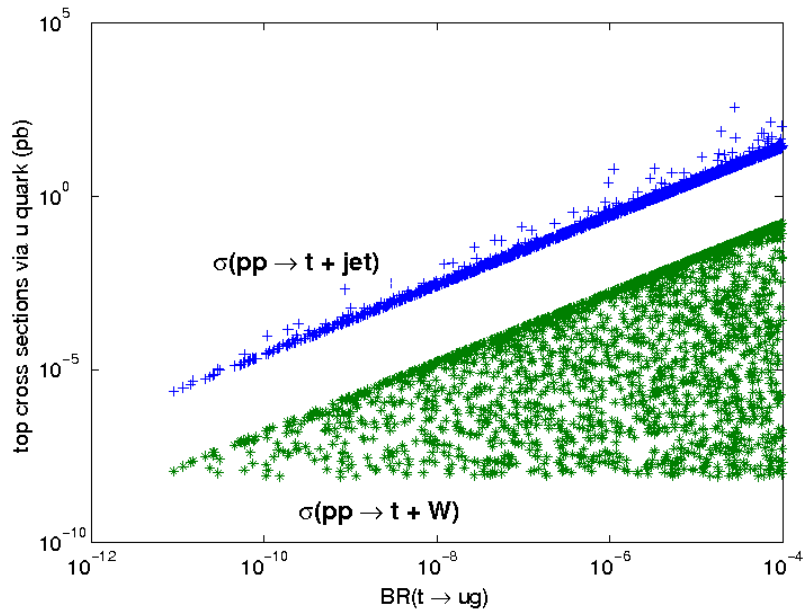}} 
\caption{Cross
sections for the processes $p\,p \rightarrow t \,+\, \rm{jet}$
(crosses) and $p\,p \rightarrow t \, + \, W$ (stars) via an $u$
quark, as a function of the branching ratio $BR(t \rightarrow g \,u)$. }
 \label{top:fig:sig_t_u}
\end{figure}

In Fig.~\ref{top:fig:sig_t_u} the cross sections for the processes $p \, p
\rightarrow t\, + \, \rm{jet}$ and $p \,p \rightarrow t\, +\, W$ via a $u$
quark versus the branching ratio $BR(t \rightarrow g \, u)$ are shown. This
plot was obtained by varying the constants $\alpha$ and $\beta$ in a random
way, as described before. Each combination of $\alpha$ and $\beta$
originates a given branching ratio and a particular value for each cross
section. Obviously, another set of points may generate the same value for
the branching ratio but a different value for the cross section, which
justifies the distribution of values of $\sigma(p\,p \rightarrow
t\,+\,\rm{jet})$ and $\sigma(p\,p \rightarrow t\,+\,W)$. Values of $\alpha$
and $\beta$ for which the branching ratio varies between the SM value and
the maximum value predicted by supersymmetry were chosen.\footnote{Both
$\alpha/\Lambda^2$ and $\beta/\Lambda^2$ were varied between $10^{-6}$ and
$1$~TeV$^{-2}$.} The cross sections for top plus jet and top plus a $W$
boson production via a $c$ quark are similar to these ones although smaller
in value. Notice that the $Wt$ cross section is proportional to only one
of the couplings, which makes it  a very attractive observable -
it may allow us to impose constraints on a single anomalous
coupling (see Ref.~\cite{Ferreira:2005dr} for details).

It should be noted that single top production depends also on the
contributions of the four fermion operators. Hence, even if the
branching ratios  $BR(t \rightarrow g \, u (c))$ are very small,
there is still the possibility of having a large single top cross
section with origin in the four fermion couplings. In
Fig.~\ref{top:fig:sig_t_u} we did not consider this possibility,
setting the four-fermion couplings to zero. For a discussion on
the four-fermion couplings do see Ref.~\cite{Ferreira:2006xe}.

\begin{figure}[bt]
\epsfysize=8.5cm \centerline{\epsfbox{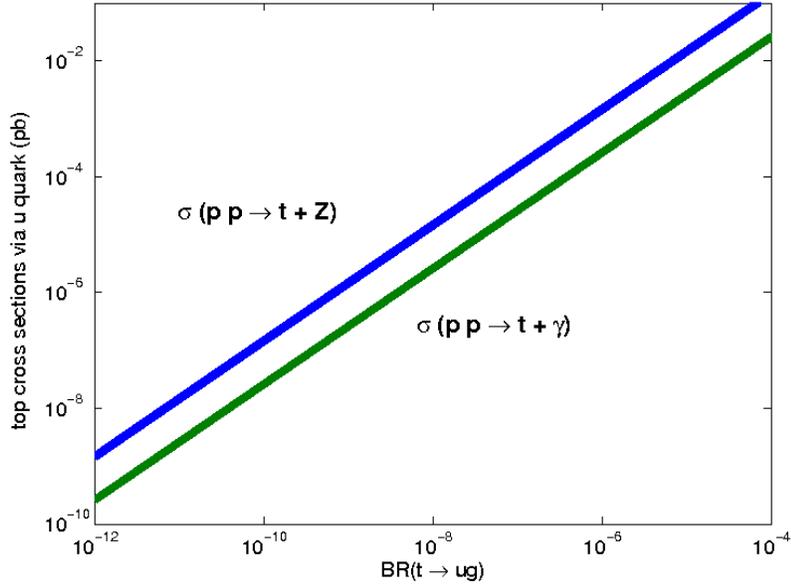}} \caption{Cross
sections for the processes $p \, p \rightarrow t \, + \,  Z$
(upper line) and $p \, p \rightarrow t \, + \,\gamma$ (lower line)
via a $u$ quark, as a function of the branching ratio $BR(t
\rightarrow g \, u)$ .} \label{top:fig:Zgamma_u}
\end{figure}

In Fig.~\ref{top:fig:Zgamma_u} the cross sections for $p \, p
\rightarrow t \, + \, Z$ and $p \, p \rightarrow t \, + \,\gamma$
via a $u$ quark, versus the branching ratio $BR(t \rightarrow g \,
u)$ are plotted. The equivalent plot with an internal $c$ quark is similar,
but the values for the cross section are much smaller. In this
plot we can see that both cross sections are very small in the
range of $\{\alpha\,\beta\}$ considered. These results imply that
their contribution will hardly be seen at the LHC, unless the
values for the branching ratio are peculiarly large.

The same, in fact, could be said for $p \, p \rightarrow t \, + \, h$. Even
for the smallest allowed SM Higgs mass, the values of the cross section for
associated top and Higgs production are very small. The same holds true for
the processes involving the anomalous couplings of the $c$ quark.

The smallness of the effects of these operators in the several
cross sections holds true, as well, for the top--anti-top channel.
In this case, even for a branching ratio $BR(t \rightarrow g \, u)
\,\simeq\,10^{-4}$, the contributions to the cross section
$\sigma(p \, p \rightarrow t \,\bar{t})$ do not exceed, in
absolute value, one picobarn. 

In conclusion, the strong FCNC effective operators are constrained in their
impact on several channels of top quark production. Namely,
Fig.~\eqref{top:fig:sig_t_u} and~\eqref{top:fig:Zgamma_u} illustrate that,
if there are indeed strong FCNC effects on the decays of the top quark,
their impact will be more significant in the single top plus jet production
channel. It is possible, according to these results, to have an excess in
the cross section $\sigma(p \, p \rightarrow t \, + \,\mbox{\rm{jet}})$
arising from new physics described by the operators we have considered
here, at the same time obtaining results for the production of a top quark
alongside a gauge and Higgs boson, or for $t\bar{t}$ production, which are
entirely in agreement with the SM predictions. This reinforces the
conclusion that the cross section for single top plus jet production is an
important probe for the existence of new physics beyond that of the SM. It
is a channel extremely sensitive to the presence of that new physics, and
boasts a significant excess in its cross section, whereas many other
channels involving the top quark remain unchanged. Nevertheless, it may
still be possible to use some of these unchanged channels, such as top plus
$W$ production, to constrain the $\beta$ parameters, through the study of
asymmetries such as $\sigma(p \, p \rightarrow t\,W^-)\, - \, \sigma(p \, p
\rightarrow \bar{t} \,W^+)$.

\subsection{Higgs boson FCNC decays into top quark in a general two-Higgs 
doublet model}
 \label{top:2hdm}

The branching ratios for FCNC Higgs boson decays are at the level of
$10^{-15}$, for Higgs boson masses of a few hundred GeV. In this section,
the FCNC decays of Higgs bosons into a top quark in a general
two-Higgs-doublet model (2HDM) are considered. In this model, the Higgs
FCNC decays branching ratios can be substantially enhanced and perhaps can
be pushed up to the visible level, particularly for $h^0$ which is the
lightest $CP$-even spinless state in these models~\cite{Gunion:1989we}.
We compute the maximum branching ratios and the number of FCNC
Higgs boson decay events at the LHC. The most favorable mode for production
and subsequent FCNC decay is the lightest $CP$-even state in the type II
{2HDM}, followed by the other $CP$-even state, if it is not very heavy,
whereas the $CP$-odd mode can never be sufficiently enhanced. The present
calculation shows that the branching ratios of the $CP$-even states may
reach $10^{-5}$, and that several hundred events could be collected in the
highest luminosity runs of the LHC. Some strategies to use these FCNC
decays as a handle to discriminate between 2HDM and supersymmetric Higgs
bosons are also pointed out.

Some work in relation with the 2HDM Higgs bosons FCNCs has already been
performed~\cite{Grzadkowski:1990sm,Bejar:2000ub}, and in the context of the
MSSM~\cite{Guasch:1999jp,Bejar:2004rz,Guasch:1997kc,Bejar:2005kv}.
In this work the production of any 2HDM Higgs boson
($h=h^0,H^0,A^0$) at the LHC is computed and analyzed, followed by the 
one-loop FCNC decay 
$h\to tc$. The maximum production rates
of the combined cross section,
\begin{eqnarray}
  \sigma(pp\to h \to tc)&
  \equiv&
  \sigma(pp\to h X)BR(h\to tc)\nonumber \, ,\\
      BR(h\to t c)&\equiv&\frac{\Gamma(
          h\to t\bar c+ \bar t c)}{\sum_i \Gamma(h\to X_i)} \, ,
    \label{top:eq:hqq-def}
\end{eqnarray}
takes into account the restrictions from the experimental determination
of $b\to s\gamma$ branching ratio ($m_{H^\pm} \gtrsim 
350\UGeV$~\cite{Gambino:2001ew}), from
perturbativity arguments ($0.1\lesssim\tan\beta\lesssim 60$, where
$\tan\beta$ is the ratio of the vacuum expectation values of each
doublet), from the custodial symmetry ($|\delta\rho^{\rm
 2HDM}|\lesssim 0.1\%$) and from unitarity of the Higgs couplings.
In this section a summarized explanation of the numerical analysis is 
given. For further details see Refs.~\cite{Bejar:2003em,Bejar:2000ub}.

\begin{figure}[tbp]
\begin{center}
        \includegraphics[width=11cm]{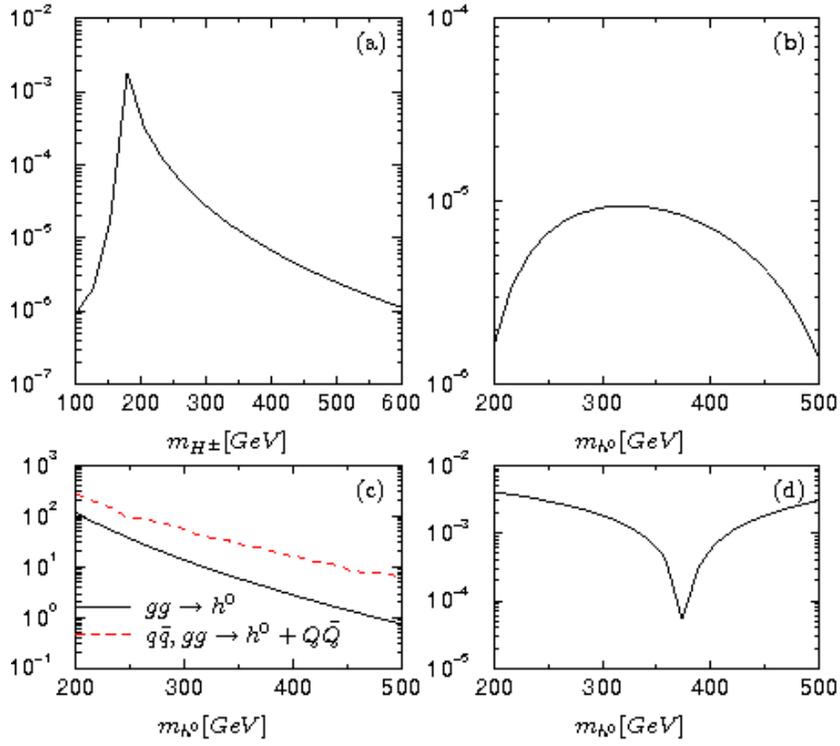}

\vspace*{10mm}
        \includegraphics[width=10cm]{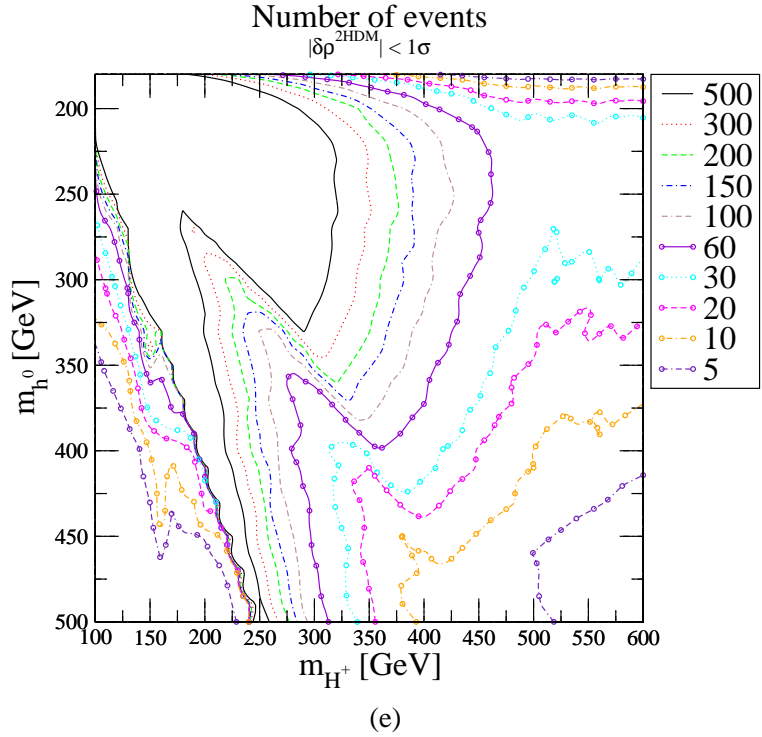}\\
{(e)}
\end{center}
    \caption{{(a)} $BR(h^0\to t c)$ versus $m_{H^\pm}$ for type~II 2HDM; 
{(b)}
      Idem, versus $m_{h^0}$; {(c)} The production cross section (in 
$\Upb$)
      of $h^0$ at the LHC versus its mass; {(d)} $\delta\rho^{\rm
        2HDM}$ versus $m_{h^0}$, for a fixed value of the other 
parameters.
      {(e)} Contour lines in the $(m_{H^\pm},m_{h^0})$-plane for 
the maximum number of light $CP$-even Higgs FCNC events $h^0 \to t c $ 
produced at the LHC for $100\Ufb^{-1}$ of integrated luminosity.}
      \label{top:fig:4figslhnumber}
\end{figure}

The full one-loop calculation of $BR(h\to t c)$ in the
type II 2HDM, as well as of the LHC production rates of these FCNC
events were included. It is considered that $BR(h \to t c)$ in the 
type I 2HDM is essentially small (for all $h$), and that these decays 
remain always invisible. The basic definitions in the general 2HDM 
framework can be found in Ref.~\cite{Gunion:1989we}.

The calculations were performed with the help of the numeric and
algebraic programs {\tt FeynArts}, {\tt FormCalc} and {\tt
LoopTools}~\cite{Kublbeck:1990xc,Hahn:1998yk,FAFCuser}. A parameter scan
of the production rates over the 2HDM parameter space in the
$(m_{H^\pm},m_{h^0})$-plane was done, keeping $\tan\beta$
fixed.

In Fig.~\ref{top:fig:4figslhnumber}~a-b, the $BR(h^0 \to t c)$ for the
lightest $CP$-even state (type II 2HDM) is shown. The $BR$ is sizeable, up
to $10^{-5}$, for the range allowed from $b\to s\gamma$.  In
Fig~\ref{top:fig:4figslhnumber}c the production cross sections explicitly
separated (the gluon-gluon fusion at one-loop and the $h^0 q \bar q$
associated production at the tree level~\cite{Spira:1997cb,Spira:1997dg}) are
presented.  The control over $\delta\rho^{\rm 2HDM}$ is displayed in
Fig.~\ref{top:fig:4figslhnumber}d.

In practice, to better assess the possibility of detection at the LHC,
one has to study the production rates of the FCNC events. A systematic
search of the regions of parameter space with the maximum number of FCNC
events for the light $CP$-even Higgs is presented in the form of contour
lines in Fig.~\ref{top:fig:4figslhnumber}e. The dominant FCNC region for
$h^0 (H^0)$ decay is where $\tan\alpha$ ($\alpha$ is the rotation angle
which diagonalises the matrix of the squared masses of the $CP$-even
scalars) is large (small), $\tan\beta$ is large and $m_{h}\ll m_{A^0}$,
with a maximum value up to few hundred events. As for the $CP$-odd state
$A^0$, it plays an important indirect dynamical role on the other decays
through the trilinear couplings, but its own FCNC decay rates never get
a sufficient degree of enhancement due to the absence of the relevant
trilinear couplings.

One should notice that in many cases one can easily distinguish whether 
the enhanced FCNC events stem from the dynamics of a general, 
unrestricted, 2HDM model, or rather from some supersymmetric mechanisms 
within the MSSM. In the 2HDM case the $CP$-odd modes $A^0\to t c$ are 
completely hopeless whereas in the MSSM they can be 
enhanced~\cite{Guasch:1999jp,Curiel:2002pf, Demir:2003bv,Bejar:2004rz}. 
Nevertheless, different ways to discriminate these rare events are discussed 
in Ref.~\cite{Bejar:2000ub}. 

The FCNC decays of the Higgs bosons into top quark final states are a 
potentially interesting signal, exceeding $1$~fb for $m_{H^+}$ up to 
$400\UGeV$ (Fig.~\ref{top:fig:4figslhnumber}e). This however, is a small 
cross section once potentially important backgrounds are considered, 
such as $Wjj$ and SM single top production. A careful study of the 
backgrounds for this process should be carried out. If it were possible 
to fully reconstruct the top, then there might be hope to observe a 
distinctive Higgs bump in the $tc$ channel~\cite{Burdman:1999sr}.

\subsection{Single top  production by direct SUSY FCNC interactions}
\label{sec:singletop}

FCNC interactions of top quarks can provide an important indirect probe
for new SUSY processes. For instance, the MSSM Higgs boson FCNC decay rates
into top-quark final states, e.g. $H^0,A^0\rightarrow t{\bar c}+{\bar
t}c$, can be of order $10^{-4}$ (see section~\ref{top:2hdm} and
Refs.~\cite{Bejar:2004rz, Demir:2003bv, Curiel:2002pf, Curiel:2003uk,
Heinemeyer:2004by}), while in the SM $BR(H\rightarrow t\,\bar{c})\sim
10^{-13}$-$10^{-16}$ (depending on the Higgs mass)~\cite{Bejar:2003em}.
There also exists the possibility to produce
$t{\bar c}$ and ${\bar t}c$ final states without Higgs bosons or any
other intervening particle~\cite{Guasch:2006hf,Eilam:2006rb}. In this
section it will be shown that the FCNC gluino interactions in the MSSM 
can
actually be one efficient mechanism for direct FCNC production of top
quarks~\cite{Guasch:2006hf}.

In general, in the MSSM we expect terms of the form gluino--quark--squark
or neutralino--fermion--sfermion, with the quark and squark having the same
charge but belonging to different flavours. In the present study only the
first type of terms, which are expected to be dominant, are considered.  A
detailed lagrangian describing these generalized SUSY--QCD interactions
mediated by gluinos can be found, e.g. in Ref.~\cite{Guasch:1999jp}. The
relevant parameters are the flavour-mixing coefficients $\delta_{ij}$. In
contrast to previous studies~\cite{Liu:2004bb}, in the present work, these
parameters are only allowed in the LL part of the $6\times 6$ sfermion mass
matrices in flavour-chirality space. This assumption is also suggested by
RG arguments~\cite{Duncan:1983iq,Duncan:1984vt}. Thus, if $M_{\rm LL}$ is
the LL block of a sfermion mass matrix, $\delta_{ij} \,(i\neq j)$ is
defined as follows: $(M_{\rm
LL})_{ij}=\delta_{ij}\,\tilde{m}_i\,\tilde{m}_j$, where $\tilde{m}_i$ is
the soft SUSY-breaking mass parameter corresponding to the LH squark of
$i$th flavour~\cite{Guasch:1999jp}. The parameter
$\delta_{23}$ is the one relating the $2$nd and $3$rd
generations (therefore involving the top quark physics) and it is
the less restricted one from the phenomenological point of view, being
essentially a free parameter ($0<\delta_{23}<1$). Concretely, we have two
such parameters, $\delta_{23}^{(t) {\rm LL}}$ and $\delta_{23}^{(b)  {\rm
LL}}$, for the up-type and down-type LL squark mass matrices respectively.
The former enters the process under study whereas the latter enters
$BR(b\to s \gamma)$, observable that we use to restrict our predictions on
$t\bar{c}+\bar{t}c$ production. Notice that $\delta_{23}^{(b) {\rm LL}}$ is
related to the parameter $\delta_{23}^{(t) {\rm LL}}$ because the two LL
blocks of the squark mass matrices are precisely related by the CKM
rotation matrix $K$ as follows: $({\cal M}_{\tilde{u}}^2)_{LL}=K\,({\cal
M}_{\tilde{d}}^2)_{LL}\,K^{\dagger}$~\cite{Gabbiani:1996hi,Misiak:1997ei}.

\begin{figure}
\begin{center}
\resizebox{!}{6.3cm}{\includegraphics*{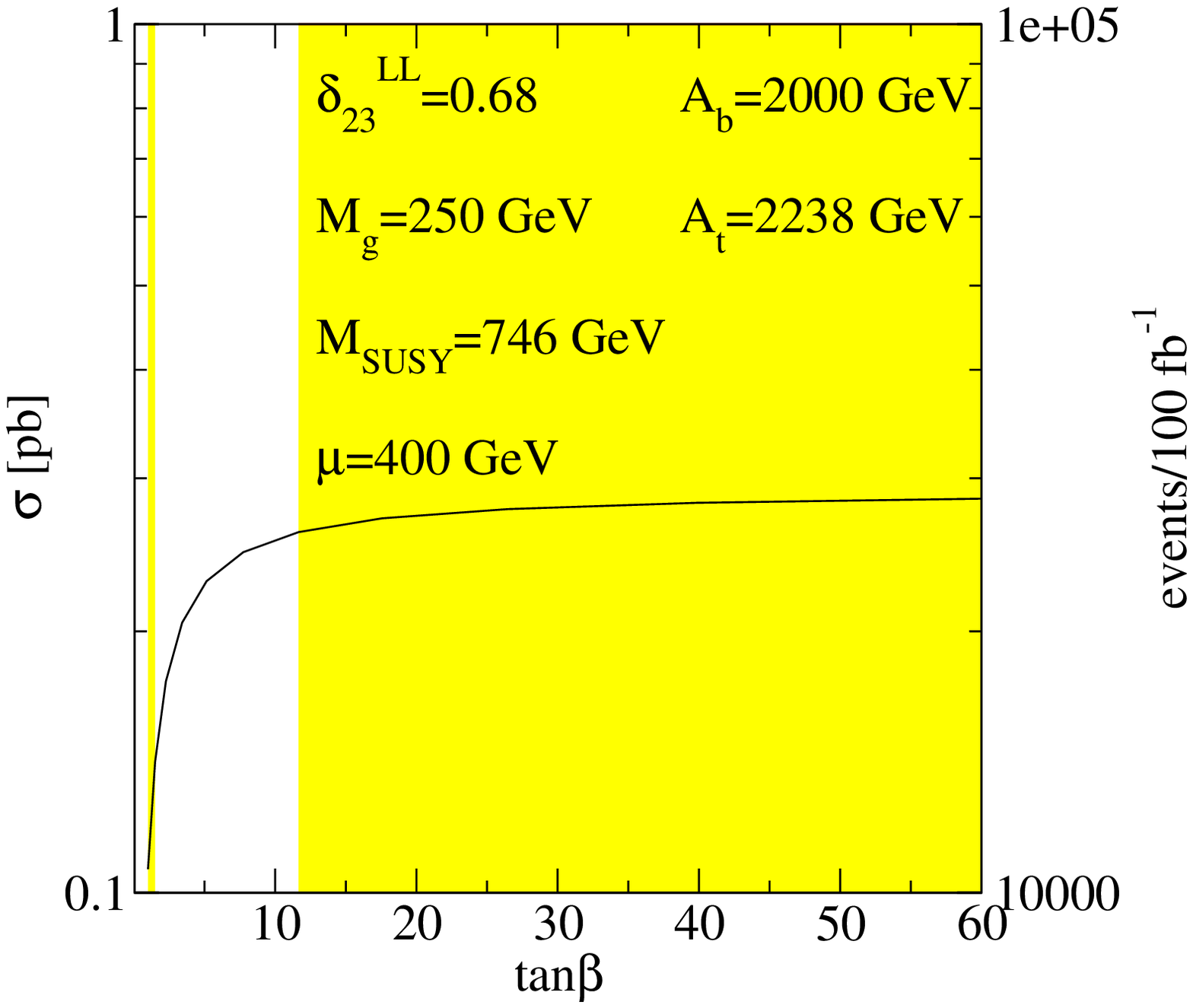}} 
\hspace*{0.5cm}
\resizebox{!}{6.3cm}{\includegraphics*{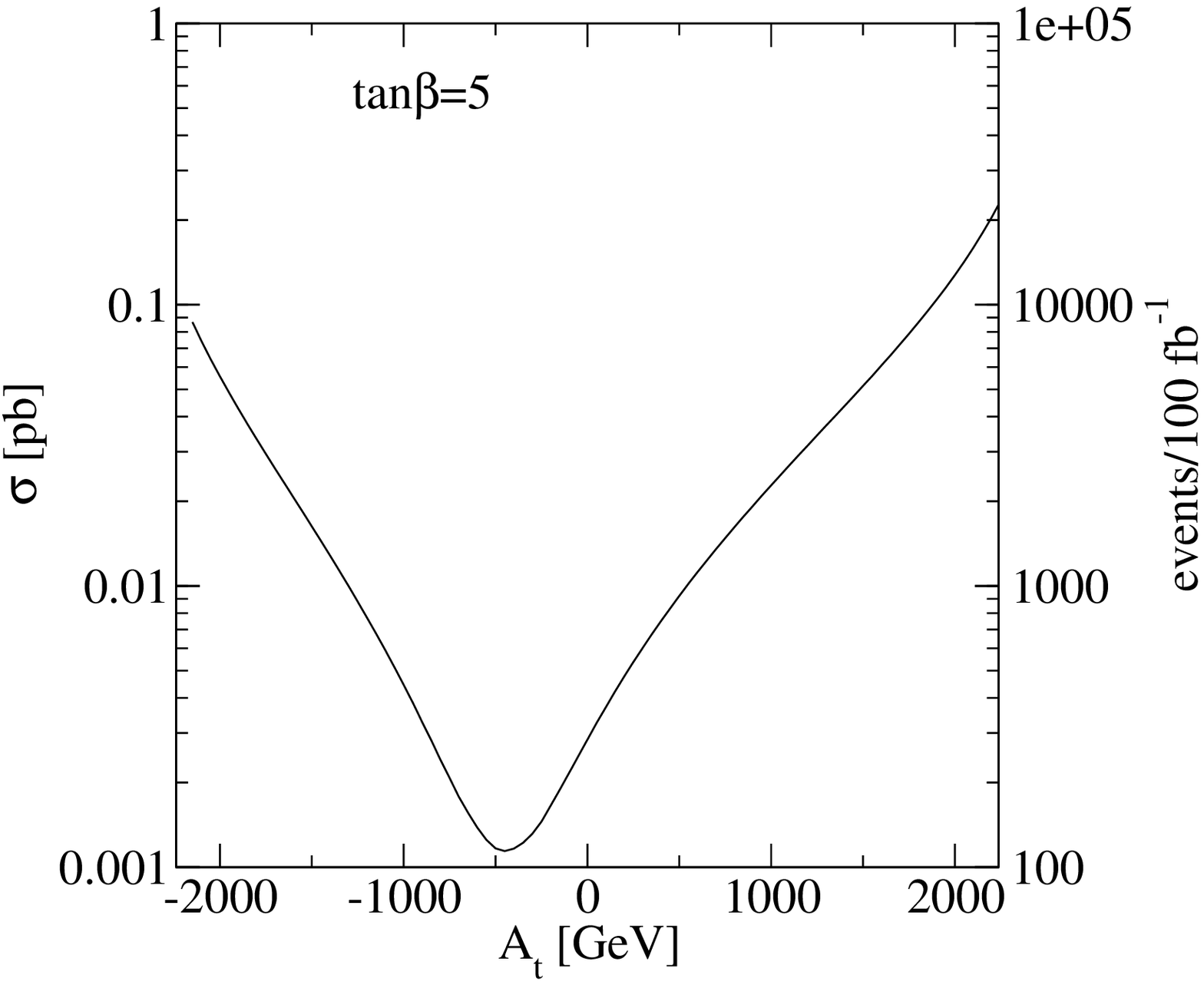}}
\end{center}
\caption{
$\sigma_{tc}$ (in ${\rm pb}$) and number of events
per $100\,{\rm fb}^{-1}$ of integrated luminosity at the LHC, as a
function of $\tan\beta$ (left) and $A_t$ (right) for the given 
parameters. The shaded
region is excluded by the experimental limits on $BR (b\to s 
\gamma)$.\label{top:fig:tbAt} }
\end{figure}

The calculation of the full one-loop SUSY--QCD cross section
$\sigma_{tc}\equiv\sigma (pp\rightarrow t\bar{c})$ using standard
algebraic and numerical packages for this kind of
computations~\cite{Hahn:2005qi,FAFCuser} has been performed. 
The typical diagrams contributing are gluon-gluon triangle loops (see
Ref.~\cite{Guasch:2006hf}for more details). In order to simplify
the discussion it will be sufficient to quote the general form of the
cross section:
\begin{equation}
\label{top:sigmatc} 
\sigma_{tc} \sim
\left(\delta_{23}^{(t)LL}\right)^2 \, \frac{m_t^2
({A_t}-\mu/\tan\beta)^2}{M_{\rm SUSY}^4} \, \frac{1}{m_{\tilde{g}}^2}\,.
\end{equation} 
Here $A_t$ is the trilinear top-quark coupling, $\mu$ the higgsino mass
parameter, $m_{\tilde{g}}$ is the gluino mass and $M_{\rm SUSY}$ stands
for the overall scale of the squark masses~\cite{Guasch:2006hf}. The
computation of $\sigma_{tc}$ together with the branching ratio $BR(b\to s
\gamma)$ in the MSSM was performed, in order to respect the experimental
bounds on $BR(b\to s \gamma)$. Specifically, $BR (b\to s
\gamma)=(2.1$-$4.5)\times 10^{-4}$ at the $3\sigma$ level is
considered~\cite{Eidelman:2004wy}. 

\begin{figure}[tb]
\begin{center}
\resizebox{!}{6.3cm}{\includegraphics*{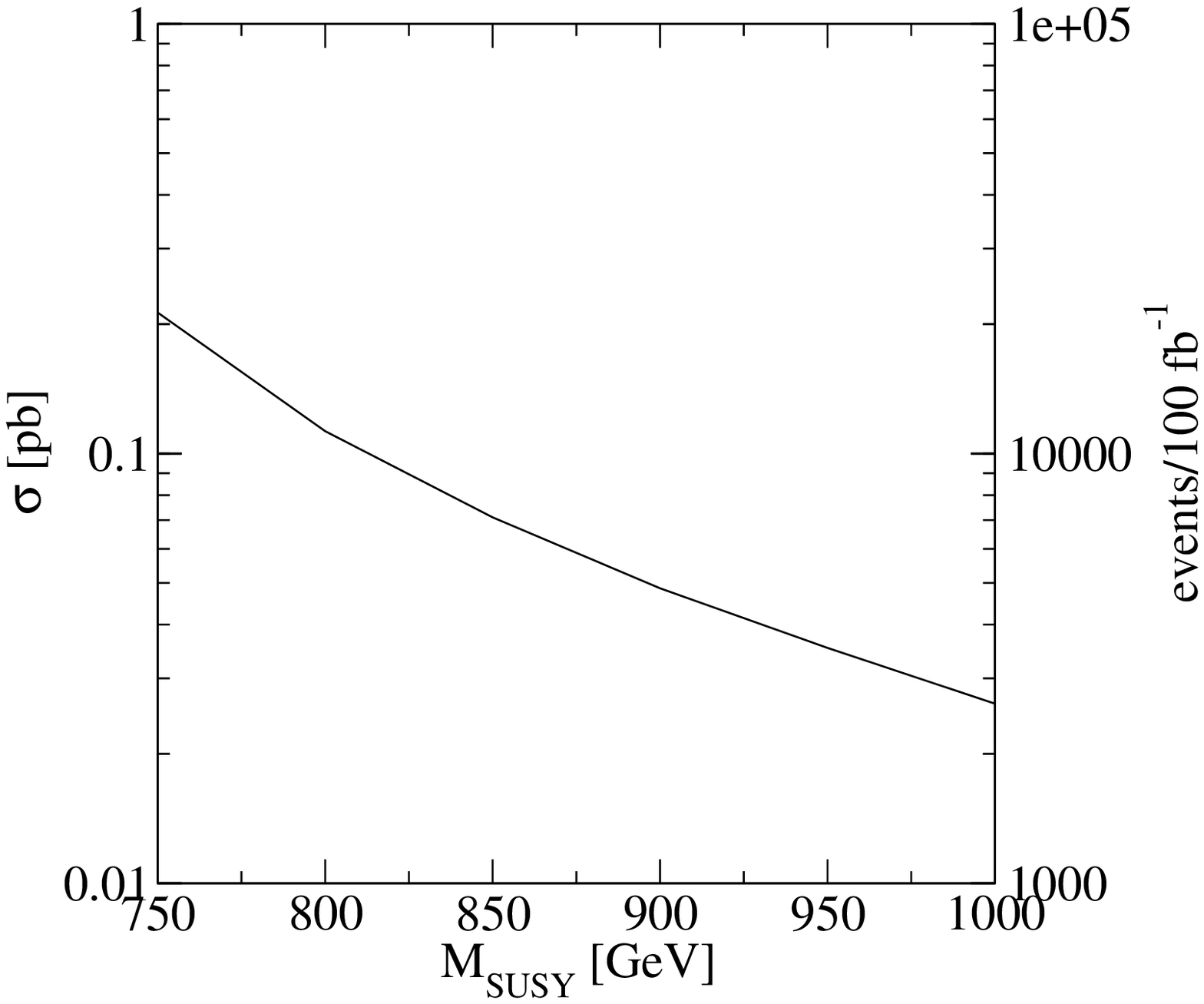}}
\hspace*{0.5cm}
\resizebox{!}{6.3cm}{\includegraphics*{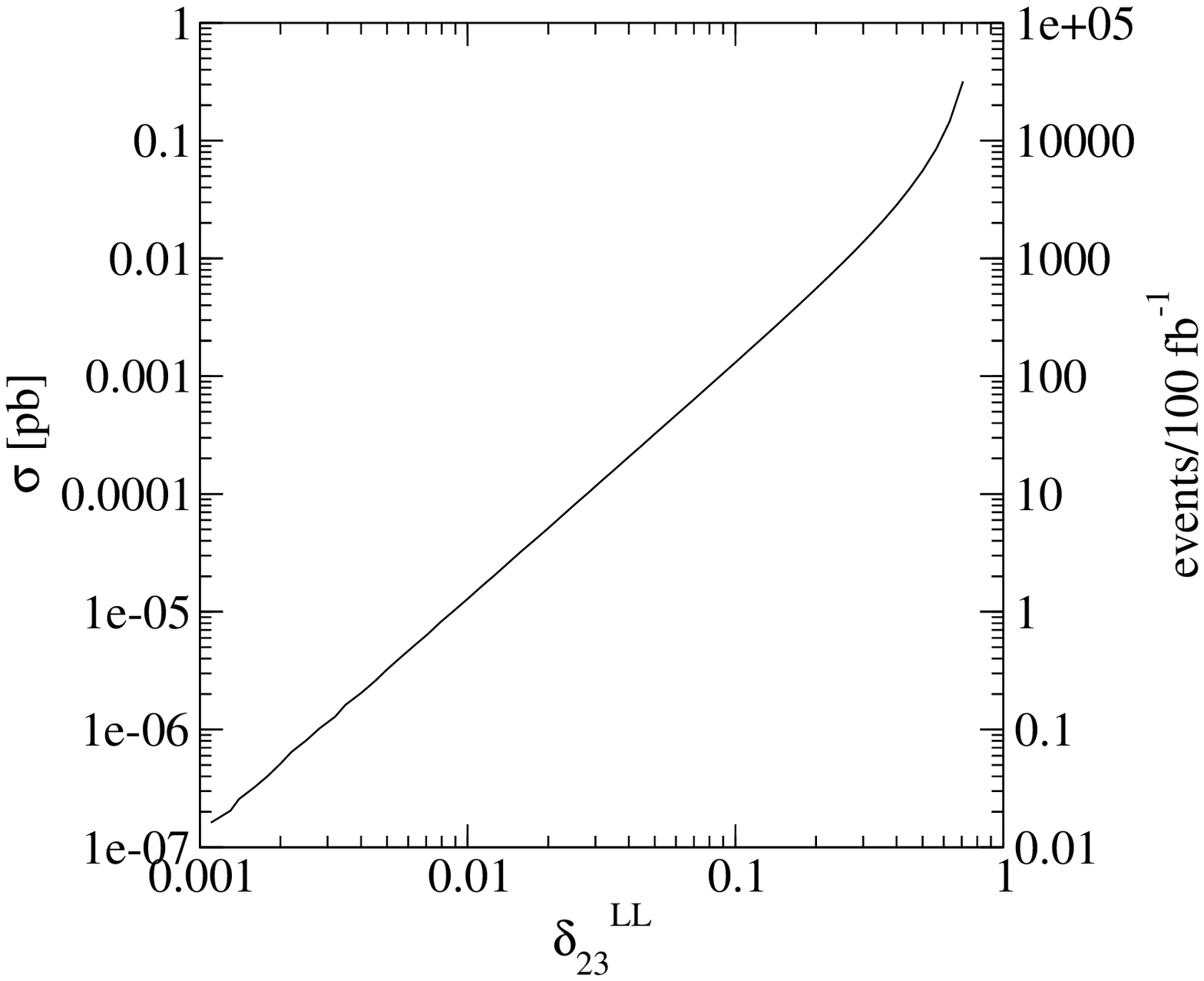}}
\end{center}
\caption{
$\sigma_{tc}$ (in ${\rm pb}$) and number of events
per $100\,{\rm fb}^{-1}$ of integrated luminosity at the LHC, as a
function of $M_{\rm SUSY}$ (left) and $\delta_{23}^{\rm LL}$ (right).
\label{top:fig:Msusyd23} }
\end{figure}

\begin{figure}[tb]
\begin{center}
\resizebox{!}{6.3cm}{\includegraphics*{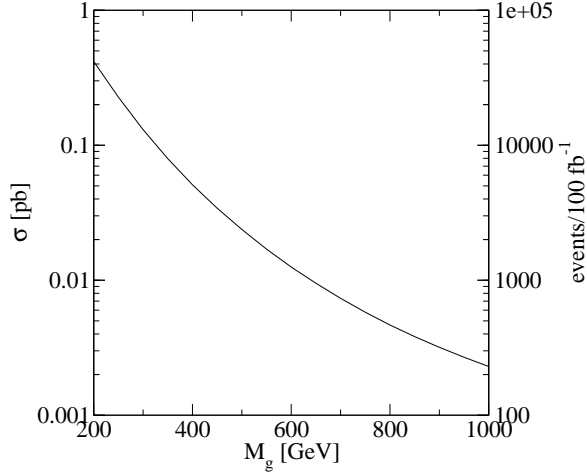}}
\end{center}
\caption{
$\sigma_{tc}$ (in ${\rm pb}$) and number of events
per $100\,{\rm fb}^{-1}$ of integrated luminosity at the LHC, as a
function of $m_{\tilde{g}}$.
\label{top:fig:Mg} }
\end{figure}

In Figs.~\ref{top:fig:tbAt}, \ref{top:fig:Msusyd23} and \ref{top:fig:Mg}
the main results of this analysis are presented. It can be seen that
$\sigma_{tc}$ is very sensitive to $A_t$ and that it decreases with
$M_{SUSY}$ and $m_{\tilde{g}}$. As expected, it increases
with $\delta_{23}^{\rm LL}\equiv \delta_{23}^{(t) {\rm LL}}$. At the
maximum of $\sigma_{tc}$, it prefers $\delta_{23}^{\rm LL}=0.68$. The
reason stems from the correlation of this maximum with the $BR(b\to s
\gamma)$ observable. At the maximum, $2\sigma_{tc}\simeq 0.5\,{\rm pb}$,
if we allow for relatively light gluino masses $m_{\tilde{g}}=250\,$ GeV
(see Fig.~\ref{top:fig:Mg}). For higher $m_{\tilde{g}}$ the cross section
falls down fast; at $m_{\tilde{g}}=500\,$ GeV it is already $10$ times
smaller. The total number of events per $100\,{\rm fb}^{-1}$ lies between
$10^4$-$10^5$ for this range of gluino masses. The fixed values of the
parameters in these plots lie near the values that provide the maximum of
the FCNC cross section. The dependence on $\mu$ is not shown, but it
should be noticed that it decreases by $\sim 40\%$ in the allowed range
$\mu=200$-$800\,$ GeV. Values of $\mu>800\,$ GeV are forbidden by $BR
(b\to s \gamma)$. Large negative $\mu$ is also excluded by the
experimental bound considered for the lightest squark mass,
$m_{\tilde{q}_1}\lesssim 150\,$ GeV; too small $|\mu|\lesssim 200\,$ GeV
is ruled out by the chargino mass bound $m_{\chi^{\pm}_1}\leq 90\,$ GeV.
The approximate maximum of $\sigma_{tc}$ in parameter space has been
computed using an analytical procedure as described in
Ref.~\cite{Guasch:2006hf}.

Finally, it should be noticed that $t\bar{c}$ final states can also be
produced at one-loop by the charged-current interactions within the SM.
This one-loop cross section at the LHC was computed, with the result
$\sigma^{\rm SM}(pp\rightarrow t\bar{c}+\bar{t}c)=7.2\times 10^{-4}\,{\rm
fb}\,.$ It amounts to less than one event in the entire lifetime of the
LHC. Consequently, an evidence for such signal above the
background would have to be interpreted as new physics.

The full one-loop SUSY--QCD cross section for the production of single 
top-quark states $t{\bar c}+{\bar t}c$ at the LHC were computed.  This 
direct production mechanism is substantially more efficient (typically a 
factor of $100$) than the production and subsequent FCNC 
decay~\cite{Bejar:2005kv,Bejar:2006hd} ($h\rightarrow t{\bar c}+{\bar 
t}c$) of the MSSM Higgs bosons $h=h^0,H^0,A^0$. It is important to 
emphasize that the detection of a significant number of $t{\bar c}+{\bar 
t}c$ states could be interpreted as a distinctive SUSY signature. It 
should be noticed however that a careful background study must be done 
for this chanel since, unlike the Higgs decay studied in the previous 
section, the kinematic distributions of the signal are not likely to 
have a very distinctive shape compared to $Wjj$ or standard model 
single-top production.

\subsection{ATLAS and CMS sensitivity to FCNC top decays}
 \label{top:fcncdecay}

Due to the high production rate for $t\bar t$ pairs and single top, the
LHC will allow either to observe top FCNC decays or to establish very
stringent limits on the branching ratios of such decays. In this section
the study of ATLAS and CMS sensitivity to top FCNC decays is presented. A
detailed description of the analysis can be found
in~\cite{Carvalho:973137, top_CMSnote-FCNC}.

Both CMS and ATLAS collaborations have investigated the $t\to q\gamma$
and $t\to qZ$ decay channels. Analyses have been optimized for searching
FCNC decays in $t\bar t$ signal, where one of the top quarks is assumed
to decay through the dominant SM decay mode ($t\to bW$) and the other is
assumed to decay via one of the FCNC modes. The $t\bar t$ final states
corresponding to the different FCNC top decay modes lead to different
topologies, according to the number of jets, leptons and photons. Only
leptonic decay channels of $Z$ and $W$ bosons are 
considered in the analysis developed by the CMS collaboration. The ATLAS
collaboration has also studied the channel corresponding to the hadronic
$Z$ decay, which is discussed elsewhere~\cite{Carvalho:973137}.

The signal is generated with {\tt TopReX}~\cite{Slabospitsky:2002ag},
while {\tt PYTHIA}~\cite{Sjostrand:2000wi} is used for background
generation and modelling of quark and gluon hadronization. The generated
events are passed through the fast (for ATLAS) and full (for CMS)
detector simulation. Several SM processes contributing as background are
studied: $t\bar t$ production, single top quark production, $ZW/ZZ/WW +
\mathrm{jets}$, $Z/W/\gamma^*+\mathrm{jets}$, $Zb\bar b$ and QCD
multi-jet production.

Although ATLAS and CMS analyses differ in some details of selection
procedure, they obtain the same order of magnitude for the FCNC
sensitivity. In both analyses, the signal is preselected by requiring the
presence of, at least, one high $p_T$ lepton (that can be used to trigger
the event) and missing energy above $20$~GeV for the ATLAS analysis and
above $25$~GeV for the CMS analysis. Additionally, two energetic central
jets from $t$ and $\bar t$ decays are required. The slight differences
in CMS and ATLAS thresholds reflect the differences in their
sub-detectors, simulation code and reconstruction algorithms.

\begin{figure}[tb]
	\centering
\includegraphics[height=6cm,width=7.9cm]{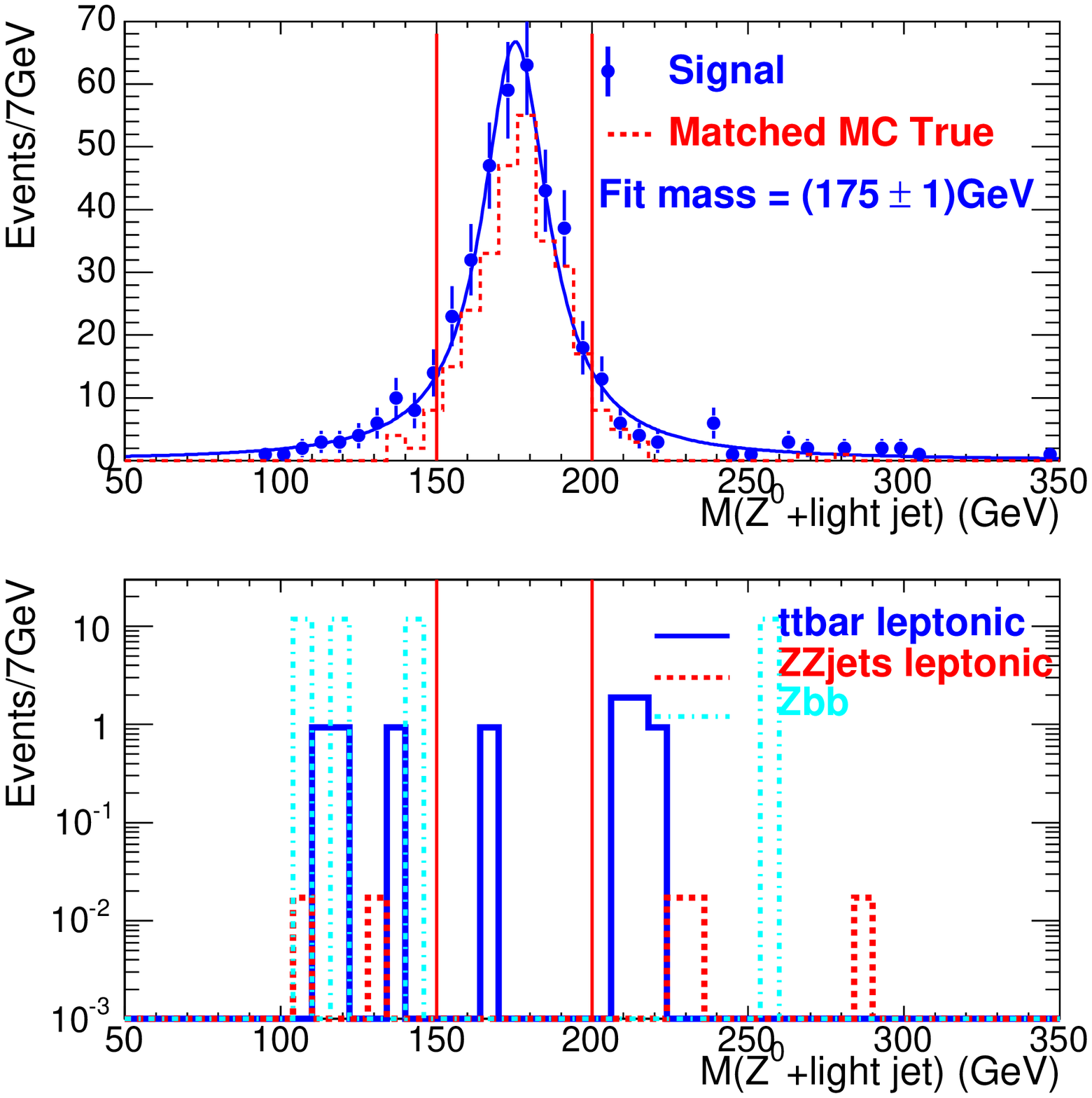} 
\hspace{0mm}
\includegraphics[height=6cm,width=7.9cm]{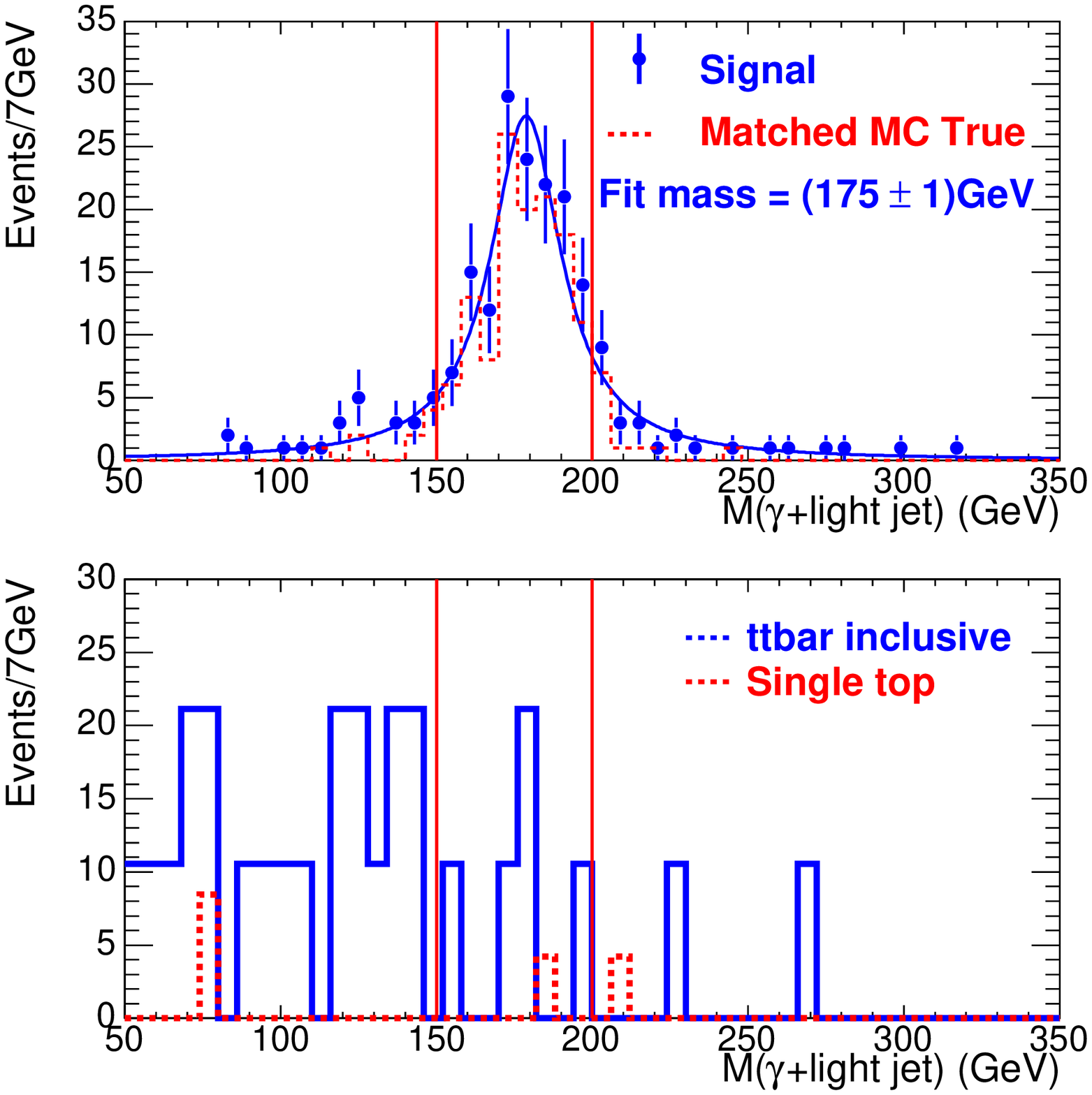}
\caption{Invariant mass plot of the FCNC top $t\to qZ$ (left) and $t\to q\gamma$ (right), as obtained in CMS after 
sequential cuts. Data are fitted with a Breit-Wigner shape and
central value is in agreement with top mass. The signal distributions obtained from reconstructed leptons and jets
matched to the corresponding generated objects are also shown (\emph{Matched MC True}).
}
\label{top:fcnc:fig:CMS_results} 
\end{figure}

The CMS analysis strongly relies on $b$-tagging capability to distinguish
the $b$-jet from SM decay and the light-jet from the anomalous one. A
series of cascade selections are applied to reduce the background. For
the $t\to q\gamma$ channel, the W boson is reconstructed requiring
the transverse mass of the neutrino and hard lepton to be less than 120
GeV and the $b$-jet is used to form a window mass 110$<m_{bW}<$ 220
GeV. The invariant mass of the light-jet and a single isolated photon
with $p_T >$ 50 GeV is bounded in the range [150,200] GeV. A final
selection of top back-to-back production ($\cos\phi(t\bar t)<-0.95$)
reduces the di-boson background. The $t\to qZ$ channel is extracted with
the search of one $Z$ (using same flavour-opposite charge leptons, which
serve as trigger and are bound to a 10 GeV window around the $Z$ mass)
and a $M_{bW}$ in the top mass region, with the same cuts of the previous
case. One hard light jet is extracted and combined with the $Z$, to
reveal the FCNC decay of a top recoiling against the one with SM decay
($\cos\phi(t\bar t)<0$). The reconstructed FCNC top invariant mass
distributions for both channels are shown in
Fig.\ref{top:fcnc:fig:CMS_results}.

\begin{figure}[tb]
	\centering
\includegraphics[height=3.5cm,width=7.9cm]{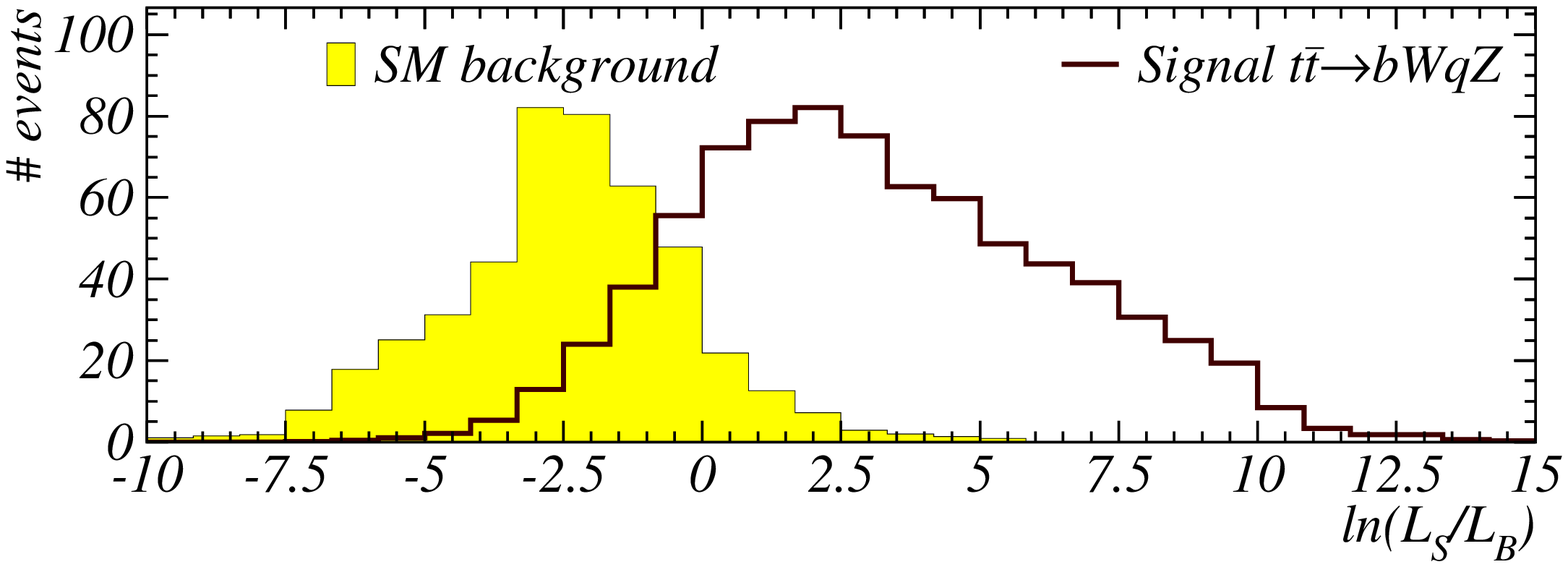} 
\hspace{0mm}
\includegraphics[height=3.5cm,width=7.9cm]{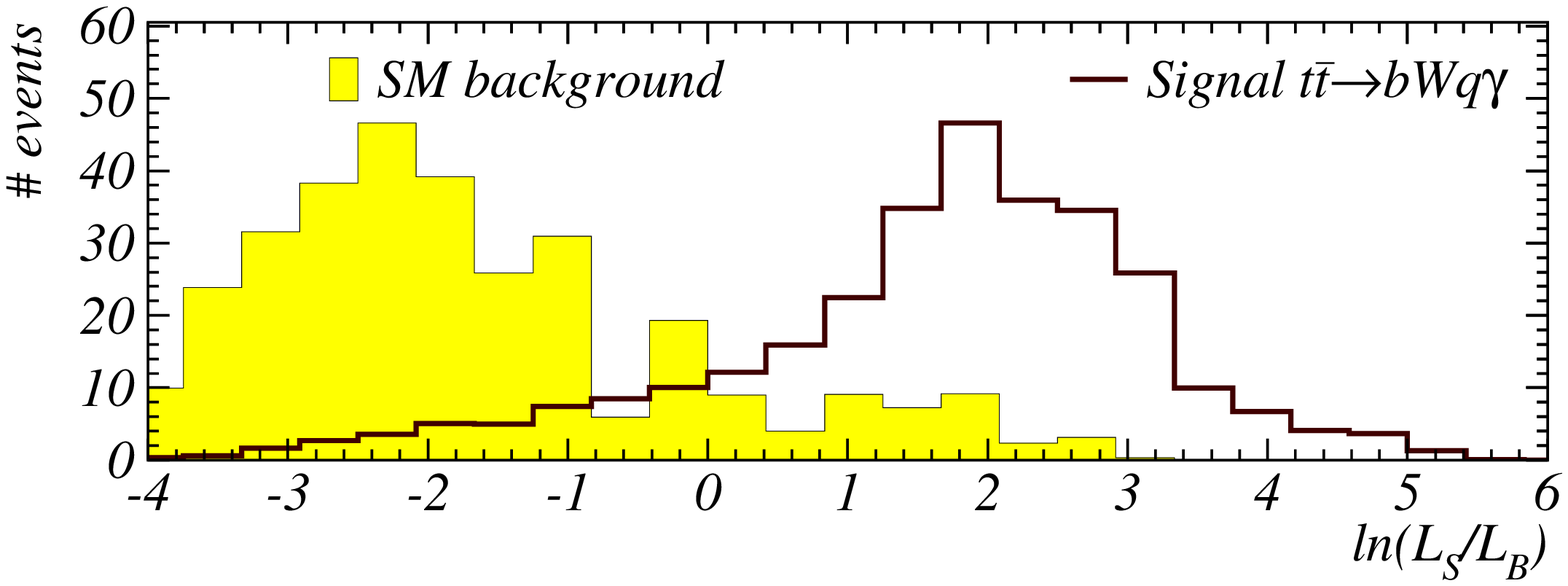}
\caption{Signal and background likelihood ratios, $L_R=\ln(L_S/L_B)$, obtained 
in 
ATLAS 
analysis 
for the $t\to qZ$ (left) and $t\to q\gamma$ (right) channels. The SM 
background (shadow region) is normalized to $L=10$~fb$^{-1}$ and the signal 
(line) is shown with 
arbitrary normalization.}
\label{top:fcnc:fig:ATLAS_results} 
\end{figure}

The ATLAS collaboration has developed a probabilistic analysis for each
of the considered top FCNC decay channels. In the $t\to qZ$ channel,
preselected events with a reconstructed Z, large missing transverse
energy and the two highest $p_T$ jets (one $b$-tagged) are used to build
a discriminant variable (likelihood ratio) $L_R=\ln(\Pi_{i=1}^n P_i^S /
\Pi_{i=1}^n P_i^B)$, where $P_i^{B(S)}$ are the signal and background
p.d.f., evaluated from the following physical distributions: the minimum
invariant mass of the three possible combinations of two leptons (only
the three highest $p_T$ leptons were considered); the transverse
momentum of the third lepton (with the leptons ordered by decreasing
$p_T$) and the transverse momentum of the most energetic non-$b$ jet.
The discriminant variables obtained for FCNC signal and the SM
background are shown in Fig.\ref{top:fcnc:fig:ATLAS_results} (left). For
the $t\to q\gamma$ channel, preselected events are required to have one
$b$-tag (amongst the two highest $p_T$ jets) and at least one photon
with transverse momentum above 75 GeV. For this channel, the likelihood
ratio is built using the p.d.f. based on the following variables:  
invariant mass of the leading photon and the non-$b$ jet; transverse
momentum of the leading photon and the number of jets. The signal and
background discriminant variables are shown in
Fig.~\ref{top:fcnc:fig:ATLAS_results} (right). For comparison with the
CMS sequential analysis, a cut on the discriminant variable
(corresponding to the best $S/\sqrt{B}$) is applied.

Once the signal efficiency ($\epsilon_S$) and the number of selected
background events ($B$) have been obtained, $BR$ sensitivities for a 
signal discovery corresponding to a
given significance can be evaluated. 
Table~\ref{top:fcnc:tab:ATLAS-CMS_Results}
reports the results of the two experiments, assuming an integrated luminosity of 
10 fb$^{-1}$, $5\sigma$ discovery level and the statistical significance ${\cal S}=2(\sqrt{B+S}-\sqrt{B})$
(a different definition for ${\cal S}$ can be found in Ref.~\cite{Carvalho:973137}).

\begin{table}[tb]
  \begin{center} 
  \caption{ATLAS and CMS results for described analysis: 
efficiency, SM background and expected branching ratios for top FCNC decays, 
assuming a $5\sigma$ significance discovery ($L=10$~fb$^{-1}$).}
    \begin{tabular}{|l||c|c|c||c|c|c|} 
    \hline
   & \multicolumn{3}{c||}{$t\to qZ$ } & \multicolumn{3}{c|}{$t\to q\gamma$ }\\ 
    \hline
     &  $\epsilon_S$ & $B$ & $BR$ ($5\sigma$) &  $\epsilon_S$ & $B$ & $BR$ ($5\sigma$)\\
     \hline
     ATLAS & 1.30\% & 0.37 & $13.0 \times 10^{-4}$ & 1.75\% & 3.13 & $1.6 \times 10^{-4}$\\
     CMS   & 4.12\% & 1.0  & $11.4 \times 10^{-4}$ & 2.12\% & 54.6 & $5.7 \times 10^{-4}$ \\
    \hline
    \end{tabular}
    \label{top:fcnc:tab:ATLAS-CMS_Results}
  \end{center}
\end{table}

\begin{figure}[tb]
	\centering
\includegraphics[width=10cm]{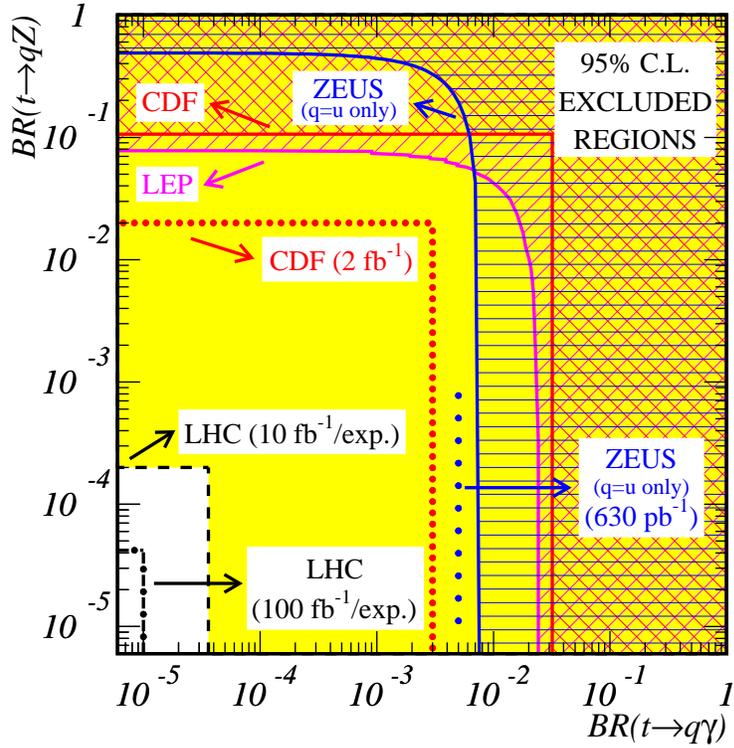} 
  \caption{The present 95\% CL limits on the $BR(t\to q\gamma)$
  versus $BR(t \to qZ)$ plane are 
  shown~\cite{CDF8888, Abe:1997fz, unknown:2003ih, Chekanov:2003yt}. The 
  expected sensitivity at the HERA  
  ($L=630$~pb$^{-1}$)~\cite{Chekanov:2003yt}, 
  Tevatron (run 
  II)~\cite{Juste:2006bs} and 
  LHC (ATLAS and CMS preliminary combination) is also represented.}
\label{top:fcnc:fig:BR_limits} 
\end{figure}

Having these two independent analyses, a preliminary combination of ATLAS
and CMS results was performed, in order to estimate the possible LHC
sensitivity to top FCNC decays. As a first attempt, the Modified
Frequentist Likelihood Method (see for example Ref.~\cite{Read:2000ru}) is
used to combine the expected sensitivity to top FCNC decays from both
experiments under the hypothesis of signal absence\footnote{For the CMS
analysis a counting experiment is used, while for the ATLAS analysis the
full shape of the discriminant variables was also taken into account.} and
an extrapolation to the high luminosity phase (100 fb$^{-1}$) is performed.
These results are showed in Table~\ref{top:fcnc:tab:ATLAS-CMS_Combination}
and indicate that a sensitivity at the level of the predictions of some new
physics models (such as SUSY) can be achieved.  The comparison with the
current experimental limits is also shown in
Fig.~\ref{top:fcnc:fig:BR_limits}. As showed, a significant improvement on
the present limits for top FCNC decays is expected at the LHC. Both
collaborations have plans to assess in detail the impact of systematic
uncertainties and improve the understanding of the detectors through
updated simulation tools. Preliminary results indicate that the effect of
theoretical systematics (as top mass, $\sigma(t\bar t)$ and parton
distribution functions) and experimental ones (such as jet/lepton energy
scale and $b$-tagging) have an impact on the limits smaller than 30\%{}. 
Thus, the order of magnitude of the results is not expected to change.

\begin{table}[bt]
\centering
    \caption{LHC 95\% CL expected limits on $t\to qZ$ 
and $t\to q\gamma$ branching ratios (ATLAS and CMS 
preliminary combination under the hypothesis of signal absence).}
    \begin{tabular}{|l||c||c|} 
    \hline
     luminosity & $BR(t\to q Z )$ & $BR(t\to q\gamma)$ \\
         \hline
     10 fb$^{-1}$ & $2.0 \times 10^{-4}$ & $3.6 \times 10^{-5}$\\
     100 fb$^{-1}$ & $4.2 \times 10^{-5}$ & $1.0 \times 10^{-5}$ \\
    \hline
    \end{tabular}
    \label{top:fcnc:tab:ATLAS-CMS_Combination}
\end{table}

\begin{figure}[tb]
	\centering
\includegraphics[height=3.5cm,width=10.2cm]{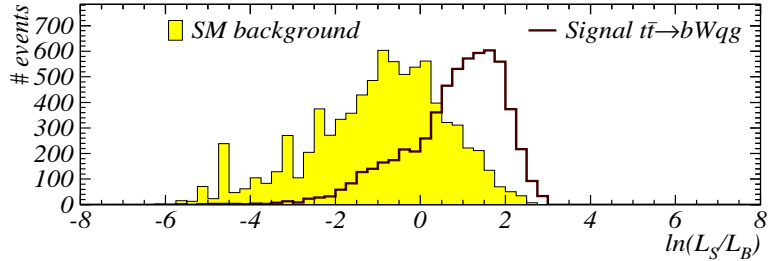} 
\caption{Signal and background likelihood ratios obtained in ATLAS analysis 
for the $t\to qg$ channel. The SM 
background (shadow region) is normalized to $L=10$~fb$^{-1}$ and the signal 
(line) is shown with 
arbitrary normalization.}
\label{top:fcnc:fig:ATLAS_qgluon} 
\end{figure}

A study of the ATLAS sensitivity to FCNC $t\to qg$ decay was also presented
in Ref.~\cite{Carvalho:973137}. In this analysis, the $t\bar t$ production
is considered, with one of the top quarks decaying into $qg$ and the other
decays through the SM decay $t\to bW$. Only the leptonic decays of the $W$
were taken into account, otherwise the final state would be fully hadronic
and the signal would be overwhelmed by the QCD background. This final state
is characterised by the presence of a high $p_T$ gluon and a light jet from
the FCNC decay, a $b$-tagged quark, one lepton and missing transverse
momentum from the SM decay. As in this topology the FCNC top decay
corresponds to a fully hadronic final state, a more restrictive event
selection is necessary. As for the $qZ$ and $q\gamma$ channels, a
probabilistic type of analysis is adopted, using the following variables to
build the p.d.f.: the invariant mass of the two non-$b$ jets with highest
$p_T$; the $b\ell\nu$ invariant mass; the transverse momenta of the $b$-jet
and of the second highest $p_T$ non-$b$ jet and the angle between the
lepton and the leading non-$b$ jet. The discriminant variables obtained for
signal and background are shown in Fig.~\ref{top:fcnc:fig:ATLAS_qgluon}.
The expected 95\% CL limit on $BR(t\to qg)$ for $L=10$~fb$^{-1}$ for
$L=10$~fb$^{-1}$ was found to be $1.3 \times 10^{-3}$. A significant
improvement on this limit should be achieved by combining the results from
$t\bar t$ production (with $t\to qg$ FCNC decay) and single top production
(see~section~\ref{top:effectivelagrangianFCNC}).

\section{New physics corrections to top quark production 
\label{top:corr_xsec}}

It is generally believed that the top quark, due to its large mass, can
be more sensitive to new physics beyond the SM than other fermions. In
particular, new processes contributing to $t \bar t$ and single top
production may be relevant. Single top processes are expected to be
sensitive to some SM extensions, such as SUSY. Another characteristic
new process could be the production in $pp$ collisions of an
$s$-channel resonance decaying to $t \bar t$. Examples of this
resonance are: (i) a spin-1 leptophobic $Z'$ boson, which would be
undetectable in leptonic decay channels; (ii) Kaluza-Klein (KK)
excitations of gluons or gravitons; (iii) neutral scalars. If these
resonances are narrow they could be visible as a mass peak over the SM
$t \bar t$ background. In such case, the analysis of $t$, $\bar t$
polarisations (in a suitable window around the peak) could provide
essential information about the spin of the resonance. If the resonance
is broad, perhaps the only way to detect it could be a deviation in $t
\bar t$ spin correlations with respect to the SM prediction. More
generally, new contributions to $t \bar t$ production which do not
involve the exchange of a new particle in the $s$ channel (including,
but not limited to, those mediated by anomalous couplings to the gluon)
do not show up as an invariant mass peak. In this case, the analysis of
the measurement of spin correlations might provide the only way to
detect new physics in $t \bar t$ production.

\subsection{Potential complementary MSSM test in single top production}

\begin{figure}[tbp]
\begin{center}
\hspace*{-20mm}
\epsfig{file=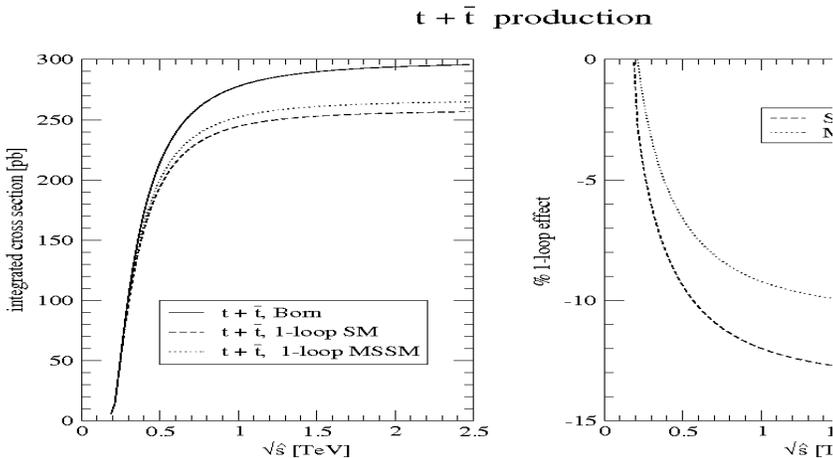, width=11cm, height=6cm}
\end{center}
\caption{Integrated cross sections for the overall $t$-channel production of a single top or antitop quark.}
\label{top:mssm_corr_fig1}
\end{figure}

At LHC, it will be possible to perform measurements of the rates of the
three different single top production processes, usually defined as
$t$-channel, associated $tW$ and $s$-channel production, with an
experimental accuracy that varies with the process.  From the most recent
analyses one expects, qualitatively, a precision of the order of 10\% for
the $t$-channel~\cite{Pumplin:2002vw}, and worse accuracies for the two
remaining processes. Numerically, the cross section of the $t$-channel is
the largest one, reaching a value of approximately 250
pb~\cite{Mahlon:1999gz}; for the associated production and the
$s$-channel one expects a value of approximately 60 pb and 10
pb~\cite{Zhang:2006cx} respectively.  For all the processes, the SM NLO
QCD effect has been computed~\cite{Campbell:2004ch,Campbell:2005bb}, and
quite recently also the SUSY QCD contribution has been
evaluated~\cite{Zhang:2006cx}.  Roughly, one finds for the $t$-channel a
relative $\sim 6$ \% SM QCD effect and a negligible SUSY QCD component;
for the associated $tW$ production a relative $\sim 10$ \% SM QCD and a
relative $\sim 6$ \% SUSY QCD effect; for the $s$-channel, a relative
$\sim$ 50 \% SM QCD and a negligible SUSY QCD component. As a result of
the mentioned calculations, one knows the relative NLO effects of
both SM and SUSY QCD. The missing part is the NLO electroweak effect. 
This has been computed for the two most relevant processes, i.e. the
$t$-channel and the associated production. The NLO calculation for the
$s$-channel is, probably, redundant given the small size of the related
cross section. It is, in any case, in progress. In this section some of
the results of the complete one-loop calculation of the electroweak effects
in the MSSM are shown for the two processes.  More precisely, eight
different $t$-channel processes (four for single top and four for single
antitop production) were considered.  These processes are defined in
Ref.~\cite{Beccaria:2006jt}. For the associated production, the process
$bg\to tW^-$ (the rate of the second process $\bar b g\to \bar t W^+$ is
the same) was considered~\cite{Beccaria:2006dt}. These calculations have
been performed using the program {\tt LEONE}, which passed three severe
consistency tests described in
Refs.~\cite{Beccaria:2006jt,Beccaria:2006dt}. For the aim of this
preliminary discussion, in this section only the obtained values of the
integrated cross sections are shown, ignoring the (known) QCD effects.
The integration has been performed from threshold to the effective centre
of mass energy ($\sqrt{\hat s}$), allowed to vary up to a reasonable
upper limit of approximately 1 TeV. Other informations are contained in
Refs.~\cite{Beccaria:2006jt,Beccaria:2006dt}.

\begin{figure}[bt]
\begin{center}
\hspace*{-15mm}
\epsfig{file=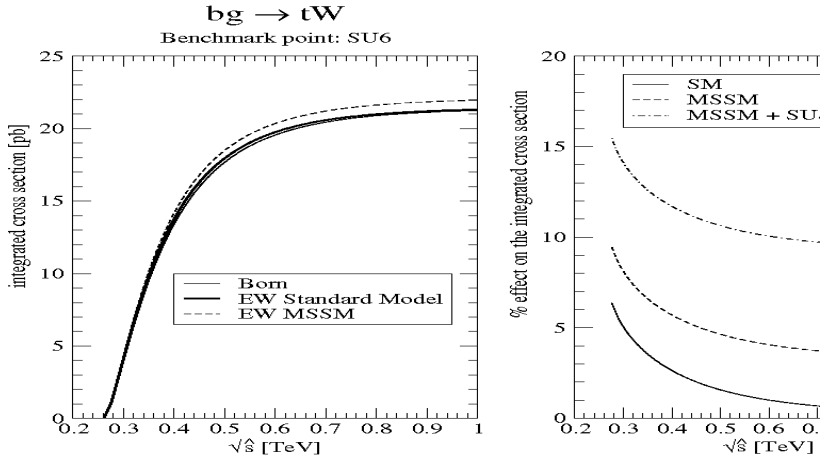, width=10.8cm, height=6cm}
\end{center}
\caption{Integrated cross sections for the associated production of a a single top quark.}
\label{top:mssm_corr_fig2}
\end{figure}

Figs.~\ref{top:mssm_corr_fig1} and \ref{top:mssm_corr_fig2} show the
obtained numerical results. In Fig.~\ref{top:mssm_corr_fig2} (right) the
discussed NLO electroweak effect was added the NLO SUSY QCD effect taken
from Ref.~\cite{Zhang:2006cx}.  From the figures the following main
conclusions can be drawn:
\begin{enumerate} 
\item The genuine SUSY effect in the $t$-channel is modest. In the most 
favourable
case, corresponding to the ATLAS DC2 point SU6~\cite{atlasdc2points}, it
reaches a value of approximately two percent.

\item The one-loop electroweak SM effect in the $t$-channel rate is large
($\sim$ 13 \%). It is definitely larger than the NLO SM QCD effect.  Its
inclusion in any meaningful computational program appears to be
mandatory.

\item The genuine SUSY effect in the associated production, if one limits
the cross section observation to relatively low (and experimentally safe
from $t\bar t$ background) energies (400-500 GeV), can be sizable. In the
SU6 point, the combined (same sign) SUSY QCD and electroweak effects can
reach a relative ten percent effect.

\item The pure electroweak SM effect in the associated production is negligible.
\end{enumerate}
From the previous remarks, one can reach the final statement that, for what
concerns the virtual NLO effects of the MSSM, the two processes $t$-channel
and associated production appear to be, essentially, complementary. In this
spirit, a separate experimental determination of the two rates might lead to
non trivial tests of the model.

\subsection{Anomalous single-top production in
warped extra dimensions}

Randall and Sundrum have proposed the use of a non-factorizable geometry
in five dimensions~\cite{Randall:1999ee} as a solution of the hierarchy
problem.  The extra dimension is compactified on an orbifold $S_1/Z_2$ of
radius $r$ so that the bulk is a slice of Anti- de Sitter 
space between two
four-dimensional boundaries. The metric depends on the five dimensional
coordinate $y$ and is given by
\begin{equation}
ds^2 = e^{-2\sigma(y)} \eta_{\mu\nu} dx^\mu dx^\nu - dy^2~,
\label{top:metric}
\end{equation} 
where $x^\mu$ are the four dimensional coordinates, $\sigma(y) = k |y|$,
with $k\sim M_P$ characterizing the curvature scale.  This metric
generates two effective scales: $M_P$ and $M_P e^{-k\pi r}$. In this way,
values of $r$ not much larger than the Planck length ($kr\simeq (11-12)$)
can be used in order to generate a scale $\Lambda_r\simeq M_Pe^{-k\pi
r}\simeq~{\rm O(TeV)}$ on one of the boundaries.

In the original Randall-Sundrum (RS) scenario, only gravity was allowed
to propagate in the bulk, with the SM fields confined to one of the
boundaries.  The inclusion of matter and gauge fields in the bulk has
been extensively treated in the literature~\cite{Goldberger:1999wh,
Pomarol:1999ad, Chang:1999nh, Grossman:1999ra, Gherghetta:2000qt,
Huber:2000ie, Davoudiasl:2000wi, Hewett:2002fe}.  The Higgs field must
be localized on or around the TeV brane in order to generate the weak
scale.  As it was recognized in Ref.~\cite{Gherghetta:2000qt}, it is
possible to generate the fermion mass hierarchy from $O(1)$ flavour
breaking in the bulk masses of fermions.  Since bulk fermion masses
result in the localization of fermion zero-modes, lighter fermions
should be localized toward the Planck brane, where their wave-functions
have an exponentially suppressed overlap with the TeV-localized Higgs,
whereas fermions with order one Yukawa couplings should be localized
toward the TeV brane.  This constitutes a theory of fermion masses, and
it has a distinct experimental signal at the LHC, as discussed below. 

Since the lightest KK excitations of gauge bosons are localized toward
the TeV brane, they tend to be strongly coupled to zero-mode fermions
localized there. Thus, the flavour-breaking fermion localization leads to
flavour-violating interactions of the KK gauge bosons, particularly with
third generation quarks. For instance, the first KK excitation of the
gluon, will have flavour-violating neutral couplings such as
$G_\mu^{a(1)}(t\gamma^\mu T^a\bar q)$, where $q=u,c$.

In this section, results of a study of the flavour-violating signals of
the top at the LHC are presented, following the work described in
Ref.~\cite{Aquino:2006vp}. The
localization of fermions in the extra dimension, and therefore their 4D
masses and their couplings to the KK gauge bosons, is determined by their
bulk masses. We choose a range of parameters that is consistent with the
observed fermion masses and quark mixing, as well as low energy flavour
and electroweak constraints. The implications for low energy flavour
physics were considered in
Refs.~\cite{Burdman:2002gr,Burdman:2003nt,Agashe:2004ay}.  The bulk
masses of the third generation quark doublet is fixed, as well as that of
the right-handed top. The following ranges were considered: $c_L^3=
[0.3,0.4]$ and $c_R^t=[-0.4,0.1]$, where the fermion bulk masses
$c^f_{L,R}$ are expressed in units of the inverse AdS radius $k$.  Since
the latter is of the order of the Planck scale, the fermion bulk mass
parameters must be naturally of order one. 

The only couplings that are non-universal in practice are those of the
$t_R$, $t_L$ and $b_L$ with the KK gauge bosons. All other fermions,
including the right-handed b quark must have localizations toward the
Planck brane in order to get their small masses.  The non-universality
of the KK gauge boson couplings leads to tree-level flavour violation.
The diagonalization of the quark mass matrix requires a change of basis
for the quarks fields. In the SM, this rotation leads to the CKM matrix
in the charged current, but the universality of the gauge interactions
results in the GIM mechanism in the neutral currents. However, since
the KK excitations of the gauge bosons are non-universal, tree-level
GIM-violating couplings will appear in the physical quark basis. 

The dominant non-universal effect is considered as coming from the
couplings of $t_R$, $t_L$ and $b_L$ to the first KK excitation of the
gluon: $g_{t_R}$, $g_{t_L}$ and $g_{b_L}$ respectively.  The $SU(2)_L$
bulk symmetry implies $g_{t_L}=g_{b_L}$.  For the considered range of
$c_L^3$ and $c_R^t$, the following results were obtained: 
\begin{equation}
g_{t_L} =
g_{b_L} = [1.0,2.8]\,g_s
\end{equation}
and 
\begin{equation}
g_{t_R} = [1.5,5]\, g_s~,
\end{equation} 
where $g_s$ is
the usual 4D $SU(3)_c$ coupling.  The light quarks as well as the
right-handed b quark have 
\begin{equation}
g_L^q = g_R^q = g_R^b \simeq -0.2\, g_s~,
\end{equation}
so they are, in practice, universally coupled, as mentioned above. 

Computing the width of the intermediate KK gluon with the range of
couplings obtained above, results in a range of $\Gamma_{\rm min.}\simeq
0.04\,M_G$ and $\Gamma_{\rm max.} \simeq 0.35\,M_G$. Then, it can be seen
that the range of values for the couplings allow for rather narrow or
rather broad resonances, two very different scenarios from the point of
view of the phenomenology. This strong coupling of the KK gluon to the
top, will also produce a $t\bar t$ resonance. Here we concentrate on the
flavour-violating signal, since the presence of a $t\bar t$ resonance
will not constitute proof of the flavour theory due to the difficulty in
identifying resonances in the light quark channels.

In the quark mass eigen-basis the left-handed up-type quarks couple to
the KK gluon through the following currents: $U_L^{tt}\,(\bar t_L T^a
\gamma_\mu t_L)$, $U_L^{tc}\,(\bar t_L T^a \gamma_\mu c_L)$ and
$U_L^{tu}\,(\bar t_L T^a \gamma_\mu u_L)$.  Similarly, the right-handed
up-type quarks couple through $U_R^{tt}\,(\bar t_R T^a\gamma_\mu t_R)$,
$U_R^{tc}\,(\bar t_R T^a \gamma_\mu c_R)$ and $U_R^{tu}\,(\bar t_R
T^a\gamma_\mu u_R)$. Here, $U_L$ and $U_R$ are the left-handed and
right-handed up-type quark rotation matrices responsible for the
diagonalization of the Yukawa couplings of the up-type quarks. In what
follows 
\begin{equation}
U_L^{tu} \simeq V_{ub}\simeq 0.004~, 
\end{equation}
will be conservatively 
assumed, and
$U_R^{tc}$ and $U_R^{tu}$ will be taken as free parameters. Since no
separation of charm from light jets is assumed, we define
\begin{equation}
U_R^{tq}\equiv
\sqrt{(U_R^{tc})^2 + (U_R^{tu})^2}~, 
\end{equation} 
and the sensitivity of
the LHC to this parameter for a given KK gluon mass is studied.

These flavour-violating interactions could be directly 
observed by the s-channel production of the first KK excitation of the gluon
with its subsequent decay to a top and a charm or up quark. For instance, at the 
LHC we could have the reaction 
\begin{equation}
p p\to G_\mu^{a(1)} \to t q~,
\label{top:singletop}
\end{equation}
with $q=u,c$. 
Thus, the Randall-Sundrum scenario with bulk matter predicts anomalous 
single top production at a very high invariant mass, which is determined by the 
mass of the KK gluon. 

In order to reduce the backgrounds, only the semi-leptonic decays of the 
top quarks were considered:
$   p p  \to t \bar{q}\; (\bar{t} q) \to b \ell^+ \nu_\ell \bar{q} \;
(\bar{b} \ell^- \bar\nu_\ell q) $, 
where $\ell =e$ or $\mu$, and $q=u,c$.  Therefore, this signal exhibits
one $b$-jet, one light jet, a charged lepton and missing transverse
energy. There are many SM backgrounds for this process. The dominant one
is $pp\to W^\pm jj \to \ell^\pm \nu jj$ where one of the light jets is
tagged as a $b$-jet. There is also $W^\pm b \bar{b} \to \ell^\pm \nu b
\bar{b}$ where one of the $b$-jets is mistagged; single top production
via $W-$gluon fusion and s-channel $W^*$, and $t\bar t$ production at
high invariant mass, mostly dominated by the flavour-conserving KK gluon
decays.

Initially, the following jet and lepton acceptance cuts were imposed: 
$p_T^j > 20 \hbox{ GeV}$, $| y_j | < 2.5$, 
$p_{T}^\ell \geq 20 \; {\rm GeV}$,  
$|y_\ell|\leq 2.5$, 
$\Delta R_{\ell j}\geq 0.63$, 
$\Delta R_{\ell \ell}\geq 0.63$,  
where $j$ can be either a light or a $b$-jet.
In order to further reduce the background the following additional cuts
were also imposed:

\begin{enumerate}
\item The invariant mass of the system formed by the lepton, the b
  tagged jet and the light jet was required to be within a window 
\begin{equation}
  M_{G^{(1)}} - \Delta \le M_{bj\ell} \le M_{G^{(1)}} + \Delta
\label{top:cuts2}
\end{equation} 
around the first KK excitation of the gluon mass. This cut ensures that
the selected events have large invariant masses, as required by the large
mass of the s-channel object being exchanged. The values of $\Delta$ used
in this study are presented in Table~\ref{top:t1}.

\item The transverse momentum of the light jet was required to be larger 
than
$p_{\rm cut}$, i.e.,
\begin{equation}
    p_{j~light} \ge p_{\rm cut}
\label{top:cuts3}
\end{equation}
Since the light jet in the signal recoils against the top forming with it a large invariant mass, 
it tends to be harder than the jets occurring in the background.
We present in Table~\ref{top:t1} the values for $p_{\rm cut}$ used in our 
analysis.

\item The invariant mass of the charged lepton 
and the $b$-tagged jet was also required to
be {\em smaller} than 250 GeV:
\begin{equation}
  M_{b\ell} \le 250 \hbox{ GeV}~.
\label{top:cuts4}
\end{equation}
This requirement is always passed by the signal, but eliminates a sizable
fraction of the $Wjj$ background. It substitutes for the full top
reconstruction when the neutrino momentum is inferred, which is not
used here. 
 
\end{enumerate}

\begin{table}
\begin{center}
\caption{Cuts used in the analysis (see text for details).}
\begin{tabular}{|c|c|c|}
\hline
$M_{G^{(1)}}$ (TeV)      &    $\Delta$ (GeV)   & $p_{\rm cut}$ (GeV)
\\
\hline
1     &     120   &  350
\\
\hline
2     &     250   &  650
\\
\hline
\end{tabular}
\label{top:t1}
\end{center}
\end{table}

In Table~\ref{top:t2} the cross sections for signal and backgrounds for
$M_{G}=1~$TeV and $2~$TeV are presented. The main sources of backgrounds
are $Wjj$ and $t\bar t$ production.  The signal is obtained for $U_R^{tq}
= 1$ and neglecting the contributions from left-handed final states,
corresponding to $U_L^{tq}=0$.  Regarding the choice of bulk masses,
these are fixed to obtain the minimum width which, as mentioned above,
can be as small as $\Gamma_G\simeq 0.04\, M_G$.~\footnote{The study of
broader resonances is left for future work.} 

\begin{table}[h]
\begin{center}
\caption{Signal and background cross sections for a KK gluon of $M_{G} = 1$~TeV and 
$2~$TeV, 
after the successive 
application of the cuts defined in (\ref{top:cuts2}), (\ref{top:cuts3}) 
and 
(\ref{top:cuts4}). 
  Efficiencies and b tagging probabilities are already included. 
$U_R^{tq}=1$ was used.}
\begin{tabular}{|c||c|c|c||c|c|c|}
\hline
 Process &  \multicolumn{3}{c||}{$M_G = 1~$TeV} & \multicolumn{3}{c|}{$M_G = 2~$TeV}
\\
\hline
& $\sigma ~-$  (\ref{top:cuts2}) & $\sigma ~-$ (\ref{top:cuts3})
& $\sigma ~-$ (\ref{top:cuts4})
& $\sigma ~-$  (\ref{top:cuts2}) & $\sigma ~-$ (\ref{top:cuts3})
& $\sigma ~-$ (\ref{top:cuts4})
\\
\hline
$pp\to tj$    &  148 fb   & 103 fb & 103 fb &  5.10 fb   & 2.18  fb& 2.18 fb
\\
\hline
$pp \to Wjj$                    &  243 fb   &  42.0 fb & 21.0 fb  & 25.4 fb   & 3.79 fb & 0.95 fb
\\
\hline
$pp \to Wbb$                  &  11.1 fb  & 4.07 fb & 3.19 fb & 0.97 fb & 0.45 fb & 0.06  fb
\\
\hline
$pp \to tb$                       &  1.53 fb  & 0.70 fb & 0.61 fb & 0.04 fb  & 0.02 fb & 0.02 fb
\\
\hline
$pp \to t \bar{t}$               &  44.4 fb  & 15.1 fb & 14.2 fb  & 1.60 fb  & 0.29 fb & 0.24 fb
\\
\hline
$Wg$ fusion                  &  32.0 fb  & 5.23 fb & 5.23 fb  & 1.20 fb  & 0.10 fb & 0.10 fb
\\
\hline
\end{tabular}
\label{top:t2}
\end{center}
\end{table}

\begin{table}[h]
\begin{center}
\caption{Reach in $U_R^{tq}$ for various integrated luminosities. }
\begin{tabular}{|c|c|c|c|}
\hline
$M_G~$ [TeV] & $30 fb^{-1}$ & $100 fb^{-1}$ & $300 fb^{-1}$
\\
\hline
1            & 0.24         & 0.18          & 0.14
\\
\hline
2            & 0.65         & 0.50          & 0.36
\\
\hline
\end{tabular}
\label{top:reach}
\end{center}
\end{table}

In order to evaluate the reach of the LHC, a significance of $5\,\sigma$
for the signal over the background is required. For a given KK gluon mass
and accumulated luminosity, this can be translated into a reach in the
flavour-violating parameter $U_R^{tq}$ defined above. This is shown in
Table~\ref{top:reach}.  It can be seen that the LHC will be sensitive to
tree-level flavour violation for KK gluon masses of up to at least
$2~$TeV, probing a very interesting region of values for $U_R^{tq}$. The
reach can be somewhat better if we allow for the reconstruction of the
momentum of the neutrino coming from the $W$ decay, which typically
reduces the $Wjj$ background more drastically.

Finally, we should point out that a very similar signal exists in Topcolor-assisted 
Technicolor~\cite{Hill:1994hp}, where the KK gluon is replaced by the Topgluon, 
which has FCNC interactions with the third generation quarks~\cite{Buchalla:1995dp}. 
The main difference between these two, is that the latter is typically a broad
resonance, whereas the KK gluon could be a rather narrow one, as it was shown above.

\subsection{Non-standard contributions to $t \bar t$ production 
\label{top:nonstandcont_tt}}

In $t \bar t$ events the top quarks are produced unpolarised at the tree level.
However, the $t$ and $\bar t$ spins
are strongly correlated, which allows to construct asymmetries using the
angular distributions of their decay products. These spin asymmetries
are dependent on the top spin. For the decay $t \to W^+ b \to \ell^+ \nu
b,q \bar q' b$, the angular distributions of $X=\ell^+,\nu,q,\bar
q',W^+,b$, in the top quark rest frame are given by
\begin{equation}
\frac{1}{\Gamma} \frac{d\Gamma}{d \cos \theta_X} = \frac{1}{2} (1+\alpha_X \cos
\theta_X )
\label{top:ec:tdist}
\end{equation}
with $\theta_X$ being the angle between the three-momentum of $X$ in the
$t$ rest frame and the top spin direction. In the SM the spin analysing
power ($\alpha_X$) of the top decay products are $\alpha_{\ell^+} =
\alpha_{\bar q'} = 1$, $\alpha_\nu = \alpha_q = -0.32$, $\alpha_{W^+} = -
\alpha_b = 0.41$ at the tree level~\cite{Jezabek:1994qs} ($q$ and $q'$
are the up- and down-type quarks, respectively, resulting from the $W$
decay). For the decay of a top antiquark the distributions are the same,
with $\alpha_{\bar X} = - \alpha_X$ as long as $CP$ is conserved in the
decay. One-loop corrections modify these values to $\alpha_{\ell^+} =
0.998$, $\alpha_{\bar q'} = 0.93$, $\alpha_\nu = -0.33$, $\alpha_q =
-0.31$, $\alpha_{W^+} = - \alpha_b = 0.39$~\cite{Czarnecki:1990pe,
Brandenburg:2002xr, Bernreuther:2004jv}. We point out that in the
presence of non-vanishing $V_R$, $g_L$ or $g_R$ couplings the numerical
values of the constants $\alpha_X$ are modified, but the functional form
of Eq.~(\ref{top:ec:tdist}) is maintained. We have explicitly calculated
them for a general $CP$-conserving $Wtb$ vertex within the narrow width
approximation. Explicit expressions can be found in
Ref.~\cite{Aguilar-Saavedra:2006fy}.  Working in the helicity basis the
double angular distribution of the decay products $X$ (from $t$) and
$\bar X'$ (from $\bar t$) can be written as a function of the relative
number of like helicity minus opposite helicity of the $t \bar t$ pairs
($C$)~\cite{Stelzer:1995gc} that measures the spin correlation between
the top quark and antiquark. Its actual value depends to some extent on
the PDFs used and the $Q^2$ scale at which they are evaluated. Using the
CTEQ5L PDFs~\cite{Lai:1999wy} and $Q^2 =\hat s$, (where $\hat s$ is the
partonic centre of mass energy), we find $C = 0.310$. At the one loop
level, $C = 0.326 \pm 0.012$~\cite{Bernreuther:2004jv}.

Using the spin analysers $X$, $\bar X'$ for the respective decays of $t$, 
$\bar t$, one can define the asymmetries
\begin{equation}
A_{X \bar X'} \equiv \frac{N(\cos \theta_X \cos \theta_{\bar X'} > 0) 
- N(\cos \theta_X \cos \theta_{\bar X'} < 0)}
{N(\cos \theta_X \cos \theta_{\bar X'} > 0) +
N(\cos \theta_X \cos \theta_{\bar X'} <0)} \,,
\end{equation}
whose theoretical value  is
\begin{equation}
A_{X \bar X'} = \frac{1}{4} C \alpha_X \alpha_{\bar X'} \,.
\label{top:ec:Ath}
\end{equation}
The angles $\theta_X$, $\theta_{\bar X'}$ are measured using as spin axis
the parent top (anti)quark momentum in the $t \bar t$ CM system. If $CP$
is conserved in the decay, for charge conjugate decay channels we have
$\alpha_{X'} \alpha_{\bar X} = \alpha_X \alpha_{\bar X'}$, so the
asymmetries $A_{X' \bar X} = A_{X \bar X'}$ are equivalent. Therefore, we
can sum both channels and drop the superscripts indicating the charge,
denoting the asymmetries by $A_{\ell \ell'}$, $A_{\nu \ell'}$, etc. In
semileptonic top decays we can select as spin analyser the charged
lepton, which has the largest spin analysing power, or the neutrino, as
proposed in Ref.~\cite{Jezabek:1994zv}. In hadronic decays the jets
corresponding to up- and down-type quarks are very difficult to
distinguish, and one possibility is to use the least energetic jet in the
top rest frame, which corresponds to the down-type quark 61\% of the
time, and has a spin analysing power $\alpha_j = 0.49$ at the tree level.
An equivalent possibility is to choose the $d$-jet by its angular
distribution in the $W^-$ rest frame~\cite{Mahlon:1995zn}. In both
hadronic and leptonic decays the $b$ ($\bar b$) quarks can be used as
well.

\begin{figure}[tb]
\begin{center}
\begin{tabular}{c c}
\epsfig{file=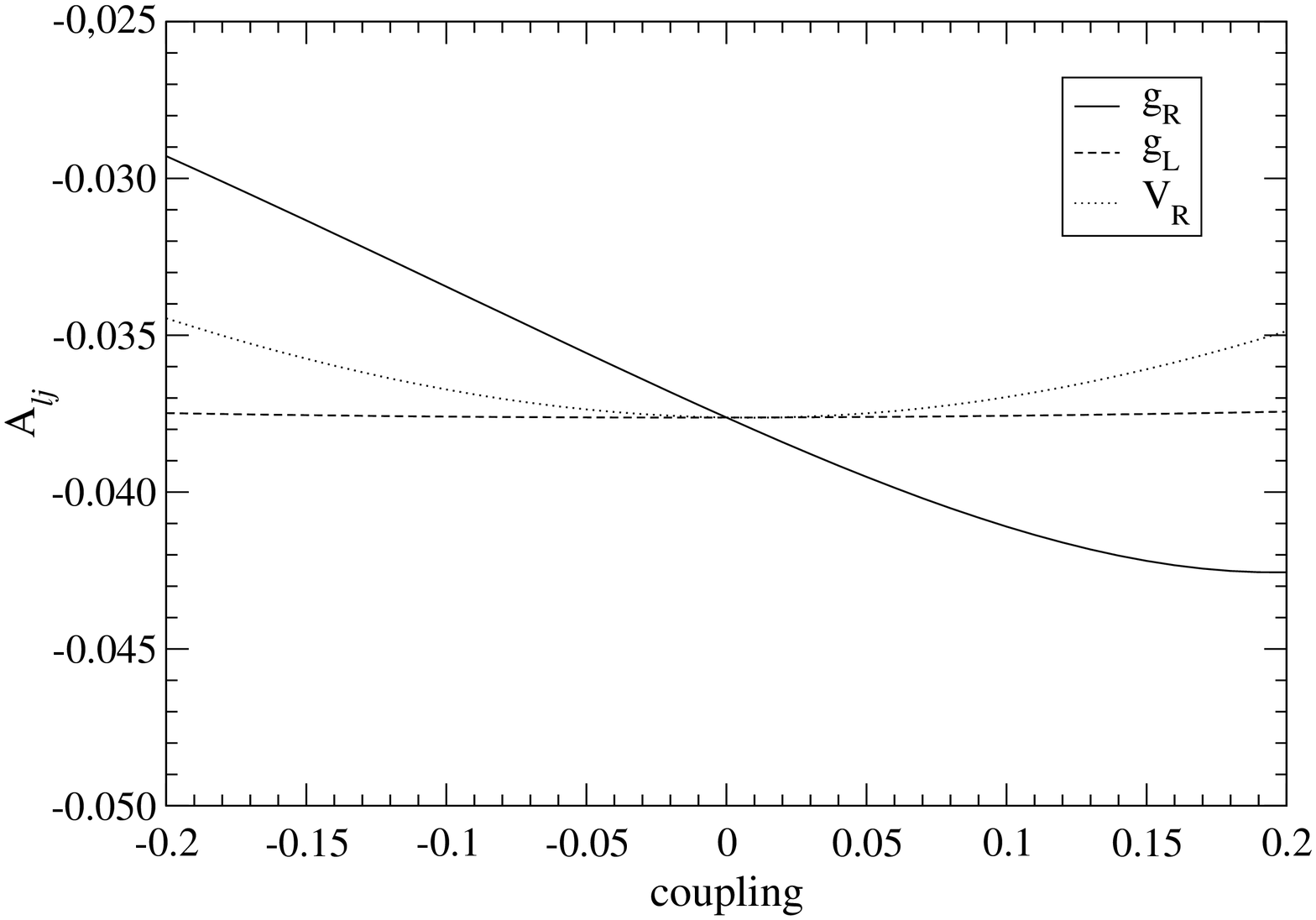,height=5.3cm,clip=} &
\epsfig{file=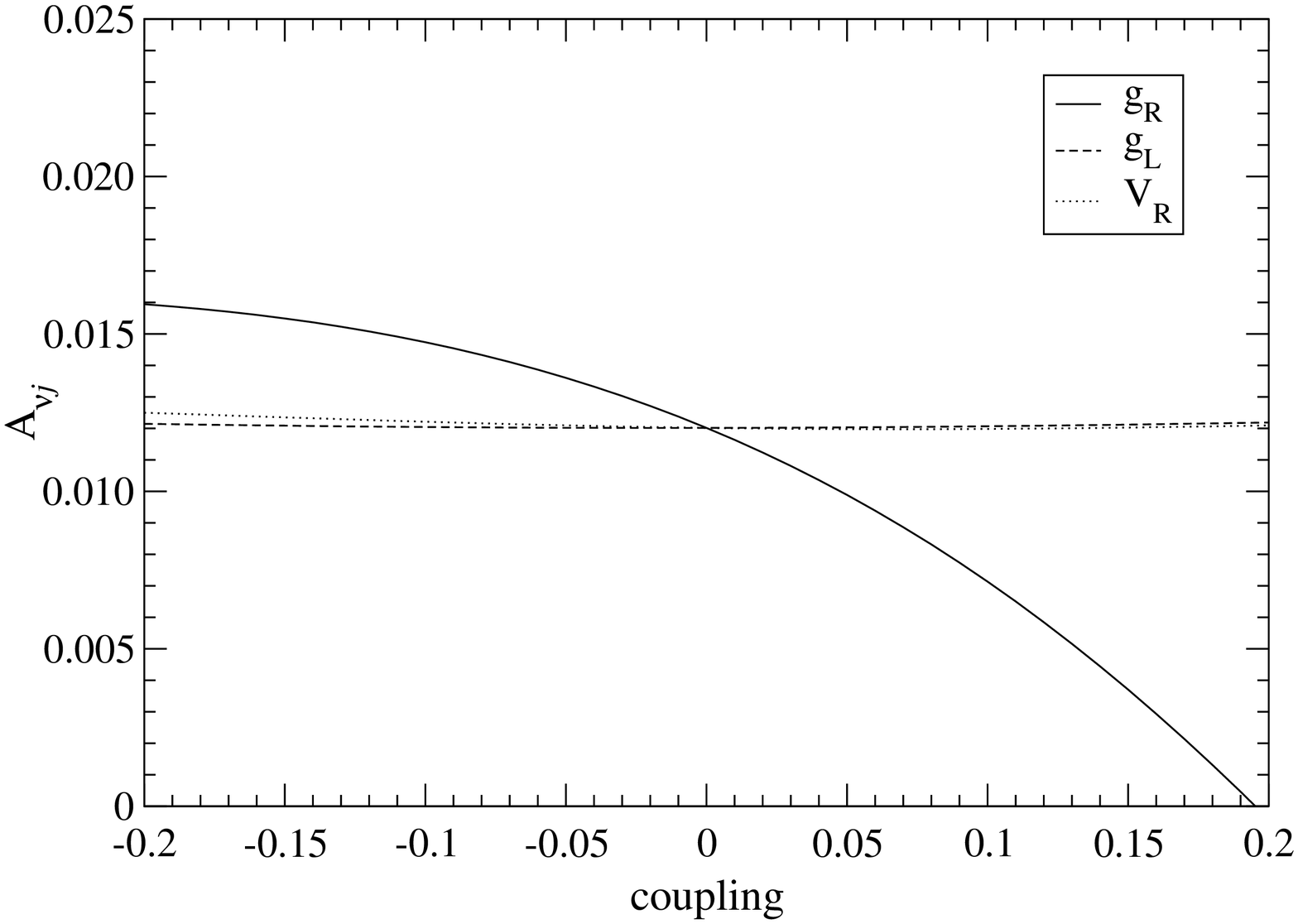,height=5.3cm,clip=} \\[0.5cm]
\epsfig{file=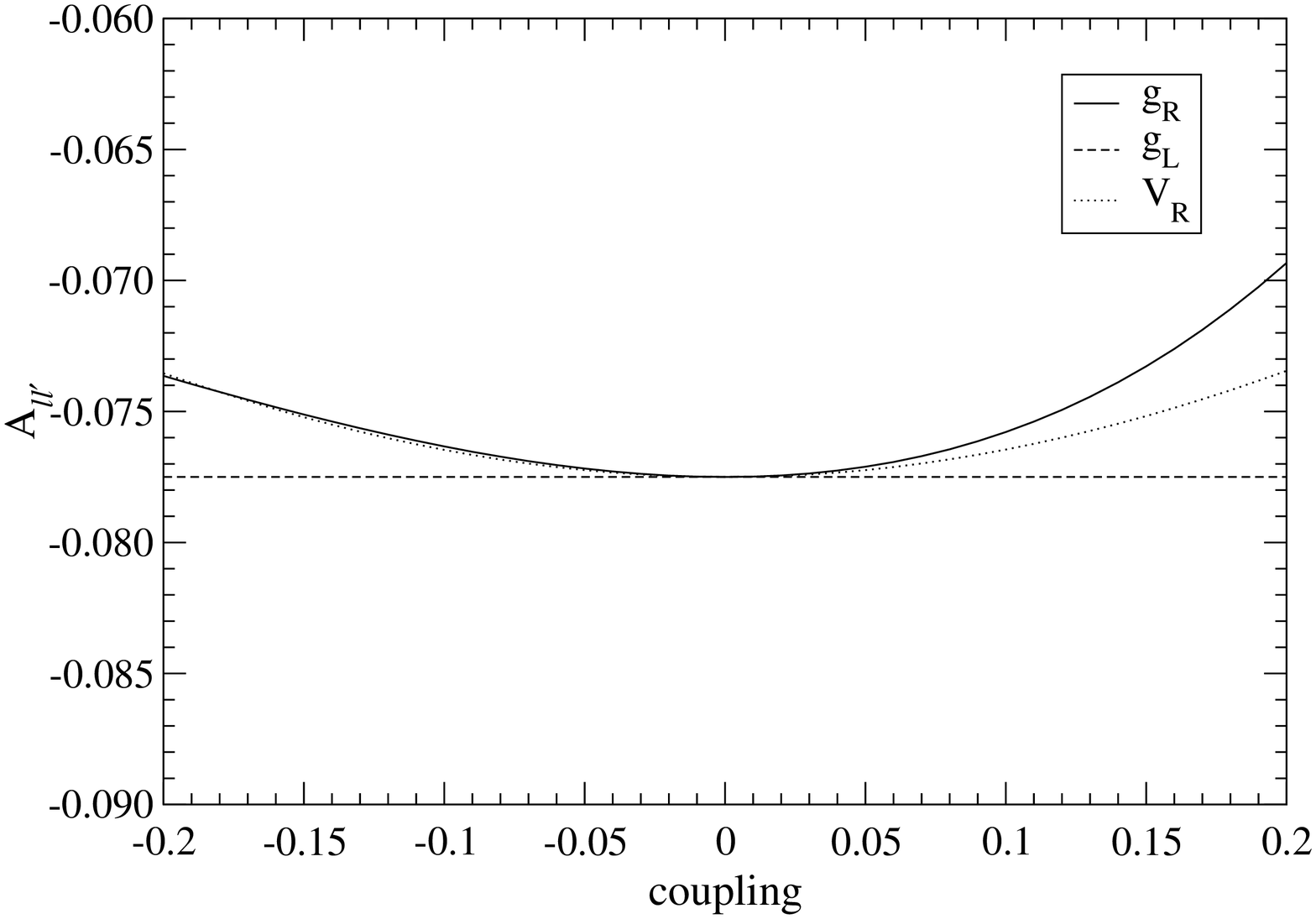,height=5.3cm,clip=} &
\epsfig{file=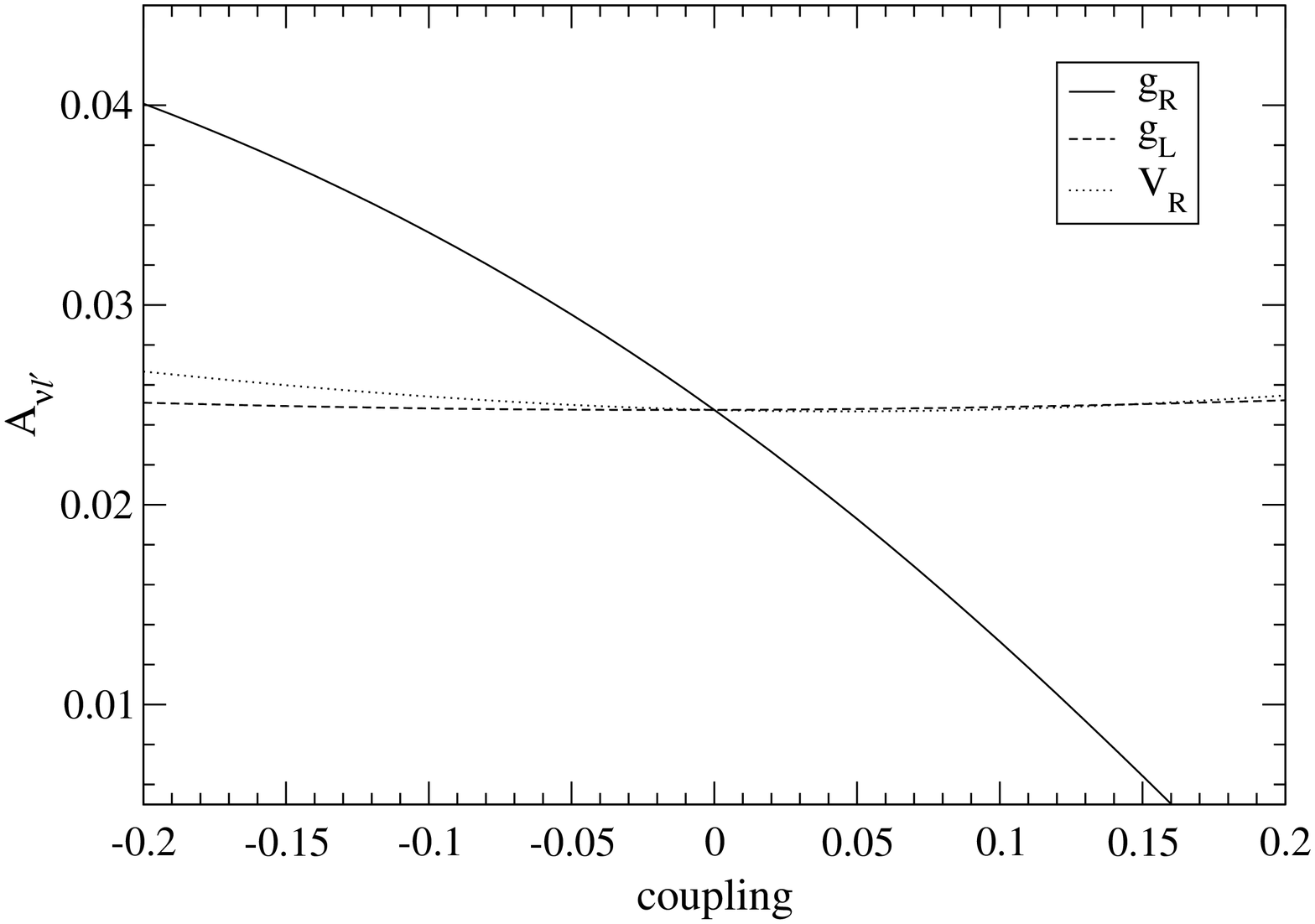,height=5.3cm,clip=} \\[0.5cm]
\epsfig{file=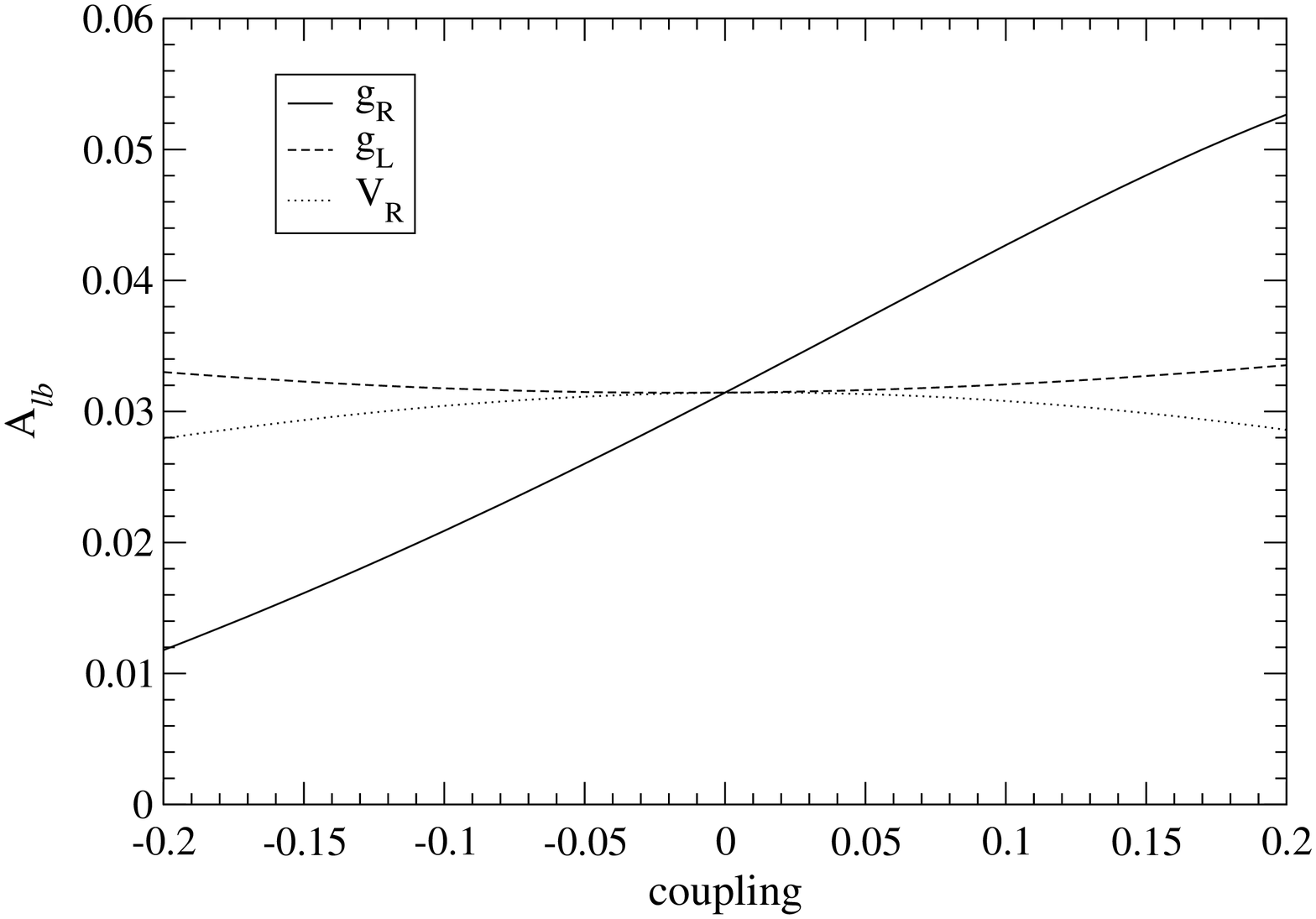,height=5.3cm,clip=} &
\epsfig{file=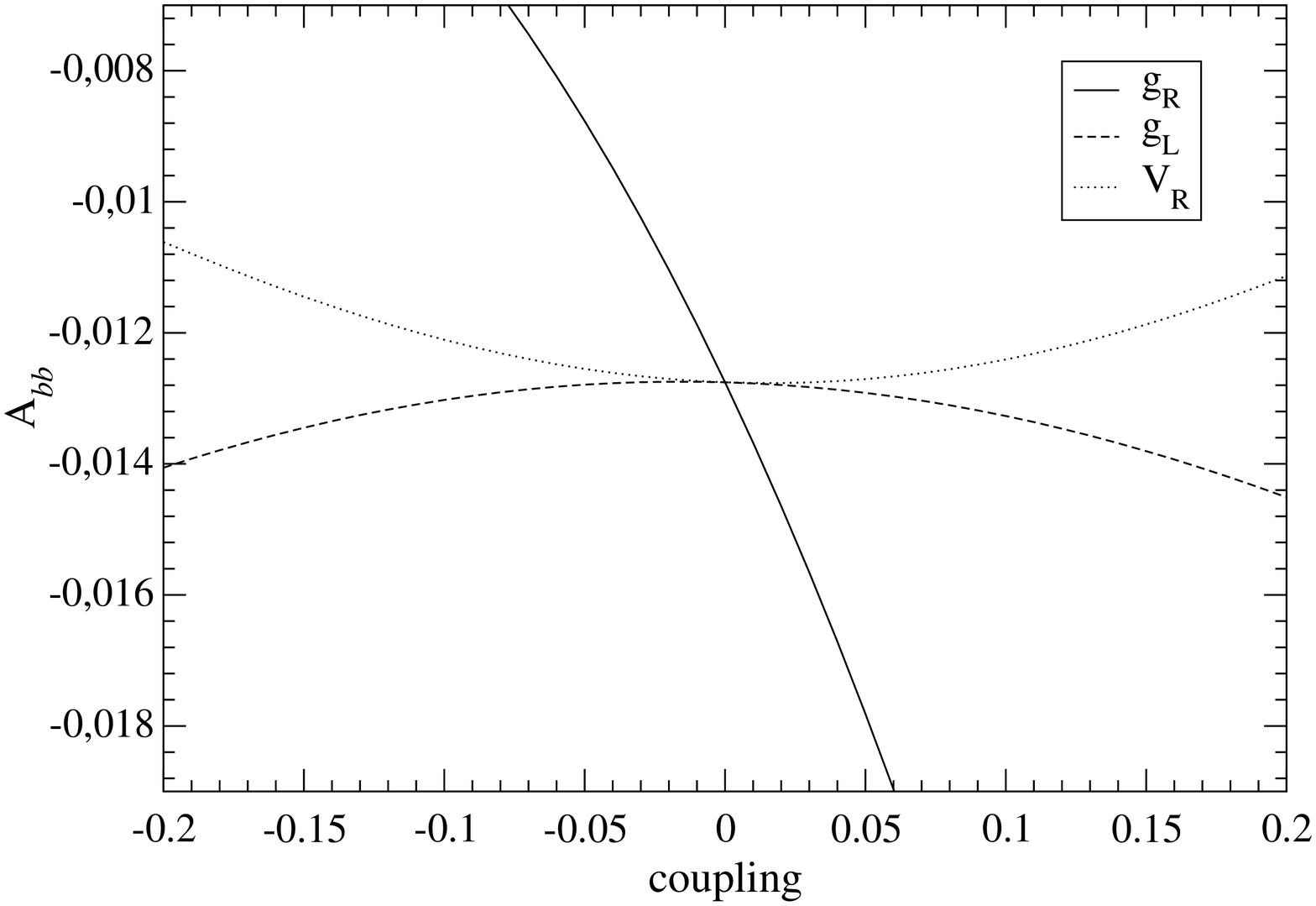,height=5.3cm,clip=} 
\end{tabular}
\caption{Dependence of several spin correlation asymmetries
on the couplings $g_R$, $g_L$ and $V_R$, for the $CP$-conserving case.}
\label{top:fig:aspin1}
\end{center}
\end{figure}

In the lepton $+$ jets decay mode of the $t \bar t$ pair, $t \bar t \to
\ell \nu b jj \bar b$ we choose the two asymmetries $A_{\ell j}$, $A_{\nu
j}$, for which we obtain the SM tree-level values $A_{\ell j} = -0.0376$,
$A_{\nu j} = 0.0120$. With the precision expected at
LHC~\cite{Hubaut:2005er, Aguilar-Saavedra:952732}, the measurements
$A_{\ell j} \simeq -0.0376 \pm 0.0058$, $A_{\nu j} \simeq 0.0120 \pm
0.0056$ are feasible ($L=10$~fb$^{-1}$).  The dependence of these
asymmetries on anomalous $Wtb$ couplings is depicted in
Fig.~\ref{top:fig:aspin1} from Ref.~\cite{Aguilar-Saavedra:2006fy}.  In
the di-lepton channel $t \bar t \to \ell \nu b \ell' \nu \bar b$ the
asymmetries $A_{\ell \ell'}$, $A_{\nu \ell'}$, whose SM values are
$A_{\ell \ell'} = -0.0775$, $A_{\nu \ell'} = 0.0247$, are selected. The
uncertainty in their measurement can be estimated from
Refs.~\cite{Hubaut:2005er, Aguilar-Saavedra:952732}, yielding $A_{\ell
\ell'} = -0.0775 \pm 0.0060$ and $A_{\nu \ell'} = 0.0247 \pm 0.0087$.
Their variation when anomalous couplings are present is shown in
Fig.~\ref{top:fig:aspin1}. We also plot the asymmetries $A_{lb}$,
$A_{bb}$, which can be measured either in the semileptonic or di-lepton
channel. Their SM values are $A_{lb} = 0.0314$, $A_{bb} = -0.0128$, but
the experimental sensitivity has not been estimated. It is expected
that it may be of the order of 10\% for $A_{lb}$, and worse for $A_{bb}$.
The determination of the correlation factor $C$ from these asymmetries
would eventually give
\begin{eqnarray}
A_{\ell \ell'} & \rightarrow & C = 0.310 \pm 0.024 ~\text{(exp)} 
 ~^{+0.}_{-0.0043} ~(\delta V_R) 
 ~^{+1 \times 10^{-5}}_{-3 \times 10^{-6}} ~(\delta g_L)
 ~^{+7 \times 10^{-6}}_{-0.0004} ~(\delta g_R)
 \,, \notag \\
A_{\ell j} & \rightarrow & C = 0.310 \pm 0.045 ~\text{(exp)} 
 ~^{+0.}_{-0.0068} ~(\delta V_R)
 ~^{+0.0001}_{-0.0008} ~(\delta g_L)
 ~^{+0.0004}_{-0.0009} ~(\delta g_R)
  \,.
\label{top:ec:Cresul}
\end{eqnarray}
The first error quoted corresponds to the experimental systematic and
statistical uncertainty. The other ones are theoretical uncertainties
obtained varying the anomalous couplings, one at a time. The confidence
level corresponding to the intervals quoted is 68.3\%. The numerical
comparison of the different terms in Eq.~(\ref{top:ec:Cresul}) also shows
that $A_{\ell j}$ and $A_{\ell \ell'}$ are much less sensitive to
non-standard top couplings than observables independent of the top spin
(see section~\ref{top:anomalouscoup}).

It is also interesting to study the relative distribution of one spin
analyser from the $t$ quark and other from the $\bar t$. Let $\varphi_{X
\bar X'}$ be the angle between the three-momentum of $X$ (in the $t$ rest
frame) and of $\bar X'$ (in the $\bar t$ rest frame). The angular
distribution can be written as~\cite{Bernreuther:2004jv}:
\begin{equation}
\frac{1}{\sigma} \frac{d\sigma}{d \cos \varphi_{X \bar X'}} =
\frac{1}{2} (1+D \, \alpha_X \alpha_{\bar X'} \cos \varphi_{X \bar X'}) \,,
\label{top:ec:sgldist}
\end{equation}
with $D$ a constant defined by this equality. From simulations, the
tree-level value $D = -0.217$ is obtained, while at one loop $D = -0.238$
\cite{Bernreuther:2004jv}, with a theoretical uncertainty of $\sim 4$\%. 
Corresponding to these distributions, the following asymmetries can be 
built:
\begin{equation} \tilde A_{X \bar X'} \equiv \frac{N(\cos \varphi_{X \bar
X'} > 0) - N(\cos \varphi_{X \bar X'} < 0)}{N(\cos \varphi_{X \bar X'} >
0) + N(\cos \varphi_{X \bar X'} < 0)} = \frac{1}{2} D \alpha_X
\alpha_{\bar X'} \,. 
\end{equation} 
For charge conjugate decay channels
the distributions can be summed, since $\alpha_{X'} \alpha_{\bar X} =
\alpha_X \alpha_{\bar X'}$ provided $CP$ is conserved in the decay. The
dependence of these asymmetries $\tilde A_{X \bar X'}$ on anomalous
couplings is (within the production $\times$ decay factorisation
approximation) exactly the same as for the asymmetries $A_{X \bar X'}$
defined above. Simulations are
available for $A_{\ell j}$ and $A_{\ell \ell'}$, whose theoretical SM
values are $A_{\ell j} = 0.0527$, $A_{\ell \ell'} = 0.1085$. The
experimental precision expected~\cite{Aguilar-Saavedra:952732,
Hubaut:2005er} is $A_{\ell j} \simeq 0.0554 \pm 0.0061$, $A_{\ell \ell'}
\simeq 0.1088 \pm 0.0056$. This precision is better than for $A_{\ell j}$
and $A_{\ell j}$, respectively, but still not competitive in the
determination of the $Wtb$ vertex structure.\footnote{Except for the 
case
of fine-tuned cancellations, see Ref.~\cite{Aguilar-Saavedra:2007rs}.}
Instead, we can use them to test top spin correlations. From these
asymmetries one can extract the value of $D$, obtaining
\begin{eqnarray}
A_{\ell \ell'} & \rightarrow & D = -0.217 \pm 0.011 ~\text{(exp)} 
 ~^{+0.0031}_{-0.} ~(\delta V_R) 
 ~^{+2 \times 10^{-6}}_{-8 \times 10^{-6}} ~(\delta g_L)
 ~^{+0.0003}_{-0.} ~(\delta g_R)
 \,, \notag \\
A_{\ell j} & \rightarrow & D = -0.217 \pm 0.024 ~\text{(exp)} 
 ~^{+0.0047}_{-0.} ~(\delta V_R) 
 ~^{+0.0006}_{-9 \times 10^{-6}} ~(\delta g_L)
 ~^{+0.0004}_{-6 \times 10^{-5}} ~(\delta g_R)  \,.
\label{top:ec:Dresul}
\end{eqnarray}
The errors quoted correspond to the experimental systematic and
statistical uncertainties, and the variation when one of the anomalous
couplings is allowed to be nonzero. From Eqs.~(\ref{top:ec:Cresul}) and
(\ref{top:ec:Dresul}) it is clear that the measurement of spin
correlations is a clean probe for new $t \bar t$ production processes,
independently of possible anomalous $Wtb$ couplings.  This is possible
because the sensitivity of spin correlation asymmetries to top anomalous
couplings is much weaker than for helicity fractions and related
observables, discussed in section~\ref{top:anomalouscoup}.

\chapter{Flavour violation in  supersymmetric models}
{\small M.~Klasen, N.~Krasnikov, T.~Lari, W.~Porod, and A.~Tricomi}
\label{chap:susy}

\section{Introduction}

The SM explains successfully the observed flavour violating phenomena
except that
for the observation in the neutrino sector one has to extend it
 by introducing either right-handed neutrinos or additional scalars. 
This implies that 
extensions of the SM with additional flavour structures are severely 
constrained by the wealth of existing data in the flavour sector.
Supersymmetry contains, as we will see below, various sources of additional
flavour structures. Therefore, the question arises if
there can still be large flavour violating effects in the production
and decays of supersymmetric particles 
despite the stringent existing constraints.

Every supersymmetric model is characterized by a K\"ahler potential,
the superpotential
$W$ and the corresponding soft SUSY breaking Lagrangian
(see e.g.~\cite{Martin:1997ns} and refs. therein). The first
describes the gauge interaction and the other two Yukawa interactions
and flavour violation. As the K\"ahler potential in general does not
contain flavour violating terms we will not discuss it further.
The most general superpotential containing only the SM fields and
being compatible with its gauge symmetry
 $G_{SM}=SU(3)_c\times SU(2)_L\times U(1)_Y$ is given  as
\cite{Weinberg:1981wj,Sakai:1981pk}:
\begin{eqnarray}
W &=& W_{MSSM} + W_{R_p \hspace{-3mm}/ } \,\,\,\, ,
\label{WG1:eq:superpot} \\
W_{MSSM} &=& h_{ij}^E \hat L_i \hat H_d \hat E^c_j
           + h_{ij}^D \hat Q_i \hat H_d \hat D^c_j
           + h_{ij}^U \hat H_u \hat Q_i \hat U^c_j
           - \mu \hat H_d \hat H_u \, ,
 \label{WG1:eq:mssmsuper} \\
W_{R_p \hspace{-3mm}/} &=& \frac{1}{2} \lambda_{ijk} \hat L_i \hat L_j \hat E^c_k
         + \lambda'_{ijk} \hat L_i \hat Q_j \hat  D^c_k
         + \frac{1}{2} \lambda''_{ijk} \hat  U^c_i \hat  D^c_j \hat  D^c_k
         + \epsilon_i \hat L_i \hat H_u \, ,
\label{WG1:eq:rpvsuper1}
\end{eqnarray}
where $i,j,k=1,2,3$ are generation indices. 
$ \hat L_i$ ($ \hat Q_i$) are the lepton (quark) $SU(2)_L$ doublet superfields.
$\hat E^c_j$ ($\hat D^c_j,\hat U^c_j$)
are the electron (down- and up-quark) $SU(2)_L$ singlet superfields. 
$h_{ij}^E$, $h_{ij}^D$, $h_{ij}^U$,
$\lambda_{ijk}$,
$\lambda_{ijk}'$, and $\lambda_{ijk}''$ are dimensionless
Yukawa couplings, whereas the $\epsilon_i$ are dimensionful 
mass parameters. Gauge invariance implies
that the first term in $W_{R_p \hspace{-3mm}/}$\,\,\, is anti-symmetric in
$\{i,j\}$ and the third one is anti-symmetric in $\{j,k\}$.
Equation~(\ref{WG1:eq:rpvsuper1})
thus contains $9+27+9+3=48$ new terms beyond those of the Minimal
Supersymmetric Standard Model (MSSM). 
At the level of the superpotential one can actually rotate  the
$(\hat H_d, \hat L_i)$ by an $SU(4)$ transformation, 
so that the $\epsilon_i$ can be set to zero. However, as discussed below, 
this cannot be done simultaneously for the corresponding
soft SUSY breaking terms and, thus, we keep them for the moment as
free parameters.
 The soft SUSY breaking potential is given by
\begin{eqnarray}
V_{soft} &=& V_{MSSM, soft} + V_{R_p \hspace{-3mm}/ \,\,\,,soft } \,\,\,\, ,
\label{WG1:eq:soft} \\
V_{MSSM} &=& M^2_{L,ij} \tilde L_i \tilde L_j^*
     +  M^2_{E,ij} \tilde E_i \tilde E_j^*  + M^2_{Q,ij} \tilde Q_i \tilde Q_j^*
     +  M^2_{U,ij} \tilde U_i \tilde U_j^* +  M^2_{D,ij} \tilde D_i \tilde D_j^* 
\nonumber \\ &&+ M^2_d H_d H^*_d + M^2_u H_u H^*_u  - (\mu B  H_d  H_u + h.c.)
\nonumber \\ &&
           + ( T_{ij}^E \tilde L_i  H_d \tilde E_j
           + T_{ij}^D \tilde Q_i H_d \tilde D_j
           + T_{ij}^U H_u \tilde Q_i \tilde U_j + h.c.)
           \, ,
 \label{WG1:eq:mssmsoft} \\
V_{R_p \hspace{-3mm}/\,\,\,,soft} &=& 
       \frac{1}{2} T^\lambda_{ijk} \tilde L_i \tilde L_j \tilde E^*_k
         + T^{\lambda'}_{ijk} \tilde L_i \tilde Q_j \tilde  D^*_k
         + \frac{1}{2} T^{\lambda''}_{ijk} \tilde  U_i \tilde  D_j \tilde  D_k
         + \epsilon_i B_i \tilde L_i  H_u  + h.c. \,  
\label{WG1:eq:rpvsoft1}
\end{eqnarray}
The mass matrices $M^2_F$ ($F=L,E,Q,U,D$) are $3\times 3$ hermitian matrices,
whereas the $T^F$ are general $3\times 3$ and 
$3\times 3\times 3$ complex tensors.
Obviously, the $T^\lambda_{ijk}$ ($T^{\lambda''}_{ijk}$) have to be antisymmetric
in the first (last) two indices due to gauge invariance.
In models, where the flavour violating terms are neglected, the $T_{ij}^F$ terms
are usually decomposed into the following products 
$T_{ij}^F = A_{ij}^F h_{ij}^F$
and analogously for the trilinear terms.

The simultaneous appearance of lepton and baryon number breaking terms
leads in general to a phenomenological catastrophe if all involved particles
have masses of the order of the electroweak scale: rapid proton decay
\cite{Weinberg:1981wj,Sakai:1981pk}.
To avoid this problem a discrete multiplicative symmetry,
called  R-parity ($R_p$),
had been invented \cite{Farrar:1978xj} which can be written
as
\begin{eqnarray}
R_p= (-{\bf 1})^{3B+L+2S} \, ,
\label{WG1:eq:rparity}
\end{eqnarray}
where $S$ is the spin of the corresponding particle. For
all superfields of MSSM, the SM field has $R_p=+1$ and its
superpartner has $R_p=-1$, e.g.~the electron has $R_p=+1$
and the selectron has $R_p=-1$. In this way all terms in \eq{WG1:eq:rpvsuper1}
are forbidden and one is left with the superpotential given in
\eq{WG1:eq:mssmsuper}.
To prohibit proton decay it is not necessary to forbid both type of terms
but it is sufficient to forbid either the lepton or the baryon number
violating terms (see e.g.~\cite{Ibanez:1991hv,Dreiner:2005rd}), 
e.g.~the baryon number
terms can be forbidden by baryon triality \cite{Dreiner:2006xw}.
Another possibility would be to break lepton number and  thus
R-parity spontaneously as discussed below. This requires, however,
an enlargement of the particle content.

\subsection{The MSSM with R-parity conservation}

The existence of the soft SUSY breaking terms implies that fermions
and sfermions cannot be rotated by the same rotation matrices from the
electroweak basis to the mass eigenbasis. It is very convenient to work
in the super-CKM basis for the squarks and to assume that $h^E$ is diagonal
and real which can be done without loss of generality. In this way
the additional flavour violation in the sfermion sector is most apparent. 
In this way, the additional flavour violation is encoded in the mass matrices
of the sfermions which read as (see also section 16 of \cite{Allanach:2006fy}):
\begin{equation}
M^2_{\tilde f} = \left(
\begin{array}{cc}
M^2_{LL} &  M^{2\dagger}_{RL} \\
M^2_{RL} &  M^2_{RR} \\
\end{array} \right)~,
\label{WG1:eq:sleptonmass}
\end{equation}
where the entries are $3 \times 3$ matrices. They are given by
\begin{eqnarray}
M^2_{LL} & = & K^\dagger {\hat{M}_Q}^2 K + m^2_{u} + D_{u\,LL} \, ,\\
M^2_{RL} & = &  v_d  {\hat T}^U  - \mu^* m_u \cot\beta \, \\
 M^2_{RR} & = & {\hat M}^2_U + m^2_{u} + D_{u\,RR} 
\end{eqnarray}
for $u$-type squarks in the basis 
$(\tilde u_L, \tilde c_L, \tilde t_L, \tilde u_R, \tilde c_R, \tilde t_R)$.
$K$ is the $CKM$ matrix and we have defined
\begin{eqnarray}
{\hat M_Q}^2 \equiv V^\dagger_d M^2_{\tilde Q} V_d
\end{eqnarray}
where $V_d$ is the mixing matrix for the left $d$-quarks. ${\hat T}^U$ and 
${\hat M}^2_U$ are
given by a similar transformation involving the mixing matrix for left- and
right-handed u-quarks. The same type of notation will be kept below
for $d$ squarks and sleptons.
Finally,
 the $D$-terms are given by
 \begin{equation}
 D_{f\,LL,RR} =  \cos 2\beta \, M_Z^2 
   \left(T_f^3 - Q_f \sin^2\theta_W \right) \iddrei \,.
\label{WG1:eq:dterm}
\end{equation}
The entries for $d$-type squarks read in the basis
$(\tilde d_L, \tilde s_L, \tilde b_L, \tilde d_R, \tilde s_R, \tilde b_R)$.
as
\begin{eqnarray}
M^2_{LL} & = & {\hat{M}_Q}^2  + m^2_{d} + D_{d\,LL} \, ,\\
M^2_{RL} & = &  v_u  {\hat T}^D  - \mu^* m_d \tan\beta \, \\
 M^2_{RR} & = & {\hat M}^2_D + m^2_{d} + D_{d\,RR} \, .
\end{eqnarray}
For the charged sleptons one finds in the basis
$(\tilde e_L,\tilde\mu_L,\tilde\tau_L,\tilde e_R,\tilde\mu_R,\tilde\tau_R)$
\begin{eqnarray}
M^2_{LL} & = & {\hat M}_L^2  + m^2_{l} + D_{l\,LL} \, ,\\
M^2_{RL} & = &  v_u  {\hat T}^E  - \mu^* m_l \tan\beta \, ,\\
 M^2_{RR} & = & {\hat M}^2_E + m^2_{l} + D_{l\,RR} \, .
\label{WG1:eq:sleptonmassRR}
\end{eqnarray}
Assuming that there are only left-type sneutrinos one finds for them
in the basis $(\tilde\nu_{eL},\tilde\nu_{\mu L},\tilde\nu_{\tau L})$
the mass matrix
\begin{eqnarray}
M^2_{LL} & = & {\hat M}_L^2  + D_{\nu\,LL} \, .
\end{eqnarray}

For sleptons the relevant interaction Lagrangian, e.g.~not considering the
slepton Higgs or slepton gauge boson interactions, for the studies below
is given  in terms of mass
eigenstates by:
\begin{eqnarray}
\label{WG1:eq:CoupChiSfermion}
{\mathcal L} &=& \bar \ell_i ( c^L_{ikm} P_L + c^R_{ikm} P_R)
           \tilde \chi^0_k \tilde \ell_m  
    + \bar{\ell_i} (d^L_{ilj} P_L + d^R_{ilj} P_R)
           \tilde \chi^-_l \tilde{\nu}_j
+\bar\nu_i e^R_{ikj}P_R \tilde \chi^0_k \tilde\nu_j
\nonumber \\
 &+& \bar{\nu_i} f^R_{ilm} P_R \tilde \chi^+_l \tilde{\ell}_m
+{\rm h.c.}~.
\end{eqnarray}
The specific forms of the couplings $c^L_{ikm}$, $c^R_{ikm}$,
$d^L_{ilj}$, $d^R_{ilj}$, $e^R_{ikj}$ and $f^R_{ilm}$ can be found in
\cite{Chung:2003fi}. The first two terms in \eq{WG1:eq:CoupChiSfermion}
give rise to the LFV signals studied here, whereas the 
last one will give rise to the SUSY background
because the neutrino flavour cannot be discriminated in high energy collider
experiments. In particular the following decays are of primary interest:
\begin{eqnarray}
\tilde l_j &\to& l_i \tilde \chi^0_k \label{eq:SleptonDecays1} \\
\tilde  \chi^0_k &\to&\tilde l_j l_i \\
\tilde  \chi^0_k &\to& l_j l_i \tilde  \chi^0_r 
\label{eq:SleptonDecays3}
\end{eqnarray}
Several studies for these decays have been performed assuming
either specific high-scale models
or specifying the LFV parameters at the low scale (see for instance refs.\
\cite{Hisano:1995cp,Donoghue:1983mx,Hall:1985dx,Borzumati:1986qx,%
Gabbiani:1988za,Hagelin:1992tc,Barbieri:1994pv,Gabbiani:1996hi,Hisano:1996qq,%
Hisano:1997tc,Krasnikov:1995qq,Hisano:1998wn,Nomura:2000zb,Guchait:2001us,%
Porod:2002zy,Carvalho:2002jg,Demir:2003bv,Porod:2004vi,Deppisch:2003wt,%
Hamaguchi:2004ne,Oshimo:2004wv,Paradisi:2005fk,Arkani-Hamed:1996au,%
Arkani-Hamed:1997km,Bartl:2007ua}).

Performing Monte Carlo studies on the parton level it has been shown
that LHC can observe SUSY LFV by studying
the LFV decays of the second neutralino
$\tilde\chi^0_2$ arising from cascade decays of gluinos
and squarks, i.e. $\tilde\chi^0_2\to \tilde\ell \ell'\to \ell' \ell''\tilde\chi^0_1$:
signals of SUSY LFV can be extracted despite considerable
backgrounds and stringent experimental bounds on flavour violating lepton
decays in case of two generation mixing in either the right or left slepton
sector in the mSUGRA model
\cite{Agashe:1999bm,Hinchliffe:2000np,Hisano:2002iy}. The $\tilde e_R-\tilde \mu_R$
mixing case was studied in \cite{Agashe:1999bm,Hisano:2002iy} and the
$\tilde \mu_L-\tilde \tau_L$ mixing case in \cite{Hinchliffe:2000np}.

In the (s)quark sector one has to analogue decays as the ones given in
Eqs.~(\ref{eq:SleptonDecays1})--(\ref{eq:SleptonDecays3}). In addition there are
decays into charginos and gluinos if kinematically allowed. Flavour
effects in these decays has first been discussed in \cite{Hurth:2003th}.
There it has been shown that one can have large effects in squark and
gluino decays despite stringent constraints from B-meson physics
as discussed in the WG2 chapter. In addition, flavour mixing
in the squark sector can induce flavour violating decays of Higgs
bosons as e.g.~$H^0 \to b s$ \cite{Curiel:2003uk}.

In the discussion we have considered so far models where the parameters
are freely given at the electroweak scale. The fact that no flavour violation
in the quark sector has been found beyond SM expectations has led to the
development of the concept of minimal flavour violation (MFV). The basic idea is
that at a given scale the complete flavour information is encoded in the Yukawa
couplings \cite{D'Ambrosio:2002ex}, 
e.g.~that in a GUT theory the parameters at the GUT scale are
given by $M^2_F = M^2_0 \iddrei$ and $T_F = A_0 h_U$ with $M_0$ and
$A_0$ being a real and a complex number, respectively. In such models
it has been shown that the branching ratios for
flavour violating squark decays are very small and most likely not 
observable at LHC \cite{Lunghi:2006uf}. A similar concept has
been developed for (s)leptons 
\cite{Cirigliano:2005ck,Davidson:2006bd}. In contrast to the
squark sector one has large mixing effects in the neutrino sector which
can lead to observable effects in the slepton sector at future collider
experiments
\cite{Deppisch:2004pc} and section 5.2.3~of the WG3 chapter.

\subsection{The MSSM with broken R-parity}

Recent neutrino experiments have shown that neutrinos are massive particles
which mix among themselves (for a review see e.g.~\cite{Maltoni:2004ei}).
In contrast to leptons and quarks, neutrinos need not
 be Dirac particles but can be Majorana particles. In the latter case
the Lagrangian contains a mass term which violates explicitly lepton number
by two units.
This motivates one to allow the lepton number breaking terms in the
superpotential in particular as they automatically imply the existence
of massive neutrinos without the need of introducing right-handed
neutrinos and explaining their mass hierarchies
\cite{Hall:1983id}. The $\lambda''$ terms can still be forbidden by a
discrete symmetry such as baryon triality \cite{Allanach:2003eb}.

Let us briefly comment on the number of free parameters  before discussing
the phenomenology in more detail.
The last term in \eq{WG1:eq:rpvsuper1},
$ \hat L_i  \hat H_u$, mixes the lepton and the Higgs superfields.
In supersymmetry $\hat L_i$ and $\hat H_d$ have the same gauge and Lorentz
quantum numbers and we can redefine them by a rotation in $(\hat H_d, \hat L_i)$.
 The terms
$\epsilon_i  \hat L_i  \hat H_u$ can then be rotated to zero in the
superpotential  \cite{Hall:1983id}. However, there are still
the corresponding terms  in the  soft supersymmetry breaking Lagrangian
\begin{eqnarray}
 V_{R_p \hspace{-3mm} / \,\,\,\, ,soft} = B_i \epsilon_i \tilde L_i H_u
\label{WG1:eq:bisoft}
\end{eqnarray}
 which can only
be rotated away if $ B_i = B$ and $M^2_{H_d} = M^2_{L,i}$  \cite{Hall:1983id}.
 Such an alignment of the superpotential terms with
the soft breaking terms is not stable under the renormalization group equations
\cite{deCarlos:1996du}.
Assuming an alignment at the unification scale, the resulting
effects are small \cite{deCarlos:1996du} except for neutrino masses
\cite{deCarlos:1996du,Nardi:1996iy,Hempfling:1995wj,Nilles:1996ij,Hirsch:2000ef}.
Models containing only bilinear terms do not introduce
trilinear terms as can easily be seen from the fact that bilinear terms
have dimension mass whereas the trilinear are dimensionless. For this
reason we will keep in the following explicitly the bilinear terms in
the superpotential.
These couplings induce decays of the LSP violating lepton number, e.g.
\begin{eqnarray}
\begin{array}{rclll}
\tilde \nu &\to& q \bar{q}   \, , &
            l^+ l^-  \, , & \nu \bar{\nu} \\
 \tilde l &\to& l^+ \nu \, , &
           q \bar{q}' \\
\tilde \chi^0_1 &\to& W^\pm l^\mp \, , & Z \nu_i \\
\tilde \chi^0_1 &\to& l^\pm q \bar{q}'\, , & q \bar{q}  \nu_i
 \, , & l^+ l^-  \nu_i
\end{array}
\end{eqnarray}

How large can the branching ratio for those decay modes be? To answer
this question one has to take into account existing constraints on
R-parity violating parameters from low energy physics. As most of
them are given in terms of trilinear couplings,
we will work for this particular considerations
in the  ``$\epsilon$-less'' basis, e.g.~rotate away the
bilinear terms in the superpotential \eq{WG1:eq:rpvsuper1}.
Therefore, the trilinear couplings get additional contributions.
 Assuming, without
loss of generality, that the lepton and down type Yukawa couplings are
diagonal
they are given to leading order in $\epsilon_i/\mu$ as
\cite{Allanach:2003eb,Dreiner:2003hw,AristizabalSierra:2004cy}:
\begin{equation}
\lambda'_{ijk} \to \lambda'_{ijk} + \delta_{jk} h_{d_k} \frac{\epsilon_i}{\mu}
\label{WG1:eq:efflamp}
\end{equation}
and
\begin{eqnarray}
&&\lambda_{ijk} \to \lambda_{ijk} +  \delta \lambda_{ijk}  , \hskip42mm
\label{WG1:eq:efflam} \\
&&\delta \lambda_{121} = h_{e} \frac{\epsilon_2}{\mu}, \hskip5mm
\delta \lambda_{122} = h_{\mu} \frac{\epsilon_1}{\mu}, \hskip5mm
\delta \lambda_{123} = 0 \nonumber \\ \nonumber
&&\delta \lambda_{131} = h_{e} \frac{\epsilon_3}{\mu}, \hskip5mm
\delta \lambda_{132} = 0, \hskip11mm
\delta \lambda_{133} = h_{\tau} \frac{\epsilon_1}{\mu} \\ \nonumber
&&\delta \lambda_{231} = 0, \hskip11mm
\delta \lambda_{232} = h_{\mu} \frac{\epsilon_3}{\mu}, \hskip5mm
\delta \lambda_{233} = h_{\tau} \frac{\epsilon_2}{\mu} \\ \nonumber
\end{eqnarray}
where we have used the fact that neutrino physics requires
$|\epsilon_i / \mu| \ll 1$ \cite{Hirsch:2000ef}.
An essential point to notice is that the additional contributions
in \eqs{WG1:eq:efflamp}{WG1:eq:efflam} follow the hierarchy
dictated by the down quark and charged lepton masses of the
standard model.

A comprehensive list of bounds on various R-parity violating parameters
can be found in \cite{Allanach:1999ic}. However, there the recent
data from neutrino experiments like Super-Kamiokande \cite{Ashie:2005ik},
SNO \cite{Aharmim:2005gt} and KamLAND \cite{Araki:2004mb}
are not taken into account. These experiments yield strong bounds
on trilinear couplings involving the third generation
\cite{Abada:2000xr,Borzumati:2002bf}. In addition also the sneutrino
vevs are constrained by neutrino data \cite{Abada:2000xr,Hirsch:2000ef}.
Most of the
trilinear couplings have a bound of the order
$(10^{-2} - 10^{-1}) \cdot m_{\tilde f}/(100$ GeV) where $m_{\tilde f}$ is the
mass of the sfermion in the process under considerations. The cases with
stronger limits are: $|\lambda'_{111}| \lsim O(10^{-4})$ due to neutrino-less
double beta decay and
$|\lambda_{i33}| \simeq 5 |\lambda'_{i33}| \simeq  O(10^{-4})$ due to
neutrino oscillation data. Moreover, neutrino oscillation data imply
$|\mu^2 (v^2_1 + v^2_2 + v^2_3) /\det({\mathcal M}_{\chi^0})| \lsim 10^{-12}$
 where $v_i$ are the sneutrino vevs and $\det({\mathcal M}_{\chi^0})$ is the
determinant of the MSSM neutralino mass matrix.

There exists a vast literature on the effects of R-parity violation at LHC
\cite{Gonzalez-Garcia:1991ap,Dreiner:1991pe,Dreiner:1993ba,Baer:1996wa%
,Bartl:1996cg,Dreiner:2000vf,Yin:2001cy}. However,
in most of these studies, in particular those considering trilinear couplings
only, very often the existence of a single coupling
has been assumed. However, such an assumption is only valid
 at a given scale as renormalization effects
imply that additional couplings are present when going to a different scale
via RGE evolution. Moreover, very often the bounds stemming from neutrino
physics are not taken into account or are out-dated (e.g. assuming an MeV
tau neutrino). Last but not least one should note, that also in this
class of models there are potential dark matter candidates, e.g.~a
very light gravitino 
\cite{Borgani:1996ag,Takayama:2000uz,Hirsch:2005ag,Buchmuller:2007ui}.

Recently another class of models with explicitly broken R-parity has been
proposed where the basic idea is that the existence of 
right handed neutrino superfields is the source of the
$\mu$-term of the MSSM as well as the source or neutrino masses
\cite{Lopez-Fogliani:2005yw}.
In this case the superpotential contains only trilinear terms. Beside
the usual Yukawa couplings of the MSSM the following couplings are present:
\begin{eqnarray}
W_{\nu^C} =  h_{ij}^\nu \hat H_u \hat L_i \hat \nu^c_j
- \lambda_i  \hat \nu^c_i \hat H_d \hat H_u
+ \frac{1}{3} \kappa^{ijk}  \hat \nu^c_i  \hat \nu^c_j \hat \nu^c_k
\end{eqnarray}
Note, that (i) the second and third term break R-parity and that
the sneutrino fields play the role of the gauge singlet field of the 
Next to Minimal Supersymmetric Standard Model (NMSSM)
\cite{Nilles:1982dy,Frere:1983ag,Derendinger:1983bz,Veselov:1985gd}.

\subsection{Spontaneous R-parity violation}
\label{WG1:sec:rpspon}

Up to now we have only considered explicit R-parity violation keeping
the particle content of the MSSM. In the case that one enlarges
the spectrum by gauge singlets one can obtain models where lepton number
and, thus, R-parity is broken spontaneously together with $SU(2) \otimes U(1)$
\cite{Aulakh:1982yn,Ross:1984yg,Masiero:1990uj,Romao:1992vu,Shiraishi:1993di}.
A second possibility to break R-parity spontaneously
 is to enlarge the gauge symmetry
\cite{Huitu:1994zm}.

The most general superpotential terms involving the  MSSM superfields
in the presence of the  $SU(2) \otimes U(1)$
singlet superfields $({\widehat \nu^c}_i,
\widehat{S}_i,{\widehat \Phi})$
carrying a conserved lepton number assigned as $(-1, 1,0)$, respectively,
is given as~\cite{Romao:1992zx}
\begin{eqnarray} \nonumber
{\cal W} &\hskip-4mm=\hskip-4mm& \varepsilon_{ab}\Big(
h_U^{ij}\widehat Q_i^a\widehat U_j\widehat H_u^b
+h_D^{ij}\widehat Q_i^b\widehat D_j\widehat H_d^a
+h_E^{ij}\widehat L_i^b\widehat E_j\widehat H_d^a
+h_{\nu}^{ij}\widehat L_i^a\widehat \nu^c_j\widehat H_u^b 
\!- {\hat \mu}\widehat H_d^a\widehat H_u^b
\!- h_0 \widehat H_d^a\widehat H_u^b \widehat\Phi \Big)  \\
& &+\hskip 5mm   h^{ij} \widehat\Phi \widehat\nu^c_i\widehat S_j +
M_{R}^{ij}\widehat \nu^c_i\widehat S_j
+ \frac{1}{2}M_{\Phi} \widehat\Phi^2 +\frac{\lambda}{3!} \widehat\Phi^3
\label{WG1:eq:Wsuppot}
\end{eqnarray}
The first three terms together with the $ {\hat \mu}$ term define the
R-parity conserving MSSM, the terms in the second line only involve the
$SU(2) \otimes U(1)$ singlet superfields
$({\widehat \nu^c}_i,\widehat{S}_i,{\widehat \Phi})$,
while the remaining terms couple
the singlets to the MSSM fields. For completeness we note that
lepton number is fixed via the
Dirac-Yukawa $h_\nu$ connecting the right-handed neutrino superfields
to the lepton doublet superfields. For simplicity we assume
in the discussion below that only one generation of
$({\widehat \nu^c}_i,\widehat{S}_i)$ is present.

The presence of singlets in the model is essential in order to drive
the spontaneous violation of R-parity and electroweak symmetries in a
phenomenologically consistent way. As in the case of explicit R-parity
violation all sneutrinos obtain a vev beside the Higgs bosons as well
as the $\tilde S$ field and the singlet field $\Phi$. For completeness
we want to note that in the limit where all sneutrino vevs vanish and
all singlets carrying lepton number are very heavy one obtains the
NMSSM as an effective theory.
The spontaneous breaking of R-parity also entails the spontaneous
violation of total lepton number. This implies that one of the neutral
CP--odd scalars, which we call Majoron $J$ and which is approximately given by
the imaginary part of
\begin{equation}
\frac{\sum_i v_{i}^2}{Vv^2} (v_u H^0_u - v_d H^0_d) +
\sum_i \frac{v_{i}}{V} \tilde{\nu_{i}}
+\frac{v_S}{V} S
-\frac{v_R}{V} \tilde{\nu^c}
\label{WG1:eq:maj}
\end{equation}
remains massless, as it is the Nambu-Goldstone boson associated to the
breaking of lepton number. $v_R$ and $v_S$ are the vevs of $\tilde \nu^c$
and $\tilde S$, respectively and $V=\sqrt{v_R^2 + v^2_S}$.
Clearly, the presence of these additional
singlets enhances further the number of neutral scalar and pseudo-scalar
bosons. Explicit formulas for the mass matrices of
scalar and pseudo-scalar bosons
 can be found e.g.~in \cite{Hirsch:2004rw}.

The case of an enlarged gauge symmetry can be obtained for example
in left-right symmetric models, e.g.~with the gauge group
$SU(2)_L \times SU(2)_R \times U(1)_{B-L}$ \cite{Huitu:1994zm}.
 The corresponding
superpotential is given by:
\begin{eqnarray}
W & = & h_ {\phi Q} \widehat Q_{L}^{T}i\tau_2 \widehat \phi  \widehat Q_{R}^c
+ h_{ \chi Q} \widehat Q_{L}^{T} i\tau_2\widehat
\chi  \widehat Q_{R}^c \nonumber \\
&&+h_ {\phi L} \widehat L_{L}^{T}i\tau_2 \widehat \phi  \widehat L_{R}^c
+h_{ \chi L} \widehat L_{L}^{T} \widehat i\tau_2 \widehat\chi
 \widehat L_{R}^c
+h_{\Delta} \widehat L_{R}^{cT} i\tau_2
{\widehat \Delta }  \widehat L_{R}^c \nonumber\\
&& + \mu_1 {\rm Tr} (i\tau_2 \widehat \phi^T i\tau_2 \widehat \chi )
+\mu_2  {\rm Tr} (\widehat \Delta \widehat \delta ) ,\label{WG1:eq:sponpot2}
\end{eqnarray}
where the Higgs sector consists of two triplet and two bi-doublet Higgs
superfields with the following
$SU(2)_L \times SU(2)_R \times U(1)_{B-L}$ quantum numbers:
\begin{eqnarray}
\label{WG1:eq:higgses}
&&\widehat\Delta =\left( \begin{array}{cc}
\widehat\Delta^-/\sqrt{2} & \widehat\Delta^{0}\\
 \widehat\Delta^{--}&-\widehat\Delta^{-}/\sqrt{2} \end{array} \right)
\sim ({\bf 1,3,}-2),\nonumber\\
&&\widehat\delta =\left( \begin{array}{cc}
\widehat\delta^{+}/\sqrt{2}& \widehat\delta^{++}\\
 \widehat\delta^{0} &-\widehat\delta^{+}/\sqrt{2} \end{array} \right)
\sim ({\bf 1,3,}2),
\,\,\,\,\,\,\,\,\,\,\,\, \nonumber\\
&& \widehat\phi =\left( \begin{array}{cc}\widehat\phi_1^0&
\widehat\phi_1^+\\\widehat\phi_2^-&\widehat\phi_2^0
\end{array} \right) \sim ({\bf 2,2,}0),
\,\,\,\,\,\,\,\,\,\,\,\,
\widehat\chi =\left( \begin{array}{cc}\widehat\chi_1^0&
\widehat\chi_1^+\\\widehat\chi_2^-&\widehat\chi_2^0
\end{array} \right)  \sim ({\bf 2,2,}0).
\end{eqnarray}
Looking at the decays of the Higgs bosons, one has to distinguish
two scenarios: (i) Lepton number is  gauged  and  thus
 the  Majoron  becomes 
the  longitudinal  part  of  an  additional
 neutral  gauge  boson.  (ii) The Majoron
remains a physical particle in the spectrum. 
In the case of the enlarged gauge group there are  additional doubly
charged Higgs bosons $H^{--}_i$
which have lepton number violating couplings. In $e^- e^-$ collisions they
can be produced according to
\begin{eqnarray}
e^- e^- \to H^{--}_i
\end{eqnarray}
and have decays of the type
\begin{eqnarray}
 H^{--}_i &\to& H^-_j \, H^-_k \\
 H^{--}_i &\to& l^-_j l^-_k \\
 H^{--}_i &\to& \tilde l^-_j \tilde l^-_k \, .
\end{eqnarray}
 In addition there exist doubly charged
charginos which can have lepton flavour violating decays:
\begin{eqnarray}
 \tilde \chi^{--}_i &\to& \tilde l^-_j l^-_k \,.
\end{eqnarray}

\subsection{Study of supersymmetry at the LHC}

If Supersymmetry exists at the electroweak scale, it could hardly escape 
detection at the LHC. In most R-parity conserving models, the production cross 
section is expected to be dominated by the pair production of coloured 
states (squarks and 
gluinos). These decay to lighter SUSY particles and ultimately to the LSP 
(Lightest Supersymmetric Particle). If this is stable and weakly interacting, 
as implied by R-parity conservation and cosmological arguments, it leaves the 
experimental apparatus undetected. The supersymmetric events are thus expected 
to show up at the LHC as an excess over SM expectations of events with several 
hard hadronic jets and missing energy. The LHC center of mass of 14~TeV 
extends the search for SUSY particles up to squark and gluino masses 
of 2.5 to 3~TeV~\cite{ATLASTDR,Abdullin:1998pm}. 

If squarks and gluinos are lighter than 1~TeV, as implied 
by naturalness arguments, this signature would be observed with high 
statistical significance already during the first year of running at the 
initial LHC luminosity of 
$2 \; 10^{33} \; \mbox{cm}^{-2} \mbox{s}^{-1}$~\cite{Tovey:2002jc}.
In practice, discovery would be achieved as soon as a good understanding 
of the systematics on Standard Model rates at the LHC is obtained. 

A significant part of the efforts in preparation for the LHC startup is 
being spent in the simulations of the new physics potential. We give below  
a brief overview of these studies, dividing them in three categories:
inclusive searches of the non-SM physics, measurement of SUSY particle masses, 
and measurements of other properties of SUSY particles, such as their spin 
or the flavour structure of their decays. 

\subsection{Inclusive searches}

In these studies, the typical discovery strategy 
consists in searching for an excess of events with a given topology. 
A variety of final state signatures has been considered. Inclusive 
searches have 
mainly be carried out in the framework of mSUGRA, which has five independent 
parameters specified at high energy scale: the common gaugino mass 
$m_{1/2}$, the common scalar mass $m_0$, the common trilinear coupling 
$A_0$, the ratio of the vacuum expectations values of the two Higgs 
doublets $\tan \beta$ and the sign of the Higgsino mixing parameter $\mu$. 
The masses and decay branching ratios of the SUSY particles are then 
computed at the electroweak scale using the renormalization group equations, 
and used as input to the LHC simulation codes. 
 
For each point of a grid covering the mSUGRA parameter space, 
signal events are generated at parton level and handed over to the 
parametrized detector simulation. The main Standard Model background 
sources are simulated, where the most relevant are processes with an hard
neutrino in the final state ($t \bar t$, $W+$jets, $Z+$jets). Multi-jet 
QCD is also relevant because its cross section is several orders of magnitude
larger than SUSY. However it is strongly suppressed by the requirement 
of large transverse missing
energy and it gives a significant contribution only to the final 
state search channels without isolated leptons. The detailed detector 
simulation, much more demanding in terms of computing CPU power, 
validates the results with parametrized detector simulations for 
the Standard Model backgrounds and selected points in the mSUGRA parameter 
space.  

\begin{figure}[t]
\setlength{\unitlength}{1mm}
\begin{picture}(150,64)
\put(25,0){\epsfig{figure=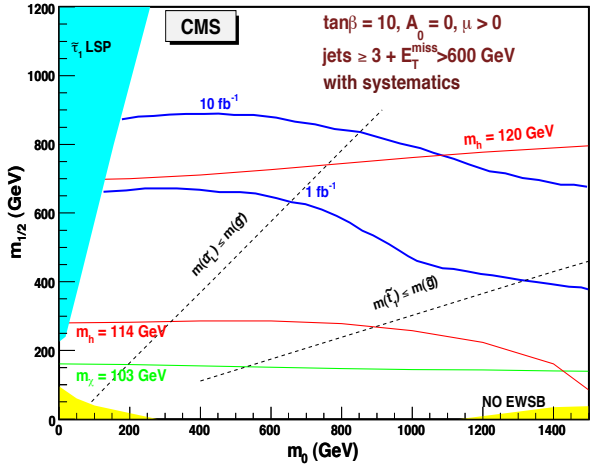
    ,height=70mm,width=95mm}}
\end{picture}
\caption{ CMS 5$\sigma$ discovery 
potential using multi-jets and missing transverse energy final 
state~\cite{DellaNegra:942733a}.}
\label{WG1:fig:reach}
\end{figure}

Cuts on missing transverse energy, the transverse momentum of jets, and other 
discriminating variables are optimized to give the best statistical significance
for the (simulated) observed excess of events. For each integrated luminosity
the regions of the parameter space for which the statistical significance 
exceeds the conventional discovery value of 5$\sigma$ are then displayed. 
An example is shown in \fig{WG1:fig:reach} for the CMS  
experiment~\cite{DellaNegra:942733a} with similar results for 
ATLAS~\cite{Tovey:2002jc}. A slice of 
the mSUGRA parameter space is shown, for fixed $\tan \beta = 10$, $A=0$ and 
$\mu > 0$. The area of parameter space favoured by naturalness arguments
can be explored with an integrated luminosity of only $1 \mbox{fb}^{-1}$. 

Although these results were obtained in the context of mSUGRA, 
the overall SUSY reach in terms of squark and 
gluino masses is very similar for most R-parity conserving models, 
provided that the LSP mass is much lower than the squark and gluino 
masses. This has been shown to be the case for GMSB 
and AMSB models~\cite{Barr:2002ex} and even the 
MSSM~\cite{Aurenche:2001ys}. 

\subsection{Mass measurement}

A first indication of the mass scale of the SUSY particles produced 
in the pp interaction will probably be obtained measuring the "effective 
mass", which is the scalar sum of transverse missing energy and the 
$p_T$ of jets and leptons in the event. Such a distribution is expected to 
have a peak correlated with the SUSY mass scale. The correlation is 
strong in mSUGRA, 
and still usable in the more general MSSM~\cite{Tovey:2002jc}. 

The reconstruction of the mass spectrum of Supersymmetric particles
will be more challenging. Since SUSY particles would be produced in 
pairs, there are two undetected LSP 
particles in the final state, which implies that mass peaks can not be 
reconstructed from invariant mass combinations, unless the 
mass of the LSP itself is already known.

The typical procedure consists in choosing a particular decay chain, 
measuring invariant mass combinations and looking for kinematical 
minima and maxima. Each kinematical endpoint is a function of the
masses of the SUSY particles in the decay chain. If enough endpoints
can be measured, the masses of all the SUSY particles involved in the 
decay chain can be obtained. Once the mass of the LSP is known, mass 
peaks can be reconstructed.

After reducing the SM background very effectively through hard missing
transverse energy cuts, the main background for this kind of
measurements usually comes from supersymmetric events in which the
desired decay chain is not present or was not identified correctly by
the analysis. For this reason, these studies are made using data
simulated for a specific point in SUSY parameter space, for which all
Supersymmetric production processes are simulated.

The two body decay chain $\chi^0_2 \rightarrow \tilde l^{\pm} l^{\pm} 
\rightarrow l^{\pm} l^{\pm} \chi^0_1$ is particularly promising, as it 
leads to a very sharp edge in the distribution of the invariant mass of 
the two leptons, which measures:
\begin{eqnarray}
m^2_{edge}(\ell\ell) = \frac{(m^2_{\tilde \chi^0_2} - m^2_{\tilde\ell_i})
                       ( m^2_{\tilde\ell_i} - m^2_{\tilde \chi^0_1})}
                      {m^2_{\tilde\ell_i}}~.
\label{WG1:eq:egde}
\end{eqnarray}

The basic signature of this decay chain are two opposite-sign, same-flavour 
(OSSF) leptons; but two such leptons can also be produced by other 
processes. If the two leptons are
independent of each other, one would expect equal amounts of OSSF leptons and
OSOF leptons (i.e combinations $e^{+}\mu^{-}$, $e^{-}\mu^{+}$). Their
distributions should also be identical, and this allows to remove the
background contribution for OSSF by subtracting the OSOF events.\\

\begin{figure}[t]
\begin{center}
\setlength{\unitlength}{1mm}
\begin{picture}(150,54)
\put(-5,-12){\epsfig{figure=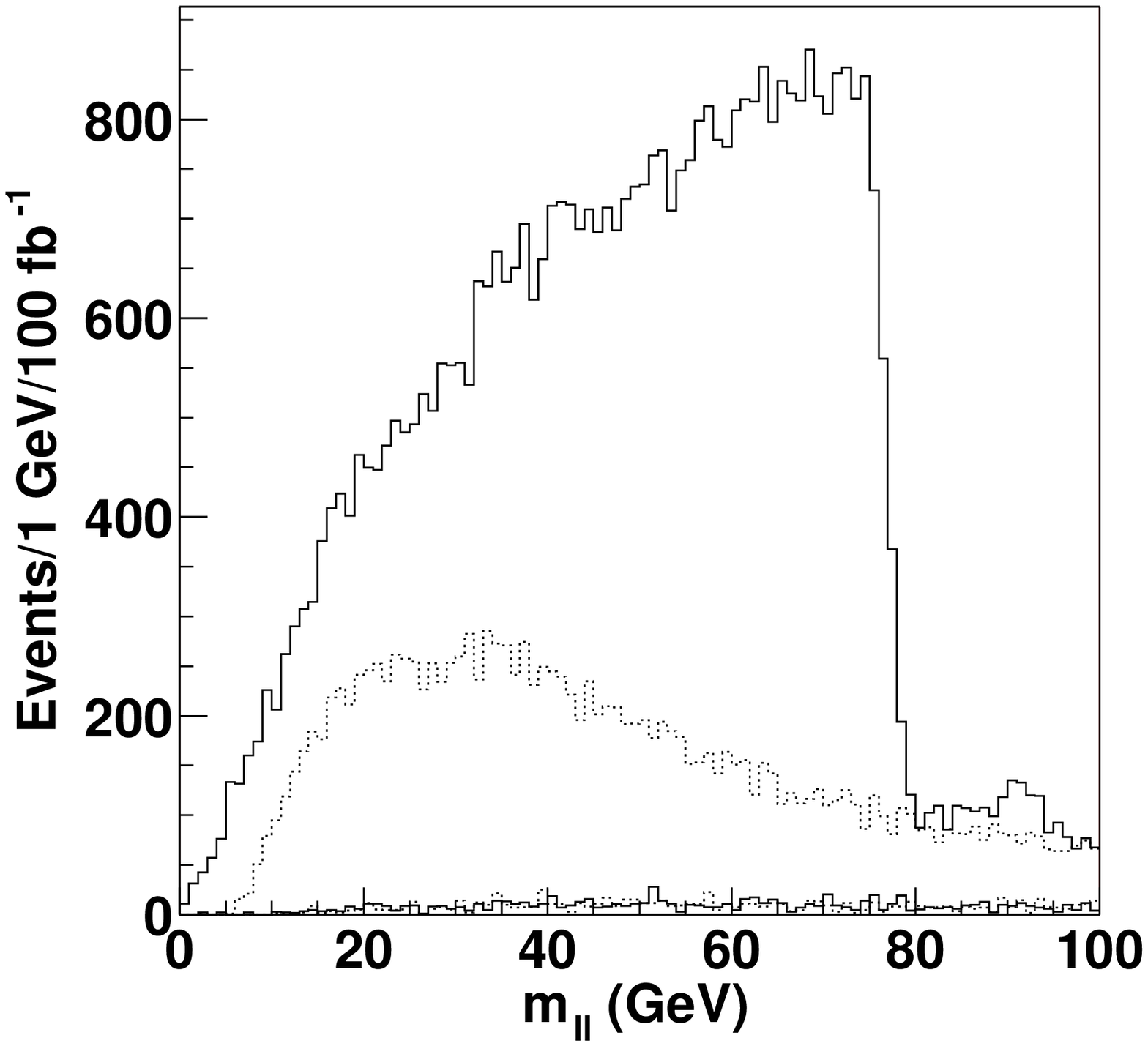
    ,height=70mm,width=85mm}}
\put(15,40){\small OSSF}  
\put(18.5,39.3){\vector(2,-3){4.4}}
\put(30,21.3){\small OSOF}  
\put(33.5,20.3){\vector(-2,-3){2.4}}
\put(20,7){\small SM}  
\put(23.5,6.3){\vector(2,-3){2.4}}
\put(75,-12){\epsfig{figure=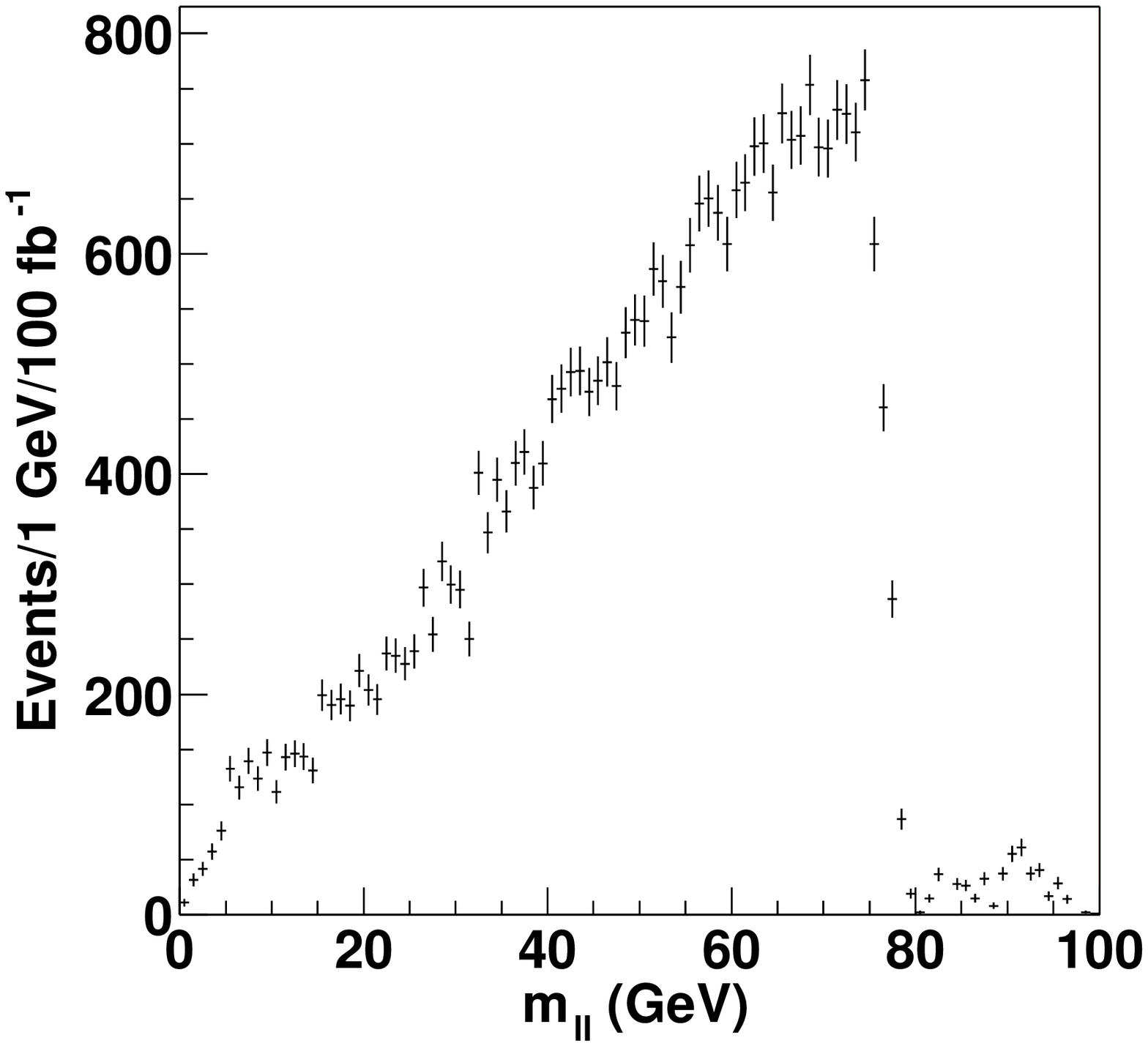
                     ,height=70mm,width=85mm}}
\end{picture}
\end{center} 
\caption{ Effect of subtracting background leptons, for
the mSUGRA benchmark point SPS1a and 
an integrated luminosity of 100 $\mbox{fb}^{-1}$. In the left plot: the curves
represent OSSF leptons, OSOF leptons and the SM contribution. In the
right plot, the flavour subtraction OSSF-OSOF have been plotted: the triangular
shape of the theoretical expectation is reproduced.}
\label{WG1:eq:dilepSPS1A}
\end{figure}

\Bfig{WG1:eq:dilepSPS1A} shows the invariant mass of the two leptons obtained 
for SPS1A point~\cite{Allanach:2002nj} with 100~$\mbox{fb}^{-1}$ of simulated 
ATLAS data~\cite{SPS1}. The Standard Model 
background is clearly negligible. The
real background consists of other SUSY processes, that are
effectively removed by the OSOF subtraction.\\

\begin{figure}[t]
\begin{center}
\setlength{\unitlength}{1mm}
\begin{picture}(150,54)
\put(-5,-12){\epsfig{figure=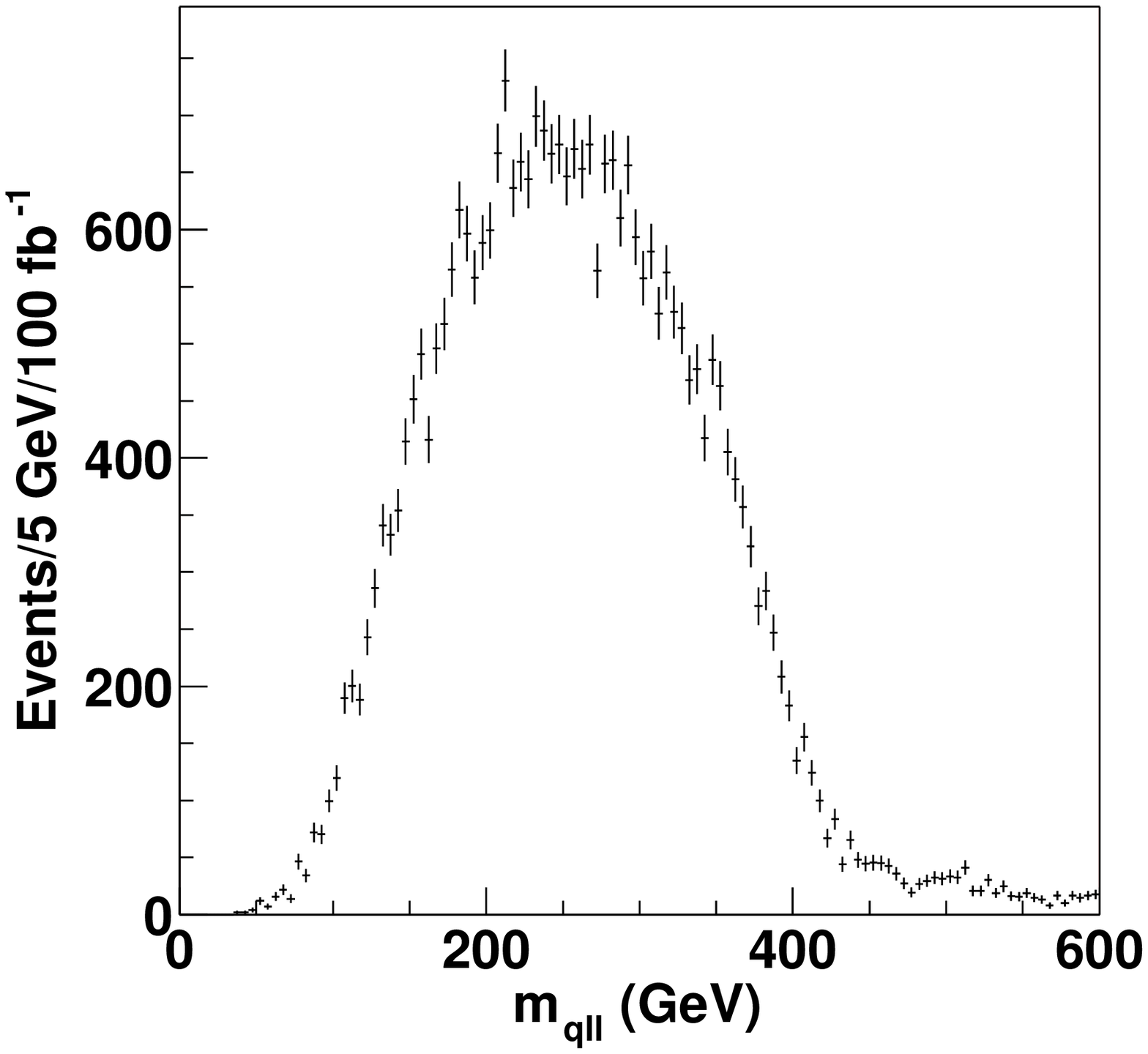
    ,height=70mm,width=85mm}}
\put(75,-12){\epsfig{figure=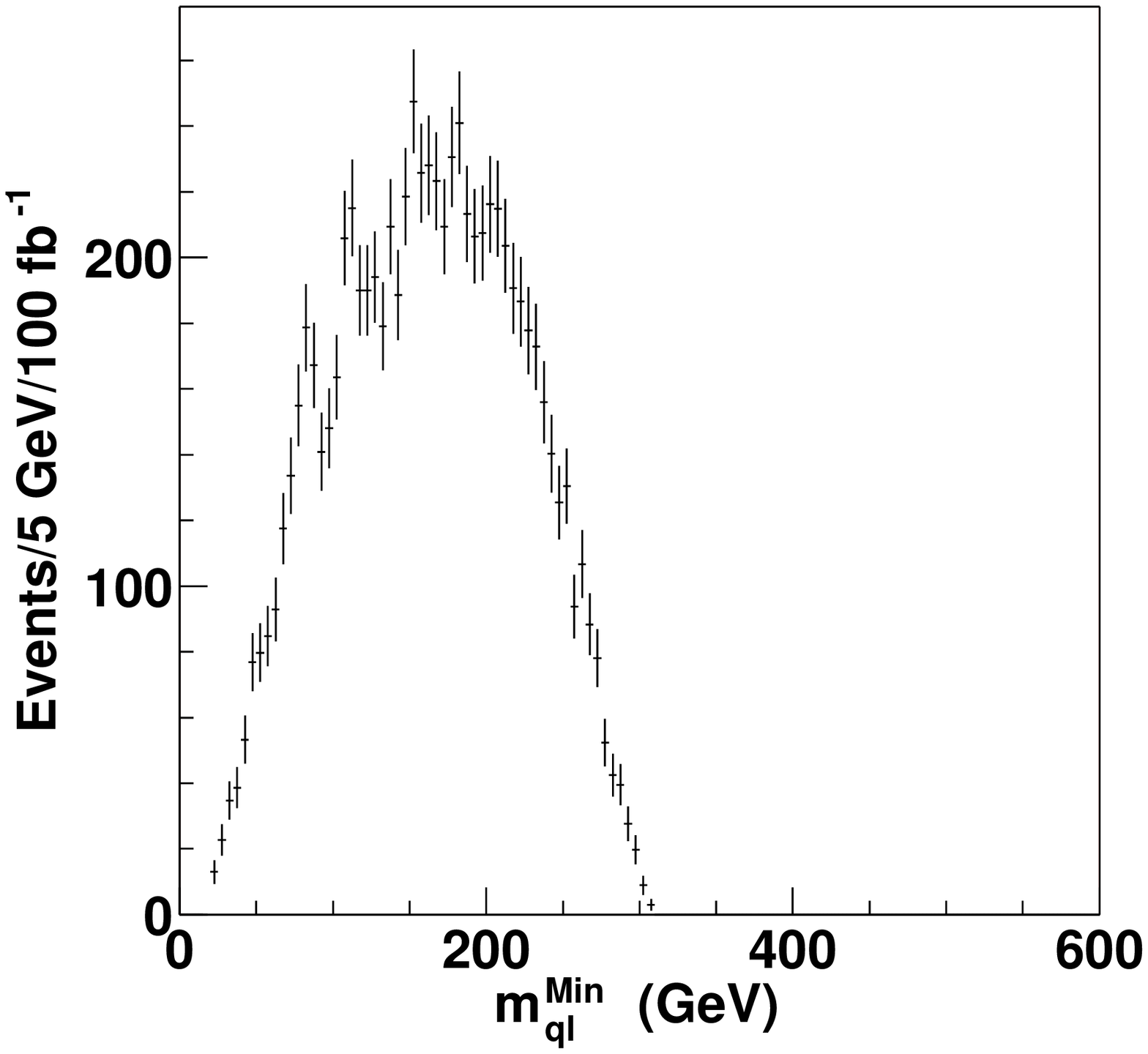
                     ,height=70mm,width=85mm}}
\end{picture}
\end{center}
\caption{ Invariant mass distributions
with kinematical endpoints, for an integrated luminosity of 100 
$\mbox{fb}^{-1}$. In the
left plot for $qll$ combination, in the right plot for the maximum of $ql$
combination.}
\label{WG1:fig:SPS1Acomb}
\end{figure}

Several other kinematical edges can be obtained using various invariant  
mass combinations involving jets and leptons. Two of such distributions are 
reported in  \fig{WG1:fig:SPS1Acomb} for the point SPS1a and 100~$\mbox{fb}^{-1}$
of ATLAS simulated data~\cite{SPS1}. Five endpoints, each providing 
a constraints on the mass of four particles, can be measured. The 
masses of the supersymmetric particles present in the decay chain 
(the left-handed squark, the right-handed sleptons, and the two 
lightest neutralinos) can thus be measured with an error between 
3\% (for the squark) and 12\% (for the lightest neutralino) for 
100~$\mbox{fb}^{-1}$of integrated luminosity.

\begin{figure}[t]
\begin{center}
\setlength{\unitlength}{1mm}
\begin{picture}(150,65)
\put(-6,-5){\epsfig{figure=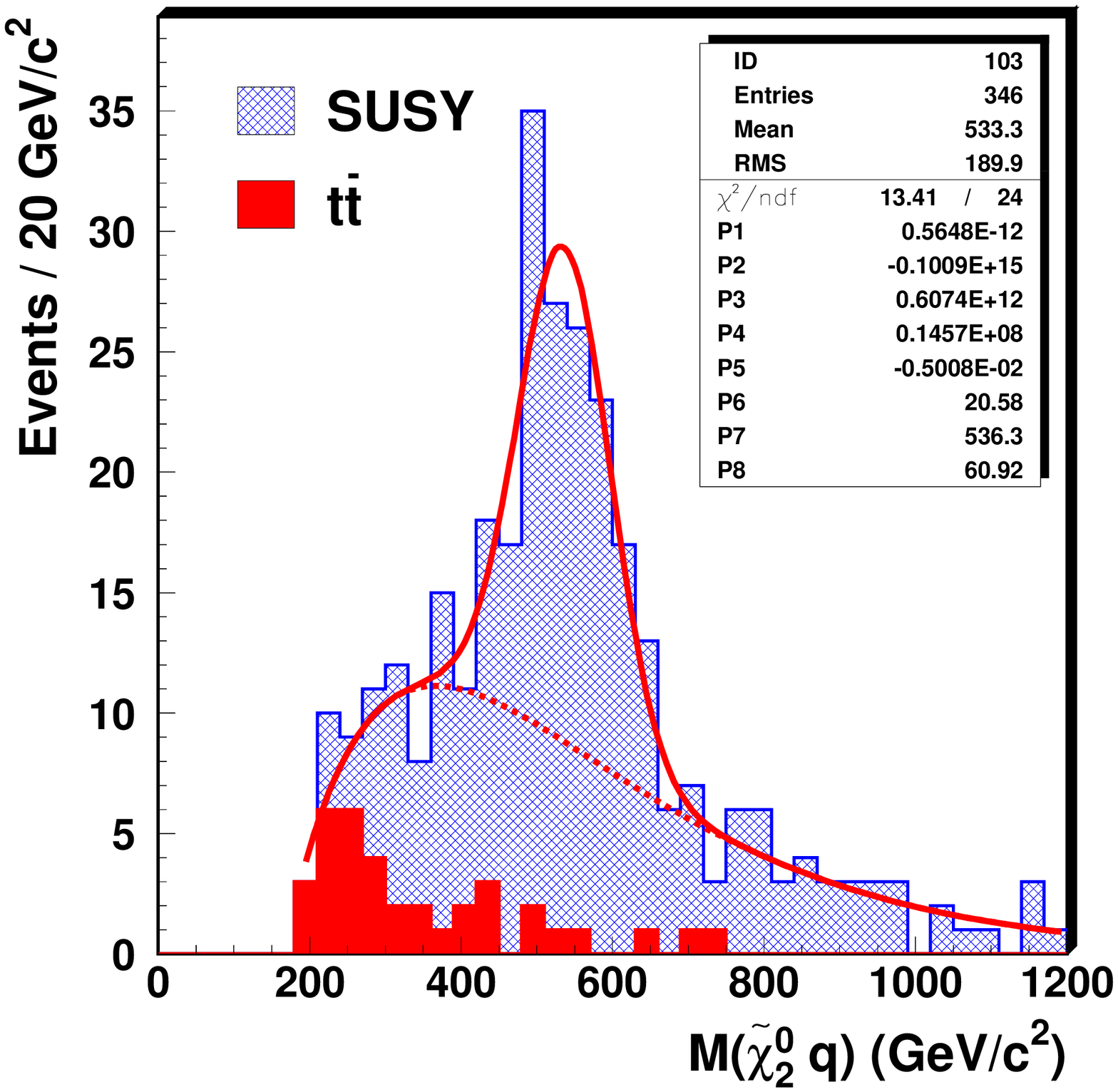
    ,height=70mm,width=52mm}}
\put(47,-5){\epsfig{figure=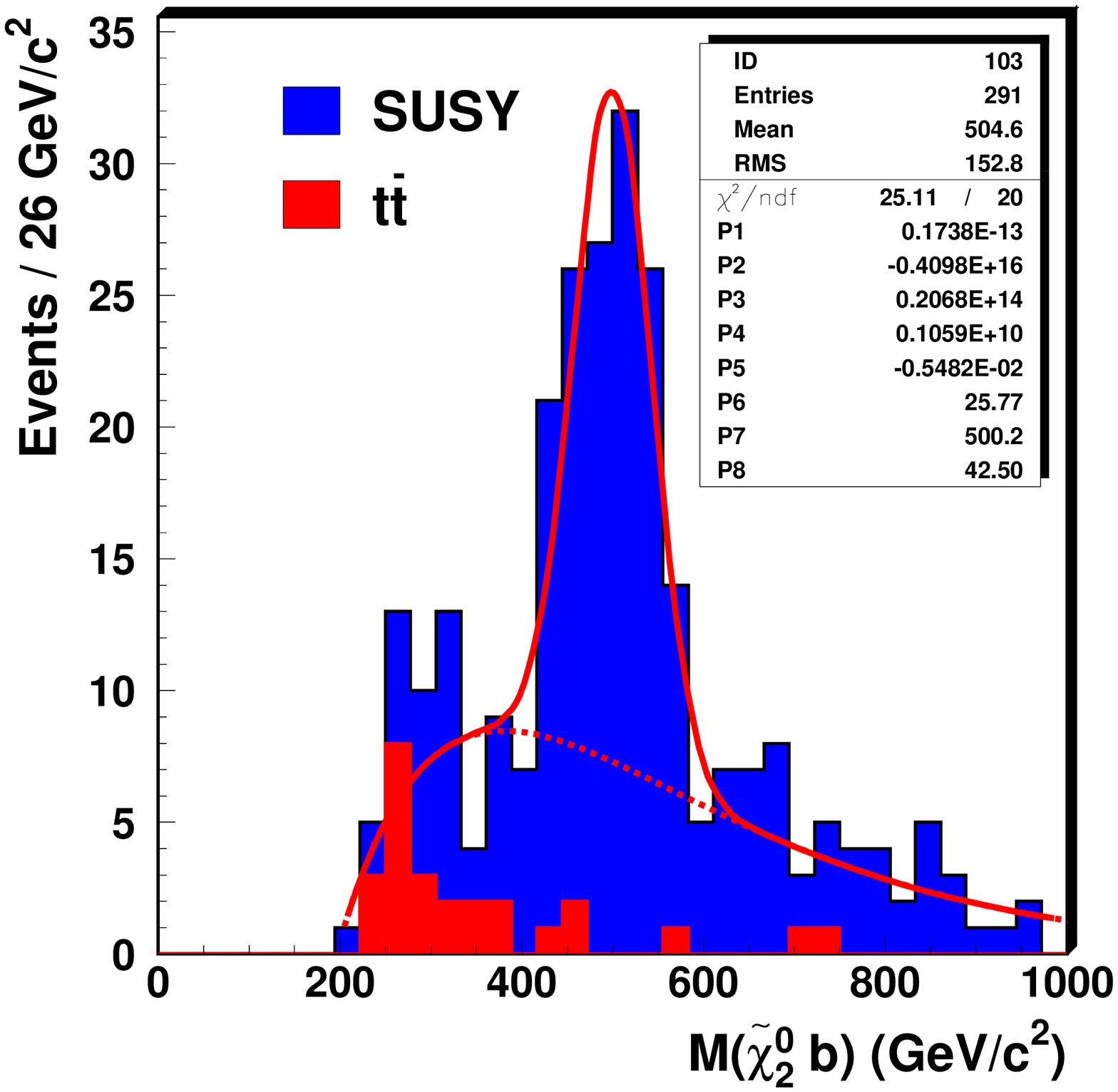
                     ,height=70mm,width=52mm}}
\put(100,-5){\epsfig{figure=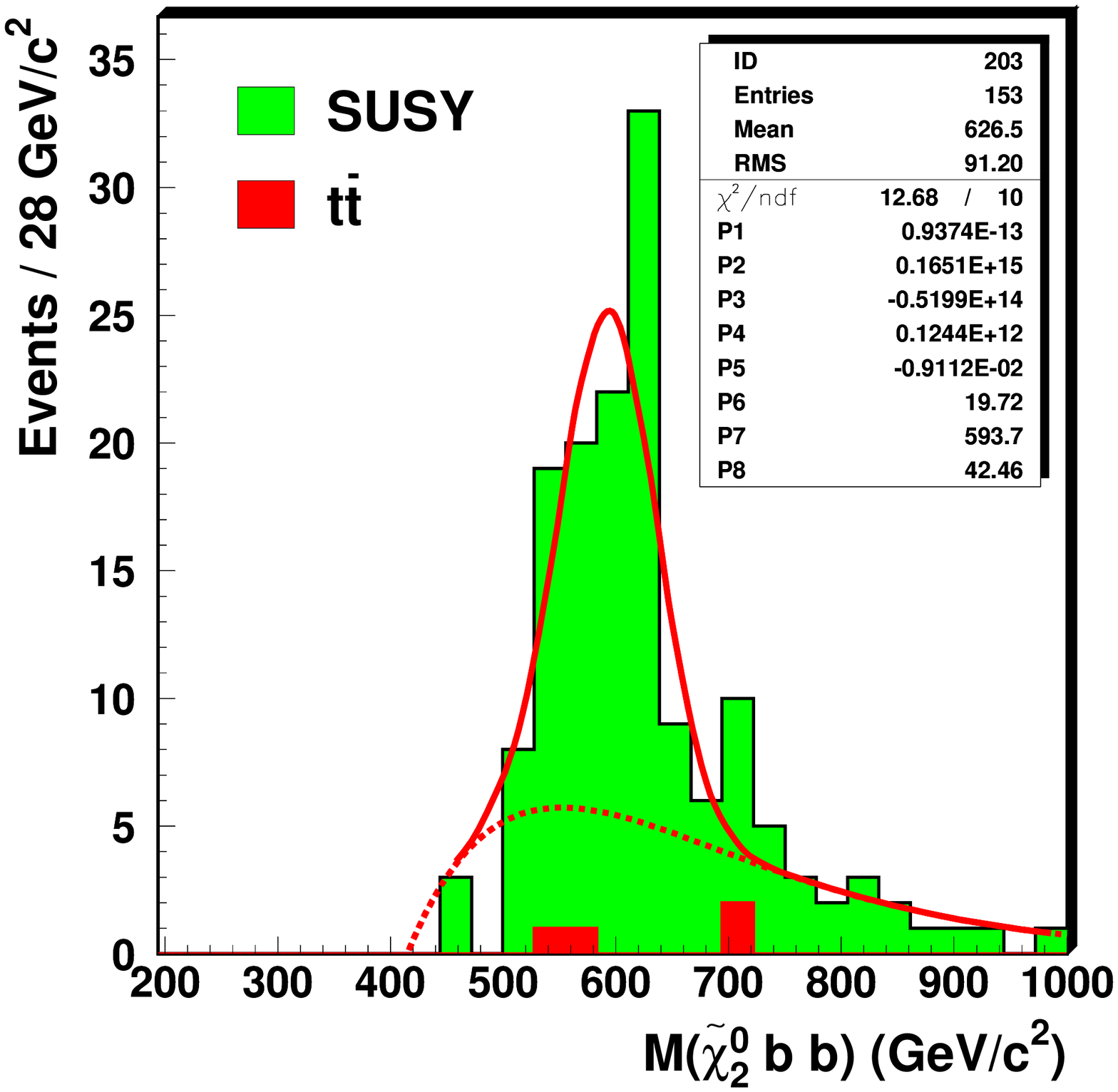
                     ,height=70mm,width=52mm}}
\end{picture}
\end{center}
\caption{Invariant mass peaks for squark (left), 
sbottom (middle) and gluino (right) at point B. The picture has 
been obtained using the parametrized simulation of the CMS detector. 
The integrated luminosity is 1~$\mbox{fb}^{-1}$ for the squarks and 
10~$\mbox{fb}^{-1}$ for the other mass peaks.}
\label{WG1:fig:susyexpintro5} 
\end{figure}

For lepton pairs with an invariant mass near the kinematical endpoint, 
the relation
\begin{equation}
p_\mu (\tilde \chi^0_2) = \left(1-\frac{m_{\tilde \chi^0_1}}{m_{ll}} \right)
 p_\mu (ll)
\end{equation}
can be used to get the four-momentum of the $\chi^0_2$, provided that the 
mass of the lightest neutralino has already been measured. This four-vector
can then be combined with that of hadronic jets to measure the gluino and 
squark masses. In \fig{WG1:fig:susyexpintro5} the gluino and squark mass 
peaks obtained with CMS parametrized simulation are reported for another 
mSUGRA benchmark point, called point B \cite{Chiorboli:2004tc}, 
which is defined by
$m_0 = 100$~GeV, $m_{1/2} = 250$~GeV, A=0, $\mu > 0$, $\tan \beta = 10$. 

Several other techniques to reconstruct the masses of Supersymmetric 
particles have been investigated by the ATLAS and CMS collaborations. 
Here we will only mention a few other possibilities:

\begin{itemize}

\item At large $\tan \beta$ the decays into third generation leptons 
are dominant. The $\tau^+ \tau^-$ kinematic endpoint is still measurable 
using the invariant mass of the tau visible decay products, but the 
expected precision is worse than that achievable with electrons and muons.

\item The right handed squark often decays directly in the LSP. 
$\tilde q_R \tilde q_R \rightarrow q \chi^0_1 q \chi^0_1$ events can be 
used to reconstruct the mass of this squark. A similar technique can 
be used to measure the mass of left-handed sleptons which decay directly 
into the LSP.

\end{itemize}

For the point SPS1a and an integrated luminosity of 300~$\mbox{fb}^{-1}$
ATLAS expects to be able to measure at least 13 mass 
relations~\cite{SPS1}. The constraints which 
can be put on the SUSY parameter space and on the relic density of 
neutralinos using these measurements are 
discussed in Ref.~\cite{Nojiri:2005ph}. 

\subsection{Flavour studies}

Most studies by the LHC collaborations have focused on the discovery 
strategies and the measurement of the masses of SUSY particles. 
However, the possibility to measure other properties of the 
new particles, such as their spin or the branching ratios of flavour 
violating decays, has also been investigated.

The measurement of the spin is interesting because it allows to 
confirm the supersymmetric nature of the new particles. This measurement
was investigated in Ref.~\cite{Barr:2004ze,Barr:2005dz} and it is also 
discussed later in this chapter.

In the hadronic sector, the experiments are not able to discriminate 
the flavour of quarks of the first two generations. Hence the only 
possibility for flavour studies relies on $b$-tagging techniques. 
In this report, the possibility to measure kinematical endpoints 
involving the scalar top is discussed. The scalar bottom masses 
may also be measured at the LHC.

The leptonic sector is more favourable from the experimental point 
of view, as the flavour of the three charge leptons can be identified 
accurately by the detectors with relatively low backgrounds. This allows 
the possibility to test the presence of decays violating lepton flavour. 
This possibility was already discussed in early 
studies~\cite{Hinchliffe:2000np,Bityukov:1997ck,Bityukov:1998va}
and it is investigated in a few contributions to this report.


\section{Effects of lepton flavour violation on di-lepton invariant mass spectra}

In this section we discuss the effect of lepton flavour violation (LFV) on
di-lepton invariant mass spectra in the decay chains
\begin{eqnarray}
\tilde \chi^0_2 \to \tilde \ell^+_i \ell^-_j \to \ell^+_k \ell^-_j \tilde \chi^0_1~.
\label{WG1:eq:chi2dec}
\end{eqnarray}
In these events one studies the invariant di-lepton mass spectrum $d N
/ d m(\ell \ell)$ with $m(\ell\ell)^2 = (p_{\ell^+} +
p_{\ell^-})^2$. Its kinematical endpoint is used in combination with
other observables to determine masses or mass differences of
sparticles \cite{Paige:1996nx,Bachacou:1999zb,Allanach:2000kt}.

\begin{figure}[t]
\begin{center}
\setlength{\unitlength}{1mm}
\begin{picture}(150,54)
\put(0,0){\epsfig{figure=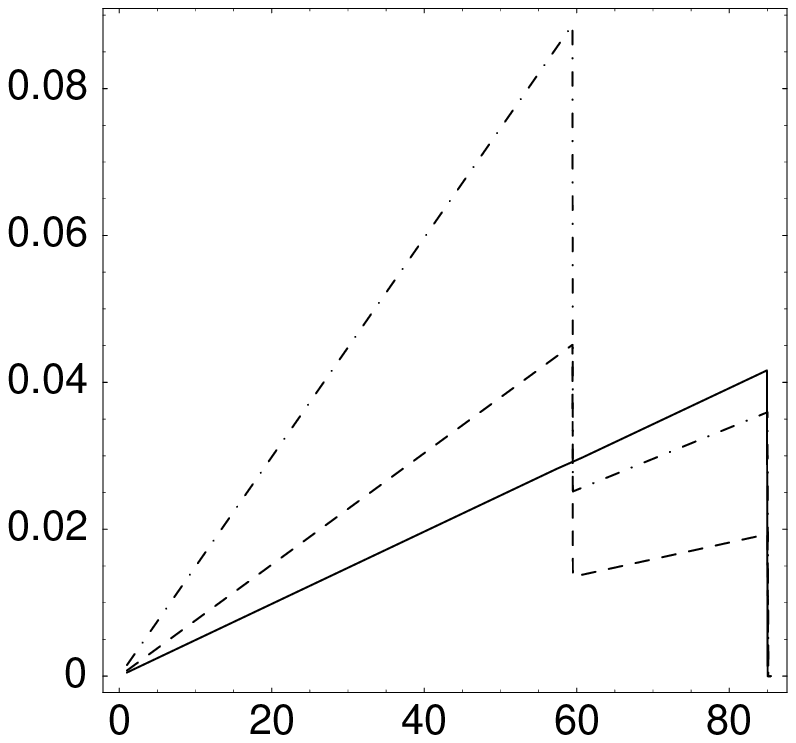,height=50mm,width=70mm}}
\put(-2,51){\small \mbox{\bf (a)}  
{$100 \Gamma^{-1}_{tot}
  d \, \Gamma(\tilde \chi^0_2 \to \ell^\pm_i \ell^\mp_j \tilde \chi^0_1) 
               / d \, m(\ell^\pm_i \ell^\mp_j)$~[GeV$^{-1}$]}}

\put(45,-2){\mbox{\small $m(\ell^\pm_i \ell^\mp_j)$~[GeV]}}
\put(32,35){\mbox{$e\tau$}}
\put(30,19){\mbox{$\mu\tau$}}
\put(60,26){\mbox{$e\mu$}}
\put(82,0){\epsfig{figure=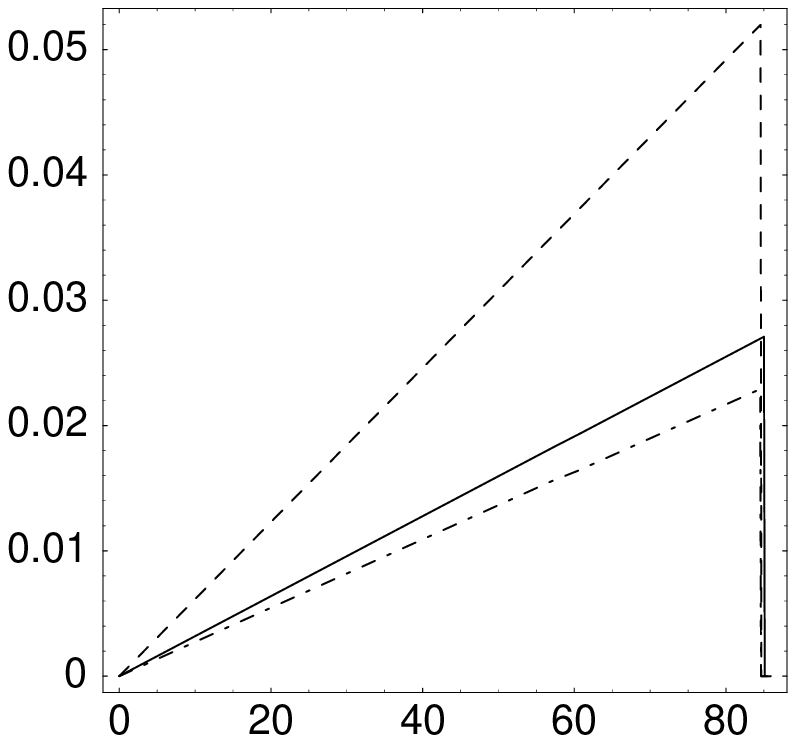,height=50mm,width=70mm}}
\put(78,51){\mbox{\small {\bf (b)} $100 \Gamma^{-1}_{tot}
  d \, \Gamma(\tilde \chi^0_2 \to \ell^+ \ell^- \tilde \chi^0_1) 
               / d \, m(\ell^+ \ell^-)$~[GeV$^{-1}$]}}
\put(126,-1){\mbox{$m(\ell^+ \ell^-)$~[GeV]}}
\put(122,34){\mbox{$\mu^+\mu^-$}}
\put(132,41){\mbox{$e^+ e^-$}}
\put(130,24){\mbox{$\mu^+\mu^-$}}
\put(131,11){\mbox{$e^+ e^-$}}
\put(131,13){\line(-2,3){2.5}}
\end{picture}
\end{center}
\caption{Invariant mass spectra 
$100\Gamma^{-1}_{tot} d \Gamma(\tilde \chi^0_2 \to \ell_i \ell_j \tilde \chi^0_1) 
               / d m(\ell_i \ell_j)$ versus $m(\ell_i \ell_j)$.
In {\bf (a)} we show
the ``flavour violating'' spectra summed over charges in the LFV case for
the SPS1a' scenario:
 $e^\pm \mu^\mp$ (full line), $e^\pm \tau^\mp$ (dashed dotted line)
and $\mu^\pm \tau^\mp$ (dashed line) and in 
{\bf (b)}  we show the ``flavour conserving'' spectra: 
$e^+ e^-$ (dashed line) and $\mu^+ \mu^-$ (dashed line) are
for the LFC case in the SPS1a' scenario, and 
$e^+ e^-$ (dashed dotted line) and 
$\mu^+ \mu^-$ (full line) are for the LFV case in the SPS1a' scenario.}
\label{WG1:fig:dilepton}
\end{figure}

\begin{table}
\caption{Relevant on-shell parameters for the SPS1a' 
\cite{Aguilar-Saavedra:2005pw} scenario.}
\begin{center}
\begin{tabular}{l|l||l|l||l|l}
\hline
$\tan\beta$ & 10        & $M_{L,11}=M_{L,22}$ & 184~GeV   &
   $M_{E,33}$ & 111~GeV \\
$M_1$       & 100.1~GeV & $M_{L,33}$          & 182.5~GeV &
   $A_{11}$ &  -0.013~GeV \\
$M_2$       & 197.4~GeV & $M_{E,11}$          & 117.793~GeV &
   $A_{22}$ & -2.8~GeV \\
$\mu$       & 400~GeV   & $M_{E,22}$          & 117.797~GeV &
  $A_{33}$  & -46~GeV \\ \hline
\end{tabular}
\end{center}
\label{WG1:tab:sps1a}
\end{table}
Details on the parameter dependence of flavour violating decays can be found
for example in ref.~\cite{Bartl:2005yy}. 
As an example  the study point SPS1a' 
\cite{Aguilar-Saavedra:2005pw} is considered which has a
 relatively light spectrum of charginos/neutralinos and
sleptons with the three lighter charged sleptons being mainly
$\tilde\ell_R$: $m_{\tilde \chi^0_1}=97.8$~GeV,
$m_{\tilde \chi^0_2}=184$~GeV, $m_{\tilde e_1}=125.3$~GeV,
$m_{\tilde \mu_1}=125.2$~GeV, $m_{\tilde\tau_1}=107.4$~GeV.
The underlying parameters are given in \tab{WG1:tab:sps1a}, where 
$M_1$ and $M_2$ are the $U(1)$ and $SU(2)$
gaugino mass parameters, respectively. 
In this example the flavour off-diagonal elements of
$M^2_{E,\alpha\beta}$ ($\alpha \ne \beta$)
 in \eq{WG1:eq:sleptonmassRR} are expected to
give the most important contribution to the LFV decays of the lighter
charginos, neutralinos and sleptons. We therefore discuss LFV only in
the right slepton sector. 
To illustrate the effect of LFV on these spectra, in
\fig{WG1:fig:dilepton} we present invariant mass distributions for various
lepton pairs taking the following LFV parameters: $M^2_{E,12} =
30$~GeV$^2$, $M^2_{E,13} = 850$~GeV$^2$ and $M^2_{E,23} =
600$~GeV$^2$, for which we have
$(m_{\tilde\ell_1},m_{\tilde\ell_2},m_{\tilde\ell_3}) =(106.4,125.1,126.2)$~GeV.
These parameters are chosen such that large LFV $\tilde\chi^0_2$ 
decay branching ratios are possible consistently with the experimental 
bounds on the rare lepton decays, for which we obtain:
 BR$(\mu^- \to e^- \gamma)=9.5 \times 10^{-12}$,
BR$(\tau^- \to e^- \gamma)=1.0 \times 10^{-7}$ 
and BR$(\tau^- \to \mu^- \gamma)=5.2 \times 10^{-8}$.
We find for the $\tilde\chi^0_2$ decay branching ratios:
BR$(e\mu)=1.7$\%, BR$(e\tau)=3.4$\%, BR$(\mu\tau)=1.8$\%, BR$(e^+ e^-)=1$\%, 
BR$(\mu^+\mu^-)=1.2$\%, BR$(\tau^+\tau^-)=51$\% with BR$(\ell_i\ell_j)\equiv 
{\rm BR}(\tilde\chi^0_2\to \ell_i\ell_j\tilde\chi^0_1)$. Note, that
we have summed here over all contributing sleptons.

In \fig{WG1:fig:dilepton}a) we show the flavour violating spectra
$(100/\Gamma_{tot}) d \Gamma(\tilde \chi^0_2 \to \ell^\pm_i \ell^\mp_j \tilde \chi^0_1) 
/ d m(\ell^\pm_i \ell^\mp_j)$ versus $m(\ell^\pm_i \ell^\mp_j)$ for the final states
$\mu \tau$, $e \tau$ and $e \mu$.
In cases where the final state contains a $\tau$-lepton, 
one finds two sharp edges. The
first one at $m \simeq 59.4$~GeV is due to an intermediate 
$\tilde \ell_1 (\sim \tilde \tau_R)$ 
and the second one at $m \simeq 84.6$~GeV is due to intermediate states of
the two heavier sleptons $\tilde\ell_2~(\sim \tilde\mu_R)$ and  
$\tilde\ell_3~(\sim \tilde e_R)$ with $m_{\tilde\ell_2}\simeq m_{\tilde\ell_3}$.
 The position of the 
edges can be expressed in terms of the neutralino and intermediate slepton 
masses \cite{Paige:1996nx}, see \eq{WG1:eq:egde}.
In the case of the $e \mu$ spectrum the first edge is practically invisible
because the branching ratios of ${\tilde \chi}^0_2$ into $\tilde\ell_1 \, e$
and $\tilde\ell_1 \, \mu$ are tiny for this example~\cite{Bartl:2005yy}.
Note that the rate for the $e \tau$ final state is largest in this case because 
$|M^2_{E,13}|$ is larger than the other LFV parameters.

In \fig{WG1:fig:dilepton}(b) we show the ``flavour conserving'' spectra for the
final states with $e^+ e^-$ and $\mu^+ \mu^-$. The dashed line corresponds to 
the flavour conserving case where $M^2_{E,ij}=0$ for $i\ne j$.
LFV causes firstly a reduction of the height of the end point peak.
Secondly, it induces a difference between 
the $\mu^+ \mu^-$ and $e^+ e^-$ spectra
because the mixings among the three slepton generations are in 
general different from each other.
The peaks at $m \simeq 59.4$~GeV in the $\mu^+ \mu^-$ and $e^+ e^-$ spectra
are invisible as in the $e\mu$ spectrum as the branching ratios of the
corresponding flavour violating decays are small.
As for the $\tau^+ \tau^-$ spectrum we remark that the height of the peak
(due to the intermediate $\tilde\ell_1$ ($\sim \tilde\tau_R$))
in the $\tau^+ \tau^-$ spectrum gets reduced by about 5\% and that the
contributions due to the intermediate $\tilde\ell_{2,3}$ are invisible.
Moreover, the peak position gets shifted to a smaller value by about 2.7 GeV 
since the mass of the intermediate $\tilde\ell_1$ gets reduced by 1 GeV 
compared to the flavour conserving case. 

It is interesting to note that in the LFV case the rate of the channel 
$e\tau$ can be larger than those of the channels with the 
same flavour, $e^+ e^-$ and $\mu^+ \mu^-$. Moreover,
by measuring all di-lepton spectra for the flavour violating as well as
flavour conserving channels, one can make an important cross check
of this LFV scenario:
the first peak position of the lepton flavour violating spectra 
(except the $e\mu$ spectrum) must coincide
with the end point of the $\tau^+ \tau^-$ spectrum and the second
peak must coincide with those of the $e^+ e^-$ and $\mu^+ \mu^-$
spectra.

Up to now  the di-lepton mass spectra
taking SPS1a' as a specific example has been  investigated in detail.
Which requirements must  other scenarios fulfill to obtain observable 
double-edge structures? Obviously the kinematic condition
$m_{\tilde\chi^0_s}> m_{\tilde\ell_i,\tilde\ell_j}>m_{\tilde\chi^0_r}$
must be fulfilled and sufficiently many  $\tilde\chi^0_s$ must be produced.
In addition there should be two sleptons contributing in a sizable 
way to the decay $\tilde\chi^0_s\to \ell' \ell''\tilde\chi^0_r$
and, of course, the corresponding branching ratio has to be large enough
to be observed. For this the corresponding LFV entries in the slepton
mass matrix have to be large enough. Moreover, also the mass difference between
the two contributing sleptons has to be sufficiently large so that the
difference of the positions of the two peaks is larger than the experimental 
resolution. In mSUGRA-like scenarios, which are
characterized by a common mass $m_0$ for the scalars and
a common gaugino mass $m_{1/2}$ at the GUT scale, the kinematic requirements 
(including the positions of the peaks) are fulfilled in the regions
of parameter space where 
$m_0^2 \lsim 0.4~m^2_{1/2}$ and $\tan\beta\gsim 8$.
The first condition provides for right sleptons lighter 
than the $\tilde\chi^0_2$
and the second condition ensures that the mass difference between
$\tilde\tau_1$ and the other two right sleptons is sufficiently large.
In the region where $m_0^2 \lsim 0.05~m^2_{1/2}$ also the left sleptons
are lighter than $\tilde\chi^0_2$, giving the possibility of additional 
structures in the di-lepton mass spectra. 

Details on background processes will be presented in the subsequent
sections, where studies by the two experiments ATLAS and CMS are
presented. Here we give a brief summary of the expected dominant
background.  The largest SM background is due to $t \bar{t}$
production. There is also SUSY background due to uncorrelated leptons
stemming from different squark and gluino decay chains. The resulting
di-lepton mass distributions will, however, be smooth and decrease
monotonically with increasing di-lepton invariant mass as was
explicitly shown in a Monte Carlo analysis in
\cite{Hinchliffe:2000np,Hisano:2002iy}.  It was also shown that the single edge
structure can be observed over the smooth background in the $e\mu$ and
$\mu\tau$ invariant mass distributions.  Therefore the novel
distributions as shown in \fig{WG1:fig:dilepton}, in particular the
characteristic double-edge structures in the $e\tau$ and $\mu\tau$
invariant mass distributions, should be clearly visible on top of the
background.  Note that the usual method for background suppression, by
taking the sum $N(e^+ e^-)+N(\mu^+ \mu^-)-N(e^\pm \mu^\mp)$, is not
applicable in the case of LFV searches. Instead one has to study the
individual pair mass spectra. Nevertheless, one can expect that these
peaks will be well observable \cite{Hinchliffepc}.


\section{Lepton flavour violation in the long-lived stau NLSP scenario}

Supersymmetric scenarios can be roughly classified into
two main classes, depending on the nature of the lightest
supersymmetric particle (LSP). The most popular
choice for the LSP is the neutralino, although scenarios
with superweakly interacting LSP, such as the gravitino
or the axino, are also compatible with all the collider
experiments and cosmology. Here, we would like to concentrate
on the latter class of scenarios, focusing for definiteness
on the case with gravitino LSP.

Under the assumption of universality of the soft-breaking scalar,
gaugino and trilinear soft terms at a high-energy scale,
the so-called constrained MSSM,  
the next-to-LSP (NLSP) can be either the lightest neutralino or the stau.
If R-parity is conserved, the  NLSP can only decay into
the gravitino and Standard Model particles, with a decay rate
very suppressed by the gravitational interactions.
As a result, the NLSP can be very long lived, with
lifetimes that could be as long as seconds, minutes
or even longer, mainly depending on the gravitino mass. When the NLSP
is the lightest neutralino, the signatures for LFV
are identical to the case with neutralino LSP, 
which have been extensively discussed in the literature~\cite{Krasnikov:1995qq,Arkani-Hamed:1996au,Hisano:1998wn,Agashe:1999bm,Nomura:2000zb,Guchait:2001us,Porod:2002zy,Deppisch:2003wt}. On the other hand, when
the NLSP is a stau, the signatures could be very different. 
In this note we discuss possible signatures and 
propose  strategies to look for
LFV in future colliders in scenarios where the gravitino (or the axino)
is the LSP and the stau is the NLSP~\cite{Hamaguchi:2004ne,Ibarra:2006sz}.

Motivated by the spectrum of the constrained MSSM
we will assume that the NLSP is mainly a 
right-handed stau, although it could have some admixture 
of left-handed stau or other leptonic flavours, and will be
denoted by $\widetilde\tau_1$. We will also assume that next 
in mass in the supersymmetric  spectrum are the right-handed 
selectron and smuon,  denoted by $\widetilde e_R$  
and $\widetilde \mu_R$  respectively, also with a 
very small admixture of left-handed states and
some admixture of stau. Finally, we will assume that
next in mass are the lightest neutralino and the rest of 
SUSY particles. Schematically, the spectrum reads:
\begin{equation}
m_{3/2}<m_{\widetilde\tau_1}<m_{\widetilde e_R,\;\widetilde\mu_R}<
m_{\chi^0_1}, 
m_{\widetilde e_L,\widetilde\mu_L}, m_{\widetilde\tau_2}...
\label{WG1:eq:spectrum-stauNLSP}
\end{equation}

In this class of scenarios, staus could be long lived
enough to traverse several layers of the vertex detector
before decaying, thus being detected as a heavily ionizing
charged track. This signature is very distinctive and is
not produced by any Standard Model particle, hence
the observation of heavily ionizing charged tracks would give
strong support to this scenario and would allow
the search for LFV essentially 
without Standard Model backgrounds. 

Long lived staus could even be stopped in the detector 
and decay at late times,
producing very energetic particles that would spring
from inside the detector. Recently, prospects of collecting staus and
detecting their decay products in future colliders have been 
discussed~\cite{Hamaguchi:2004df,Feng:2004yi}. At the LHC, cascade
decays of squarks and gluinos could produce of the order of $10^6$ staus
per year if the sparticle masses are close to the present experimental
limits~\cite{Beenakker:1996ch}.
Among them, ${\mathcal O}(10^3$--$10^4)$ staus could be collected by
placing 1--10 kton massive material around the LHC detectors.  On the
other hand, at the ILC up
to ${\mathcal O}(10^3$--$10^5)$ staus could be collected and studied.

If there is no LFV, the staus
could only decay into taus and gravitinos,
$\widetilde \tau\rightarrow \tau \psi_{3/2}$. If on the contrary
LFV exists in nature, some of the staus could decay into electrons
and muons. Therefore the detection of very energetic electrons and muons 
coming from inside the detector would constitute a signal of lepton flavour
violation. 

There are potentially two sources of background in this analysis.
First, in certain regions of the SUSY parameter space selectrons or smuons
could also be long lived, and the electrons and muons from their
flavour conserving decays could be mistaken for electrons and muons
coming from the lepton flavour violating  decay of the stau. However,
if flavour violation is
large enough to be observed in these experiments, the selectron
decay channel $\widetilde e\rightarrow\widetilde \tau \;e\; e$ is very 
efficient. Therefore,
selectrons (and similarly, smuons) are never long lived 
enough to represent an important
source of background. It is remarkable the interesting double role
that flavour violation plays in this experiment, both as object
of investigation and as crucial ingredient for the success
of the experiment itself. 

A second source of background for this analysis are the muons
and electrons from tau decay, that could be mistaken for muons
and electrons coming from the lepton
flavour violating decays ${\widetilde \tau}\rightarrow \mu \; \psi_{3/2}$,
${\widetilde \tau}\rightarrow e\; \psi_{3/2}$. Nevertheless, 
this background can be distinguished from the signal
by looking at the energy spectrum: the leptons from the flavour
conserving tau decay present a continuous energy spectrum, in stark
contrast with the leptons coming from the two body gravitational
decay, whose energies are sharply peaked at $E_0 =
(m^2_{{\widetilde \tau},\,{\widetilde e}}+ m^2_{\mu,\,e}-m^2_{3/2})/(2
m_{{\widetilde \tau},\,{\widetilde e}})$.  
It is easy to check that only a very small
fraction of the electrons and muons from the tau decay have energies
close to this cut-off energy.  For instance, for the typical 
energy resolution of an electromagnetic calorimeter,
$\sigma\simeq 10\%/\sqrt{E({\rm GeV})}$, only
$2\times 10^{-5}$ of the taus with energy 
$E_0\sim 100$GeV will produce electrons
with energy $\simeq E_0$, within the energy 
resolution of the detector, which could be mistaken for electrons coming
from the LFV stau decay. Therefore, 
for the number of NLSPs that can be typically trapped at
the LHC or the ILC, the number of electrons or muons from this source
of background turns out to be negligible in most instances.

Using this technique, we have estimated that at the LHC or at 
the future Linear
Collider it would be possible
to probe mixing angles in the slepton sector
down to the level of 
$\sim 3\times 10^{-2}~(9\times 10^{-3})$ at 90\% confidence 
level if $3\times 10^3~(3\times 10^4)$  staus are
collected~\cite{Hamaguchi:2004ne}. A different technique,
that does not require to stop the staus, was proposed
in \cite{Ibarra:2006sz} for the case of an $e^+ e^-$ or $e^-e^-$ linear
collider.


\section{Neutralino decays in models with broken R-parity}
\label{WG1:sec:rpbi}

In supersymmetric models neutrino masses can be explained
intrinsically supersymmetric, namely the breaking of R-parity.  The
simplest way to realize this idea is to add the bilinear terms of
$W_{R_p \hspace{-3mm}/}$ to the MSSM superpotential $W_{\rm MSSM}$
(see \eqs{WG1:eq:mssmsuper}{WG1:eq:rpvsuper1}):
\begin{eqnarray}
W = W_{\rm MSSM} + \epsilon_i \hat L_i \hat H_u
\label{WG1:eq:model}
\end{eqnarray}
For consistency one has also to add the corresponding bilinear terms
to soft SUSY breaking (see \eqs{WG1:eq:mssmsoft}{WG1:eq:rpvsoft1}) which
induce small vevs for the sneutrinos. These vevs in turn induce a
mixing between neutrinos and neutralinos, giving mass to one neutrino
at tree level. The second neutrino mass is induced by loop effects
(see
\cite{Romao:1999up,Hirsch:2000ef,Diaz:2003as} and references
therein). The same parameters that induce neutrino masses and mixings
are also responsible for the decay of the lightest supersymmetric
particle (LSP). This implies that there are correlations between
neutrino physics and LSP decays
\cite{Mukhopadhyaya:1998xj,Porod:2000hv,Hirsch:2002ys,Hirsch:2003fe,%
AristizabalSierra:2004cy}.

 In particular, the neutrino mixing angles 
\begin{eqnarray}
\tan^2 \theta_{atm} \simeq \left( \frac{\Lambda_2}{\Lambda_3}\right)^2 \, , \,
U^2_{e3} \simeq \frac{|\Lambda_1|}{\sqrt{\Lambda_2^2+\Lambda_3^2}}\, , \,
\tan^2 \theta_{sol} \simeq
 \left( \frac{\tilde\epsilon_1}{\tilde \epsilon_2}\right)^2
\end{eqnarray}
can be related to ratios of couplings and branching ratios, for example
\begin{eqnarray}
 \tan^2 \theta_{\rm atm} \simeq \left| \frac{\Lambda_2}{\Lambda_3} \right|^2
   \simeq \frac{BR({\tilde \chi}^0_1 \to \mu^\pm W^\mp)}
               {BR({\tilde \chi}^0_1 \to \tau^\pm W^\mp)}
   \simeq  
 \frac{BR({\tilde \chi}^0_1 \to \mu^\pm \bar{q} q')}
      {BR({\tilde \chi}^0_1 \to \tau^\pm \bar{q} q' )},
\label{WG1:eq:corr}
\end{eqnarray}
in the case of a neutralino LSP. Here $\Lambda_i = \epsilon_i v_d + \mu v_i$,
$v_i$ are the sneutrino vevs and $v_d$ is the vev of $H^0_d$; 
$\tilde \epsilon_i = V_{ij} \epsilon_j$ where $ V_{ij}$ is the neutrino
mixing matrix at tree level which is given as a function of the $\Lambda_i$.
Details on the neutrino masses and mixings can be found in 
\cite{Hirsch:2000ef,Diaz:2003as}.

\begin{figure}[t]
\setlength{\unitlength}{1mm}
\begin{picture}(150,120)
\put(1,63){\mbox{\epsfig{figure=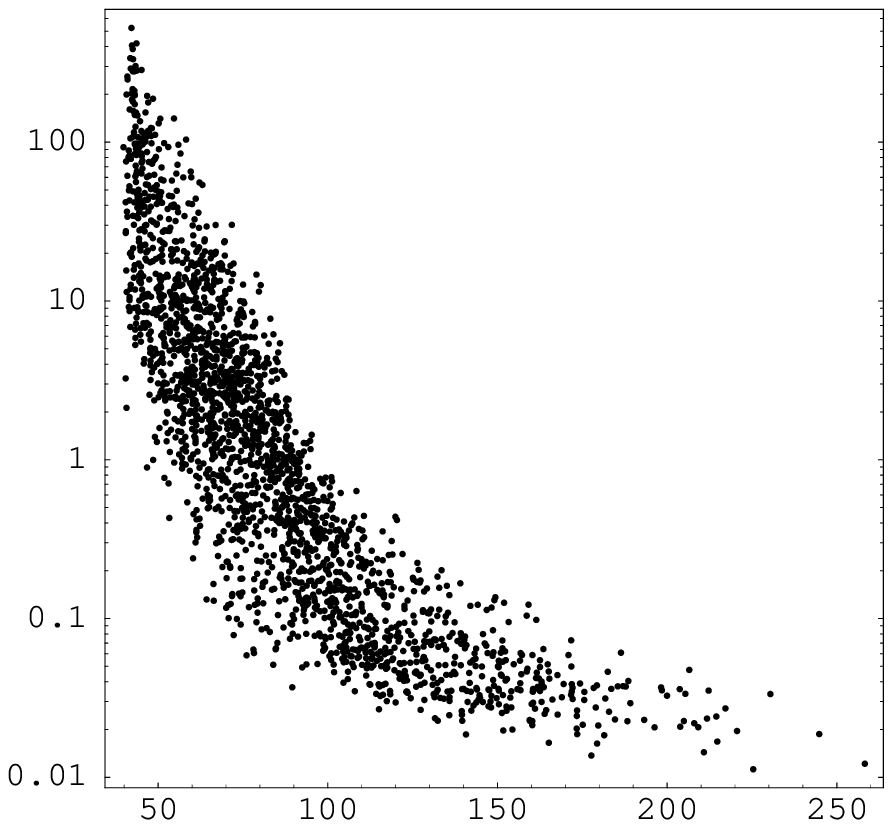,
height=5.2cm,width=7.cm}}}
\put(0,117){\makebox(0,0)[bl]{{ a)}}}
\put(5,116){\makebox(0,0)[bl]{{\small $L({\tilde \chi}^0_1)$~[cm]}}}
\put(74,60){\makebox(0,0)[br]{{ $m_{{\tilde \chi}^0_1}$~[GeV]}}}
\put(82,62){\mbox{\epsfig{figure=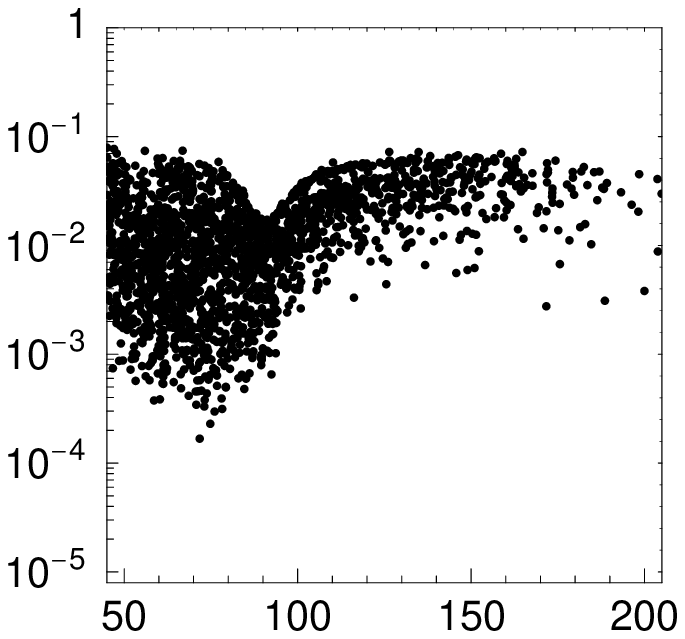,
height=5.5cm,width=7.2cm}}}
\put(82,117){\makebox(0,0)[bl]{{ b)}}}
\put(87,116){\makebox(0,0)[bl]{{\small 
        BR(${\tilde \chi}^0_1 \to \sum_{i,j,k} \nu_i \nu_j \nu_k$)}}}
\put(150,60){\makebox(0,0)[br]{{$m_{{\tilde \chi}^0_1}$~[GeV]}}}
\put(0,0){\mbox{\epsfig{figure=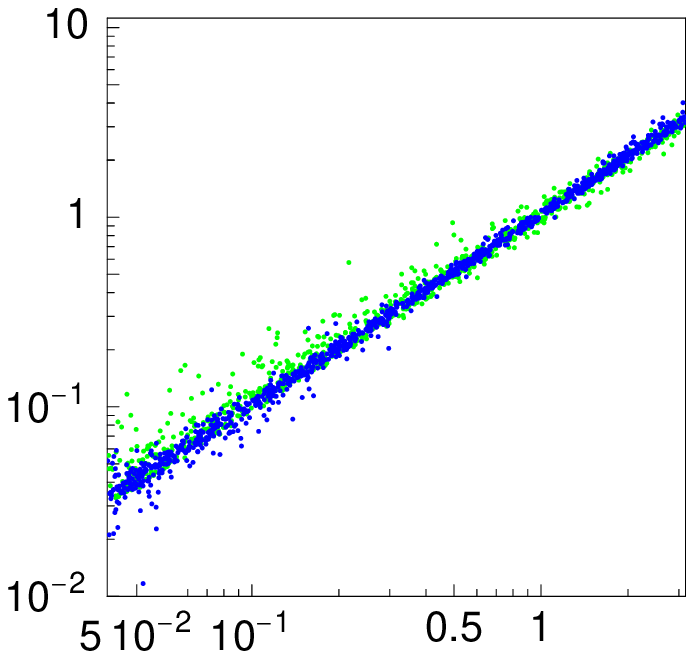,
      height=5.5cm,width=7.2cm}}}
\put(0,56){\makebox(0,0)[bl]{{ c)}}}
\put(5,55){\makebox(0,0)[bl]{{\small  BR$(\tilde \chi^0_1 \to  \mu q' \bar q)/$
BR$(\tilde \chi^0_1 \to\tau q' \bar q)$}}}
\put(74,-2){\makebox(0,0)[br]{{ $\tan^2(\theta_{atm})$}}}
\put(82,1){\mbox{\epsfig{figure=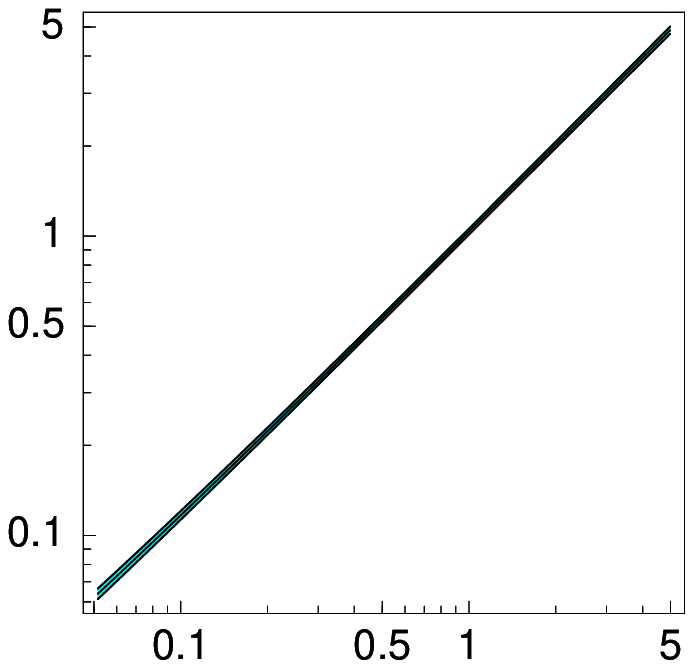,
     height=5.3cm,width=7.2cm}}}
\put(82,56){\makebox(0,0)[bl]{{ d)}}}
\put(87,55){\makebox(0,0)[bl]{{\small BR$(\tilde \chi^0_1 \to  \mu q' \bar q)/$
BR$(\tilde \chi^0_1 \to\tau q' \bar q)$}}}
\put(150,-2){\makebox(0,0)[br]{{$\tan^2(\theta_{atm})$}}}
\end{picture}
\caption[]{Various neutralino properties:  a) Neutralino decay length
and b) invisible neutralino
 branching ratio summing over all neutrinos as a function of 
$m_{{\tilde \chi}^0_1}$; c) 
BR$(\tilde \chi^0_1 \to  \mu q' \bar q)/
$BR$(\tilde \chi^0_1 \to\tau q' \bar q)$ scanning over the SUSY parameter and
 d) 
BR$(\tilde \chi^0_1 \to  \mu q' \bar q)/$
BR$(\tilde \chi^0_1 \to\tau q' \bar q)$ for 10\% variations around a
 fixed SUSY point as a function of $\tan^2(\theta_{atm})$.}
\label{WG1:fig:NeutProp}
\end{figure}

The smallness of the $R$-parity violating couplings which is required by the
neutrino data implies that the production and decays of the SUSY particles
proceed as in the MSSM with conserved $R$-parity except that the LSP
decays. There are several  predictions for the LSP properties 
discussed in the literature above. 
Here we discuss various important examples pointing out
generic features. The first observation is, that the smallness
of the couplings can lead to finite decay lengths of the LSP which
are measurable at LHC. As an example we show in \fig{WG1:fig:NeutProp}a
the decay length of a neutralino LSP as a function of its mass. The SUSY
parameters have been varied such that collider constraints as well as
neutrino data are fulfilled. This is important for LHC as a secondary vertex
for the neutralino decays implies that the neutralino decay products
can be distinguished from the remaining leptons and jets within a cascade
of decays. A first attempt to use this to establish the predicted
correlation between neutralino decays and neutrino mixing angles
has been presented in \cite{Porod:2004rb}. The finite decay length
can also be used to  enlarge the reach of colliders for SUSY searches as
has been shown in ref.~\cite{deCampos:2005ri} for the Tevatron
and in ref.~\cite{deCampos:2007bn} for the LHC. 
The fact, that the decay products of the neutralino can be identified
via a secondary vertex is important for the check if the predicted
correlations indeed exist. As an example we show in 
\fig{WG1:fig:NeutProp}c and d the ratio 
BR$(\tilde \chi^0_1 \to  \mu q' \bar q)/$
BR$(\tilde \chi^0_1 \to\tau q' \bar q)$
as  a function of the atmospheric neutrino mixing angle $\tan^2(\theta_{atm})$.
In \Bfig{WG1:fig:NeutProp}c a general scan is performed over the SUSY 
parameter space yielding a good correlation whereas in 
\fig{WG1:fig:NeutProp}d the situation is shown if one assumes that
the underlying SUSY parameters are known with an accuracy of 10\%.
The branching ratios themselves are usually of order 10\%.

It is usually argued that broken R-parity implies that the missing energy
signature of the MSSM is lost. This is not entirely correct if 
$R$-parity is broken via lepton number breaking as in the model discussed
here. The reason is that neutrinos are not detected at LHC or ILC.
This implies that the missing energy signature still is there although somewhat
reduced. However, there are still cases where the LSP can decay
completely invisible: $\tilde \chi^0_1 \to 3 \nu$ or 
$\tilde \nu_i \to \nu_j \nu_k$. In \Bfig{WG1:fig:NeutProp}b we see
that the decay branching ratio for $\tilde \chi^0_1 \to 3 \nu$ can go up
to several per-cent. In the sneutrino case is at most per-mille 
\cite{AristizabalSierra:2004cy}. If one adds trilinear $R$-parity
breaking couplings to the model, then these branching ratios will be reduced.
In models with spontaneous breaking of $R$-parity the situation can
be quite different, e.g.~the invisible modes can have in total nearly
100\% branching ratio \cite{Hirsch:2006di}.

\begin{figure}[t]
\setlength{\unitlength}{1mm}
\begin{picture}(150,60)
\put(0,0){\mbox{\epsfig{figure=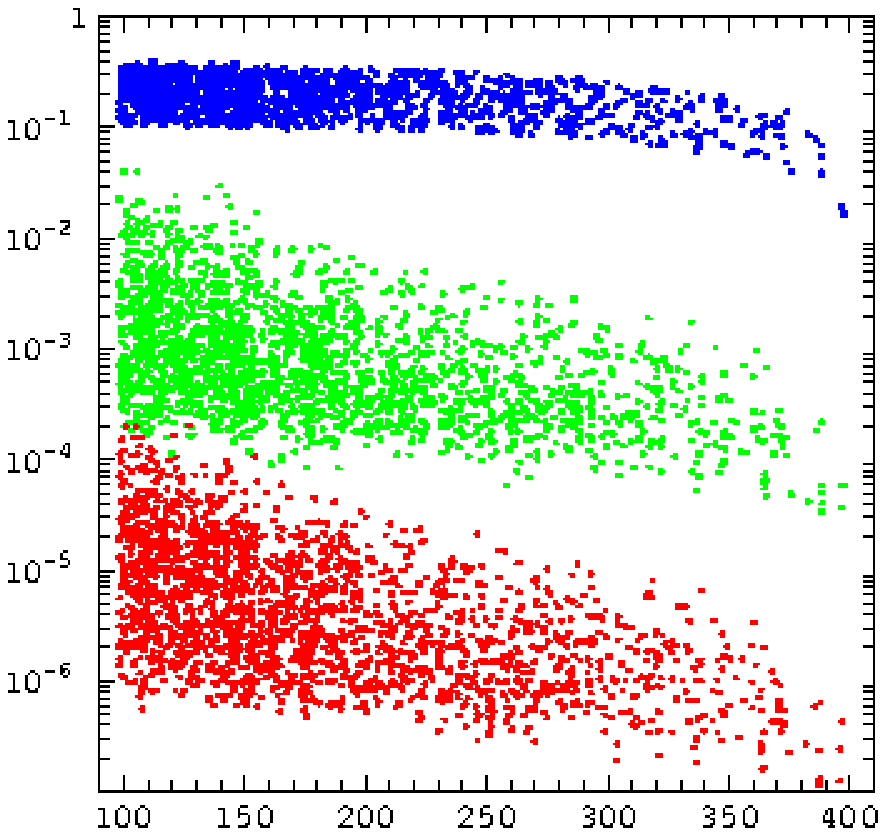,
      height=5.5cm,width=7.2cm}}}
\put(0,56){\makebox(0,0)[bl]{{ a)}}}
\put(5,55){\makebox(0,0)[bl]{{\small Decay length ($\tilde e$, $\tilde \mu$,
 $\tilde \tau$ ) [cm]}}}
\put(69,48){\makebox(0,0)[br]{{$L(\tilde e)$ }}}
\put(69,32){\makebox(0,0)[br]{{$L(\tilde \mu)$ }}}
\put(69,14){\makebox(0,0)[br]{{$L(\tilde \tau)$ }}}
\put(74,-2){\makebox(0,0)[br]{{$m_{\tilde l}$ }}}
\put(82,-28){\mbox{\epsfig{figure=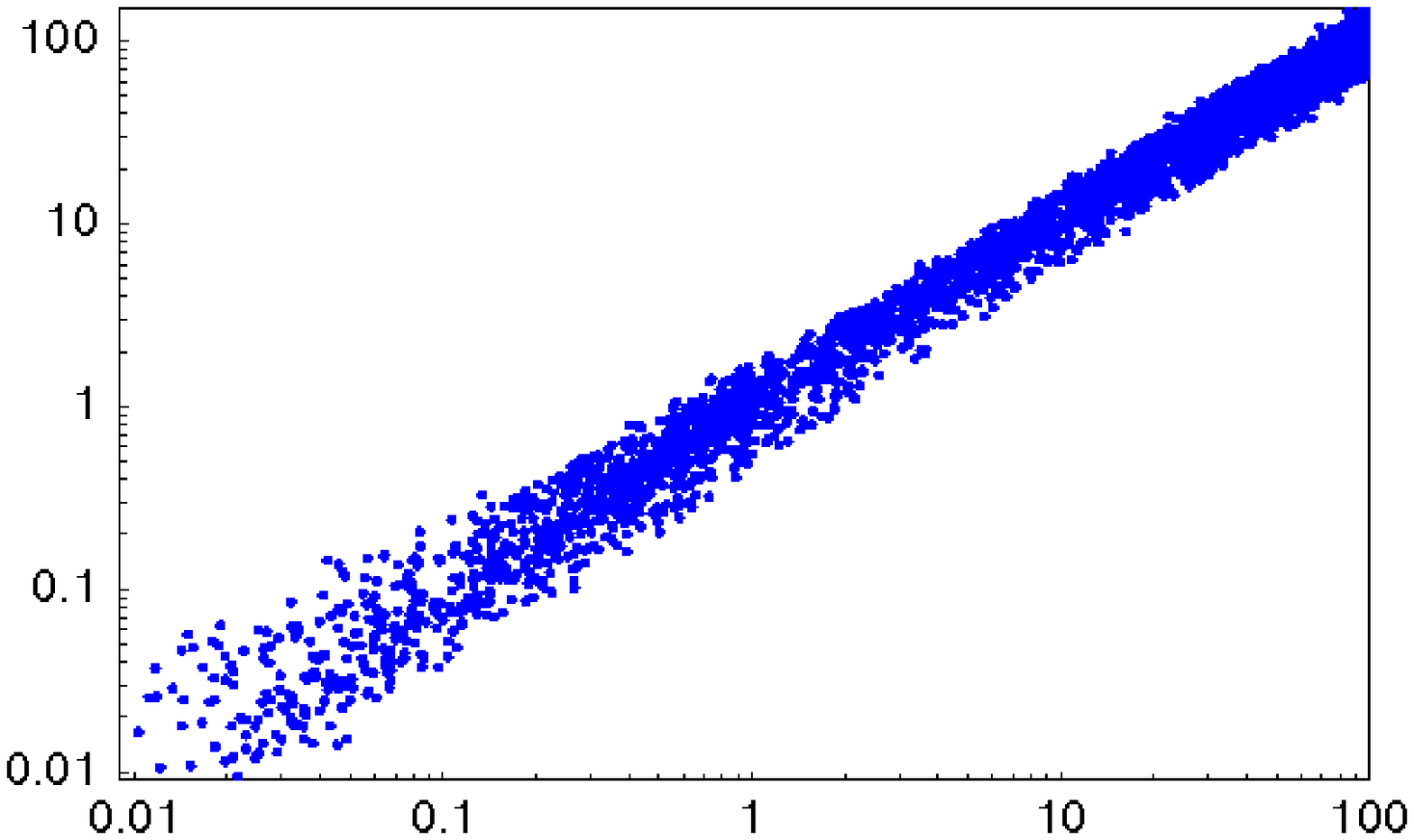,
     height=11.3cm,width=7.cm}}}
\put(82,56){\makebox(0,0)[bl]{{ b)}}}
\put(87,55){\makebox(0,0)[bl]{{\small 
BR$(\tilde \tau_1 \to  e \nu)/$
BR$(\tilde \tau_1 \to \mu \nu)$}}}
\put(150,-2){\makebox(0,0)[br]{{$\tan^2(\theta_{sol})$}}}
\end{picture}
\caption[]{Various slepton properties:  a) decay lengths
 as a function of  $m_{\tilde l}$ and b)
BR$(\tilde \tau_1 \to  e \nu)/$
BR$(\tilde \tau_1 \to \mu \nu)$ as a function of $\tan^2(\theta_{sol})$.}
\label{WG1:fig:SlepProp}
\end{figure}

As a second example, we present in \fig{WG1:fig:SlepProp}a the decay
lengths slepton LSPs as they motivated in GMSB models. Also in this
case we have performed a generous scan of the SUSY parameter space.
One sees that the sleptons have different decay lengths which is again
useful to distinguish the various 'flavours'. However, at LHC
it might be difficult to separate smuons from staus in this scenario.
provided this is possible one could measure for example the correlation
between stau decay modes and the solar neutrino mixing angle as
shown  in \fig{WG1:fig:SlepProp}b.


\section{Reconstructing neutrino properties from collider
    experiments in a Higgs triplet neutrino mass model}
\label{slep:triplet}

In the previous section the neutrino masses are solely due to
$R$-parity violation and the question arises how the situation changes
if there are additional sources for neutrino masses. Therefore a model
is considered where Higgs triplets give additional contributions to the
neutrino masses. It can either be obtained as a limit of spontaneous $R$-parity
breaking models discussed in \sect{WG1:sec:rpspon} 
or as  the supersymmetric extension of the original triplet model of
neutrino mass~\cite{Schechter:1980gr} with additional bilinear
$R$-parity breaking terms~\cite{Hall:1983id,Ross:1984yg,Ellis:1984gi}.
The particle content is that of the MSSM augmented by a pair of Higgs triplet
superfields, $\widehat{\Delta}_u$ and $\widehat{\Delta}_d$, with
hypercharges $Y=+2$ and $Y=-2$, and lepton number $L=-2$ and $L=+2$,
respectively. The superpotential of this model is then given by a sum
of three terms,
\begin{eqnarray} \label{WG1:eq:higgstreesuper}
W&=&W_{\mathrm{MSSM}} + \epsilon_i \hat L_i \hat H_u +W_{\Delta} \\
W_{\Delta}&=&\mu_{\Delta}\widehat{\Delta}_u\widehat{\Delta}_d+h_{ij}
\widehat{L}_i\widehat{L}_j\widehat{\Delta}_u
\label{WG1:eq:triplet-superpotential}
\end{eqnarray}
 Additional details of the model can be found in
ref.~\cite{AristizabalSierra:2003ix}. 
From the analytical study of the Higgs sector, it is possible to show that the
Higgs triplet vevs are suppressed by two powers of the BRPV parameters, 
as already emphasized in Ref.~\cite{Ma:2002xt}. 

The nonzero vevs of this model ($v_u\equiv\langle H_u^0\rangle,\, v_d\equiv
\langle H_d^0\rangle,\, v_i\equiv
\langle\widetilde{\nu}_i\rangle,\,\langle\Delta^0_u\rangle$ and  $\langle\Delta^0_u\rangle$) 
produce a mixing between neutrinos, gauginos and Higgsinos. 
For reasonable ranges of
parameters, atmospheric neutrino physics is determined by the BRPV 
parameters, whereas the solar neutrino mass scale
depends mostly on the triplet Yukawa couplings and the triplet mass. 
This situation is different from the one in the model with only BRPV, where 
the solar mass scale is generated by radiative corrections to neutrino masses, 
thus requiring $\epsilon^2/\Lambda\sim\mathcal{O}(0.1-1)$.
Now, as the solar mass scale is generated by the Higgs triplet, $\epsilon_i$ can be 
smaller.
Using the experimentally measured values of $\tan^2\theta_{\textsc{atm}} \simeq 1$ 
and $\sin^22\theta_{\textsc{chooz}} \ll 1$ one can find a simple formula for the 
solar angle in terms of the Yukawa couplings $h_{ij}$ of the triplet Higgs boson to the doublet leptons, which is approximately given by
\begin{equation}
\tan(2\theta_{\textsc{sol}})\simeq\frac{-2\sqrt{2}(h_{12} - h_{13})}
{-2h_{11}+h_{22}+h_{33} -2h_{23}}\equiv x
\label{WG1:eq:x}
\end{equation}

One of the characteristic features of the triplet model of neutrino
mass is the presence of doubly charged Higgs bosons $\Delta^{--}_u$.
At LHC, the doubly charged Higgs boson can be produced in different
processes, such as:
(a) It can be singly produced via vector boson fusion or via the
fusion of a singly charged Higgs boson with either a vector boson or
another singly charged Higgs boson; its production cross section is
$\sigma(WW,WH,HH\to\Delta)=\Unit{(10-1.5)}{fb}$ for a triplet mass of
$M_{\Delta}=(300-800)\UGeV$, assuming the triplet vev to be
$9$~GeV~\cite{Azuelos:2005uc,Huitu:1996su}. However, the triplet VEV
is of order eV in this model, thus suppressing this production mechanism.
(b) It can be doubly produced via a Drell-Yan process, with $\gamma/Z$ exchange in the
$s$-channel; 
its production cross section is $\sigma(q\bar{q}\to\gamma/Z\to\Delta\Delta)=\Unit{(5-0.05)}{fb}$ for a triplet mass of $M_{\Delta}=(300-800)\UGeV$~\cite{Dion:1998pw}.
(c) It can be singly produced with a singly charged Higgs boson, with
the exchange of $W$ in the $s$-channel; its production cross section
is $\sigma(q\bar{q}'\to W\to\Delta H)=\Unit{(35-0.3)}{fb}$ for a
triplet mass of $M_{\Delta}=(300-800)\UGeV$, where some splitting
among the masses of the doubly and singly charged Higgs bosons is
allowed~\cite{Akeroyd:2005gt,Dion:1998pw}.
Assuming a luminosity of $\mathcal{L}=\Unit{100}{(fb\,$\cdot$ year)}^{-1}$ for the LHC, 
the number of events for the above mentioned production processes is $\mathcal{O}(10^3-10^1)$ per year, depending on the Higgs triplet mass.

\begin{figure}[t]
\begin{center}
\includegraphics[width=75mm,height=50mm]{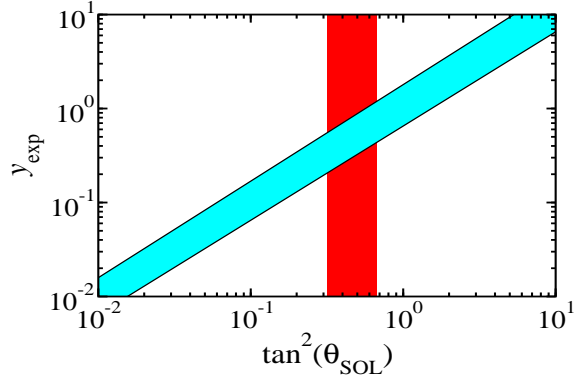}
\end{center}
\caption{Ratio of doubly charged Higgs boson leptonic decay branching ratios
 (assuming a $10\%$ uncertainty)
indicated by the variable $y_{\mathrm{exp}}$ of
\eqs{WG1:eq:yratio}{WG1:eq:x-exp} versus the solar neutrino mixing
angle. The vertical band indicates current $3\sigma$ allowed range. }
\label{WG1:fig:tansqsolvsyexp01}
\end{figure}

The most remarkable feature of the present model is that the decays of 
the doubly charged Higgs bosons can be a perfect tracer of the
solar neutrino mixing angle. 
Considering \eq{WG1:eq:x} and taking into account that the leptonic
decays of the doubly charged Higgs boson are proportional to
$h_{ij}^2$, we construct the following ratio
\begin{equation}
  \label{WG1:eq:x-exp}
x_{\mathrm{exp}}\equiv\frac{-2\sqrt{2}(\sqrt{\mathrm{BR}_{12}}-\sqrt{\mathrm{BR}_{13}})}{-2\sqrt{2 \mathrm{BR}_{11}}+\sqrt{2 \mathrm{BR}_{22}}+\sqrt{2 \mathrm{BR}_{33}}-2\sqrt{\mathrm{BR}_{23}}}
\end{equation}
with $\mathrm{BR}_{ij}$ denoting the measured branching ratio for the process 
($\Delta_u^{--} \to l_i^-l_j^-$). 
\Bfig{WG1:fig:tansqsolvsyexp01} shows the ratio $y_{\mathrm{exp}}$ of the 
leptonic decay branching ratios of the doubly charged Higgs boson
versus the solar neutrino mixing angle.
The ratio of
doubly charged Higgs boson decay branching ratios is
specified by the variable
\begin{equation}
  \label{WG1:eq:yratio}
y_{\mathrm{exp}}\equiv\tan^2\left(\frac{\arctan(x_{\mathrm{exp}})}{2}\right)  
\end{equation}
where, 
for the determination of $x_{\mathrm{exp}}$, 
a $10\%$ uncertainty in the measured branching ratios has been assumed and the triplet
mass has been fixed at $M_{\Delta_u}=\Unit{500}{GeV}$. 
As can be seen from the figure, there is a very strong correlation
between the pattern of Higgs triplet decays and the solar neutrino mixing angle. 
The $3\sigma$ range permitted by current solar and
reactor neutrino data 
 (indicated by the vertical band in   
\fig{WG1:fig:tansqsolvsyexp01}) fixes a minimum value for $y_{\mathrm{exp}}$, 
thus requiring minimum values for the off-diagonal leptonic decay
channels of the doubly charged Higgs triplet. If $\mathrm{BR}_{23}=0$,
at least either $\mathrm{BR}_{12}$ or $\mathrm{BR}_{13}$
must be larger than $0.5$. On the other hand, if $\mathrm{BR}_{23}\ne 0$, then
at least one of the  off-diagonal branching ratios 
must be larger than $0.2$.

As  in  \sect{WG1:sec:rpbi}, 
the decay pattern of a neutralino LSP
is predicted in terms of the atmospheric neutrino mixing angle
The main difference is, that
the $\epsilon_i$ can be smaller in this model compared to the previous
one. This implies that the main decay mode BR(${\tilde \chi}^0_1 \to \nu b\bar{b}$)
gets reduced \cite{Porod:2000hv} and the branching ratios into the final
states 
$l q q'$ ($l=e,\mu\tau$) increase.


\section{SUSY (s)lepton flavour studies with ATLAS}
\label{WG1:sec:atlasslep}

In this section main features of Monte Carlo studies for slepton
masses and spin measurements are presented as well as a study of
slepton non-universality. As a reference model the SPS1a point is
taken \cite{SPS1}, which is derived from the following high scale
parameters: $m_{0}$= 100 GeV, $m_{1/2}=$ 250 GeV, $A_{0}=$- 100 GeV,
$\tan{\beta}=10$ and sign$(\mu)=+$, where $m_{0}$ is a common scalar
mass, $m_{1/2}$ a common gaugino mass, $A_{0}$ a common trilinear
coupling, $\tan\beta$ the ratio of the Higgs vacuum expectation
values.

 Sleptons are produced either directly in pairs 
$\tilde{l}^{+}\tilde{l}^{-}$ or indirectly from decays
of heavier charginos and neutralinos (typical mode $\;\tilde{\chi}_{2}^{0}\,\rightarrow\, \tilde{l}_{R}l\;$). They can decay according to:  
$\;\tilde{l}_{R}\, \rightarrow \,l\, \tilde{\chi}_{1}^{0}\; $, 
$\;\tilde{l}_{L}\, \rightarrow \,l\, \tilde{\chi}_{1}^{0}\;$, 
$\;\tilde{l}_{L}\, \rightarrow \,l\, \tilde{\chi}_{2}^{0}\;$,
$\;\tilde{l}_{L}\, \rightarrow \,\nu\, \tilde{\chi}_{1}^{\pm}\;$. 
At the end of every SUSY decay chain is undetectable lightest
neutralino $\tilde{\chi}_{1}^{0}$ and kinematic endpoints in the
invariant mass distributions are measured rather than the mass
peaks. Kinematic endpoints are the function of SUSY masses which can
be extracted from the set of endpoint measurements. Fast simulation
studies  of left squark
cascade decay
$\;\tilde{q}_{L}\,\rightarrow\,\tilde{\chi}_{2}^{0}\,q\,\rightarrow\,
\tilde{l}^{\pm}_{R}\,l^{\mp}\,q \,\rightarrow\,l^{+}\,l^{-}\,q\,\;\tilde{\chi}_{1}^{0} \;\;
(l=e,\,\mu)\;$ were performed in
refs.~\cite{Gjelsten:2004ki,SPS1}. Events with two same flavour and
opposite sign (SFOS) leptons, at least 4 jets with
$\;p_{T}>150,100,50,50\; \text{GeV}\;$, and effective mass
$\;M_{eff}=\Sigma_{i=1}^{4}p_{T}(jet)+ \ETmi > 600\;
\text{GeV}$ and missing transverse energy $\ETmi > \max(100\,{\rm GeV},
\;0.2M_{eff})\;$ were selected.  Flavour subtraction
$\,e^{+}e^{-}\,+\,\mu^{+}\mu^{-}\,-\,e^{\pm}\mu^{\mp}\,$ was
applied. After the event selection, SM background
becomes negligible and significant part of SUSY background is
removed. Few kinematic endpoints 
were reconstructed and fitted \cite{Gjelsten:2004ki}:
 the maximum of the distribution of the
dilepton invariant mass $\,M_{ll}^{max}\,$, the maximum and the
minimum of the distribution of the $\,M(llq)\,$ invariant mass
$\,M_{llq}^{max}\,$ and $\,M_{llq}^{min}\,$, the maximum of the
distribution of the lower of the two $l^{+}q,\,l^{-}q$ invariant
masses $\,(M_{lq}^{low})^{max}\,$ and the maximum of the distribution
of the higher of the two $l^{+}q,\,l^{-}q$ invariant masses
$\,(M_{lq}^{high})^{max}\,$. From this set of endpoint measurements
and by taking into account statistical fit error and systematic error
on the energy scale ($1\%$ for jets and $0.1\%$ for leptons), SUSY
masses $m_{\tilde{q}_{L}}=540$ GeV,
$m_{\tilde{\chi}_{2}^{0}}=177$ GeV, $m_{\tilde{l}_{R}}=143$
GeV and $m_{\tilde{\chi}_{1}^{0}}= 96$ GeV were
extracted with a 6 GeV resolution for squarks and 4 GeV for
non-squarks ($L=300\;\text{fb}^{-1}$).

Few experimentally challenging points in the mSUGRA parameter space
constrained by the latest experimental data (see
Ref.~\cite{Ellis:2003cw}) were recently selected and studied by using
full Geant4 simulation.  Preliminary full simulation studies of left
squark cascade decay for the bulk point, the coannihilation point and
the focus point are reported (see
Ref.~\cite{Borjanovic:2005jr}). Events with two SFOS leptons are
selected and flavour subtraction
$\,e^{+}e^{-}\,+\,\mu^{+}\mu^{-}\,-\,e^{\pm}\mu^{\mp}\,$ was
applied. The bulk point ($m_{0}= 100$ GeV,
$m_{1/2}=300 $ GeV, $A_{0}=-300$ GeV,
$\tan{\beta}=6\,$, sign$(\mu)=+\,$ ) is a typical mSUGRA point where
easy SUSY discovery is expected. The endpoints $M_{ll}^{max}\,$,
$M_{llq}^{max}\,$, $M_{llq}^{min}\,$, $\,(M_{lq}^{high})^{max}\,$
and $\,(M_{lq}^{low})^{max}\,$ were reconstructed for integrated
luminosity $L=5\;\text{fb}^{-1}$.  The coannihilation point (
$\,m_{0}=70 $ GeV, $\,m_{1/2}=350 $ GeV,
$A_{0}=0$ GeV, $\tan{\beta}=10\,$, sign$(\mu)=+\,$ ) is
challenging due to the soft leptons present in the final state. The
decay of the second lightest neutralino to both left and right
sleptons is open: $\,\tilde{\chi}^{0}_{2} \rightarrow
\tilde{l}_{L,R}l\,$. The endpoints $M_{ll}^{max}\,$,
$M_{llq}^{max}\,$, $\,(M_{lq}^{high})^{max}\,$ and
$\,(M_{lq}^{low})^{max}\,$ were reconstructed for integrated
luminosity $L=20\;\text{fb}^{-1}$.  The focus point ($\,m_{0}=3550\;
\text{GeV}\,$, $\,m_{1/2}=300$ GeV, $A_{0}=0\;
\text{GeV}\,$, $\tan{\beta}=10\,$, sign$(\mu)=+\,$) predicts
multi-TeV squark and slepton masses. Neutralinos decay directly to
leptons: $\,\tilde{\chi}_{3}^{0}\;\rightarrow
\;l^{+}\,l^{-}\,\tilde{\chi}_{1}^{0}\,$,
$\;\;\tilde{\chi}_{2}^{0}\;\rightarrow
\;l^{+}\,l^{-}\,\tilde{\chi}_{1}^{0}\, $ and dilepton endpoints
$M_{ll}^{max}$ were reconstructed for $L=7\;fb^{-1}$.  All
reconstructed endpoints are at the expected positions.

In the case of direct slepton production where both sleptons decay to
lepton and the first lightest neutralino
$\;\tilde{l}_{L}\tilde{l}_{L}/\tilde{l}_{R}\tilde{l}_{R}\,\rightarrow
\,l^{+}\,l^{-}\,\tilde{\chi}_{1}^{0}\,\tilde{\chi}_{1}^{0}\;$, there
are no endpoints in the invariant mass distributions because of two
missing final state particles. It is possible to estimate slepton mass
by using variable transverse mass $M_{T2}=min
_{E_{T}^{miss}=E_{T1}^{miss}+E_{T2}^{miss}}
\left\{max \left\{m_{T}^{2}(p_{T}^{l1},E_{T1}^{miss} ),m_{T}^{2}(p_{T}^{l2},E_{T2}^{miss}) 
\right\} \right\}\,$ (see Ref.~\cite{Lester:1999tx}). 
The endpoint of the stransverse mass distribution is a function of
mass difference between slepton and the first lightest neutralino
$\tilde{\chi}_{1}^{0}$. In the case of mSUGRA point SPS1a, fast
simulation studies (see Ref.~\cite{SPS1}) show that by using
stransverse mass left slepton mass $m_{\tilde{l}_{L}}=202$ GeV can be
estimated with the resolution of 4 GeV ($L=100\;\text{fb}^{-1}$).

Left squark cascade decays
 $\;\;\tilde{q}_{L}\,\rightarrow\,\tilde{\chi}_{2}^{0}\,q\,\rightarrow\,
\tilde{l}^{\pm}_{L,R}\,l^{near(\mp)}\,q \,\rightarrow\,l^{far(\pm)}\,l^{near(\mp)}\,q\,\;\tilde{\chi}_{1}^{0}\;\;$ are
very convenient for the supersymmetric particles' spin measurement
(see Ref.~\cite{Barr:2004ze}). Due to slepton and squark spin-0 and
neutralino $\tilde{\chi}_{2}^{0}$ spin-1/2, invariant mass of quark
and first emitted (`near') lepton $M(ql^{near(\pm)})$ is charge
asymmetric. The asymmetry is defined as $A=(s^{+}-s^{-})/(s^{+}+s^{-}
),\;s^{\pm}=(d \sigma)( d M(ql^{near(\pm)}))\,$. Asymmetry
measurements are diluted by the fact that it is usually not possible
to distinguish the first emitted (`near') from the second emitted
(`far') lepton. Also, squark and anti-squark have opposite asymmetries
and are experimentally indistinguishable, but LHC is proton-proton
collider and more squarks than anti-squarks will be produced. Fast
simulation studies of few points in the mSUGRA space
\cite{Barr:2004ze,Goto:2004cp} show asymmetry distributions not
consistent with zero, which is the direct proof of the neutralino
spin-1/2 and slepton spin-0.  In the case of point SPS1a, non-zero
asymmetry may be observed with $30\;\text{fb}^{-1}$.
 
For some of the points in mSUGRA space , mixing between left and right
smuons is not negligible. Left-right mixing affects decay branching
ratios $\;\tilde{\chi}^{0}_{2} \rightarrow \tilde{l}_{R}l\;$ and
charge asymmetry of invariant mass distributions from left squark
cascade decay.  For the point SPS1a with modified $\tan(\beta)=20$,
fast simulation studies \cite{Goto:2004cp} show that
different decay branching ratios for selectrons and smuons can be
detected at LHC for $300\;\text{fb}^{-1}$.

Fast simulation studies show that SUSY masses can be extracted by
using kinematic endpoints and stransverse mass. Preliminary full
simulation analysis show that large number of mass relations can be
measured for leptonic signatures with few $\text{fb}^{-1}$ in
different mSUGRA regions. What is still needed to be studied more
carefully are: acceptances and efficiencies for electrons and muons,
calibration, trigger, optimization of cuts against SM background and
fit to distributions.  The asymmetry distributions are consistent with
neutralino spin-1/2 and slepton spin-0. Different branching ratios for
selectron and smuon, caused by smuon left-right mixing, can be
detected at ATLAS.


\section{Using the $l^{+}l^{-} ~+~ \ETmi ~+$ jet~veto
          signature for slepton detection}
\label{WG1:sect:CMSsleptons}

The aim of this section, which is based on Ref.~\cite{Andreev:2006sq},
is to study the possibility of detecting sleptons at CMS.
 Note the previous related papers where the sleptons detection was
studied at the level of a toy detector
\cite{delAguila:1990yw,Baer:1993ew,CMS_TN_96-059,Andreev:2004qq,Bityukov:1997ck}
whereas we perform a full detector simulation.

{\tt ISASUSY}~7.69~\cite{Paige:2003mg} was used for
the calculation of coupling constants
and cross sections in the leading order approximation for SUSY processes.
For the calculation of the next-to-leading order corrections
to the SUSY cross sections the {\tt PROSPINO}
code \cite{Beenakker:1996ed} was used. Cross sections of the background events
were calculated
with {\tt PYTHIA} 6.227 \cite{Sjostrand:2000wi} and {\tt CompHEP} 4.2pl
\cite{Pukhov:1999gg}. For
considered backgrounds the NLO corrections are known and they were
used.  Official data sets production were used for the study of CMS
test point LM1 and backgrounds ${\rm t \bar t}$, ZZ, WW, Wt, Z${\rm b
\bar b}$, DY2e, DY2$\tau$, where DY denotes Drell-Yan processes.  For
WZ, DY2$\mu$ and W+jet backgrounds the events were generated with
{\tt PYTHIA} 6.227.
The detector simulation
and hits production were made with full CMS simulation \cite{OSCAR},
digitized and reconstructed  \cite{ORCA}.
The DY2$\mu$ and W+jet
backgrounds were simulated with fast simulation \cite{PTDR1_FAMOS}.

Jets were reconstructed using an iterative cone algorithm with cone
size 0.5 and their energy was corrected with the GammaJet calibration.
The events are required to pass the Global Level 1 Trigger (L1),
the High Level Trigger (HLT) and at least one of the following triggers:
single electron, double electron, single muon, double muon.
The CMS fast simulation  was used for the determination of the
sleptons discovery plot.

As discussed in the previous section, sleptons can be either
produced at LHC directly via the Drell-Yan mechanism or in cascade
decays of squarks and gluinos.
The slepton production and decays described previously  lead to
the signature
with the simplest event topology: $ two \, ~leptons + \ETmi +
 ~jet~veto$. This signature arises for both direct
and indirect slepton pair production.
In the case of indirectly produced sleptons not only the event topology with
two
leptons but with single, three and four leptons is possible.
Besides, indirect
slepton production from decays of squarks and gluino through charginos,
neutralinos can lead to the event topology
$ two ~leptons + \ETmi + (n \geq 1) ~jets $.

Close to the optimal cuts are:
\begin{itemize}
\item[a.] for leptons:

\begin{itemize}
\item {
  $p_T$ - cut on leptons ($p_T^{lept} > 20$ GeV, $|\eta| < 2.4 $)
and lepton isolation within ${\Delta}R<0.3$ cone containing calorimeter
cells and tracker;}

\item 
  effective mass of two opposite-sign and the same-flavour leptons is
  outside the ($M_Z-15$~GeV, $M_Z+10$~GeV) interval; 
\item $\Phi(l^{+}l^{-})  < 140^{\circ}$  cut on angle between two leptons; 
\end{itemize}

\item[b.] for $\ETmi$ :

\begin{itemize}
\item { $\ETmi > 135$~GeV cut on missing $\ET$; }
\item { $\Phi (\ETmi, ll) > 170^{\circ}$
 cut on relative azimuthal angle between dilepton and $\ETmi$;}
\end{itemize}

\item[c.] for jets:
\begin{itemize}
\item { jet veto cut: $N_{jet} = 0 $ for a $E^{jet}_T > 30$~GeV
(corrected jets) threshold in the pseudorapidity interval $|\eta| < 4.5 $. }
\end{itemize}

\end{itemize}

The Standard Model (SM) backgrounds are:
${\rm t \bar t}$, WW, WZ, ZZ, Wt, Z${\rm b \bar b}$, DY, W+jet.
The main contributions come from WW and ${\rm t \bar t}$ backgrounds.
There are also internal SUSY backgrounds which arise from
$\tilde{q}\tilde{q}$, $\tilde{g}\tilde{g}$ and $\tilde{q}\tilde{g}$
productions and subsequent cascade decays with jets outside the
acceptance or below the threshold.
Note that when we are interested in new physics discovery
we have to compare the calculated number of
SM background events $N_{SMbg}$ with new physics signal events
$N_{new~physics} = N_{slept} + N_{SUSYbg}$,
so SUSY background events increase the discovery potential
of new physics.

For the point LM1 with the  set of cuts for an integral
luminosity $\mathcal{L}$~$= 10~fb^{-1}$
the number of signal events (direct sleptons plus sleptons from
chargino/neutralino decays) is
$N_{S} = 60$, whereas the number of SUSY background events is
$N_{SUSYbg} = 4$ and the number of SM background events  is $N_{SMbg} = 41$.
The total signal efficiency is $1.16 \cdot 10^{-4}$
and the background composition is $1.32 \cdot 10^{-6}$ of the total ttbar,
$1.37 \cdot 10^{-5}$
of the total WW, $4 \cdot 10^{-6}$ of the total WZ, $4.4 \cdot 10^{-5}$ of the total ZZ,
$8.1 \cdot 10^{-6}$ of the total Wt, 0 of the total Zbb, DY, W+jet.

\begin{figure}[t]
  \begin{center}
    \resizebox{12cm}{!}{\includegraphics{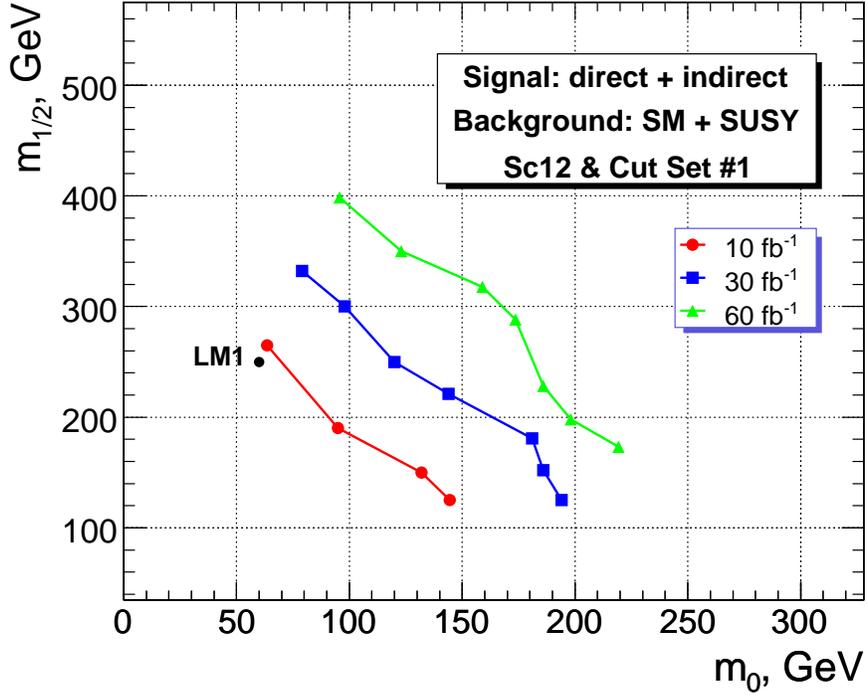}}
      \caption{Discovery plot ($\tan\beta = 10$, $sign( \mu ) = +$,
        $ A = 0$) for final states with $l^+l^{-}$, 
        missing transverse energy and a jet veto.}
      \label{WG1:fig:slept_fig23}
  \end{center}
\end{figure}

The SUSY background is rather small compared to the signal,
so we can assume $N_S~=~$
$N_{direct~sleptons} + N_{chargino/neutralino} + N_{SUSY bg} =  64$.
This corresponds to  significances
$S_{c12} = 7.7$ and  $S_{cL} = 8.3$ where the quantity $S_{c12}$ is defined
in ref.~\cite{Bityukov:1998ju} and $S_{cL}$ in refs.~\cite{Cousins:2004jc,
CMS_Note_2005-004}.
Taking into account the systematic
uncertainty of 23\% related to inexact knowledge of backgrounds
leads to the decrease of significance $S_{c12}$ from 7.7 to 4.3.
The ratio of the numbers of background events from two different channels
$N(e^+e^- + \mu^+\mu^-) /$ $N(e^\pm \mu^\mp)$ = 1.37 will be used to keep the
backgrounds under control.
The CMS discovery plot for $two~leptons + \ETmi + ~jet~veto$
signature is presented in \fig{WG1:fig:slept_fig23}.

\section{ Using the $e^{\pm}\mu^{\mp}~+~\ETmi$ signature 
          in the search for supersymmetry and lepton flavour
          violation in neutralino decays}

The aim of this section  based on Ref.~\cite{Andreev:2006sd} 
is the study of the possibility to detect SUSY
and LFV using
the $e^{\pm}\mu^{\mp} ~+ ~\ETmi$ signature at CMS. The details
concerning the simulations are the same as described in 
\sect{WG1:sect:CMSsleptons}.

The SUSY production $ pp \rightarrow \tilde{q}\tilde{q}^{'}, \tilde{g}
\tilde{g}, \tilde{q}\tilde{g}$ with subsequent decays
leads to the event topology $e^{\pm}\mu^{\mp} ~+ ~\ETmi$.
In the MSSM with lepton flavour conserving neutralino decays into leptons
$\tilde{\chi}^0_{2,3,4} \rightarrow l^+l^- \tilde{\chi}^0_1$ do not
contribute to this signature and contribute only to
$l^{+}l^{-} ~+ ~\ETmi$ signature (here $l=e$ or $\mu$). The main backgrounds
contributing to the $e^{\pm}\mu^{\mp}$ events
are: ${\rm t \bar t}$, ZZ, WW, WZ, Wt, Z${\rm b \bar b}$, DY2$\tau$, Z+jet.
It has been found that ${\rm t \bar t}$ background is the biggest one and
it gives  more than 50\% contribution to the total background.

Our set of cuts is the following:
\begin{itemize}
\item {
  $p_T$ - cut on leptons ($p_T^{lept} > 20$ GeV, $|\eta | < 2.4$)
and lepton isolation within ${\Delta}R<0.3$ cone.}
\item { $\ETmi > 300$~GeV cut on missing $\ET$. }
\end{itemize}
For integrated luminosity $\mathcal{L}$~$ = ~10$~fb$^{-1}$ the number of background
events with this set of cuts
is $N_B = 93$. The results for various CMS study points at
this luminosity are presented in \tab{WG1:tab:fvnino_results}.

\begin{table}[t]
\caption{ Number of signal events and significances
$S_{c12}$\cite{Bityukov:1998ju} and $S_{cL}$\cite{Cousins:2004jc,
CMS_Note_2005-004} for $\mathcal{L}$~$~=~10$~fb$^{-1}$.}
\label{WG1:tab:fvnino_results}
\begin{center}
\begin{tabular}{c|rrr}
\hline
\hline
~~Point~~ & ~$N$ events & ~~~$S_{c12}$ & ~~~$S_{cL}$ \\
\hline
LM1 & 329 & 21.8 & 24.9  \\
LM2 &  94 &  8.1 &  8.6  \\
LM3 & 402 & 25.2 & 29.2  \\
LM4 & 301 & 20.4 & 23.1  \\
LM5 &  91 &  7.8 &  8.3  \\
LM6 & 222 & 16.2 & 18.0  \\
LM7 &  14 &  1.4 &  1.4  \\
LM8 & 234 & 16.9 & 18.8  \\
LM9 & 137 & 11.0 & 11.9  \\
\hline
\hline
\end{tabular}
\end{center}
\end{table}

At point LM1 the signal over background ratio is 3
and the signal efficiency is $6 \cdot 10^{-4}$.
The background composition is $9.5 \cdot 10^{-6}$ of the total ttbar, $3.4 \cdot 10^{-6}$
of the total WW, $4 \cdot 10^{-6}$ of the total WZ, $3.2 \cdot 10^{-6}$ of the total Wt,
$2.2 \cdot 10^{-6}$ of the total Z+jet, 0 of the total ZZ, Z${\rm b \bar b}$, DY2$\tau$.

\begin{figure}[t]
\centering
\resizebox{12cm}{!}{\includegraphics{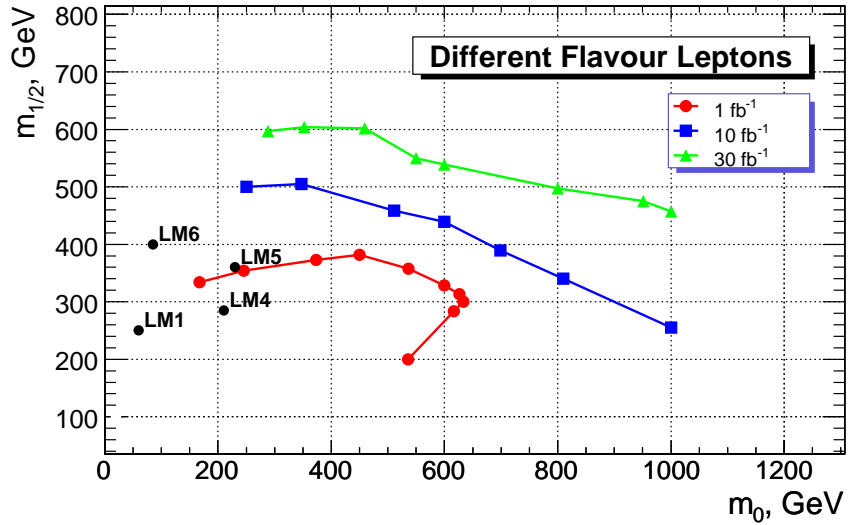}}
\caption{Discovery plot ($\tan\beta = 10$, $sign( \mu ) = +$,
$ A = 0$) for the luminosities $\mathcal{L}$~$~=~1,~10,~30$~fb$^{-1}$
for the $e^{\pm}\mu^{\mp} ~+ ~\ETmi$ signature.
    }
 \label{WG1:fig:fvnino_fig:9}
\end{figure}

The CMS discovery plot for the $e^{\pm}\mu^{\mp} ~+ ~\ETmi$
signature is presented in \fig{WG1:fig:fvnino_fig:9}.

It has been shown in refs.\
\cite{Krasnikov:1994hr,Arkani-Hamed:1996au,Krasnikov:1996np}  that
it is possible to look for lepton flavour violation at supercolliders
through the production and decays of the sleptons.  For LFV at the LHC
one of the most promising processes is the LFV decay of the second
neutralino \cite{Agashe:1999bm,Hisano:2002iy} $~\tilde{\chi}^0_2
\rightarrow \tilde{l}l \rightarrow \tilde{\chi}^0_1~ll^{'}$, where the
non zero off-diagonal component of the slepton mass matrix leads to
the different flavours for the leptons in the final state.  By using
the above mode, LFV in $\tilde{e} ~-~ \tilde{\mu}$ mixing has been
investigated in refs.~\cite{Agashe:1999bm,Hisano:2002iy} at a parton
model level and a toy detector simulation. Here we study the perspectives of 
LFV detection in CMS on the basis of full simulation of both signal and
background. To be specific, we study the point LM1. We
assume that the LFV is due to nonzero mixing of right-handed smuon and
selectron. The signal of the LFV $\tilde{\chi}^0_2$ decay is two
opposite-sign leptons ($e^+\mu^{-}$ or $e^{-}\mu^{+}$) in the final
state with a characteristic edge structure.  In the limit of lepton
flavour conservation, the process $ \tilde{\chi}^0_2 \rightarrow
\tilde{l}l \rightarrow ll\tilde{\chi}^0_1$ has an edge structure for
the distribution of the lepton-pair invariant mass $m_{ll}$ and the
edge mass $m^{max}_{ll}$ is expressed by the slepton mass
$m_{\tilde{l}}$ and the neutralino masses $m_{\tilde{\chi}^0_{1,2}}$
as follows:
\begin{equation}
(m^{max}_{ll})^2 = m^2_{\tilde{\chi}^0_2}(1 - \frac{m^2_{\tilde{l}}}
{m^2_{\tilde{\chi}^0_2  }})(1 - \frac{m^2_{\tilde{\chi}^0_1}}
{m^2_{\tilde{l}}  }   )
\end{equation}

The SUSY background for the LFV comes from uncorrelated leptons
from different squark or gluino decay chains. The SM background
comes mainly from
\begin{equation}
t\bar{t} \rightarrow b W b W \rightarrow bl bl^{'}\nu \nu^{'}
\end{equation}

Drell-Yan background from $pp \rightarrow \tau\tau \rightarrow e\mu\dots$
is negligible.
It should be stressed that for the signature with $e^{\pm}\mu^{\mp}$
in the absence of the LFV we
do not have the edge structure for the distribution on the invariant mass
$m_{inv}(e^{\pm}\mu^{\mp})$. As the result of the LFV the edge structure
for $e^{\pm}\mu^{\mp}$ events arises too. Therefore the signature of the LFV is
the existence of an edge structure in the $e^{\pm}\mu^{\mp}$ distribution.
The rate for a flavour violating decay is
\begin{equation}
BR({\tilde \chi}^0_2 \rightarrow e^{\pm}\mu^{\mp}{\tilde \chi}^0_1) =
 \kappa BR({\tilde \chi}^0_2 \rightarrow  e^{+}e^{-}{\tilde \chi}^0_1,
 \mu^{+}\mu^{-}{\tilde \chi}^0_1),
\end{equation}
where:
\begin{equation}
BR({\tilde \chi}^0_2 \rightarrow e^+e^-{\tilde \chi}^0_1, \mu^{+}\mu^{-}{\tilde \chi}^0_1)
= BR({\tilde \chi}^0_2 \rightarrow e^+e^-{\tilde \chi}^0_1)
+ BR({\tilde \chi}^0_2 \rightarrow {\mu}^{+}\mu^{-}{\tilde \chi}^0_1),
\end{equation}
\begin{equation}
\kappa = 2x \sin^{2}\theta \cos^{2} \theta,
\end{equation}
\begin{equation}
x = \frac{\Delta m^{2}_{\tilde{e}\tilde{\mu}}}
{\Delta m^2_{\tilde{e}\tilde{\mu}}
+ \Gamma^{2}},
\end{equation}
\begin{equation}
BR({\tilde \chi}^0_2  \rightarrow e^{\pm}\mu^{\mp}) =
BR({\tilde \chi}^0_2  \rightarrow e^{+}\mu^{-}) +
BR({\tilde \chi}^0_2  \rightarrow e^{-}\mu^{+}).
\end{equation}

Here $\theta$ is the mixing angle between $\tilde{e}_R $ and $\tilde{\mu}_R$
and $\Gamma $ is the sleptons decay width.
The parameter $x$ is the measure of the quantum interference effect.
There are some limits on $\tilde{e} - \tilde{\mu}$ mass splitting from
lepton flavour violating processes but they are not
very strong.

\begin{figure}[t]
\resizebox{8cm}{!}{\includegraphics{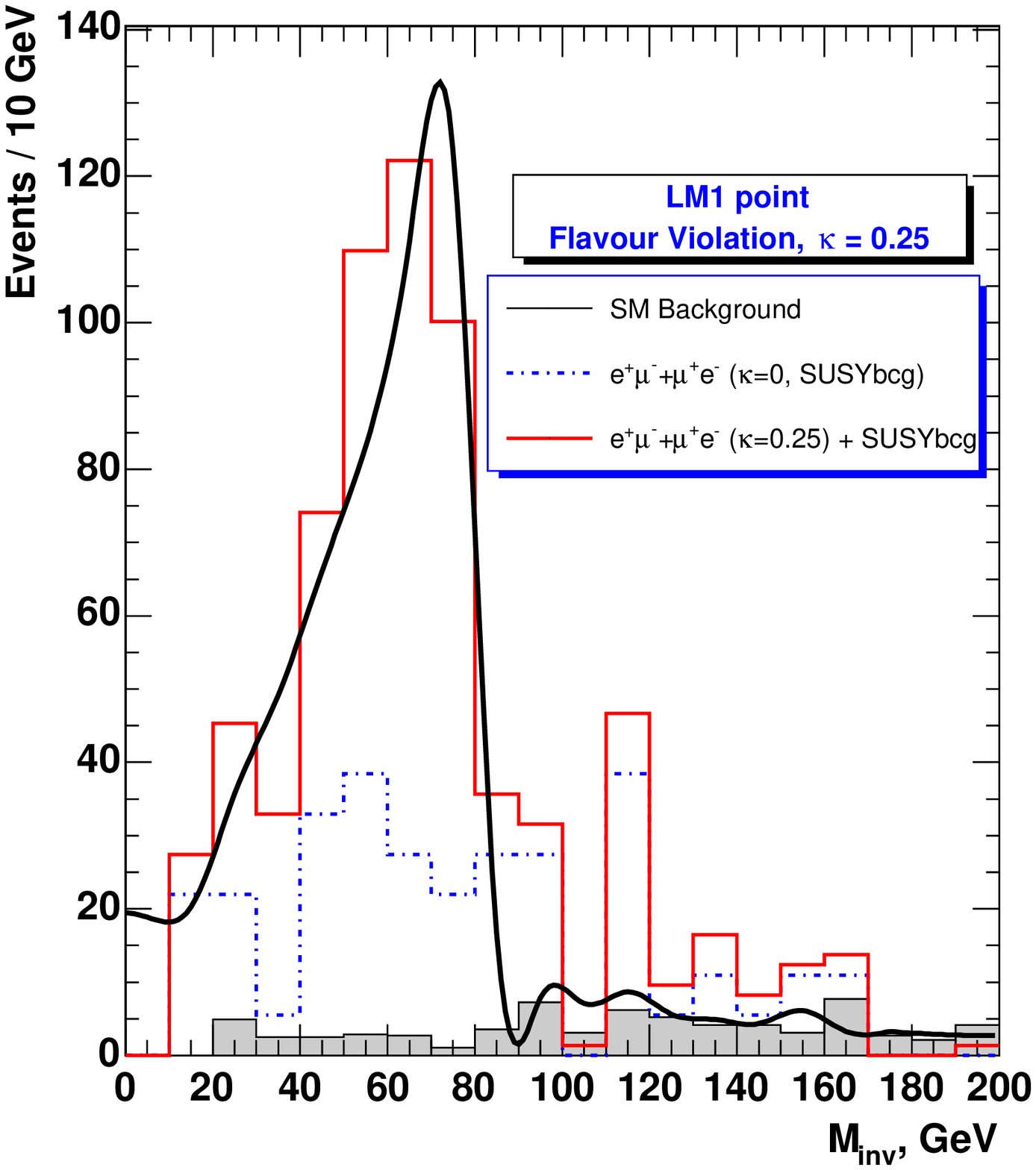}}
\resizebox{8cm}{!}{\includegraphics{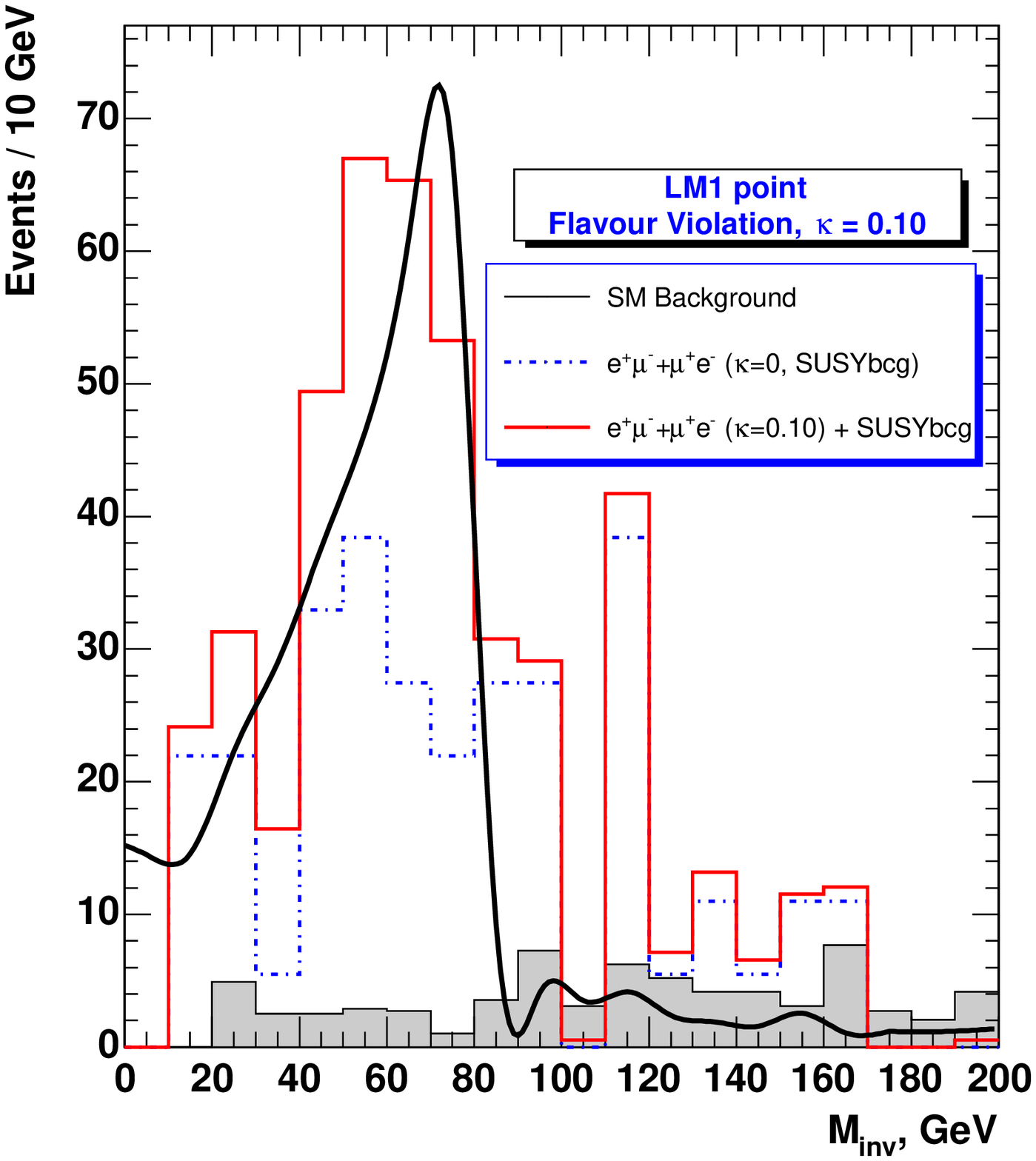}}
\caption{The distribution of dilepton invariant mass after selection of two
isolated $e^{\pm} \mu^{\mp}$ leptons with $p_T^{lept} > 20$~GeV and
$\ETmi > 300$~GeV for flavour violation parameter
$ k = 0.25 $ (left) and $ k = 0.1 $ (right). The superimposed curves are fits
to the invariant mass distribution for the case of 100\% LFV. }
\label{WG1:fig:fvnino_fig:11}
\end{figure}

For $\kappa = 0.25$, $\kappa = 0.1$   the distributions of the number of
$e^{\pm}\mu^{\mp}$ events on the invariant mass $m_{inv}(e^{\pm}\mu^{\mp})$
(see \fig{WG1:fig:fvnino_fig:11})
clearly demonstrates the existence of the edge
structure\cite{Baer:1994nr}, i.e. the existence of the lepton flavour violation in
neutralino decays.
It appears that for the point LM1 the use of an additional cut
\begin{equation}
m_{inv}(e^{\pm}\mu^{\mp}) < 85\,\,\mathrm{GeV}
\end{equation}
reduces both the SM and SUSY backgrounds and increases the
discovery potential in the LFV search.
For the point LM1 we found that in the assumption of exact knowledge of
the background (both the SM and SUSY backgrounds) for the integrated
luminosity $\mathcal{L}$~$ = 10$~fb$^{-1}$ it would be possible to detect LFV
at $5 \sigma$ level in ${\tilde \chi}^0_2$
decays for $\kappa  \geq 0.04 $.


\section{Neutralino spin measurement with ATLAS}

Charge asymmetries in invariant mass
distributions containing leptons can be used 
to prove that the neutralino spin is 1/2.
This is based on a method \cite{Barr:2004ze} which allows to choose between
different hypotheses for spin assignment, and to discriminate SUSY
from an Universal Extra Dimensions (UED) model apparently
mimicking low energy SUSY
 \cite{Datta:2005zs,Smillie:2005ar}.
For this the decay chain
\begin{equation}
\tilde{q}_{L}\,\rightarrow\,\tilde{\chi}_{2}^{0}\,q\,\rightarrow\,
\tilde{l}_{L,R}^{\pm}\,l^{\mp}\,q
\,\rightarrow\,l^{+}\,l^{-}\,q\,\;\tilde{\chi}_{1}^{0} \; \label{sqldecay}
\end{equation}
will be used.
In the following, the first lepton (from ${\tilde \chi}^0_2$ decay) 
is called {\it near}, and the one from slepton decay is called {\it far}. 

In the MSSM, squarks and sleptons are spin-0 particles and their
decays are spherically symmetric, differently from the
$\tilde{\chi}_2^0$ which has spin 1/2.  A charge asymmetry is expected
in the invariant masses $m(ql^{near(\pm)})$ formed by the quark and
the near lepton. Also $m(ql^{far})$ shows some small charge asymmetry
\cite{Datta:2005zs,Smillie:2005ar}, but it is not always possible to
distinguish experimentally near from far lepton, thus leading to
dilution effects when measuring the $m(ql^{near(\pm)})$ charge asymmetry.

In the cascade decay (\ref{sqldecay}), the asymmetry in the
corresponding $m(\bar{q}l)$ charge distributions is the same as the
asymmetry in $m(ql)$ from $\tilde{q}_{L}$ decay, but with the opposite
sign \cite{Richardson:2001df}. Though it is not possible to
distinguish $q$ from $\bar{q}$ at a $pp$ collider like the LHC, more
squarks than anti-squarks will be produced. Here only
electrons and muons are considered for analysis.

Two mSUGRA points were selected for analysis \cite{Biglietti:2007xx}: 
SU1, in the
stau-coannihilation region ($m_{0}$= 70 GeV, $m_{1/2}$= 350 GeV,
$A_{0}$=0 GeV, $tan{\beta}$=10, $sgn{\mu}$=+) and SU3, in the
bulk region ($m_{0}$= 100 GeV, $m_{1/2}$= 300 GeV, $A_{0}$=-300 GeV,
$tan{\beta}$=6, $sgn{\mu}$=+).  In SU1 (SU3) LO cross section for all
SUSY is 7.8 pb (19.3 pb), and the observability of charge asymmetry is
enhanced by $\sim$5 ($\sim$2.5) in $\tilde{q}$/$\bar{\tilde{q}}$
production yield.

In the SU1 point, owing to a small mass difference between
$\tilde{\chi}_{2}^{0}$ and $\tilde{l}_{L}$ (264 GeV and 255 GeV,
respectively), the near lepton has low $p_T$ in the
$\tilde{\chi}_{2}^{0} \rightarrow \tilde{l}_{L}\,l$ decay, while the
small mass difference between $\tilde{l}_{R}$ and
$\tilde{\chi}_{1}^{0}$ (155 GeV and 137 GeV, respectively), implies
low values for far lepton's $p_T$ in $\tilde{\chi}_{2}^{0} \rightarrow
\tilde{l}_{R}\,l$ decay.  As a consequence, near and far leptons are
distinguishable.  Decay (\ref{sqldecay}) represents $\sim 1.6\%$ of
all SUSY production. From the three detectable particles
$l^{+},\,l^{-},\,q$ (where the quark hadronizes to a jet) in the final state
of the $\tilde{q}_{L}$ decay (\ref{sqldecay}) four invariant masses
are formed: $m(ll)$, $m(qll)$, $m(ql^{near})$ and $m(ql^{far})$. Their
kinematic maxima are given by: $m(ll)^{max}$ = 56 GeV
($\tilde{l}_{L}$), 98 GeV ($\tilde{l}_{R}$), $m(qll)^{max}$ = 614 GeV
($\tilde{l}_{L}$, $\tilde{l}_{R}$), $m(ql^{near})^{max}$ = 181 GeV
($\tilde{l}_{L}$), 583 GeV ($\tilde{l}_{R}$) and $m(ql^{far})^{max}$ =
329 GeV ($\tilde{l}_{R}$), 606 GeV ($\tilde{l}_{L}$).  In the SU3
point, only the decay $\tilde{\chi}^{0}_{2}\rightarrow
\tilde{l}_{R}^{\pm}\,l^{\mp}$ is allowed (3.8\% of all SUSY
production). The endpoints for $m(ll)$, $m(qll)$, $m(ql^{near})$ and
$m(ql^{far})$ are 100, 503, 420 and 389 GeV, respectively.

Events were generated with {\tt HERWIG} 6.505 \cite{Corcella:2000bw}. 
SUSY samples corresponding to integrated luminosities of 100 fb$^{-1}$ for SU1 
and 30 fb$^{-1}$ for SU3 were analysed.
Also the most relevant SM processes have been also 
studied, i.e. $t\bar{t}$ + jets, $W$ + jets and $Z$ + jets 
backgrounds were produced with {\tt Alpgen} 2.0.5 \cite{Mangano:2002ea}.
Events were passed through a parametrized simulation of ATLAS 
detector, {\tt ATLFAST} \cite{Richter-Was:1998xx}.

In order to separate SUSY signal from SM background these {\it preselection} cuts were applied:
\begin{itemize}
\item[$\bullet$] missing transverse energy $E_{T}^{miss}\,>\,100$ GeV,
\item[$\bullet$] 4 or more jets with transverse momentum $p_{T}(j_1)>100$ GeV 
and $p_{T}(j_2,j_3,j_4)>50$ GeV.
\item[$\bullet$] exactly two SFOS leptons ($p_{T}^{lepton}>6$ GeV for 
SU1, and $p_{T}^{lepton}>10$ GeV for SU3).
\end{itemize}
At this selection stage, few invariant masses are formed:
the dilepton invariant mass $m(ll)$, the lepton-lepton-jet invariant mass $m(jll)$,
and the lepton-jet invariant masses $m(jl^+)$ and $m(jl^-)$, where $l^\pm$ are 
the leptons and $j$ is one of the two most energetic jets in the event.
Subsequently
\begin{itemize}
\item[$\bullet$] $m(ll)< 100\;\text{GeV}$, 
$m(jll)<615 \;\text{GeV}$ (for SU1) \ or \ 
$m(jll)<500 \;\text{GeV}$ (for SU3)
\end{itemize}
is required. In SU1, the decays (\ref{sqldecay}) with $\tilde{l}_{L}$
or $\tilde{l}_{R}$ are distinguished asking for $m(ll)<57\;\text{GeV}$
or $57\; {GeV} <m(ll)<100\;\text{GeV}$, respectively.  For SU1, in the
decay (\ref{sqldecay}) with $\tilde{l}_{L}$, the near (far) lepton is
identified as the one with lower (higher) $p_T$, and vice versa for
the decay (\ref{sqldecay}) with $\tilde{l}_{R}$.  The efficiencies and
signal/background ratios after all the cuts described so far, when
applied on SUSY and SM events, are shown in Table \ref{eff_tab}.
\begin{table}[t]
\caption{Efficiencies and S/B ratios for SUSY signal and background (SU1, SU3)
 and for SM background.}
 \label{eff_tab}
\begin{center}
\begin{tabular}{|c|c|c|c|c|}
\hline
 & Efficiency (SU1)& S/B (SU1)& Efficiency (SU3)& S/B (SU3)\\ \hline
Signal & (17.0 $\pm$ 0.3) \% & / & (20.0 $\pm$ 0.3)\% & / \\ \hline 
SUSY Background & (0.94 $\pm$ 0.01)\% & 0.33 & (0.75 $\pm$ 0.01)\% & 1 \\ \hline
$t\bar{t}$ & (2.69 $\pm$ 0.02) 10$^{-4}$ & 0.18 & (3.14 $\pm$ 0.02) 10$^{-4}$ & 0.9 \\ \hline
$W$ & (1.4 $\pm$ 0.9) 10$^{-5}$ & $\sim$16 & (0.4 $\pm$ 0.4) 10$^{-5}$ & $\sim$300 \\ \hline
$Z$ & (1.1 $\pm$ 0.3) 10$^{-5}$ & $\sim$12 & (0.9 $\pm$ 0.2) 10$^{-5}$ & $\sim$100 \\ \hline
\end{tabular}
\end{center}
\end{table}
Further background reduction is applied by subtracting statistically in
invariant mass distributions events with two opposite flavour opposite
sign (OFOS) leptons: $e^{+}e^{-} + \mu^{+}\mu^{-} - e^{\pm}\mu^{\mp}$
(SFOS-OFOS subtraction). This reduces SUSY background by about a factor of
2 and makes SM events with uncorrelated leptons compatible with zero.

\begin{figure}[t]
\begin{center}
\vspace{4.5cm}
\includegraphics{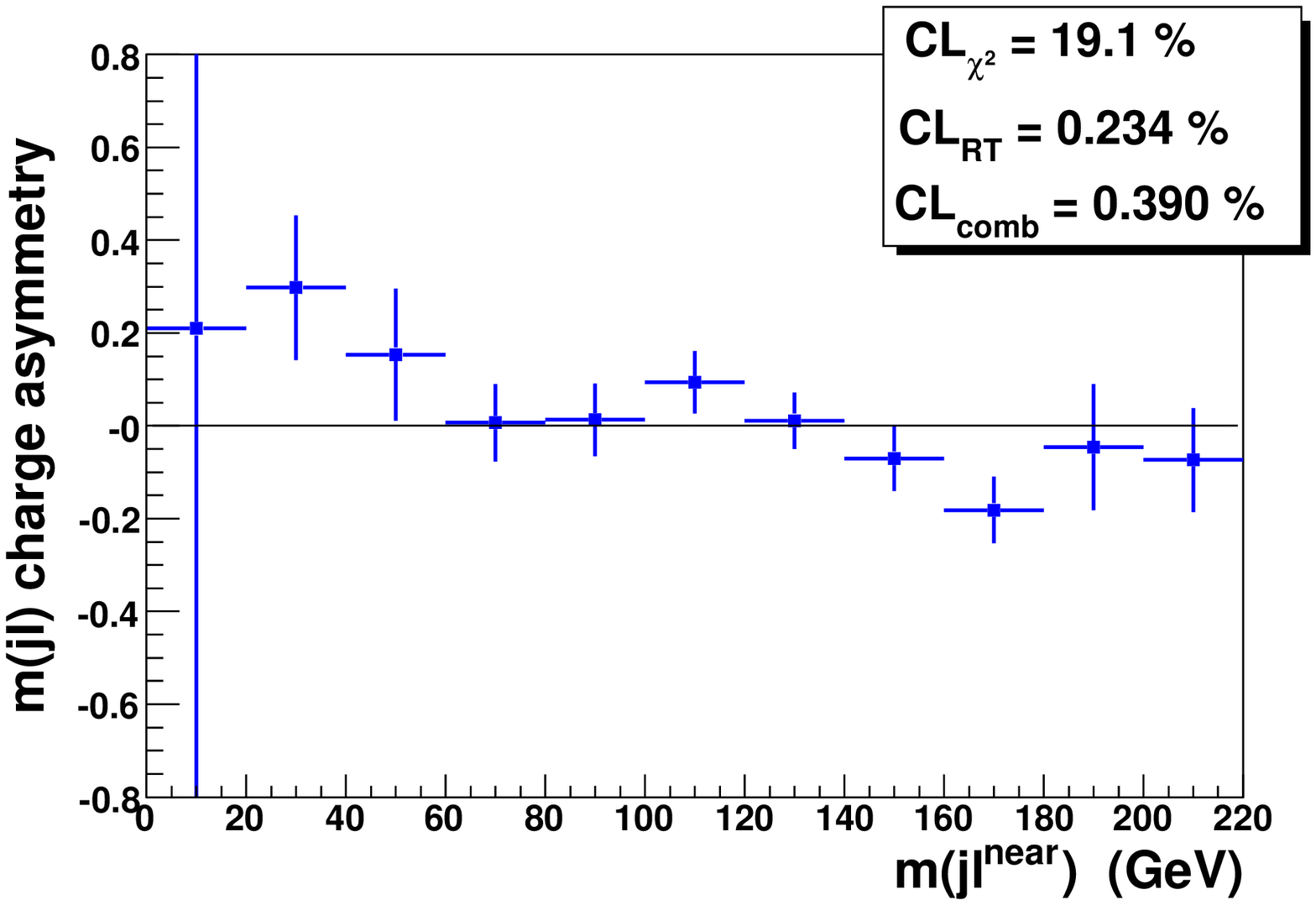}
\includegraphics{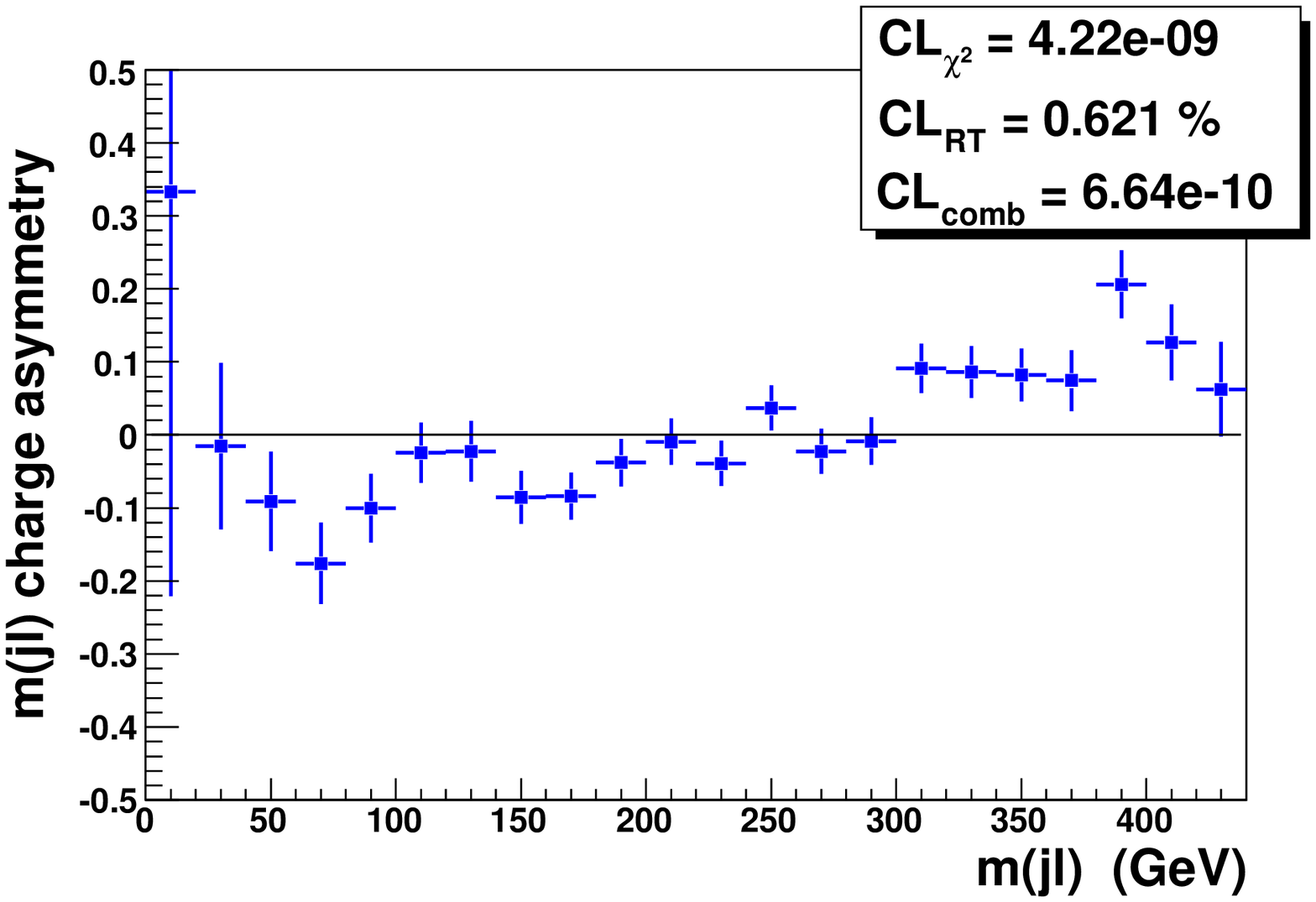}
\caption{Charge asymmetries for lepton-jet invariant masses after SFOS-OFOS subtraction. 
Left: using the near lepton from the chain involving $\tilde{l}_L$ for the
 SU1 point.
Right: using both near and far leptons for the SU3 point.}
\label{asym_rec_lq}
\end{center}
\end{figure}

Charge asymmetries of $m(jl)$ distributions have been computed after
SFOS-OFOS subtraction in the ranges
$\left[ 0, 220 \right]$ GeV for SU1 (only for the decay (\ref{sqldecay}) with
$\tilde{l}_L$ and near lepton) and $\left[ 0, 420 \right]$ GeV for SU3.
Two methods have been applied to detect the presence of a non-zero charge asymmetry:
\begin{itemize}
\item[$\bullet$] a non parametric $\chi^2$ test with respect to a constant 
0 function, giving confidence level CL$_{\chi^2}$,
\item[$\bullet$] a {\it Run Test} method \cite{Frodesen:1979fy} providing a
confidence level CL$_{RT}$ for the hypothesis of a zero charge asymmetry.
\end{itemize}
The two methods are independent and are not influenced by the actual shape of 
charge asymmetry. Their probabilities can be combined \cite{Frodesen:1979fy} providing 
a final confidence level CL$_{comb}$. 
In \fig{asym_rec_lq} charge asymmetries are reported for $m(jl^{near})_L$ 
in SU1 and for $m(jl)$ in SU3.
With 100 fb$^{-1}$, in SU1 CL$_{comb}$ is well below 1\%, while for
SU3 30 fb$^{-1}$ are enough to get a CL$_{comb}\sim$10$^{-9}$.
Different sources of background and possible systematic effects have
been investigated for SU1 and SU3 samples and the obtained confidence
levels are reported in Table \ref{tablesys} (letters {\bf b.} to {\bf
f.}), compared to the final SUSY selected sample (letter {\bf
a.}). They refer to: selected OFOS lepton pairs ({\bf b.}), SFOS
background SUSY events ({\bf c.}), SFOS and OFOS selected SM
background events ({\bf d.} and {\bf e.}, respectively) and events
with $m(jl)$ formed with a wrong jet ({\bf f.}). Anyway, confidence
levels are much higher than the final selected SUSY sample.
\begin{table}[t]
\caption{Confidence levels for the two methods described in the text, separately and
combined, obtained on $m(jl)$ distributions for the final selected samples and
for various sources of background/systematics.}
\label{tablesys}
\begin{center}
\begin{tabular}{|l|c|c|c|c|c|c|}
\hline
\ \ \ \ \ \ \ \ \ \ \ Analysed & \multicolumn{3}{|c|}{SU1 selection} & \multicolumn{3}{|c|}{SU3 selection} \\ \cline{2-7}
\ \ \ \ \ \ \ \ \ \ \ \ \ sample & CL$_{\chi^2}$ & CL$_{RT}$ & CL$_{comb}$ & CL$_{\chi^2}$ & CL$_{RT}$ & CL$_{comb}$ \\ \hline
{\bf a.} SUSY SFOS-OFOS & 19.1\% & 0.234\% & 0.390\% & 4.22$\cdot 10^{-9}$ & 0.621\% & 6.64$\cdot 10^{-10}$ \\ \hline
{\bf b.} SUSY OFOS & 57.1\% & 92.1\% & 86.4\% & 19.3\% & 93.3\% & 48.9\% \\ \hline
{\bf c.} SUSY SFOS bkg & 30.7\% & 24.0\% & 26.6\% & 53.5\% & 30.9\% & 46.2\% \\ \hline
{\bf d.} SM SFOS bkg & 21.4\% & 24.0\% & 20.3\% & 61.3\% & 84.1\% & 85.7\% \\ \hline
{\bf e.} SM OFOS bkg & 73.8\% & 50.0\% & 73.7\% & 95.5\% & 30.9\% & 65.5\% \\ \hline
{\bf f.} SUSY wrong jet & 62.8\% & 50.0\% & 67.8\% & 19.7\% & 15.9\% & 14.0\% \\ \hline
\end{tabular}
\end{center}
\end{table}

It is observed that the evidence with a 99\% confidence level for a 
charge asymmetry needs at least 100 fb$^{-1}$ in the case of SU1, 
while even less than 10 fb$^{-1}$ would be needed for SU3
\cite{Biglietti:2007xx}.

\section{SUSY Higgs-boson production and decay}

Flavour-changing neutral current (FCNC) interactions of neutral Higgs bosons
are extremely suppressed in the Standard Model (SM). In the SM, one finds 
$\mathcal{B}(H_{\text{SM}}\to bs) \approx 4\times 10^{-8}$ for
$m_{H_{\text{SM}}}= 114$ GeV \footnote{%
   In the following, $\mathcal{B}(H\to b s)$ denotes the sum of the Higgs 
   branching ratios into $b\bar s$ and $\bar bs$.  The Higgs boson 
   $H$ stands for that of the SM, $H_{\text{SM}}$, or one of 
   those of the MSSM, $H^0$ or $A^0$.}.
For the neutral MSSM Higgs bosons the 
ratios could be of $\mathcal{O}(10^{-4}$--$10^{-3})$.
Constraints from $b\to s\gamma$
data reduce these rates, though~\cite{Bejar:2004rz,Bejar:2005kv,Bejar:2006hd,Curiel:2003uk}.
The FCNC decays
$BR(t\rightarrow H_{\rm SM}\,c)$ or $BR(H_{\rm SM}\rightarrow t\,c)$ are of
the order $10^{-14}$ or less~\cite{Mele:1998ag,Bejar:2000ub,Bejar:2003em,%
Bejar:2004rz}, hence $10$ orders of magnitude below other more conventional
(and relatively well measured) FCNC processes like $b\rightarrow s\gamma$\,
\cite{Eidelman:2004wy}. The detection of Higgs FCNC interactions would be
instant evidence of new physics. The Minimal Supersymmetric Standard Model
(MSSM) introduces new sources of FCNC interactions mediated by the
strongly-interacting sector\footnote{For description of these interactions
see e.g.\ Refs.\ \cite{Gabbiani:1996hi,Misiak:1997ei,Guasch:1999jp} and
references therein.}. They are produced by the misalignment of the quark
mass matrix with the squark mass matrix, and the main parameter
characterizing these interactions is the non-flavour-diagonal term in the
squark-mass-matrix, which we parametrize in the standard fashion
\cite{Gabbiani:1996hi,Misiak:1997ei} as
$
  (M^2)_{ij} = \delta_{ij} \tilde{m}_i \tilde{m}_j\ (i \neq j)
$,
$\tilde{m}_i $ being the flavour-diagonal mass-term of the $i$-flavour
squark. Since there are squarks of different chiralities, there are
different $\delta_{ij}$ parameters for the different chirality
mixings.

\subsection{SUSY Higgs-boson flavour-changing neutral currents at the LHC}

Some work in relation with the MSSM Higgs-boson FCNCs has already been
performed \cite{Guasch:1999jp,Bejar:2004rz,Guasch:1997kc,Guasch:1999ve,%
Bejar:2001sj,Curiel:2002pf,Demir:2003bv,Curiel:2003uk,Heinemeyer:2004by,%
Aguilar-Saavedra:2004wm,Hahn:2005qi,Bejar:2005kv,Bejar:2006hd}. Here,
we compute and analyze the production of any MSSM Higgs boson ($h=h^0,H^0,
A^0$) at the LHC, followed by the one-loop FCNC decay $h\rightarrow
b\,s$ or $h\rightarrow t\,c$, and we find the maximum production rates
of the combined cross-section
\begin{equation}
  \sigma(pp\to h\to q\,{q'})
  \equiv
  \sigma(pp\to h X)BR(h\to q\,{q'})\ \ , \ \ 
      BR(h\to q\,{q'})\equiv\frac{\Gamma(h\to
          q\,\bar{q'}+\bar{q}\,q')}{\sum_i \Gamma(h\to X_i)},
    \label{eq:hqq-def}
\end{equation}
$qq'$ being a pair of heavy quarks $(qq'\equiv bs$ or $tc)$, taking into
account the restrictions from the experimental determination of
$B(b\to s\gamma)$~\cite{Eidelman:2004wy}. For other signals  of SUSY FCNC at
the LHC, without Higgs-boson couplings, see \chapt{chap:top}{sec:singletop}
and Ref.~\cite{Guasch:2006hf}. For comparison of the same signal in non-SUSY
models see \chapt{chap:top}{top:2hdm} 
and refs.~\cite{Bejar:2000ub,Bejar:2003em}.
Here we assume flavour-mixing only among the
left squarks, since these mixing terms are expected to be the
largest ones by Renormalization Group analysis~\cite{Duncan:1983iq}.

In the following  we give a summarized explanation of the computation,
for further details see Refs.\cite{Bejar:2005kv,Bejar:2004rz}. 
We include the full one-loop SUSY-QCD contributions to the FCNC partial
decay widths $\Gamma(h\to q\,{q'})$ in the observable of Eq.\ (\ref{eq:hqq-def}).
 The Higgs sector parameters (masses and CP-even mixing angle
  $\alpha$) have been treated using the leading $m_t$ and $m_b\tan\beta$
  approximation  to the one-loop result~\cite{Yamada:1994kj,Chankowski:1994er,Dabelstein:1995hb,Dabelstein:1995js}. 
 The Higgs-boson total decay widths $\Gamma(h\to X)$ are computed
  at leading order, including all the relevant channels.
 The MSSM Higgs-boson production cross-sections  have been computed
using the programs  \texttt{HIGLU 2.101} and \texttt{PPHTT
1.1}\,\cite{Spira:1995mt,Spira:1997dg,Spira:1995rr}. 
We have used the leading order
approximation for all channels. The QCD renormalization scale is
set to the default value for each program. We
have used the set of CTEQ4L PDF\cite{Lai:1996mg}.
For the constraints on the FCNC parameters, we use $BR(b\to s\gamma)=(2.1-4.5)\times
10^{-4}$ as the experimentally allowed range within three standard
deviations~\cite{Eidelman:2004wy}. We also require that the sign of the $b\to s\gamma$
amplitude is the same as in the SM~\cite{Gambino:2004mv}\footnote{This
    constraint automatically excludes the \textit{fine-tuned}
    regions of Ref.~\cite{Bejar:2004rz}.}.
Running quark masses $m_q(Q)$ and strong coupling constants
$\alpha_s(Q)$ are used throughout, with the renormalization scale set
to the decaying Higgs-boson mass in the decay processes. These
computations have been implemented in the computer code
\texttt{FchDecay}~\cite{FCHDECAY} 
(see also \chapt{chap:tools}{sec:tools:FchDecay}).
Given this setup, we have performed a Monte-Carlo
maximization \cite{Brein:2004kh} of the
cross-section in Eq.\ (\ref{eq:hqq-def}) over the MSSM parameter space, keeping
the parameter $\tan\beta$ fixed and under the simplification that the squark
and gluino soft-SUSY-breaking parameter masses are at the same scale,
$ 
m_{\tilde{q}_{L,R}}=m_{\tilde{g}}\equiv M_{\mathrm{SUSY}}
$. 

%
\begin{figure}
 \begin{center}
  \includegraphics[width=52mm]{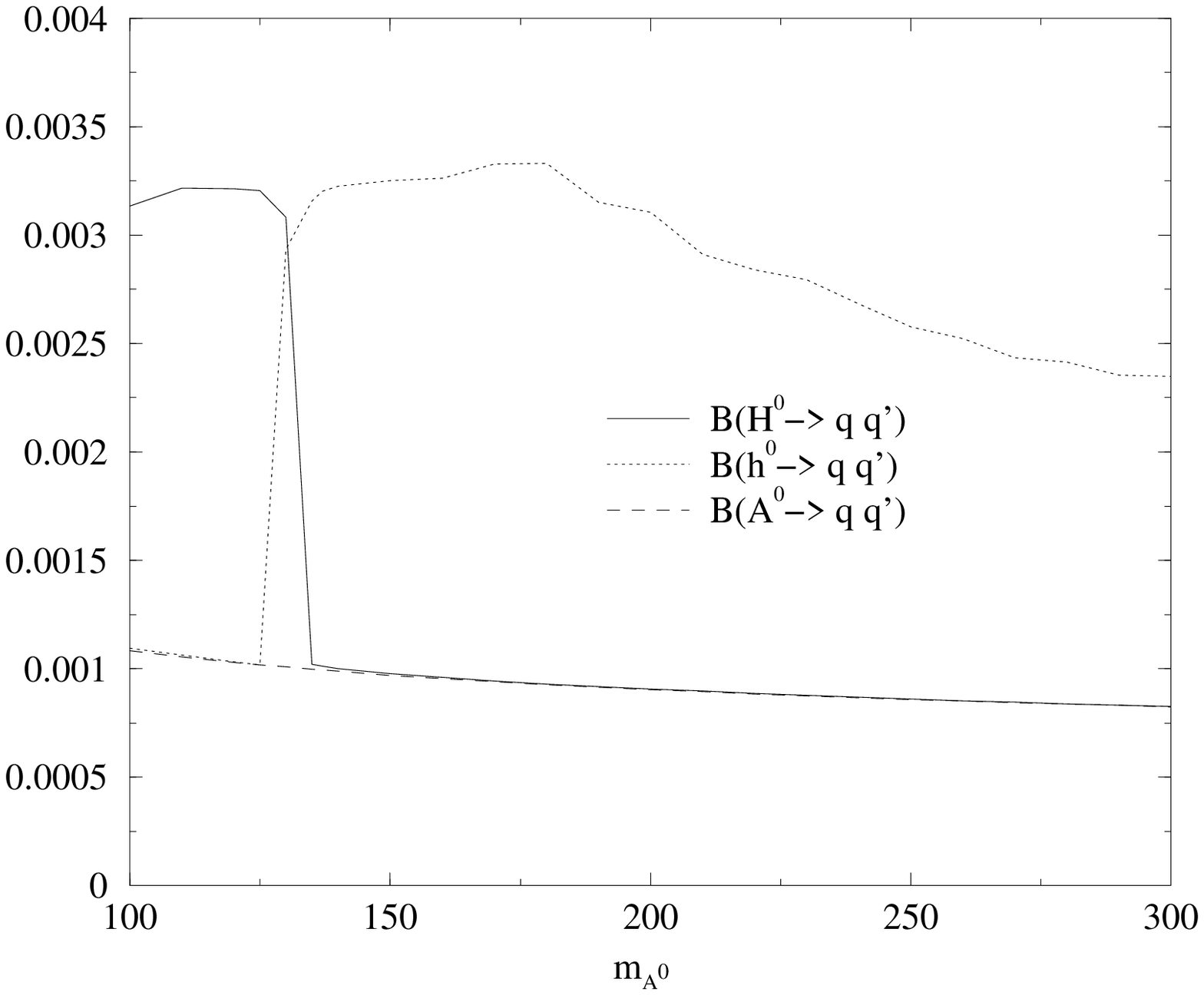}
  \includegraphics[width=49mm]{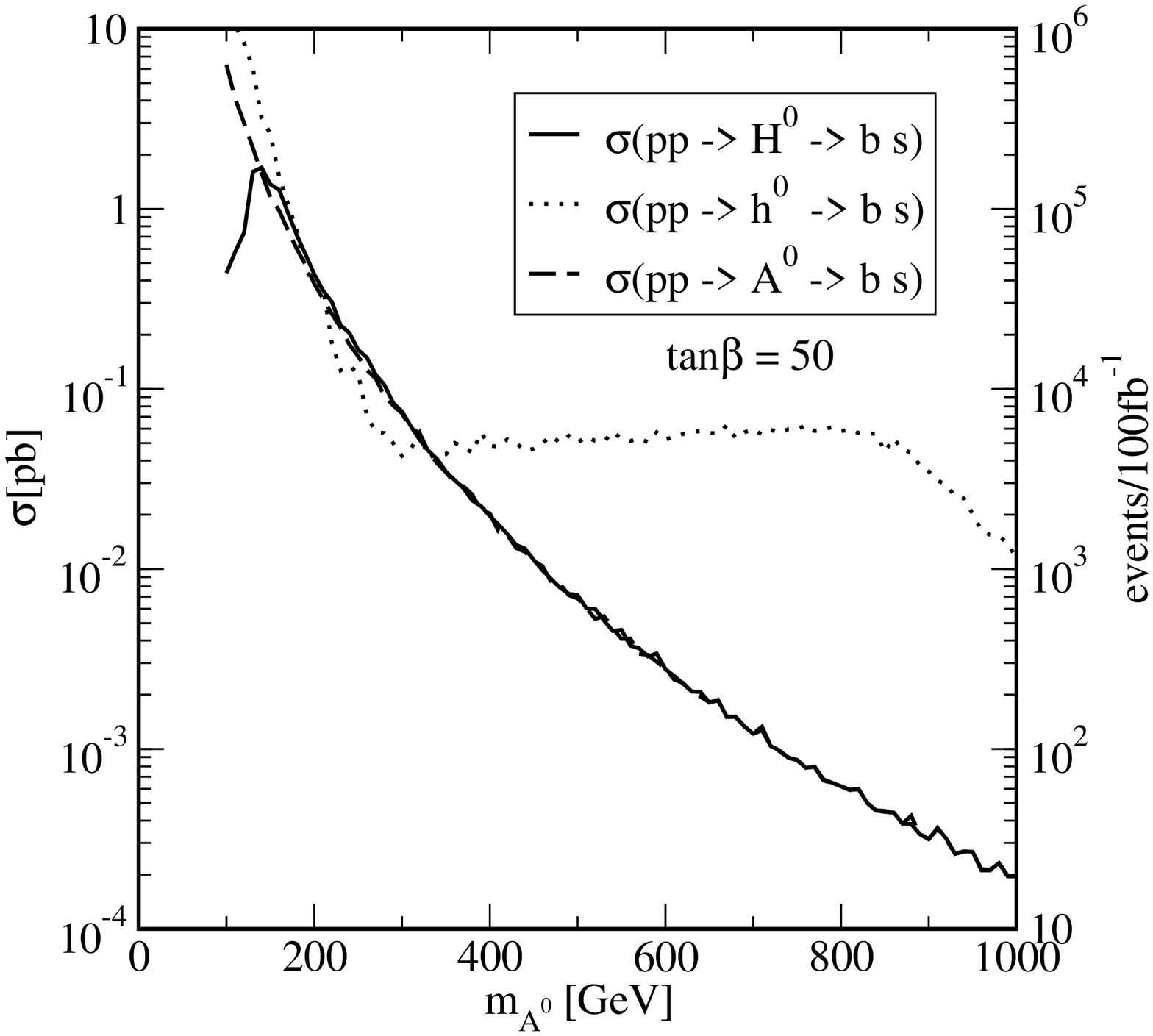}
  \includegraphics[width=49mm]{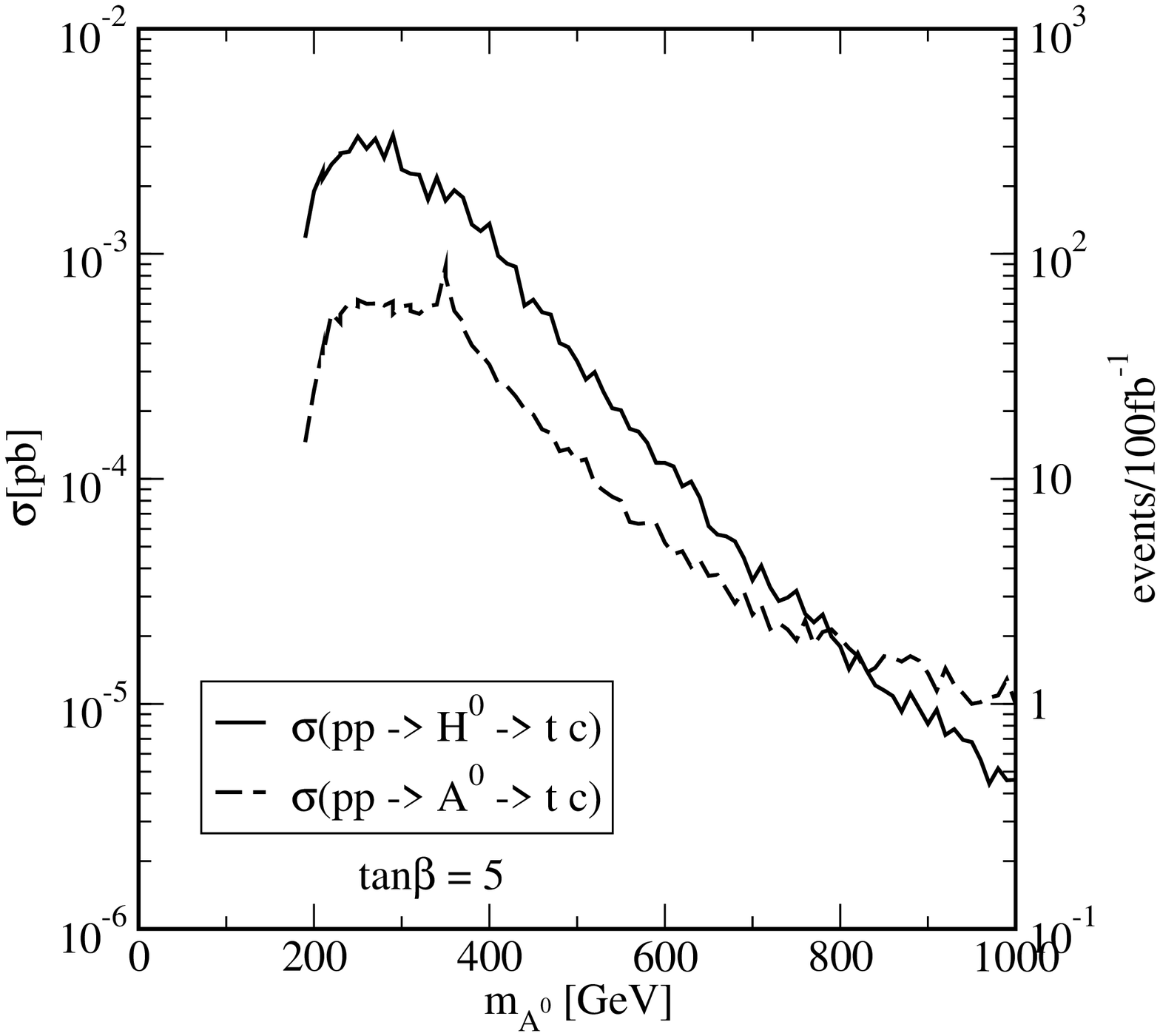}
  \caption{Left: The maximum value of $BR(h\to bs)$ as a function of $m_{A^0}$ for $\tan\beta=50$. 
  Centre: Maximum SUSY-QCD contributions to $\sigma(pp\rightarrow h\rightarrow b\,s)$ as a function of $m_{A^0}$ for $\tan\beta=50$.
  Right: Maximum SUSY-QCD contributions to $\sigma(pp\rightarrow h\rightarrow t\,c)$ as a function of $m_{A^0}$ for $\tan\beta=5$.}
\label{fig:hbsma0}
 \end{center}
\end{figure}
%

%
\begin{table}
 \begin{center}
  \caption{Top: Maximum values of $BR(h\to bs)$ and corresponding SUSY parameters for $m_{A^0}=200\UGeV$ and $\tan\beta=50$.
  Centre: Maximum value of $\sigma(pp\to h\to b\,s)$ and corresponding SUSY parameters for $m_{A^0}=200\UGeV$ and $\tan\beta=50$.
  Bottom: Maximum value of $\sigma(pp\to h\to t\,c)$ and corresponding SUSY parameters for $m_{A^0}=300\UGeV$ and $\tan\beta=5$.}
\label{tab:maximnowindow}
  \begin{tabular}{|c||c|c|c|}
  \hline $h$ & $H^0$ & $h^0$ & $A^0$ \\
  \hline\hline $BR(h\to bs)$ & $9.1\times 10^{-4}$ & $3.1\times 10^{-3}$ & $9.1\times 10^{-4}$\\
  \hline $\Gamma(h\to X)$ & $11.2 \UGeV$ & $1.4\times 10^{-3} \UGeV$ & $11.3 \UGeV$ \\
  \hline $\delta_{23}$ & $10^{-0.43}$& $10^{-0.8}$ & $10^{-0.43}$\\\hline $M_{\mathrm{SUSY}}$ & $1000 \UGeV$ &  $975 \UGeV$ & $1000 \UGeV$ \\
  \hline $A_b$ & $-1500 \UGeV$ & $-1500 \UGeV$ & $-1500 \UGeV$\\
  \hline $\mu$ & $-460 \UGeV$ & $-1000 \UGeV$ & $-460 \UGeV$ \\
  \hline $BR(b\to s\gamma)$ &  $4.49\times 10^{-4}$ &  $4.48\times 10^{-4}$ &$4.49\times 10^{-4}$ \\
  \hline
  \end{tabular}
  \begin{tabular}{|c||c|c|c|}
        \hline
        $h$ &  $H^0$ & $h^0$ & $A^0$ \\\hline\hline
        $\sigma(pp\to h\to b\,s)$ &  $0.45\Upb$ & $0.34\Upb$ & $0.37\Upb$ \\\hline
        events/$100\Ufb^{-1}$ & $4.5\times10^4$ & $3.4\times 10^4$ & $3.7\times10^4$\\\hline
        $BR(h\to bs)$ & $9.3\times 10^{-4} $& $2.1\times 10^{-4} $& $8.9\times10^{-4} $ \\\hline
        $\Gamma(h\to X)$ & $10.9\UGeV$ & $1.00\UGeV$ & $11.3\UGeV$
        \\\hline
        $\delta_{23}$ & $10^{-0.62}$ & $10^{-1.32}$ & $10^{-0.44}$ \\\hline
        $m_{\PSQ}$ & $990\UGeV$ &  $670\UGeV$ & $990\UGeV$ \\\hline
        $A_b$ & $-2750\UGeV$ & $-1960\UGeV$ & $-2860\UGeV$ \\\hline
        $\mu$ & $-720\UGeV$ & $-990\UGeV$ & $-460\UGeV$ \\\hline
        $BR(b\to s\gamma)$ & $4.50\times 10^{-4}$ & $4.47\times 10^{-4}$ & $4.39\times
        10^{-4}$ \\\hline
    \end{tabular}
    \begin{tabular}{|c||c|c|c|}
        \hline
        $h$ &  $H^0$ & $A^0$ \\\hline\hline
        $\sigma(pp\to h\to t\,c)$ &  $2.4\times 10^{-3}\Upb$ & $5.8\times 10^{-4}\Upb$ \\\hline
        events/$100\Ufb^{-1}$ & 240 & 58  \\\hline
        $BR(h\to tc)$ & $1.9\times 10^{-3} $& $5.7\times 10^{-4} $\\\hline
        $\Gamma(h\to X)$ & $0.41\UGeV$ & $0.39\UGeV$ \\\hline
        $\delta_{23}$ & $10^{-0.10}$ & $10^{-0.13}$ \\\hline
        $m_{\PSQ}$ & $880\UGeV$ & $850\UGeV$ \\\hline
        $A_t$ & $-2590\UGeV$ & $2410\UGeV$ \\\hline
        $\mu$ & $-700\UGeV$ & $-930\UGeV$ \\\hline
        $BR(b\to s\gamma)$ & $4.13\times 10^{-4}$ & $4.47\times 10^{-4}$ \\\hline
    \end{tabular}
 \end{center}
\end{table}

It is enlightening to look at the approximate leading expressions to
understand the qualitative trend of the results. The SUSY-QCD contribution
to the $b\to s\gamma$ amplitude can be approximated to
\begin{equation}\label{bsgamma}
A^{SQCD}(b\to s\gamma)\sim \delta_{23}\,{m_b(\mu-A_b\tan\beta)}/{M_{\mathrm{SUSY}}^2}\,,
\end{equation}
whereas the MSSM Higgs-boson FCNC effective couplings behave as
\begin{equation}
g_{h q\bar{q'}}   \sim \delta_{23} \frac{-\mu \,
    m_{\tilde{g}}}{M_{SUSY}^2} \left\{ \begin{array}{cc}
 \sin(\beta-\alpha_{\rm eff})& (H^0)\\
    \cos(\beta-\alpha_{\rm eff}) & (h^0)\\
 1& (A^0)
  \end{array}\right..
\label{eq:approxleading}
\end{equation}
The different structure of the amplitudes in Eqs.\ (\ref{bsgamma}) and 
(\ref{eq:approxleading}) allows us to obtain an appreciable FCNC
Higgs-boson decay rate, while the prediction for $BR(b\to s\gamma)$ stays inside
the experimentally allowed range.

For the analysis of the bottom-strange production channel, we study
first the Higgs-boson branching ratio in Eq.\ (\ref{eq:hqq-def}).
Fig.~\ref{fig:hbsma0} (left) shows the maximum value of $BR(h\to bs)$ 
as a function of the pseudoscalar Higgs-boson mass $m_{A^0}$. We observe
that fairly large values of $BR(h^0\to bs)\sim 0.3\%$ are
obtained. Tab.\ \ref{tab:maximnowindow} (top) shows the actual values of the
maximum branching ratios and the parameters that provide them for each
Higgs boson. Let us discuss first the general trend, which is valid
for all studied processes: The maximum is attained at large $M_{\mathrm{SUSY}}$ and
moderate $\delta_{23}$. The SUSY-QCD contribution to $b\to s\gamma$ in Eq.\
(\ref{bsgamma})
decreases with $M_{\mathrm{SUSY}}$, therefore to keep $BR(b\to s\gamma)$ in the allowed range
when $M_{\mathrm{SUSY}}$ is small, it has to be compensated with a low value of
$\delta_{23}$, providing a small FCNC effective
coupling in Eq.\ (\ref{eq:approxleading}). On the other hand, at large $M_{\mathrm{SUSY}}$
the second factor in Eq.\ (\ref{bsgamma}) decreases, allowing a larger
value of $\delta_{23}$. Thus, the first factor in
Eq.\ (\ref{eq:approxleading}) grows, but the second factor in
Eq.\ (\ref{eq:approxleading}) stays fixed (provided that $|\mu|\sim
M_{\mathrm{SUSY}}$), overall providing a larger value of the effective coupling. 
On the other hand, a too large value of $\delta_{23}$ has to be compensated
by a small value of $|\mu|/M_{\mathrm{SUSY}}$ in Eq.\ (\ref{bsgamma}), provoking a
reduction in Eq.\ (\ref{eq:approxleading}). In the end, the balance of
the various interactions involved produces the results of
Tab.~\ref{tab:maximnowindow} (top). 

The maximum value of the branching ratio for the lightest Higgs-boson
channel is obtained in the \textit{small $\alpha_{\rm eff}$
  scenario}~\cite{Carena:1999bh,Carena:2002qg}. In this scenario the
coupling of bottom quarks to $h^0$ is extremely suppressed. The
large value of $BR(h^0\to bs)$ is obtained because the total decay width
$\Gamma(h^0\to X)$ in the denominator of Eq.\ (\ref{eq:hqq-def}) tends to
zero (Fig.~\ref{fig:hbsma0}, top), and not because of a large FCNC
partial decay width in its  numerator \cite{Bejar:2004rz}. 

The leading production channel of
$h^0$ at the LHC at high $\tan\beta$ is the associated production with bottom quarks, and
therefore the $h^0$ production will be suppressed when $BR(h^0\to bs)$ is
enhanced. We have to perform a combined analysis of the full
process in Eq.\ (\ref{eq:hqq-def}) to obtain the maximum production rate of FCNC
Higgs-boson meditated events at the LHC. Fig.\ \ref{fig:hbsma0} (centre)
shows the result of the maximization of the
production cross-section~(\ref{eq:hqq-def}). The central column of
Tab.~\ref{tab:maximnowindow} (center) shows that when performing the combined
maximization $\Gamma(h^0\to X)$ has a much larger value, and
therefore the maximum of the combined cross-section is not obtained in
the \textit{small $\alpha_{\rm eff}$ scenario}. The number of expected
events at the LHC is around 50,000 events/100$\Ufb^{-1}$. While it is a
large number, the huge $b$-quark background at the LHC will most likely
prevent its detection. Note, however, that the maximum FCNC branching
ratios are around $10^{-4}$--$10^{-3}$, which is 
 at the same level as the already measured $BR(b\to s\gamma)$.

The numerical results  for the $tc$ channel are similar to
the $bs$ channel, so we  focus mainly on the differences.
Figure~\ref{fig:hbsma0} (right) shows the maximum value of the
production cross-section $\sigma(pp\to h\to t\,c)$ as a function of $m_{A^0}$.
 Only the heavy neutral Higgs bosons contribute to this
channe and we obtain
 a maximum of  $\sigma^{\rm max}(pp\to h\to tc)\simeq
10^{-3}-10^{-2}\Upb$, which means several hundreds events per
100$\Ufb^{-1}$ at the LHC.
 Due to the single top quark signature they
should be easier to detect than the $bs$ channel, providing the key to a
new door to study  physics beyond the Standard Model.
It is now an
experimental challenge to prove that
these events can be effectively be separated from the background.

The single top-quark FCNC signature can also be produced in other
processes, like the direct production 
(see \chapt{chap:top}{sec:singletop} and Ref.~\cite{Guasch:2006hf}), or other
models, like the two-Higgs-doublet model (see \chapt{chap:top}{top:2hdm} and
Refs.~\cite{Bejar:2000ub,Bejar:2003em}). In Table~\ref{tab:comparison}
we make a schematic comparison of these different modes. The two modes
available in SUSY models probe different parts of the
parameter space. While the maximum of the direct production is larger,
it decreases quickly with the  mass, in the end, at 
$M_{\mathrm{SUSY}}=m_{\tilde{g}}\sim 800 \UGeV$ both channels have a similar production
cross-section. As for the comparison with the two-Higgs-doublet model,
the maximum for this later model is obtained in a totally different
parameter set-up than the SUSY model: large $\tan\beta$, large $m_{A^0}$, large
splitting among the Higgs-boson masses, and extremal values of the
CP-even Higgs mixing angle $\alpha$ (large/small $\tan\alpha$ for
$h^0/H^0$). The first two conditions would produce a small value for the
production in SUSY models, while the last two conditions are not
possible in the SUSY parameter space. Then, the detection of a FCNC $tc$
channel at the LHC, together with some other hint on the parameter space
(large/small $\tan\beta$, $m_{A^0}$) would give a strong indication (or
confirmation) of the underlying physics model (SUSY/non-SUSY) chosen by nature.

%
\begin{table}
 \begin{center}
\caption{Comparison of several FCNC top-charm production cross-sections
  at the LHC, for $\sigma^{SUSY}(pp\to h \to tc)$ [this work, and Refs.~\cite{Bejar:2005kv,Bejar:2004rz}],
direct production $\sigma^{SUSY}(pp\to tc)$ (\chapt{chap:top}{sec:singletop} 
and Ref.\cite{Guasch:2006hf}), 
and two-Higgs-doublet model $\sigma^{2HDM}(pp\to h \to tc)$ 
(\chapt{chap:top}{top:2hdm} and
Refs.~\cite{Bejar:2000ub,Bejar:2003em}).}
\label{tab:comparison}
\begin{tabular}{|c||c|c|c|}
\hline
Parameter & SUSY $h\to tc$ & Direct Production & 2HDM $h\to tc$ \\
\hline
Maximum cross-section & $10^{-2}-10^{-3}\Upb$ & $1\Upb$ & $5\times 10^{-3}\Upb$ \\\hline
$\tan\beta$     & Decreases fast & insensitive      & Increases fast \\\hline
$m_{A^0}$     & Decreases fast & insensitive      & Prefers large \\\hline
$M_{\mathrm{SUSY}}$  & Prefers large    & Decreases fast & -- \\\hline
$A_t$     & insensitive    & very sensitive   & --  \\\hline
$\delta_{23}$ & Moderate      & Moderate         & -- \\\hline
Preferred Channel & $H^0$ & -- & $H^0/h^0$ \\\hline
Higgs mass splitting & Given (small) & -- & Prefers large\\\hline
\end{tabular}
\end{center}
\end{table}

\providecommand{\href}[2]{#2}

\subsection{$H\to b\bar s$ and $B$-physics in the MSSM with NMFV}

Here we summarize the results from a phenomenological analysis of the
general constraints on flavour-changing neutral Higgs decays $H\to b\bar s,
s\bar b$, set by bounds from $b\to s\gamma$ on the flavour-mixing parameters in
the squark mass matrices of the MSSM with non-minimal flavour violation
(NMFV) and 
compatible with the data from $B\to X_s\mu^+\mu^-$, assuming
first one and then several types of flavour mixing contributing at a
time~\cite{Hahn:2005qi}.  
Details of the part of the soft-SUSY-breaking Lagrangian responsible for 
the non-minimal squark family mixing and of the 
parametrization of the flavour-non-diagonal squark mass matrices 
are given in~\cite{Hahn:2005qi,FAFCuser} (see also 
\chapt{chap:tools}{sec:tool:feynarts} for a brief description). 
Previous analyses of bounds on SUSY flavour-mixing
parameters from $b\to s\gamma$~\cite{Borzumati:1999qt,Besmer:2001cj,Besmer:2001yd} have shown the importance of the
interference effects between the different types of flavour
violation~\cite{Gabbiani:1996hi,Misiak:1997ei}. 

We define the dimensionless flavour-changing parameters
$(\delta^{u}_{ab})_{23}$ $(ab = \mathrm{LL},\mathrm{LR},\mathrm{RL},\mathrm{RR})$ from the 
flavour-off-diagonal elements of the squark mass matrices 
in the following way (see~\cite{Hahn:2005qi,FAFCuser}),
\begin{equation}
\begin{aligned}
\label{eq:FCparam}
\Delta_\mathrm{LL}^u &\equiv (\delta_{\mathrm{LL}}^u)_{23}\, 
M_{\tilde L,c}\, M_{\tilde L,t}\,,\quad &
\Delta_\mathrm{LR}^u &\equiv (\delta_{\mathrm{LR}}^u)_{23}\, 
M_{\tilde L,c}\, M_{\tilde R,t}\,,\\
\Delta_\mathrm{RL}^u &\equiv (\delta_{\mathrm{RL}}^u)_{23}\, 
 M_{\tilde R,c}\, M_{\tilde L,t}\,,\quad &
\Delta_\mathrm{RR}^u &\equiv (\delta_{\mathrm{RR}}^u)_{23}\, 
M_{\tilde R,c}\, M_{\tilde R,t}\,,
\end{aligned}
\end{equation}
and analogously for the down sector $(\{u,c,t\} \rightarrow
\{d,s,b\})$. For simplicity, we take the same values for the
flavour-mixing parameters in the up- and down-squark sectors:
$(\delta_{ab})_{23}\equiv(\delta_{ab}^u)_{23}=(\delta_{ab}^d)_{23}$.  
The expression for 
the branching ratio $BR(B\to X_s\gamma)$ to NLO is taken 
from \cite{Kagan:1998bh,Kagan:1998ym}. Besides, we assume a common value 
for the soft SUSY-breaking
squark mass parameters, $M_{\text{SUSY}}$, 
and all the various trilinear parameters to be
universal, $A\equiv A_t = A_b = A_c = A_s$~\cite{Hahn:2005qi}.  
These parameters and the
$\delta$'s will be varied over a wide range, subject only to the
requirements that all the squark masses be heavier than 100 GeV, 
$|\mu| > 90$ GeV and $M_2 > 46$ GeV~\cite{Eidelman:2004wy}.  
We have chosen as a reference the following set of parameters:
\begin{equation}
\label{eq:numparameters}
\begin{gathered} 
M_{\text{SUSY}} = 800\text{ GeV}\,, \quad
M_2 = 300\text{ GeV}\,, \quad
M_1 = \frac 53 \frac{s_W^2}{c_W^2} M_2\,, \\
A   = 500\text{ GeV}\,, \quad
m_A = 400\text{ GeV}\,, \quad
\tan\beta = 35\,, \quad
\mu = -700\text{ GeV}\,.
\end{gathered}
\end{equation} 
We have modified the MSSM model file of {\tt FeynArts} to include
general flavour mixing, and added $6\times 6$ squark mass and mixing
matrices to the {\tt FormCalc} evaluation. Both extensions are
publicly
available~\cite{Hahn:2000kx,Hahn:2001rv,Hahn:1998yk,FAFCuser}.  The
masses and total decay widths of the Higgs bosons were computed with
{\tt
FeynHiggs}~\cite{Hahn:2005cu,Heinemeyer:1998yj,Heinemeyer:1998np,FHPage}.

Next we derive the maximum values of $BR(H^0 \to b s)$ compatible with
$BR(B\to X_s\gamma)_{\text{exp}} = (3.3\pm 0.4)\times 10^{-4}$
\cite{Chen:2001fj,Aubert:2002pb} within three standard deviations by varying the
flavour-changing parameters of the squark mass matrices.
The results for the $A^0$ boson are very similar and we do not show them 
separately.

\begin{figure}
 \begin{center}
 \includegraphics[width=0.49\textwidth]{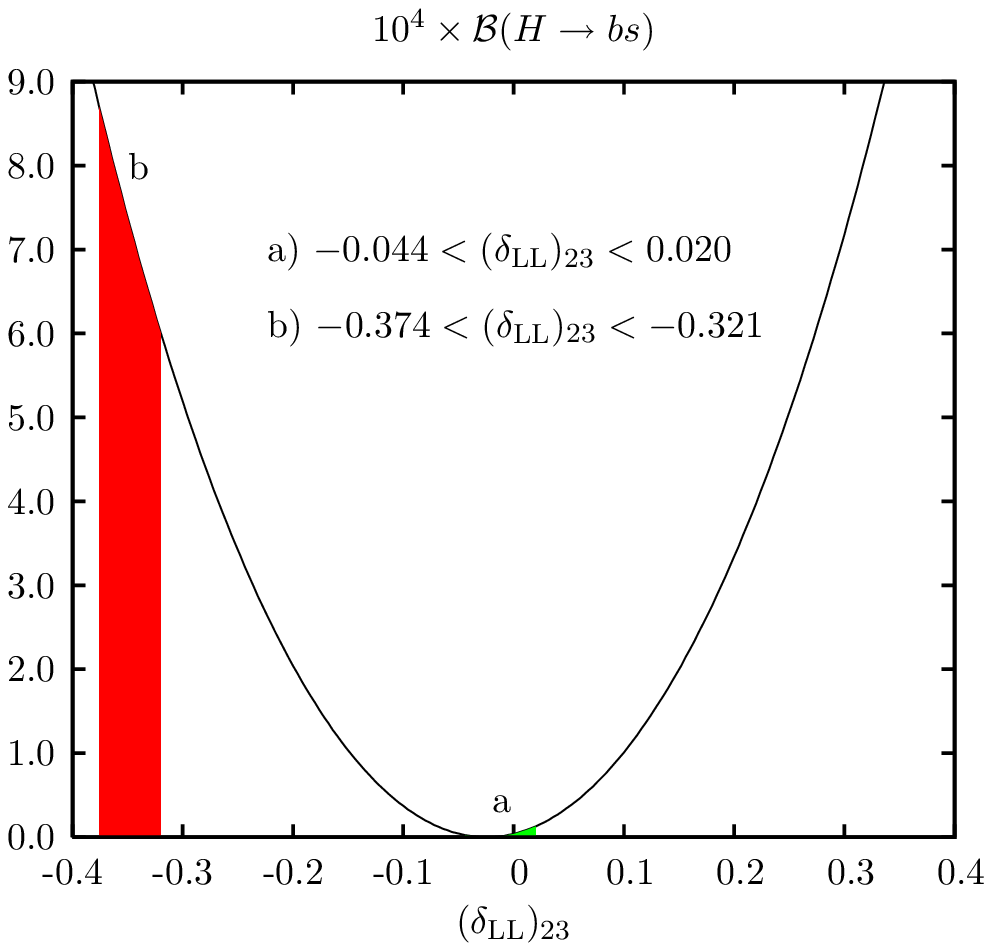}
 \includegraphics[width=0.49\textwidth]{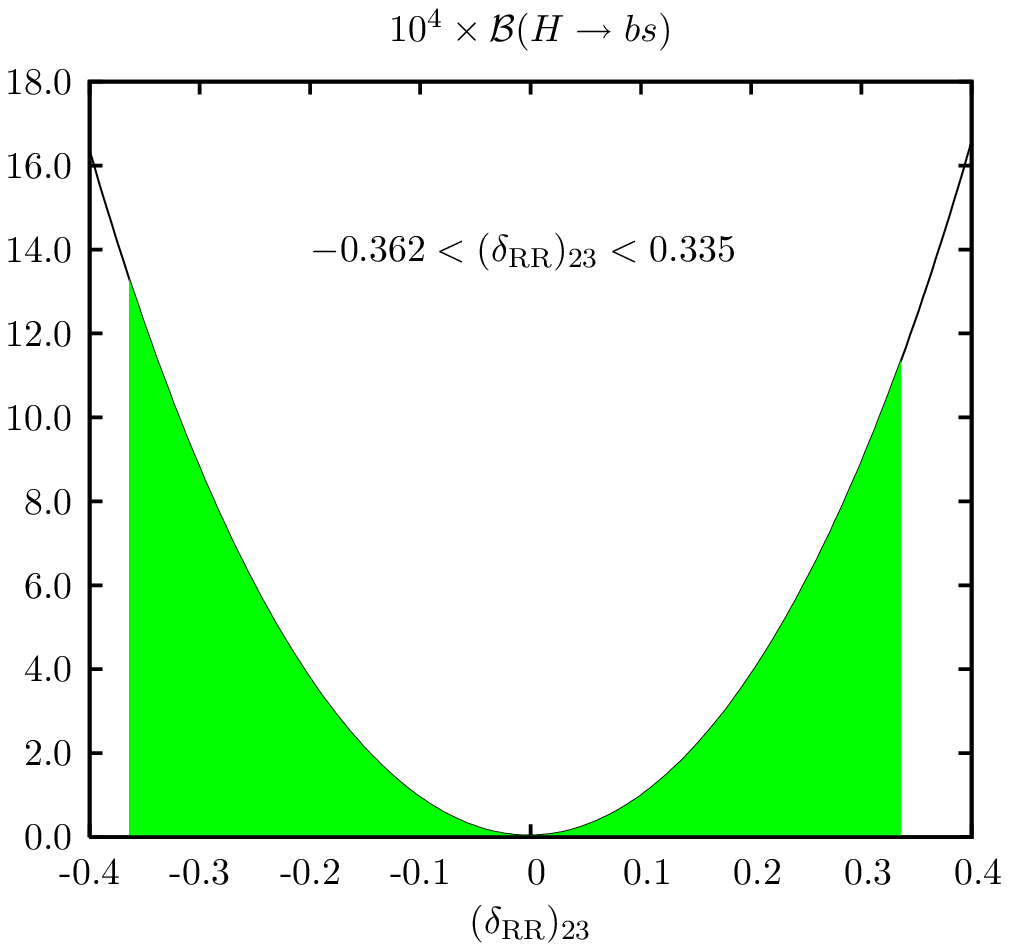} 
 \caption{$BR(H^0\to b s)$ as a function of
 $(\delta_{\mathrm{LL},\mathrm{RR}})_{23}$.  The allowed intervals of these 
 parameters determined from $b\to s\gamma$ are
 indicated by coloured areas. The red (dark-shaded) areas are disfavoured by
 $B\to X_s\mu^+\mu^-$.}
\label{fig:Hbs}
 \end{center}
\end{figure}
As a first step, we select one possible type of flavour violation in the
squark sector, assuming that all the others vanish.  The interference
between different types of flavour mixing is thus ignored.  
We found that the flavour-off-diagonal
elements are independently constrained to be at most $(\delta_{ab})_{23}\sim
10^{-3}$--$10^{-1}$.  As expected~\cite{Gabbiani:1996hi,Misiak:1997ei,Borzumati:1999qt,Besmer:2001cj,Besmer:2001yd}, the bounds on
$(\delta_{\mathrm{LR}})_{23}$ are the strongest, 
$(\delta_{\mathrm{LR}})_{23} \sim 10^{-3}$--$10^{-2}$. The
data from $B\to X_s\mu^+\mu^-$ further constrain the parameters 
$(\delta_{\mathrm{LL}})_{23}$
and $(\delta_{\mathrm{LR}})_{23}$, the others remaining untouched.
The allowed intervals for the corresponding flavour-mixing parameters 
thus obtained are given in~\cite{Hahn:2005qi}.
For our reference point (\ref{eq:numparameters}) we find that the
largest allowed value of $BR(H^0\to b s)$, 
of $\mathcal{O}(10^{-3})$ or $\mathcal{O}(10^{-5})$, 
is induced by $(\delta_{\mathrm{RR}})_{23}$ or 
$(\delta_{\mathrm{LL}})_{23}$, respectively (see~Fig.\ref{fig:Hbs}).
These are the flavour-changing parameters least
stringently constrained by the $b\to s\gamma$ data.  
$BR(H^0\to b s)$ can reach
$\mathcal{O}(10^{-6})$ if induced by $(\delta_{\mathrm{LR}})_{23}$ or 
by $(\delta_{\mathrm{RL}})_{23}$, the most stringently 
constrained flavour-changing parameter. Because of the restrictions imposed by 
$b\to s\gamma$, $BR(H^0\to b s)$
depends very little on $(\delta_{\mathrm{LR}})_{23}$ and 
$(\delta_{\mathrm{RL}})_{23}$. 

\begin{figure}
\begin{center}
\includegraphics[width=0.32\textwidth]{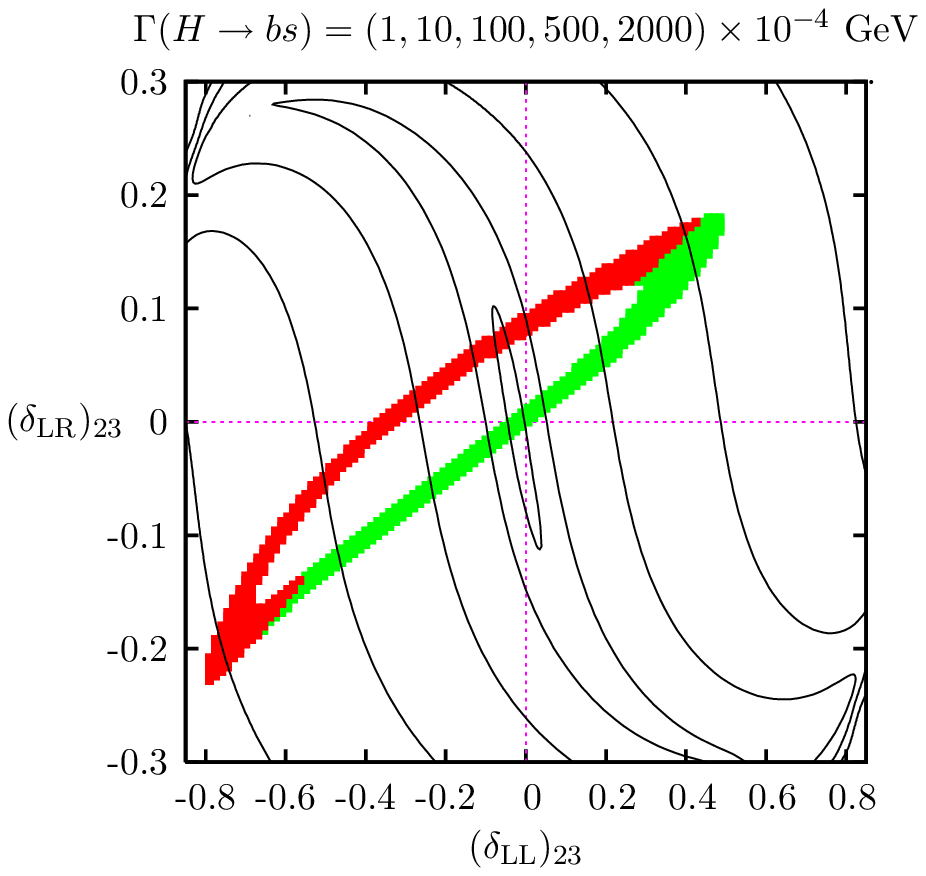}
\includegraphics[width=0.32\textwidth]{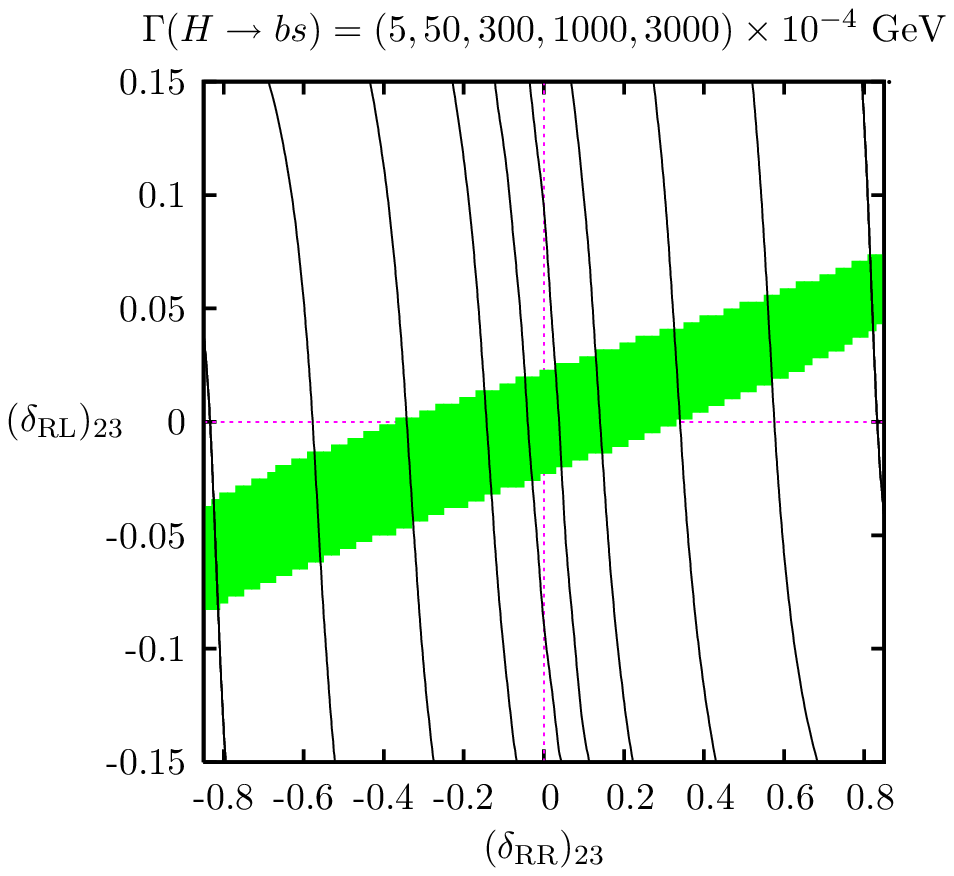}
\includegraphics[width=0.32\textwidth]{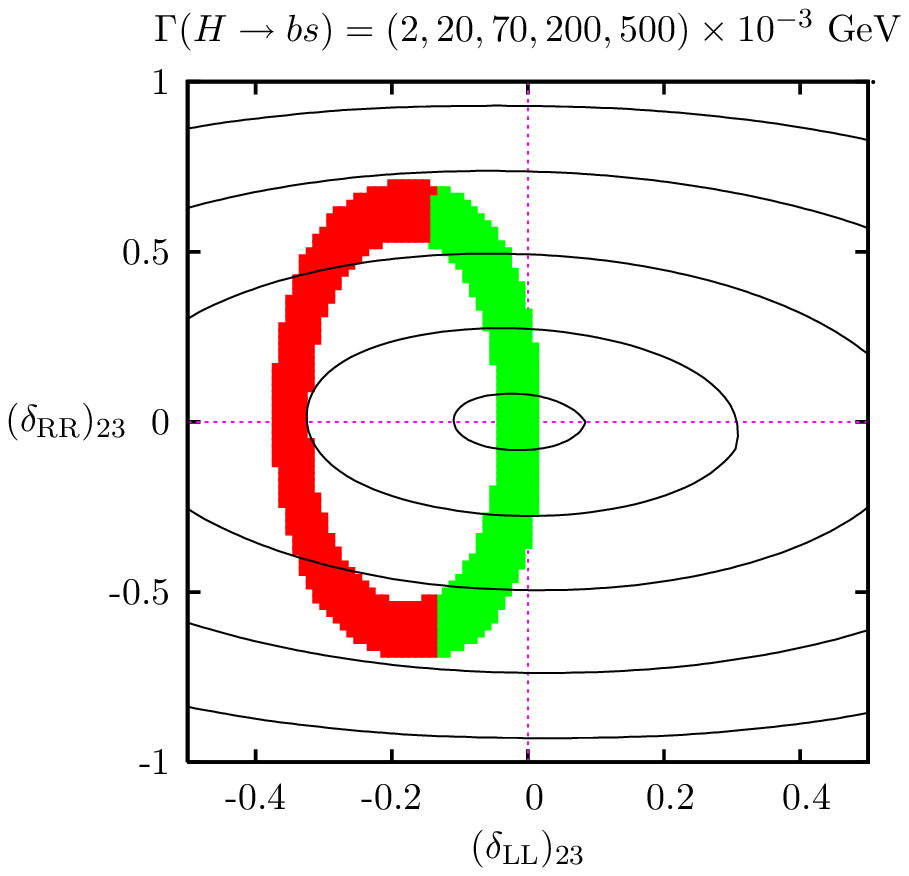}
\end{center}
\vspace*{-3ex}
\caption{Contours of constant $\Gamma(H^0\to b s)$ in 
various planes of $(\delta_{ab})_{23}$.  
The coloured
bands indicate regions experimentally allowed by $B\to X_s\gamma$. The red
bands show regions disfavoured by $B\to X_s\mu^+\mu^-$.}
\label{fig:planes}
\end{figure}

Then, we investigate the case
when two off-diagonal elements of the squark mass matrix contribute
simultaneously. Indeed, we performed the analysis for all
possible combinations of two of the four dimensionless parameters
(\ref{eq:FCparam}). The full results are given in~\cite{Hahn:2005qi}. 
Fig.~\ref{fig:planes} displays part of the results for our parameter set
(\ref{eq:numparameters}).  Contours of constant
$\Gamma(H^0\to b s)\equiv \Gamma(H^0\to b\bar s) + \Gamma(H^0\to s\bar b)$
are drawn
for various combinations $(\delta_{ab})_{23}$--$(\delta_{cd})_{23}$ 
of flavour-mixing
parameters, which we shall refer to as ``$ab$--$cd$ planes'' for short
in the following.  The coloured bands represent regions experimentally
allowed by $B\to X_s\gamma$. The red bands are regions disfavoured by
$B\to X_s\mu^+\mu^-$. The bounds on $(\delta_{\mathrm{LR}})_{23}$, 
the best constrained for only one non-zero
flavour-off-diagonal element, are dramatically relaxed when
other flavour-changing parameters contribute simultaneously.  Values of
$(\delta_{\mathrm{LR}})_{23} \sim 10^{-1}$ are allowed. 
As shown in~Fig.~\ref{fig:planes}, large although
fine-tuned values of $(\delta_{\mathrm{LL}})_{23}$ and 
$(\delta_{\mathrm{LR}})_{23}$ combined are not excluded by
$b\to s\gamma$, yielding e.g. 
$\Gamma(H^0\to b s)_{\text{max}} = 0.25\text{ GeV}
\text{ for } (\delta_{\mathrm{LR}})_{23}=-0.22,  
(\delta_{\mathrm{LL}})_{23}=-0.8$. 
This translates to branching ratios compatible with experimental data 
of $BR(H^0\to b s)_{\text{max}}\sim 10^{-2}$.~\footnote{Here we have 
used the total 
width of $\Gamma(H\to X)\approx 26$ GeV, $H = H^0, A^0$, for the point
(\ref{eq:numparameters}) in the MSSM with MFV.}
It also occurs for the $\mathrm{RL}$--$\mathrm{RR}$ case. 
The combined effects of $\mathrm{RR}$--$\mathrm{LL}$ lead to 
$\Gamma(H^0\to b s)_{\text{max}} = 0.12\text{ GeV}
\text{ for } (\delta_{\mathrm{RR}})_{23} = 0.65, 
(\delta_{\mathrm{LL}})_{23} = \pm0.14\,,$ leading to
$BR(H^0\to b s)_{\text{max}}\sim 10^{-2}$. 

\section{Squark/gaugino production and decay}

Non-minimal flavour violation (NMFV) arises 
in the MSSM from a possible misalignment
between the rotations diagonalizing the quark and squark sectors. It is
conveniently parametrized in the super-CKM basis by non-diagonal entries in
the squared squark mass matrices $M^2_{\tilde{Q}}$, $M^2_{\tilde{U}}$, and
$M^2_{\tilde{D}}$ and the trilinear couplings $A_u$ and $A_d$. Squark
mixing is expected to be the largest for the second and third generations due to
the large Yukawa couplings involved. In addition, stringent experimental
constraints for the first generation are imposed by precise measurements of
$K^0-\bar{K}^0$ and $D^0-\bar{D}^0$ mixing. Furthermore, direct searches of
flavour violation depend on the possibility of flavour tagging, established
experimentally only for heavy flavours. We therefore consider here only
mixings of second- and third-generation squarks and follow the conventions
of Ref.\ \cite{Heinemeyer:2004by}.

\subsection{Flavour-violating squark- and gaugino-production at the LHC}

We impose mSUGRA [$m_0$, $m_{1/2}$, $A_0$, $\tan\beta$, and sgn$(\mu)$]
parameters at a large (grand unification) scale and use two-loop
renormalization group equations and one-loop finite corrections as
implemented in the computer program {\tt SPheno} 2.2.2 \cite{Porod:2003um} to
evolve them down to the electroweak scale. At this point, we generalize the
squark mass matrices by including non-diagonal terms $\Delta_{ij}$. The
scaling of these terms with the SUSY-breaking scale $M_{\rm SUSY}$ implies
a hierarchy $\Delta_{\rm LL}\gg\Delta_{\rm LR,RL}\gg\Delta_{\rm RR}$
\cite{Brax:1995up}. We therefore take $\Delta_{\rm LR,RL}=\Delta_{\rm RR}=
0$, while $\Delta_{\rm LL}^t=\lambda^t M_{\tilde{L}_t}M_{\tilde{L}_c}$ and
$\Delta_{\rm LL}^b=\lambda^b M_{\tilde{L}_b} M_{\tilde{L}_s}$, and assume
for simplicity $\lambda=\lambda^t=\lambda^b$. The squark mass matrices are
then diagonalized, and constraints from low-energy and electroweak precision
measurements are imposed on the corresponding theoretical observables,
calculated with the computer program {\tt FeynHiggs} 2.5.1
\cite{Heinemeyer:1998yj}.

Flavour-changing neutral-current (FCNC) $B$-decays and $B^0-\bar{B}^0$ mixing
arise in the SM only at the one-loop level. These processes are therefore
particularly sensitive to non-SM contributions entering at the same order in
perturbation theory and have been intensely studied at $B$-factories. The
most stringent constraints on SUSY-loop contributions in minimal and
non-minimal flavour violation come today from the inclusive $b\to s\gamma$
decay rate as measured by BaBar, Belle, and CLEO, ${\rm BR}(b\to s\gamma)=
(3.55\pm 0.26)\times 10^{-4}$ \cite{Yao:2006px}, which affects directly the
allowed squark mixing between the second and third generation
\cite{Hahn:2005qi}.

Another important consequence of NMFV in the MSSM is the generation of large
splittings between squark-mass eigenvalues. The splitting within isospin
doublets influences the $Z$- and $W$-boson self-energies at zero-momentum
$\Sigma_{Z,W}(0)$ in the electroweak $\rho$-parameter $\Delta\rho=\Sigma_Z
(0)/M_Z^2-\Sigma_W(0)/M_W^2$ and consequently the $W$-boson mass $M_W$ and
the squared sine of the weak mixing angle $\sin^2\theta_W$. The latest
combined fits of the $Z$-boson mass, width, pole asymmetry, $W$-boson and
top-quark mass constrain new physics contributions to $\Delta\rho$ to
$T=-0.13\pm0.11$ or $\Delta\rho=-\alpha T=0.00102\pm0.00086$
\cite{Yao:2006px}.

A third observable sensitive to SUSY loop-contributions is the anomalous
magnetic moment $a_\mu=(g_\mu-2)/2$ of the muon, for which recent BNL data
and the SM prediction disagree by $\Delta a_\mu=(22\pm10)\times 10^{-10}$
\cite{Yao:2006px}. In our calculation, we take into account the SM and MSSM
contributions up to two loops \cite{Heinemeyer:2003dq,Heinemeyer:2004yq}.

For cosmological reasons, we require the lightest SUSY particle (LSP) to be
electrically neutral. We also calculate, albeit for minimal flavour violation
($\lambda=0$) only, the cold dark matter relic density using the computer
program DarkSUSY \cite{Gondolo:2004sc} and impose a limit of $0.094<
\Omega_c h^2<0.136$ at 95\% (2$\sigma$) confidence level. This limit has
recently been obtained from the three-year data of the WMAP satellite,
combined with the SDSS and SNLS survey and Baryon Acoustic Oscillation data
and interpreted within a more general (11-parameter) inflationary model
\cite{Hamann:2006pf}. This range is well compatible with the older,
independently obtained range of $0.094 <\Omega_c h^2<0.129$
\cite{Ellis:2003cw}.

\begin{figure}
 \centering
 \includegraphics[width=0.4\columnwidth]{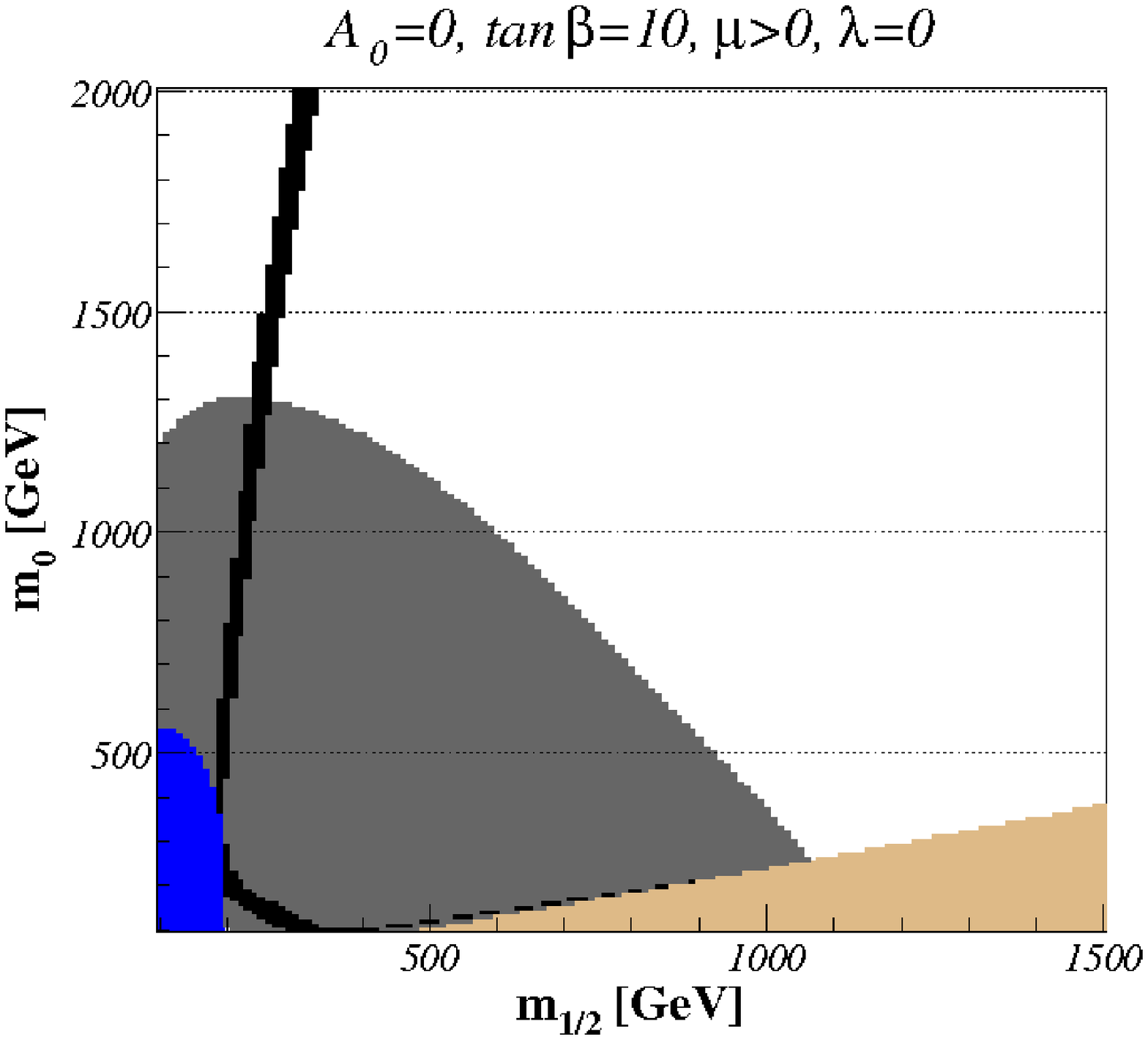}
 \includegraphics[width=0.4\columnwidth]{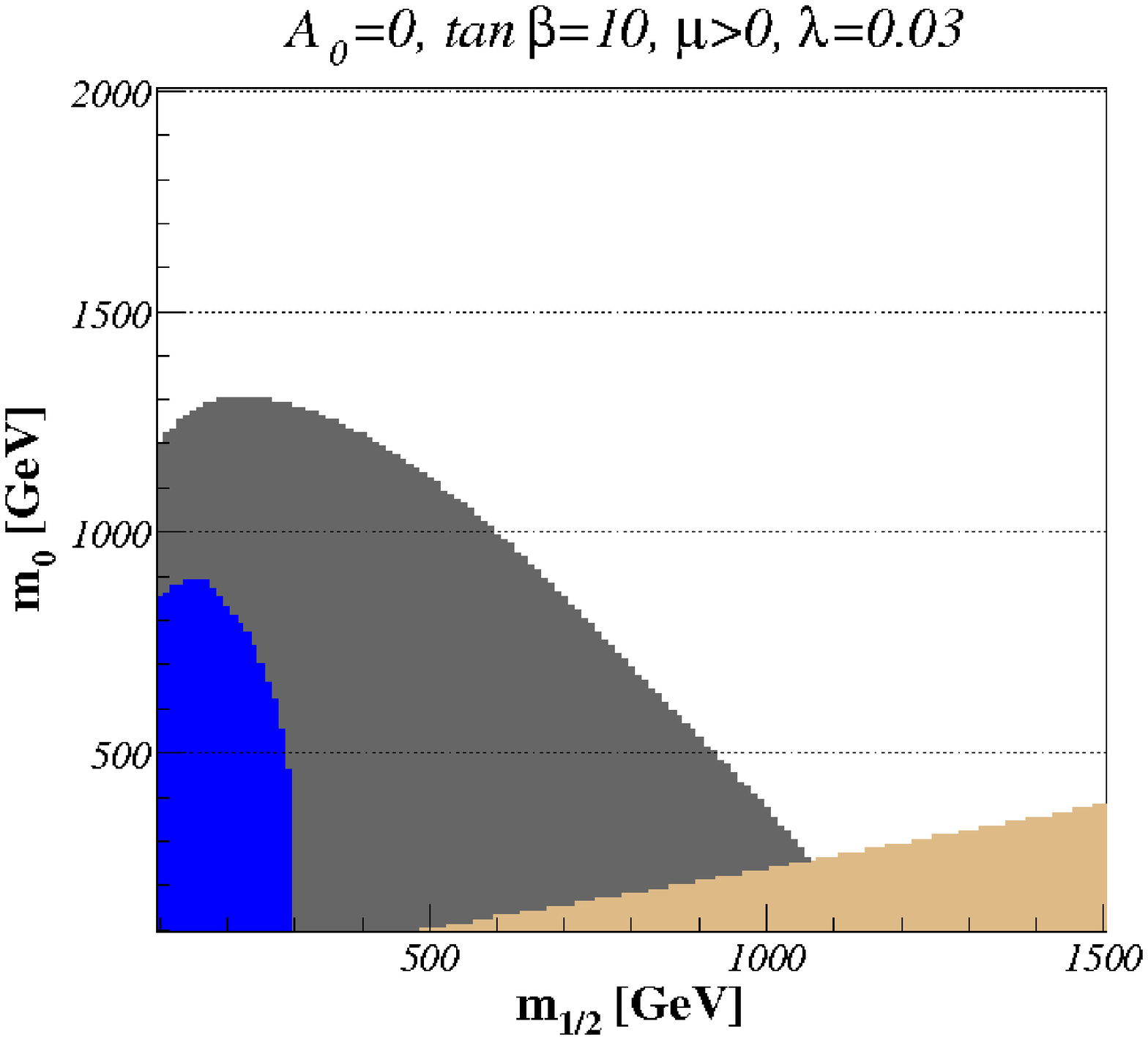}
 \includegraphics[width=0.4\columnwidth]{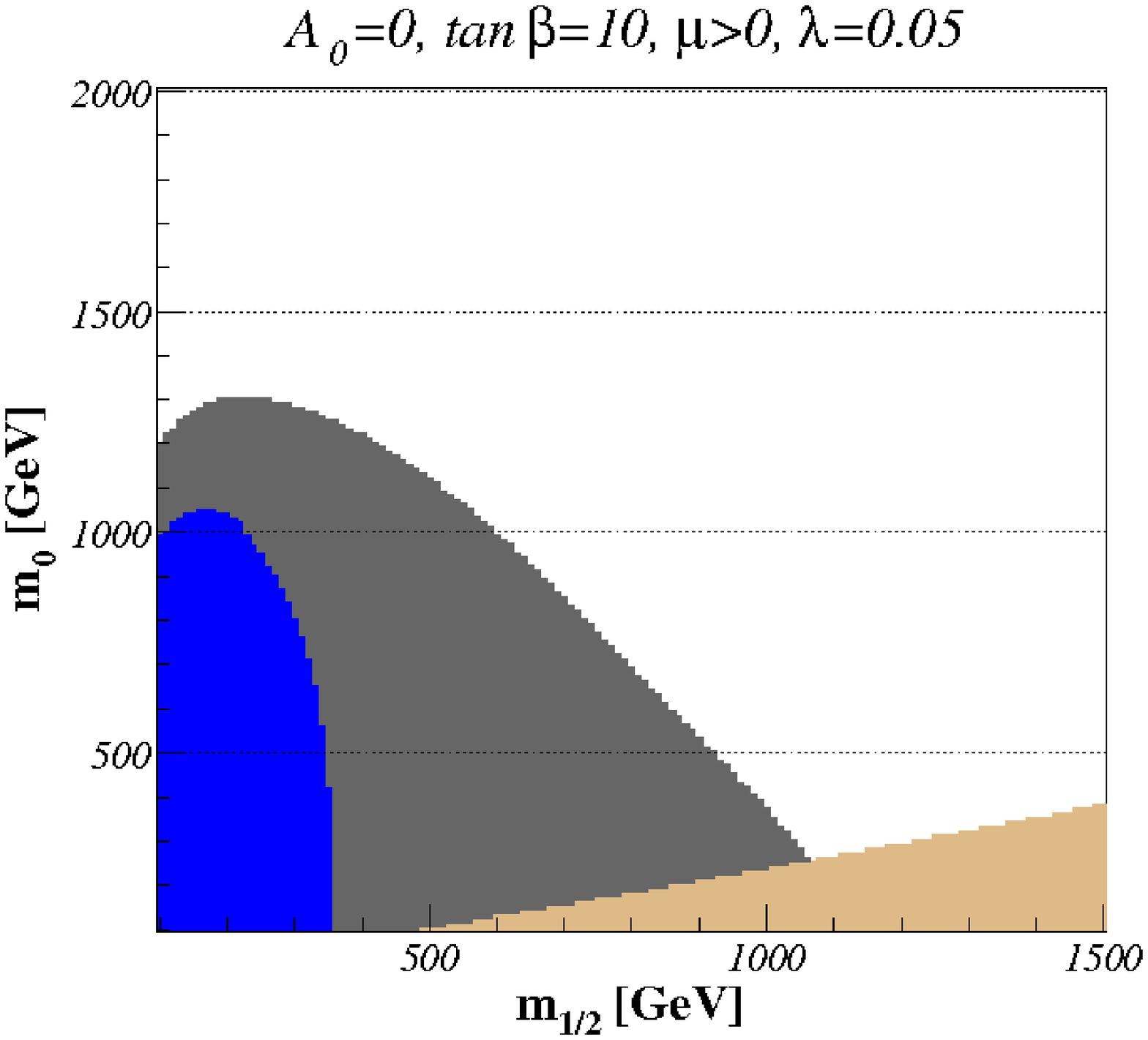}
 \includegraphics[width=0.4\columnwidth]{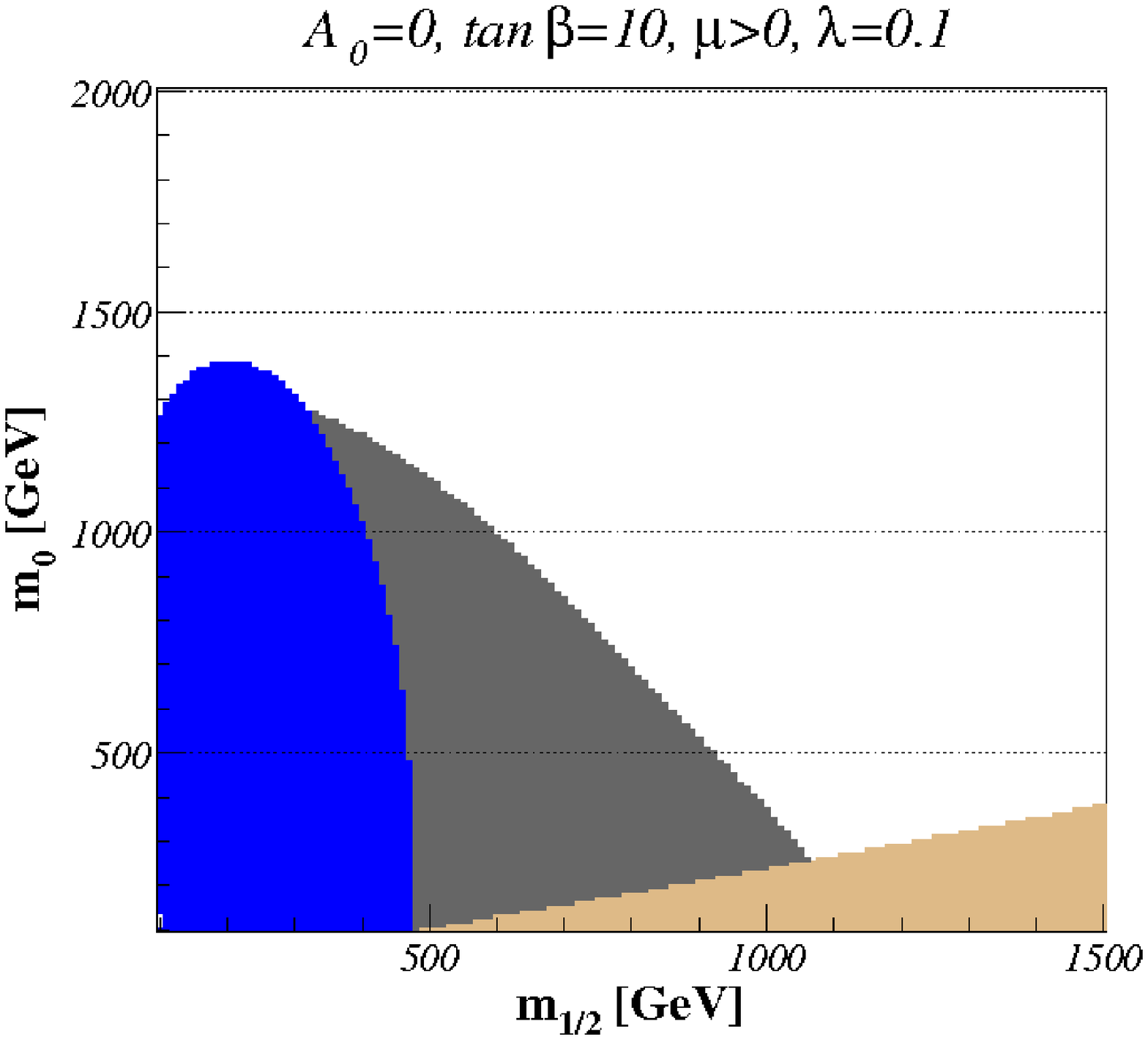}
 \caption{$a_\mu$ (grey) and WMAP (black) favored as
 well as $b\to s\gamma$ (blue) and charged LSP (orange) excluded regions of
 mSUGRA parameter space in minimal ($\lambda=0$) and non-minimal
 ($\lambda>0$) flavour violation.}
\label{fig:bfhk01}
\end{figure}

Typical scans of the mSUGRA parameter space with $\tan\beta=10$, $A_0=0$
and $\mu>0$ and all experimental limits imposed at the $2\sigma$ level are
shown in Fig.\ \ref{fig:bfhk01}. Note that $\mu<0$ is disfavored by
%
%
$g_\mu-2$ data, while $\Delta\rho$ only constrains the parameter space
outside the mass regions shown here. In minimal flavour-violation, light
SUSY scenarios such as the SPS 1a benchmark point ($m_0=100$ GeV,
$m_{1/2}=250$ GeV) \cite{Allanach:2002nj} are favored 
 $g_\mu-2$ data. The dependence on the trilinear coupling $A_0$
($-100$ GeV for SPS 1a, $0$ GeV in our scenario) is extremely weak.

\begin{figure}
 \centering
 \includegraphics[width=0.4\columnwidth]{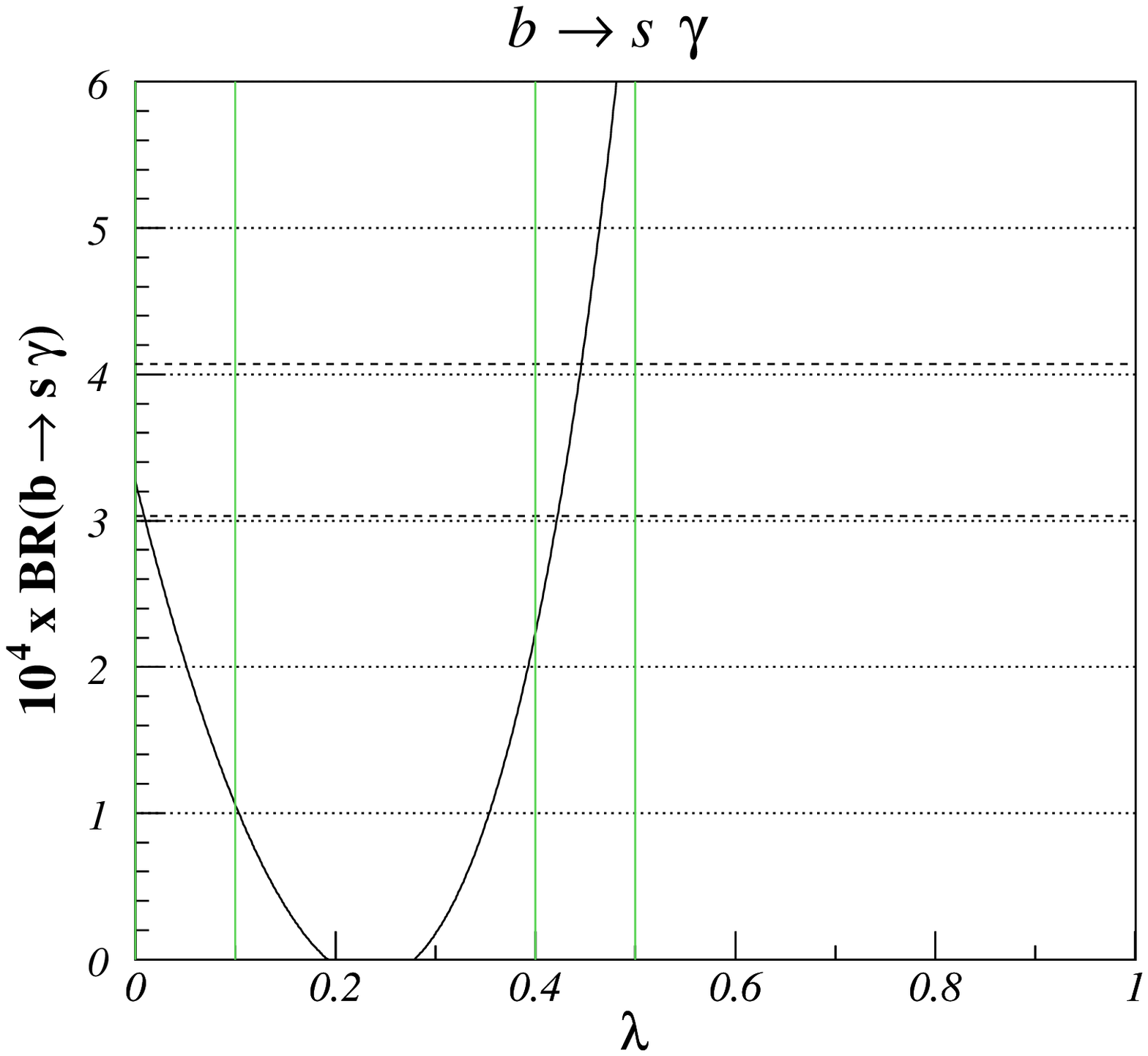}
 \includegraphics[width=0.4\columnwidth]{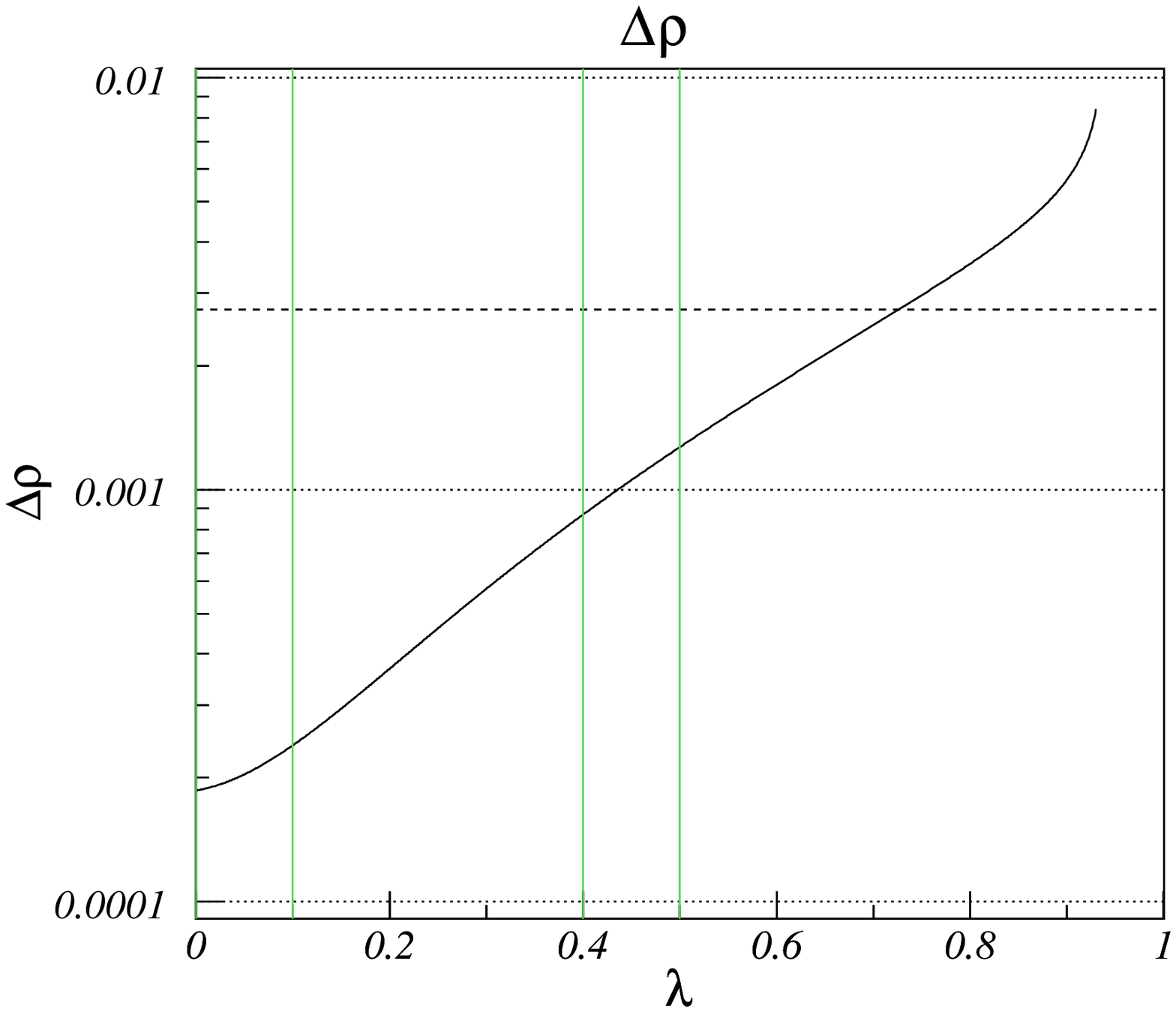}
 \includegraphics[width=0.4\columnwidth]{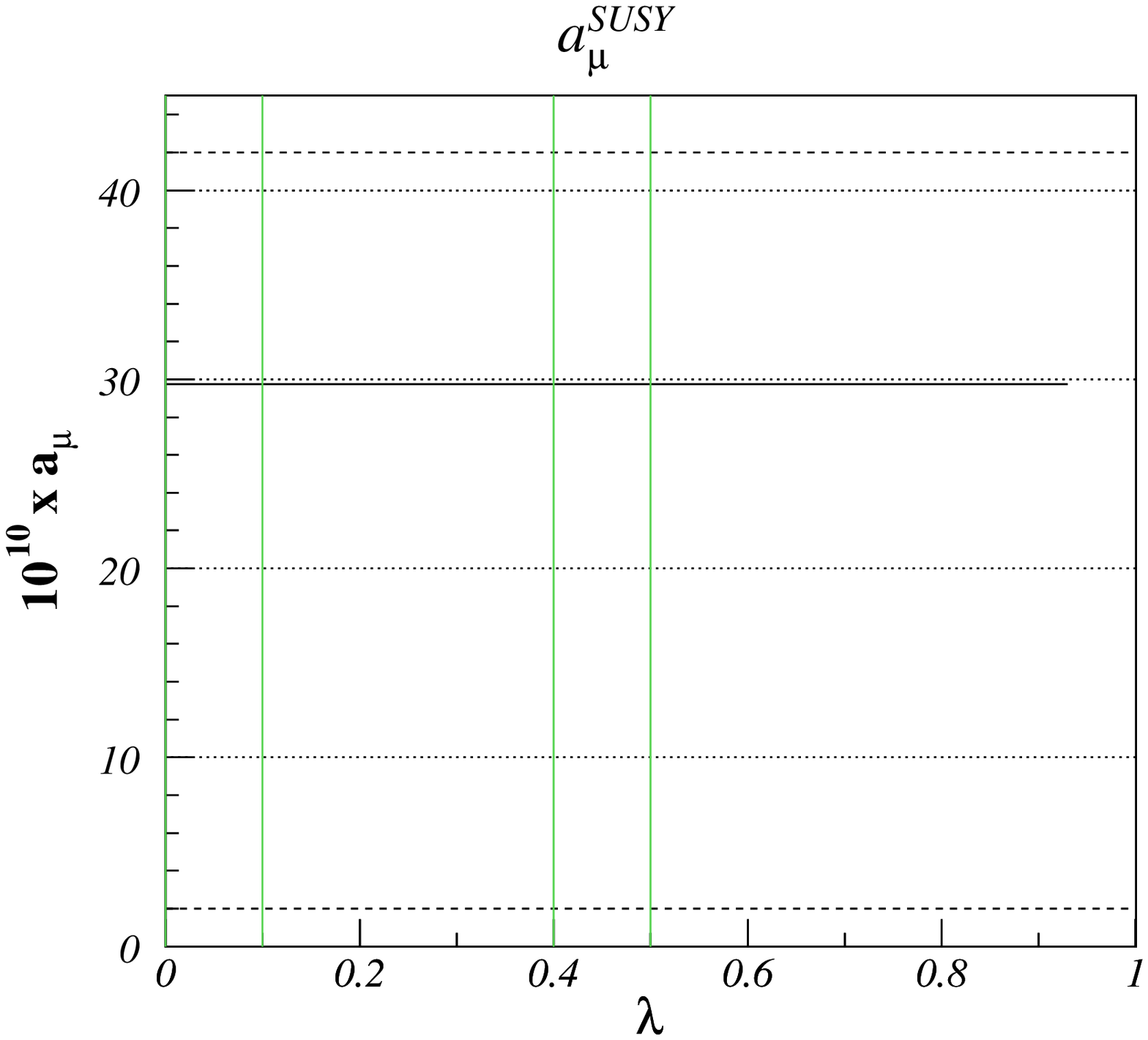}
 \includegraphics[width=0.4\columnwidth]{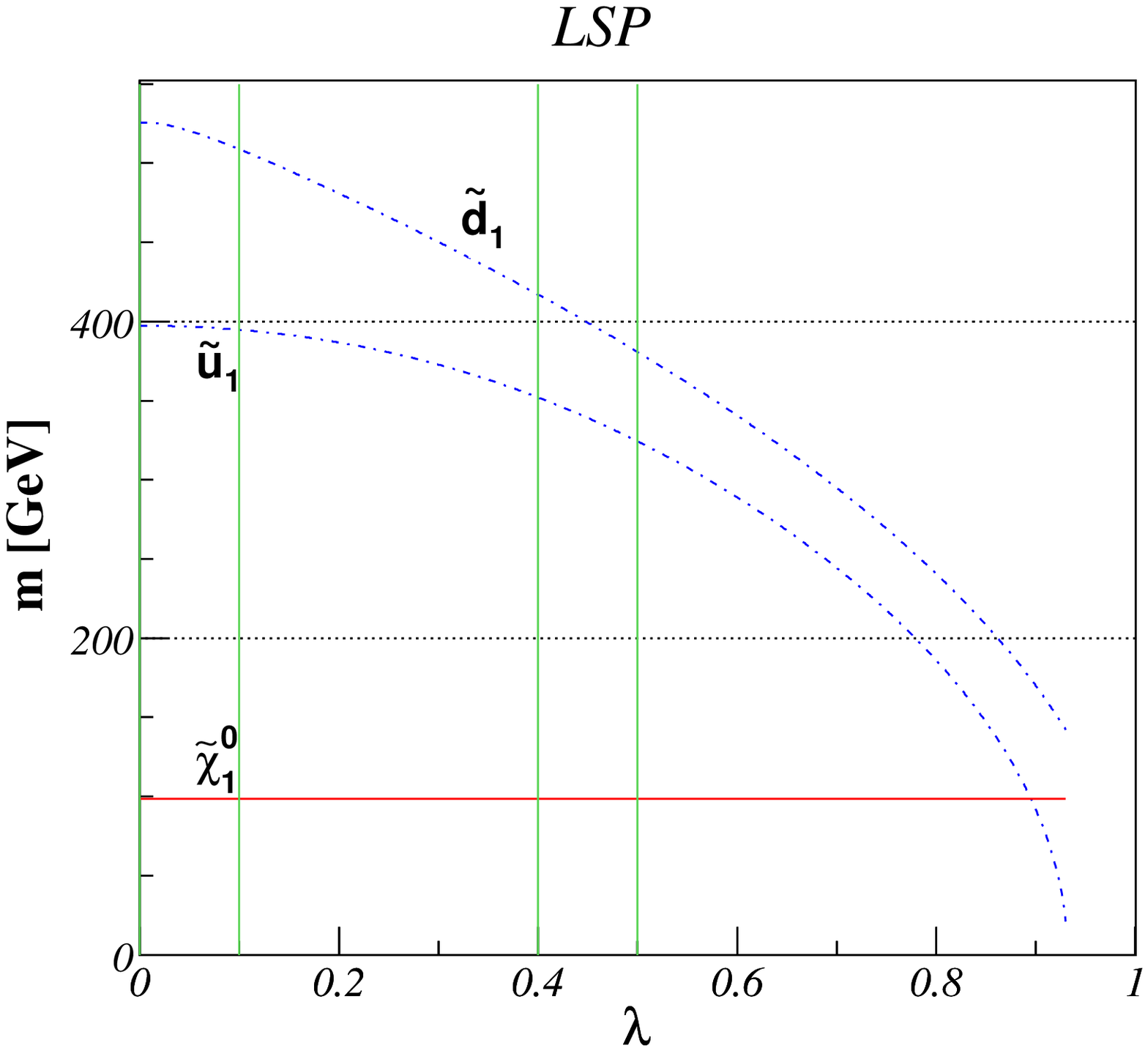}
 \caption{Dependence of the precision variables
 BR$(b\to s\gamma)$, $\Delta\rho$, and $a_\mu$ and the lightest SUSY
 particle masses on the NMFV parameter $\lambda$.}
\label{fig:bfhk02}
\end{figure}

In Fig.\ \ref{fig:bfhk02} we show for our (slightly modified) SPS 1a
%
%
benchmark point the dependence of the electroweak precision variables and
the lightest SUSY particle masses on the NMFV parameter $\lambda$,
indicating by dashed lines the ranges allowed experimentally within two
standard deviations. It is interesting to see that for this benchmark point,
not only the region close to minimal flavour violation ($\lambda<0.1$) is
allowed, but that there is a second allowed region at $0.4<\lambda<0.5$.

\begin{figure}
 \centering
 \includegraphics[width=\columnwidth]{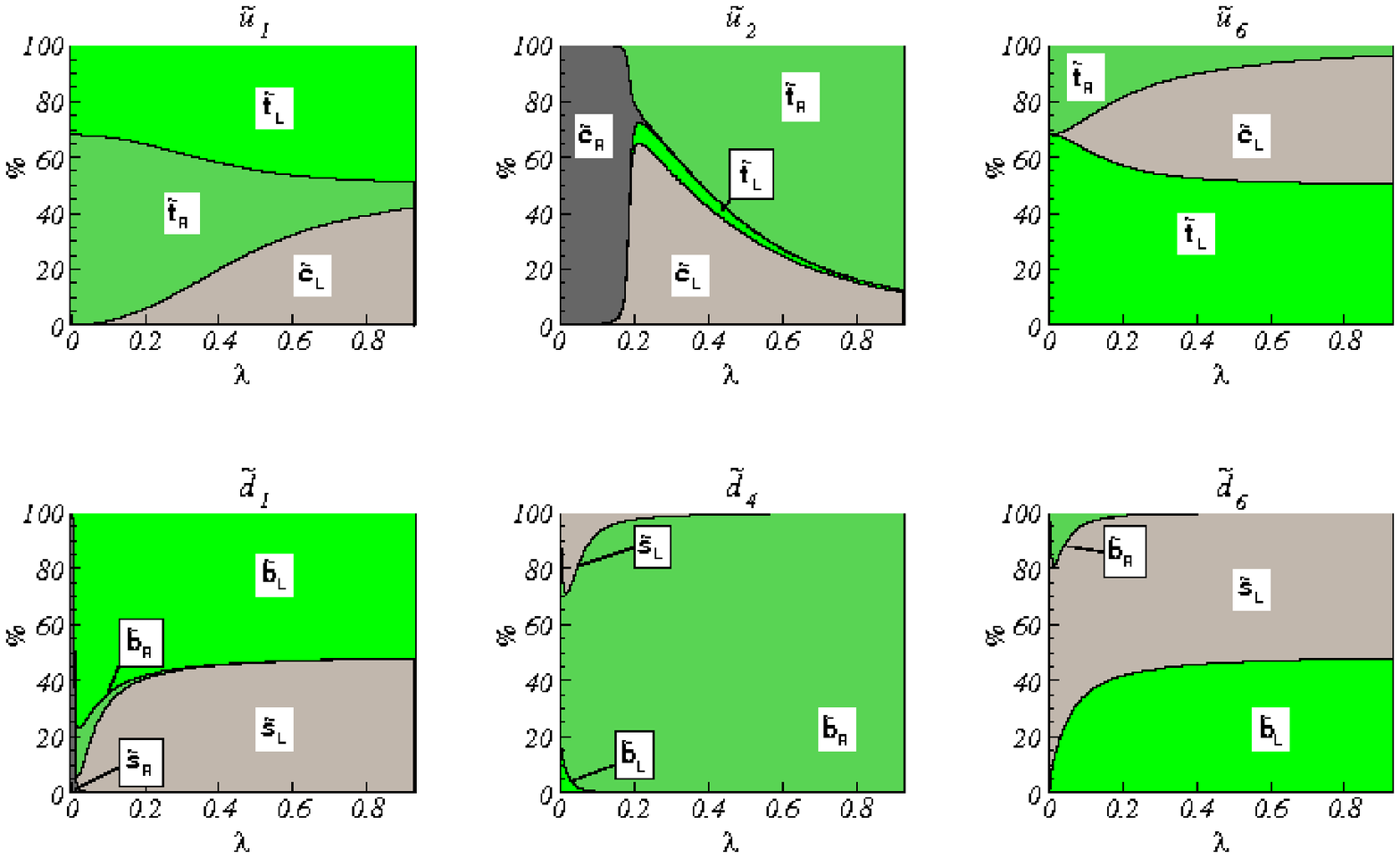}
 \caption{Decomposition of the chirality (L,R) and flavour
 (c,t and s,b) content of the lightest ($\tilde{q}_1,\tilde{q}_2$) and
 heavier ($\tilde{q}_4,\tilde{q}_6$) up- ($q=u$) and down-type ($q=d$)
 squarks on the NMFV parameter $\lambda$.}
\label{fig:bfhk03}
\end{figure}

Next, we study in Fig.\ \ref{fig:bfhk03} the chirality and flavour
%
%
decomposition of the light (1,2) and heavy (4,6) squarks, which changes
mostly in a smooth way, but sometimes dramatically in very small intervals
of $\lambda$. In particular, the second allowed region at larger $\lambda$
has a quite different flavour and chirality mixture than the one at small
$\lambda$.

\begin{figure}
 \centering
 \includegraphics[width=0.49\columnwidth]{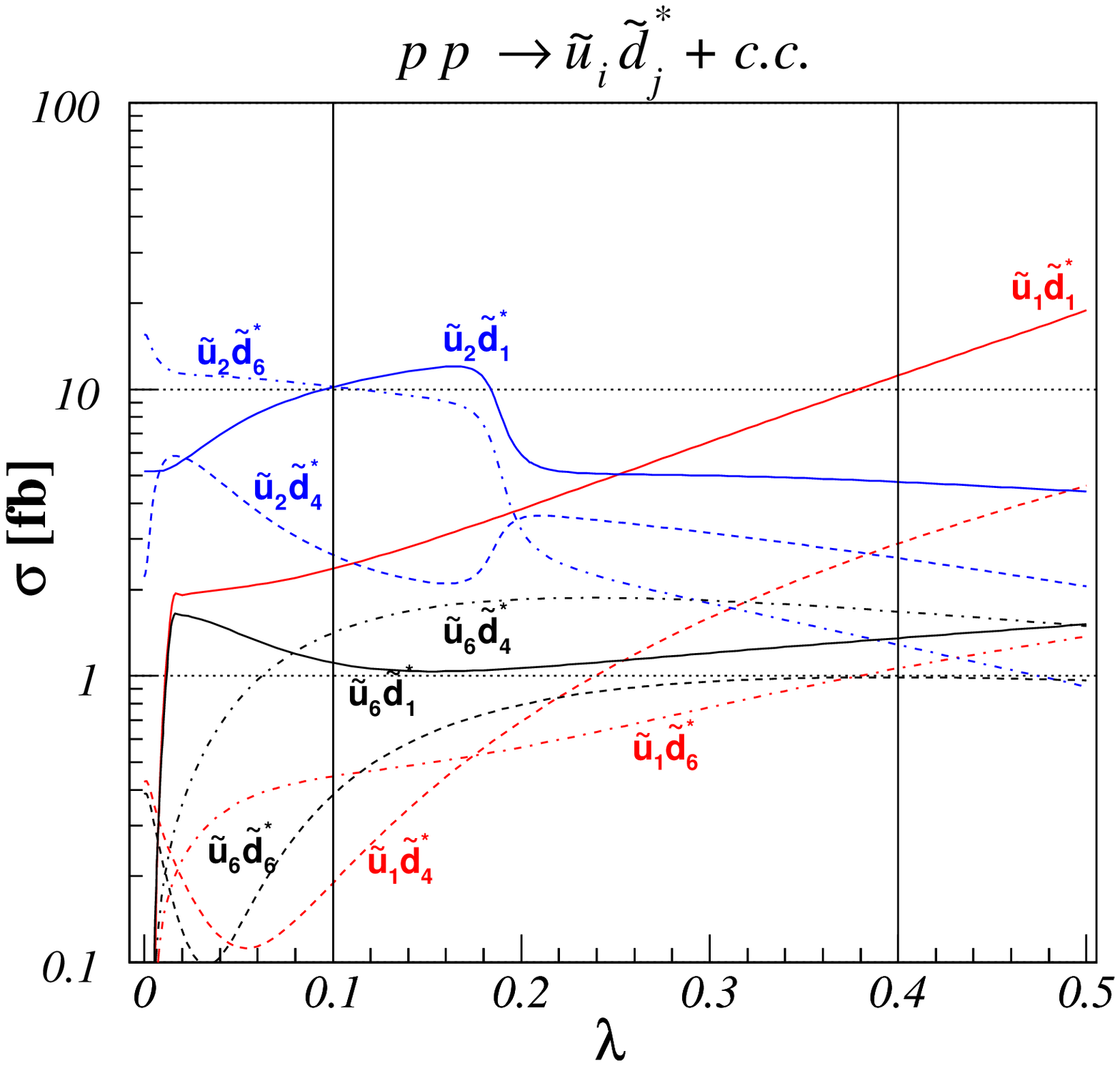}~~
 \includegraphics[width=0.49\columnwidth]{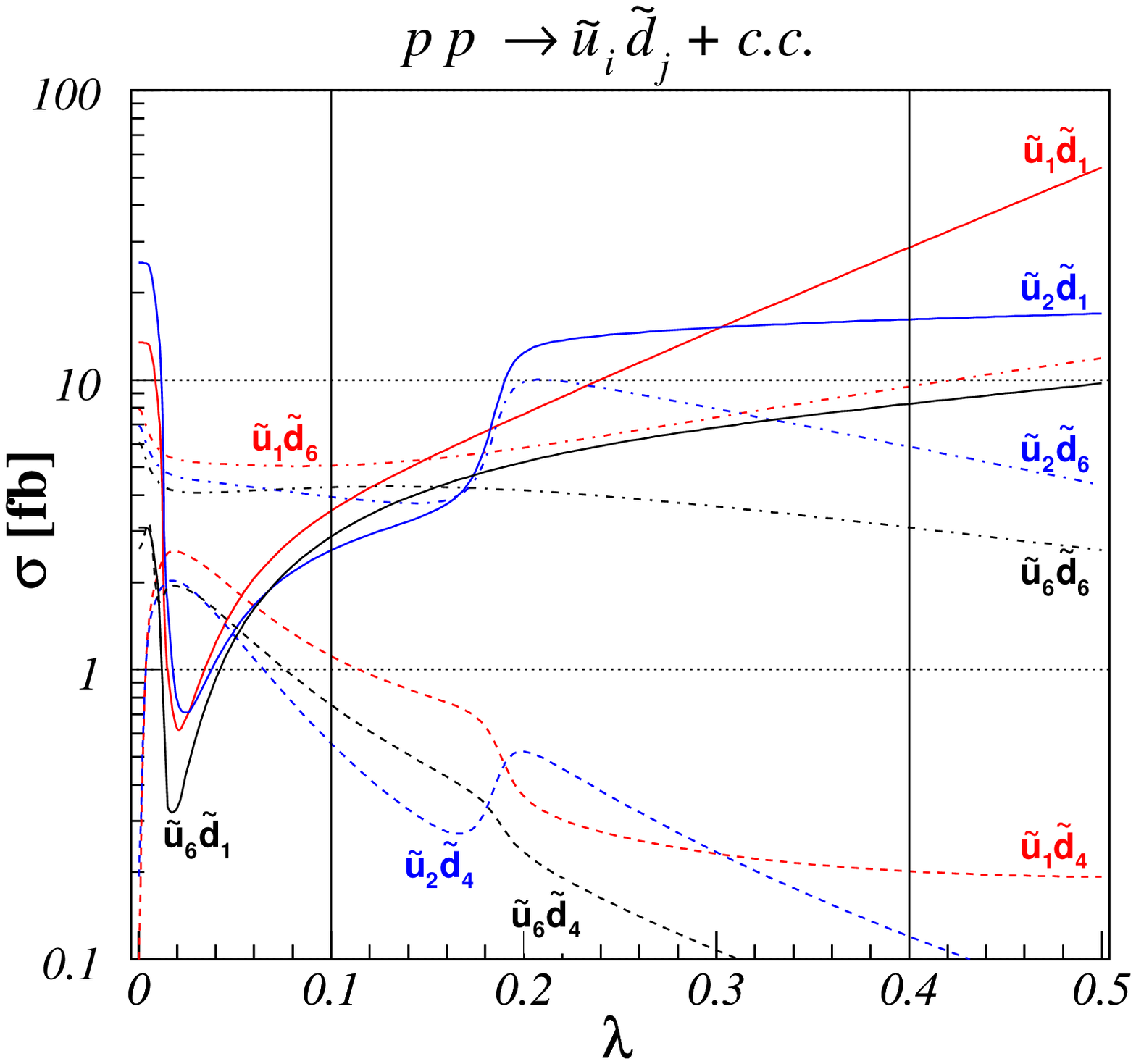}\\[1mm]
 \includegraphics[width=0.49\columnwidth]{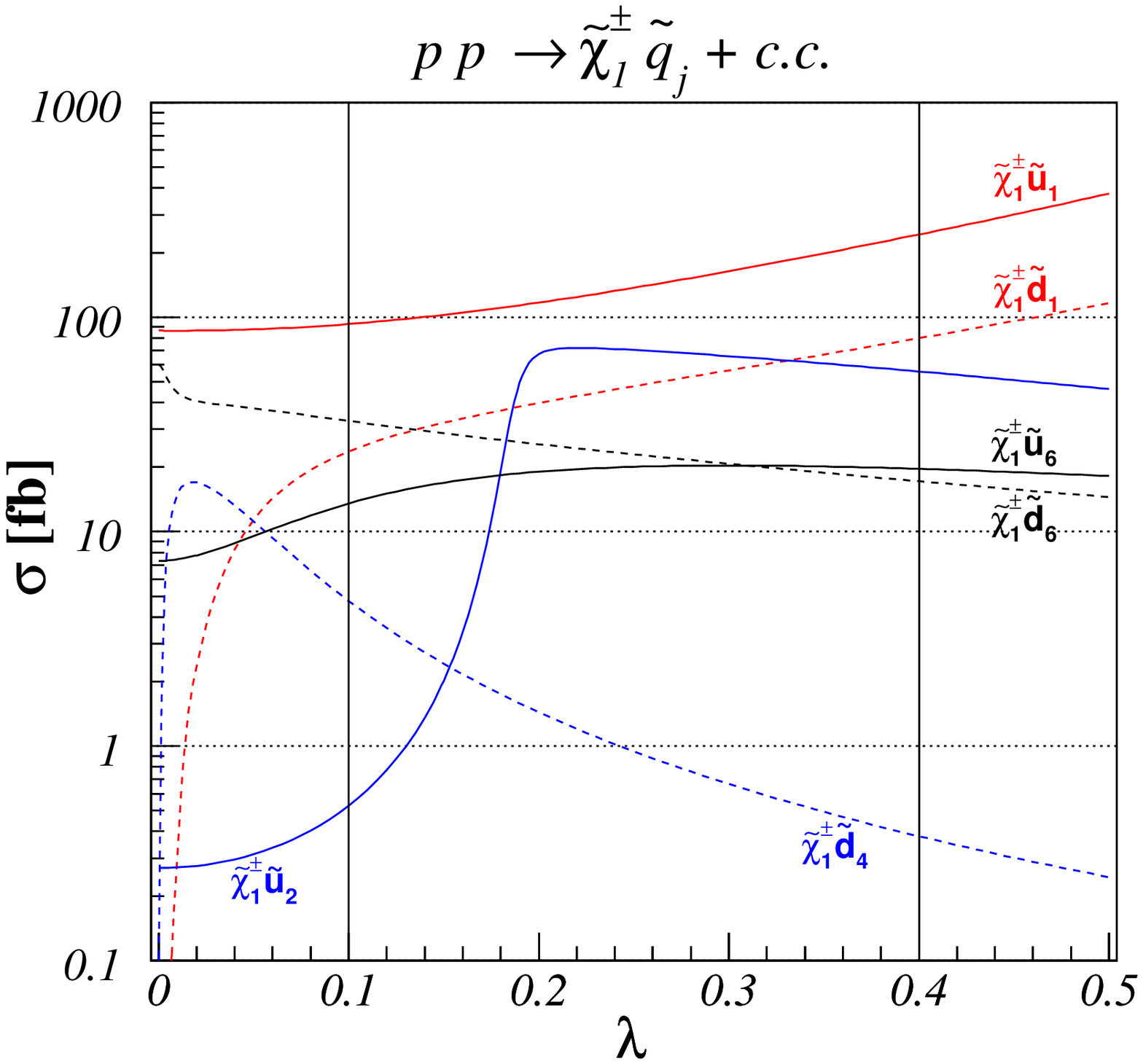}~~
 \includegraphics[width=0.49\columnwidth]{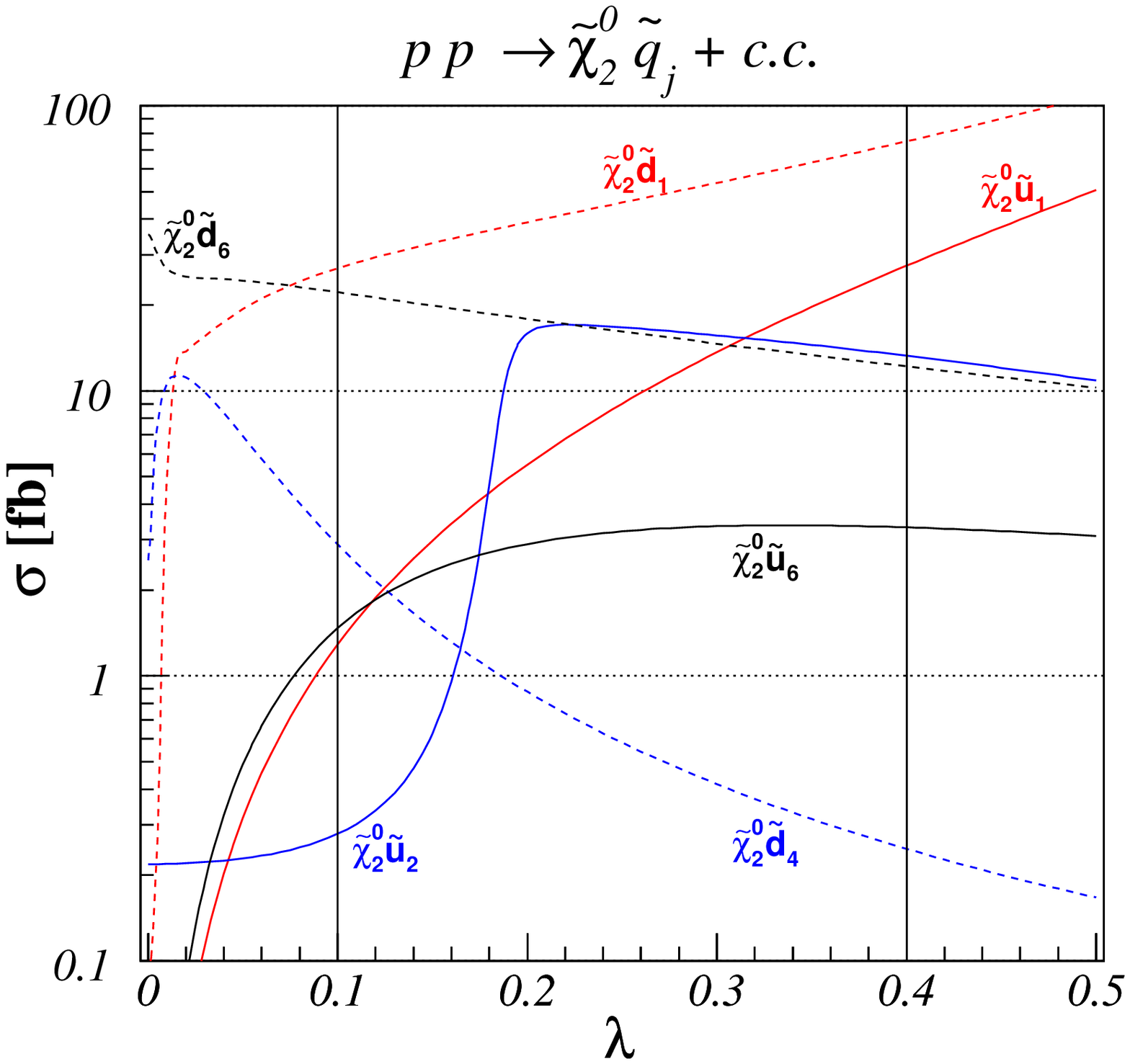}
 \caption{Representative sample of squark and gaugino
 production cross sections at the LHC in NMFV.}
\label{fig:bfhk04}
\end{figure}

The main result of our work is the calculation of all electroweak (and
strong) squark and gaugino production channels in NMFV SUSY \cite{bfhk07}.
We show in Fig.\ \ref{fig:bfhk04} a small, but representative sample of
%
%
these production cross sections: charged squark-antisquark pair production,
non-diagonal squark-squark pair production, as well as chargino-squark and
neutralino-squark associated production. The two $b\to s\gamma$ allowed
regions ($\lambda<0.1$ and $0.4<\lambda<0.5$) are indicated by vertical
lines. Note that NMFV allows for a top-flavour content to be produced from
non-top initial quark densities and for right-handed chirality content to be
produced from strong gluon or gluino exchanges. The cross sections shown
here are all in the fb range and lead mostly to experimentally identifiable
heavy-quark (plus missing transverse-energy) final states.

In conclusion, we have performed a search in the NMFV-extended mSUGRA
parameter space for regions allowed by electroweak precision data as well as
cosmological constraints. In a benchmark scenario similar to SPS 1a, we find
two allowed regions for second- and third-generation squark mixing,
$\lambda<0.1$ and $0.4<\lambda<0.5$, with distinct flavour and chirality
content of the lightest and heaviest up- and down-type squarks. Our
calculations of NMFV production cross sections at the LHC demonstrate
that the corresponding squark (anti-)squark pair production channels and the
associated production of squarks and gauginos are very sensitive to the NMFV
parameter $\lambda$. For further details see Ref.\ \cite{Bozzi:2007me}.


\subsection{Flavour-violating squark and gluino decays}

In the study of squark decays two general scenarios can be distinguished
depending on the hierarchy within the SUSY spectrum:
\begin{itemize}
\item $m_{\tilde g} > m_{\tilde q_i}$ ($q=d,u$; $i=1,\dots,6$):  In 
  this case the gluino will mainly decay according to
  \begin{eqnarray}
    \tilde g \to d_j \, \tilde d_i \,,\,\,\,    \tilde g \to u_j \, \tilde u_i 
  \end{eqnarray}
  with $d_j = (d,s,b)$ and $u_j = (u,c,t)$ followed by squark decays
  into neutralino and charginos
  \begin{eqnarray}
   \tilde u_i \to u_j \tilde \chi^0_k \, , \,  d_j \tilde \chi^+_l \,, \,\,\,  
   \tilde d_i \to d_j \tilde \chi^0_k \, , \,  u_j \tilde \chi^-_l \,\,.
   \end{eqnarray}
  In addition there can be decays into gauge- and Higgs bosons if
  kinematically allowed: 
  \begin{eqnarray}
   \tilde u_i &\to& Z \tilde u_k\,, \,\,  H^0_r \tilde u_k\,, \,\,
       W^+ \tilde d_j\,, \,\,  H^+ \tilde d_j \\
   \tilde d_i &\to&  Z \tilde d_k\,, \,\,  H^0_r \tilde d_k\,, \,\,
       W^- \tilde u_j\,, \,\,  H^- \tilde u_j
  \end{eqnarray}
  where $H^0_r = (h^0, H^0, A^0)$, $k < i$, $j=1,\dots,6$.
  Due to the fact, that there is left-right mixing in the
  sfermion mixing, one has flavour changing neutral decays into $Z$-bosons
  at tree-level. 
\item $m_{\tilde g} < m_{\tilde q_i}$ ($q=d,u$; $i=1,\dots,6$): In 
  this case the squarks decay mainly into a gluino:
  \begin{eqnarray}
   \tilde u_i \to u_j \tilde g \,, \,\,\,
   \tilde d_i \to d_j \tilde g \,\,
  \end{eqnarray}
  and the gluino decays via three-body decays and loop-induced two-body
  decays into charginos and neutralinos
  \begin{eqnarray}
    \tilde g \to d_j \, d_i \, \tilde \chi^0_k \,,\,\, 
                   u_j \, u_i \, \tilde \chi^0_k \,,\,
    \tilde g \to u_j \, d_i \, \tilde \chi^\pm_l \,,\,
    \tilde g \to g  \, \tilde \chi^0_k
  \end{eqnarray}
  with $i,j=1,2,3$, $l=1,2$ and $k=1,2,3,4$. The first two decay
  modes 
contain 
  states with quarks of different generations.
\end{itemize}
Obviously, the flavour mixing final states of the decays listed above
are constrained by the fact that all observed phenomena in rare
meson decays are consistent with the SM predictions. Nevertheless,
one has to check how large the branching ratios for the flavour
mixing final states still can be. One also has to study the impact of such final
states on discovery of SUSY as well as the determination of the underlying
model parameters.

For simplicity  we restrict ourselves 
to the mixing
between second and third generation of (s)quarks. 
We will take the so--called SPA point SPS1a' \cite{Aguilar-Saavedra:2005pw} 
as a specific example
which is specified by the mSUGRA parameters
$m_0=70$~GeV, $m_{1/2}=250$~GeV, $A_0=-300$~GeV,
$\tan\beta=10$ and $\rm{sign}(\mu)=1$. We have checked that main 
features
discussed below are also present in other study points, e.g.~$I''$ and
$\gamma$ of \cite{DeRoeck:2005bw}.
 At the electroweak scale (1 TeV) 
one gets the following data with the SPA1a' point:  
$M_2=193$~GeV, $\mu=403$~GeV, $m_{H^+}=439$~GeV and
$m_{\tilde g}=608$~GeV.
We have used the program {\tt SPheno} \cite{Porod:2003um} for the calculation.

It has been shown, that in Minimal Flavour Violating scenarios the
flavour changing decay modes are quite small \cite{Lunghi:2006uf}.
To get sizable flavour changing decay branching ratios, 
we have added the flavour mixing parameters as given in  
Table~\ref{squarkstab:par}; the resulting up-squark masses in GeV are
in ascending order: 315, 488, 505, 506, 523 and 587 [GeV] whereas the
resulting down-squark masses are 457, 478, 505, 518, 529, 537 [GeV]. 
This point is a random, but also typical one out of 20000 points 
fulfilling the constraints derived from the experimental measurements
 of the following three key observables of the $b \rightarrow s$ sector: 
$b\to s \gamma$, $\Delta M_{B_s}$
and $b \to s l^+ l^-$. For the calculation we have used the formula
given in \cite{Cho:1996we,Hurth:2003dk},  
for $b\to s \gamma$, the formula for $\Delta M_{B_s}$
given in  \cite{Buras:2002vd} 
and the formula for $b \to s l^+ l^-$ given in  \cite{Cho:1996we,Huber:2005ig}.
Note, that we have included all contributions mediated by chargino,
neutralino and gluino loops as we depart here considerably from 
Minimal Flavour Violation. 
The most important branching ratios for gluino and squark decays are given
in Table~\ref{squarkstab:br}.
 In addition the following branching ratios are larger
than 1\%, namely  BR($\tilde u_6 \to \tilde d_1 W$)=8.9\% and
 BR($\tilde u_6 \to \tilde d_2 W$)=1.8\%. We have not displayed the branching
ratios of the first generation nor the ones of the gluino into first generation.

\begin{table}
\begin{center}
\caption{Flavour violating parameters in GeV$^2$  which are
             added to the SPS1a' point. The
         corresponding values for the low energy observables are
        BR$(b\to s \gamma) =3.8 \cdot 10^{-4}$,
        $|\Delta(M_{B_s})| = 19.6$ ps$^{-1}$ and
        BR$(b\to s \mu^+ \mu^-) =1.59 \cdot 10^{-6}$.}
\label{squarkstab:par}
\begin{tabular}{|ccc|cccc|}
\hline
$M^2_{Q,23}$ & $M^2_{D,23}$ & $M^2_{U,23}$ & $v_u A^u_{23}$ 
 & $v_u A^u_{32}$  & $v_d A^d_{23}$ & $v_d A^d_{32}$ \\ \hline
 -18429 & -37154 & -32906 & 28104 & 16846 & 981 & -853 \\
\hline
\end{tabular}
\end{center}
\end{table}

\begin{table}
\begin{center}
\caption{Branching ratios (in \%) for squark and gluino decays for the point
 specified in 
         Table~\ref{squarkstab:par}. Only branching ratios larger than 1\% are
 shown.}
\label{squarkstab:br}
\begin{tabular}{|c|cc|cc|cc|ccc|}
\hline
          & $\tilde \chi^0_1 c$ &  $\tilde \chi^0_1 t$ & 
              $\tilde \chi^0_2 c$ &  $\tilde \chi^0_2 t$ & 
              $\tilde \chi^+_1 s$ &  $\tilde \chi^+_1 b$ & 
                $\tilde \chi^+_2 b$ &
           $\tilde u_1 Z^0$ &   $\tilde u_1 h^0$ \\ 
$\tilde u_1$ & 1.4                & 16.8 &
                               &   & 
                & 81.1    &
                              &    &      \\  
$\tilde u_2$ & 9.1               &   &
               21.0                & 3.6 &
               42.9 & 14.3  &  &
               5.3                & 1.3  \\
$\tilde u_3$ & 20.9                &   &
               21.9                 &   &
               47.5 & 1.1    &     &
               1.9               & 5.5 \\
$\tilde u_6$ & 1.5                &  2.7  &
               1.6                &  3.7  &
               4.0 &  14.1  
                               & 14.2  &
             39.2                 & 5.2  \\ 
\hline
\hline
          & $\tilde \chi^0_1 s$ &  $\tilde \chi^0_1 b$ & 
              $\tilde \chi^0_2 s$ &  $\tilde \chi^0_2 b$ & 
               $\tilde \chi^0_3 b$ & $\tilde \chi^0_4 b$ &
              $\tilde \chi^-_1 c$ &  $\tilde \chi^-_1 t$ & 
              $\tilde u_1  W^-$   \\ 
$\tilde d_1$ & 1.4                & 5.7 &
               2.7                & 2.8 & 
                              &     &
              6.5                & 28.1   &
               27.3    \\  
$\tilde d_2$ & 4.2               & 2.9  &
               6.3                & 17.8 &
                              &     &
               13.4                & 18.8 &
               34.8    \\
$\tilde d_4$ & 1.8                &   &
               23                 & 3.7  &
                              &     &
               41.5               & 5.8  &
               20.0                   \\
$\tilde d_6$ & 77.3                &  15.9  &
              4.6                &  3.7  &
                  2.4  & 2.4  &
             7.7                 & 5.1  &
             40.                     \\ 
\hline
\hline
 & $\tilde d_1 s$ & $\tilde d_1 b$ & $\tilde d_2 s$ & $\tilde d_2 b$ 
 & $\tilde d_3 d$ & $\tilde d_4 s$ & $\tilde d_5 d$ & $\tilde d_6 s$
 & $\tilde d_6 b$ \\ 
$\tilde g$ & 3.4 & 12.8 & 5.5 & 7.5 & 8.2 & 5.8 & 5.1 & 2.1 & 2.2 \\
\cline{2-10}
 & $\tilde u_1 c$ & $\tilde u_1 t$ & $\tilde u_2 c$ & $\tilde u_3 c$ 
 & $\tilde u_4 u$ & $\tilde u_5 u$ &  & & \\
 & 1.2 & 14 & 8.8 & 7.9 & 8.2 & 5.5 &  & & \\
\hline
\end{tabular}
\end{center}
\end{table}

It is clear from Table~\ref{squarkstab:br} that all listed particles have large
flavour changing decay modes. This clearly has an impact on the discovery
strategy of squarks and gluinos as well as on the measurement of the underlying
parameters. For example, in mSUGRA points without flavour mixing one
 finds usually that
the left-squarks of the first two generations as well as the right squarks
have similar masses. Large flavour mixing implies that there is a considerable
mass splitting as can be seen by the numbers above. Therefore, the assumption
of nearly equal masses should be reconsidered if sizable flavour changing
decays are discovered in squark and gluino decays.

An important part of the decay chains considered for SPS1a' and nearby points
are $\tilde g \to b \tilde b_j \to b \bar{b} \tilde \chi^0_k$ which are
used to determine the gluino mass as well as the sbottom masses or at least
their average value if these masses are close. In the analysis the existence
of two b-jets has been assumed, which need not to be the case as shown in
the example above. Therefore, this class of analysis should be re-done
requiring only one b-jet + one additional non b-jet to study the impact
of flavour mixing on the determination of these masses. 

Similar conclusions hold for  the variable $M^w_{tb}$ defined in 
\cite{Hisano:2003qu}. For this variable one considers final states containing
$b \tilde \chi^+_1$. In our example, three u-type squarks 
contribute with branching
ratios larger than 10\% in contrast to assumption that only the two stops 
contribute. The influence of the additional state requires for a sure
a detailed Monte Carlo study which should be carried out in the future.


\section{Top squark production and decay}

Supersymmetric scenarios with a particularly light stop have been recently
considered as potential candidates to provide a solid explanation of the
observed baryon asymmetry of the Universe~\cite{Cline:2006ts}.
Independently of this proposal, measurements of the process of
stop-chargino associated production at LHC have  been considered as a rather
original way of testing the usual assumptions about the Supersymmetric CKM matrix~\cite{Fuks}. 
In a very recent paper~\cite{Beccaria:2006wz}, 
the latter associated production process has been studied in some
detail for different choices of the SUSY benchmark points, trying to
evidentiate and to understand an apparently strong $\tan\beta$ dependence of
the  production rates. As a general feature of that study,
the values of the various rates appeared, typically, below the one $pb$ size, to
be compared with the (much) bigger rates of the stop-antistop process (see e.g. \cite{Beenakker:1997ut}). 


\subsection{Associated stop-chargino production at LHC: A light stop scenario test}

Given the possible relevance of an experimental
determination, it might be opportune to perform a more detailed
study of the
production rate size in the special light stop scenario, where one expects
that the numerical value is as large as possible. Here we
present the results of this study, performed at the simplest Born level
given the preliminary nature of the investigation.

The starting point is the expression of the differential cross section,
estimated at Born level in the c.m. frame of the incoming pair of the
partonic process $bg\to \chi^-_i\,\widetilde{t}_j$. Its detailed expression 
has been derived and discussed in~\cite{Beccaria:2006wz}.
The associated c.m. energy distribution
(at this Born level identical to the final invariant mass distribution) is 
\begin{eqnarray}
{d\sigma(pp\to \tilde{t_a}\chi^-_i+X)\over d\hat{s}}&=&
{1\over S}~\int^{\cos\theta_{max}}_{\cos\theta_{min}}
d\cos\theta\, L_{bg}(\tau, \cos\theta)
{d\sigma_{bg\to  \tilde{t_a}\chi^-_i}\over d\cos\theta}(\hat{s}),
\end{eqnarray}
where $\sqrt{\hat{s}}$ and $\sqrt{S}$ are the parton and total $pp$
c.m.~energies, respectively, $\tau=\hat{s}/S$, and $L_{bg}$ is the
parton process luminosity that we have evaluated using the parton
distribution functions from the Heavy quark CTEQ6
set~\cite{Kretzer:2003it}.  The rapidity and angular integrations are
performed after imposing a cut $p_T \ge 10$ GeV.

\begin{figure}
\setlength{\unitlength}{1mm}
\begin{picture}(100,55)
\put(20,0){\mbox{\epsfig{figure=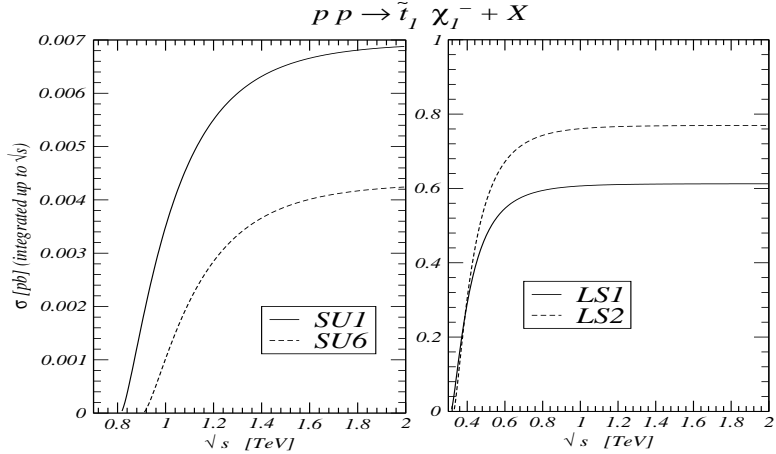,height=6cm,width=12cm}}}
\end{picture}
\caption{Integrated cross sections for the process $pp\to \widetilde{t}_1\,\chi^-_1 + X$ at the four MSSM points SU1, SU6, LS1, LS2.}
\label{fig:tchi1}
\end{figure}

For a preliminary analysis, we have considered the total cross section
(for producing the lightest stop-chargino pair), defined as the
integration of the distribution from threshold to a final energy
$\sqrt{s}$ left as free variable, generally fixed by experimental
considerations. To have a first feeling of the size of this quantity,
we have first estimated it for two pairs of sensible MSSM benchmark
points.  The first pair are the ATLAS Data Challenge-2 points SU1, SU6
whose detailed description can be found in~\cite{atlasdc2points}. 
The second pair
are the points LS1, LS2 introduced in \cite{Beccaria:2006dt}. These
points are typical {\em light SUSY} scenarios and in particular share
a rather small threshold energy $m_{\widetilde{t}}+m_{\chi}$ which
appears to be a critical parameter for the observability of the
considered process.  The main difference between SU1 and SU6 or LS1
and LS2 is the value of $\tan\beta$ (larger in SU6 and LS2).  The
results are shown in Fig.\ \ref{fig:tchi1}.  As one sees, the various
rates are essentially below the one $pb$ size, well below the expected
stop-antistop values.

\begin{figure}
\setlength{\unitlength}{1mm}
\begin{picture}(100,55)
\put(20,0){\mbox{\epsfig{figure=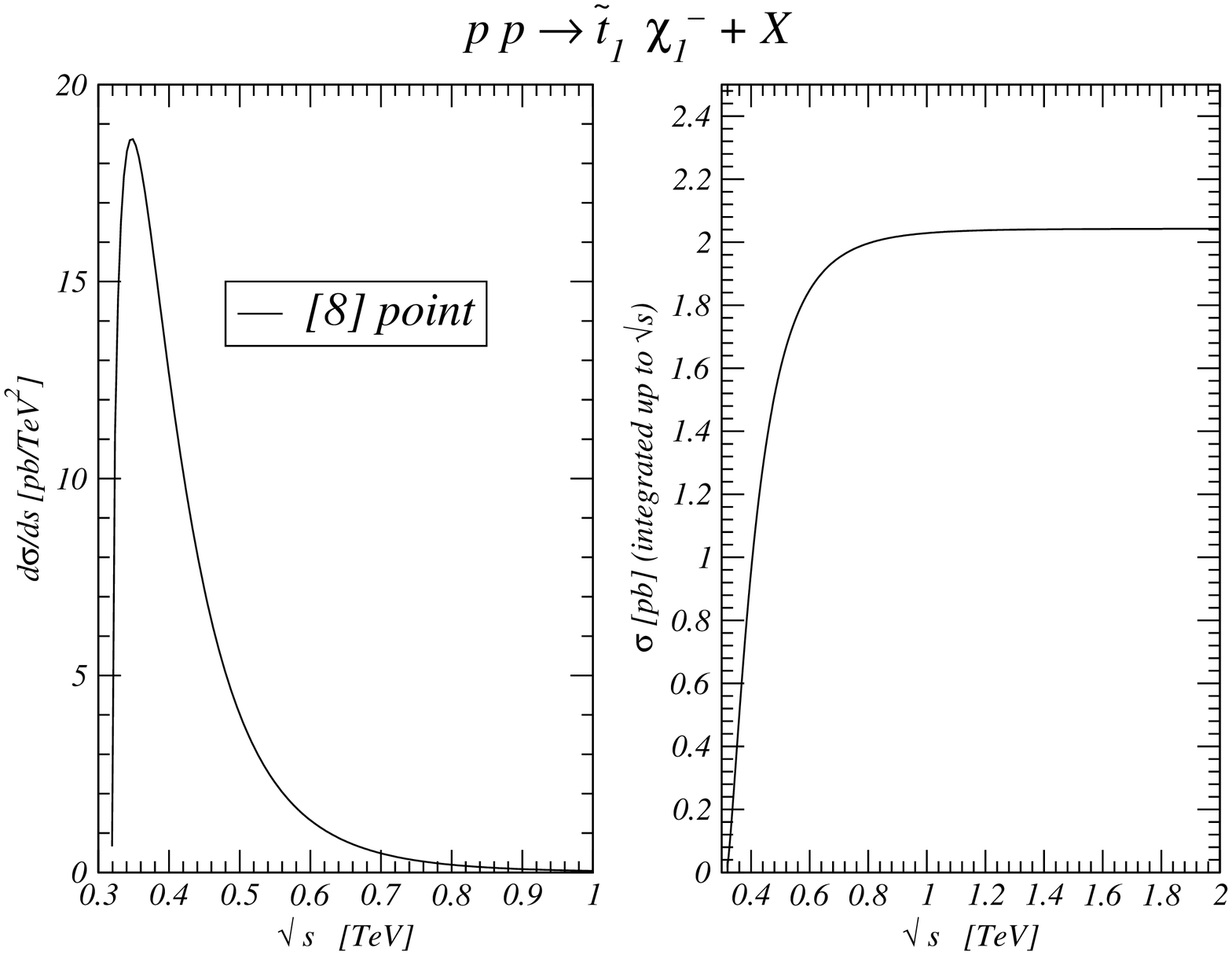,height=6cm,width=12cm}}}
\end{picture}
\caption{Distribution $d\sigma/ds$ and integrated cross sections for the process $pp\to \widetilde{t}_1\,\chi^-_1 + X$ at the point LST2.}
 \label{fig:tchi2}
\end{figure}

In the previous points, no special assumptions about the value of the
stop mass were performed, hence keeping a conservative attitude. One
sees, as expected, that the bigger rate values correspond to the
lighter stop situations (LS1 and LS2). In this spirit, we have
therefore considered a different MSSM point where the final stop is
particularly light.  More precisely, we have concentrated our analysis
on the point LST2, introduced and discussed in Section~\ref{sec:lightstop} and
characterized by the MSSM parameters (we list the relevant ones at
Born level)
\begin{equation}
M_1 = \frac{5}{3}\,\tan^2\theta_{\rm W}\,M_2 = 110\,{\rm GeV},\ \mu = 300\,{\rm GeV},\ \tan\beta = 7,\ 
\widetilde{t}_1 \simeq \widetilde{t}_R,\ m_{\widetilde{t}_1} = 150\,{\rm GeV},
\end{equation}
and consistent with the cosmological experimental bounds on the relic
density. Now the threshold energy is even smaller than in the previous
examples. The integrated cross section, shown in Fig.\ \ref{fig:tchi2}
reaches a maximum of about 2 pb, that might be detected by a dedicated
experimental search.


\subsection{Exploiting gluino-to-stop decays in the light stop scenario}
\label{sec:lightstop}

To achieve a strong first-order electroweak phase transition in the
MSSM, the lighter of the two stops, $\tilde t_1$, has to be lighter
than the top quark~\cite{Delepine:1996vn,Carena:1997ki,Cline:1998hy,
Laine:1998vn,Balazs:2004bu}. Assuming a stable ${\tilde \chi}^0_1$ LSP, there
hence exists a very interesting parameter region with a small
${\tilde \chi}^0_1$--${\tilde t}_1$ mass difference, for which (i) coannihilation with
${\tilde t}_1$~\cite{Boehm:1999bj,Ellis:2001nx} leads to a viable neutralino
relic density and (ii) the light stop decays dominantly into
$c\tilde\chi^0_1$~\cite{Hikasa:1987db}.

In this case, stop-pair production leads to \mbox{2 $c$-jets $+\not{\!\!E}_T$}, 
a signal which is of very limited use at the LHC.
One can, however, exploit~\cite{Kraml:2005kb} gluino-pair production
followed by gluino decays into stops and tops: 
since gluinos are Majorana particles, 
they can decay either into $t\tilde t_1^*$ or $\bar t\tilde t_1^{}$;
pair-produced gluinos therefore give same-sign top quarks in half of the 
gluino-to-stop decays. 
Here note that in the light stop scenario, $\tilde g\to t\tilde t_1^*$
(or $\bar t\tilde t_1^{}$) has practically 100\% branching ratio. 
With ${\tilde t}_1\to c{\tilde \chi}^0_1$, $t\to b W$, and the $W$'s decaying leptonically,
this leads to a signature of two $b$-jets plus two same-sign leptons 
plus jets plus missing transverse energy:
\begin{equation}
    pp \to \tilde g\tilde g\to 
    bb\,l^+l^+\: ({\rm or}\: \bar b\bar b\, l^-l^-) 
    + {\rm jets\:} + \not\!\!E_T\,.
\label{eq:bbllsignature}
\end{equation}
In~\cite{Kraml:2005kb} we performed a case study for the `LST1'
parameter point with $m_{{\tilde \chi}^0_1}=105$~GeV, $m_{{\tilde
t}_1}=150$~GeV, $m_{\tilde g}=660$~GeV and showed that the signature
Eq.~(\ref{eq:bbllsignature}) is easily extracted from the
background. In this contribution, we focus more on the stop
coannihilation region and discuss some additional issues.

We define a benchmark point `LST2' in the stop coannihilation region
by taking the parameters of LST1 and lowering the stop mass to
$m_{{\tilde t}_1}=125$~GeV.  We generate signal and background events
equivalent to $30~\text{fb}^{-1}$ of integrated luminosity and perform
a fast simulation of a generic LHC detector as described
in~\cite{Kraml:2005kb}. The following cuts are then applied to extract
the signature of Eq.~(\ref{eq:bbllsignature}):
\begin{itemize}
  \item require two same-sign leptons ($e$ or $\mu$) with
  $p^{\mathrm{lep}}_T>20$~GeV;
  \item require two $b$-tagged jets with $p^{\mathrm{jet}}_T>50$~GeV;
  \item missing transverse energy $\not{\!\!E}_T > 100$~GeV;
  \item demand two combinations of the two hardest leptons and $b$-jets \\
  that give invariant masses $m_{bl}<160$~GeV, consistent with a top quark.
\end{itemize}
This set of cuts emphasizes the role of the same-sign top quarks in
our method, and ignores the detectability of the jets initiated by the
${\tilde t}_1$ decay.  Table~\ref{tab:sstcuts} shows the effect of the cuts
on both the signal and the backgrounds.
Detecting in addition the (soft) $c$-jets from the 
${\tilde t}_1\to c{\tilde \chi}^0_1$ decay, together with the exess in 
events with \mbox{2 $c$-jets $+\not{\!\!E}_T$} from stop-pair production, 
can be used to strengthen the light stop hypothesis. 
A reasonable $c$-tagging efficiency would be very helpful in this case.

\begin{table}\begin{center}
\caption{
Number of events at LST2 left after cumulative cuts for
$30~\text{fb}^{-1}$ of integrated luminosity. ``2lep, 2$b$'' means two
leptons with $p^{\mathrm{lep}}_T>20$~GeV plus two $b$-jets with
$p^{\mathrm{jet}}_T>50$~GeV. ``2$t$'' is the requirement of two tops
(i.e.\ $m_{bl}<160$~GeV), and ``SS'' that of two same-sign leptons.}
\label{tab:sstcuts}
\begin{tabular}{ll|rrrrrrr} \hline
Cut         &      & 2lep, 2$b$ & $\not{\!\!E}_T$ & 2$t$ & SS \\ \hline
Signal:     & $\tilde g \tilde g$ &  1091 &  949 &  831 & 413 \\
Background: & SM                  & 34224 & 8558 & 8164 &  53 \\
            & SUSY                &   255 &  209 &  174 &  85 \\
\hline
\end{tabular}\end{center}
\end{table}

\begin{figure}
  \centerline{%
  \includegraphics[width=.49\textwidth]{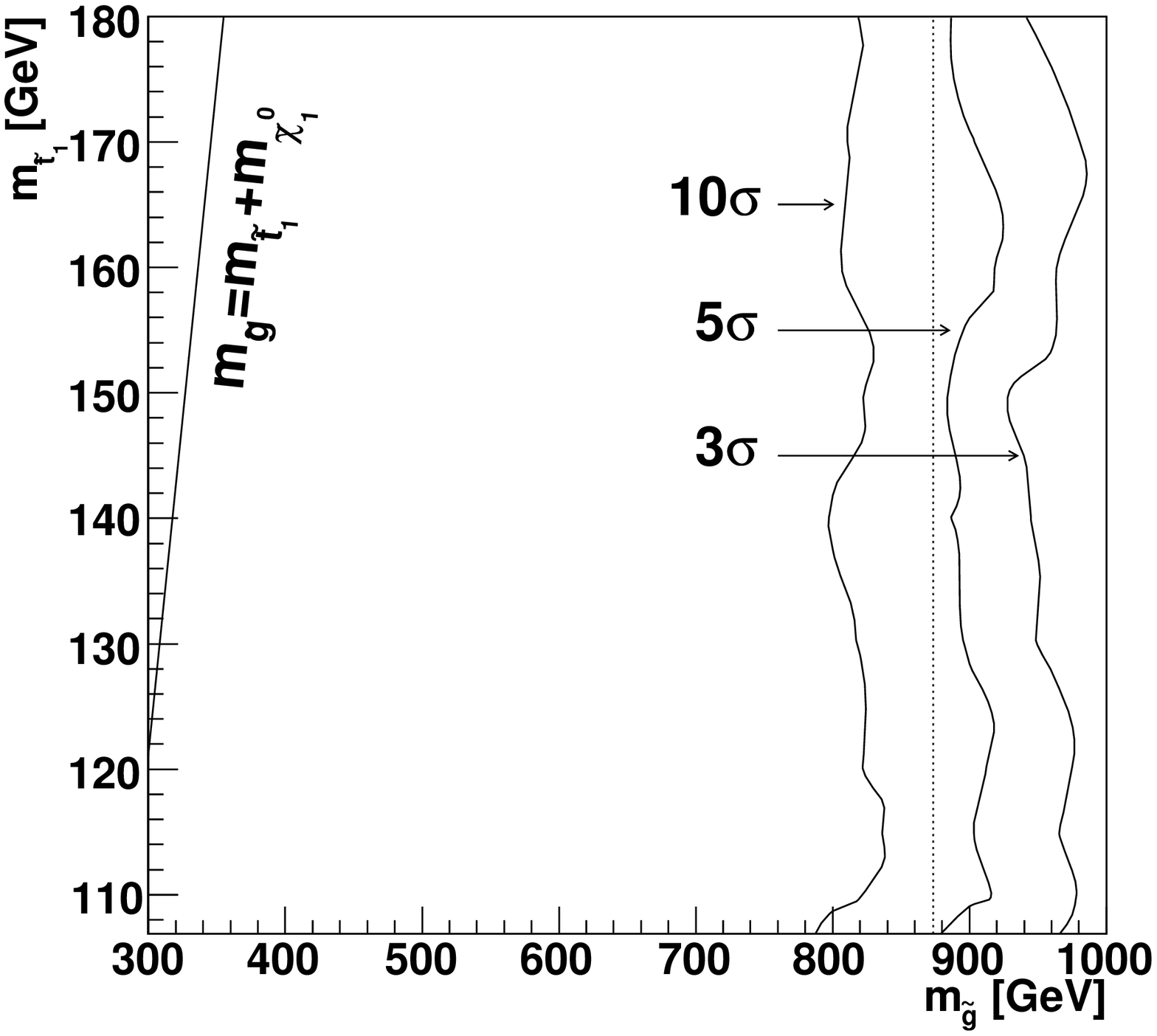}
  \includegraphics[width=.49\textwidth]{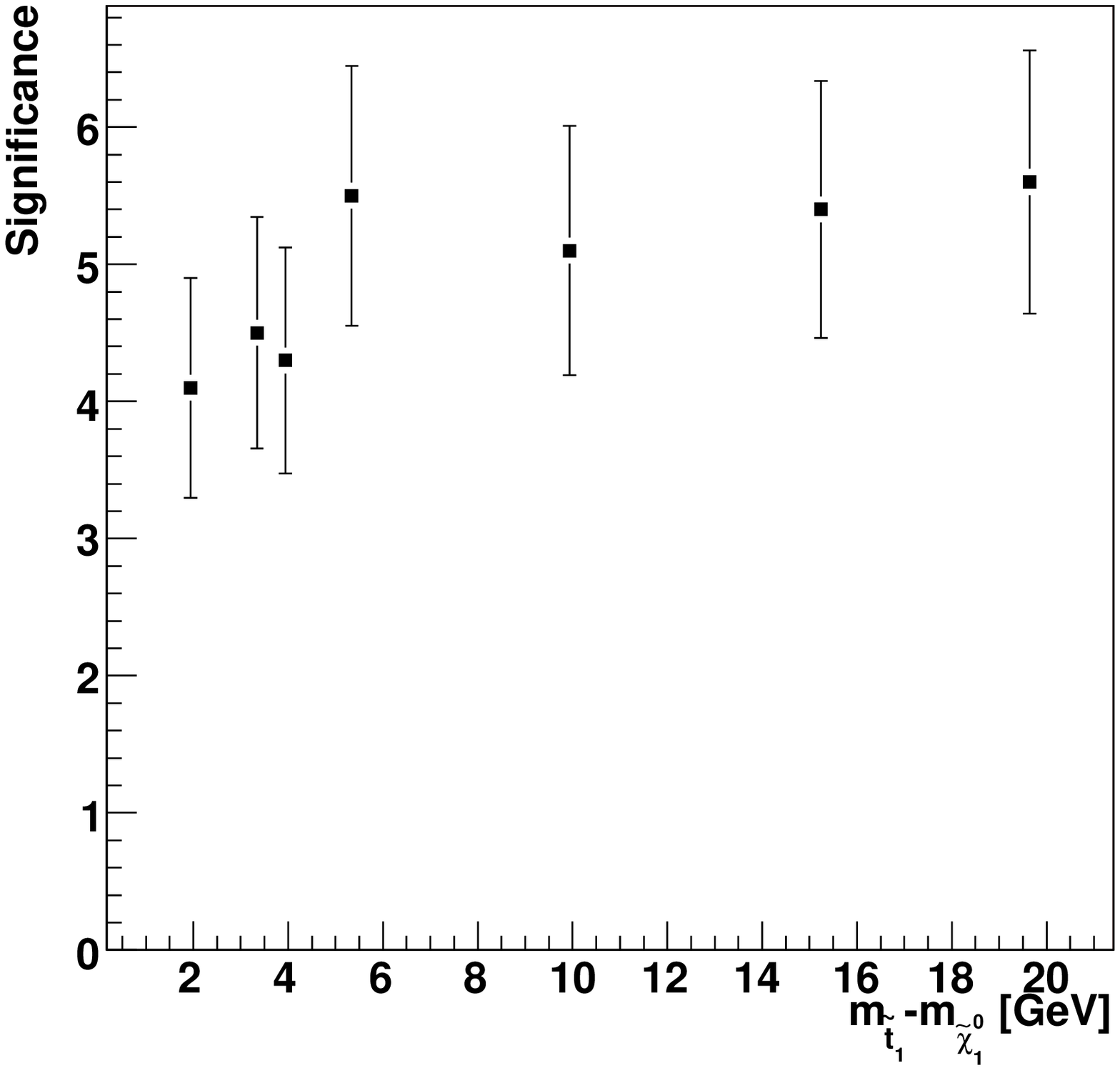}}
  \caption{
  Reach for the signature of Eq.~(\ref{eq:bbllsignature}) in the
  gluino--stop mass plane (left) and significance as a function of
  stop--neutralino mass difference with $m_{\tilde g}=900$~GeV
  (right).}
\label{fig:sstreach}
\end{figure}

To demonstrate the robustness of the signal, we show in
Fig.~\ref{fig:sstreach} (left) contours of $3\sigma$, $5\sigma$ and
$10\sigma$ significance\footnote{We define significance as
$S/\sqrt{B}$, where $S$ and $B$ are the numbers of signal and
background events.} in the ($m_{\tilde g}$, $m_{{\tilde t}_1}$) plane.  
For comparison we also show as a dotted line the result of a
CMS study \cite{DellaNegra:942733a}, which found a reach down to $1$~pb in terms
of the total cross section for same-sign top production. In
Fig.~\ref{fig:sstreach} (right), we show the decreasing significance
for $m_{\tilde g}=900$~GeV, as the stop--neutralino mass difference
goes to zero.
To be conservative, both panels in Fig.~\ref{fig:sstreach} assume that
all squarks other than the ${\tilde t}_1$ are beyond the reach of the LHC;
$\tilde q\tilde q$ and $\tilde g\tilde q$ production would increase the signal
through $\tilde q\to \tilde g q$ decays (provided $m_{\tilde q}>m_{\tilde g}$)
while adding only little to the background; see
\cite{Kraml:2005kb,Kraml:2006ca} for more detail.

The usual way to determine SUSY masses in cascade decays is through    
kinematic endpoints of the invariant-mass distributions of the SM 
decay products, see e.g. \cite{Hinchliffe:1996iu,Bachacou:1999zb,
Allanach:2000kt,Lester:2001zx}. 
In our case, there are 
four possible endpoints: $m_{bl}^{\max}$, $m_{bc}^{\max}$,
$m_{lc}^{\max}$ and $m_{blc}^{\max}$, of which the first simply gives
a relationship between the masses of the $W$ and the top, and the
second and third are linearly dependent, so that we are left with
three unknown masses and only two equations. Moreover, because of the
information lost with the escaping neutrino the distributions of
interest all fall very gradually to zero. 

\begin{figure}
  \centerline{\includegraphics[height=.26\textheight]{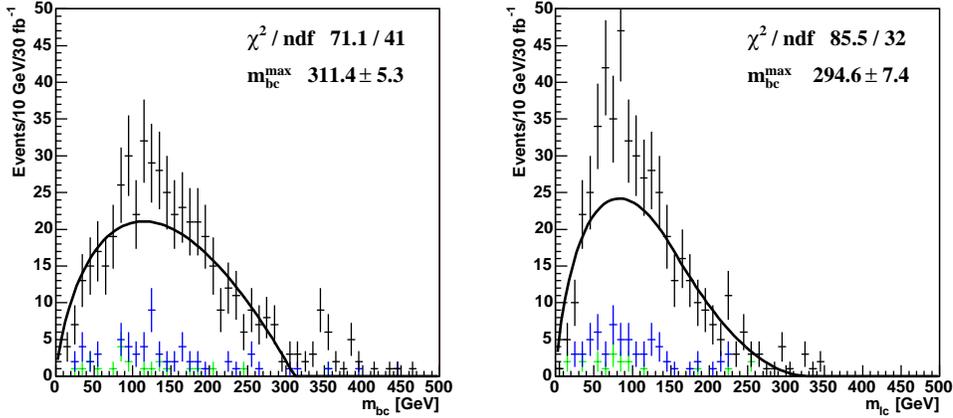}}
  \caption{
  Invariant-mass distributions $m_{bc}$ (left) and $m_{lc}$ (right)
  with $20\%$ $c$-tagging efficiency after $b$-tagging (black with
  error bars) and best fit for LST2.
  Also shown are the contributions from the SM
  background (green) and the SUSY background (blue).
}
\label{fig:im_ctag}
\end{figure}

In order to nevertheless get some information on the ${\tilde \chi}^0_1$, 
${\tilde t}_1$ and $\tilde g$ masses, we fit the whole $m_{bc}$ and $m_{lc}$
distributions~\cite{Kraml:2005kb,Miller:2005zp} and not just the
endpoints.  This requires, of course, the detection of the jets
stemming from the ${\tilde t}_1$ decay. For small 
$m_{{\tilde t}_1}-m_{{\tilde \chi}^0_1}$ these
are soft, so we demand two jets with $p^{\mathrm{jet}}_T<50$~GeV in
addition to the cuts listed above. The results
of the fits for LST2, assuming 20\% $c$-tagging
efficiency,~\footnote{When one or none of the remaining jets are
$c$-tagged we pick the $c$-jets as the hardest jets with
$p^{\mathrm{jet}}_T<50$~GeV.} are shown in Fig.~\ref{fig:im_ctag}. The
combined result of the two distributions is $m_{bc}^{\max}=305.7\pm
4.3$, as compared to the nominal value of $m_{bc}^{\max}\simeq
299$~GeV.

As mentioned above, the gluino-pair production leads to 
50\% same-sign (SS) and 50\% opposite-sign (OS) top-quark pairs, and  
hence $R=N(SS)/N(SS+OS)\simeq 0.5$ with $N$ denoting the number of 
events. In contrast, in the SM one has $R\lsim 0.01$. 
This offers a potential test of the 
Majorana nature of the gluino. The difficulty is that the number 
of OS leptons is completely dominated by the $t\bar t$ background. 
This can easily be seen from the last two rows of Table~\ref{tab:sstcuts}: 
$R\sim 0.5$ (0.02) for the signal (backgrounds) as expected; signal and 
backgrounds combined, however, give $R\sim 0.06$. 
A subtraction of the $t\bar t$ background as described in 
Section~\ref{sec:LariPolesello} 
may help to extract $R(\tilde g\tilde g)$. 


\subsection{A study on the detection of a light stop squark with the ATLAS
 detector at the LHC}
\label{sec:LariPolesello}

We present here an exploratory study of a benchmark model in which the 
stop quark has a mass of 137~GeV, and 
the two-body decay of the stop squark into a chargino and a $b$ quark
is open. We address in detail
the ability of the ATLAS experiment to separate the stop signal from 
the dominant Standard Model backgrounds.

For the model under study\cite{Allanach:2006fy} 
all the masses of the first two generation squarks
and sleptons are set at 10~TeV, and the gaugino masses are related by the  
usual gaugino mass relation $M_1:M_2=\alpha_1:\alpha_2$.
The remaining parameters are thus defined:
$$
M_1=60.5~{\rm GeV} \; \; \; \mu=400~{\rm GeV} \; \; \; \tan\beta=7 
\; \; \; M_3=950~{\rm GeV}
$$
$$
m(Q_3)=1500~{\rm GeV} \; \; \; m({\tilde t}_R)=0~{\rm GeV} 
\; \; \; m({\tilde b}_R)=1000~{\rm GeV}  \; \; \; 
A_t=-642.8~{\rm GeV}
$$
The resulting relevant masses are $m({\tilde t}_1)=137$~GeV, 
$m({\tilde \chi}^\pm_1)=111$~GeV, $m({\tilde\chi_1^0})=$58~GeV. 
The ${\tilde t}_1$ decays with 100\% BR into 
\mbox{${\tilde \chi}^\pm_1 b$}, and ${\tilde \chi}^\pm_1$ decays with 
100\% BR into an 
off-shell $W$ and ${\tilde \chi}^0_1$. The final state signature is therefore 
similar to the one for $t\bar{t}$ production: 2 $b$-jets, $E_T^{miss}$ 
and either
2 leptons ($e,\mu$) (4.8\% BR) or 1 lepton and 2 light jets (29\% BR).\par 
The signal cross-section, calculated at 
NLO with the {\tt PROSPINO}\cite{Beenakker:1997ut} program is 412~pb. \\
We analyze here the semi-leptonic channel, where only one of the two 
${\tilde t}_1$ legs has a lepton in the final state. 
We apply the standard cuts for the search of the semileptonic top
channel as applied in \cite{ATLASTDR},  but with softer requirements 
on the kinematics:
\begin{itemize}
\item
One and only one isolated lepton ($e$, $\mu$), $p_T^l>20$~GeV.
\item
$E_T^{miss}>20$~GeV.
\item
At least four jets $P_T(J_1,J_2)>35$~GeV, $P_T(J_3,J_4)>25$~GeV.
\item
Exactly two jets in the events must be tagged as $b$-jets, anda congratulacin 	
they both must have \mbox{$p_T>20$~GeV}.
The standard ATLAS b-tagging efficiency of 60\% for a rejection
factor of 100 on light jets is assumed.
\end{itemize}

A total of 600k SUSY events were generated using {\tt HERWIG} 6.5
\cite{Corcella:2000bw, Moretti:2002eu}, 
1.2M $t\bar{t}$ events using {\tt PYTHIA} 6.2\cite{Sjostrand:2001yu}.
The only additional background considered for this exploratory 
study was the associated production of a W boson with two $b$ jets
and two non-$b$ jets, with the W decaying into $e$ or $\mu$.
This is the dominant background for top 
searches at the LHC. For this process, we generated 
60k events using {\tt Alpgen}\cite{Mangano:2002ea}. 
The number of events generated corresponds to $\sim 1.8$~fb$^{-1}$. 
The generated events are then passed through {\tt ATLFAST},
a parametrized simulation of the ATLAS detector \cite{Richter-Was:683751}.\\
After the selection cuts the efficiency for the $t\bar{t}$ background is 
3.3\%\footnote{The emission of additional hard jets at higher 
orders in the QCD interaction can increase the probability that 
the $t\bar t$ events satisfy the requirement of 4 jets. The 
cut efficiency is observed to increase by about 20\% if MC@NLO is 
used to generate the $t \bar t$ background. We do not expect such an effect 
to change the conclusions of the present analysis, but future studies should 
take it into account.},
for $Wbbjj$ 3.1\%, and for the signal 0.47\%, yielding a background 
which is $\sim$15 times higher than the signal.\\

\begin{figure}
\begin{center}
\includegraphics[width=0.4\textwidth,height=5cm]{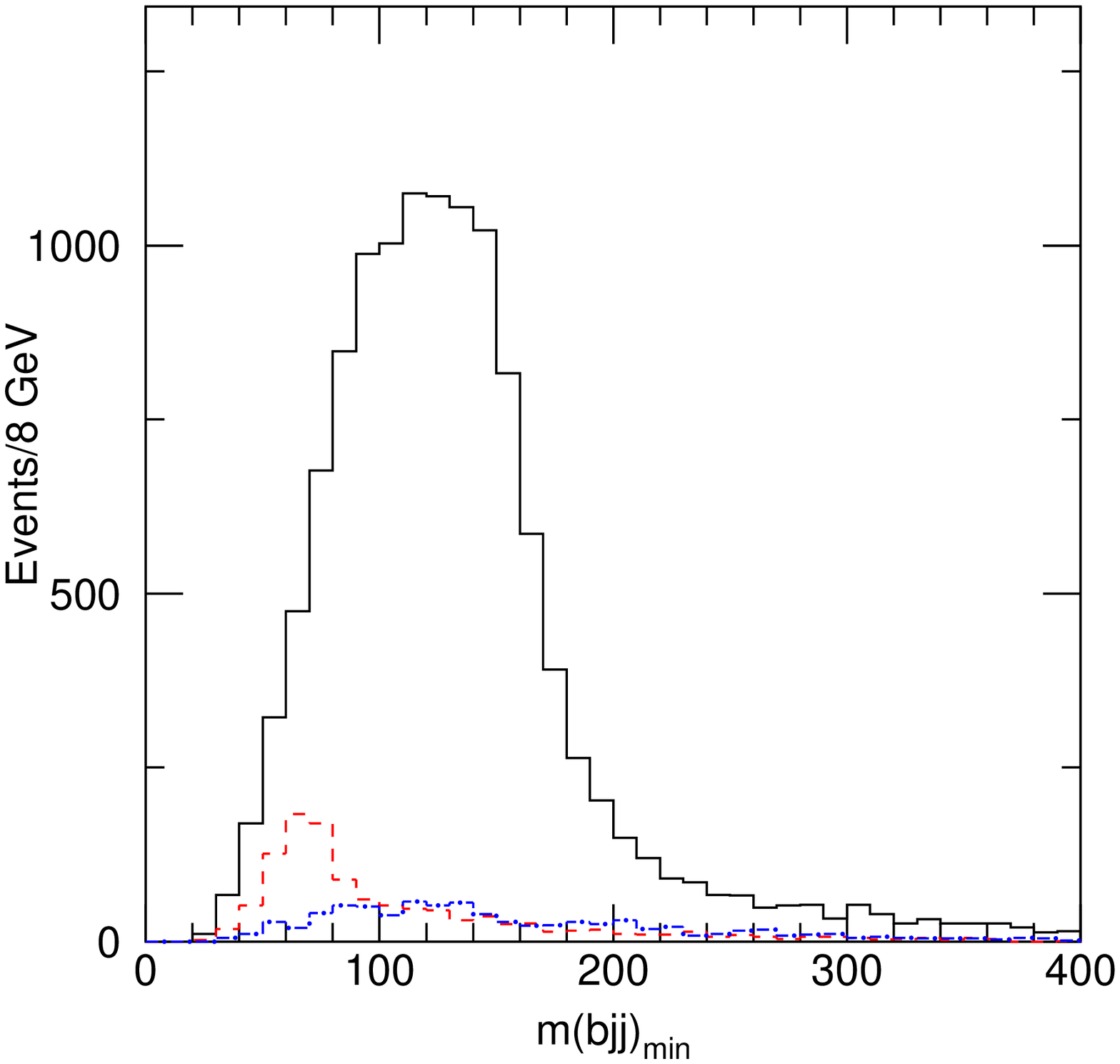}
\includegraphics[width=0.4\textwidth,height=5cm]{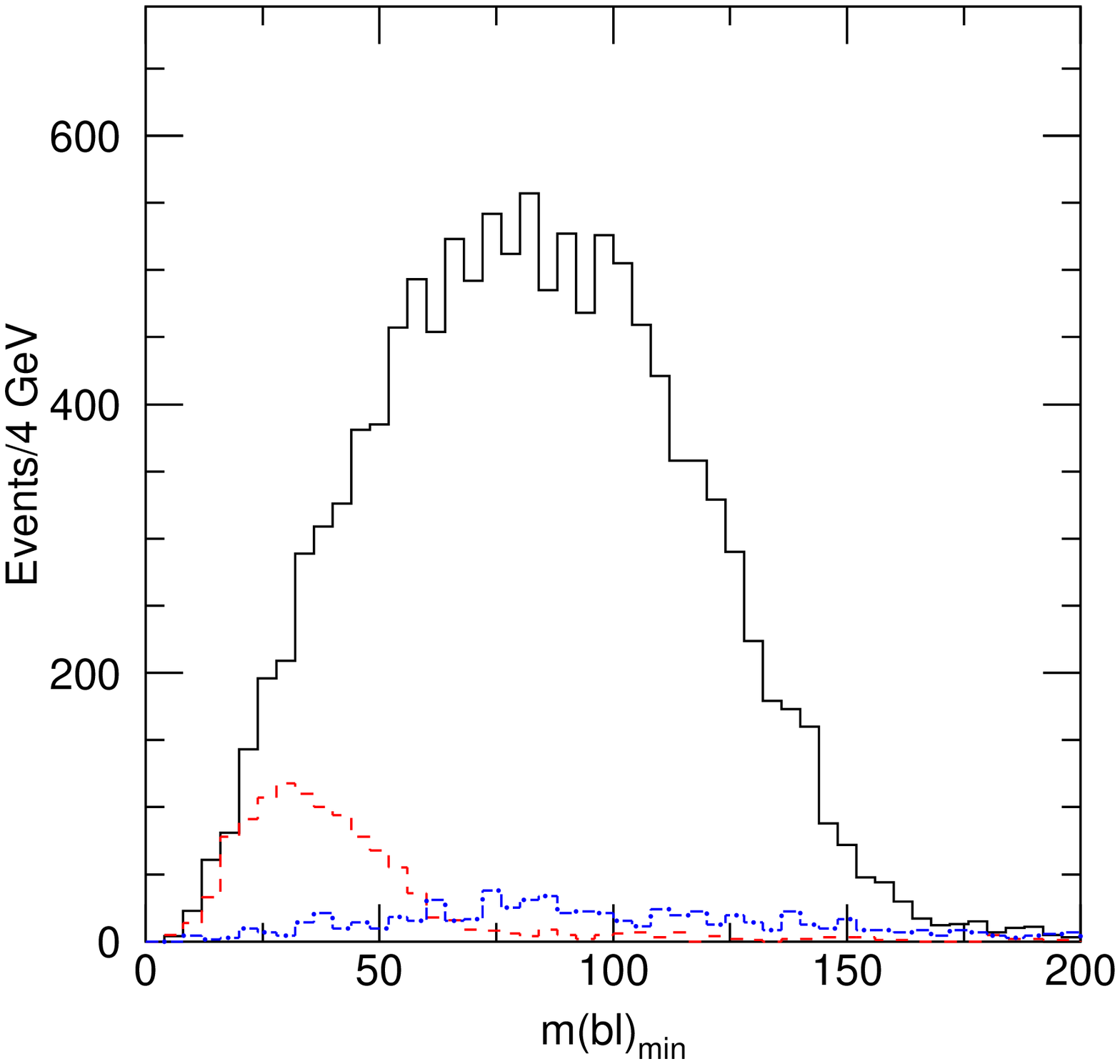}
\caption { 
Left: Distributions of the minimum $bjj$ invariant mass 
for top background (full black line), $Wbb$ background (dot-dashed blue line),
signal (dashed red line).
Right:  Distributions of the minimum $bl$ invariant mass 
for top background (full black line), $Wbb$ background (dot-dashed blue line),
signal (dashed red line).
}
\label{fig:mbjjmin}
\end{center}
\end{figure}

An improvement of the signal/background ratio can be obtained using 
the minimum invariant mass of all the non-b jets with 
$p_T>25$~GeV. This distribution peaks near the value of the W mass for 
the top background, 
whereas the invariant mass for the signal should be smaller than 54 GeV,
which is the mass difference between the ${\tilde \chi}^\pm$ and the 
${\tilde \chi}^0_1$. Requiring $m(jj) < 60$~GeV improves the 
signal/background ratio to 1/10, with a loss of a bit
more than half the signal. 
We show in the left plot of Figure~\ref{fig:mbjjmin}
after this cut the distributions for the variable
$m(bjj)_{min}$, i.e. the invariant mass for the combination a b-tagged
jet and the two non-b jets yielding the minimum invariant 
mass. If the selected jets are from the decay of the stop, this
invariant mass should have an end point at $\sim$79~GeV, whereas the 
corresponding end-point should be at 175~GeV for the top background.
The presence of the stop signal is therefore visible as a shoulder 
in the distribution as compared to the pure top contribution.
A significant contribution from $Wbbjj$ is present, without a particular
structure. 
Likewise, the variable $m(bl)_{min}$ has an end point at $\sim$66 GeV
for the signal and at 175~GeV for the top background, as shown in 
Figure~\ref{fig:mbjjmin}, and the same shoulder structure is observable. 
We need therefore to predict precisely the shape of the 
distributions for the top background in order to subtract it
from the experimental distributions and extract the signal distributions.\par 

The top background distributions can be estimated
from the data themselves by exploiting the fact that we select 
events where one of the $W$ from the top decays into two jets
and the other decays into lepton neutrino.
One can therefore select two pure top samples, with minimal 
contribution from non-top events by applying separately hard cuts
on each of the two legs.
\begin{itemize}
\item
Top sample 1: the best reconstructed $bl\nu$ invariant mass is within 
15 GeV of 175~GeV, and $(m_{\ell b})_{min}>60$~GeV in order to 
minimize the contribution from the stop signal. The neutrino 
longitudinal momentum is calculated by applying the $W$ mass constraint.
\item
Top sample 2:  the best reconstructed $bjj$ mass is within 10~GeV 
of 175~GeV. 
\end{itemize} 
We assume here that we will be able to predict the $Wbb$ background
through a combination of Monte Carlo and the study of $Zbb$ production 
in the data, and we subtract this background both from the 
observed distributions and from the Top samples. 
More work is required to assess the uncertainty on 
this subtraction. Given the fact 
that this background is smaller than the signal, and it has a significantly
different kinematic distribution, we expect that a 10-20\% uncertainty
on it will not affect the conclusions of the present analysis.\par

For Top sample 1, the top selection is performed by applying severe 
cuts on the lepton leg, it can therefore be expected that
the  minimum $bjj$ invariant mass distribution, which is built from 
jets from the decay of the hadronic side be essentially unaffected 
by the top selection cuts. This has indeed be verified to be the 
case~\cite{Allanach:2006fy}. The $m(bjj)$ distribution from Top sample 1 is 
then normalized to the observed distribution in the high mass region, 
where no signal is expected, and subtracted from it.
A similar procedure is followed for the $m(bl)$ distribution: the top 
background is estimated using Top sample 2, normalized to the 
observed distribution in the high-mass region, and subtracted from it.
\begin{figure}
\begin{center}
\includegraphics[width=0.4\textwidth,height=5cm]{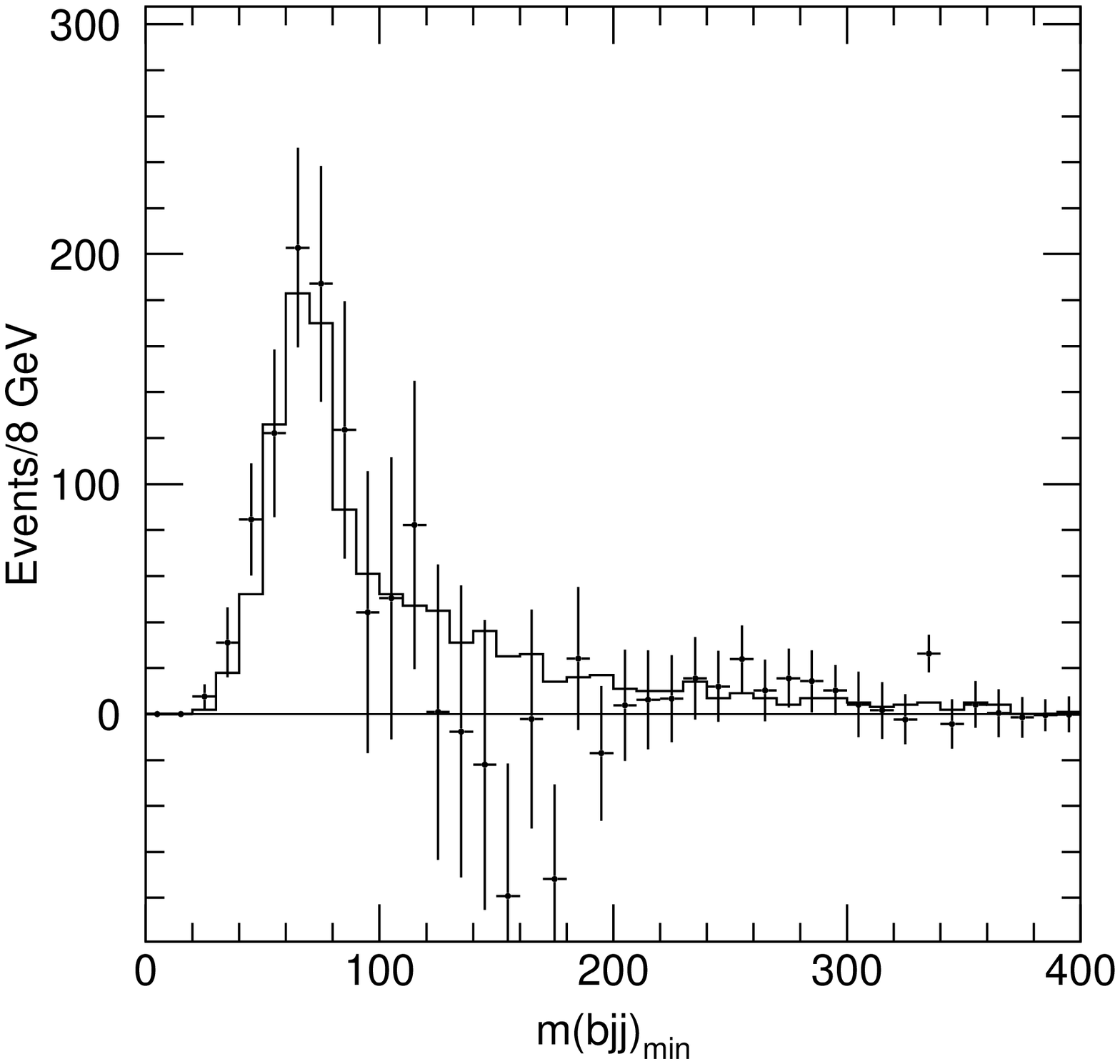}
\includegraphics[width=0.4\textwidth,height=5cm]{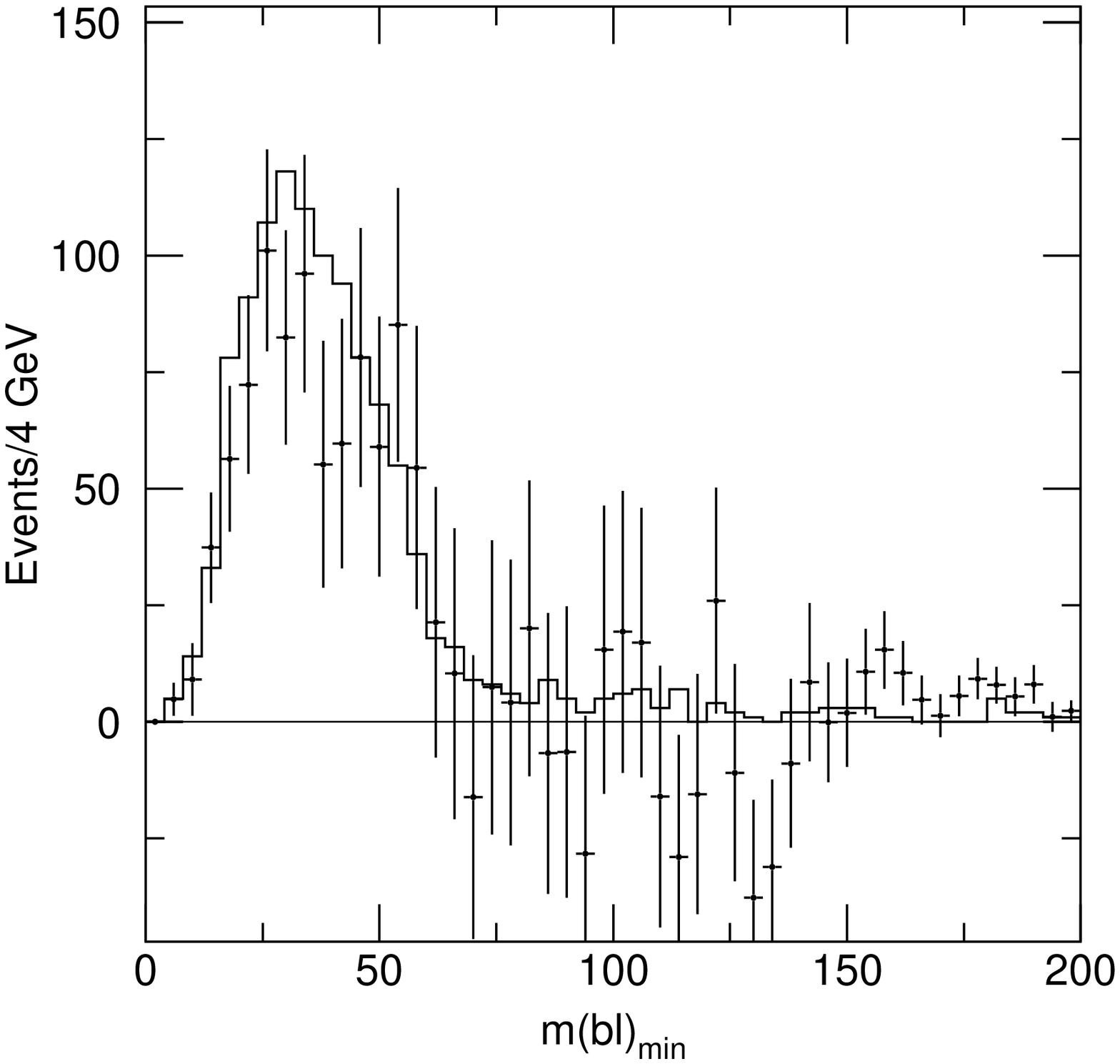}
\caption{
Left: distribution of the minimum $bjj$ invariant mass
after the subtraction procedure (points with errors)
superimposed to the original signal distribution (full line).
Right: distribution of the minimum $bl$ invariant mass
after the subtraction procedure (points with errors)
superimposed to the original signal distribution (full line).
}
\label{fig:mbsig}
\end{center}
\end{figure}
The results are shown in Figure~\ref{fig:mbsig}, with superimposed
the corresponding distributions for the signal. 
As discussed above, we have subtracted the $Wbb$ background from the observed
distributions. \par
For both variables the true and measured distributions for the signal are 
compatible, showing the goodness of the background subtraction technique, 
and the expected kinematic structure is observable, 
even with the very small statistics generated for this analysis, 
corresponding to little more than one month of data taking at the
initial luminosity of \mbox{$10^{33}$~cm$^{-1}$s$^{-1}$}.\par 
Further work, outside the scope of this initial exploration, is 
needed on the evaluation of the masses of the involved sparticles
through kinematic studies of the selected sample

A preliminary detailed analysis of a SUSY model with a stop squark
lighter than the top quark decaying into a chargino and a $b$-jet
was performed.
It was shown that for this specific model after simple
kinematic cuts a signal/background ratio of $\sim$1/10  can be achieved.
A new method, based on the selection of pure top samples to subtract
the top background was demonstrated. Through this method it is possible
to observe the kinematic structure of the stop decays, and thence to
extract a measurement of the model parameters. This analysis can yield
a clear signal for physics beyond the SM for just $1-2~$fb$^{-1}$, and
is therefore an excellent candidate for early discovery at the LHC.


\subsection{Stop decay into right-handed sneutrino LSP}

Right-handed neutrinos offer the possibility to accommodate
neutrino masses.  In supersymmetric models this implies the
existence of right-handed sneutrinos.  Right-handed sneutrinos are
expected to be as light as other supersymmetric particles 
\cite{Gopalakrishna:2006kr,deGouvea:2006wd} if the
neutrinos are either Dirac fermions or if the lepton-number breaking scale is
at (or below) the SUSY breaking scale , assumed to be
around the electroweak scale. Depending on the mechanism of SUSY
breaking, the lightest right-handed sneutrino $\tilde N_R$ may be the lightest
supersymmetric particle (LSP).  We consider in the following such a scenario
focusing on the case where the right-handed stop is the next to lightest
SUSY particle assuming $R$-parity conservation. Details on the
model and other scenarios can be found in 
\cite{Gopalakrishna:2006kr,deGouvea:2006wd}.

 As the right-handed
neutrino has a mass around 100 GeV, the neutrino Yukawa couplings $Y_N$ must be
very small to accommodate neutrino data: $Y_N \sim 10^{-6}$ ($Y_N \sim 10^{-12}$)
 in the case of Majorana neutrinos (Dirac neutrinos).
 This has as immediate consequence that 
if the SUSY breaking sneutrino trilinear ``A-term'' is also proportional to $Y_N$, the
left-handed and right-handed sneutrinos hardly mix independent of neutrino 
physics because the left-right mixing term
is proportional to $Y_N$. Decays into $\tilde N_R$ will give tiny
decay widths as $Y_N$ is the only coupling of $\tilde N_R$. For this
reason, all decays of supersymmetric particles are as in the usual MSSM, 
but for the NLSP whose life-time can be long since it can only decay into 
the $\tilde N_R$.
 In the case of a stop NLSP the
dominant decay mode is 
$\tilde t_1 \to  b \, \ell^+ \, \tilde N_R$, followed by CKM suppressed ones
into $s$ and $d$ quarks.
In the limit where mixing effects for stops and charginos are neglected
the corresponding matrix element squared in the rest frame of the stop
reads as:
\begin{eqnarray}
|T_{fi}|^2 \sim \frac{4 |Y_t|^2 |Y_N|^2 M_{\tilde t_R}^2 E_b E_l}
                     {\left((p_{\tilde t_R} - k_b)^2 - M_{\tilde H}^2 \right)^2}
\frac{\left(1+\cos{\theta_{b \ell}}\right)}{2} \ .
\label{TfistR.EQ}
\end{eqnarray}
where we have assumed that the right-handed stop $\tilde t_R$ is the lightest 
stop and $\tilde H$ is the Higgsino, $E_b$ ($E_\ell$) 
is the energy of the b-quark (lepton), $\theta_{b \ell}$ is the angle
between the fermions.
The complete formula can be found in \cite{deGouvea:2006wd}. The last factor
in Eq.~(\ref{TfistR.EQ}) implies that the $b$-quark and the lepton
have a tendency to go in the same direction.

\begin{figure}[t]
\setlength{\unitlength}{1mm}
\begin{picture}(170,80)
\put(0,76){\mbox{a)}}
\put(5,80){\epsfig{figure=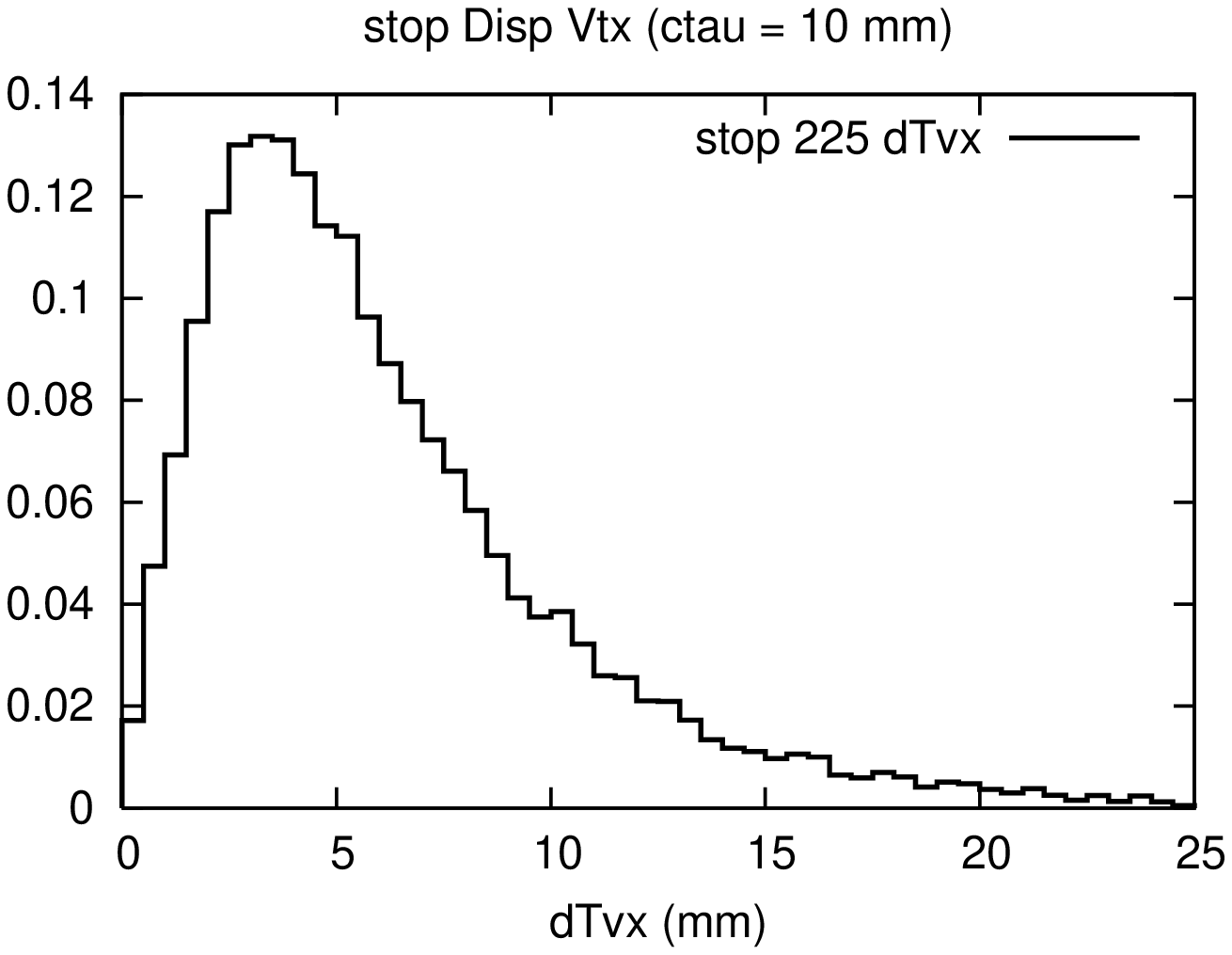,height=75mm,width=40mm,angle=270}}
\put(78,76){\mbox{b)}}
\put(80,80){\epsfig{figure=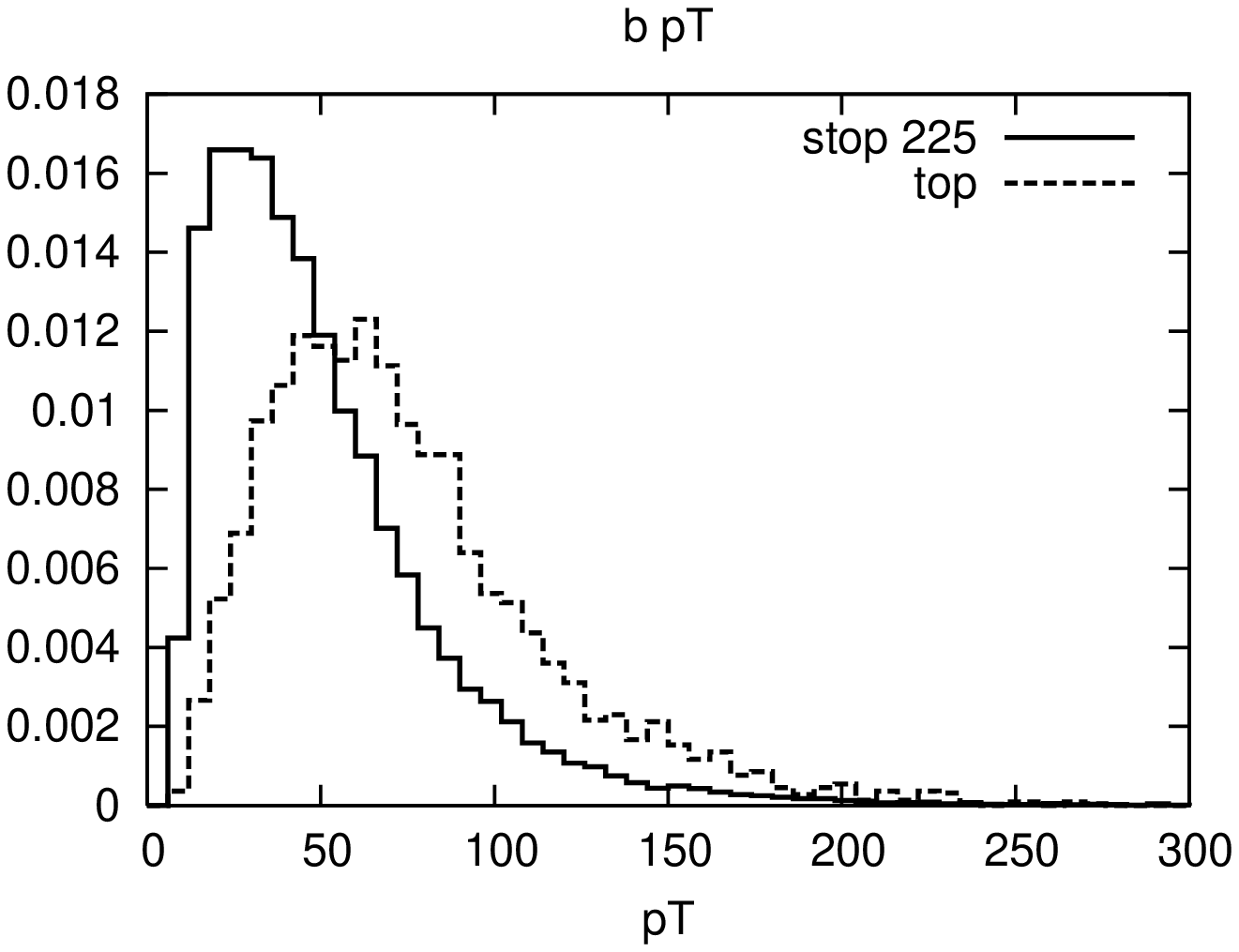,height=75mm,width=40mm,angle=270}}
\put(0,36){\mbox{c)}}
\put(5,40){\epsfig{figure=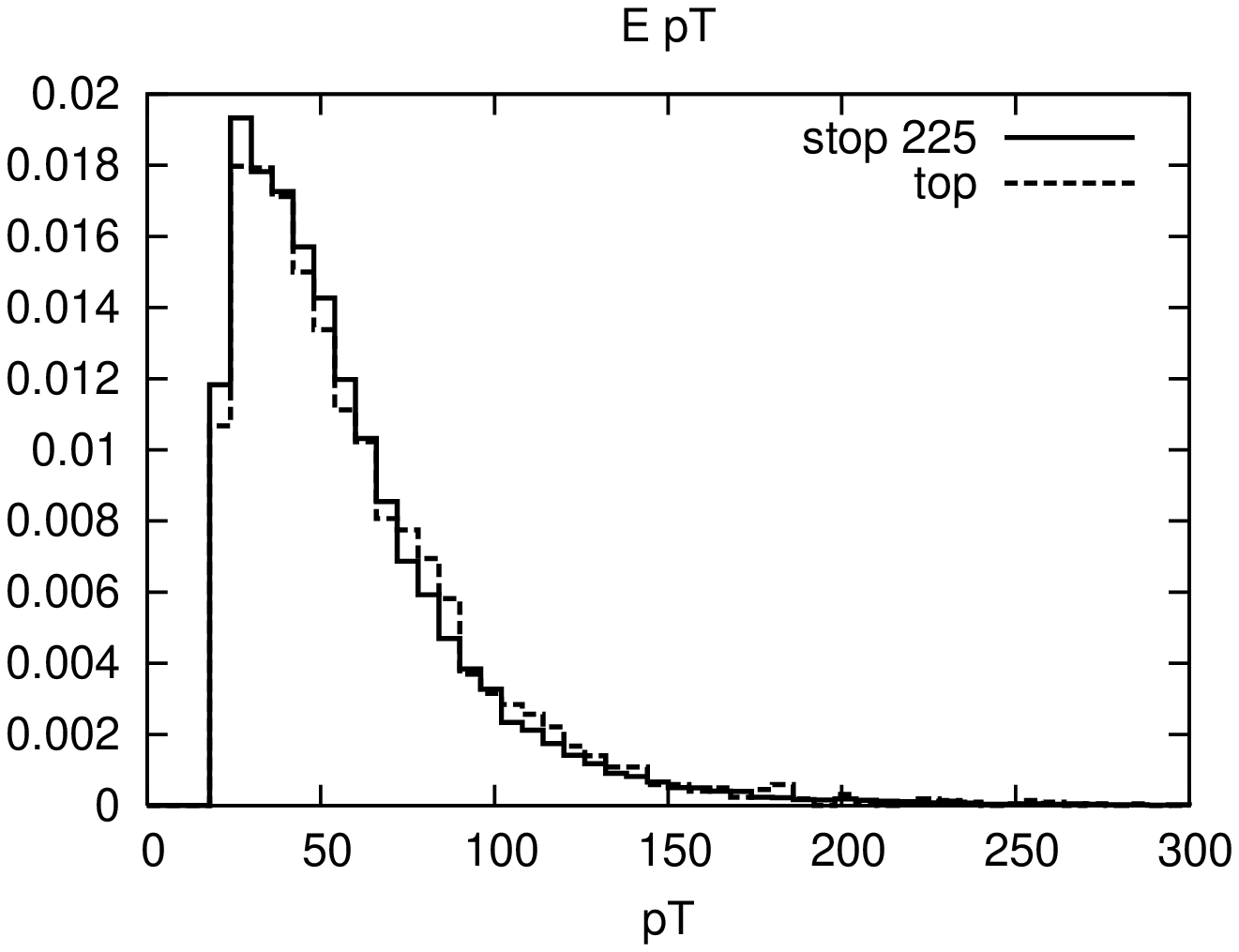,height=75mm,width=40mm,angle=270}}
\put(78,36){\mbox{d)}}
\put(80,40){\epsfig{figure=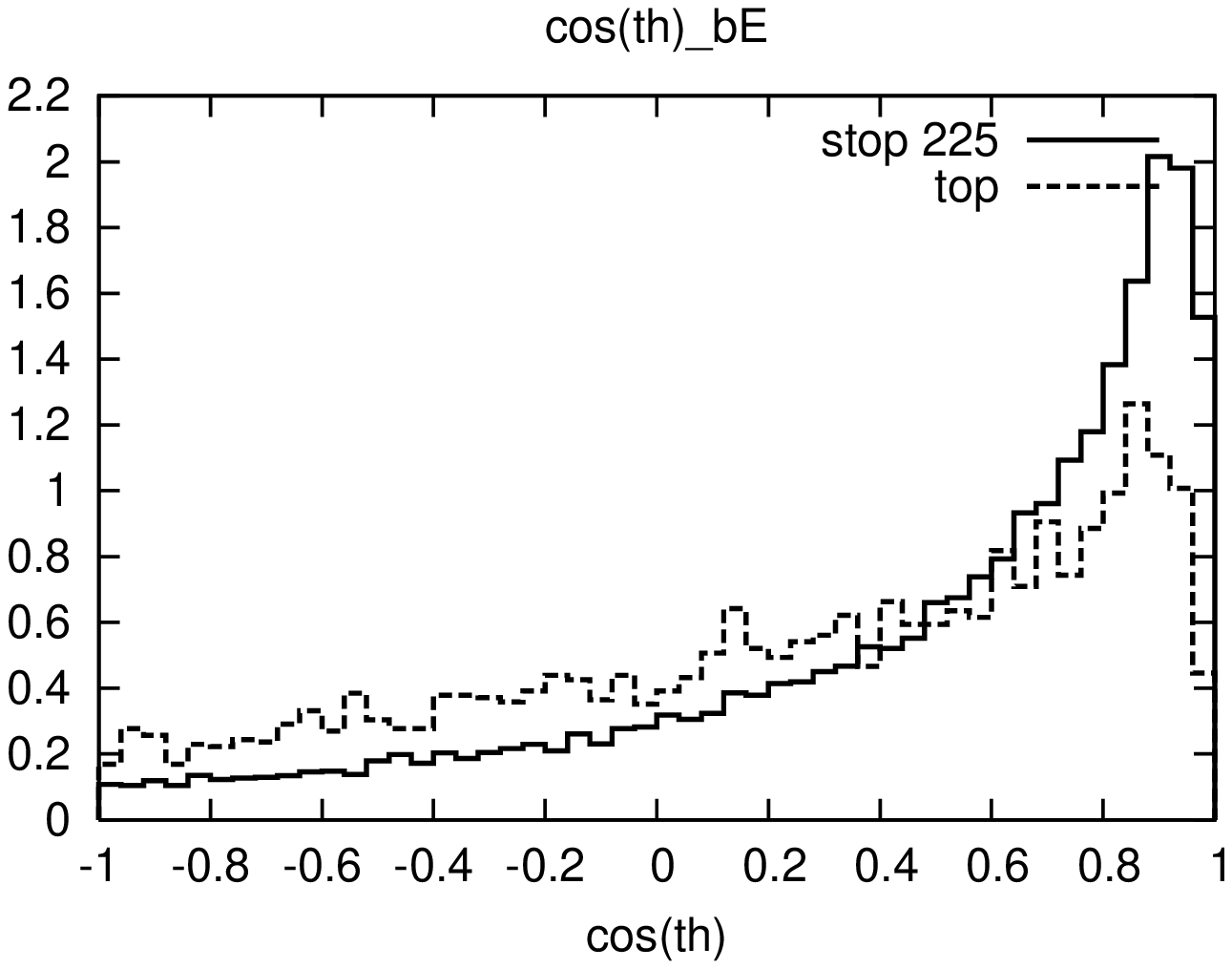,height=75mm,width=40mm,angle=270}}
\end{picture}
\caption{Distributions of stop and top decays: 
a) the transverse displacement of the stop (in mm), 
b)   $p_T$ of the $b$-quark
c) $p_T$ of the charged lepton and
d) $\cos\theta_{b\ell}$, the angle between
the 3-momenta ${\bf k}_b$ and $\bf k_\ell$.
}
\label{fig:distributions}
\end{figure}

In the following we summarize the results of a Monte Carlo study at the
parton level
\cite{deGouvea:2006wd} using {\tt PYTHIA} 6.327 \cite{Sjostrand:2003wg}.
We have taken $M_{\tilde t_R}=$ 225 GeV, $M_{\tilde N_R}=$ 100 GeV,
$M_{\tilde H} =$ 250~GeV and $Y_N = 4 \cdot 10^{-6}$ resulting
in a mean decay length of 10~mm. Note, that the stop will hadronize before
decaying. However, we have neglected the related effects in this study.
We have only considered direct stop pair production, 
and neglected 
stops from cascade decays, e.g.~$\tilde g \to t \tilde t_R$. 
The signal is
 $p p (\bar{p}) \rightarrow {\tilde t_R} {\tilde t_R}^*
\rightarrow b \ell^+ \bar{b} \ell^- + E^{\rm miss}_T$. The dominant physics background
 is  top quark pair production: 
$p p (\bar{p}) \rightarrow t \bar t \rightarrow b W^+ \bar{b} W^- 
\rightarrow b \ell^+ \bar{b} \ell^- + E^{\rm miss}_T$, 
where the missing energy is due to
neutrinos in the final state. We have imposed the
following ''Level 1'' cuts: 
(i) fermion rapidities: $|\eta_{\ell}|<2.5$, $|\eta_{b}|<2.5$
(ii) $p_{T}$ cuts ${p_{T}}_{\ell}>20~$GeV, ${p_{T}}_{b}>10~$GeV and
(iii) isolation cut
      $R_{b\ell}\equiv (\phi_b - \phi_\ell)^2 + (\eta_b - \eta_\ell)^2 >0.4$.

Figure~\ref{fig:distributions} shows various distributions for
stop and top decays.
Figure~\ref{fig:distributions}a) depicts the resulting transverse
displacement after including the boost of the stop. If it decays
before exiting the tracking subsystem, a displaced vertex may be
reconstructed through the stop decay products' 3-momenta meeting away
from the primary interaction point.  On each side, the $b$-quark
itself leads to an additional displaced vertex, and its 3-momentum
vector can be reconstructed from its decay products. In combination
with the 3-momentum of the lepton, the stop displaced vertex can be
determined.  In order to reveal the displaced vertex, one must require
either the $b$-quark or the charged lepton 3-momentum vector to miss
the primary vertex.  Since a pair of stops is produced, we would
expect to discern two displaced vertices in the event (not counting
the displaced vertices due to the b-quarks).
 Such an event with two displaced
vertices, from each of which originates a high $p_T$ $\ell$ and
$b$-quark might prove to be the main distinguishing
characteristics of such a scenario. A cut on the displaced vertex
will be very effective to separate stop events from the top
background provided one can efficiently explore such cuts.
We anticipate that NLSP stop searches may turn out to be physics-background
free in such a case.

If the stop displaced vertex cannot be efficiently resolved, one will
have to resort to more conventional analysis methods. In the remainder
 we explore various kinematical distributions for both
the signal ($\tilde t_R$ pair production) and the physics
background ($t$ pair production), obtained after imposing the level~1
cuts given above. Figures~\ref{fig:distributions}b) and c) depict the
$p_T$ spectra of the produced fermions.  
The $p_T$ of the $b$-quark from the $225~$GeV stop peaks at a lower
value compared to the top quark background, and therefore accepting
them at high efficiency for $p_T \lesssim 40~$GeV will be very helpful
in maximizing the signal acceptance.  The signal and background shapes
are quite similar and no simple set of $p_T$ cuts can be made in order
to significantly separate signal from background. 

Fig.~\ref{fig:distributions}d) depicts the distribution of $\cos\theta_{b \ell}$,
the angle between the 3-momenta ${\bf k}_b$ and ${\bf k}_\ell$, for both
the signal and background.  It is important to appreciate that, by
default, {\tt PYTHIA} generates stop decays into the 3-body final state
according only to phase-space, ignoring the angular dependence of the
decay matrix element. We have reweighted {\tt PYTHIA} events to include the
correct angular dependence in the decay matrix element. Consistent
with the expectation from Eq.~(\ref{TfistR.EQ}), we see for the signal
that the distribution peaks for the $b$-quark and charged lepton
3-momenta aligned, unlike the background.
 It is unfortunate that the isolation level~1 cut
on the leptons removes more signal events than background
events. Relaxing this constraint as much as practical would help in
this regard.
Additional work will be necessary to include also the effect of
spin correlations in top production and top decays so that
information from the quantities ${\bf k}_b\cdot{\bf k}_{\bar b}$ and
$\bf k_{\ell^+}\cdot\bf k_{\ell^-}$ can be exploited.

Assuming efficiencies of $\epsilon_b=0.5$ and $\epsilon_l=0.9$ for
$b$-quark and lepton identification, respectively, it has been
shown in \cite{deGouvea:2006wd} that stops with masses up to
500~GeV can be detected at the 5$\sigma$ level for an integrated
luminosity of 10 fb$^{-1}$ if $\ell=e,\mu$ even if the
displaced vertex signature is not used. Clearly, the situation will be worse in the
case of $\ell=\tau$. Provided one can exploit the displaced vertex
information, we expect a considerable improvement as we could not identify
any physics background.
 Further studies are planned to investigate the questions we
have touched upon here.


\section{ SUSY Searches at $\sqrt{s}=14$ TeV with CMS}

This section summarizes the recent results on SUSY searches reported at 
\cite{CMSTDR}. In the context of this work 
we refer to the generalized  classification of  models of new physics according 
to  how they affect flavour physics:
\begin{itemize}
\item {\bf CMFV:}  Constrained Minimal Flavour Violation \cite{Buras:2003jf}
models: in these models the only
source of quark flavour violation is the CKM matrix.
Examples include minimal supergravity models with low or moderate
tan$\,\beta$, and models with a universal large extra dimension.
\item {\bf MFV:}  Minimal 
Flavour Violation \cite{D'Ambrosio:2002ex} models: a set of CMFV models 
with some new relevant operators that contribute to flavour transitions. 
Examples include SUSY models with large
tan$\,\beta$.
\item {\bf NMFV:} Next-to-Minimal Flavour Violation \cite{Agashe:2005hk}
models: they involve third generation quarks and help to solve the flavour
problems that appear in frameworks such as Little Higgs, topcolour,
and RS models.
\item {\bf GFV:} General Flavour
Violation \cite{Foster:2005kb} models; they provide with new sources of flavour 
violation. These  include most of the MSSM parameter space, and almost any BSM model
before  flavour constraints are considered.
\end{itemize}

A useful discussion on these models can be found in \cite{lykken:2006gs}.
The SUSY searches that are summarized here fall in the category 
of MFV (mSUGRA specifically) and all results are obtained with the detailed 
Geant-4 based CMS  simulation. In the context of this workshop and in collaboration 
with the theory community we try to also move towards interpretation within NMVF models 
(see e.g  contributions by R. Cavanaugh and O. Buchmueller in this volume).
Note that since the squarks and sleptons can have significant flavour changing 
vertices and be complex, the connection to collider physics can be 
subtle indeed, the main implication being that the superpartners 
cannot be too heavy and that  larger tan$\,\beta$ is favored 
-- with no direct  signature in general. For interpretations of
recent Tevatron and B-factory results the interested reader can refer for example 
to  \cite{Carena:2006ai}, \cite{Isidori:2006qy}, \cite{Ligeti:2006pm}, \cite{Becher:2006pu} 
and to relevant contributions at this workshop.

The SUSY search path has been described in the past years as a 
successive approximation of serial steps that move from inclusive to 
more exclusive measurements as follows:
\begin{itemize}
\item Discovery: using canonical inclusive searches
\item Characterization: putting together a picture given the channels
  that show excess and ratios of the observed objects
  (e.g. multi-leptons, photons (GMSB), ratio of same sign leptons to  opposite ones, 
ratios of positive pairs to negative, departure from lepton universality, third generation excesses).
\item Reconstruction: in canonical dark matter LSP SUSY the final
  state contains two neutralinos hence there is no direct mass peak
  due to the missing transverse energy in the event. The kinematics of
  the intermediate decays provide however a multitude of endpoints and
  edges that might provide mass differences and help orient towards the right mass hierarchy. 
\item Measurement of the underlying theory: the classical SUSY solving
  strategy involves more mass combinations, more decay chains, mass peaks 
and once the LSP mass is known the determination of  the mass hierarchies,
particles' spins, and eventually the model parameters. An outstanding
  question remains as to how many simple measurements do we need  to
  ``nail'' the theory? Remember that we did not need to measure all
  the Standard Model particles and their properties in order to measure the Standard Model.
\end{itemize}

In the past three years the ``inclusive'' and ``exclusive'' {\it modus
quaestio questio} have been approached in coincidence and in many
works that range in exploitation strategy from statistical methods to
fully on-shell description of unknown models and inclusion of
cosmological considerations such as in \cite{Lester:2005je},
\cite{Allanach:2007qj}, \cite{Baltz:2004aw}, \cite{Nojiri:2005ph} and
\cite{Arkani-Hamed:2007fw}, to mention but a few.

It is rather safe to claim that the program of discovery and
characterization will be  (much) more convoluted than the one 
described in the  serial steps above. Realistic studies of 
kinematic edges across even the ``simple'' mSUGRA parameter 
space show that this is a difficult job and it will take a lot of work 
and wisdom to do it right. Endpoint analyses by definition involve particles 
which are very soft in some reference frame and non trivial issues  of 
acceptance need to be considered.

Some of most recent  SUSY searches  at CMS \cite{CMSTDR}, proceed in the following channels:
\begin{itemize}
\item canonical inclusive 
  \begin{itemize} 
    \item  multijets+$\ETmi$
      \item $\mu$+jets+$\ETmi$
        \item same-sign dimuon + $\ETmi$
\item opposite-sign same flavour dielectron and dimuon +$\ETmi$
\item opposite-sign same flavour hadronic ditau +$\ETmi$
\item trileptons at high $m_0$
\end{itemize}
\item  higher reconstructed object inclusive 
\begin{itemize}
\item $Z^{0}$ + $\ETmi$
\item hadronic top + $\ETmi$
\item $h(\rightarrow b\bar{b})$+ $\ETmi$
\end{itemize}
\item flavour violating 
\begin{itemize}
\item opposite-sign different flavour $e\mu$  FV neutralino decays 
(contributing to this workshop also)
\end{itemize}
\end{itemize}

\begin{figure}[ht]
\begin{center}
    \epsfig{figure=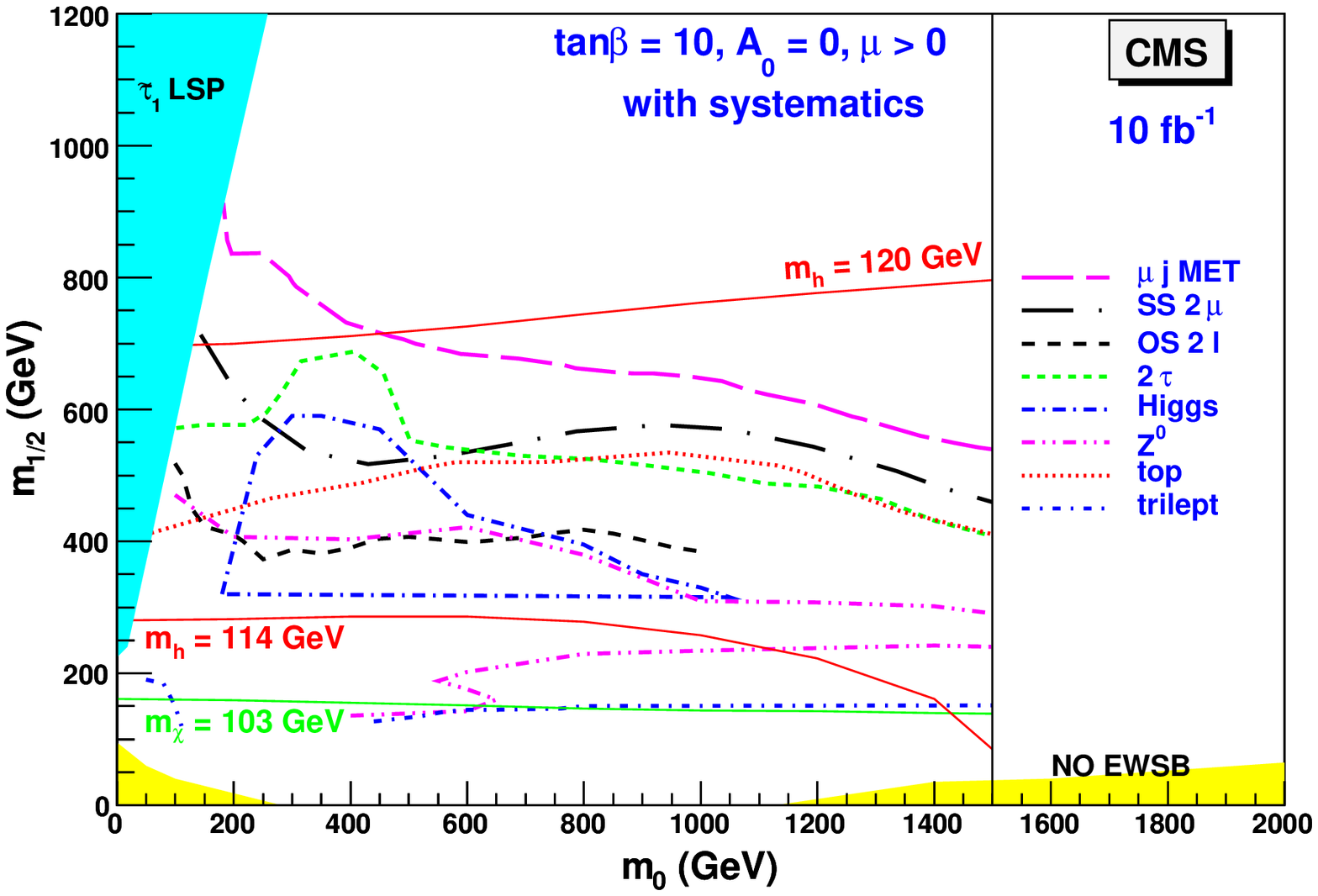, width=9cm,height=5cm, angle=0.}
    \epsfig{figure=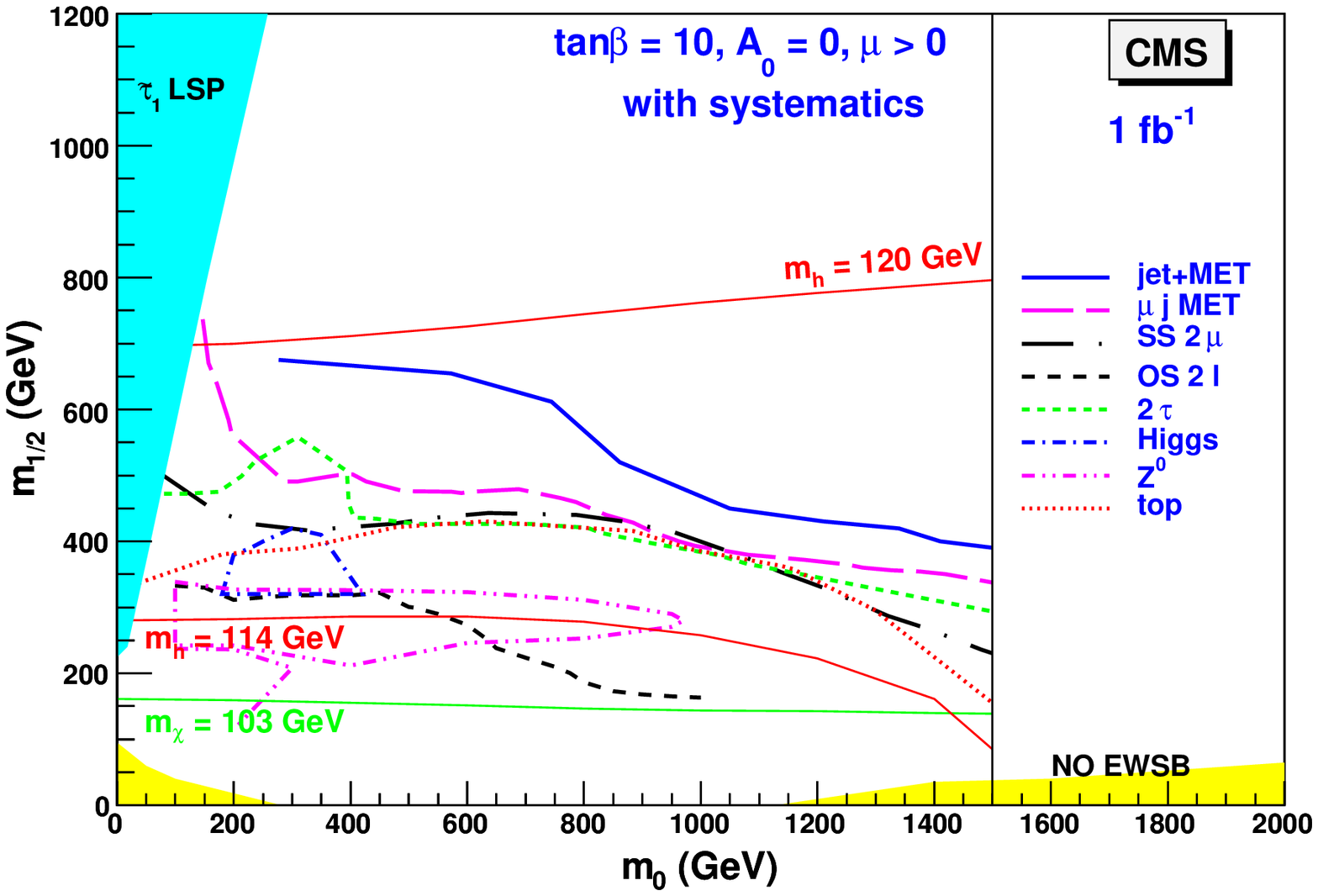, width=9cm,height=5cm, angle=0.}
\end{center}
    \caption{5$\sigma$ reach for 1 fb$^{-1}$ and 10 fb$^{-1}$  at CMS in different channels.}
    \label{fig:susyreach}
\end{figure}

The attempt is to have an as model-independent signature-based search strategy 
with educated input from theory. The interpretation of the search results
are given in the context and parameter space of mSUGRA but re-interpreting 
them in different models is possible. All of the searches are including detector 
systematic uncertainties and a scan that provides the  5$\sigma$ reach 
in the mSUGRA parameter space is derived for 1$fb^{-1}$ and 10$fb^{-1}$ as 
shown in figure \ref{fig:susyreach}
The details of the analyses and individual search results can be found at
\cite{CMSTDR}.

\chapter{Non-supersymmetric Standard Model extensions}
\label{chap:NS}
{\small J. A. Aguilar Saavedra and G. {\"U}nel}

\section{Introduction}
Although the Standard Model (SM) has seemingly survived many stringent tests
offered by both precision measurements and direct searches, it has a 
number of shortcomings. The most unpleasant one is the ``instability'' of 
the Higgs boson mass with respect to radiative corrections, known as the 
hierarchy problem. If the SM is assumed valid up
to a high scale $\Lambda$ of the order of the Planck mass, radiative 
corrections to $M_h$ from top quark loops are of order $\delta M_h^2 \sim 
\Lambda^2$, {\em i.e.} much larger than $M_h$ which is expected to be of 
the order of the electroweak scale. The requirement that $M_h$ and 
$\delta M_h$ are of the same order would imply a cutoff 
(and hence new physics at) $\Lambda \sim 1-2$ TeV.
Some other aspects of the SM that make it unappealing as the 
ultimate theory of fundamental interactions (excluding gravity) are:
\begin{itemize}
\item the lack of simplicity of the gauge structure,
\item the large hierarchy of fermion masses and quark mixings, and the large
number of apparently free parameters necessary to describe them,
\item the source of baryogenesis, which can not be explained by the amount of CP
 violation present in the SM,
\item the unknown mechanism behind the neutrino mass generation 
(neutrinos can have Dirac masses simply with the introduction of right-handed
fields, 
but present limits $m_\nu \sim 1$ eV require unnaturally small Yukawa couplings).
\end{itemize}

Therefore, the SM is believed to be the low-energy limit of a more 
fundamental theory. Several arguments suggest that this theory may 
manifest itself at energies not much higher than the electroweak 
scale, and give support to the hope that LHC will provide signals 
of new physics beyond the SM.

This chapter deals with non-supersymmetric candidate theories as 
extensions to the SM. Among the most frequently studied ones, 
the following ones can be mentioned.
\begin{enumerate}

\item Grand unified theories (GUTs). In these models the
SM gauge group $\text{SU}(3)_c \times \text{SU}(2)_L \times \text{U}(1)_Y$
is embedded into a larger symmetry group, which is recovered at a higher scale. 
They predict the existence of new fermions ({\em e.g.} $Q=-1/3$ singlets) 
and new gauge bosons (especially $Z'$), which may be at the reach of LHC.

\item Little Higgs models. They address the hierarchy problem with the
introduction of extra gauge symmetries and extra matter which stabilise
the Higgs mass up to a higher scale $\Lambda \sim 10$ TeV. In particular, 
the top quark loop contribution to the Higgs mass is partially cancelled with
the introduction of a $Q=2/3$ quark singlet $T$.

\item Theories with extra dimensions. The various extra-dimensional models avoid
the hierarchy problem by lowering the Planck scale in the higher dimensional
theory, and some of them can explain the large hierarchies between fermion
masses. The observable effect of the additional dimension is the appearance of
``towers'' of Kaluza-Klein (KK) excitations of fermions and bosons, with
increasing masses. Depending on the model, the lightest modes can have a mass
around the TeV scale and thus be produced at LHC.
\end{enumerate}
It should be stressed that these SM extensions, sometimes labelled as
``alternative theories'' do not exclude supersymmetry
(SUSY). In fact, SUSY in its minimal versions does not address some of the open
questions of the SM. One example is the motivation behind the apparent gauge
coupling unification. The
renormalisation-group evolution of the coupling constants strongly suggests
that they unify at a very high scale $M_\text{GUT} \sim 10^{15}$ GeV, and that
the SM gauge group
is a subgroup of a larger one, {\em e.g.} $\text{SO}(10)$, $\text{E}_6$ or other
possibilities. Thus, SUSY can naturally coexist with GUTs. Another example of
complementarity is SUSY + Little Higgs models. If SUSY is broken at the TeV
scale or below, it may give dangerous contributions to flavour-changing neutral
(FCN) processes and electric dipole moments (EDMs). These contributions must be
suppressed with some (well justified or not) assumption, like minimal
supergravity (MSUGRA) with real parameters. These problems are alleviated if
SUSY is broken at a higher scale and, up to that scale, the Higgs mass is
stabilised by another mechanism, as it happens in the Little Higgs theories.

With the forthcoming LHC, theories beyond the SM will be tested  directly  
through the searches of the new particles, and indirectly, with measurements of the
deviations from SM precision variables. Instead of studying the different SM
extensions and their additional spectrum separately, we follow a
phenomenological/experimental approach. Thus, this chapter is organised
according to the new particles which are expected to be produced.
Section \ref{sec_new_quarks} reviews the searches for the new  
quarks and section \ref{sec_new_leptons} for new heavy neutrinos. Studies for
new gauge bosons are collected in sections \ref{sec_new_zp} and
\ref{sec_new_wp},
and in section \ref{sec_new_scalars} some new scalar signals are
presented. Detailed information about the SM extensions predicting these new
particles is not included in this report for brevity (although the text is as
self-contained as possible). Instead, the interested reader is encouraged to
refer to the original papers and dedicated reviews (see for instance
\cite{Langacker:1980js,Schmaltz:2005ky,Perelstein:2005ka,Hewett:2002hv,Csaki:2004ay}).
The observation of these new particles would prove, or at least provide
hints, for the proposed theories. In this case,
the identification of the underlying theory might be possible with the
measurement of the couplings, production and decay modes of the new
particle(s). Alternatively, the non-observation of the predicted signals would
disprove the models or impose lower bounds for their mass scales.

\section{New quarks}
 \label{sec_new_quarks}

At present, additional quarks are not required neither by experimental data nor
by the
consistency of the SM. But on the other hand they often appear in 
Grand Unified Theories \cite{Hewett:1988xc,Frampton:1999xi},
Little Higgs
models \cite{Arkani-Hamed:2001nc,Arkani-Hamed:2002qy,Perelstein:2005ka},
Flavour Democracy\cite{Fritzsch:1989qm}
and models with extra
dimensions \cite{Mirabelli:1999ks,DelAguila:2001pu,Csaki:2004ay}. 
Their existence is not experimentally excluded but their mixing, mainly with the
lightest SM
fermions, is rather constrained. They can lead to various indirect effects at
low energies, and their presence could explain experimental deviations
eventually found, for instance in $CP$ asymmetries in $B$ decays.
They can also enhance flavour-changing processes, especially those involving the
top quark. These issues have been dealt with in other chapters of this report.
Here we are mainly concerned with their direct production and detection at LHC.

\begin{figure}[htb]
\begin{center}
\epsfig{file=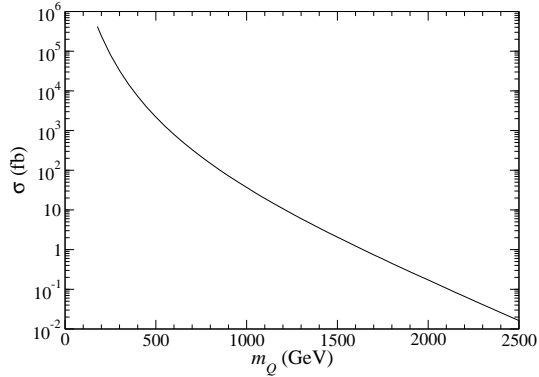,height=5.0cm,clip=}
\caption{Tree-level cross section for pair production of heavy quarks $Q$ in
$pp$ collisions at $\sqrt{s}=14$~TeV, $gg,q \bar q \to Q \bar Q$.
CTEQ5L PDFs are used.}
\label{fig:mQ-sigma}
\end{center}
\end{figure}

New quarks share the same electromagnetic and strong interactions of standard
quarks, and thus they can be produced at LHC by $q \bar q$ annihilation and
gluon fusion in the same way as the top quark, with a cross section which only
depends on their mass, plotted in
Fig.~\ref{fig:mQ-sigma}. Depending on
their electroweak mixing with the SM fermions, they can be produced singly
as well \cite{Han:2003wu,delAguila:2004sj,Sultansoy:2006rx}.
Their decay always takes place through electroweak interactions or interactions with scalars,
and the specific decay modes available depend on the particular SM extension
considered. Let us consider a SM extension with $N$ ``standard'' chiral
generations (left-handed doublets and right-handed singlets under
$\mathrm{SU}(2)_L$), plus $n_u$ up-type and $n_d$ down-type singlets under this
group.\footnote{Anomaly cancellation requires that the number of lepton
generations is also $N$. For $N > 3$ this implies additional neutrinos heavier
than $M_Z/2$ to agree with the $Z$ invisible width measurement at LEP. On the
other hand, quark singlets can be introduced alone, since they do not contribute to
anomalies \cite{Frampton:1999xi}.}
While (left-handed) $\mathrm{SU}(2)_L$ doublets couple to the $W^\pm$ and $W^3$
bosons, singlet fields do not. 
The Lagrangian in the weak eigenstate basis reads
\begin{eqnarray}
\mathcal{L}_W & = & -\frac{g}{\sqrt 2} \left[ \bar u'_L \gamma^\mu d'_L \right]
W_\mu^\dagger + \mathrm{h.c.} \,, \notag \\
\mathcal{L}_Z & = & -\frac{g}{2 c_W} \left[ \bar u'_L \gamma^\mu u'_L 
-\bar d'_L \gamma^\mu d'_L -2 s_W^2 J_\mathrm{EM} \right] Z_\mu \,, 
\label{ec:NS:lagrW}
\end{eqnarray}
where $(u',d')_L$ are the $N$ doublets under $\mathrm{SU}(2)_L$
and $J_\mathrm{EM}$ is
the electromagnetic current which includes all (charged) quark fields.
The number of mass eigenstates
with charges $2/3$ and $-1/3$ is $N_u \equiv N+n_u$, $N_d \equiv N+n_d$,
respectively. 
The resulting
weak interaction Lagrangian in the mass eigenstate basis is
\begin{eqnarray}
\mathcal{L}_W & = & -\frac{g}{\sqrt 2} \left[ \bar u_L \gamma^\mu V d_L \right]
W_\mu^\dagger + \mathrm{h.c.} \,, \notag \\
\mathcal{L}_Z & = & -\frac{g}{2 c_W} \left[ \bar u_L \gamma^\mu X^u u_L 
-\bar d_L \gamma^\mu X^d d_L -2 s_W^2 J_\mathrm{EM} \right] Z_\mu \,, 
\label{ec:NS:lagrQS}
\end{eqnarray}
where $u_{L,R}$, $d_{L,R}$ are column vectors of dimensions $N_u$, $N_d$, and
$J_\mathrm{EM}$ is the electromagnetic current (including all mass eigenstates).
The $N_u \times N_d$ matrix $V$ (not necessarily square) is the generalisation
of the $3 \times
3$ CKM matrix. The matrices $X^u = V V^\dagger$, $X^d = V^\dagger V$ have
dimensions $N_u \times N_u$ and $N_d \times N_d$, respectively.
In case that $n_u > 0$ the up-type mass eigenstates are mixture of weak
eigenstates with different isospin, and thus the matrix $X^u$ is not necessarily
diagonal.
In other words, $V$ is not unitary (but its $3 \times 3$ submatrix involving SM
quarks is almost unitary), what prevents the GIM mechanism from
fully operating.
Analogous statements hold for the down sector. Therefore, models with
quark singlets can have tree-level flavour-changing neutral (FCN) couplings
to the $Z$ boson. These couplings are suppressed by the mass of the new mass
eigenstates, {\em e.g.} $X_{tc} \sim m_t m_c / m_T^2$ (with $T$ a new
charge $2/3$ quark), what forbids dangerous FCN currents in the down sector but
allows for observable effects in top physics.

As within the SM, its extensions with extra quarks typically have one Higgs
doublet which breaks the electroweak symmetry and originates the fermion masses.
The surviving scalar field $h$ couples to the chiral fields (through Yukawa
couplings) but not to the weak eigenstate isosinglets.
In the mass eigenstate basis, the scalar-quark interactions read
\begin{eqnarray}
\mathcal{L}_h & = & - \frac{g}{2 M_W} \left[ \bar u_R \mathcal{M}^u X^u u_L 
+ \bar d_R \mathcal{M}^d X^d d_L \right] h + \mathrm{h.c.} \,,
\label{ec:NS:Hterm}
\end{eqnarray}
with $\mathcal{M}^u$, $\mathcal{M}^d$ the diagonal mass matrices, of
dimensions $N_u \times N_u$ and $N_d \times N_d$.
SM extensions with extra quarks usually introduce further scalar fields, {\em
e.g.} in $\text{E}_6$ additional scalar singlets are present, but with VEVs
typically much higher than the mass scale of the new quarks and small mixing
with $h$. Also, in supersymmetric versions of $\text{E}_6$
there are two Higgs doublets, in which case the generalisation of
Eq.~(\ref{ec:NS:Hterm}) involves two scalar fields and the ratio of their VEVs
$\tan \beta$. However, the main phenomenological features of these models can be
described with the minimal scalar sector and Lagrangian in
Eq.~(\ref{ec:NS:Hterm}). (Of course, this does not preclude that with
appropriate
but in principle less natural choices of parameters one can build models with a
completely different behaviour.) In particular, from Eq.~(\ref{ec:NS:Hterm}) it
follows that FCN  interactions with scalars have the same strength as the ones mediated
by the $Z$ boson, up to mass factors. Note also that Eq.~(\ref{ec:NS:Hterm})
does not contradict the fact that new heavy mass eigenstates, which are mainly
$\mathrm{SU}(2)_L$ singlets, have small Yukawa couplings. For example, with an
extra $Q=2/3$ singlet the Yukawa coupling of the new mass eigenstate $T$ is
proportional to $m_T X_{TT} \simeq m_T |V_{Tb}|^2 \sim m_t/m_T^2$ (see also
section \ref{subsec_q_2_3} below).

More general extensions of the SM quark sector include right-handed
fields transforming non-trivially under $\mathrm{SU}(2)_L$. The simplest of such
possibilities is the presence of
additional isodoublets $(T,B)_{L,R}$. Their interactions are described with the
right-handed analogous of the terms in Eqs.~(\ref{ec:NS:lagrQS}) and a
generalisation of Eq.~(\ref{ec:NS:Hterm}). From the point
of view of collider phenomenology, their production and decay takes place
through the same channels as fourth generation or singlet quarks (with
additional gauge bosons there would be additional modes). However, the
constraints from low energy processes are much more stringent, since mixing with
a heavy isodoublet $(T,B)_{L,R}$ can induce right-handed charged currents among
the
known quarks, which are absent in the SM. An example of this kind is a $W t_R
b_R$ interaction, which would give a large contribution to the radiative decay
$b \to s \gamma$ (see \chapt{chap:top}{top:const_bfac}).

A heavy quark $Q$ of either charge can decay to a lighter quark $q'$ via charged
currents, or to a lighter quark $q$ of the same charge via FCN
couplings if they are nonzero. The partial widths for these decays are
\cite{Aguilar-Saavedra:2006gw}
\begin{align}
\Gamma(Q \to W^+ q') & = \frac{\alpha}{16 s_W^2} |V_{Qq'}|^2
  \frac{m_Q}{M_W^2} \lambda(m_Q,m_{q'},M_W)^{1/2} \nonumber \\
  & \times  \left[ 1 + \frac{M_W^2}{m_Q^2}
  - 2  \frac{m_{q'}^2}{m_Q^2} - 2  \frac{M_W^4}{m_Q^4}  + \frac{m_t^4}{m_Q^4}
  + \frac{M_W^2 m_{q'}^2}{m_Q^4} \right] \,, \nonumber \\
\Gamma(Q \to Z q) & = \frac{\alpha}{32 s_W^2 c_W^2} |X_{Qq}|^2
  \frac{m_Q}{M_Z^2} \lambda(m_Q,m_q,M_Z)^{1/2} \nonumber \\
  & \times  \left[ 1 + \frac{M_Z^2}{m_Q^2}
  - 2  \frac{m_q^2}{m_Q^2} - 2  \frac{M_Z^4}{m_Q^4}  + \frac{m_t^4}{m_Q^4}
  + \frac{M_Z^2 m_q^2}{m_Q^4} \right] \,, \nonumber \\
\Gamma(Q \to h q) & = \frac{\alpha}{32 s_W^2} |X_{Qq}|^2
 \frac{m_Q}{M_W^2} \lambda(m_Q,m_q,M_h)^{1/2} \nonumber \\
  & \times  \left[ 1 + 6 \frac{m_q^2}{m_Q^2} - \frac{M_h^2}{m_Q^2} 
  + \frac{m_q^4}{m_Q^4} - \frac{m_q^2 M_h^2}{m_Q^4} \right] \,,
\label{ec:NS:pwidQ}
\end{align}
with
\begin{eqnarray}
\lambda (m_Q,m,M) & \equiv & (m_Q^4 + m^4 + M^4
- 2 m_Q^2 m^2 - 2 m_Q^2 M^2 - 2 m^2 M^2)
\end{eqnarray}
a kinematical function. (The superscripts $u$, $d$ in the FCN couplings
$X_{Qq}$ may be dropped when they are clear from the context.)
Since QCD and electroweak production processes are the
same for $4^\mathrm{th}$ generation and exotic quarks, their decays provide the
way to distinguish them. For quark singlets the neutral current decays
$Q \to Zq$ are possible, and
kinematically allowed (see below). Moreover, for a doublet of SM quarks
$(q,q')$ of the same generation one has $\Gamma(Q \to Zq) \simeq 1/2 \;
\Gamma(Q \to W q')$, for $m_Q \gg m_q, m_{q'},M_Z, M_W$. Depending on the Higgs
mass, decays $Q \to h q$ may be kinematically allowed as well, with a partial
width $\Gamma(Q \to hq) \simeq 1/2 \; \Gamma(Q \to W q')$ for $m_Q$ much larger
than the other masses involved. Both FCN decays, absent for $4^\mathrm{th}$
generation heavy
quarks,\footnote{For $4^\mathrm{th}$ generation quarks neutral decays can take
place radiatively, and can have sizeable branching ratios if tree-level charged
current decays are very suppressed, see section~\ref{sec:NS:4gensc}~.}
provide clean final states in which new quark singlets could be discovered, in
addition to the charged current decays present in all cases.
If the new quarks mix with the SM sector through right-handed interactions with
the SM gauge and Higgs bosons, the decays are the same as in
Eq.~(\ref{ec:NS:pwidQ}) but replacing $V_{Qq'}$ and $X_{Qq}$ by their
right-handed analogues. If the new quarks are not too heavy, the chirality
of their interactions can be determined by measuring angular or energy
distributions of the decay products. For instance, in a decay $T \to W^+ b \to
\ell \nu b$ the charged lepton angular distribution in $W$ rest frame (or its
energy distribution in $T$ rest frame) an be used to probe the $WTb$ interaction
(see the discussion after Eq.~(\ref{ec:NS:dist-el}) below,
and also \chapt{chap:top}{top:wtb_anom_coup}).

Searches at Tevatron have
placed the 95\% CL limits $m_{B} \geq 128$ GeV \cite{Yao:2006px} (in charged
current decays, assuming 100\% branching ratio), $m_{b'} \geq 199$
GeV~\cite{Affolder:1999bs} (assuming BR($b' \to Z b$)=1), where $b'$ is a charge $-1/3$ quark. 
If {\em a priori} assumptions on $b'$ decays are not made, limits
can be found on the branching ratios of these two channels \cite{delphibprime}
(see also \cite{Oliveira:2003hf,Oliveira:2003gc}).
In particular, it is found that for $b'$ quarks with masses $\sim 100$ GeV
near the LEP kinematical limit there are
some windows in parameter space where $b'$ could
have escaped discovery.
For a charge $2/3$ quark $T$, the present Tevatron bound is $m_T \geq 258$ GeV
\cite{cdf-T} in charged current decays $T \to W^+ b$, very close to the
kinematical limit
$m_t + M_Z$ where decays $T \to Zt$ are kinematically possible.
The prospects for LHC are reviewed in the following.

\subsection{Singlets: charge $2/3$}
 \label{subsec_q_2_3}

\begin{figure}[htb]
\begin{center}
\epsfig{file=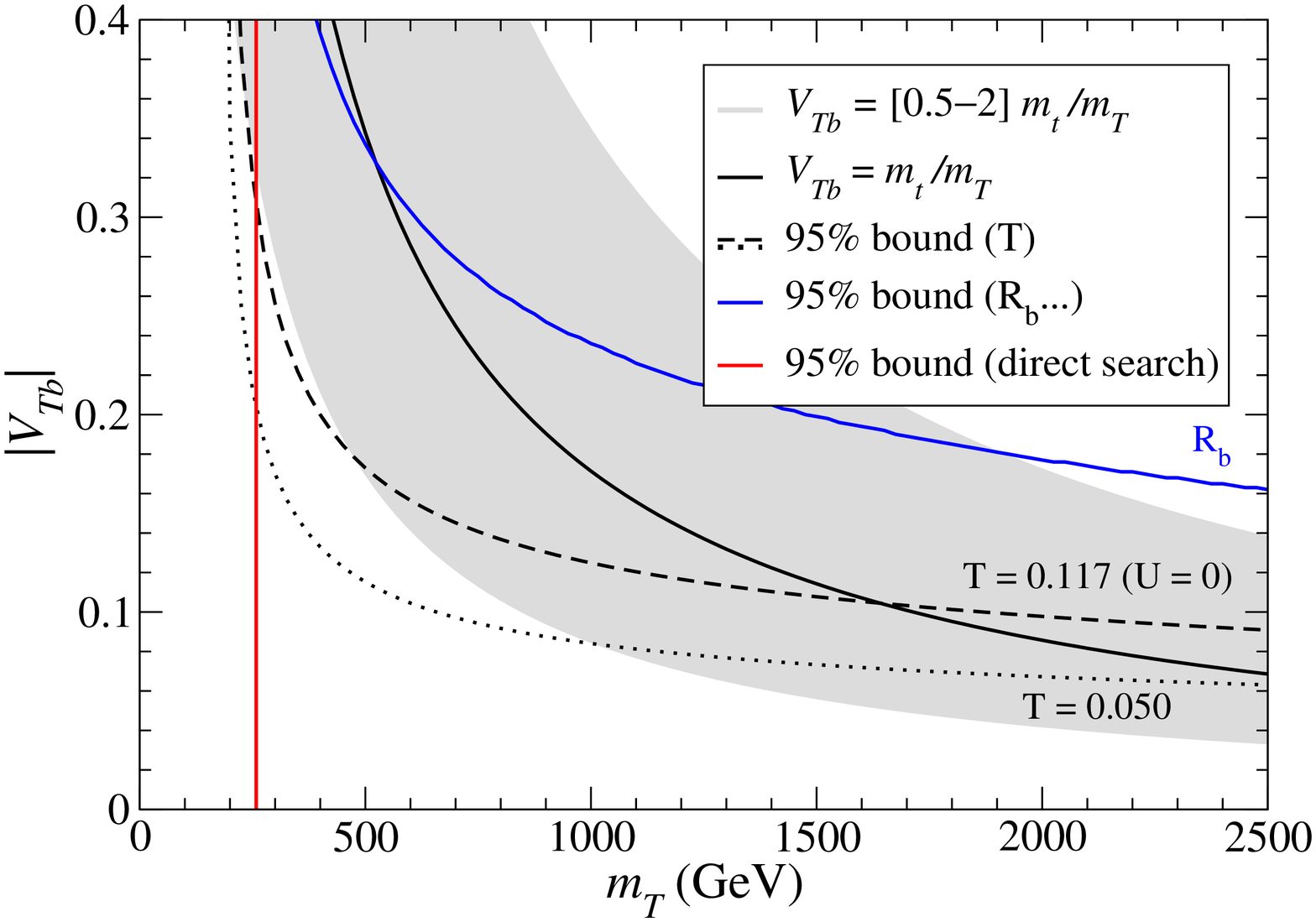,height=5.5cm,clip=}
\caption{95\% CL bounds on $|V_{Tb}|$ from the $\mathrm{T}$ parameter and from
$R_b$, and
values derived from the relation $V_{Tb} = m_t / m_T$.}
\label{fig:NS:VTblim}
\end{center}
\end{figure}

A new up-type singlet $T$ is expected to couple preferrably to the third
generation, due to the large mass of the top quark. The CKM matrix element
$V_{Tb}$ is expected to be of order $m_t / m_T$, although for $T$ masses at the
TeV scale or below
the exact relation $V_{Tb} = m_t / m_T$ enters into conflict with latest
precision electroweak data. In particular, the most stringent constraint comes
from the $\mathrm{T}$ parameter \cite{Aguilar-Saavedra:2002kr}. The most
recent values \cite{Yao:2006px} $\mathrm{T} = -0.13 \pm 0.11$
(for $\mathrm{U}$ arbitrary),
$\mathrm{T} = -0.03 \pm 0.09$ (setting $\mathrm{U}=0$) imply the 95\% CL
bounds $\mathrm{T} \leq 0.05$, $\mathrm{T} \leq 0.117$, respectively. The
resulting limits on $|V_{Tb}|$ are plotted in Fig.~\ref{fig:NS:VTblim},
including for completeness the limit from $R_b$ (plus other correlated
observables like $R_c$, the FB asymmetries and coupling parameters)
and the bound on $m_T$ from direct searches. The mixing values
obtained from the relation $V_{Tb} = \lambda m_t/m_T$ are also displayed, for
$\lambda = 1$ (continuous line) and $\lambda = 0.5-2$ (gray band).
In this class of
models the new contributions to $\mathrm{U}$ are very small, so it is sensible
to use the less restrictive bound $\mathrm{T} \leq 0.117$. Even in this case,
mixing angles $V_{Tb} = m_t/m_T$ seem too large for $T$ lighter than 1.7 TeV. Of
course, the importance of the bound $\mathrm{T} \leq 0.117$, and indirect bounds
in general, must not be neither overemphasised nor neglected. Additional new
particles present
in these models also contribute to $\mathrm{T}$ and can cancel the contribution
from the new quark. But this requires fine-tuning for lower $T$ masses and/or
larger $V_{Tb}$ mixings.

The main decays of the new quark are $T \to W^+ b$,
$T \to Z t$, $T \to h t$, with partial widths given by Eqs.~(\ref{ec:NS:pwidQ}).
Their characteristic features are:
\begin{enumerate}
\item[(i)] $T \to W^+ b$: The decays $W \to \ell^+ \nu$, $\ell=e,\mu$
originate very energetic charged leptons, not only due to the large
$T$ mass but also to spin effects \cite{Aguilar-Saavedra:2006gv}: for large
$m_T$ the charged leptons are emitted more towards the $W$ flight direction.

\item[(ii)] $T \to Z t$: The leptonic decays $Z \to \ell^+ \ell^-$ produce a
very clean final state, although with a small branching ratio.

\item[(iii)] $T \to h t$: For a light Higgs, its decay $h \to b \bar b$ and the
decay
of the top quark give a final state with three $b$ quarks, which can be tagged
to reduce backgrounds. They have an additional interest as they can produce
Higgs bosons with a large cross section
\cite{delAguila:1989rq,Aguilar-Saavedra:2006gw}.
\end{enumerate}

\subsubsection{Discovery potential}
\label{sec:NS:Tdisc}

\begin{figure}[htb]
\begin{center}
\begin{tabular}{cc}
\epsfig{file=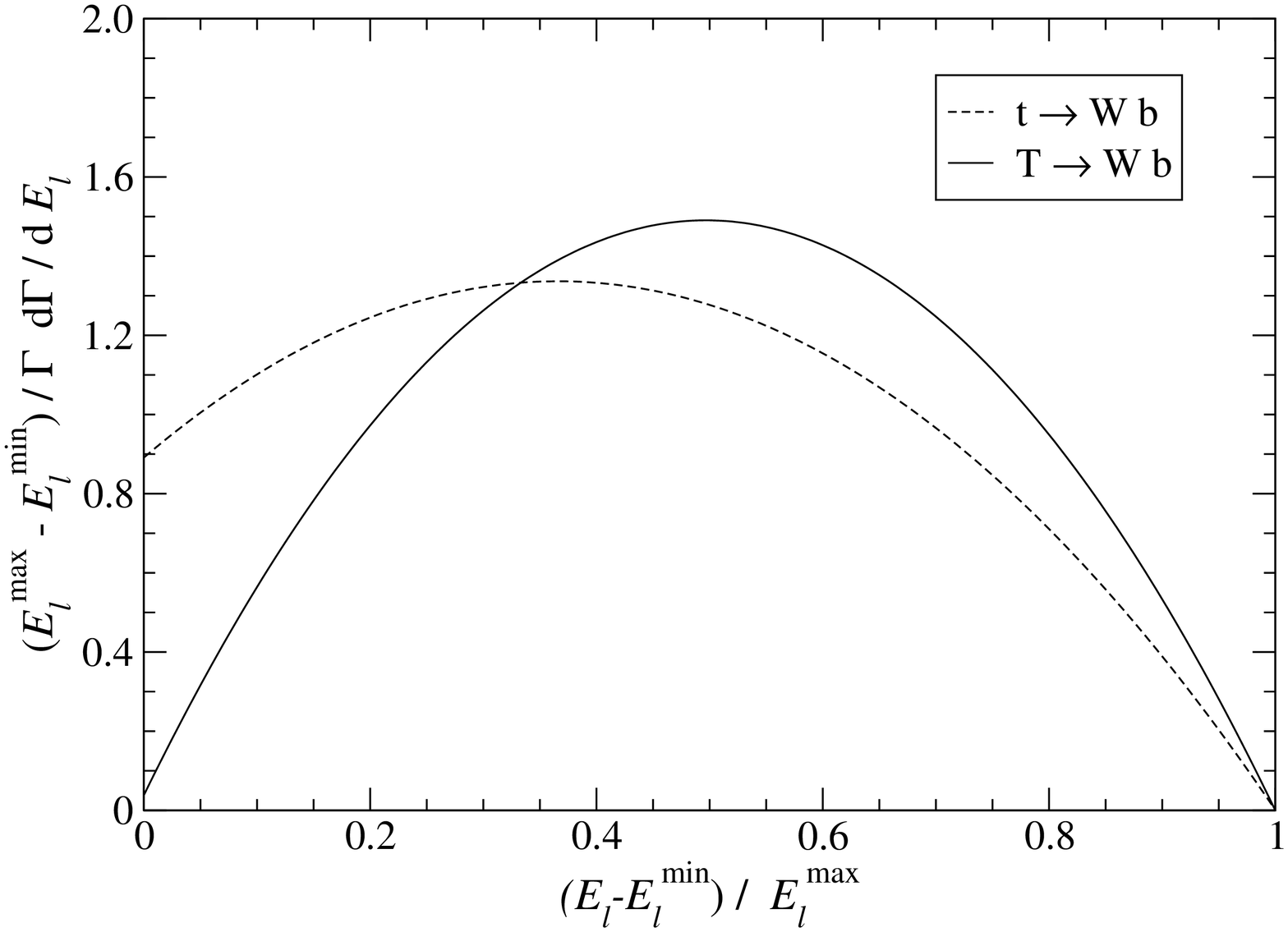,height=5.5cm,clip=} &
\epsfig{file=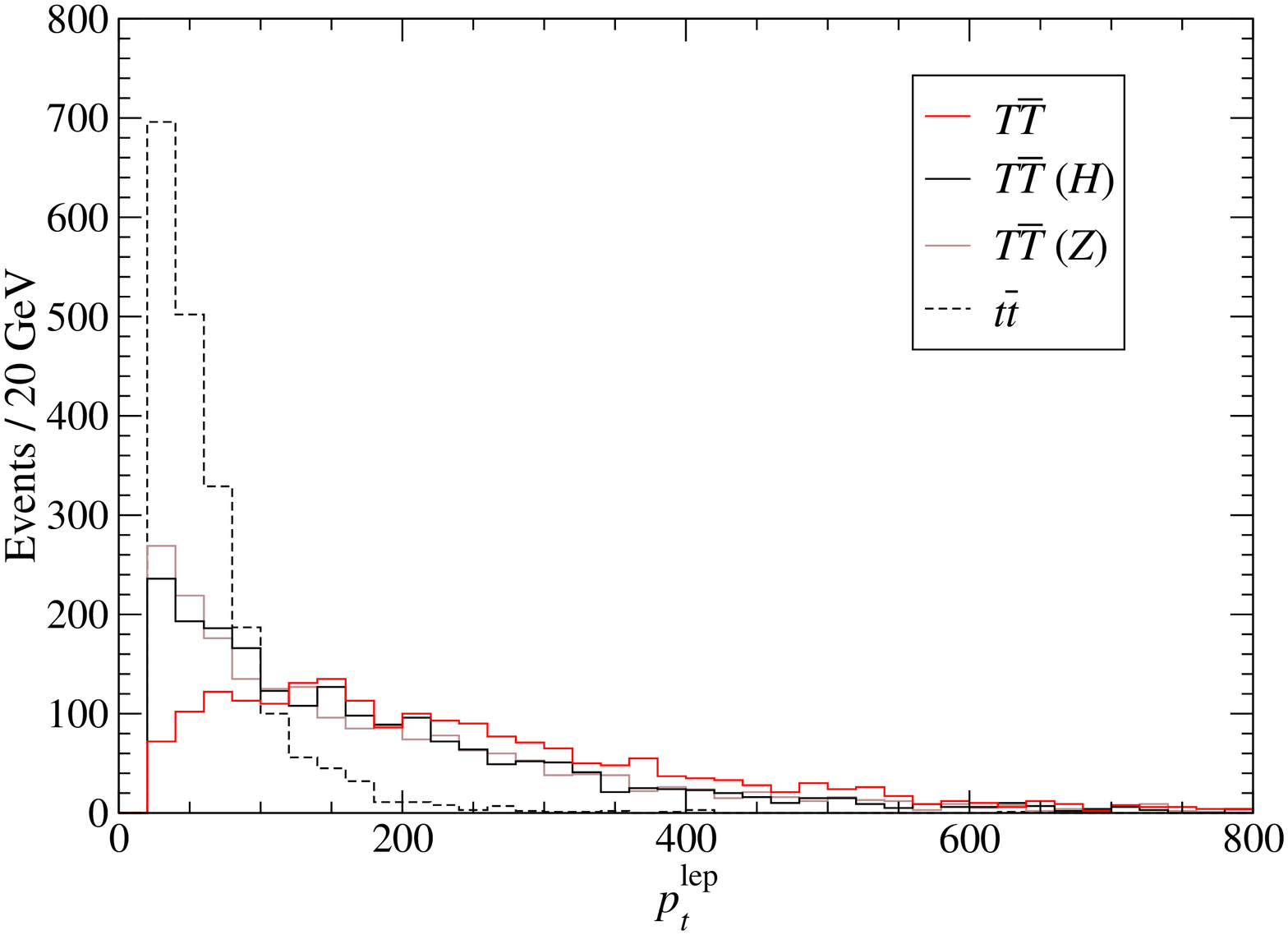,height=5.5cm,clip=} 
\end{tabular}
\caption{Left: Normalised energy distributions of the charged
lepton from $t$ and $T$ semileptonic decays, in $t$ ($T$) rest frame, taking
$m_T = 1$ TeV. Right:
the resulting transverse momentum distribution in laboratory frame
for the processes in Eqs.~(\ref{ec:NS:TTtoWW}) (right)}
\label{fig:NS:PTlepWW}
\end{center}
\end{figure}

In $T$ pair production  the largest $m_T$ reach is provided by the mode
$T \bar T \to W^+ b W^- \bar b$ and subsequent semileptonic decay of the $W^+
W^-$ pair, plus additional contributions from other decay modes giving
the same signature plus additional jets or missing energy
\cite{Aguilar-Saavedra:2005pv,Aguilar-Saavedra:2006gv}
\begin{align}
& T \bar T \to W^+ b \, W^- \bar b \to \ell^+ \nu b \, \bar q q'
\bar b \,,  \nonumber \\
& T \bar T \to W^+ b \, h \bar t / h t \, W^- \bar b 
   \to W^+ b \, W^- \bar b \, h \to \ell^+ \nu b \, \bar q q' \bar b 
   \; b \bar b / c \bar c \,, \nonumber \\
& T \bar T \to W^+ b \, Z \bar t / Z t \, W^- \bar b 
   \to W^+ b \, W^- \bar b \, Z \to \ell^+ \nu b \, \bar q q' \bar b 
   \; q'' \bar q'' / \nu \bar \nu \,,
\label{ec:NS:TTtoWW}
\end{align}
These signals are characterised by one energetic charged lepton, two $b$ jets
and at least two additional jets. Their main backgrounds are top pair and
single top production and $W/Z b \bar b$ plus jets. Charged leptons originating
from $T \to W b \to \ell \nu b$ decays are much more energetic than those from
$t \to W b \to \ell \nu b$, as it has been stressed above. The charged lepton  
energy distribution in $T$ ($t$) rest frame reads
\begin{eqnarray}
\frac{1}{\Gamma} \frac{d \Gamma}{d E_\ell} & = & \frac{1}{(E_\ell^\text{max} -
E_\ell^\text{min})^3} \left[ 3 (E_\ell - E_\ell^\text{min})^2 \, F_R
+ 3 (E_\ell^\text{max} - E_\ell)^2 \, F_L \right. \notag \\[1mm]
& & \left. + 6 (E_\ell^\text{max} - E_\ell) (E_\ell - E_\ell^\text{min}) \, F_0
\right] \,,
\label{ec:NS:dist-el}
\end{eqnarray}
with $F_i$ the $W$ helicity fractions (see \chapt{chap:top}{top:wtb_anom_coup}),
which satisfy
$F_L+F_R+F_0 = 1$. For the top quark they are $F_0 = 0.703$, $F_L = 0.297$,
$F_R \simeq 0$, while for $T$ with a mass of 1 TeV they are $F_0 = 0.997$,
$F_L = 0.013$, $F_R \simeq 0$. It must be pointed out that for large $m_T$,
$F_0 \simeq 1$ even when right-handed $WTb$ interactions are included; thus,
the chirality of this vertex cannot be determined from these observables.
The maximum and minimum energies depend on the mass of the decaying fermion,
and are 
$E_\ell^\text{min} = 18.5$ GeV, $E_\ell^\text{max} = 87.4$ GeV for $t$, and
$E_\ell^\text{min} = 3.2$ GeV, $E_\ell^\text{max} = 500$ GeV for $T$
(with $m_T$=1~TeV).
The resulting energy distributions are presented in Fig.~\ref{fig:NS:PTlepWW}
(left) for the same $T$ mass of 1~TeV.
The larger mean energy in the rest frame of the parent quark is
reflected in a larger transverse momentum $p_T^\text{lep}$ in laboratory frame,
as it can be observed in
Fig.~\ref{fig:NS:PTlepWW} (right). For the second and third decay channels in
Eq.~(\ref{ec:NS:TTtoWW}), denoted by $(h)$ and $(Z)$ respectively,
the tail of the distribution is less pronounced. This is so because the charged
lepton originates from $T \to W b \to \ell \nu b$ only half of the times, and
the rest comes from $t \to W b \to \ell \nu b$ and is less energetic.

\begin{figure}[htb]
\begin{center}
\begin{tabular}{cc}
\epsfig{file=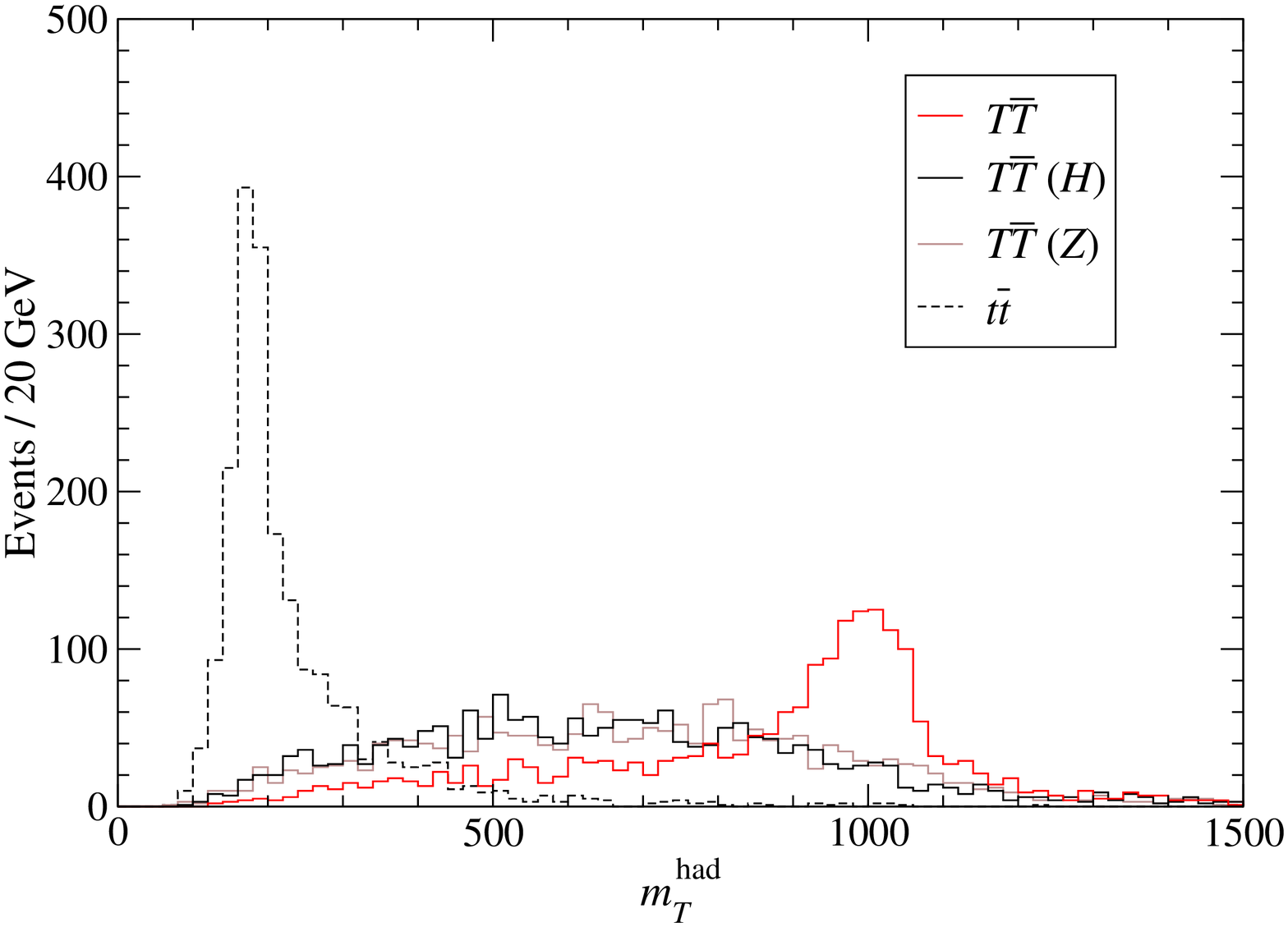,height=5.5cm,clip=} &
\epsfig{file=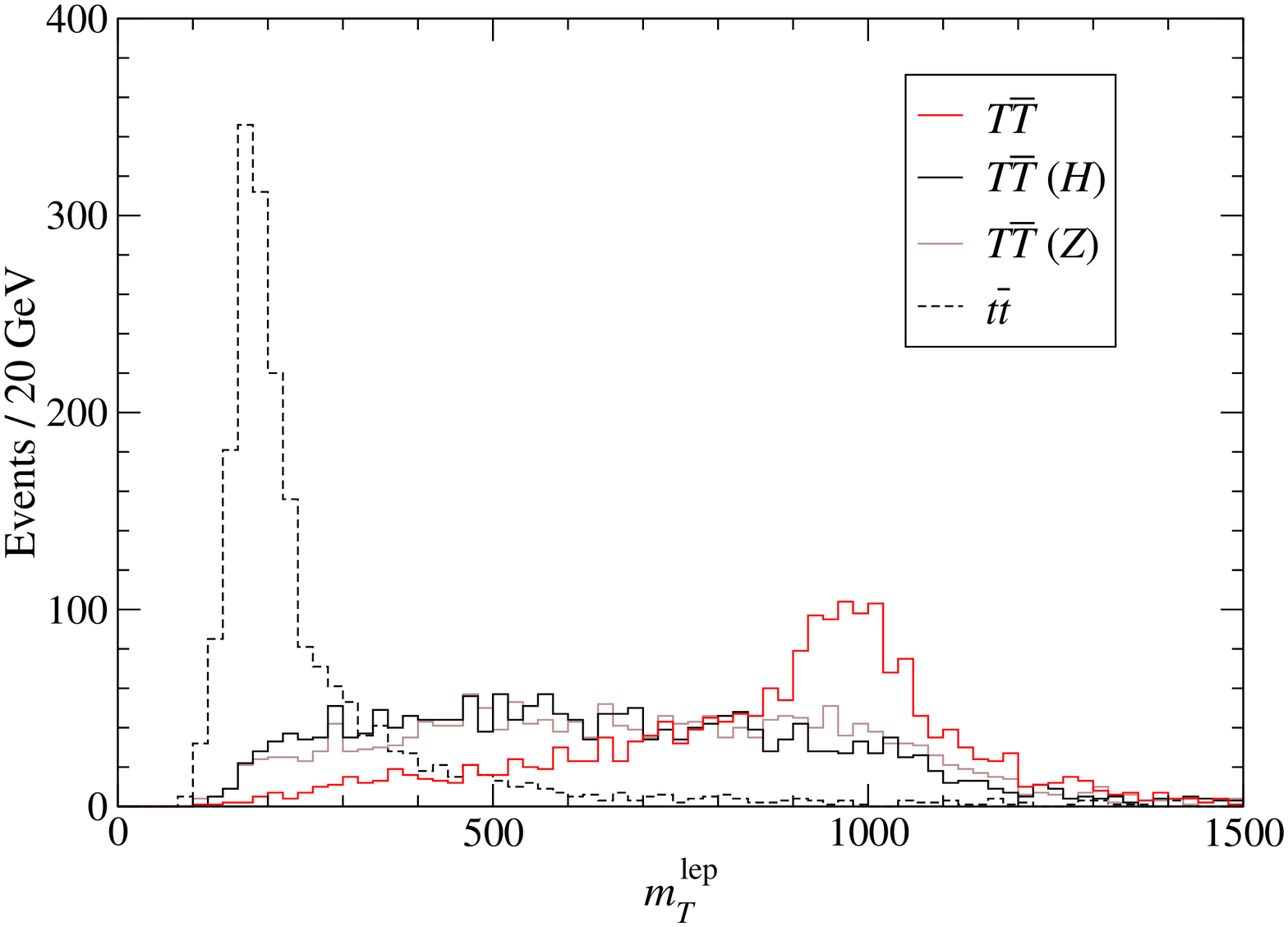,height=5.5cm,clip=} 
\end{tabular}
\caption{Reconstructed masses of the heavy quarks decaying hadronically (left)
and semileptonically (right), for the processes in Eqs.~(\ref{ec:NS:TTtoWW})
with $m_T = 1$ TeV, and their main background $t \bar t$.}
\label{fig:NS:mTrec1}
\end{center}
\end{figure}

Background is suppressed by requiring large
transverse momenta of the charged lepton and the jets, and with the heavy quark
mass reconstruction. The reconstructed masses of the heavy quarks decaying
hadronically ($m_T^\text{had}$) and semileptonically ($m_T^\text{lep}$)
are shown in Fig.~\ref{fig:NS:mTrec1}. For the leading decay mode
$T \bar T \to W^+ b W^- \bar b$ these distributions have a peak around the true
$m_T$ value, taken here as 1 TeV, but for the additional signal contributions
the events spread over a wide range. Thus, kinematical cuts on $p_T^\text{lep}$,
$m_T^\text{had}$, $m_T^\text{lep}$ considerably reduce the extra signal
contributions.

\begin{figure}[htb]
\begin{center}
\epsfig{file=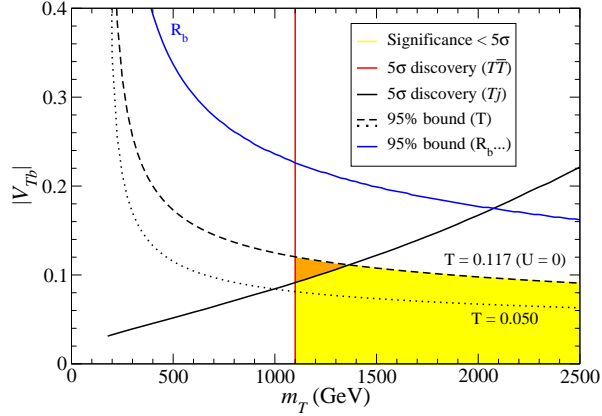,height=5.5cm,clip=}
\caption{Estimated $5\sigma$ discovery limits for a new charge $2/3$ quark
$T$ in $T \bar T$ and $Tj$ production.}
\label{fig:mTreach}
\end{center}
\end{figure}

The estimated $5\sigma$ discovery limits for 300 fb$^{-1}$ can be summarised in
Fig.~\ref{fig:mTreach}. They also include the results from $Tj$ (plus $\bar T j$)
production, where the decay $T \to W^+ b$ (or $\bar
T \to W^- \bar b$) also gives the highest sensitivity for large $T$ masses
\cite{Azuelos:2004dm}. The $m_T$ reach in $T \bar T$ production is independent
of $V_{Tb}$, but the $Tj$ cross section scales with
$|V_{Tb}|^2$, and thus the sensitivity of the latter process
depends on $V_{Tb}$. $T$ masses on the left of the vertical line
can be seen with $5\sigma$ in $T \bar T$ production. Values of $m_T$ and
$V_{Tb}$ over the solid curve can be seen in $Tj$ production. 
The latter discovery limits have been obtained by rescaling the results for
$m_T = 1$ TeV in Refs.~\cite{Azuelos:2004dm,Costanzo}. The 95\% CL bounds
from the $\mathrm{T}$ parameter (for $\mathrm{U}=0$ and $\mathrm{U}$ arbitrary)
ares represented by the dashed and dotted lines, respectively. Then, the yellow
area (light grey in print) represents the parameter region where the new quark
cannot be discovered with $5\sigma$ , and the orange triangle (dark gray) the
parameters for which it can be discovered in single but not in pair production.

Several remarks are in order regarding these results. The limits shown for
$T \bar T$ and $Tj$ only include the channel $T \to W^+ b$ (with additional
signal contributions giving the same final state in the former case). In both
analyses the evaluation of backgrounds, {\em e.g.} $t \bar t$,
does not include higher order processes with extra hard jet radiation:
$t \bar tj$, $t \bar t2j$, etc.
These higher order $t \bar t nj$ contributions may be important in
the large transverse momenta region where the new quark
signals are searched.
Systematic uncertainties in the background
are not included either, and they lower the significance with respect to the
values presented here. On the other hand, additional $T$ decay
channels can be included and the event selection could be refined, {\em e.g.} by
a probabilistic method, so that the
limits displayed in Fig.~\ref{fig:mTreach} are not expected to be significantly
degraded when all of these improvements are made in the analysis.

\subsubsection{Higgs discovery from $T$ decays}

Apart from the direct observation of the new quark, another exciting
possibility is to discover the Higgs boson from $T$ decays
\cite{delAguila:1989rq,Aguilar-Saavedra:2006gw}. Very recent
results from CMS have significantly lowered the expectations for the discovery 
of a light Higgs boson in $t \bar t h$ production, with $h \to b \bar b$. This
decrease is due to a more careful calculation of the $t \bar t n j$ background,
and to the
inclusion of systematic uncertainties \cite{CMSTDR}. As a result, a light Higgs
is impossible to see in this process, with a statistical significance of only
$\sim 0.47\sigma$ for 30 fb$^{-1}$~of luminosity and $M_h = 115$ GeV. But if a
new quark $T$ exists with a moderate mass, its pair production and decays
\begin{align}
& T \bar T \to W^+ b \, h \bar t / h t \, W^- \bar b 
   \to W^+ b \, W^- \bar b \, h \to \ell^+ \nu b \, \bar q q' \bar b 
   \; b \bar b / c \bar c \,, \nonumber \\
& T \bar T \to h t \, h \bar t \to W^+ b \, W^- \bar b \, h h 
  \to \ell^+ \nu b \, \bar q q' \bar b
 \;  b \bar b / c \bar c \;  b \bar b / c \bar c \,, \nonumber \\
& T \bar T \to Z t \, h \bar t / h t \, Z \bar t 
\to W^+ b \, W^- \bar b \, h Z \to \ell^+ \nu' b \, \bar q q' \bar b
 \;  b \bar b / c \bar c \; q'' \bar q'' / \nu \bar \nu
\label{ec:NS:TTtoH}
\end{align}
provide an additional source of Higgs bosons with a large cross section
(see Fig.~\ref{fig:mQ-sigma}) and a total branching ratio close to
$1/2$. The final state is the same as in $t \bar t h$ production with
semileptonic decay: one charged lepton, four or more $b$-tagged
jets and two non-tagged jets. The main backgrounds are $t \bar t n j$ production
with two $b$ mistags and $t \bar t b \bar b$ production.
The inclusion of higher order ($n > 2$) contributions is relevant
because of their increasing efficiency for larger $n$ (the probability to have
two $b$ mistags grows with the jet multiplicity). 
The larger transverse momenta involved for larger $n$ also make higher order
processes more difficult to suppress with respect to the $T \bar T$ signal.
Lower order contributions ($n<2$) are important as well, due to pile-up.
The method followed to evaluate top pair production plus jets is to calculate
$t \bar t nj$
for $n=0,\dots,5$ with {\tt Alpgen}\ \cite{Mangano:2002ea} and use 
{\tt PYTHIA}\ 6.4~\cite{Sjostrand:2006za} to include soft jet radiation, using
the MLM matching prescription \cite{MLM} to avoid double counting. 

\begin{figure}[htb]
\begin{center}
\begin{tabular}{cc}
\epsfig{file=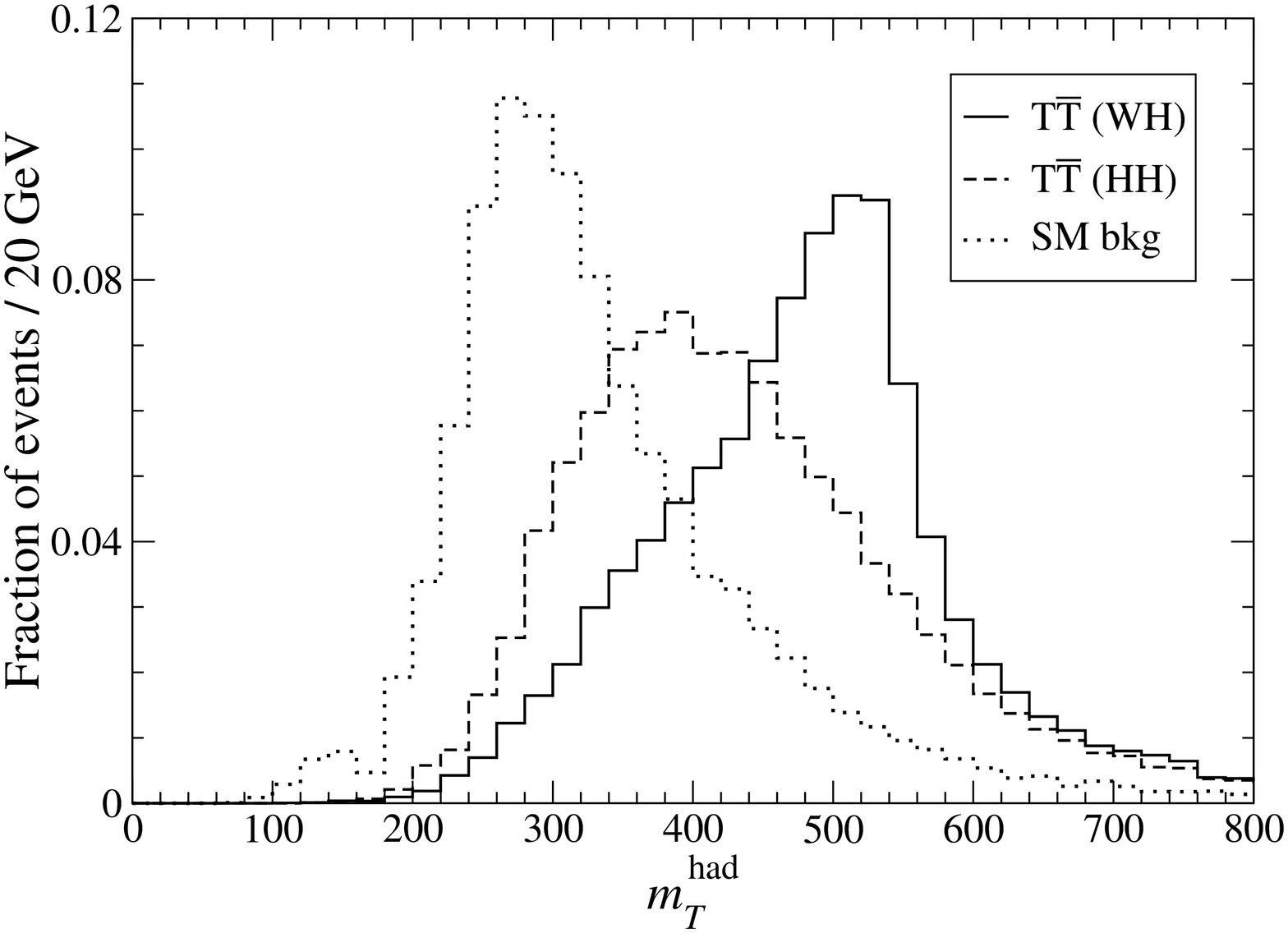,height=5.2cm,clip=} &
\epsfig{file=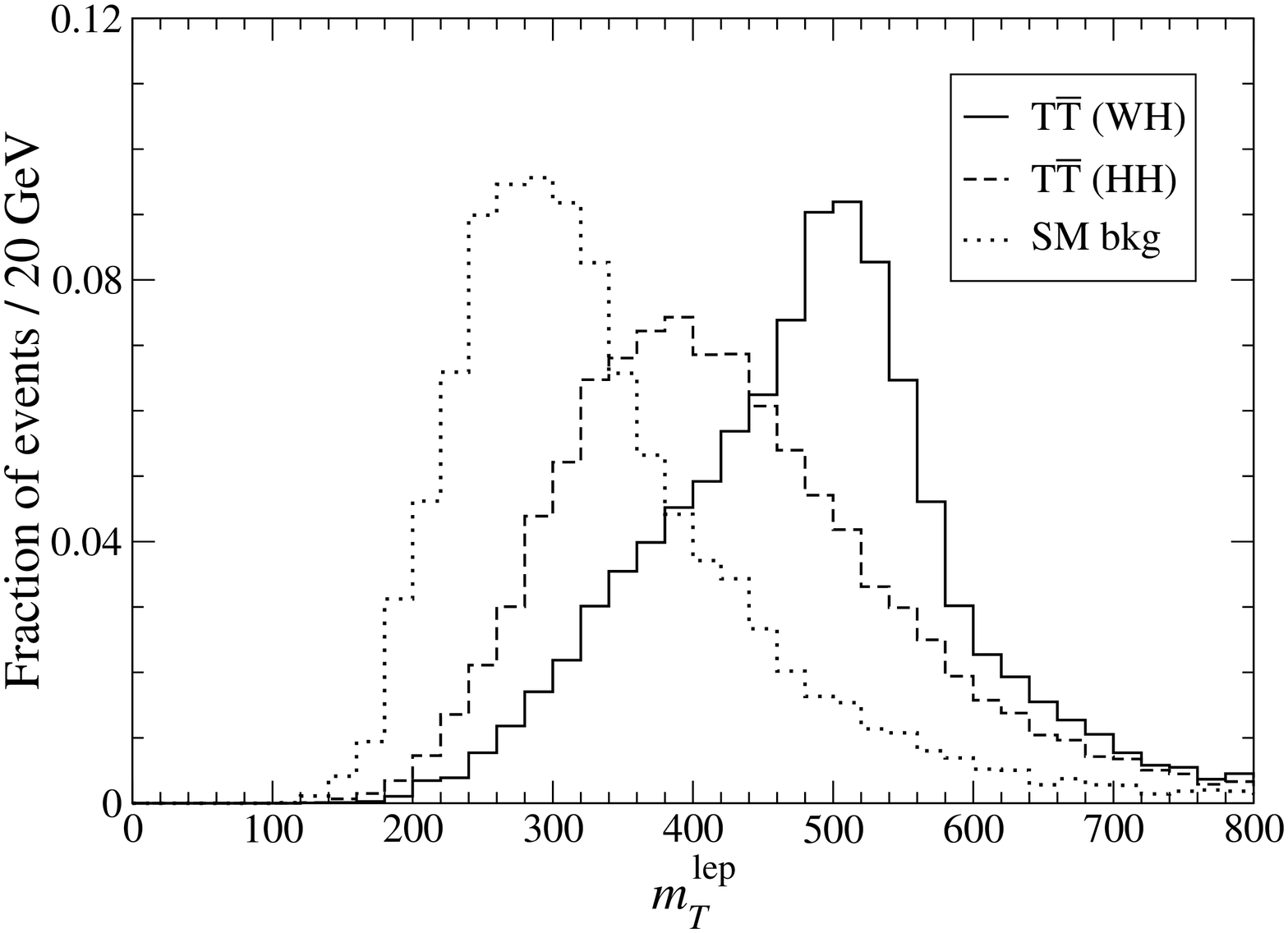,height=5.2cm,clip=} \\
\epsfig{file=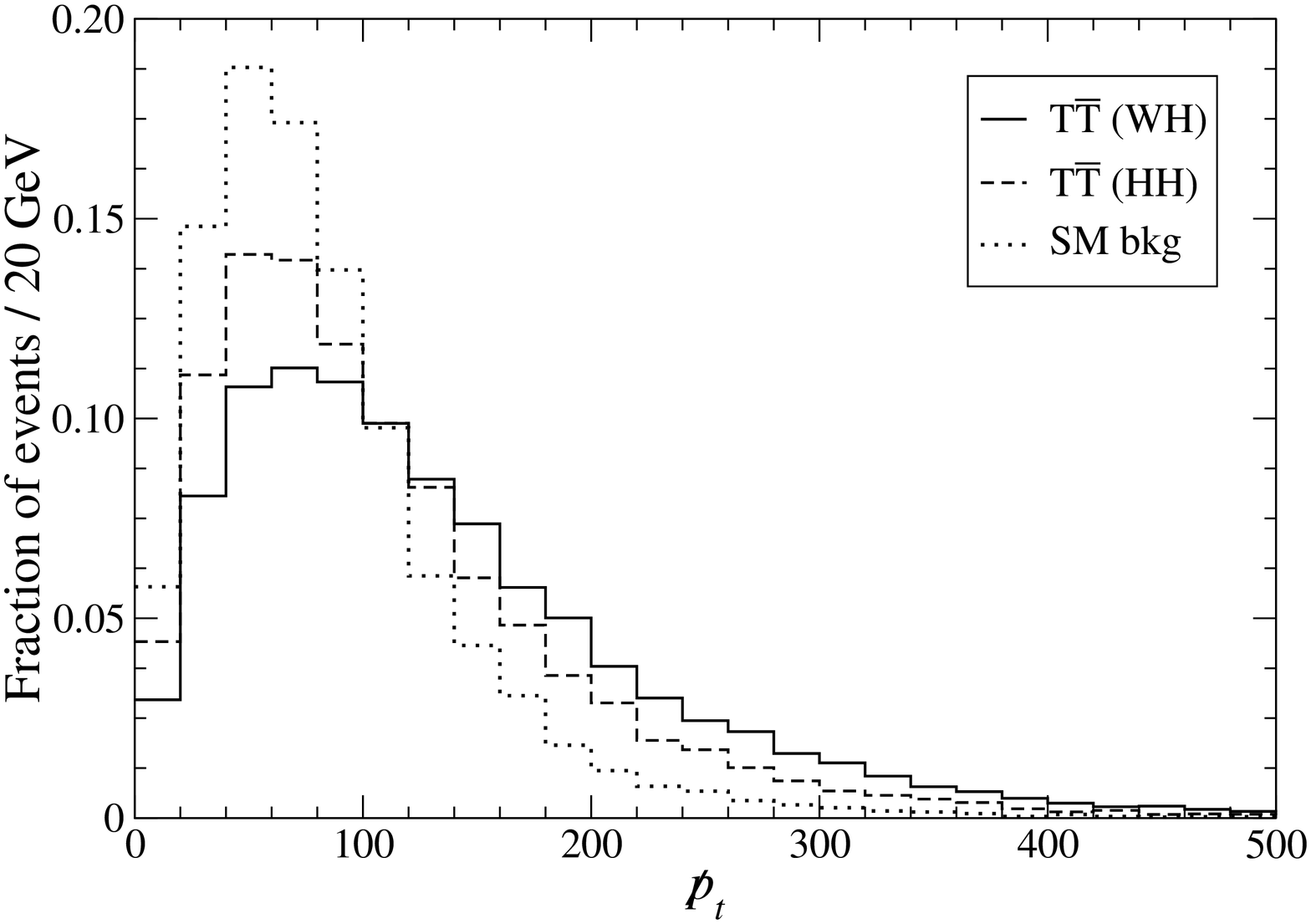,height=5.2cm,clip=} &
\epsfig{file=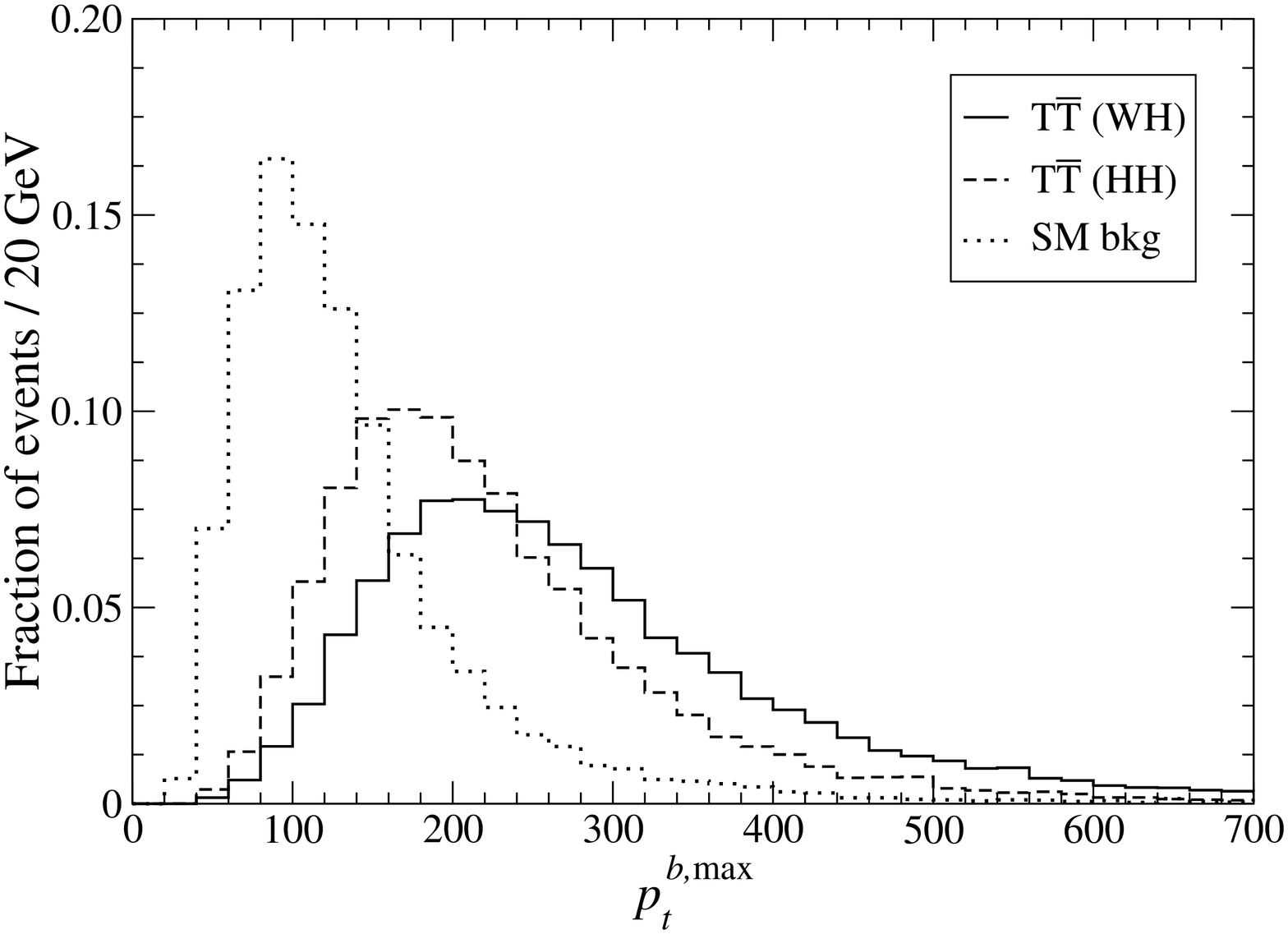,height=5.2cm,clip=} 
\end{tabular}
\caption{Several useful variables to discriminate between heavy quark signals
and background for $T \bar T$ production in 4$b$ final states:
heavy quark reconstructed masses ($m_T^\text{had}$, $m_T^\text{lep}$),
missing energy ($\ptmiss$), and maximum $p_t$ of the
$b$-tagged jets ($p_t^{b,\text{max}}$).
The main signal processes (first two
ones in Eqs.~(\ref{ec:NS:TTtoH}) are denoted by $Wh$, $hh$, respectively.}
\label{fig:NS:lik}
\end{center}
\end{figure}

Background suppression is challenging because the higher-order $t \bar t nj$
backgrounds are less affected by large transverse momentum requirements.
Moreover, the
signal charged  leptons are not so energetic, and cannot be used to discriminate
signal and background as efficiently as in the previous final state.
Background is suppressed with a likelihood method.
Signal and background likelihood functions $L_S$, $L_B$ can be built.
using as variables several 
transverse momenta and invariant masse, as those shown in
Fig.~\ref{fig:NS:lik},
as well as angles and rapidities of final state particles. 
(Additional details can be found in Ref.~\cite{Aguilar-Saavedra:2006gw}.)
Performing cuts on these and
other variables greatly improves the signal observability.
For a luminosity of 30 fb$^{-1}$, the statistical significances obtained for
the Higgs signals in final states with four, five
and six $b$ jets are \cite{Aguilar-Saavedra:2006gw}
\begin{align}
4\; b\, \hbox{jets}: & \quad 6.43 \sigma \,, \notag \\
5\; b\, \hbox{jets}: & \quad 6.02 \sigma \,, \notag \\
6\; b\, \hbox{jets}: & \quad 5.63 \sigma \,,
\end{align}
including a 20\% uncertainty in the background.
Additional backgrounds like electroweak $t \bar t b \bar b$ production,
$t \bar t c \bar c$ (QCD and electroweak) and $W/Z b \bar b$ plus jets are
smaller but have also been included. 
The combined significance is 
$10.45\sigma$, a factor of 25 larger than in $t \bar t h$ production alone.
Then, this process offers  a good opportunity to
quicky discover a light Higgs boson (approximately with 8 fb$^{-1}$) in final
states containing a charged lepton and four or
more $b$ quarks. These figures are conservative, since additional signal
processes $T \bar T n j$ have not been included in the signal evaluation.
The decay channels in Eqs.~(\ref{ec:NS:TTtoH}) also provide the best discovery
potential for $m_T$ relatively close to the electroweak scale. For $m_T = 500$ 
GeV, as assumed here, $5\sigma$ discovery of the new quark could be
possible with 7 fb$^{-1}$.

\subsection{Singlets: charge $-1/3$}
Down-type iso-singlet quark arise in the $E_{6}$ GUT models \cite{Hewett:1988xc}. 
These models postulate that the group structure of the SM, 
$\mathrm{SU}_{C}(3)\times \mathrm{SU}_{W}(2)\times \mathrm{U}_{Y}(1)$, originates from the breaking of 
the $\mathrm{E}_{6}$ GUT scale down to the electroweak scale, and thus 
extend each SM family by the addition of one isosinglet down type quark. 

Following the literature, the new quarks are denoted by letters $D$, $S$, and $B$.
The mixings between these and SM down type quarks is responsible for
the decays of the new quarks. In this study, the intrafamily mixings of the new quarks are assumed
to be dominant with respect to their inter-family mixings. In addition,
as for the SM hierarchy, the $D$ quark is taken to be the lightest
one. The usual CKM mixings, represented by superscript $\theta$,
are taken to be in the up sector for simplicity of calculation (which
does not affect the results). Therefore, the Lagrangian relevant for
the down type isosinglet quark, $D$, becomes a simplification of equation set
\ref{ec:NS:lagrQS}. It can explicitely be written as: 

\begin{eqnarray}
{\cal L_D} & = & \frac{\sqrt{4\pi\alpha_{em}}}{2\sqrt{2}\sin\theta_{W}}\left[\bar{u}^{\theta}
\gamma_{\alpha}\left(1-\gamma_{5}\right)d\cos\phi+\bar{u}^{\theta}\gamma_{\alpha}\left(1-\gamma_{5}
\right)D\sin\phi\right]W^{\alpha}\label{lagrangian}\\
 & - & \frac{\sqrt{4\pi\alpha_{em}}}{4\sin\theta_{W}}\left[\frac{\sin\phi\cos\phi}{\cos\theta_{W}}\bar{d}
 \gamma_{\alpha}\left(1-\gamma_{5}\right)D\right]Z^{\alpha}\nonumber \\
 & - & \frac{\sqrt{4\pi\alpha_{em}}}{12\cos\theta_{W}\sin\theta_{W}}\left[\bar{D}\gamma_{\alpha}\left(4\sin^
{2}\theta_{W}-3\sin^{2}\phi(1-\gamma_{5})\right)D\right]Z^{\alpha}\nonumber \\
 & - & \frac{\sqrt{4\pi\alpha_{em}}}{12\cos\theta_{W}\sin\theta_{W}}\left[\bar{d}\gamma_{\alpha}\left(4\sin^{2}\theta_{W}-3\cos^{2}
\phi(1-\gamma_{5})\right)d\right]Z^{\alpha}\;+\hbox{h.c.}\nonumber 
\end{eqnarray}

The measured values of $V_{ud}$,$V_{us}$,$V_{ub}$ constrain the
$d$ and $D$ mixing angle $\phi$ to $|\sin\phi|$ $\leq0.07$~
assuming the squared sum of row elements of the new $3\times4$ CKM
matrix equal unity (see\cite{Eidelman:2004wy} and references therein for CKM matrix
related measurements). The total decay width and the contribution
by neutral and charged currents were already estimated in \cite{Cakir:1997xxx}.
As reported in this work, the $D$ quark decays through a $W$ boson
with a branching ratio of 67\% and through a $Z$ boson with a branching
ratio of 33\%. If the Higgs
boson exists, in addition to these two modes, $D$ quark might also
decay via the $D\rightarrow h\, d$ channel which is available due
to $D-d$ mixing. The branching ratio of this channel for the case 
of $m_{h}=120$~GeV and $\sin\phi=0.05$ is calculated to be about
25\%, reducing the branching ratios of the neutral
and charged channels to 50\% and 25\%, respectively \cite{Andre:2003wc,Sultansoy:2006cw}.

\subsubsection{The discovery potential}

\begin{figure}[htb]
\noindent \begin{center}
\includegraphics[scale=0.415]{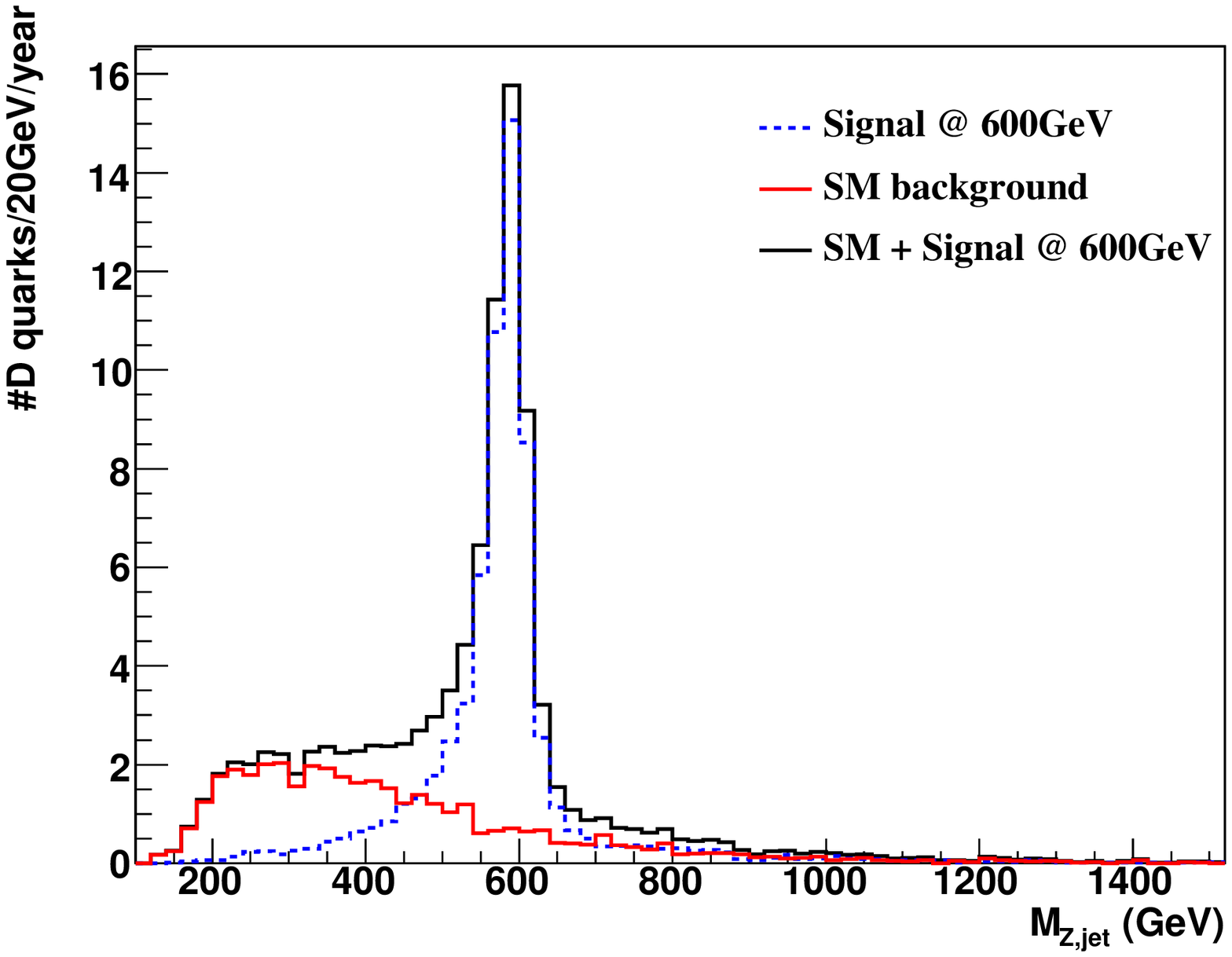}\includegraphics[scale=0.415]{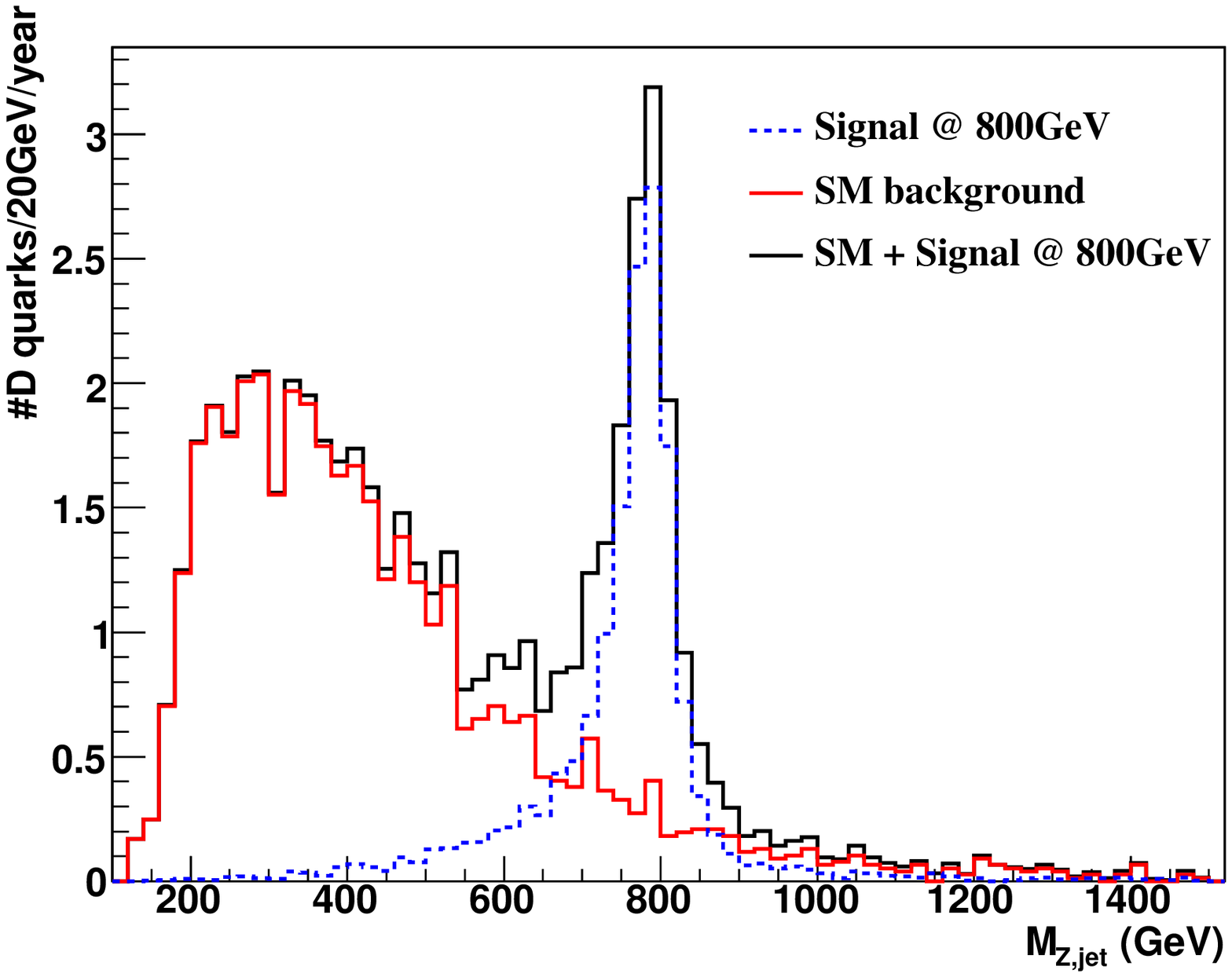}
\includegraphics[scale=0.415]{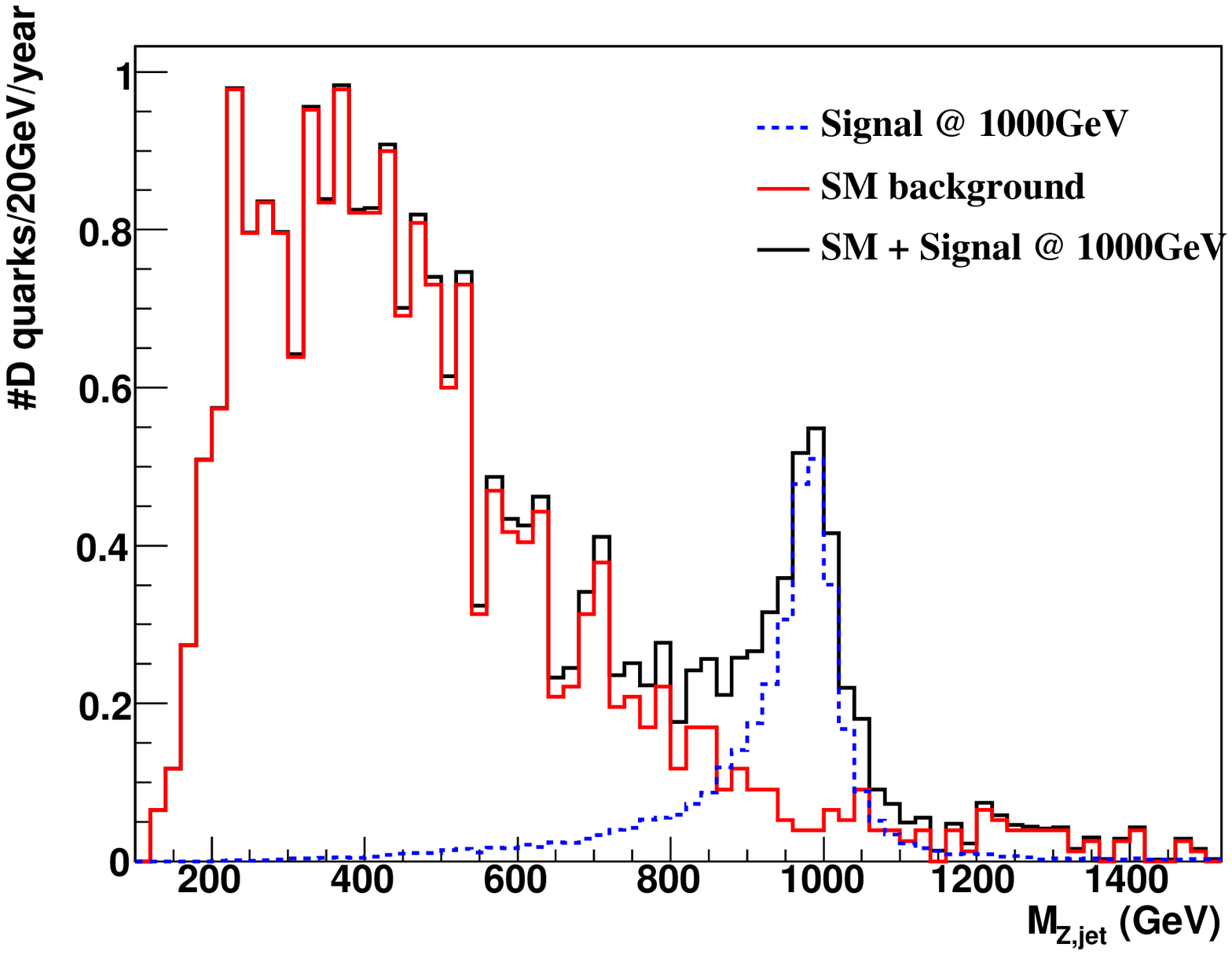}\includegraphics[scale=0.415]{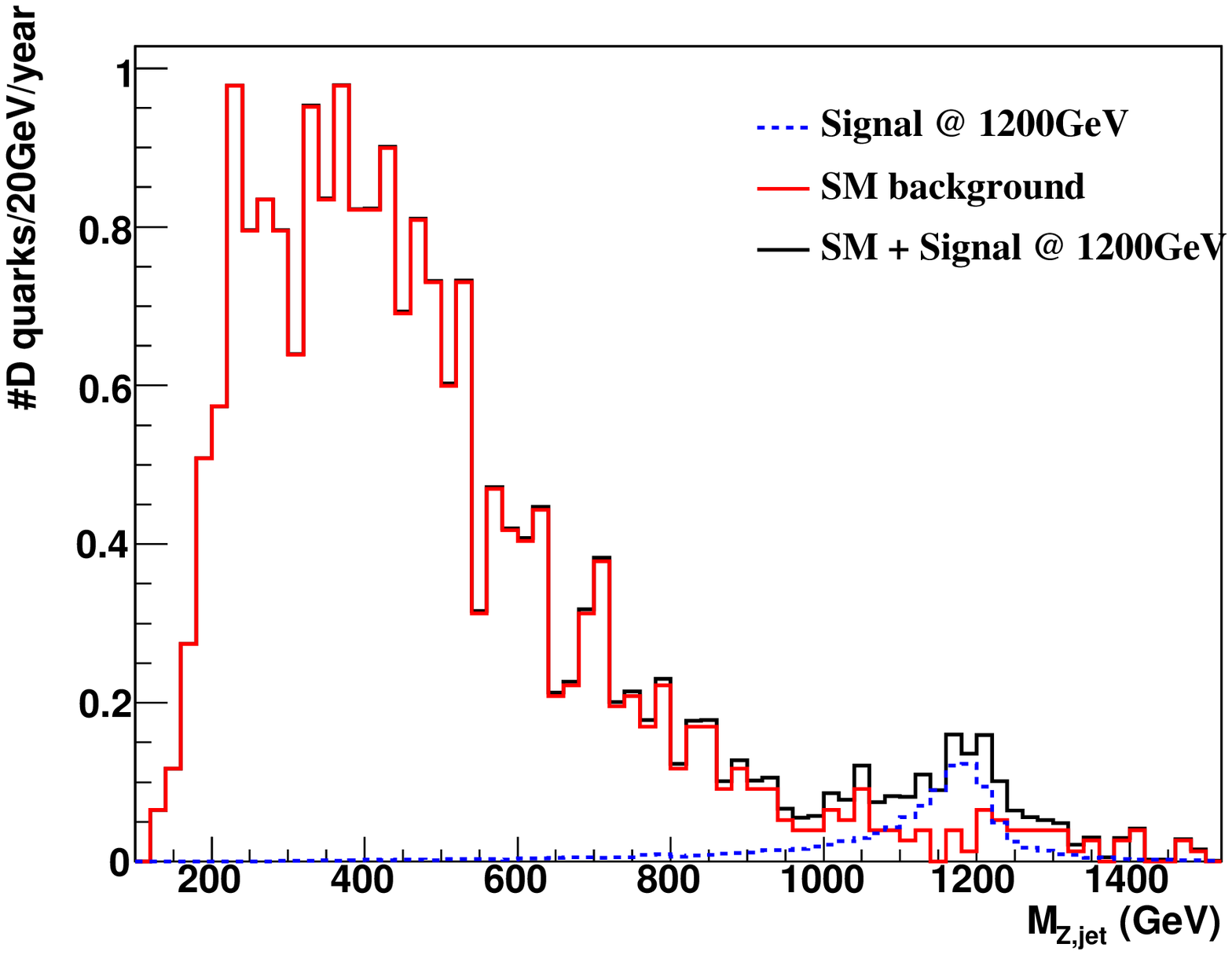}
\end{center}
\caption{Combined results for possible signal observation at $M_{D}=$ 600,
800, 1000, 1200~GeV . The reconstructed $D$ quark mass and the relevant
SM background are plotted for a luminosity of 100 fb$^{-1}$ which
corresponds to one year of nominal LHC operation. The dark line shows
the signal and background added, the dashed line is for signal only
and the light line shows the SM background.}
 \label{fig:Combined-results}
\end{figure}

The discovery potential of the lightest isosinglet quark has been investigated using the pair production channel
which is quasi-independent of the mixing angle $\phi$.
The main tree level Feynman diagrams for the pair production of $D$
quarks at LHC are gluon fusion, and $q-\overline{q}$ annihilation. The
$gD\overline{D}$ and $\gamma D\overline{D}$ vertices are the same
as their SM down quark counterparts. The modification to the $Zd\overline{d}$
vertex due to $d-D$ mixing can be neglected due to the small value
of $\sin\phi$.

The Lagrangian in Eq. (\ref{lagrangian}) was implemented
into tree level event generators, {\tt CompHEP}\ 4.3 \cite{Boos:2004kh}
and {\tt MadGraph}\ 2.3 \cite{Stelzer:1994ta}. 
The impact of uncertainties in parton distribution
functions (PDFs) \cite{Pumplin:2002vw}, is calculated by
using different PDF sets, to be less than 10\% for $D$ quark mass values
from 400 to 1400$\;$GeV. 
For example at $m_{D}=800$~GeV and $Q^{2}=m_{Z}^{2}$, the cross
section values are 450~fb ({\tt CompHEP}, CTEQ6L1) and 468~fb ({\tt CompHEP}, CTEQ5L)
versus 449~fb ({\tt MadGraph}, CTEQ6L1) and 459~fb ({\tt MadGraph}, CTEQ5L) with
an error of about one percent in each calculation.  The largest contribution to the total cross section comes
from the gluon fusion diagrams for $D$ quark masses below 1100~GeV,
while for higher $D$ quark masses, contributions from $s$-channel
$q\bar{q}$ annihilation subprocesses becomes dominant. For these computations,
$q\bar{q}$ are assumed to be only from the first quark family since,
the contribution to the total cross section from $s\bar{s}$ is about
10 times smaller and the contribution from $c\bar{c}$ and $b\bar{b}$
are about 100 times smaller. The $t$-channel diagrams mediated by
$Z$ and $W$ bosons, 
which are suppressed by the small value of $\sin\phi$ (for example
0.4 fb at $m_{D}=800$~GeV) were also included in the signal generation.
The isosinglet quarks being very heavy are expected to immediately
decay into SM particles. The cleanest signal can be obtained from both
$D$s decaying via a $Z$ boson. Although it has the smallest branching ratio, 
the 4 lepton and 2 jet final state offers the possibility
of reconstructing the invariant mass of $Z$ bosons and thus of 
both $D$ quarks.  The high transverse momentum of the jets coming from
the $D$ quark decays can be used to distinguish the signal events
from the background. 

The $D$ quarks in signal events were made to decay in {\tt CompHEP}\ into SM particles. 
The final state particles for both signal and background events were fed into
{\tt PYTHIA}\ version
6.218 \cite{Sjostrand:2000wi} for initial and final state radiation, as well
as hadronisation using the {\tt CompHEP}\ to {\tt PYTHIA}\ and {\tt MadGraph}\ to {\tt PYTHIA}\
interfaces
provided by {\tt ATHENA}\ 9.0.3 (the ATLAS offline software framework).
To incorporate the detector effects, all event samples were processed
through the ATLAS fast simulation tool, {\tt ATLFAST}\ \cite{Froidevaux:682460},
and the final analysis has been done using physics objects that it
produced. The cases of 4 muons, 4 electrons and 2 electrons plus 2 muons were 
separately treated to get the best reconstruction efficiency.
As an example, table \ref{tab:emu-cut-efficiencies} 
gives the selection efficiencies for the mixed lepton case at $m_D=800$~GeV.

\begin{table}[!h]
\caption{The individual selection cut efficiencies $\epsilon$ for one $Z\rightarrow ee$
and one $Z\rightarrow\mu\mu$ sub-case. The subscript $\ell$ represents
both electron and muon cases.}
 \label{tab:emu-cut-efficiencies}
\begin{center}\begin{tabular}{c|cccccc}
channel& $N_{\ell}$& $M_{Z}$& $P_{T,\ell}$& $N_{jet}$& $P_{T,jet}$& $\epsilon_{combined}$\tabularnewline
\hline
cut& =4 & =90$\pm$20~GeV & $\mu(e)>40(15)$~GeV& $\geq$2 & $\geq$100~GeV& \tabularnewline 
$\epsilon$ Signal& 0.44& 0.94& 0.71& 1& 0.93& 0.28\tabularnewline 
$\epsilon$ Background& 0.35& 0.97& 0.34& 0.95& 0.10& 0.011\tabularnewline
\end{tabular}\end{center}
\end{table}

\begin{figure}[htb]
\begin{center}\includegraphics[%
  bb=0bp 0bp 520bp 520bp,
  scale=0.43]{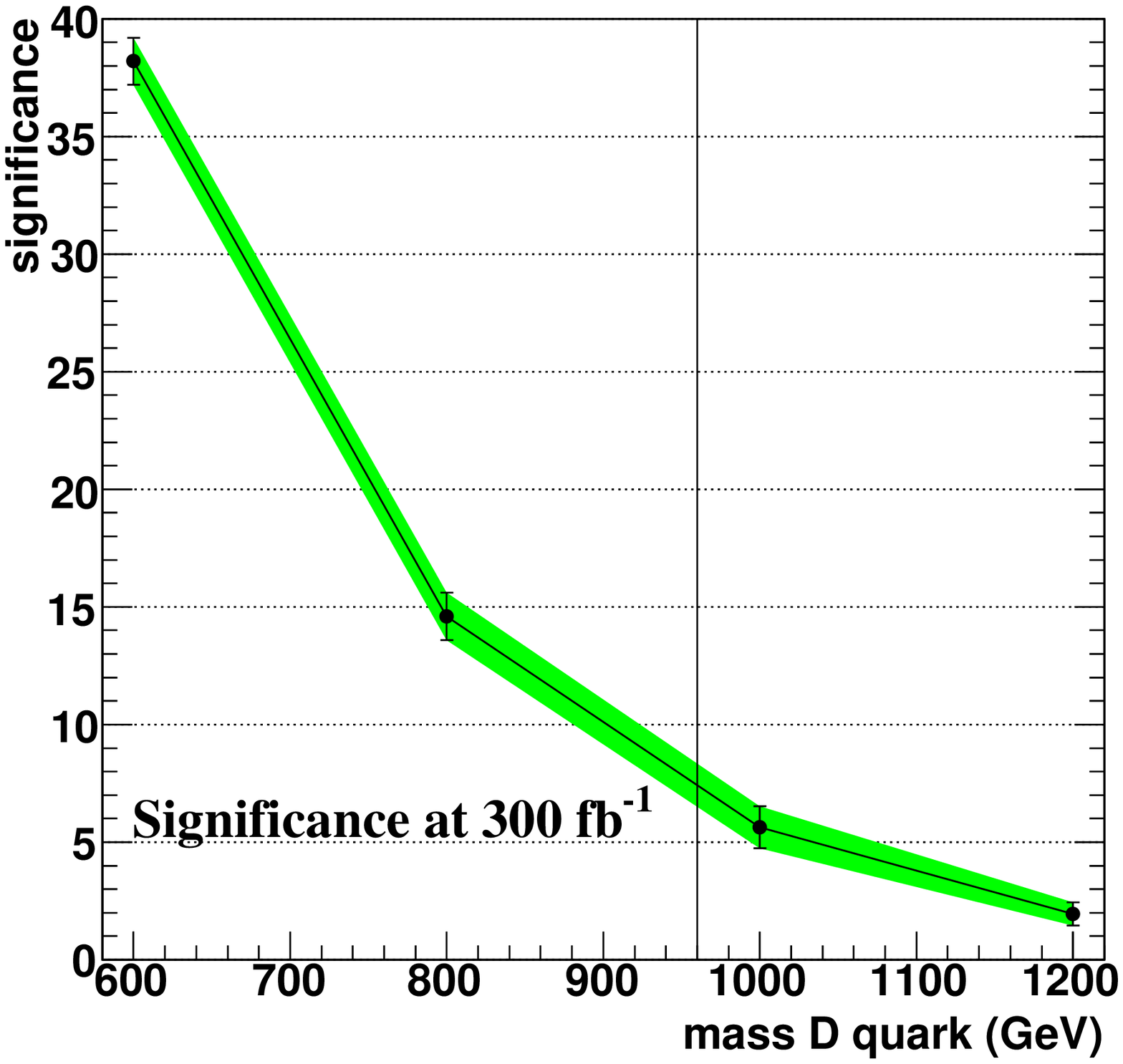}\includegraphics[%
  scale=0.40]{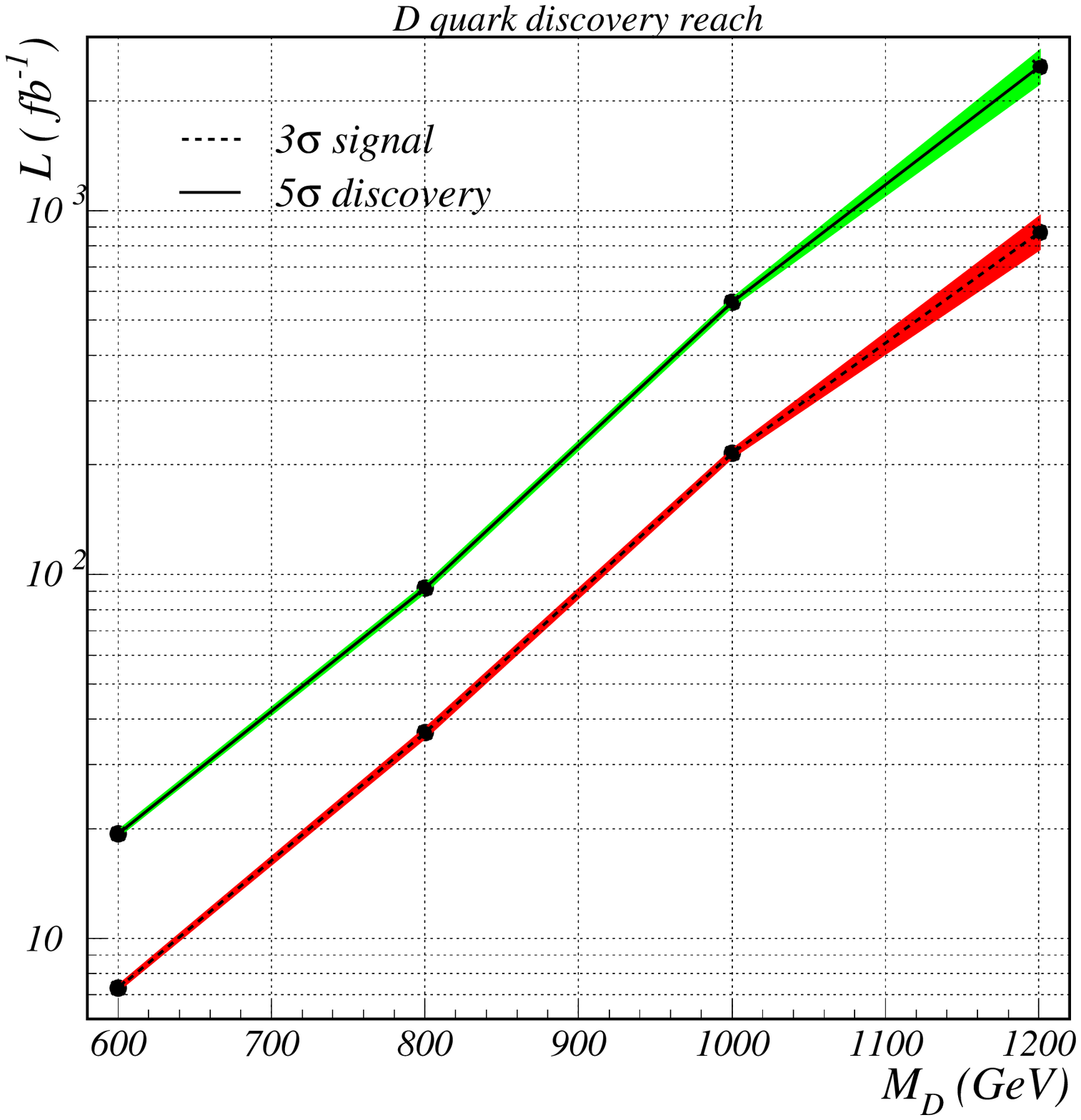}\end{center}
\caption{On the left: the expected statistical significance after 3 years
of running at nominal LHC luminosity assuming Gaussian statistics.
The vertical line shows the limit at which the event yield drops below
10 events. On the right: the integrated luminosities for 3 $\sigma$ observation
and 5 sigma discovery cases as a function of $D$ quark mass. The
bands represent uncertainties originating from finite
MC sample size.}
 \label{cap:Statistical-significance-NC}
\end{figure}

Using the convention of defining a running accelerator year as $1\times10^{7}$seconds,
one LHC year at the design luminosity corresponds to 100~fb$^{-1}$.
For one such year worth of data, all the signal events are summed
and compared to all SM background events as shown in Fig.~\ref{fig:Combined-results}.
It is evident that for the lowest of the considered masses, the studied
channel gives an easy detection possibility, whereas for the highest
mass case ($M_{D}$=1200~GeV) the signal to background ratio is of
the order of unity. For each $D$ quark mass value that was considered, a Gaussian is
fitted to the invariant mass distribution around the $D$ signal peak
and a polynomial to the background invariant mass distribution. The
number of accepted signal ($S$) and background ($B$) events are
integrated using the fitted functions in a mass window whose width
is equal to $2\sigma$ around the central value of the fitted Gaussian.
The significance is then calculated at each mass value as $S/\sqrt{B}$,
using the number of integrated events in the respective mass windows.
The expected signal significance for three years of nominal LHC luminosity
running is shown in Fig. \ref{cap:Statistical-significance-NC} left
hand side. The shaded band in the same plot represents the systematic
errors originating from the fact that for each signal mass value,
a finite number of Monte Carlo events was generated at the start of
the analysis and the surviving events were selected from this event
pool. For $M_{D}=600$~GeV, ATLAS could observe the $D$ quark with a significance more than 3
$\sigma$ before the end of the first year of low luminosity running (10~fb$^{-1}$/year)
whereas to claim discovery with 5 $\sigma$ significance, it would need
about 20~fb$^{-1}$ integrated luminosity. For $M_{D}=1000$~GeV,
about 200~fb$^{-1}$ integrated luminosity is necessary for a 3 $\sigma$
signal observation claim.

\subsubsection{The mixing angle to SM quarks} 
\begin{figure}[htb]
\begin{center}
\begin{centering}\includegraphics[scale=0.4]{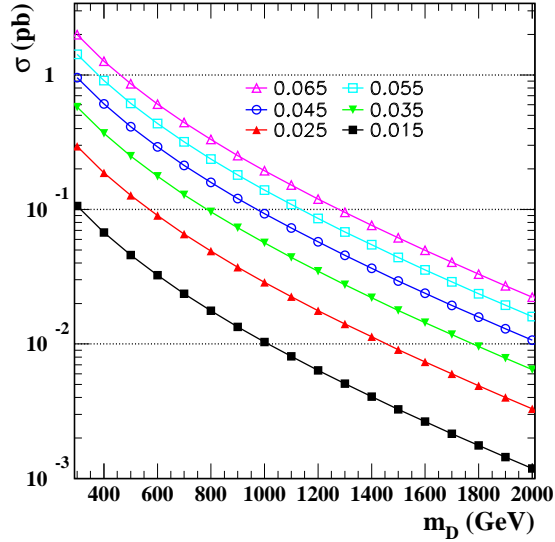}
\end{centering}
\caption{Cross section in single D production as a function of D quark mass
for different $\sin\phi$ values.}
\label{fig:single-Cross-sections}
\end{center}
\end{figure}

This section addresses the discovery of the isosinglet quarks via their
jet associated single production at the LHC and the measurements of the mixing angle
between the new and the SM quarks. The current upper limit on $\phi$
is $|\sin\phi|<0.07$, allowed by the known errors on the 
CKM matrix elements assuming unitarity of its extended version\cite{Mehdiyev:2006tz}.   
However, in this work, a smaller thus a more conservative value, $\sin\phi=0.045$, 
was considered for the calculation of the cross sections and decay widths. 
For other values of $\sin\phi$, both of these two quantities can be scaled with a $\sin^{2}\phi$ dependence.
For both the signal and the background studies, the contributions from sea quarks were
also considered. The used parton distribution function was CTEQ6L1
and the QCD scale was set to be the mass of the $D$ quark for both
signal and background processes. The cross section for single production
of the $D$ quark for its mass up to 2 TeV and for various mixing
angles is given in  Fig.~\ref{fig:single-Cross-sections}. 
The main tree level signal processes are originating from the valance quarks
exchanging $W$ or $Z$ bosons via the $t$ channel. The remaining processes originating
from the sea quarks contribute about 20 percent to the total signal
cross section. 

Although the work in this section is at the generator level, various parameters
of the ATLAS detector \cite{Armstrong:1994it} such as the barrel calorimeter
geometrical acceptance, minimum angular distance for jet separation
and minimum transverse momentum for jets\cite{Jenni:616089} were taken
into account. Five mass values (400, 800,1200, 1500 and 2000~GeV)
were studied to investigate the mass dependence of the discovery potential
for this channel. The cuts common to all considered mass values are:

\begin{eqnarray*}
P_{Tp} & > & 15\,\hbox{GeV}\\
|\eta_{p}| & < & 3.2\\
|\eta_{Z}| & < & 3.2\\
R_{p} & > & 0.4\\
M_{Zp} & = & M_{D}\pm20\,\hbox{GeV}
\end{eqnarray*}
where $p$ stands for any parton; $R$ is the cone separation angle
between two partons; $\eta_{p}$ and $\eta_{Z}$ are pseudorapidities
of a parton and $Z$ boson respectively; and $P_{Tp}$ is the parton
transverse momentum.   For each mass case, the optimal cut value is found
by maximizing the significance ($S/\sqrt{B}$) and it is used for
calculating the effective cross sections presented in Table \ref{tab:Expected-events}.
To obtain the actual number of events for each mass value, the $e^+\,e^-$
and $\mu^+\,\mu^-$ decays of the $Z$ boson were considered for simplicity
of reconstruction. The last 3 rows of the same table contain the expected
number of reconstructed events for both signal and background for
100 fb$^{-1}$ of data taking. Although the lepton identification and
reconstruction efficiencies are not considered, one can note that
the statistical significance at $m_{D}=$1500~GeV, is above 5$\sigma$
after one year of nominal luminosity run.

\begin{figure}[htb]
\begin{centering}\includegraphics[scale=0.45]{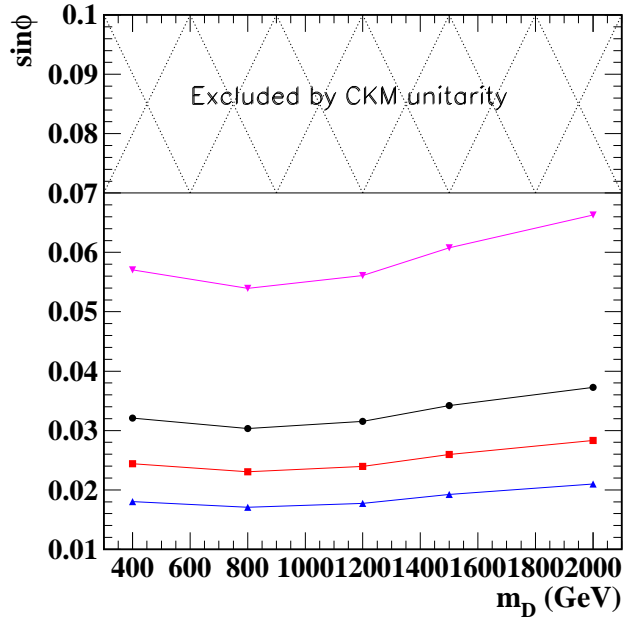}\par\end{centering}
\caption{3$\sigma$ exclusion curves for 10, 100, 300, 1000 fb $^{-1}$ integrated
luminosities are shown from top to down.}
 \label{cap:Determination-of-sinphi}
\end{figure}

\begin{table}
\caption{The signal and background effective cross sections before the $Z$
decay and after the optimal cuts, obtained by maximixing the $S/\sqrt{B}$, 
together with the $D$ quark width in GeV for each considered mass. The number
of signal and background events also the signal siginificance were calculated
for an integrated luminosity of 100 fb$^{-1}$. }
\begin{centering}\begin{tabular}{c|ccccc}
$M_{D}$(GeV)& 400& 800& 1200& 1500& 2000\tabularnewline
\hline
$\Gamma$(GeV)& 0.064& 0.51& 1.73& 3.40& 8.03\tabularnewline
Signal (fb)& 100.3& 29.86& 10.08& 5.09& 1.92\tabularnewline
Background (fb)& 2020& 144& 18.88& 6.68& 1.36\tabularnewline
optimal $p_{T}$ cut& 100& 250& 450& 550& 750\tabularnewline
Signal Events& 702& 209& 71& 36& 13.5\tabularnewline
Background Events& 14000& 1008& 132& 47& 9.5\tabularnewline
Signal significance ($\sigma$)& 5.9& 6.6& 6.1& 5.2& 4.37\tabularnewline
\end{tabular}\label{tab:Expected-events}\par\end{centering}
\end{table}

The single production discovery results given in Table \ref{tab:Expected-events}
can be used to investigate the mixing angle. 
In the event of a discovery in the single production case,
the mixing angle can be obtained directly. If no discoveries are made,
then the limit on the cross section can be converted to a limit curve
in the $D$ quark mass vs mixing angle plane. Therefore 
the angular reach for a 3$\sigma$ signal is calculated by extrapolating to other
$\sin\phi$ values. Figure~\ref{cap:Determination-of-sinphi} gives
the mixing angle versus $D$-quark mass plane and the 3$\sigma$ reach
curves for different integrated luminosities ranging from 10 fb$^{-1}$
to 1000 fb$^{-1}$, which correspond to one year of low luminosity
LHC operation and one year of high luminosity super-LHC operation
respectively. The hashed region in the same plot is excluded using
the current values of the CKM matrix elements. One should note that,
this channel allows reducing the current limit on $\sin\phi$ by half
in about 100 fb$^{-1}$ run time. The process of single production of the $\mathrm{E}_{6}$ isosinglet quarks
could essentially enhance the discovery potential if $\sin\phi$ exceeds
0.02. For example, with 300 fb$^{-1}$ integrated luminosity, the
3$\sigma$ discovery limit is $m_{D}=2000\,$ GeV, if $\sin\phi=0.03$.
It should also be noted that for pair production the 3$\sigma$ discovery
limit was found to be about 900~GeV, independent of $\sin\phi$. If
ATLAS discovers an 800~GeV $D$ quark via pair production, single
production will give the opportunity to confirm the discovery and
measure the mixing angle if $\sin\phi>0.03.$ The FCNC decay channel
analysed in this paper is specific for isosinglet down type quarks
and gives the opportunity to distinguish it from other models also
involving additional down type quarks, for example the fourth SM family.

\subsubsection{The impact on the Higgs searches}
The origin of the masses of SM particles is explained by using the
Higgs Mechanism. The Higgs mechanism can also be preserved in $\mathrm{E}_{6}$
group structure as an effective theory, although other alternatives
such as dynamical symmetry breaking are also proposed \cite{Hosotani:1983xw,McInnes:1989ft}.
On the other hand, the origin of the mass of the new quarks ($D,\, S,\, B$)
should be due to another mechanism since these are isosinglets. 
However, the mixing between $d$ and $D$ quarks will lead to decays of the latter involving
$h$ after spontaneous symmetry breaking (SSB). To find these decay
channels, the interaction between the Higgs field and both down type
quarks of the first family should be considered before SSB. 
After SSB, the Lagrangian for the interaction between $d,\; D$ quarks, and the Higgs boson
becomes :
\begin{eqnarray}
{\cal L}_{h}  & = & \frac{m_{D}}{\nu}\sin^{2}\phi_{L}\bar{D}Dh\label{eq:DDh}\\
 & - & \frac{{\sin{\phi_{L}}\cos{\phi_{L}}}}{2\nu}\bar{{D}}\left[(1-\gamma^{5})\, m_{D}+(1+\gamma^{5})\,
 m_{d}\right]\, d\, h\nonumber \\
 & - & \frac{{\sin{\phi_{L}}\cos{\phi_{L}}}}{2\nu}\bar{{d}}\left[(1+\gamma^{5})\, m_{D}+(1-\gamma^{5})\,
 m_{d}\right]\, D\, h\nonumber \\
 & + & \frac{m_{d}}{\nu}\cos^{2}\phi_{L}\bar{d}dh\nonumber 
\end{eqnarray}
where $\nu=\eta/\sqrt{2}$ and $\eta=246$ GeV is the vacuum expectation
value of the Higgs field. It is seen that the
$D$~quark has a narrow width and becomes even narrower with decreasing
values of $\phi$ since it scales through a $\sin^{2}\phi$ dependence. 
The relative branching ratios for the decay of the $D$~quark depend
on both the $D$~quark and the Higgs mass values. For example, at
the values of $D$~quark mass around 200~GeV and the Higgs mass
around 120$\;$GeV: BR($D\rightarrow Wu$)$\sim$60\%, BR($D\rightarrow hd$)$\sim$12\%,
BR($D\rightarrow Zd$)$\sim$28\%, whereas as the $D$~quark mass
increases the same ratios asymptotically reach 50\%, 25\% and 25\%
respectively. As the Higgs mass increases from 120~GeV, these limit
values are reached at higher $D$~quark masses. 

\begin{figure}[htb]
\begin{centering}\includegraphics[scale=0.4]{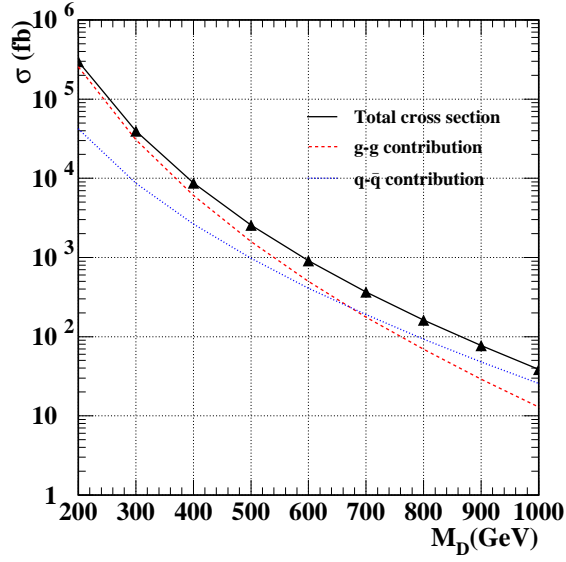}\par\end{centering}
\caption{Pair production of $D$ quarks at LHC computed at tree level with CTEQ6L1 
and QCD scale set at the mass of the $D$ quark.}
\label{fig:D-Pair-prod}
\end{figure}

Depending on the masses of the $D$ quark and the Higgs boson itself,
the $\mathrm{E}_{6}$ model could  boost the overall Higgs production at
the LHC. This boost is particularly interesting for the Higgs hunt,
one of the main goals of the LHC experiments. For example, if the
$D$~quark mass is as low as 250~GeV, the pair production cross
section at the LHC becomes as high as $10^{5}$ fb$^{-1}$, which
is enough to compensate for the relatively small Higgs branching ratio
of 17\%, as can be seen in Fig.~\ref{fig:D-Pair-prod}. In the low
mass range considered in this section (from 115 up to 135~GeV), the
branching ratio $h\rightarrow b\overline{b}$ is about 70\% \cite{Armstrong:1994it}.
Table \ref{tab:expected-final-states} lists the decays involving
at least one Higgs boson and the expected final state particles associated
with each case. Although the case involving the $Z$ is more suitable
from the event reconstruction point of view, the focus will be on
the last row, which has the highest number of expected Higgs events
per year.

\begin{table}
\caption{For pair production of D quarks, the decay channels involving the
Higgs particle. The branching ratios and the number of expected Higgs
particles are calculated assuming $m_{h}$=120~GeV and $m_{D}$=250
(500) GeV.}
 \label{tab:expected-final-states}
\begin{center}
\begin{tabular}{cc|ccc}
$D_{1}$ & $D_{2}$ & BR & \#expected Higgs/100fb$^{-1}$ & expected final state\tabularnewline
\hline
$D\rightarrow h\, j$ & $D\rightarrow h\, j$ & 0.029 (0.053) & 0.58$\times10^{6}$ (2.65$\times10^{4}$ ) & $2j\;4j_{b}$\tabularnewline
\hline
$D\rightarrow h\, j$ & $D\rightarrow Z\, j$ & 0.092 (0.120)  & 0.92$\times10^{6}$ (3.01$\times10^{4}$) & $2j\;2j_{b}\;2 \ell $\tabularnewline
\hline
$D\rightarrow h\, j$ & $D\rightarrow W\, j$ & 0.190 (0.235) & 1.9$\times10^{6}$ (6.04$\times10^{4}$) & $2j\;2j_{b}\; \ell \; E_{T,miss}$\tabularnewline
\end{tabular}

\end{center}
\end{table}

The full Lagrangian also involving the Higgs interaction
has been implemented in a tree level event generator, {\tt CompHEP}\ 4.4.3 \cite{Boos:2004kh},
to investigate the possibility of detecting the Higgs particle and
reconstructing it from $b$-jets. Assuming a light Higgs boson of mass 120~GeV, 
four mass values for the $D$~quark have been taken as examples: 250~GeV, 500~GeV,
750~GeV, and 1000~GeV.  10$\,$000 signal events were produced for each mass value under consideration
with the $W\, h\, j\, j$ final states using the CTEQ6L1 PDF set \cite{Pumplin:2002vw}.
The generator level cuts on the partons, guided by the performance of the ATLAS detector, are listed as:

\begin{eqnarray*}
|\eta_{p}| & \leq & 3.2\quad,\\
p_{T\,p} & \geq & 15\, GeV\quad,\\
R_{p} & > & 0.4
\end{eqnarray*}

where $\eta_{p}$ is the pseudo-rapidity for the partons giving rise
to jets; $p_{T\,, p}$ is the transverse momentum of the partons; and
$R_{p}$ is the angular separation between the partons. The imposed
maximum value of $\eta$ requires the jets to be in the central region
of the calorimeter where the jet energy resolution is optimal. The
imposed lower value of $p_{T}$ ensures that no jets that would eventually
go undetected along the beam pipe are generated at all. The imposed
lower value of $R$ provides good separation between the two jets
in the final state. Using the interface provided by {\tt CPYTH}\ 2.3 \cite{Belyaev:2000wn},
the generated particles are processed with {\tt ATHENA}\
11.0.41, which uses {\tt PYTHIA}\ \cite{Sjostrand:2000wi} for
hadronization and {\tt ATLFAST}\ \cite{Froidevaux:682460} for fast detector
response simulation.  However,
one should note that the reconstructed $b$-jet energy and momenta
were re-calibrated like in \cite{Armstrong:1994it} to have a good match
between the mean value of the reconstructed Higgs mass and its parton
level value.

As for the background estimations, all the SM interactions giving the 
$W^{\pm} b\, b\, j\, j$
final state have been computed in another tree level generator, {\tt MadGraph}\
2.1. \cite{Stelzer:1994ta}, using the same parton level cuts and parton distribution functions.
The SM background cross section is calculated to be 520 $\pm$11~pb.
The reasons for using two separate event generators, their compatibility,
and their relative merits have been discussed elsewhere \cite{Mehdiyev:2006tz}.
The generated 40$\,$000 background events were also processed in
the same way using {\tt ATLFAST}\ for hadronization and calculation of detector effects.

\begin{table}
\begin{center}
\caption{Optimised event selection cuts and their efficiencies for $m_{D}=500$
GeV.}
\label{tab:Event-selection-cuts}
\begin{tabular}{cc|cc}
& cut & $\epsilon$ signal & $\epsilon$ background \tabularnewline \hline 
N-leptons&           =1 &                0.83 & 0.79 \tabularnewline 
N-jets&            $\geq4$&              0.99 & 0.99 \tabularnewline 
N-bjets&            $\geq2$&             0.33 & 0.36 \tabularnewline 
$P_{T}-bjet$&    $\geq1$ GeV&            1.00 & 1.00 \tabularnewline 
$P_{T}-$lepton&   $\geq15$ GeV&          0.95 & 0.94 \tabularnewline 
$P_{T}-jet$&       $\geq100$GeV&         0.83 & 0.69 \tabularnewline 
$\cos\theta_{bj\, bj}$&   $\geq$-0.8&    0.97 & 0.89 \tabularnewline 
$M_{j\, j}$&          $\geq90$GeV&       0.99 & 0.65 \tabularnewline 
$H_{T}$&             $\geq800$ GeV&      0.90 & 0.55 \tabularnewline 
$|m_{D1}-m_{D2}|$&   $\leq100$GeV &      0.59 & 0.37 \tabularnewline 
\end{tabular}
\end{center}
\end{table}

\begin{figure}[htb]
\begin{center}
\includegraphics[scale=0.6]{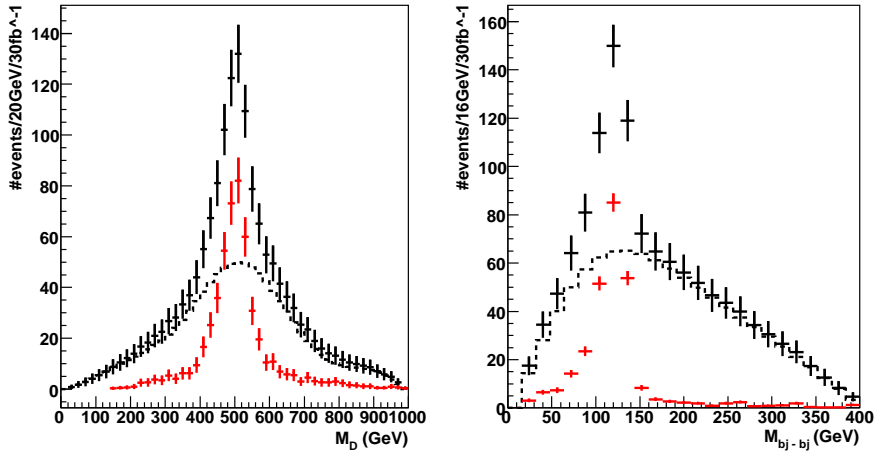}
\end{center}
\caption{Reconstructed invariant masses of the $D$ quark (left, red crosses) 
and of the Higgs boson (right, red crosses) together with the SM background (dotted lines)
and the total signal (black crosses) after 10 fb$^{-1}$ integrated luminosity. 
The mass of the $D$ quark is set to 500~GeV and 
Higgs boson to 120~GeV.}
\label{fig:Reconstructed-invariant-masses250}
\end{figure}

The selection cuts for example values for $D$ quark mass  of 500~GeV, and $h$ boson mass of 120~GeV,
are given in Table \ref{tab:Event-selection-cuts}.
The invariant mass distributions after the  selection cuts for the same example values are 
 presented in Fig.~\ref{fig:Reconstructed-invariant-masses250}
for 30 fb$^{-1}$ integrated luminosity. The signal window
for $D$ can be defined as $M_{D}\pm50$ GeV and for $h$ as $M_{h}\pm30$
GeV. The number of events for the signal (S) and the background (B)
can be summed in their signal windows 
for both signal and background cases to calculate the statistical significance
$\sigma=S/\sqrt{S+B}$. For this set of parameters, it is found that
the $D$ quark can be observed with a significance of 13.2$\sigma$
and at the same time the Higgs boson with a significance of about
9.5$\sigma$. One should note that, in the SM Higgs searches, such
a high statistical significance can only be reached with more than
3 times more data: with about 100 fb$^{-1}$ integrated luminosity.

\begin{figure}[htb]
\begin{center}
\begin{centering}\includegraphics[scale=0.4]{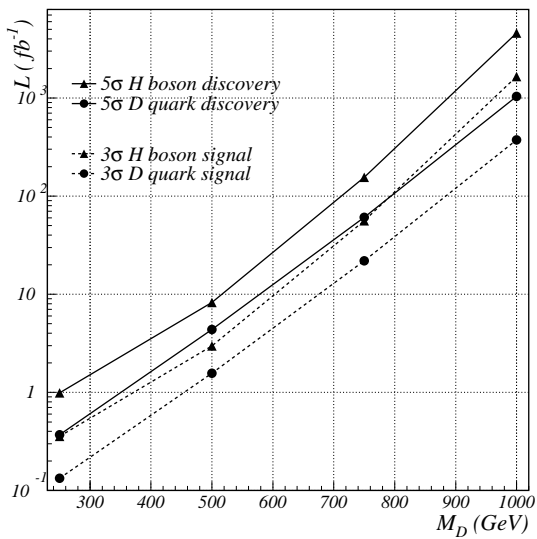}
\end{centering}
\caption{The reach of ATLAS in the Higgs search for increasing $D$ quark mass values. 
The dashed lines show the 3 $sigma$ and the solid lines show the 5 $\sigma$ reaches
of Higgs boson (triangles) and $D$ quark (circles) searches.}
\label{fig:The-reach}
\end{center}
\end{figure}

An analysis similar to the above one was
performed for the other three $D$ quark masses: 250, 750 and 1000
GeV. For each mass, the cut values were re-optimised to get the best
statistical significance in the Higgs boson search. Figure \ref{fig:The-reach} contains
the 3$\sigma$  and the 5$\sigma$ signal significance reaches
of the Higgs boson and the $D$ quark as a function of their masses.
It can be seen that, a  light Higgs boson could be discovered
with a 5$\sigma$ statistical significance using the $D\bar{D}\rightarrow hWjj$
channel within the first year of low luminosity data taking (integrated
luminosity of 10 fb$^{-1}$) if $m_{D}<500$ GeV. Under the same conditions
but with one year of design luminosity (integrated luminosity of 100
fb$^{-1})$, the 5$\sigma$ Higgs discovery can be reached if $m_{D}\leq700$
GeV. This is to be compared with the studies from the ATLAS Technical Design Report, where
the most efficient channel to discover such a light Higgs is the $h\rightarrow\gamma\gamma$
decay. This search yields about 8$\sigma$ signal significance with
100 fb$^{-1}$ integrated luminosity. The presently discussed model
could give the same significance (or more) with the same integrated
luminosity if $m_{D}<630$ GeV. Therefore, if the isosinglet quarks
exist and their masses are suitable, they will provide a considerable
improvement for the Higgs discovery potential.

%
%
\subsection{Quarks from extra dimensions: charges $-1/3$ and $5/3$}
\label{sec:NS:q1353}

Heavy quarks of charges (-1/3, 2/3, 5/3) (denoted $\tilde{q}$) are
well-motivated in  Randall-Sundrum (RS) models with custodial symmetry 
\cite{Agashe:2003zs,Agashe:2004ci,Agashe:2004bm,Agashe:2006at,Contino:2006qr}.
They are partners of the SM right-handed top quark 
and have a mass between 500 and 1500 GeV.
Their presence  can be attributed to the heaviness of the top quark.
This section studies the pair-production of heavy $Q=-1/3$ and $Q=5/3$ quarks,
which takes place through standard QCD
interactions with a cross section $\sim {\cal O}(10)$ pb for masses of several
hundreds of GeV.  
The focus is on the 4-$W$ events, which are characteristic of the decay of
new charge $-1/3$ singlets coupling to the $(t,b)_L$ doublet, in contrast with
the preceding section in which the singlet $D$ is assumed to couple to the $d$
quark. The process considered is $gg, q\overline{q}\rightarrow  \tilde{q} 
\overline{\tilde{q}} \rightarrow W^- t \ W^+  \overline{t}
  \rightarrow W^- W^+ b \ W^+  W^- \overline{b}$.
A straightforward trigger criterion for these events is that of a single,
 isolated lepton with missing $E_T$ originating from 
the leptonic decay of one of the $W$ bosons.
The remaining $W$ bosons can be reconstructed using dijet pairs.  
The goal in this analysis is to investigate the feasibility  of multi-$W$
 reconstruction and therefore identify 
$\tilde{q}$ at the LHC.  A simulation of this signal and its main background 
has been performed, and an analysis
strategy outlined which distinguishes the signal from the sizable SM
 backgrounds \cite{Dennis:2007tv}.

There can be several $\tilde{q}$-type KK quarks in the class of 
composite Higgs models under consideration, leading to the same signature. 
Typically, in the minimal models, there is one heavy quark with electric 
charge $5/3$ as well as a $Q=-1/3$ quark, decaying into $tW^+$ and $tW^-$ 
respectively, both with branching ratio essentially equal to 1. 
In addition, there is another bottom-type quark with $tW^-$ branching ratio 
$\sim 1/2$. All these $\tilde{q}$ quarks are almost degenerate in mass. 
For the present model analysis, the mass of $\tilde{q}$ is 
taken as $m_{\tilde{q}}=500$ GeV.
The Lagrangian of the model \cite{Dennis:2007tv} has been implemented into
{\tt CalcHep}\ 2.4.3 \cite{Pukhov:2004ca} for
the simulation of $\tilde{q}$ pair production and decay through the
$tW$ channel. 
The actual number of 
$4W$ events coming from the pair production and decay of the other $Q=-5/3$
 KK quarks, in a typical model, is taken into account by a multiplying factor.

\begin{figure}[!htb]
\begin{center}
\includegraphics[height=4.cm,width=7.95cm]{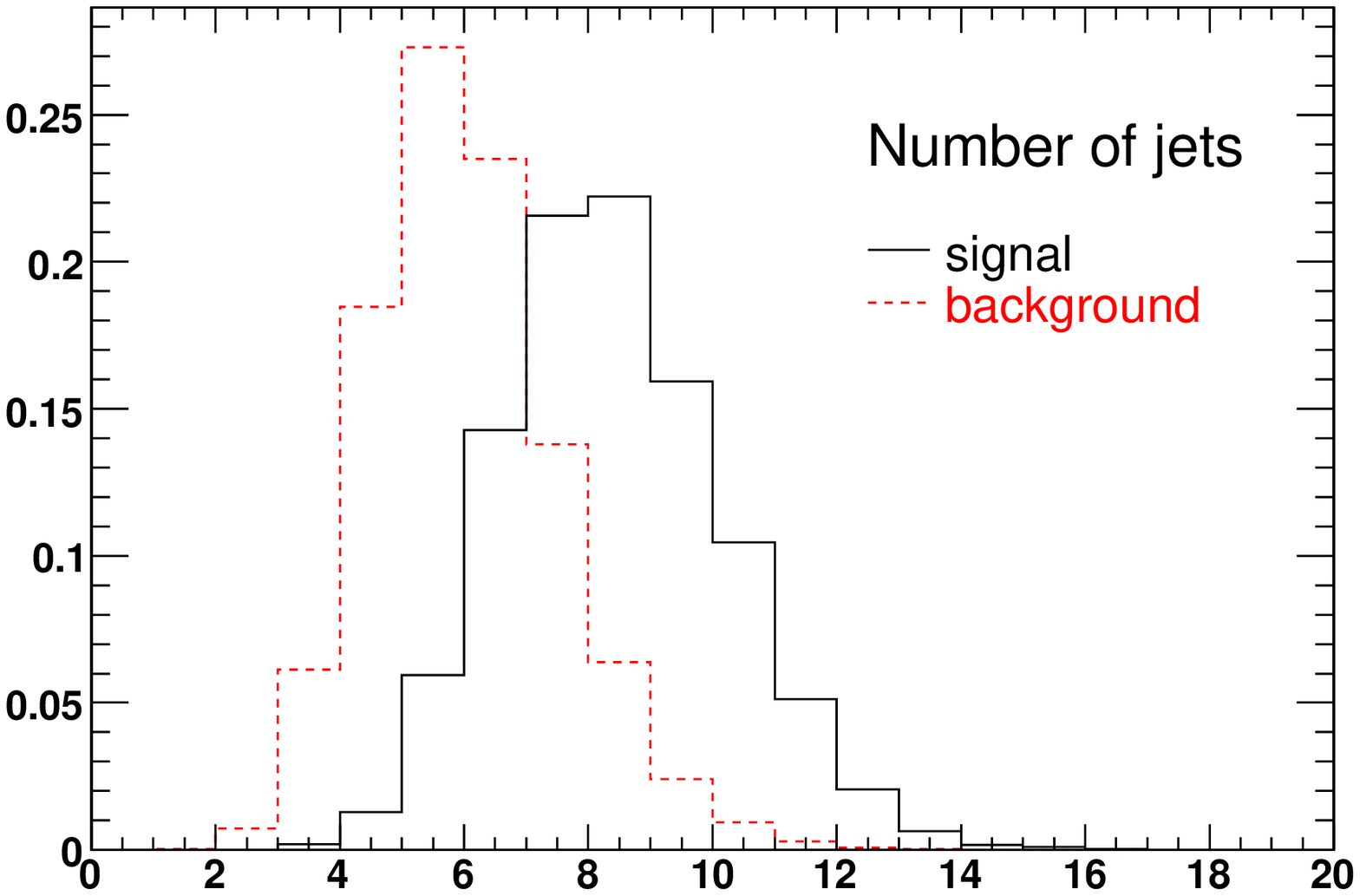}
\includegraphics[height=4.cm,width=7.95cm]{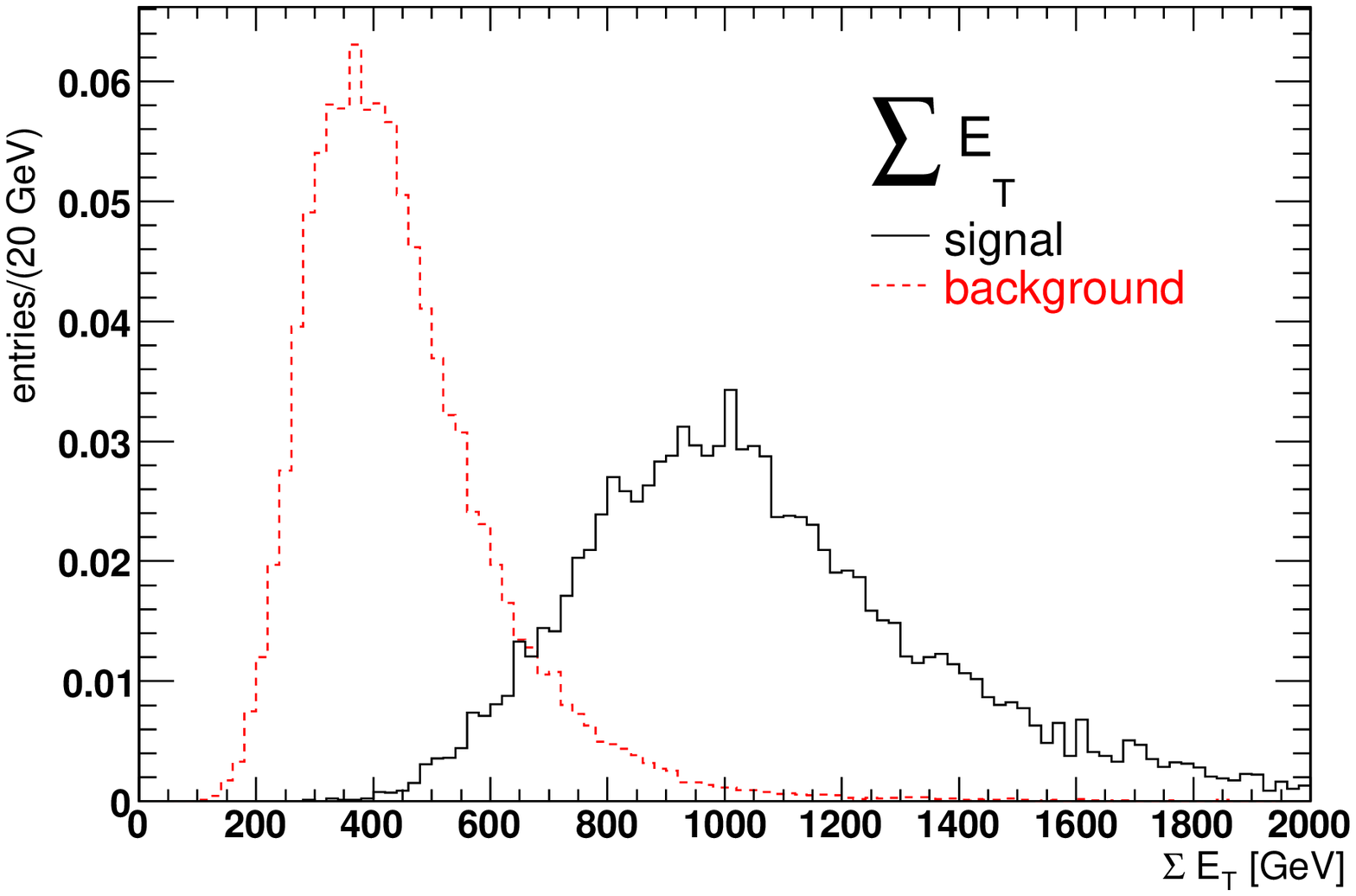}
\includegraphics[height=5.cm,width=7.9cm]{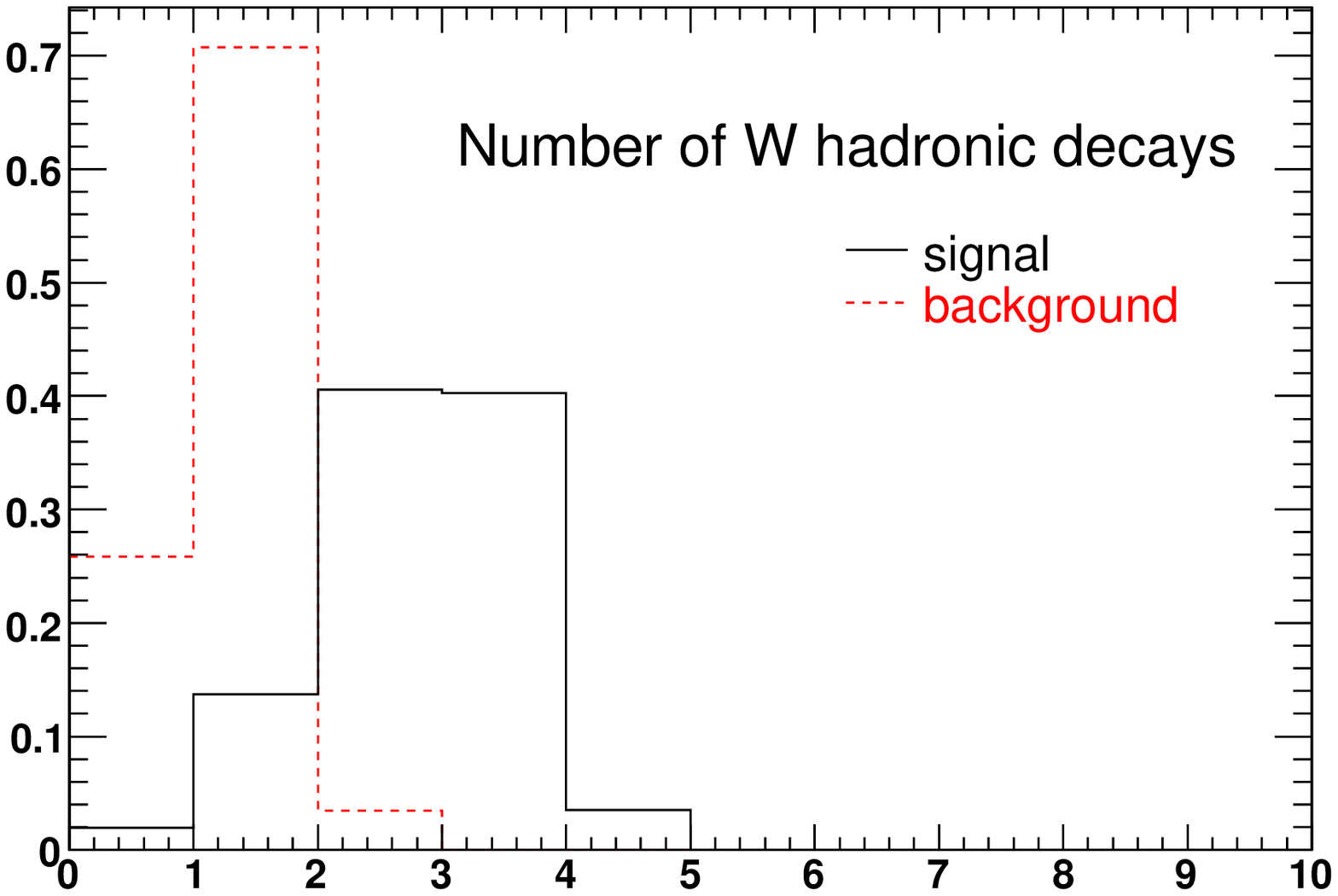}
\includegraphics[height=5.cm,width=8.cm]{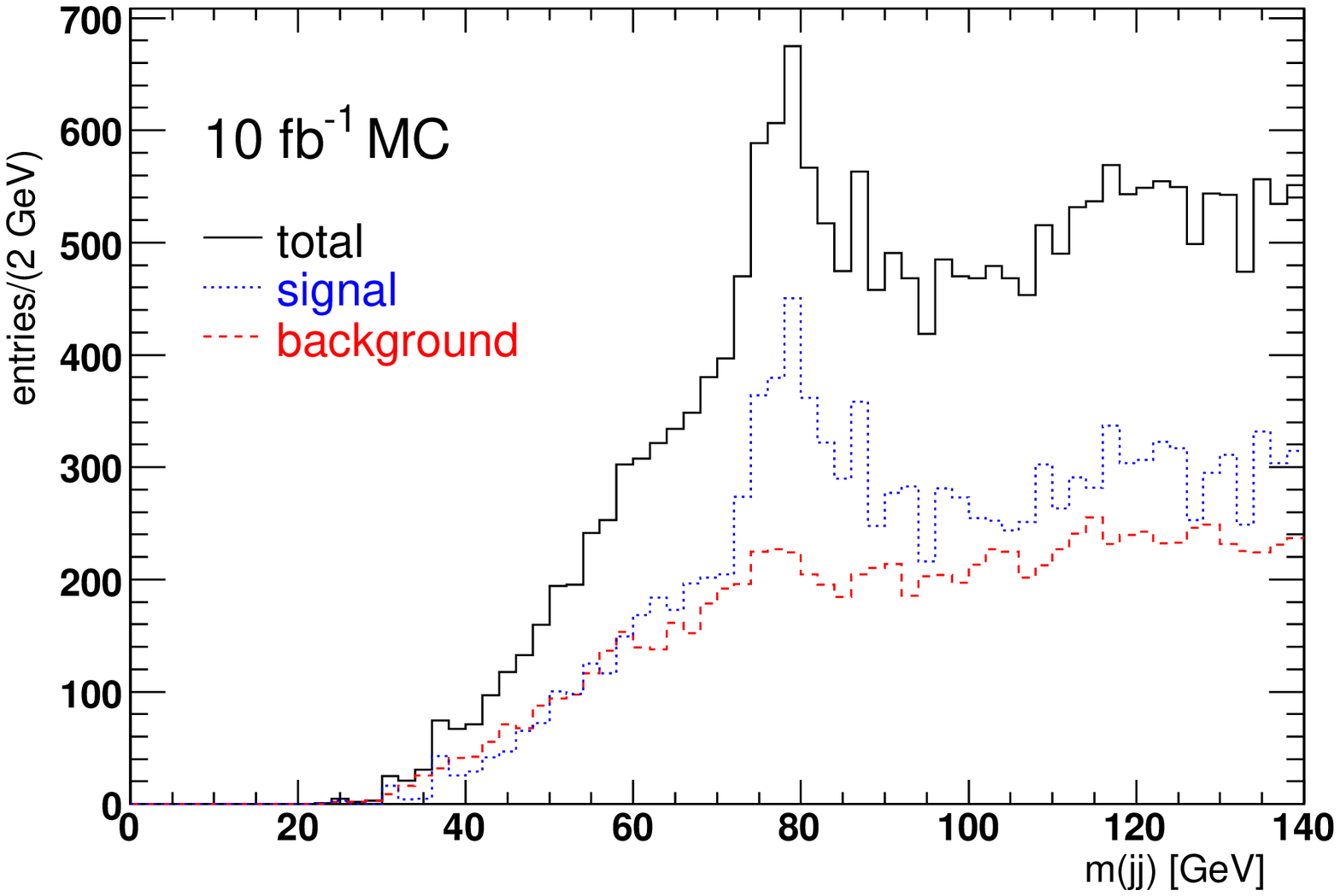}
\caption{Top left (a): Number of jets with $p_T > 20\;{\rm GeV}$. Top Right (b):  Scalar sum of $E_T$;
Bottom left (c): number of $W$'s decaying hadronically in the event. All distributions
are normalized to unit area. Bottom right (d): Dijet mass distribution after eliminating the first hadronic $W$ candidate.}
 \label{Peak}
\end{center}
\end{figure}

$t \overline{t} WW$ events from $\tilde{q}$ pair production are generated
with {\tt CalcHep}, and are further processed 
with {\tt PYTHIA}\ 6.401 \cite{Sjostrand:2006za}.
The following ``trigger'', applied to the generated events, is based on the
lepton criteria for selecting $W\rightarrow\ell\nu$ events:  
at least one electron or
muon with $p_T>25\;{\rm GeV}$ must be found within     the pseudorapidity
range $|\eta|<2.4$; then, the ``missing $E_T$'', calculated by
adding all the neutrino momenta in the event and taking the component
transverse to
the collision axis, must exceed 20 GeV.
Hadronic jets are reconstructed as they might be observed in a detector:
stable charged and neutral particles within
$|\eta|<4.9$ (the range of the ATLAS hadronic calorimeter),
excluding neutrinos, are first ranked in $p_T$ order.  Jets are seeded
starting with the highest $p_T$ tracks, with $p_T>1\;{\rm GeV}$;
softer tracks are added to the
nearest existing jet, as long as they are within $\Delta R < 0.4$
of the jet centroid, where $\Delta R = \sqrt{\Delta\phi^2 + \Delta\eta^2}$.
The number of jets with $p_T > 20\;{\rm GeV}$ is shown in Figure~\ref{Peak}a.
The signal is peaked around 8 jets.  

The two main backgrounds considered come from $t \overline{t}$ and
$t\overline{t}h$ production.  
$t \overline{t}$ leads to 2 $W$'s $+$ 2 $b$'s, with four extra jets
misinterpreted as coming from 
hadronic $W$ decays. $t\overline{t}h$ however, can lead exactly to $4W$'s and
$2b$'s when the Higgs mass is large enough. 
In this work, the Higgs mass is taken as $m_h=115$ GeV .
The background sample is dominated by $t\overline{t}$ events generated
using TopReX (version 4.11) \cite{Slabospitsky:2002ag} and {\tt PYTHIA}\
6.403, with CTEQ6L 
parton distribution functions. The small $t\overline{t}h$ contribution to the
background has been 
modeled with {\tt PYTHIA}. As expected, the background has fewer high-$p_T$ jets
than the signal, peaking  around 5 jets.

The number of $W/Z \rightarrow jj$ candidates ($N$) is counted,
ensuring that jets are used only once in each event. 
In the heavy Higgs case with a $\tilde{q}$ mass of 500 GeV, the following
sources dominate:
\begin{center}
\begin{tabular}{lll}
$N=1$:   SM $W/Z$ processes  & & $N=2$:  SM single h, $WW/WZ$, $t 
\overline{t}$ \\
$N=3$: $\tilde{q} \overline{\tilde{q}} \rightarrow tW bZ \rightarrow WWZbb$ & &
$N=4$: $\tilde{q} \overline{\tilde{q}} \rightarrow tW tW / tW bh / bh bh$
\end{tabular}
\end{center}

In order to suppress the most common ($t\overline{t}$) SM
background, the single hadronic $W$ is eliminated by searching 
for a combination of two high $p_T$ jets whose mass falls
between 70 and 90 GeV. 
The jets are combined in order of decreasing $p_T$.  If a pair is found,
it and the preceding pairs are removed; the dijet mass combinations of
the subsequent pairs are shown in Figure~\ref{Peak}d.  This procedure has been
tested on W+jet simulation to ensure that it does not sculpt the
combinatorial background distribution.  Detailed results of the $W$
reconstructions 
and consequences for $\tilde{q}$ identification are presented in
\cite{Dennis:2007tv}.
The peak obtained in the dijet mass distribution suggests that it is possible
to reach a signal significance beyond the $5\sigma$ level. Further investigation with more detailed
simulation is required to map the discovery potential for this signal at an LHC
experiment such as ATLAS, or at the ILC, and to connect the observable signal to the production
cross section.

%
%
\subsection{Fourth sequential generation}
\label{sec:4gensc}

The measurement of the $Z$ invisible width
implies well known constraints on the number of SM families
with light neutrinos. However the discoveries of neutrino masses and
mixings show that the lepton sector is richer than the traditional
SM. Moreover, some recent hints for new physics, mainly in $CP$ violation
effects in $b\to s$ transitions, might be accommodated with a fourth standard
model family \cite{Barberio:2007cr}.
A phenomenological motivation for the existence of a fourth SM family
might be attributed to the non-naturalness of the SM Yukawa couplings
which vary by orders of magnitude even among the same type fermions.
This consideration hints in the direction of accepting the SM as an
effective theory of fundamental interactions rather than of fundamental
particles. However, the electroweak theory (or SM before spontaneous
symmetry breaking) itself is a theory of massless fermions where fermions
with the same quantum numbers are indistinguishable. Therefore there
is no particular reason why the Yukawa couplings of a given type ($t$=u,d,l,$\nu$)
should be different across families. If one starts with such a unique
coupling coefficient per type $t$, for a case of $n$ families the
resulting spectrum becomes $n-1$ massless families and a single family
where all particles are massive with $m=na^{t}\eta$ where $\eta$
is the vacuum expectation value of the Higgs field. In the most simple
model, where all fermions acquire mass due to a Higgs doublet,
it is natural to also assume that the Yukawa couplings (therefore the
masses) for different types should be comparable to each other and
lie somewhere between the other couplings of EW unification:

\begin{eqnarray*}
a^{d}\approx a^{u}\approx a^{l}\approx a^{\nu} & \approx & a\\
e=g_{W}\sin\theta_{W} & <a/\sqrt{2}< & g_{Z}=g_{W}/\cos\theta_{W}\end{eqnarray*}

The measured fermion spectrum gives us a consistency check, quickly
proving that the 3rd SM family can not be the singled out heavy family
since $m_{t}\gg m_{b}\gg m_{\tau}\gg m_{\nu_\tau}\approx0$ . Therefore
if the above presented naturalness assumptions are true, not
only the reason behind the total number of families and the
lightness of the SM neutrinos is obtained but also a set of predictions for the
masses and mixings of the heavy fourth family are made through the parameterisations
and fits to the extended (4x4) CKM matrix elements.

\subsubsection{Search Scenarios}
\label{sec:NS:4gensc}

A recent detailed study~\cite{Arhrib:2006pm} of $b'$ and $t'$ decay has updated
old results done almost 20 years ago~\cite{Hou:1988yu,Hou:1988sx,Hou:1989sa,Hou:1990wz}. 
It was found that, the fourth generation while greatly enhancing FCNC top decays
(see section \ref{sec:NS:Tdisc} for heavy top searches), especially
$t\to cZ$ and $ch$, can only bring these into the borderline
($10^{-6}$--$10^{-7}$) of observability at the LHC. But the direct
search for $b'$ and $t'$ looks far more interesting. Since $t'\to bW$
always dominates $t'$ decay (unless $t'$--$b'$ mass difference
is large), hence it can be straightforwardly discovered by a ``heavy
top'' search, the focus will be on $b'$.
The search scenarios are roughly separated by kinematics, i.e. whether
$b'\to tW$ is allowed, and by pattern of quark mixing, i.e. whether
$b'\to cW$ is suppressed with respect to the neutral decay mode.

\paragraph{Case {$m_{b'}<m_{t}+M_{W}$}}

With $b'\to tW$ kinematically forbidden, it was pointed out long
ago that the phenomenology is rather rich~\cite{Hou:1988yu,Hou:1988sx}, with the
possibility of FCNC $b'\to bZ$ decay dominance, as well as the bonus
that a light Higgs could be discovered via $b'\to bh$~\cite{Hou:1989sa,Hou:1990wz}.
This can happen for light enough $b'$ when $V_{cb'}$ is small enough,
and has been searched for at the Tevatron. However, if the $b\to s$
CP violation indications are taken seriously, then $V_{cb'}\sim0.12$~\cite{Hou:2005yb}
is not small. Therefore the $b'\to cW$ channel should be kept open.
In this case, one has 3 scenarios: 

\begin{enumerate}
\item $b'\to cW$ dominance \ --- \ signature of $c\bar{c}W^{+}W^{-}$

For $V_{cb'}$ sizable, the lack of {}``charm-tagging\char`\"{} methods
that also reject $b$ makes this rather difficult. 

\item $b'\to cW$, $bZ$ (and $bh$) comparable \ --- \ signature of $\bar{c}W^{+}bZ$
(and $\bar{c}W^{+}bh$, $\bar{b}\bar{b}Zh$)

This can occur for $|V_{cb'}/V_{t'b}V_{t'b'}|\lesssim0.005$. 
The measurements on the $b'\to bg$ and $ b'->b\gamma$ neutral decays \cite{delphibprime}
can motivate this choice for the CKM matrix elements ratio.
The signature of $\bar{c}W^{+}bZ$ has never been properly studied, but
shouldn't be difficult at the LHC so long that $b'\to bZ$ branching
ratio is not overly suppressed. The possible bonus of finding
the Higgs makes this scenario quite attractive. 

\item $b'\to tW^{*}$ and $cW$, $bZ$ (and $bh$) comparable

$b'\to tW^{*}$ cannot be ignored above 230~GeV or so. This scenario
is the most complicated, but the signature of $\bar{t}W^{*+}bZ$ is
still quite tantalising. Again, one could also expect an enhancement 
to Higgs searches. One should not forget that $t\bar{c}W^{+}W^{-}$ should also
be considered. 

\end{enumerate}
Scenarios 1 and 2 form a continuum, depending on  BR$(b'\to bZ)$.

\paragraph{Case $m_{b'}>m_{t}+M_{W}$}

The $b'\to tW$ should dominate over all other
modes, except when one is still somewhat restricted by kinematics
while $V_{cb'}/V_{tb'}$ is very sizable. Therefore the two available scenarios are:

\noindent \phantom{xy} 4. $b'\to tW$ \ --- \ with a signature of $\,t\bar{t}W^{+}W^{-}$,
or $b\bar{b}W^{+}W^{-}W^{+}W^{-}$\\
\noindent \phantom{xy11} With four $W$ bosons plus two $b$-jets, the signature could be striking.

\noindent \phantom{xy} 5. $b'\to Wu \quad \hbox{or} \quad  b'\to Wc $ \ --- \ with 
signature of $W^{+}W^{-}\,j\,j$\\
\noindent \phantom{xy11} The undistinguishability of the first and second family 
quarks in the light jets makes this signature benefit from the full $b'$ branching ratio. 
Such a case is investigated in the following subsection.

It should be stressed that the standard sequential generation is considered,
hence $b'$ and $t'$ masses should be below 800~GeV from partial wave unitarity
constraints, and the mass difference between the two should be smaller or comparable to $M_{W}$. 
Scenario 4 and 5 , together with the top-like
$t'\to bW$ decay, could certainly be studied beyond 500~GeV. With
such high masses, one starts to probe strong couplings. Whether there
is an all-new level of strong dynamics~\cite{Holdom:2006mr} related to
the Higgs sector and what the Yukawa couplings would be is also a rather interesting
and different subject.

\subsubsection{A Case Study}

If the fourth family is primarily mixing with the first two families,
the dominant decay channels will be $t'\rightarrow W^{+}s(d)$
and $b'\rightarrow W^{-}c(u)$. In this case, since the light quark
jets are indistinguishable, the signature will be $W^{+}W^{-}j\, j$
for both $t'\bar{t'}$ and $b'\bar{b'}$ pair production.
According to flavour democracy, the masses of the new quarks have
to be within few GeV of each other. This is also experimentally hinted
by the value of the $\rho$ parameter's value which is close to unity
\cite{Eidelman:2004wy}. For such a mixing, both up and down type new 
quarks should
be considered together since distinguishing between $t'$ and $b'$
quarks with quasi-degenerate masses in a hadron collider seems to
be a difficult task. Moreover, the tree level pair production and decay
diagrams of the new $b'$ quarks are also valid for the $t'$ quark, 
provided $c,u$ is replaced by $s,d$.
As the model is not able to predict the masses of the new quarks,
three mass values (250, 500 and 750~GeV)  are considered as a mass scan.
The widths of the $b'$ and $t'$ quarks are proportional to
$|V_{b'u}|^{2}+|V_{b'c}|^{2}$ and $|V_{t'd}|^{2}+|V_{t's}|^{2}$ 
respectively.
Current upper limits for corresponding CKM matrix elements
are $|V_{b'u}|<0.004$, $|V_{b'c}|<0.044$, $|V_{t'd}|<0.08$ and $|V_ 
{t's}|<0.11$.
For the present case study, the common
value 0.001 is used for all four elements. As the widths of the new
quarks are much smaller than 1 GeV, this selection of the new CKM
elements has no impact on the calculated cross sections. Table
\ref{tab:The-considered-masses} gives the cross section for the 
$b'\bar{b'}$
or $t'\bar{t'}$ production processes which are within 1s\%
of each other as expected. For this reason, from this point on,
$b'$ will be considered and the results will be multiplied by
two to cover both $t'$ and $b'$ cases. Therefore in the final plots,
the notation $q_4$ is used to cover both $t'$ and $b'$.

\begin{table}
\begin{center}
\caption{The considered quark mass values and the associated width 
and pair
production cross sections at LHC.}
 \label{tab:The-considered-masses}
\begin{tabular}{c|ccc}
$M_{d4}$& 250& 500& 750\tabularnewline
\hline
$\Gamma$(GeV)& 1.00$\times10^{-5}$& 8.25$\times10^{-5}$& 2.79$ 
\times10^{-4}$\tabularnewline
$\sigma$(pb)& 99.8& $2.59$& $0.25$\tabularnewline
\end{tabular}
\end{center}
\end{table}

To estimate the discovery possibility of the fourth family quarks,
the model was implemented into a well known tree-level generator,
{\tt CompHEP}\ 4.3.3\cite{Boos:2004kh}. This tool was used to simulate 
the
pair production of the $b'$ quarks at the LHC and their subsequent
decay into SM particles. The QCD scale was set to the mass of the
$b'$ quark under study and the parton distribution function was
chosen as CTEQ6L1 \cite{Pumplin:2002vw}. The generated events
were fed into the ATLAS detector simulation and event reconstruction
framework, {\tt ATHENA}\ 11.0.41, using the interface program
{\tt CPYTH}\ 2.0.1 \cite{Belyaev:2000wn}. The partons were hadronised 
by {\tt PYTHIA}\
6.23 \cite{Sjostrand:2000wi} and the detector response was simulated by
the fast simulation software, {\tt ATLFAST}\ \cite{Froidevaux:682460}.
The decay of the pair produced $b'$ quarks result in two light jets
(originating from the quarks and/or anti-quarks of the first two SM 
families)
and two $W$ bosons. For the final state particles,
the hadronic decay of one $W$ boson and the leptonic
($e$, $\mu$) decays of the other one are considered to ease the 
reconstruction.

The direct background to the signal is from SM events yielding the 
same final state particles.
These can originate from all the SM processes which give two $W$s and 
two
non b-tagged jets. The contributions from same sign $W$
bosons were calculated to be substantially small. Some of the 
indirect backgrounds
are also taken into account. These mainly included the $t\,\bar{t}$
pair production where the $b$ jets from the decay of the top quark
could be mistagged as a light jet. Similarly the jet associated top
quark pair production ($t\,\bar{{t}}\, j\rightarrow W^{-}\, W^{+}\, b 
\,\bar{b}\, j$
) substantially contributes to the background events as the production
cross section is comparable to the pair production and only one 
mistagged
jet would be sufficient to fake the signal events. The cross section of
the next order process, namely $p\, p\to t\,\bar{t}\,2j$, was also
calculated and has been found to be four times smaller than $t\,\bar 
{t}\: j$
case: therefore it was not further investigated.

The first step of the event selection was the requirement of a single 
isolated lepton ($e$ or $\mu$) of transverse momentum 
above $15$~GeV, and at least
four jets with transverse momenta above 20 GeV. The leptonically
decaying $W$ boson was reconstructed by attributing the total missing
transverse momentum in the event to the lost neutrino, and using the
nominal mass of the $W$ as a constraint. The two-fold ambiguity in
the longitudinal direction of the neutrino was resolved by choosing
the solution with the lower neutrino energy. The four-momenta of the
third and fourth most energetic jets in the event were combined to
reconstruct the hadronically decaying $W$ boson. The invariant mass
of the combination of these jets was required to be less than $200$~GeV. 
The summary of the event selection cuts and their efficiencies for
both signal and background events are listed in Table \ref 
{tab:Efficiencies} for a quark mass of $500$~GeV.

\begin{table}
\begin{center}
\caption{Efficiencies of the selection criteria, as applied in the  
order listed,
for the $m_{q}$=500 GeV signal and the SM background.}
 \label {tab:Efficiencies}
\begin{tabular}{c|cc}
Criterion& $\epsilon$-Signal (\%)& $\epsilon$-Background (\%) 
\tabularnewline \hline
Single $e/\mu$, $p_{T}^{\ell}>15$GeV & 32& 29.1\tabularnewline
At least 4 jets, $p_{T}^{j}>20$GeV & 88.3& 94.2\tabularnewline
Possible neutrino solution& 71.3& 73.7\tabularnewline
$m_{jj}^{W}<$200~GeV& 63.5& 76.0\tabularnewline
\end{tabular}
\end{center}
\end{table}

\begin{figure}[!htb]
\begin{center}
\includegraphics[scale=0.34]{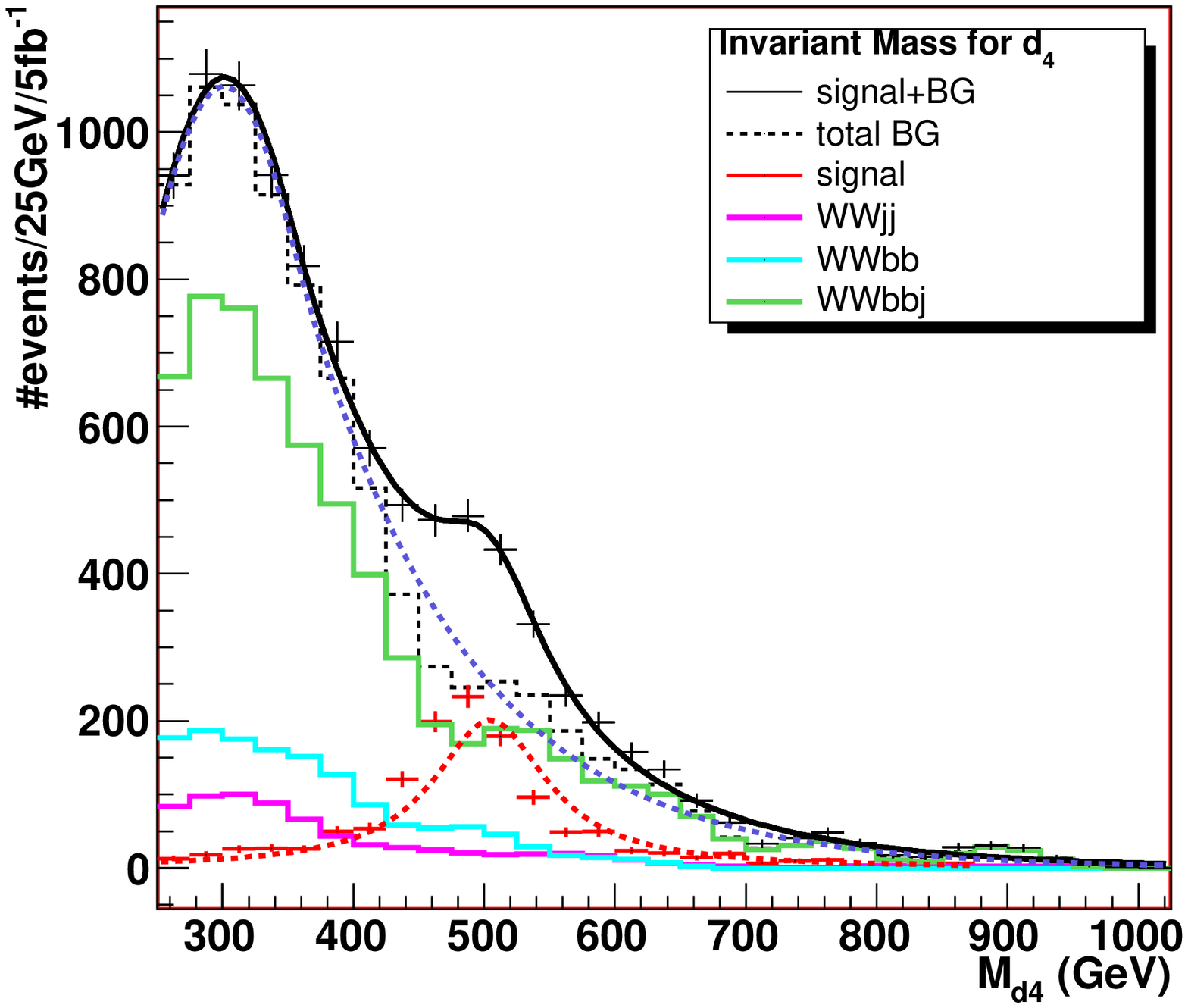}
\includegraphics[scale=0.34]{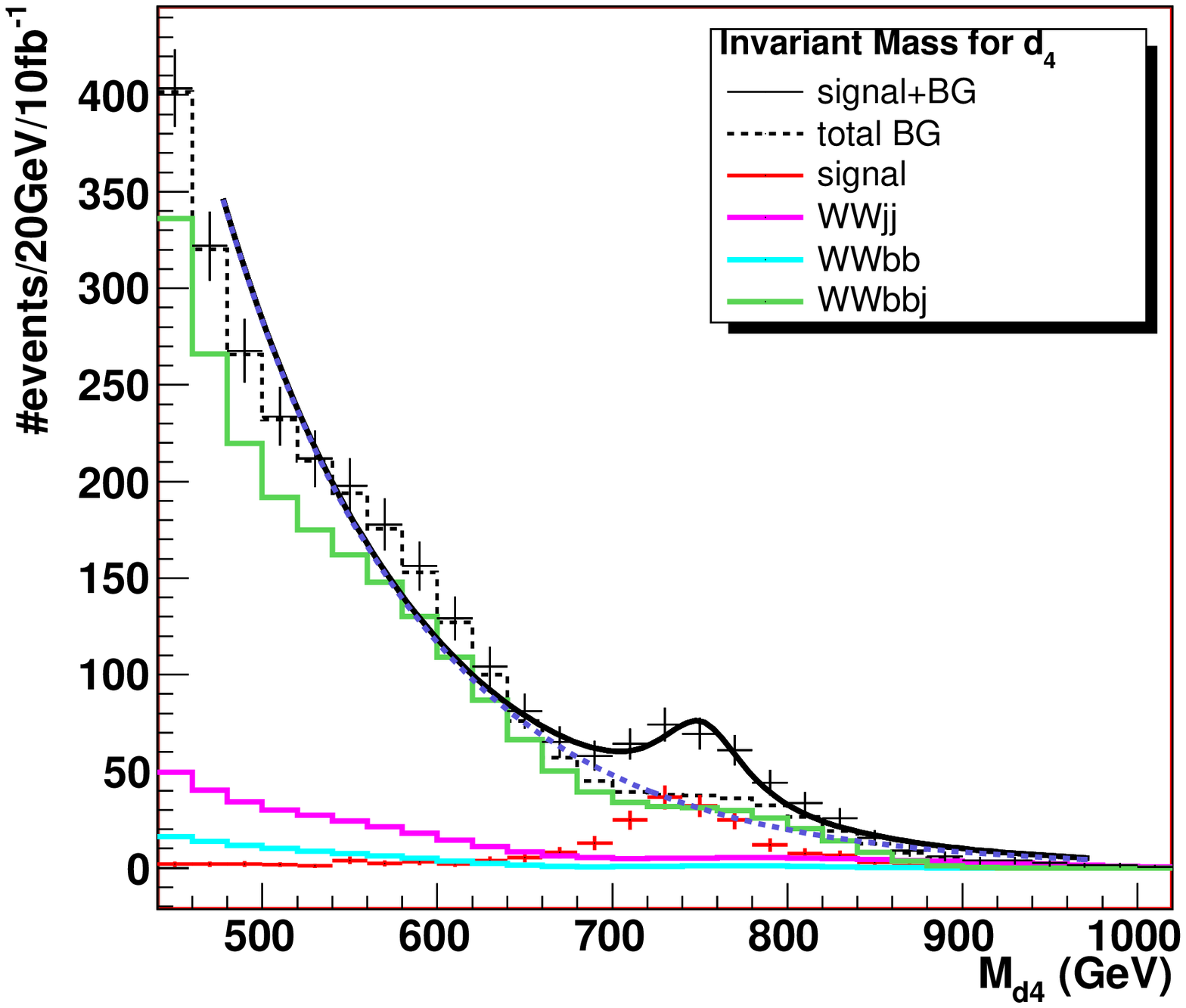}
\caption{Reconstructed signal and the stadard model background for a  
quark
of mass 500~GeV (left) and 750~GeV (right) . The colored solid
lines show SM backgrounds from various processes, the solid black
like represents the fit to the sum of background and signal events.}
\label{fig:Reconstructed}
\end{center}
\end{figure}

The surviving events were used to obtain the invariant mass of the
new quark. The $W$-jet association ambiguity was resolved by selecting
the combination giving the smallest mass difference between the two
reconstructed quarks in the same event. The results of the  
reconstruction
for quark masses of 500~GeV and 750~GeV are shown in
Fig.~\ref{fig:Reconstructed}
together with various backgrounds for integrated luminosities of 5
and 10 fb$^{-1}$ respectively. The bulk of the background in both
cases is due to $g\, g\to t\,\bar{t}\: g$ events as discussed before.

In order to extract the signal significance, an analytical function
consisting of an exponential term to represent the background and
a Breit-Wigner term to represent the signal resonance was fitted
to the total number of events in the invariant plots of Fig. \ref 
{fig:Reconstructed}.
In both cases, the fitted function is shown with the solid line, whereas
the background and signal components are plotted with dashed blue
and red lines, respectively. For the case of $m_{d_{4}}$=500~GeV,
it can be noticed that the signal function extracted from the fit
slightly underestimates the true distribution. However, using the same
fit functions and with 5 fb$^{-1}$ of data, the signal significance
is found to be 4.7$\sigma$. The significance is calculated after
the subtraction of the estimated background: the integral area around
the Breit-Wigner peak and its error are a measure of the expected number of
signal events, thus the signal significance. A similar study with the
higher mass value of 750~GeV, and with 10 fb$^{-1}$ of data gives results
with a significance of 9.4 $\sigma$.
This analysis has shown that the fourth family quarks with the 
studied mass
values can be observed at the LHC with an integrated luminosity of 10 
fb$^{-1}$.
Although these results were obtained with a fast simulation, the  
simplistic
approach in the analysis should enhance their validity.

\paragraph{Other possible Studies}

The study of $\bar{c}W^{+}bZ$ is a relatively easy one. Due to
the cleanness of the $Z\to\ell^{+}\ell^{-}$ signature, one does
not need to face $c$-jet tagging issues, and one can either have
$W\to jj$ or $W\to\ell\nu$. For the latter, the offshoot is to search
for $\bar{c}W^{+}bh$ by a $M_{b\bar{b}}$ scan with $Z$ as standard
candle. A second effort would be $\bar{t}W^{*+}bZ$, with similar
approach as above. Once experience is gained in facing $c$ as well
as $W^{*}$ (relatively soft leptons or jets, or missing $E_{T}$),
one could also consider $t\bar{c}WW^{*}$, before moving onto the
challenge of $c\bar{c}W^{+}W^{-}$. If $c$-jet tagging tools could
be developed, it could become of general use. The $t\bar{t}W^{+}W^{-}$
search for heavy $b'$ could also be pursued.

%
%
\section{New Leptons: heavy neutrinos}
 \label{sec_new_leptons}

Models with extended matter multiplets predict additional leptons, both 
charged and neutral. While heavy neutral leptons (neutrinos) can be introduced
to explain the smallness of the light neutrino masses in a natural way and the 
observed baryon asymmetry in the universe, the charged ones are not required 
by experiment. Here we concentrate on the neutral ones.

Heavy neutrinos with masses $m_N > M_Z$ appear in theories with extra dimensions
near the TeV scale and little Higgs models, in much the same way as vector-like
quarks, and in left-right models. For example, in the {\em simplest} Little
Higgs models
\cite{Schmaltz:2004de}, the matter content belongs to $\mathrm{SU}(3)$
multiplets, and the SM lepton doublets must be enlarged with one extra neutrino
$N'_{\ell L}$ per family. These extra neutrinos can get a large Dirac mass of
the order of the new scale $f \sim 1$ TeV if the model also includes
right-handed neutrinos transforming as
$\mathrm{SU}(3)$ singlets \cite{delAguila:2005yi}.
This mechanism provides a natural way of giving masses to the SM neutrinos, and
in this framework the mixing between the light leptons and the heavy neutrinos
is of order $v/\sqrt 2 f$, with $v=246$ GeV the electroweak VEV. But besides
their appearance in several specific models, heavy Majorana neutrinos are often
introduced to explain light neutrino masses via the seesaw mechanism
\cite{Minkowski:1977sc,Gell-Mann:1980vs,Yanagida:1979as,Mohapatra:1979ia}.\footnote{This
mechanism, with heavy neutrino singlets under the SM gauge group, is often referred to as
seesaw type I. Other possibilities to generate light neutrino masses are to introduce a scalar
triplet (type II seesaw, see section \ref{sec_new_scalars}) or a lepton triplet (type III).
In this section, heavy neutrinos are always assumed to be SM singlets.}
They give contributions to light neutrino masses $m_\nu$ of the order
$Y^2 v^2  / 2 m_N$, where $Y$ is a Yukawa coupling.
In the minimal seesaw realization this is the only source for
light neutrino masses, and the Yukawa couplings are assumed of order unity without any
particular symmetry. Therefore,
having $m_\nu \sim Y^2 v^2 / 2 m_N$ requires heavy masses $m_N \sim 10^{13}$ GeV
to reproduce the observed light neutrino spectrum. Additionally, the light-heavy
mixing is predicted to be $V_{\ell N} \sim \sqrt{m_\nu / m_N}$.
These ultra-heavy particles are unobservable, and thus the seesaw mechanism
is not directly testable. Nevertheless, non-minimal
seesaw models can be built, with $m_N \sim 1$ TeV or smaller, if some 
approximate flavour
symmetry suppresses the $\sim  Y^2 v^2 / 2 m_N$ contribution from seesaw
\cite{Buchmuller:1991tu,Ingelman:1993ve,Gluza:2002vs}.
These models can also provide a successful leptogenesis (see, for instance
Refs.~\cite{Pilaftsis:2003gt,Pilaftsis:2005rv,Boubekeur:2004ez,Abada:2004wn}).
Heavy neutrinos with masses near the electroweak scale can be produced at the
next generation of colliders (see Ref.~\cite{delAguila:2006dx} for a review)
if their coupling to the SM fermions and gauge bosons is not too small, or
through new non-standard interactions. The most conservative
point of view is to assume that heavy neutrinos are singlets under the SM gauge
group and no new interactions exist, which constitutes a ``minimal'' scenario in
this sense. On the other hand, with an extended gauge structure, for example
$\mathrm{SU}(2)_L \times \mathrm{SU}(2)_R \times \mathrm{U}(1)_{B-L}$
in models with left-right symmetry, additional production
processes are possible, mediated by the new $W'$ and/or $Z'$
 gauge bosons. We will discuss these possibilities in turn.

\subsection{Production of heavy neutrino singlets}
\label{sec:NS:singleN}

Heavy neutrino singlets couple to the SM fields through their mixing with the SM
neutrino weak eigenstates. The Lagrangian terms describing the interactions of
the lightest heavy neutrino (in the mass eigenstate basis) are
\begin{eqnarray}
\mathcal{L}_W & = & - \frac{g}{\sqrt 2}  \left( \bar \ell \gamma^\mu V_{\ell N}
P_L N \; W_\mu + \bar N \gamma^\mu V_{\ell N}^* P_L \ell \; W_\mu^\dagger
\right) \,, \notag \\
\mathcal{L}_Z & = & - \frac{g}{2 c_W}  \left( \bar \nu_\ell \gamma^\mu
V_{\ell N} P_L N + \bar N \gamma^\mu V_{\ell N}^* P_L \nu_\ell \right)
Z_\mu \,, \notag \\
\mathcal{L}_h & = & - \frac{g \, m_N}{2 M_W} \, \left( \bar \nu_\ell \,
V_{\ell N} P_R N + \bar N \, V_{\ell N}^* P_L \nu_\ell \right) h \,,
\label{ec:NS:lagN}
\end{eqnarray}
with $N$ the heavy neutrino mass eigenstate and $V$ the extended MNS matrix.
For Majorana $N$, the last terms in the $Z$ and $h$ interactions can be
rewritten in terms of the conjugate fields.
These interactions determine the $N$
production processes, as well as its decays. The latter can happen in the
channels $N \to W \ell$, $N \to Z \nu$, $N \to h \nu$. The partial widths
can be straightforwardly obtained from Eqs.~(\ref{ec:NS:pwidQ}) neglecting
charged lepton and light neutrino masses,
\begin{align}
\Gamma(N \to W^+ \ell^-) & =  \Gamma(N \to W^- \ell^+) \nonumber \\
& = \frac{g^2}{64 \pi} |V_{\ell N}|^2
\frac{m_N^3}{M_W^2}  
\left[ 1 -3 \frac{M_W^4}{m_N^4} + 2 \frac{M_W^6}{m_N^6} \right] \,, \nonumber
\\[0.1cm]
\Gamma_D(N \to Z \nu_\ell) & =  \frac{g^2}{128 \pi c_W^2} |V_{\ell N}|^2
\frac{m_N^3}{M_Z^2}  
\left[ 1 - 3 \frac{M_Z^4}{m_N^4} + 2 \frac{M_Z^6}{m_N^6} \right] \,, \nonumber
\\[0.2cm]
\Gamma_M(N \to Z \nu_\ell) & = 2 \, \Gamma_D(N \to Z \nu_\ell) \,, \nonumber
\\
\Gamma_D(N \to h \nu_\ell) & =  \frac{g^2}{128 \pi} |V_{\ell N}|^2
\frac{m_N^3}{M_W^2} \left[ 1 - 2 \frac{M_h^2}{m_N^2} +
\frac{M_h^4}{m_N^4} \right] \,, \nonumber
\\[0.2cm]
\Gamma_M(N \to h \nu_\ell) & = 2 \, \Gamma_D(N \to h \nu_\ell) \,.
\label{ec:NS:pwidN}
\end{align}
The subscripts $M$, $D$ refer to Majorana and Dirac heavy neutrinos,
respectively, and the lepton number violating (LNV) decay $N \to W^- \ell^+$ is only possible for a
Majorana $N$.

In the minimal seesaw the mixing angles $V_{\ell N}$ are of order
$\sqrt{m_\nu / m_N}$ (and then of order $10^{-5}$ or smaller for $m_N > M_Z$),
but in models
with additional symmetries the light-heavy mixing can be decoupled from
mass ratios \cite{Tommasini:1995ii}. Nevertheless, $V_{\ell N}$ are
experimentally constrained to be small (this fact has already been used in order
to simplify the Lagrangian above). Defining the quantities 
\begin{equation}
\Omega_{\ell \ell'} \equiv \delta_{\ell \ell'} - \sum_{i=1}^3 
V_{\ell \nu_i} V_{\ell' \nu_i}^* =  
\sum_{i=1}^3 V_{\ell N_i} V_{\ell' N_i}^*
\end{equation}
(assuming three heavy neutrinos),
limits from universality and the invisible $Z$ width imply
\cite{Bergmann:1998rg,Bekman:2002zk}
\begin{equation}
\Omega_{ee} \leq 0.0054 \,, \quad \Omega_{\mu \mu} \leq 0.0096 \,, \quad
\Omega_{\tau \tau} \leq 0.016 \,,
\label{ec:NS:Nlim1}
\end{equation}
with a 90\% confidence level (CL). In the limit of heavy neutrino masses 
in the TeV range, limits from lepton flavour violating (LFV) processes require
\cite{Tommasini:1995ii}
\begin{equation}
|\Omega_{e \mu}| \leq 0.0001 \,, \quad |\Omega_{e \tau}| \leq 0.01 \,, \quad
|\Omega_{\mu \tau}| \leq 0.01 \,.
\label{ec:NS:Nlim2}
\end{equation}
Additionally, for heavy Majorana neutrinos there are constraints on
$(V_{eN},m_N)$ from the
non-observation of neutrinoless double beta decay. These, however, may be
evaded {\em e.g.} if two nearly-degenerate Majorana neutrinos with opposite CP
parities form a quasi-Dirac neutrino.

Heavy Dirac or Majorana neutrinos with a significant coupling to the electron
can be best produced and seen at $e^+ e^-$ colliders in $e^+ e^- \to N \nu$,
which has a large cross section and whose backgrounds have moderate size
\cite{Azuelos:1993qu,Gluza:1996bz,delAguila:2005mf,delAguila:2005pf}. On the
other hand, a
Majorana $N$ mainly coupled
to the muon or tau leptons is easier to discover at a hadronic machine like LHC,
in the process $q \bar q' \to W^* \to \ell^+ N$ (plus the charge conjugate),
with subsequent decay $N \to \ell W \to \ell q \bar q'$. (Other final
states, for instance with decays $N \to Z \nu$, $N \to h \nu$, or in the
production process $p p \to Z^* \to N \nu$ have backgrounds much larger.)
Concentrating ourselves on $\ell N$ production with $N \to \ell W$,
it is useful to classify the possible signals
according to the mixing and character of the lightest heavy neutrino:
\begin{enumerate}
\item For a Dirac $N$ mixing with only one lepton flavour, the decay
$N \to \ell^- W^+$ yields a $\ell^+ \ell^- W^+$ final state, with a huge
SM background.

\item For a Dirac $N$ coupled to more than one charged lepton we can also have
$N \to \ell^{'-} W^+$ with $\ell' \neq \ell$, giving the LFV signal
$\ell^+ \ell^{'-} W^+$, which has much smaller backgrounds.

\item For a Majorana $N$, in addition to LNC signals we have LNV ones arising
from the decay $N \to \ell^{(')+} W^-$, which have small backgrounds too.
\end{enumerate}

In the following we concentrate on the case of a Majorana $N$ coupling to the
muon, which is the situation in which LHC has better
discovery prospects than ILC. The most interesting signal
is~\cite{Datta:1993nm,Panella:2001wq,Han:2006ip,delAguila:2007em}
\begin{equation}
pp \to \mu^\pm N \to \mu^\pm \mu^\pm jj \,,
\label{ec:NS:Nprod}
\end{equation}
with two same-sign muons in the final state, and at least two jets.
SM backgrounds to this LNV signal involve the production of additional leptons,
either neutrinos or charged leptons (which may be missed by the detector, thus
giving the final state in Eq.~(\ref{ec:NS:Nprod})). The main ones are $W^\pm
W^\pm nj$ and $W^\pm Z nj$, where $nj$ stands for $n=0,\dots$ additional jets
(processes with $n<2$ are also backgrounds due to the appearance of extra jets
from pile-up). The largest reducible backgrounds are $t\bar t nj$, with
semileptonic decay of the $t\bar t$ pair, and $Wb \bar b nj$, with leptonic $W$
decay. In these cases, the additional same-sign
muon results from the decay of a $b$ or $\bar b$ quark. Only a tiny fraction of
such decays produce isolated muons with sufficiently high transverse momentum
but, since the $t \bar tnj$ and $W b \bar b nj$ cross sections are so large,
these backgrounds are
much larger than the two previous ones. An important remark here is that
the corresponding backgrounds $t \bar t nj, W b \bar b nj \to e^\pm e^\pm X$
are one order of magnitude larger than the ones involving muons. The reason is
that $b$ decays produce ``apparently isolated'' electrons more often than muons,
due to detector effects. A reliable evaluation of the $e^\pm e^\pm X$ background
resulting from these processes seems to require a full simulation of the
detector. 
Other backgrounds like $Wh$ and $Zh$
are negligible, with cross sections much smaller than the ones considered,
$W/Z b \bar b$, $WZ$, $ZZ$, which give the same final states.
Note also that for this heavy neutrino mass $b \bar b nj$, which is huge,
has very different
kinematics and can be eliminated. However, for $m_N < M_W$ the heavy neutrino
signal and $b \bar b nj$ are much alike, and thus this background is the largest
and most difficult to reduce. Further details can be found in
Ref.~\cite{delAguila:2007em}.

Signals and backgrounds have been generated using {\tt Alpgen}\ (the implementation
in {\tt Alpgen}\ of heavy neutrino production is discussed in the Tools chapter).
Events are
passed through {\tt PYTHIA}\ 6.4 (using the MLM prescription for jet-parton
matching~\cite{MLM} to avoid double
counting of jet radiation) and a fast simulation of the ATLAS detector.
The pre-selection criteria used are: (i) two same-sign isolated muons with 
pseudorapidity $|\eta| \leq 2.5$ and transverse momentum $p_T$
larger than 10 GeV; (ii) no additional isolated charged leptons nor
non-isolated muons; (iii) two jets with
$|\eta| \leq 2.5$ and $p_T \geq 20$ GeV. It should be noted that requiring the
absence of
non-isolated muons reduces backgrounds involving $Z$ bosons almost by a factor
of two.

\begin{figure}[htb]
\begin{center}
\begin{tabular}{cc}
\epsfig{file=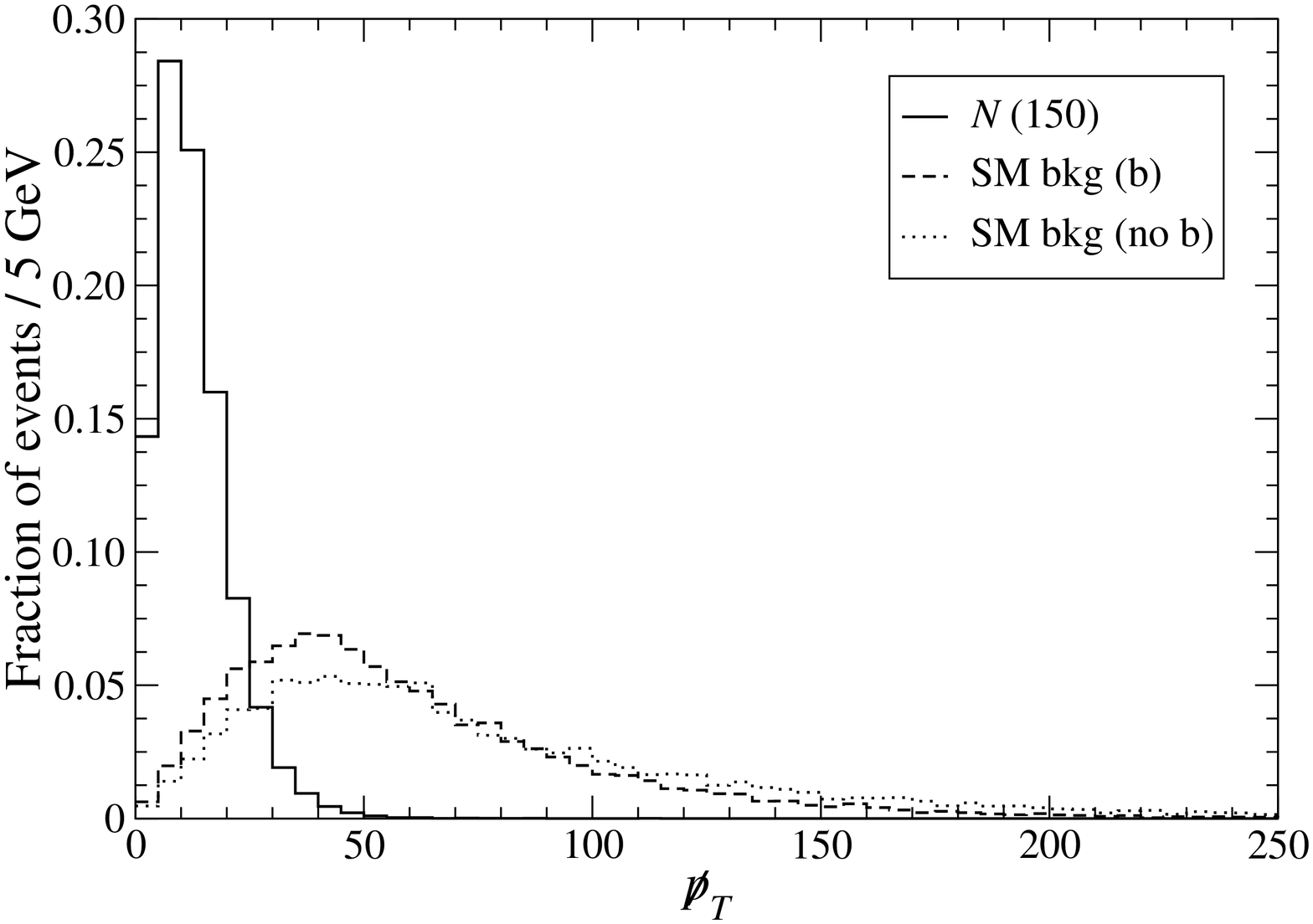,height=5.2cm,clip=} &
\epsfig{file=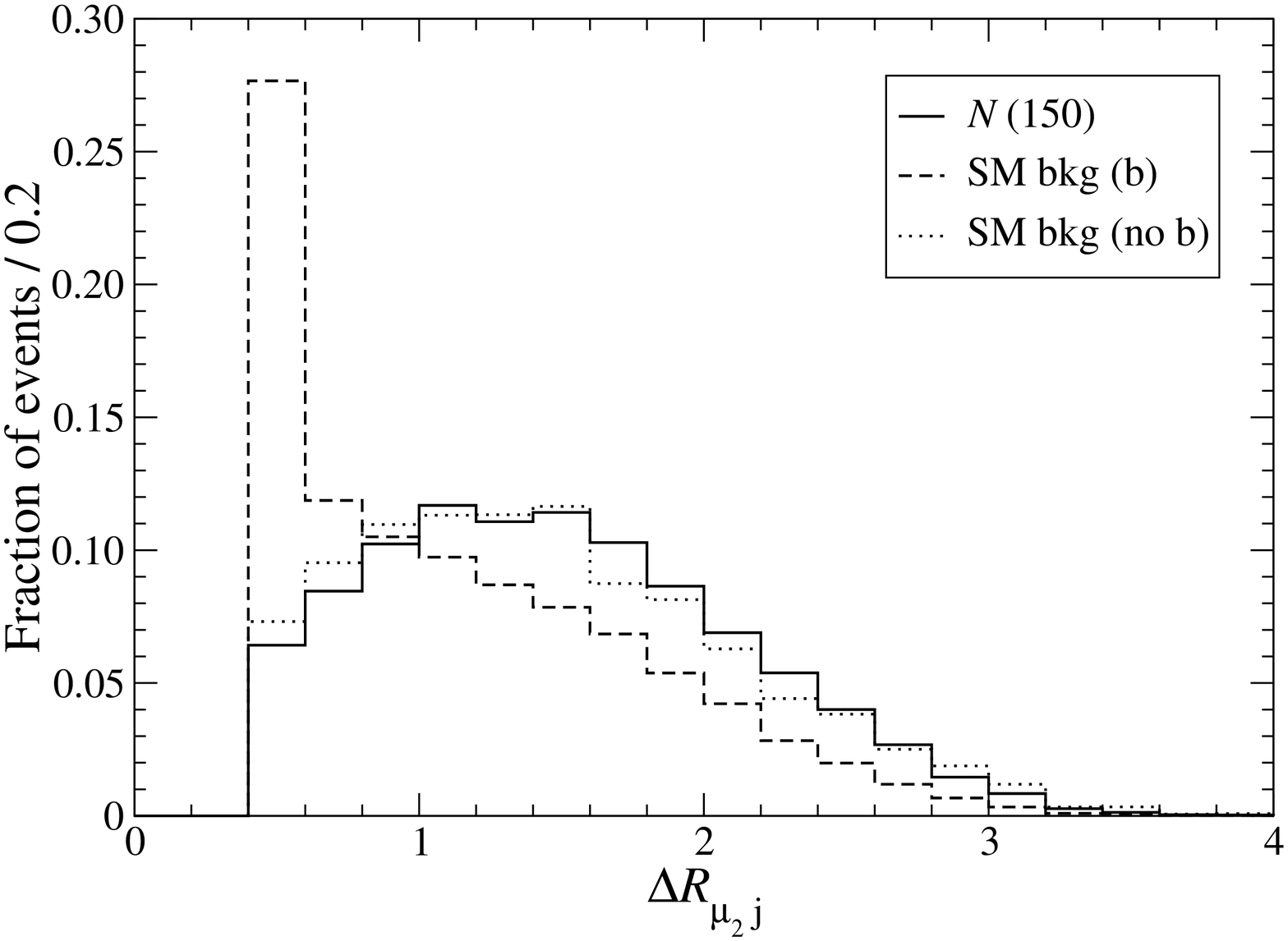,height=5.2cm,clip=} \\
\epsfig{file=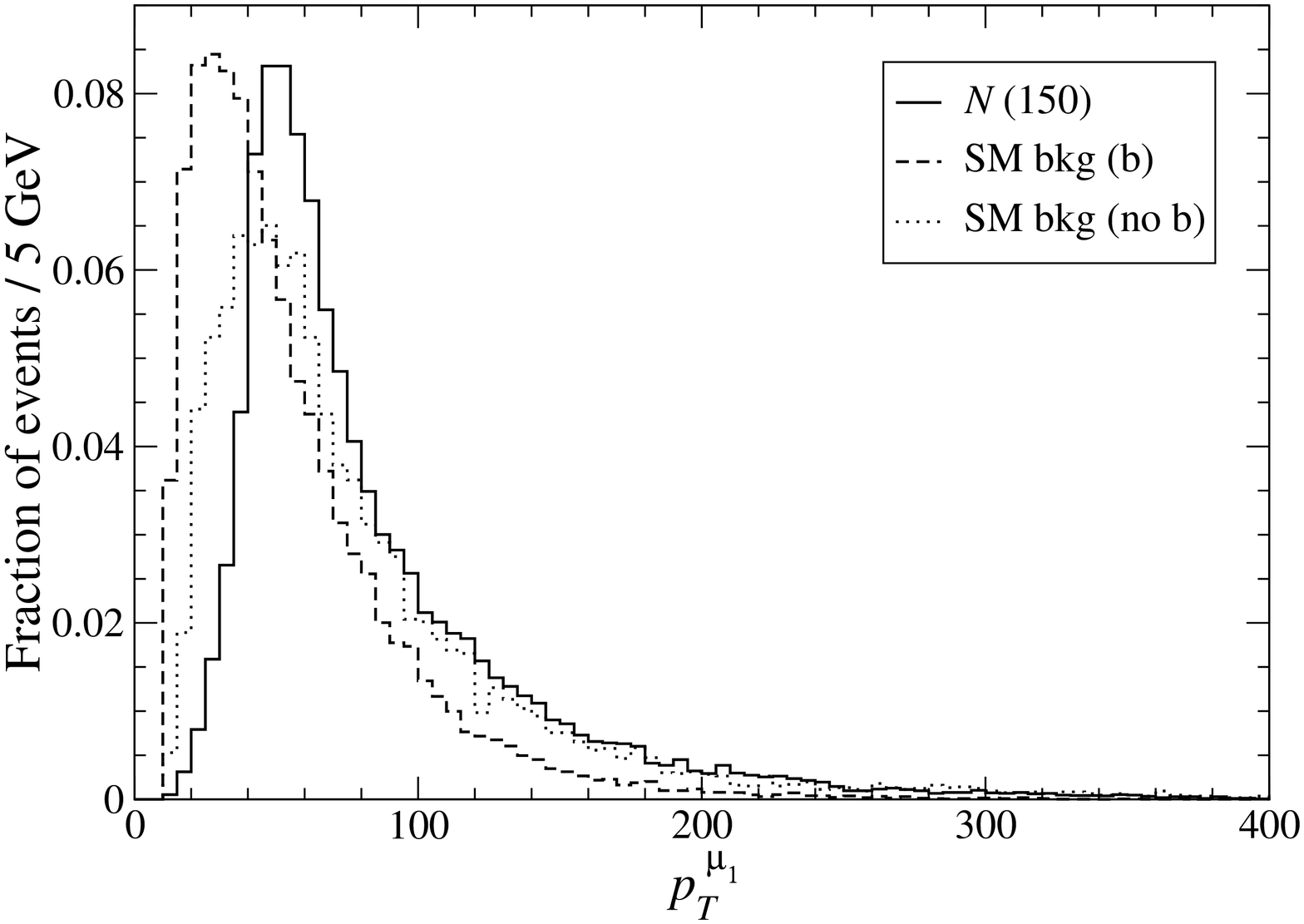,height=5.2cm,clip=} &
\epsfig{file=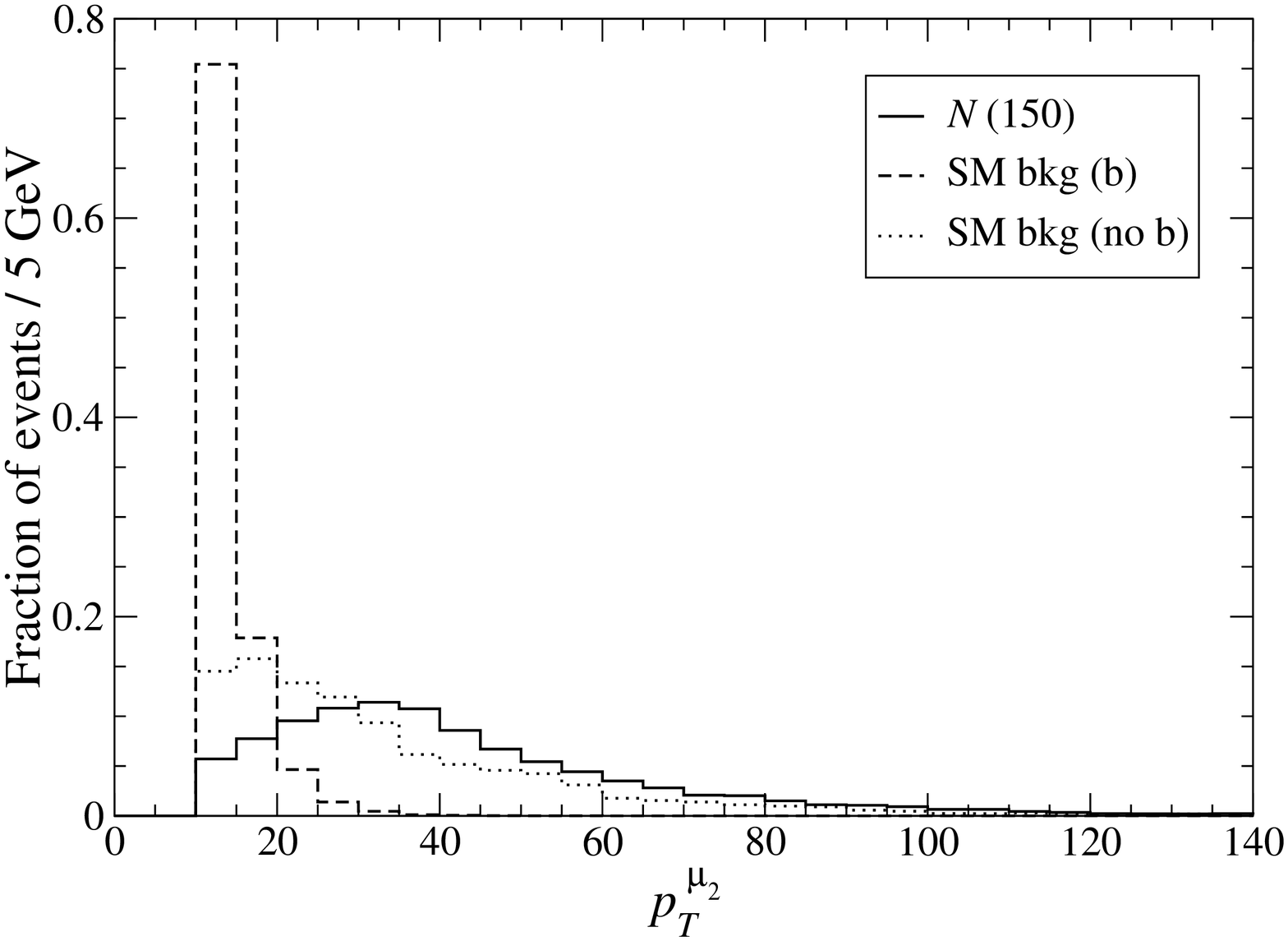,height=5.2cm,clip=} \\
\epsfig{file=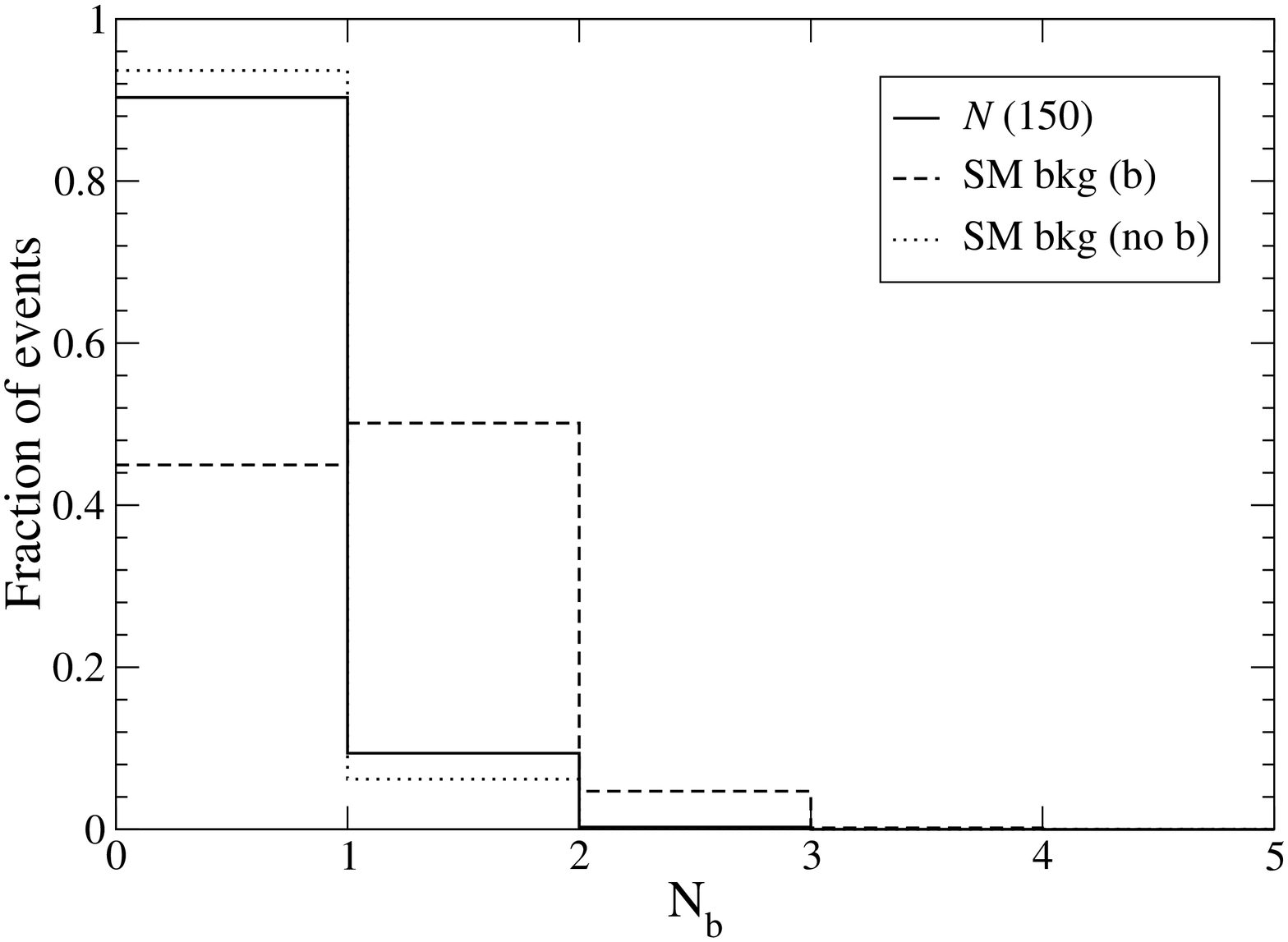,height=5.2cm,clip=} &
\epsfig{file=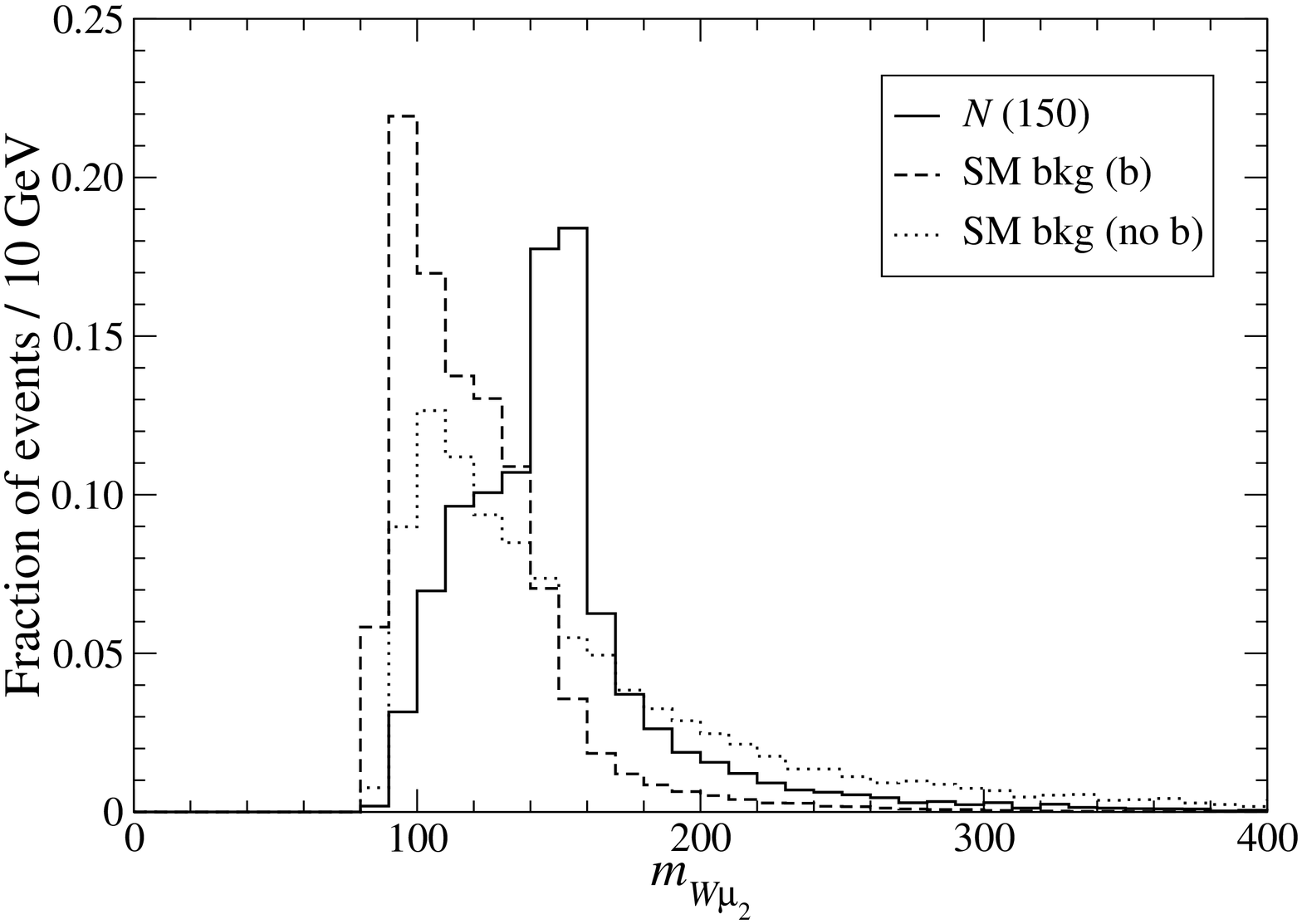,height=5.2cm,clip=}
\end{tabular}
\caption{Several useful variables to discriminate between the heavy neutrino
signal and the backgrounds, as explained in the text.}
\label{fig:NS:Nlik}
\end{center}
\end{figure}

It must be emphasised that SM backgrounds
are about two orders of magnitude larger than in previous estimations in the
literature~\cite{Han:2006ip}.
Backgrounds cannot be significantly suppressed with respect to
the heavy neutrino signal using simple cuts
on missing energy and muon-jet separation. Instead, a likelihood analysis has
been performed~\cite{delAguila:2007em}. Several
variables are crucial in order to distinguish the signal from the backgrounds:
\begin{itemize}
\item The missing momentum $\ptmiss$ (the signal does not have neutrinos in the
final state).
\item The separation between the second muon and the closest jet,
$\Delta R_{\mu_2 j}$. For backgrounds involving $b$ quarks this separation is
rather small.
\item The transverse momentum of the two muons $p_T^{\mu_1}$, 
$p_T^{\mu_2}$, ordered from higher ($\mu_1$) to lower ($\mu_2$) $p_T$. 
Backgrounds involving $b$ quarks have one muon with small $p_T$.
\item The $b$ tag multiplicity (backgrounds involving $b$ quarks often have
$b$-tagged jets).
\item The invariant mass of $\mu_2$ and the two jets which best reconstruct a
$W$ boson, $m_{W \mu_2}$.
\end{itemize}
The distribution of these variables is presented in Fig.~\ref{fig:NS:Nlik},
distinguishing three likelihood classes: the signal, backgrounds with one muon
from $b$ decays and backgrounds with both muons from $W/Z$ decays. The $b \bar b$ background can
be suppressed for $m_N \gtrsim 100$~GeV, and it is not shown.
Additional variables like jet transverse
momenta, the $\mu \mu$ invariant mass, etc. are useful, and included in the
analysis.  
Assuming a 20\% systematic uncertainty in the
backgrounds (which still has to be precisely evaluated), and taking the maximum allowed
mixing by low energy data, the following $5\sigma$ discovery limits are found:
(i) A heavy neutrino coupling only to the muon with $|V_{\mu N}|^2 = 0.0096$
can be discovered up to masses $m_N = 200$ GeV; (ii)
A heavy neutrino coupling only to the muon with $|V_{e N}|^2 = 0.0054$
can be discovered up to masses $m_N = 145$ GeV. Limits for other masses and
mixing scenarios can be found in Ref.~\cite{delAguila:2007em}.

\subsection{Heavy neutrino production from $W_R$ decays}

Models with left-right symmetry have an extended gauge structure
$\mathrm{SU}(2)_L \times \mathrm{SU}(2)_R \times \mathrm{U}(1)_{B-L}$ and, in
addition to three new gauge bosons $Z'$, $W_R^\pm$ (see
sections~\ref{sec_new_zp} and \ref{sec_new_wp}) they introduce three
right-handed neutrinos as partners of the charged leptons in $\mathrm{SU}(2)_R$
doublets $(N_\ell,\ell)_R$. The minimal scalar sector consists of a bi-doublet
and two triplets. The measurement of the $\mathrm{T}$ parameter and present
lower
bounds on the masses of the new bosons and their mixing with the $W$ and $Z$
imply the hierarchy $v_L \ll (|k_1|^2+|k_2|^2)^{1/2} \ll v_R$ among the VEVs of
the bi-doublet $k_ {1,2}$ and the triplets $v_{L,R}$. In this situation the
neutrino mass matrix exhibits a seesaw structure, heavy neutrino eigenstates $N$
are mostly right-handed and the following hierarchy
is found among the couplings of the light and heavy neutrinos to the gauge
bosons:
\begin{itemize}
\item[(i)] $\ell \nu W$ and $\ell N W_R$ are of order unity; $\ell N W$ and
$\ell \nu W_R$ are suppressed.
\item[(ii)] $\nu \nu Z$ and $NN Z'$ are of order unity; 
$\nu N Z$, $\nu N Z'$, $NN Z$ and $\nu \nu Z'$ are suppressed.
\end{itemize}

At hadron colliders the process $q \bar q' \to W_R \to \ell N$
\cite{Keung:1983uu} involves mixing
angles of order unity and only one heavy particle in the final state.
The best situation happens where $N$ is lighter than $W_R$, so that $W_R$ can be
on its mass shell and the cross section is not suppressed by an $s$-channel
propagator either. This is in sharp contrast with the analysis in the previous
subsection, in which the process $q \bar q' \to W^* \to \ell N$ is suppressed by
mixings and the off-shell $W$ propagator.

Heavy neutrino production from on-shell $W_R$ decays has been previously
described in Ref.~\cite{Datta:1992qw}, and studied in detail for the ATLAS
detector in Ref.~\cite{Ferrari:2000sp}. Here we summarise the expectations
for the CMS detector \cite{Gninenko:2003fv,Gninenko:2006br}.
Production cross sections and decay branching ratios depend on several
parameters of the model. The new coupling constant $g_R$ of $\mathrm{SU}(2)_R$
is chosen to be equal to $g_L$, as it happens {\em e.g.} in models with
spontaneous parity breaking. Mixing between gauge bosons can be safely
neglected. An additional hypothesis is that the right-handed CKM matrix equals
the left-handed one. The heavy neutrino $N$ is assumed to be lighter than $W_R$
(the other two are assumed heavier) and coupling only to the electron, with a
mixing angle of order unity.

\begin{figure}[htb]
\begin{center}
\epsfig{file=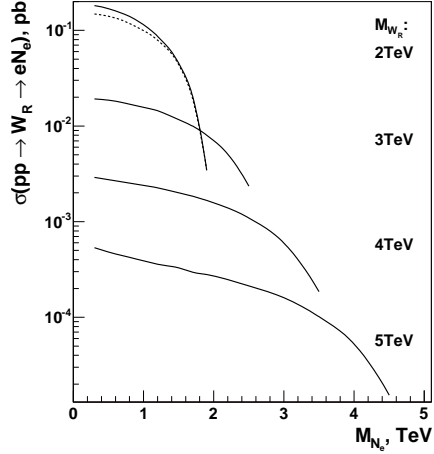,height=6cm,clip=}
\caption{Dependence of $\sigma(pp \to W_R) \times
BR(W_R \to e^\pm N)$ on the heavy neutrino mass, for different
values of $M_{W_R}$.}
\label{fig:NS:cswrd}
\end{center}
\end{figure}

For the signal event generation and calculation of cross sections,
{\tt PYTHIA}\ 6.227
is used with CTEQ5L parton distribution functions, and the model assumptions
mentioned above. The analysis is focused
on the $W_R$ mass region above 1 TeV. The signal cross section, defined as
the product of the total $W_R$ production cross section times the branching
ratio of $W_R$ decay into $e N$, is shown in Fig.~\ref{fig:NS:cswrd}
as a function of $m_N$, for several $W_R$ masses.
For the value $M_{W_R}=2$ TeV, the dashed line illustrates the decrease of the
total cross section (due to the smaller branching ratio for $W_R \to eN$)
for the case of three degenerated heavy neutrinos $N_{1-3}$, mixing with $e$,
$\mu$, $\tau$ respectively. The values $M_{W_R} = 2$ TeV,
$m_N = 500$~GeV are selected as a reference point for the detailed analysis.

\begin{figure}[t]
\begin{center}
\epsfig{file=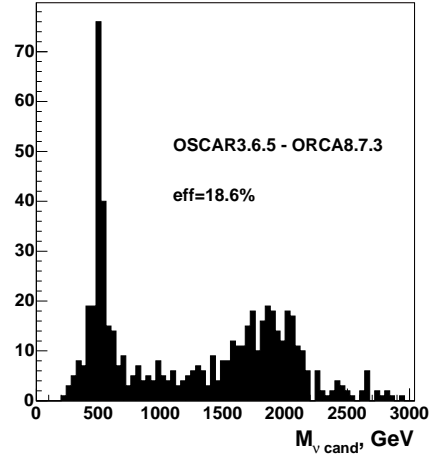,height=6cm,clip=}
\caption{Distribution of the invariant mass $m_{N_e}^{cand}$ for signal events
with a heavy neutrino with $m_N = 500$ GeV. The two possible electron
assignments are shown. The normalization is arbitrary.}
\label{fig:NS:recomnu}
\end{center}
\end{figure}

The detection of signal events is studied using the full CMS detector
simulation and reconstruction chain. For details see
Ref.~\cite{Gninenko:2006br}. The analysis proceeds through the following steps:
\begin{itemize}
\item Events with 2 isolated electrons are selected (standard isolation in the
tracker is required).
\item Events with at least 2 jets are selected. From these jets, the two ones
with the maximum $p_T$ are chosen.
\item Using the 4-momenta of the signal jet pair and the 4-momentum of a lepton,
the invariant mass $M_{ejj} = m_{N_e}^{cand}$ is calculated. Since there are
two electrons, the two $ejj$ combinations are considered. This distribution
is plotted in Fig.~\ref{fig:NS:recomnu}. The tail above 500 GeV corresponds to
a wrong choice of the electron.
\item From the 4-momenta of the jet pair and the electrons, the invariant mass
$M_{eejj} = M_{W_R}^{cand}$ is calculated.
\end{itemize}

Background is constituted by SM processes giving a lepton pair plus jets.
The production of a $Z$ boson plus jets has a large cross section, about 5
orders of magnitude larger than the signal.
In a first approximation, this process can be simulated with {\tt PYTHIA}.
This background is suppressed by a cut on the lepton pair invariant mass
$M_{ee} > 200$ GeV. In order to reduce the number of simulated events, it is
required that the $Z$ transverse momentum is larger than 20 GeV
during the simulation, and events with sufficiently high
$M_{ee}$ are pre-selected at the generator level. Another background is
$t \bar t$ production with dileptonic $W^+ W^-$ decay.
It has been checked that other decay modes do not contribute significantly.
Its cross section is about two orders of magnitude larger than the signal.
It must be pointed out that the Majorana nature of the heavy neutrino allows
to single out the LNV final state with two like-sign leptons. This does not
improve the sensitivity because, although backgrounds are smaller in this case,
the signal is reduced to one half. However, in case
of discovery comparing events with leptons having the same and opposite charges
will be an excellent cross check.

\begin{figure}[htb]
\begin{center}
\begin{tabular}{ccc}
\epsfig{file=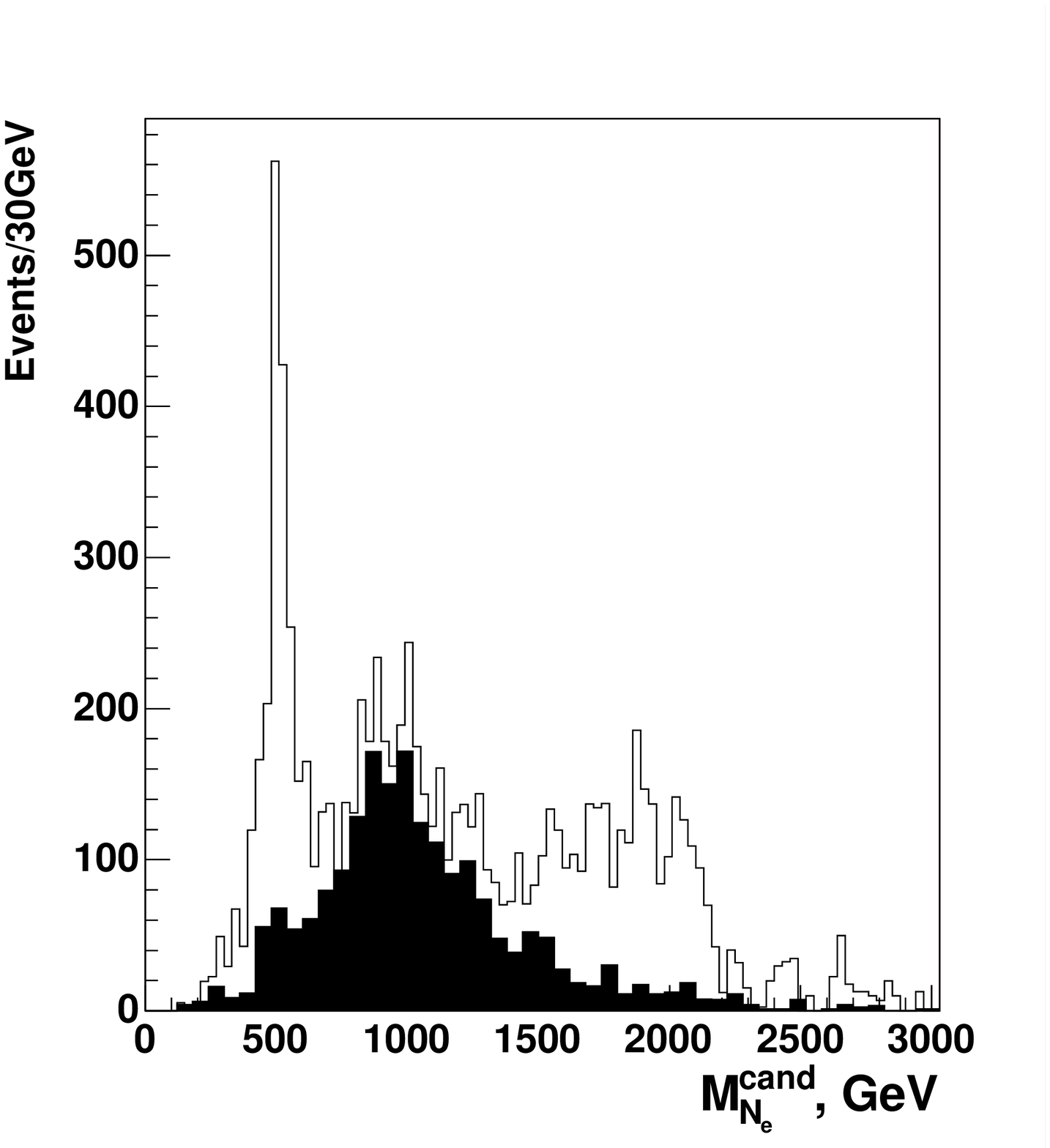,height=6cm,clip=} & \quad &
\epsfig{file=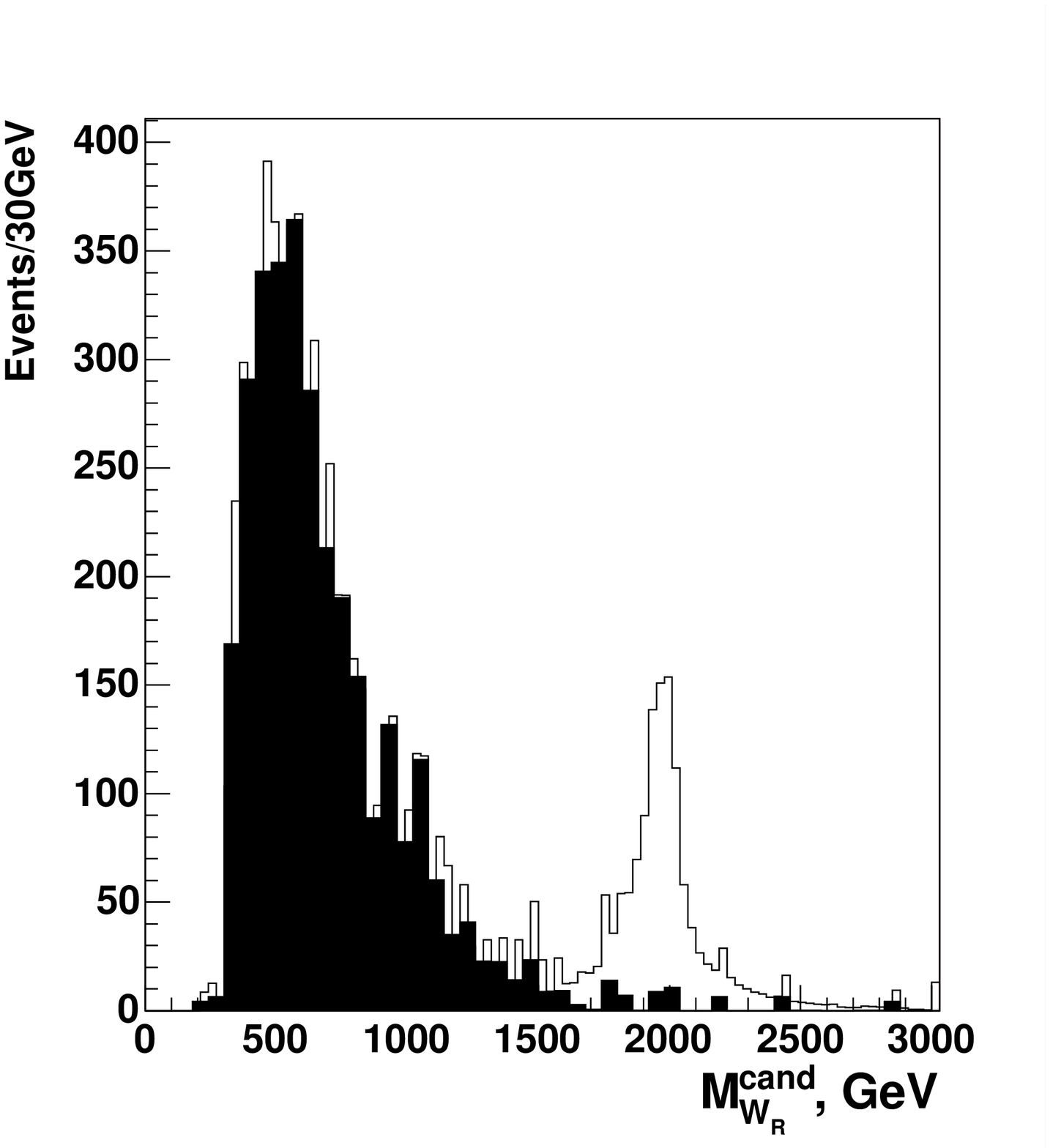,height=6cm,clip=}
\end{tabular}
\caption{Left: reconstructed heavy neutrino mass peak including the SM
background ( histogram) and background only (shaded histogram).
Right: the same for the $W_R$ mass peak. In both cases
an $eejj$ invariant mass above 1 TeV is required. The integrated luminosity
is 30~fb$^{-1}$}
\label{fig:NS:recomnubg2}
\end{center}
\end{figure}

For the values $M_{W_R} = 2$ TeV, $m_N = 500$ GeV selected
the reconstructed $N$ mass peak is well visible, though the background is
significant (comparable to the peak height).
However, if an invariant mass $M_{eejj} > 1$ TeV is required, the background
under the heavy neutrino peak drops dramatically, resulting in the mass
distribution shown in Fig.~\ref{fig:NS:recomnubg2} (left).
The reconstructed $W_R$ mass peak is shown in Fig.~\ref{fig:NS:recomnubg2}
(right).

\begin{figure}[htb]
\begin{center}
\epsfig{file=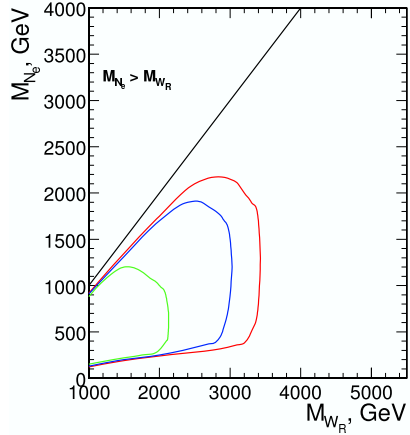,height=6cm,clip=}
\caption{CMS discovery potential for heavy Majorana neutrinos from $W_R$ decays
for integrated luminosities of 30 fb$^{-1}$ (red, outer contour),
10 fb$^{1}$ (blue, middle) and 1 fb$^{-1}$ (green, inner contour).}
\label{fig:NS:sensit}
\end{center}
\end{figure}

The discovery potential is calculated using the criterion
\cite{Bityukov:1998ju}
\begin{equation}
S=2(\sqrt{N_S+N_B}-\sqrt{N_B})\geq 5 \,,
\end{equation}
where $N_S$ and $N_B$ are the numbers of signal and background events
respectively. The discovery
limits in the $(M_{W_R}, m_N)$ plane are shown in Figure \ref{fig:NS:sensit},
for luminosities of 1, 10 and 30 fb$^{-1}$.
After three years of running at low luminosity (30 fb$^{-1}$) this process would
allow to discover $W_R$ and $N$ with masses up to 3.5 TeV and 2.3 TeV,
respectively. For $M_{W_R} = 2$ TeV and $m_N = 500$ GeV discovery could be
possible already after one month of running at low luminosity.

The influence of background uncertainties in these results is small since the
background itself is rather small and the discovery region is usually limited
by the fast drop of the signal cross section at high ratios $m_N/M_{W_R}$ or by
the fast drop of efficiency at small $m_N/M_{W_R}$.
Signal cross section uncertainties from PDFs have been estimated by taking
different PDF sets, finding changes of about 6\% in the discovery region. No
change of acceptance has been observed. Assuming a rather pessimistic value of 
6\% as the PDF uncertainty, it is easy to estimate from Fig.~\ref{fig:NS:cswrd}
that the uncertainty for the upper boundary of the discovery region is
of $1-2$\%, and for the lower boundary of $2-3$\%.

\subsection{Heavy neutrino pair production}

New heavy neutrinos can be produced in pairs by the exchange of an $s$-channel
neutral gauge boson. Since $ZNN$ couplings are quadratically suppressed,
$NN$ production is only relevant when mediated by an extra
$Z'$ boson. For example, in $\mathrm{E}_6$ grand unification both
new $Z'$ bosons and
heavy neutrinos appear. If $M_{Z'} > 2 m_N$, like-sign dilepton signals
from $Z'$ production and subsequent decay $Z' \to NN \to \ell^\pm W^\mp \ell^\pm
W^\mp$ can be sizeable. As it has been remarked before,
 like-sign dilepton signals have moderate (although not negligible) backgrounds.
These are further reduced for heavier neutrino masses, when the charged leptons
from the signal are more energetic and background can be suppressed demanding a
high transverse momentum for both leptons.

A striking possibility happens when the new $Z'$ boson is leptophobic
(see also the next section).
If the new $Z'$ does not couple to light charged leptons
the direct limits from $p \bar p \to Z' \to \ell^+ \ell^-$ searches at
Tevatron do not apply, and the $Z'$ could be relatively light,
$M_{Z'} \gtrsim 350$ GeV. A new leptophobic $Z'$ boson in this mass range could
lead to like-sign dilepton signals observable already at Tevatron.
For LHC, the $5\sigma$ sensitivity reaches $M_{Z'} = 2.5$ TeV, $m_N = 800$ GeV for
a luminosity of 30 fb$^{-1}$
\cite{delAguila:2007ua}.

To conclude this section a final comment is in order.
In the three heavy neutrino production processes
examined we have considered heavy Majorana
neutrinos which are singlets under the SM group (seesaw type I),
produced through standard or new interactions. Majorana neutrinos lead to the
relatively clean LNV signature of two like-sign dileptons, but it should be
pointed out that like-sign dilepton signals arise also in the other seesaw
scenarios: from the single production of doubly charged scalar triplets
(seesaw type II)~\cite{Huitu:1996su}, and in pair production of lepton 
triplets (seesaw type III)~\cite{Bajc:2006ia}. For this reason, like-sign
dileptons constitute an interesting final state in which to test seesaw at LHC.
Of course, additional multi-lepton signatures are characteristic of type II 
(see section~\ref{sec_new_scalars} for a discussion on scalar triplets) and
type III seesaw, and they might help reveal the nature of seesaw at LHC.

%
%
\section{New neutral gauge bosons} 
\label{sec_new_zp}

Many models beyond the SM introduce new neutral gauge bosons, generically
denoted by $Z'$. GUTs with groups larger than $\mathrm{SU}(5)$ always predict
the existence of at least one $Z'$ boson. Their mass is not necessarily of the
order of the unification scale $M_\mathrm{GUT} \sim 10^{15}$ GeV, but on the
contrary, one (or some) of these extra  bosons can be ``light'', that is, at the
TeV scale or below. Well-known examples are $\text{E}_6$ grand
unification~\cite{Hewett:1988xc} and left-right
models~\cite{Langacker:1984dc} (for reviews see
also~\cite{Leike:1998wr,Yao:2006px}). Theories with extra dimensions with gauge
bosons propagating in the bulk predict an infinite tower of KK
excitations $Z^{(n)} = Z^{(1)},Z^{(2)},\dots$,
$\gamma^{(n)} = \gamma^{(1)},\gamma^{(2)},\dots$
The lightest ones $Z^{(1)}$, $\gamma^{(1)}$, can have a mass at the TeV scale,
and a  phenomenology similar to $Z'$ gauge
bosons~\cite{Antoniadis:1994yi,Rizzo:1999en}. Little Higgs models enlarge the
$\text{SU}(2)_L \times \text{U}(1)_Y$ symmetry and introduce new gauge bosons as
well, {\em e.g.} in the littlest Higgs models based on
$[\text{SU}(2) \times \text{U}(1)]^2$ two new bosons $Z_H$, $A_H$
appear, with masses expected in the TeV range.

The production mechanisms and decay modes of $Z'$ bosons depend on their
coupling to SM fermions.\footnote{Decays to new fermions and bosons, if any, are
also possible but usually ignored in most analyses. When included they decrease
the branching ratio to SM fermions, and then they lower the signal cross
sections and discovery potential in the standard modes.}
These couplings are not fixed
even within a class of models. For example, depending on the breaking
pattern of $\text{E}_6$ down to the SM, the lightest
$Z'$ has different couplings to quarks and leptons
or, in other words, quarks and leptons have different $\mathrm{U}(1)'$
hypercharges. Three common breaking patterns are labeled as $\psi$, $\chi$ and
$\eta$, and the corresponding ``light'' $Z'$ as $Z'_\psi$, $Z'_\chi$, $Z'_\eta$.
Thus, the constraints on $Z'$ bosons, as well as the discovery
potential for future colliders refer to particular $Z'$ models.

Present limits on $Z'$ bosons result from precise measurements at the $Z$ pole
and above at LEP, and from the non-observation at Tevatron. $Z$ pole
measurements constrain the $Z-Z'$ mixing, which would induce deviations in the
fermion couplings to the $Z$. For most popular models
the mixing is required to be of order of few $10^{-3}$ \cite{Yao:2006px}
(as emphasised above, limits
depend on the values assumed for the $Z'$ couplings). Measurements
above the $Z$ pole in fermion pair and $W^+ W^-$ production set constraints on
the mass and mixing of the $Z'$. The non-observation at Tevatron in
$u \bar u,d\bar d \to Z' \to \ell^+ \ell^-$ sets lower bounds on $M_{Z'}$. For
most common models they are of the order of $700-800$~GeV
\cite{Abulencia:2006iv}, with an obvious dependence
on the values assumed for the coupling to $u$, $d$ quarks and charged leptons.
LHC will explore the multi-TeV mass region and might discover a $Z'$ with very
small luminosity, for masses of the order of 1 TeV. Below we summarise the
prospects for ``generic'' $Z'$ bosons (for example
those arising in $\text{E}_6$ and left-right models),
which couple to quarks and leptons without
any particular suppression. In this case, $u \bar u,d \bar d
 \to Z' \to e^+ e^-,\mu^+ \mu^-$ gives very clean
signals and has an excellent sensitivity to search for $Z'$ bosons
\cite{Barger:1986hd,delAguila:1986ez,Cvetic:1995zs,Godfrey:2002tn}.
Then we examine the situation when lepton couplings are suppressed, in which
case other $Z'$ decay channels must be explored.

\subsection{$Z'$ bosons in the dilepton channel}

\subsubsection{Discovery potential}

\begin{figure}[htb]
\begin{center}
\begin{tabular}{ccc}
\epsfig{file=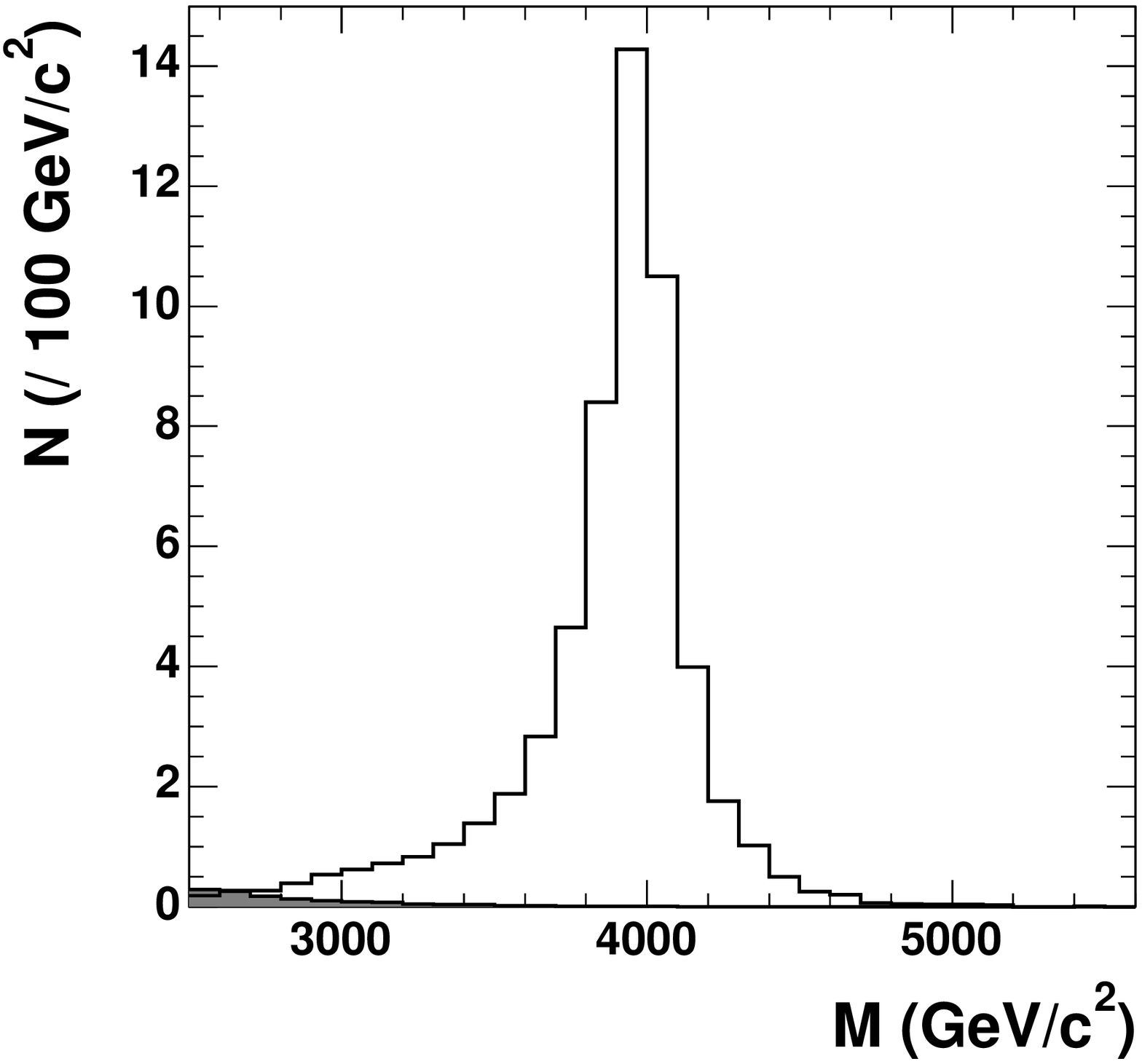,height=5cm,clip=} & \quad
\epsfig{file=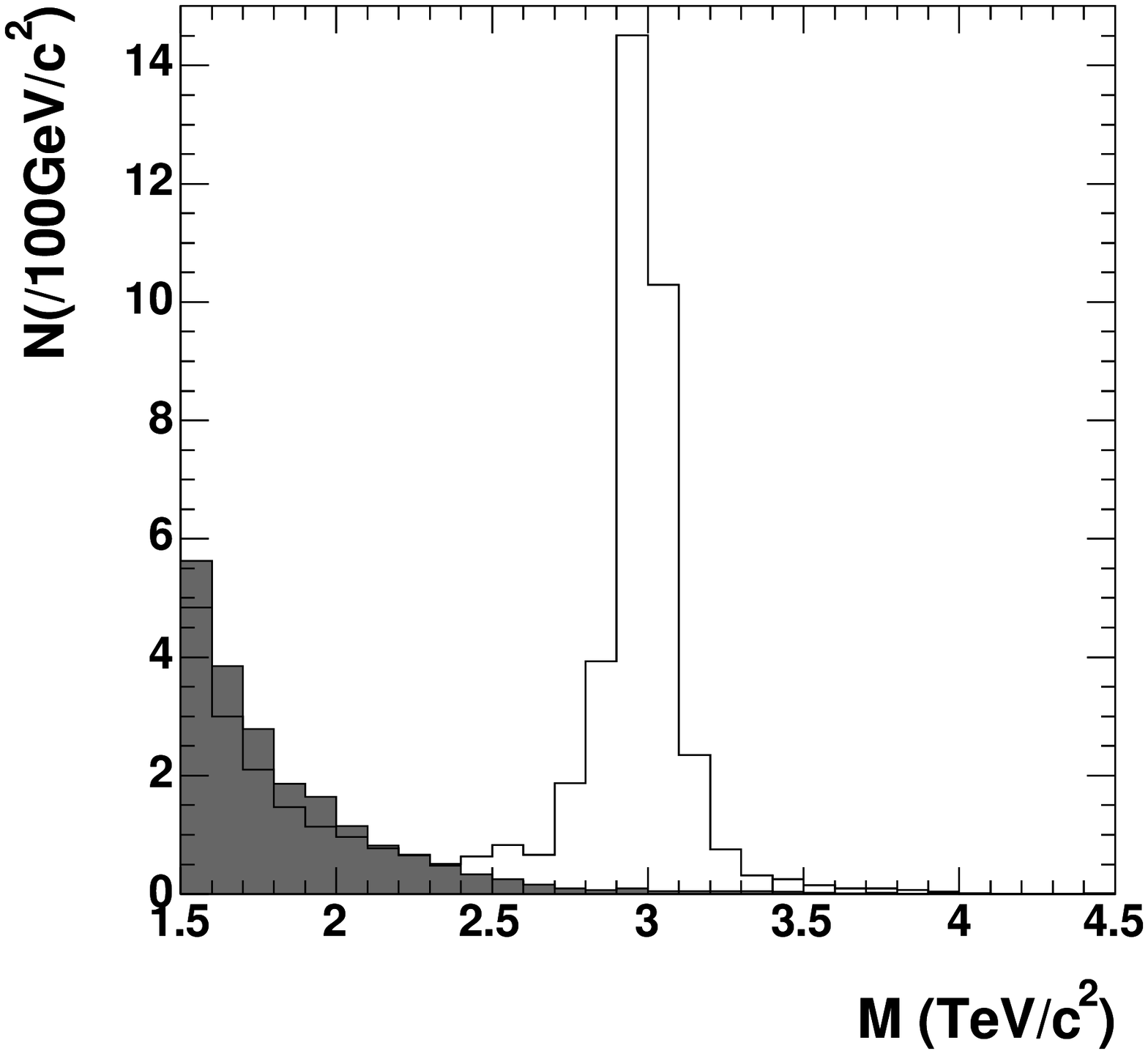,height=5.0cm,clip=}
\end{tabular}
\caption{Resonance signal (white histograms) and Drell-Yan background (shaded
histograms) for KK $Z^{(1)}/\gamma^{(1)}$ boson production with $M = 4.0$ TeV
(left), and
$Z'_\mathrm{SM}$ with $M = 3.0$ TeV (right), with an integrated luminosity of
30 fb$^{-1}$ (from CMS full simulation).}
\label{fig:NS:Zmass}
\end{center}
\end{figure}

The dilepton decay channel provides a clean signature of a $Z'$ boson.
The presence of this heavy particle would be detected
by the observation of a resonance peak in the dilepton mass 
spectrum over the SM background, the largest one coming from the Drell-Yan
process $q\bar{q} \rightarrow \gamma/Z \rightarrow \ell^+\ell^-$.
Reducible backgrounds like QCD jets and $\gamma$-jets can be suppressed 
mainly by applying isolation cuts and requirements on the energy deposited in
the hadronic calorimeter. This is illustrated in Fig.~\ref{fig:NS:Zmass}
for KK excitations of the $Z/\gamma$ and a ``reference''
$Z'_\mathrm{SM}$ (sometimes denoted as $Z'_\mathrm{SSM}$ as well) with the same
couplings as the $Z$, in the $e^+e^-$ decay channel.
These distributions have been obtained with a full simulation of the CMS
detector. More details of the analyses can be
found  in Ref.~\cite{CMS_Note_2006-083} for the $e^+e^-$ channel and in 
Refs.~\cite{CMS_Note_2006-062,CMS_Note_2005-022} for the $\mu^+ \mu^-$ channel.

\begin{figure}[htb]
\begin{center}
\epsfig{file=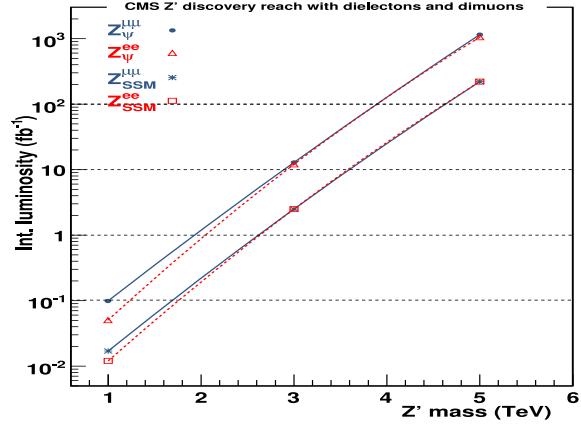,height=6cm,width=8cm,clip=}
\caption{$5\sigma$ discovery limit as a function of the resonance mass for two
examples of $Z'$ bosons, in the $e^+ e^-$ (red, dashed lines) and $\mu^+ \mu^-$
(blue, solid lines) channels  (from CMS full simulation).}
\label{fig:NS:Zdisco}
\end{center}
\end{figure}

The discovery potential is obtained using likelihood estimators
\cite{CMS_Note_2005-004}
suited for small event samples. The $e^+ e^-$ and $\mu^+ \mu^-$ channels provide
similar results, with some advantage for $e^+ e^-$ at lower $Z'$
masses. A comparison between both is given in
Fig.~\ref{fig:NS:Zdisco} for the $\text{E}_6$ $Z'_\psi$ and the reference
$Z'_\mathrm{SM}$. For masses of 1 TeV, a luminosity of 0.1 fb$^{-1}$ would
suffice to discover the $Z'$ bosons in most commonly used scenarios, such as
$Z'_\psi$, $Z'_\chi$, $Z'_\eta$ mentioned above, left-right models and
KK $Z^{(1)}/\gamma^{(1)}$. For a luminosity of 30 fb$^{-1}$, $5\sigma$
significance in the $e^+ e^-$ channel can be
achieved for masses ranging up to $3.3$ TeV ($Z'_\psi$) and 5.5 TeV
($Z^{(1)}/\gamma^{(1)}$). ATLAS studies obtain a similar sensitivity
\cite{Azuelos:2005ym}.
Theoretical uncertainties result from the poor knowledge of PDFs in the high
$x$ and high $Q^2$ domain, and from higher-order QCD and EW corrections
(K factors), and they amount to $10-20$\%. Nevertheless, measurements of real
data outside the mass peak regions will reduce this uncertainty to a large
extent.

\subsubsection{$Z'$ and implications on new physics} 

\begin{figure}[htb]
\begin{center}
\epsfig{file=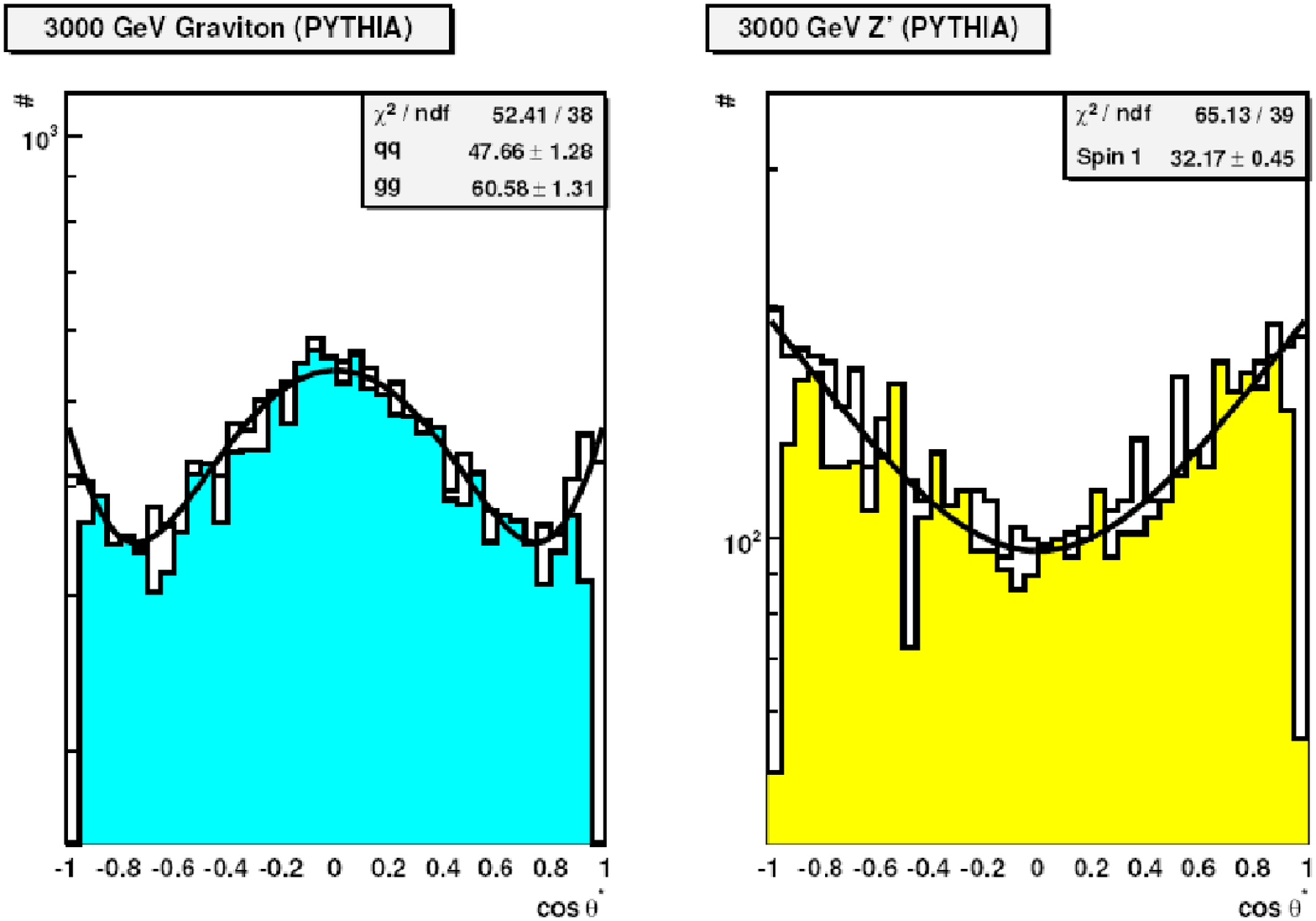,height=6.0cm,width=13.0cm}
\caption{Angular distributions for a 3 TeV graviton (left) and $Z'$ boson
(right) in the dimuon decay channel. Open histograms correspond to
generated-level data, while coloured histograms show events after full CMS
detector simulation and reconstruction. Theoretical fits to Monte Carlo data are
overlayed.}
\label{fig:NS:Zang_distr}
\end{center}
\end{figure}

Once a new resonance decaying to $\ell^+ \ell^-$ ($\ell = e,\mu$) is found,
information about the underlying theory can be extracted with the study of
angular distributions and asymmetries. The first step is the determination of
the particle spin, what can be done with the help of the
$\ell^-$ distribution in the $\ell^+ \ell^-$ rest frame
\cite{Cousins:2005pq}. Let us denote by
$\theta^*$ the angle between the final $\ell^-$ and the initial
quark.\footnote{In $pp$ collisions the quark direction is
experimentally ambiguous because the quark can originate from either proton with
equal probability. The sign ambiguity in $\cos \theta^*$ can be resolved
assuming that the overall motion of the $\ell^+ \ell^-$ system is in the
direction of the initial quark (what gives a good estimation because the
fraction of proton momentum carried by quarks is larger in average) and
taking into
account the probability for a ``wrong'' choice. Additionally, the transverse
momenta of the incoming partons is not known, and it is generally believed
that optimal results are achieved by using the Collins-Soper angle
$\theta_{\rm CS}^*$ \cite{Collins:1977iv} as the estimation for $\theta^*$.}
The $\cos \theta^*$ distribution is obviously flat for a scalar particle.
For a spin-1 particle ($\gamma$, $Z$ or $Z'$)  it is
given by
\begin{equation}
\frac{d\sigma}{d\cos \theta^*} = \frac{3}{8} [1+\cos^2 \theta^*]
+ A_\text{FB} \cos \theta^* \quad \quad (\gamma,Z,Z') \,,
\label{ec:NS:Zcos}
\end{equation}
where the coefficient of the $\cos \theta^*$ term $A_\text{FB}$ depends on the
$Z'$ couplings to quarks and leptons. (The $\cos \theta^*$ forward-backward
asymmetry is equal to this coefficient, hence our choice of notation.) For a
spin-2 graviton $G$ the corresponding distribution is
\begin{equation}
\frac{d\sigma}{d\cos \theta^*} = \frac{5}{8} [1- 3 \epsilon_q \cos^2 \theta^*
+ (\epsilon_g-4 \epsilon_q) \cos^4 \theta^* ] \quad \quad (G) \,.
\label{ec:NS:Gcos}
\end{equation}
The constants $\epsilon_q$ and $\epsilon_g$
are the relative contributions of the two processes in which gravitons
can be produced, $q \bar q \to G$ and $g g \to G$, which are fixed for a given
mass $M_G$ and depend on the PDFs.
The method in Ref.~\cite{Cousins:2005pq} uses only the even terms in
the $\cos \theta^*$ distribution (thus avoiding the dependence on the $Z'$
model
and the $\cos \theta^*$ sign ambiguity). It has been applied to the dimuon
decay channel in Ref.~\cite{Belotelov:2006bh}. Figure~\ref{fig:NS:Zang_distr}
shows the $\cos \theta^*$ distributions for a 3 TeV graviton and $Z'$.
Both distributions are rather different, and the two spin hypotheses
can be distinguished already with a relative small number of events.
Table~\ref{table:NS:Zlumi} contains, for different masses and coupling
parameters $c$ (cross sections are proportional to $|c|^2$), the integrated
luminosity required to discriminate at the $2\sigma$ level between the
spin-1 and spin-2 hypotheses. The cross sections for $Z'$ bosons are assumed to
be equal to the ones for gravitons with the given masses and $c$ values.
In the five cases the required signal is in the
range $150-200$ events, and larger for a larger number $N_B$ of background
events as one may expect. Since the production cross sections fall steeply with
the mass, the integrated luminosity required for spin discrimination increases
with $M$ (and decreases for larger $c$). Distinguishing from the
spin-0 hypothesis (a flat distribution) is harder, and requires
significantly more events than discriminating spin 2 from spin 1, as discussed 
in Ref.~\cite{Cousins:2005pq}.

\begin{table}[htb]
\caption{Number of signal events $N_S$ required to discriminate at the
2$\sigma$ level between the spin-1 and spin-2 hypotheses, in the presence of
$N_B$ background events (see the text). From full CMS detector simulation.}
\begin{center}
\begin{tabular}{ccccc}
$M$ (TeV) & $c$  & L (fb$^{-1}$) & $N_S$   & $N_B$   \\
\hline
1.0       & 0.01 & 50            & 200     & 87     \\
1.0       & 0.02 & 10            & 146     & 16      \\
1.5       & 0.02 & 90            & 174     & 41     \\
3.0       & 0.05 & 1200          & 154     & 22    \\
3.0       & 0.10 & 290           & 154     & 22    \\
\end{tabular}
\label{table:NS:Zlumi}
\end{center}
\end{table}

It should be remarked that, apart from the direct spin determination, a $Z'$ and
a graviton can be distinguished by their decay modes. Indeed, the latter can
decay to $\gamma \gamma$, and the discovery significance in this final state is
equal or better than in the electron and muon channels. On the contrary, $Z' \to
\gamma \gamma$ does not happen at the tree level.

\begin{figure}[htb]
\begin{center}
\begin{tabular}{ccc}
\epsfig{file=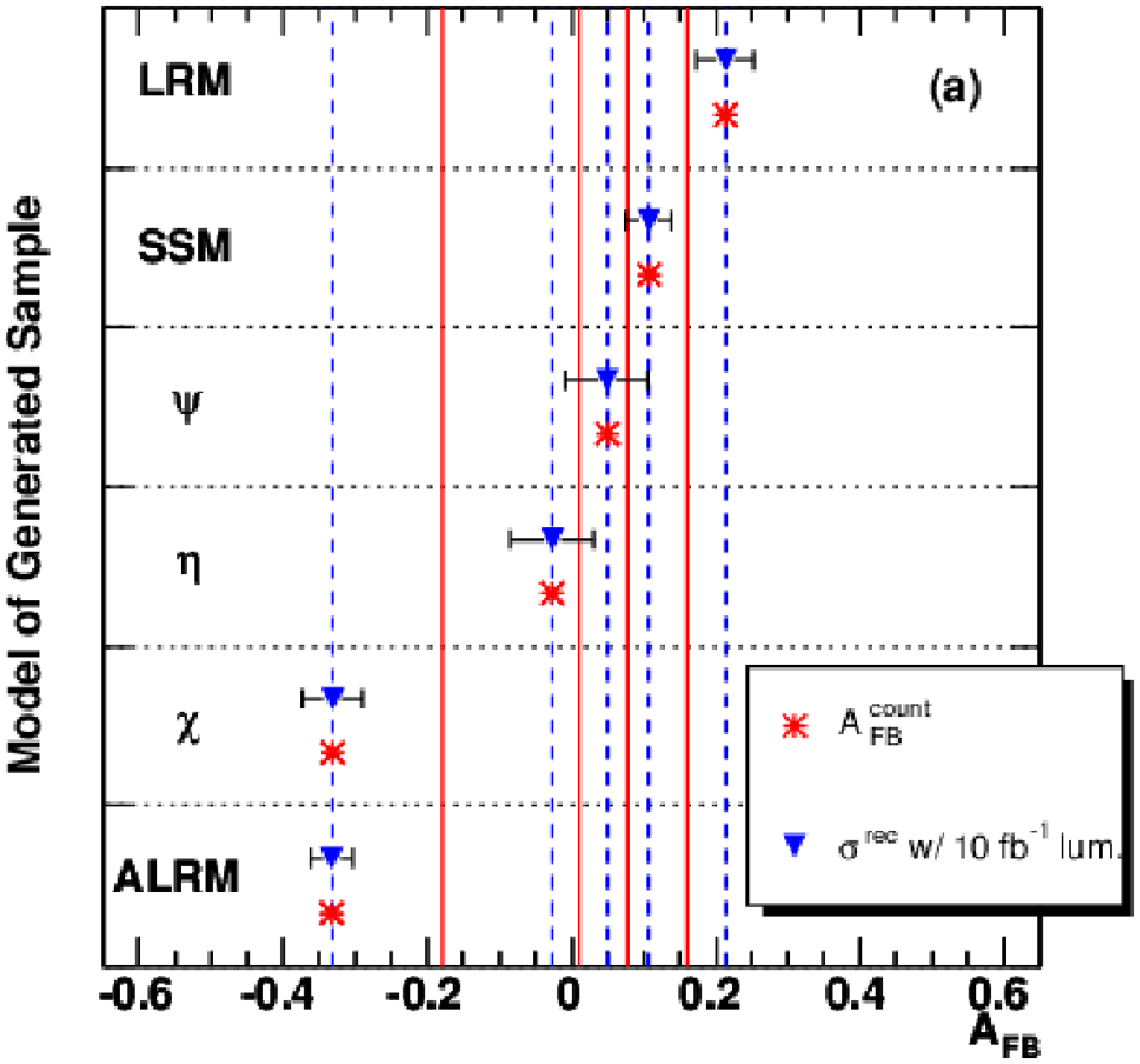,height=5.5cm,width=7.0cm,clip=} &\quad
 & 
\epsfig{file=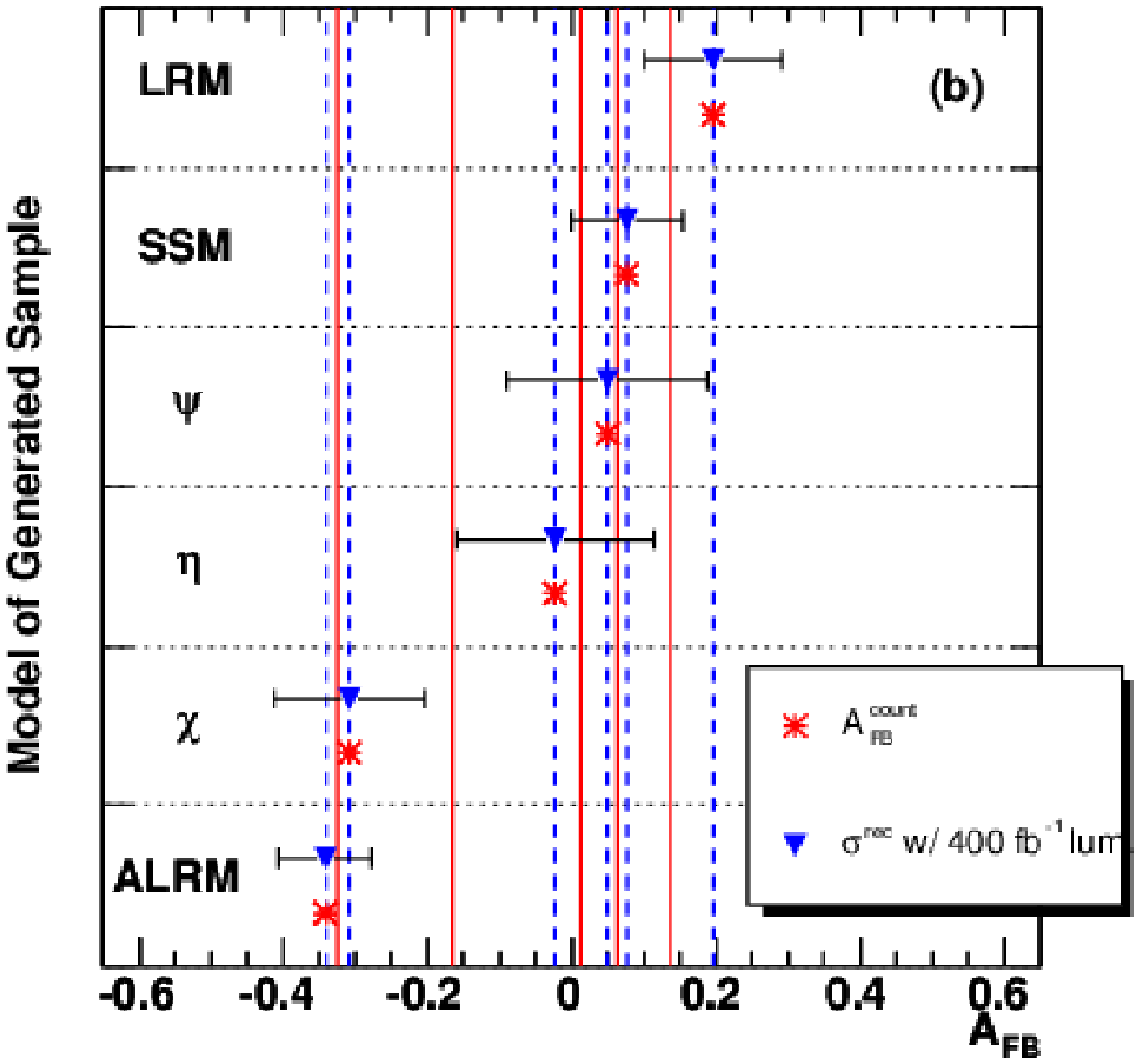,height=5.5cm,width=7.0cm,clip=}
\end{tabular}
\caption{Theoretical values $A_{\rm FB}^{\rm count}$ (dotted lines and
asterisks) and reconstructed values $A_{\rm FB}^{\rm rec}$ (triangles) of the
$A_\text{FB}$ coefficient in Eq.~(\ref{ec:NS:Zcos}), obtained for
different models (see the text), with
$M_{Z'}$ = 1 TeV (left) and  $M_{Z'}$ = 3 TeV (right). The solid vertical lines
are halfway between the adjacent values of $A_{\rm FB}^{\rm count}$. The error
bars on the $A_{\rm FB}^{\rm rec}$ triangles show the $1\sigma$ error
scaled to 10 fb$^{-1}$ (for $M_{Z'} = 1$ TeV) and 400 fb$^{-1}$ (for $M_{Z'} =
3$ TeV). Obtained from CMS full detector simulation.}
\label{fig:NS:Zp_discr}
\end{center}
\end{figure}

The various $Z'$ models are characterised by different parity-violating
$Z'$ couplings to quarks and leptons, reflected in different coefficients of the
linear $\cos \theta^*$ term in Eq.~(\ref{ec:NS:Zcos}).  This
coefficient can be measured with a technique described in
Ref.~\cite{CMS_Note_2005-022} for the dimuon decay channel. $A_{\rm FB}$ is
extracted using an unbinned maximum likelihood fit to events in a suitable
window around the $\mu^+ \mu^-$ invariant mass peak. The fit is based on a
probability density function built from several observables, including
$\cos \theta_{\rm CS}^*$ (as an estimation of the true $\cos \theta^*$).
The values obtained for $A_\text{FB}$ are shown in
Fig.~\ref{fig:NS:Zp_discr} for six different $Z'$ models: the $Z'_\psi$,
$Z'_\chi$ and $Z'_\eta$ from $\text{E}_6$ unification, a left-right model 
(LRM)~\cite{Langacker:1984dc}, an ``alternative left-right model''
(ALRM)~\cite{Ma:1986we} and the ``benchmark'' $Z'_\text{SM}$. 
With an integrated luminosity of 400 fb$^{-1}$ at CMS, one can distinguish 
between either a $Z'_\chi$ or $Z'_\text{ALRM}$ and one of the four other models
with a significance level above 3$\sigma$ up to a $Z'$ mass between 2 and 2.7
TeV. One can distinguish among the four other models up to $M_{Z'} =1-1.5$ TeV, 
whereas $Z'_\text{ALRM}$ and $Z'_\chi$ are indistinguishable for $M_{Z'} > 1$
TeV.

Additional observables, like rapidity distributions~\cite{delAguila:1993ym} or
the off-peak asymmetries~\cite{Rosner:1986cv} can be used to further
discriminate between $Z'$ models.
We finally point out that in specific models the $Z'$ boson may have other
characteristic decay channels, which would then identify the underlying theory
or provide hints towards it. One such example is the decay $Z_H,A_H \to Z h$ in
little Higgs models~\cite{Burdman:2002ns}, which could be
observable~\cite{Azuelos:2004dm}. Contrarily, in $Z'$ models from GUTs this
decay would be generically suppressed by the small $Z-Z'$ mixing, and is
unlikely to happen.

\subsubsection{$Z'$ and fermion masses}

In models which address fermion mass generation, one can go a step further and
try to relate fermion masses with other model parameters. This is the case, for
instance, of extensions of the RS \cite{Randall:1999ee}
scenario,
where the SM fields (except the Higgs boson) are promoted to bulk fields.
If the SM fermions acquire various localisations along the extra dimension,
they provide an interpretation for the large mass hierarchies among the
different flavours.
Within the framework of the RS model with bulk matter, collider
phenomenology and flavour physics are interestingly connected: the effective
4-dimensional couplings between KK gauge boson modes and SM fermions depend
on fermion localisations along the extra dimension, which are fixed
(non-uniquely) by fermion masses.\footnote{Fermion masses are determined up to
a global factor by the fermion localisations (which generate the large
hierarchies) as well as by $3 \times 3$ matrices in flavour space with entries
of order unity. Then, the relation between masses and couplings is not unique,
but involves additional parameters (four $3 \times 3$ matrices).}
Here we test the observability of KK excitations of the photon and $Z$ boson at
LHC in the electron channel, $p p \to \gamma^{(n)} / Z^{(n)} \to e^+ e^-$.
Previous estimations for RS models are given in
Ref.~\cite{Davoudiasl:2000wi}, under  
the simplificating assumption of a universal fermion location.

The fit of EW precision data typically imposes the bound $M_{KK} \gtrsim 10$
TeV \cite{Davoudiasl:2000wi,Burdman:2002gr}. However, if the EW gauge symmetry
is enlarged to ${\rm SU(2)_{L} \times SU(2)_{R} \times U(1)_{X}}$
\cite{Agashe:2003zs}, agreement of the S, T parameters is
possible for $M_{KK} \gtrsim 3$ TeV. 
The localisation of the $(t_L,b_L)$ doublet towards the TeV brane (necessary to
generate the large top quark mass) in principle generates deviations in
the $Z b_L b_L$ coupling (see also the next subsection), what can be avoided
with
a $O(3)$ custodial symmetry \cite{Agashe:2006at}. In the example presented here,
the SM quark doublets are embedded in bidoublets $(2,2)_{2/3}$ under
the above EW symmetry, as proposed in \cite{Agashe:2006at} and in contrast with
Ref. \cite{Agashe:2003zs}. Motivated by having gauge representations
symmetric between the quark and lepton sector,
the lepton doublets are embedded into bidoublets $(2,2)_0$.
This guarantees that there are no modifications of the
$Z \ell_L \ell_L$  couplings. 

\begin{figure}[htb]
\begin{center}
\epsfig{file=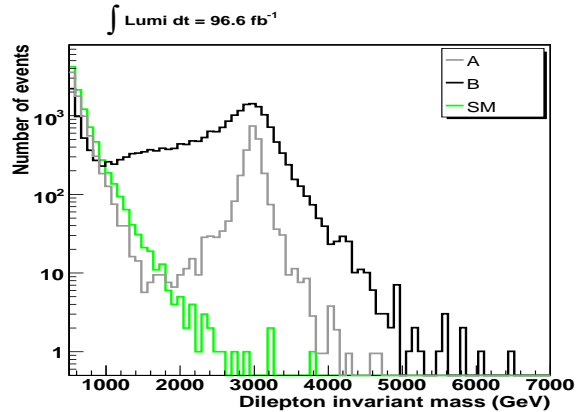,height=5.5cm,width=8cm,clip=}
\caption{Distribution of the $e^+ e^-$ invariant mass for $Z^{(n)}/\gamma^{(n)}$
production in two scenarios (A and B) for the fermion localisations
and the SM background. The number of events corresponds  
to an integrated luminosity of 96.6 fb$^{-1}$.}
\label{fig:NS:ZRS}
\end{center}
\end{figure}

The simulation of $Z^{(n)}/\gamma^{(n)}$ production \cite{Moreau_proc} is
obtained after  implementing the new processes in {\tt PYTHIA}.
Only $n=0,1,2$ are considered, since the contributions of KK excitations
with $n \geq 3$ are not significant.
The cross section depends on the fermion localisations which are clearly
model-dependent.
In Fig. \ref{fig:NS:ZRS} we show the $e^+ e^-$ invariant mass distribution
for two different fermion localisation scenarios labelled as A and B (see
Ref.~\cite{Moreau:2006np}), both with $M_{KK}=3$ TeV. These scenarios are in
agreement with all present data on quark and lepton masses
and mixings \cite{Moreau:2006np}, in the minimal SM extension where neutrinos
have Dirac masses. Furthermore, for both sets FCN processes are below the
experimental limit if $M_{KK} \gtrsim 1$ TeV. In Fig.~\ref{fig:NS:ZRS} we
observe that the signal can be easily extracted from the physical SM
background, as an excess of Drell-Yan events compared to the SM expectation. 

\subsection{$Z'$ in hadronic channels}

$Z'$ bosons with suppressed coupling to leptons (``leptophobic'' or
``hadrophilic'') have theoretical interest on their own. They were first
introduced
some time ago \cite{Chiappetta:1996km,Altarelli:1996pr,Barger:1996kr} on a
purely phenomenological basis, in an attempt
to explain reported $3.5\sigma$ and $2.5\sigma$ deviations in $R_b$ and $R_c$,
respectively, observed by
the LEP experiments at the $Z$ pole. In order to accommodate these deviations
without spoiling the good agreement for the leptonic sector, the $Z'$ couplings
to $b$, $c$ were required to be much larger than those to charged leptons, so
that the deviations in the $Zbb$, $Zcc$ couplings induced by a small $Z-Z'$
mixing were significant for quarks but not for charged leptons. As a bonus, the
introduction of leptophobic $Z'$ bosons seemed to explain an apparent excess of
jet events at large transverse momenta measured by CDF.

With more statistics available the deviations in $R_b$, $R_c$ have disappeared,
and SM predictions are now in good agreement with experiment. Nevertheless,
a $2.7\sigma$ discrepancy in $A_\text{FB}^b$ has remained until now. This
deviation might well be due to some uncontrolled systematic error. But, if one
accepts the $A_\text{FB}^b$ measurement, explaining it with new physics
contributions while keeping the good agreement for $R_b$ is quite hard.
One possibility has recently arised in the context of RS models,
where the introduction of a custodial symmetry \cite{Agashe:2006at} protects
the $Z b_L b_L$ coupling from corrections due to mixing with the $Z^{(1)}$.
$Z b_R b_R$, is allowed to receive a new contribution
from mixing, which could explain the anomaly in $A_\mathrm{FB}^b$.
Alternatively, one may allow deviations in $Z b_L b_L$ and $Z b_R b_R$, chosen
so as to fit the experimental values of $R_b$ and $A_\text{FB}^b$
\cite{Djouadi:2006rk}.
The new $Z^{(1)}$ state has a mass of $2-3$ TeV and suppressed couplings to
charged leptons. Hence, it can be produced at LHC but mainly decays
to quark-antiquark pairs.
Leptophobic $Z'$ bosons can also appear
in grand unified theories as $\mathrm{E}_6$ \cite{Babu:1996vt,delAguila:1986ez}.

\begin{figure}[htb]
\begin{center}
\epsfig{file=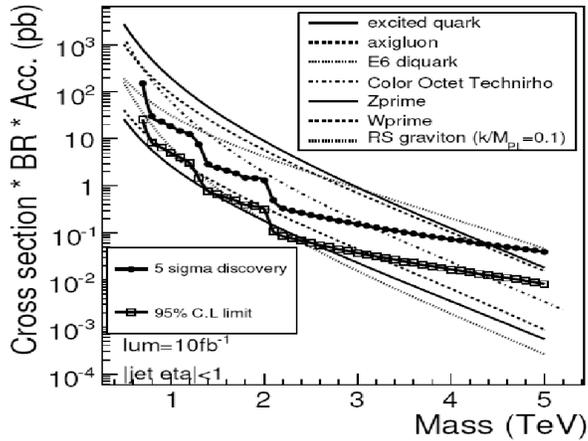,height=6cm,width=8cm,clip=}
\caption{$5\sigma$ discovery limits (circles) and 95\% upper bounds (squares)
for resonances decaying to two jets, as a function of their mass. The luminosity
is of 10 fb$^{-1}$ and a full simulation of the CMS detector is used.
The predictions of several models are also shown.}
\label{fig:NS:Zqq}
\end{center}
\end{figure}

Studies of the CMS sensitivity to narrow resonances in the dijet final states
have been performed \cite{CMS_Note_2006-070}. Experimental searches in 
the dijet channel are
challenging because of the large QCD background and the limited dijet mass 
resolution. All new particles with a natural width significantly smaller than
the  measured dijet mass resolution should all appear as a dijet mass resonance
of the same line shape in the detector. Thus, a generic analysis search has been 
developed to extract cross section sensitivities, which are compared to the
expected cross sections from different models (excited quarks, axigluons,
colorons, ${\rm E}_6$ diquarks, color octet technirhos, $W'$, $Z'$, and
RS gravitons), to determine the mass range for which we expect to be able to 
discover or exclude these models of dijet resonances.
The size of the cross section is a determining factor in whether the model
can be discovered, as illustrated in Fig.~\ref{fig:NS:Zqq} for a sequential
$Z'_\text{SM}$ and other new states. For a luminosity of 10 fb$^{-1}$ the
$Z'_\text{SM}$ signal is
about one order of magnitude below the $5\sigma$ discovery limit for all the
mass range, and a discovery is not possible. Conversely, if agreement is found
with the SM expectation, $Z'_\text{SM}$ masses between 2.1 and 2.5 TeV can be
excluded (see Fig.~\ref{fig:NS:Zqq}).

For resonances decaying to $t \bar t$ preliminary studies have been performed
in Ref.~\cite{ATLAS_Note_2006-033}. With 300 fb$^{-1}$, a 500 GeV resonance
could be
discovered for a cross section (including branching ratio to $t \bar t$) of 1.5
pb. For masses of 1 TeV and 3 TeV, the necessary signal cross sections are
650 and 11 fb, respectively.


\section{New charged gauge bosons}
 \label{sec_new_wp}

Extensions of the SM gauge group including an additional $\mathrm{SU}(2)$ factor
imply the existence of new bosons $W^{'\pm}$ (as well as an extra $Z'$ boson,
whose phenomenology has been described in  the previous section).
Two well-known examples are left-right models, in which the
electroweak gauge group is $\mathrm{SU}(2)_L \times \mathrm{SU}(2)_R \times
\mathrm{U}(1)$, and littlest Higgs models (those with group 
$[\text{SU}(2) \times \text{U}(1)]^2$). As for the neutral case, the
interactions of new $W'$ bosons depend on the specific model considered. 
For example, in left-right models the new charged bosons (commonly denoted as
$W_R$) have purely right-handed couplings to fermions, whereas in littlest
Higgs models they are purely left-handed, as the ordinary $W$ boson.
Low-energy limits are correspondingly different. 
In the former case the kaon mass difference sets
a limit on the $W_R$ mass of the order of two
TeV~\cite{Barenboim:1996nd}. This stringent limit is due to an
enhancement of the ``LR'' box diagram contribution involving $W$ and $W_R$
exchange \cite{Beall:1981ze}, compared to the ``LL'' exchange of two
charged bosons with left-handed couplings. On the other hand, in little Higgs
models (especially in its minimal versions like the littlest Higgs model)
precision electroweak data are quite constraining, and require the $W'$ masses
to be of the order of several TeV~\cite{Csaki:2003si,Hewett:2002px}.

\subsection{Discovery potential}

\begin{figure}[htbp]
\begin{center}
\epsfig{file=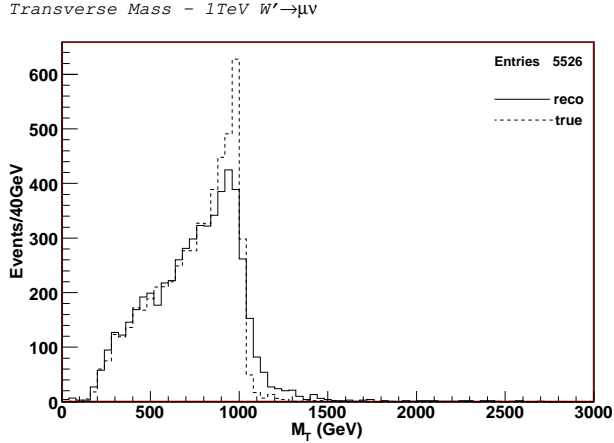,width=8cm,clip=}
\caption{Generated and reconstructed transverse mass distribution for a
simulated 1 TeV $W'$, before any detector effects and after full simulation of
the ATLAS detector.}
\label{fig:NS:mtRT}
\end{center}
\end{figure}

Most studies for $W'$ discovery potential have focused on a $W'$ boson
with SM-like couplings to fermions and $WZ$, $Wh$ decays suppressed.
The present direct limit from Tevatron is
$m_{W'} > 965$ GeV with 95\% CL \cite{D0Wprime}. 
Previous ATLAS studies have shown that a $W'$ boson
could be observed in the leptonic decay channel
$W' \to \ell \nu_\ell$, $\ell=\mu, e$, if it has a mass up to 6 TeV with
100 fb$^{-1}$ of integrated luminosity~\cite{Cousinou:682437}. 
For CMS the expectations are similar \cite{Hof:973105}.
Here the possible
detection of a $W'$ signal in the muon decay channel is investigated,
focusing on masses in the range $1-2.5$ TeV and using the full simulation 
of the ATLAS detector. 
The signal has been generated with {\tt PYTHIA}\ using CTEQ6L structure
functions. The resulting cross sections times branching ratio, as well as
the $W'$ width for various masses, are given in Table \ref{tab:NS:cxSB} (left).
The $W'$ can be identified as a smeared Jacobian peak in the transverse 
mass distribution, built with the muon transverse momentum and the
transverse missing energy $\ptmiss$. Figure \ref{fig:NS:mtRT} shows the
smearing of the edge after full simulation of the ATLAS detector.

\begin{table}[htb]
\begin{center}
\caption{Left: expected cross section times branching ratio for the
$W'\rightarrow\mu\nu$ signal, and total $W'$ width. Right: cross section times
branching ratio for the main background sprocesses.}
\begin{tabular}{lp{.05in}l} 
  \begin{tabular}{ccc}
  \multicolumn{3}{c}{Signal: $pp\rightarrow W'\rightarrow\mu\nu_{\mu}+X$} \\ 
  $m_{W'}$ (TeV) & $ \sigma\times \text{BR}$ (pb) & $ \Gamma_{tot}$ (GeV) \\ \hline
  1.0 & 3.04 & 34.7 \\
  1.5 & 0.57 & 52.6 \\
  2.0 & 0.15 & 70.5 \\
  2.5 & 0.047 & 88.5 \\ 
  \end{tabular}
&&
  \begin{tabular}{ll}
  SM Background processes & $\sigma\times \text{BR}$ (nb) \\ \hline
  $pp\rightarrow W\rightarrow\mu\nu_{\mu}+X$ & 15 \\
  $pp\rightarrow W\rightarrow\tau\nu_{\tau}\rightarrow\mu\nu_{\mu}\nu_{\tau}+X$
   & 2.6 \\
  $pp\rightarrow Z\rightarrow\mu^-\mu^++X$ & 1.5 \\
  $pp\rightarrow Z\rightarrow\tau^-\tau^+\rightarrow\mu+X$ & 0.25 \\ 
  $pp\rightarrow t\bar{t}\rightarrow WbW\bar{b}\rightarrow l\nu_l+X $ & 0.46 \\
  QCD (all di-jet processes) & $5\times10^5$ \\ 
  \end{tabular}
\end{tabular}
\label{tab:NS:cxSB}
\end{center}
\end{table}

In addition to the signal, there are contributions 
from the various SM backgrounds originating from the processes given in
Table~\ref{tab:NS:cxSB} (right). The $W$ background is irreducible, but all the
other backgrounds can be reduced applying the appropriate
selections. In Table~\ref{tab:NS:cf} the selection cuts used for the  
background rejection are shown.

\begin{figure}[htb]
\begin{center}
\epsfig{file=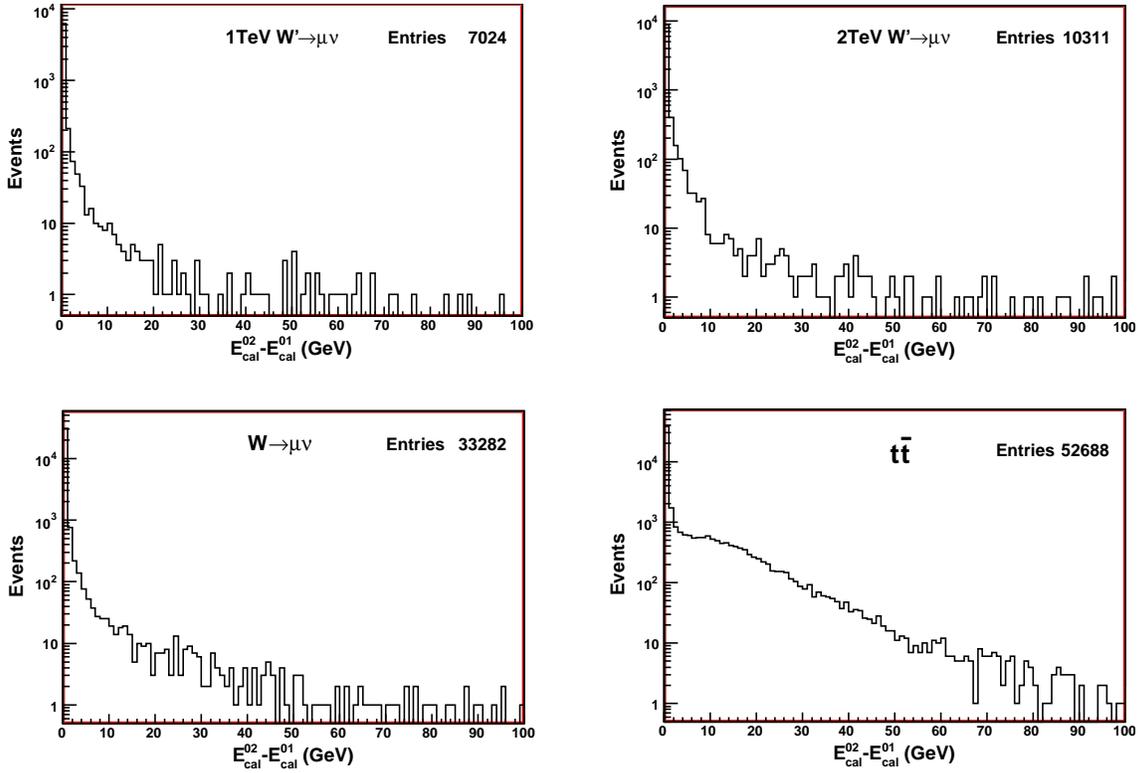,width=15cm,clip=}
\caption{Distribution of calorimetric energy contained in the difference of
two cones with $\Delta R=0.1$ and $\Delta R=0.2$ for both signal and background
events.}
\label{fig:NS:iso}
\end{center}
\end{figure}

The main signature of the signal is the presence of an energetic muon 
together with a significant missing transverse momentum in the event.
When searching for a $W'$ with mass of 1 TeV or heavier, events that contain
at least one reconstructed muon with $p_T >100$ GeV and missing transverse
momentum
$\ptmiss > 50$ GeV are selected. These cuts mainly eliminate the $t \bar t$
background, which tends to have less energetic muons, and $Z$ production,
which in general does not have significant missing energy. Muons coming from
$W'$ decays are isolated, {\em i.e.} they do not belong to a jet. Isolated
muons are identified by requiring that the calorimetric energy deposited
inside the difference of a small and a bigger cone around the muon track is
less than $E_\text{cal}^{02}-E_\text{cal}^{01} < 10$ GeV, where the cones `01'
and `02' are determined, respectively, as the ones which have
$\Delta R=0.1,0.2$.
This double cone strategy is adopted because muons from $W'$ decays are very
energetic and therefore can have significant, almost collinear radiation. 
Figure \ref{fig:NS:iso} shows the distribution of calorimetric energy contained
in the difference of the two cones for both signal and background. 
It is evident that the above cut reduces mainly the $t \bar t$ background.
Events with exactly one isolated muon are selected. $Z$ background events
contain mostly two isolated muons,  except for the cases where one of the muons
lies in a region outside the muon spectrometer ($|\eta|>2.7$) or is not
reconstructed. These cases account for about the $30\%$ of the high mass $Z$
events and remain as irreducible background. QCD and $t \bar t$ backgrounds
contain in most cases non-isolated muons coming from jets. In order to
eliminate the QCD di-jet background, which contains one jet misidentified
as a muon, events with additional high energy jets, with $p_T>200$ GeV,
are rejected (JetVeto).
The $t \bar t$ background is further reduced by applying
a $b$-jet veto cut (in ATLAS the jet tagging is done for jets with $p_T>15$
GeV).
Muons coming from cosmic rays and $b$-decays are rejected with track quality
criteria, what ensures that the muon track is well reconstructed. Specifically,
cuts are applied on the
$\chi^2$ probability over the number of degrees of freedom and the transverse
$d_0$ and longitudinal $z_0$ perigee parameters:
$Prob(\chi^2)/DoF > 0.001$, $d_0/\Delta(d_0)<10$, $z_0<300$ mm.

\begin{table}[htb]
\begin{center}
\caption{Cross-section times branching ratio to muons and relative
efficiencies after each cut. The cuts correspond to: (1) $p_T > 100$ GeV and
$\ptmiss >50$ GeV; (2) $b$-jet Veto; (3) JetVeto; (4) muon isolation and
quality.}
\begin{tabular}{c|cc|cc|cc|cc|cc}
  & \multicolumn{2}{|c}{1 TeV $W'$} & \multicolumn{2}{c|}{2 TeV $W'$}
  & \multicolumn{2}{c|}{$W$ (off-shell)} & \multicolumn{2}{c|}{$t \bar{t}$}
  & \multicolumn{2}{c}{$Z$ (off-shell)} \\
\hline
  cut & $\sigma$ (pb) & eff (\%) & $\sigma$ (pb) & eff (\%) & $\sigma$ (pb)
  & eff (\%) & $\sigma$ (pb) & eff (\%) & $\sigma$ (pb) & eff (\%) \\
1 & 2.52 & 82.8 & 0.126 & 84.0 & 2.04 & 74.4 & 8.878 & 1.93 & 0.251 & 9.89 \\ 
2 & 2.45 & 80.7 & 0.122 & 81.4 & 1.99 & 72.8 & 1.610 & 0.35 & 0.244 & 9.62 \\
3 & 2.23 & 73.3 & 0.104 & 69.4 & 1.95 & 71.1 & 0.966 & 0.21 & 0.237 & 9.34 \\
4 & 2.18 & 71.6 & 0.101 & 67.3 & 1.91 & 69.8 & 0.736 & 0.16 & 0.232 & 9.15 \\
\end{tabular}
\label{tab:NS:cf}
\end{center}
\end{table} 

\begin{figure}[htb]
\begin{center}
\epsfig{file=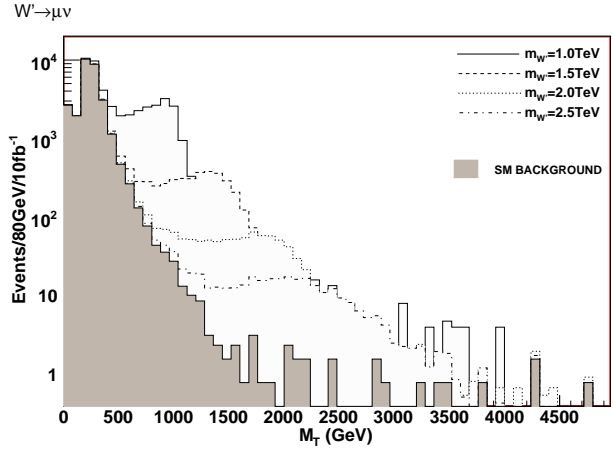,width=8cm,clip=}
\caption{Transverse mass distribution of the SM background and $W'$ signals
corresponding to different $W'$ masses, 
plotted on top of the background for an integrated luminosity of 10 fb$^{-1}$.}
\label{fig:NS:mtD}
\end{center}
\end{figure}

After the application of all the signal separation requirements the transverse 
mass distribution, shown in Fig.~\ref{fig:NS:mtD}, has been statistically 
analysed to determine the significance of the discovery. First, 
for each $W'$ mass the transverse mass interval which gives the best discovery 
significance is determined. The corresponding number of 
signal and background events for 10 fb$^{-1}$ are presented
in Table~\ref{tab:NS:nSB}. 
The minimum luminosity to have a $5\sigma$ significant discovery is 
also calculated and shown in Table~\ref{tab:NS:lum}. The significance is 
calculated assuming Poisson statistics. The errors in the luminosity 
correspond to a $5\%$ systematic uncertainty in the signal
(mainly due to the variation of PDF's) and 
a $20\%$ systematic uncertainty in the background (due to several 
different contributions). The uncertainties in the NLO corrections 
($K$ factors) are expected to influence both the signal and the background in
a similar way.
The experimental systematic uncertainties are expected to be reduced only 
after the first data taking using the control samples of $Z$ and $W$ events. 
A control sample will also be formed in the transverse mass region between
200 and 400 GeV, which will provide the final check for the systematic
uncertainties  collectively, concerning the scale as well as the shape of the
background.

\begin{table}[htb]
\begin{center}
\caption{Number of signal and background events expected for 10 fb$^{-1}$ of
integrated luminosity,  for various $W'$ masses. The best search windows in the
transverse mass distribution $(M_T)$ are also shown.}
\begin{tabular}{lcccc}
$M_{W'}$ & 1.0 TeV & 1.5 TeV & 2.0TeV & 2.5TeV \\
\hline
$M_T$ (TeV) & $0.6 - 1.7$ & $0.9 - 2.0$ & $1.2 - 2.9$ & $1.6 - 3.2$ \\
Signal Events & $15753 \pm 787$ & $3059 \pm 153$ & $603 \pm 30$
  & $225 \pm 11$ \\
SM Background Events & $469 \pm 94$ & $76 \pm 15$ & $22 \pm 5$ & $15 \pm 3$ \\ 
\end{tabular}
\label{tab:NS:nSB}
\end{center}
\end{table} 

\begin{table}[htb]
\begin{center}
\caption{Minimum luminosity required in order to have a $5\sigma$ discovery for
various $W'$ masses. $N_S$, $N_B$ stand for the number of signal and background
events, respectively, within the optimal transverse mass window.}
\begin{tabular}{cccc}
$M_{W'}$ (TeV) & Luminosity (pb$^{-1}$) & $N_S$ & $N_B$ \\
\hline
1.0 & $3.0 \pm 0.3$ & 4.7 & 0.14 \\ 
1.5 & $14.6 \pm 1.4$ & 4.5 & 0.11 \\ 
2.0 & $84 \pm 9$ & 5.1 & 0.18 \\  
2.5 & $283 \pm 31$ & 6.4 & 0.42 \\ 
\end{tabular}
\label{tab:NS:lum}
\end{center}
\end{table} 

A new $W'$ boson with SM-like couplings to fermions can be
discovered with low integrated luminosity during the initial LHC running. With
0.3 fb$^{-1}$ integrated luminosity,
a $W'$ can be discovered in the ATLAS experiment with a mass up to 2.5 TeV.
Imposing the additional requirement of observing at least 10 candidate signal
events would rise the minimum luminosity to 0.5 fb$^{-1}$.

The present study so far has been performed without pile-up 
and cavern background conditions.  Both these conditions are not 
expected to affect much the results since the initial run will be 
at very low luminosity and moreover the majority of the muons of 
the signal concentrate in the barrel region. Nevertheless, studies 
for the fake reconstruction with both kinds of background 
are under way. 

Finally, we point out that
the experimental resolution for muons with $p_T$ ranging from 0.5 to 
1~TeV is about $5-10\%$, giving an experimental width larger than the 
intrinsic width, shown in Table \ref{tab:NS:cxSB} (left). Therefore no further
attempt has been made for discriminating the underlying theory based on the
$W'$ width. However, following
the $W'$ discovery, the muonic decay channel could provide valuable information
concerning the FB asymmetry,
which in turn could be used to discriminate between various theoretical models.

%
%

\section{New scalars}
 \label{sec_new_scalars}

Additional scalars appear in theories beyond the SM to solve some of the 
problems presented in the introduction. A selection of these models and 
their goals are:
\begin{itemize}
\item{2 Higgs Doublet models: explain the origins of the CP asymmetry}
\item{Little Higgs models: solve the hierarchy problem}
\item{Babu-Zee model: explain the sources of the neutrino mass differences}
\end{itemize}

The 2 Higgs Doublet Model (2HDM) contains  two Higgs fields, one to give
mass to SM gauge bosons and the other one remaining with CP violating
terms \cite{Wu:1994ja}. The additional 2 neutral Higgs particles aim to solve the strong 
CP problem and explain the observed baryon asymmetry of the universe with
minimum impact to the SM. Such a model can be easily investigated at LHC
via either direct observation of the non-SM Higgs particles or indirectly
via the enhancement to the FCNC Higgs decays involving the top quark. The
details of such a discovery and of possible discrimination between the models can
be found in \chapt{chap:top}{top:2hdm}.

Little Higgs  models  \cite{Arkani-Hamed:2001nc,Arkani-Hamed:2002qy,Perelstein:2005ka}
aim to solve the hierarchy problem arising from the rather large 
loop corrections to the tree level Higgs mass, without imposing a symmetry
between fermions and bosons. Instead, the unwanted contributions from the loops
are removed via the same spin counterparts of the involved SM particles:
top quark, W and Z bosons and the Higgs itself. 
The coupling coefficients of these predicted particles are connected to their 
SM counterparts via the symmetries of the larger group embedding the SM gauge
group. 
Depending on the selection of the embedding group, these  models predict a variety of new particles. 
Additional charge +2/3 quarks (studied in subsection \ref{subsec_q_2_3}), 
a number of spin 1 bosons and a number of scalars, with masses less than 2, 5 and 10 TeV respectively. 
The smallest of these symmetry groups defines the Littlest Higgs model which predicts 
three nearly degenerate scalar particles with charges 2, 1 and 0. 
Experimentally, the doubly charged scalar is the most appealing one,
since its manifestation would be two like-sign leptons or $W$ bosons 
when produced singly \cite{Huitu:1996su,Maalampi:2002vx,Azuelos:2004dm},
 or two like-sign lepton pairs with equal invariant mass when produced
in pairs \cite{Akeroyd:2005gt,Azuelos:2004dm,Hektor:2007uu}.
More generally, scalar triplets appear in various type II seesaw models.
For scalar triplets in supersymmetric models see
\chapt{chap:susy}{slep:triplet}.

The Babu-Zee model, independently proposed by Zee \cite{Zee:1985id} and
Babu \cite{Babu:1988ki},  proposes a  particular radiative mass generation mechanism. 
This mechanism might help understanding the origin of neutrino masses
and mixing angles which are firmly established by the neutrino oscillation experiments 
\cite{Fukuda:1998mi,Ahmad:2002jz,Eguchi:2002dm,Apollonio:2002gd}.
The model introduces two new charged scalars $h^+$ and $k^{++}$, both singlets under $SU(2)_{L}$,
which couple only to leptons. 
Neutrino masses in this model arise at the two-loop level. 
Since present experimental neutrino data requires at least one neutrino to have a mass of the order of
${O}(0.05)$ eV \cite{Maltoni:2004ei} an estimation for
the value of neutrino masses in the model indicates that for
couplings $f$ and $h$ of order ${O}(1)$
(see Eq.~(\ref{eq:yuks})) the new scalars should have masses
in the range ${O}(0.1-1)$ TeV (see ref.
\cite{AristizabalSierra:2006gb}). The model is
therefore potentially testable at the LHC.

\subsection{Scalar triplet seesaw models}

An important open issue to be addressed in the context of little Higgs
models~\cite{Arkani-Hamed:2001nc,Arkani-Hamed:2002qy,Perelstein:2005ka}
is the origin of neutrino 
masses~\cite{Han:2005nk,delAguila:2005yi,Abada:2005rt}.
A neutrino mass generation mechanism  which naturally occurs in these
models is type II
seesaw~\cite{Cheng:1980qt,Schechter:1980gr,Gelmini:1980re},
which employs a scalar with the $\text{SU}(2)_L\times \text{U}(1)_Y $ quantum
numbers $\Phi\sim (3,2)$. The existence of such a multiplet in some 
little Higgs models~\cite{Arkani-Hamed:2002qy,Chang:2003zn} 
is a direct consequence of the global $[\text{SU}(2) \times \text{U}(1)]^2$
symmetry breaking which makes the SM Higgs 
light. 
Although $\Phi$ is predicted to be heavier than the SM Higgs boson, 
the little Higgs philosophy implies that its mass could be of order
$O(1)$~TeV. Due to its specific quantum numbers the triplet Higgs boson
only couples to the left-chiral lepton doublets $L_i\sim (2,-1)$, $i=e, \mu, \tau$,
via Yukawa interactions given by 
\begin{equation}
\mathcal{L}_\Phi=i\bar L^c_{i} \tau_2  Y_{ij} (\tau\cdot \Phi) L_{j} 
+ \text{h.c.} \,,
\label{L}
\end{equation}
where $Y_{ij}$ are Majorana Yukawa couplings. 
The interactions in Eq.~(\ref{L})
induce LFV decays of charged leptons which have not been
observed. The most stringent constraint on the Yukawa couplings comes from the 
upper limit on the tree-level decay $\mu\to eee$~\cite{Huitu:1996su,Yue:2007kv}
\begin{equation}
Y_{ee}Y_{e\mu} <3 \times 10^{-5} (M_{\Phi^{++}}/\text{TeV})^2 \,,
\end{equation}
with $M_{\Phi^{++}}$ the mass of the doubly charged scalar, constrained
by direct Tevatron searches to be
$M_{\Phi^{++}} \ge 136$~GeV \cite{Acosta:2004uj,Abazov:2004au}.
Experimental bounds on the 
tau Yukawa couplings are much less stringent.

According to Eq.~(\ref{L}), the neutral component of the 
triplet Higgs boson $\Phi^0$ couples to left-handed neutrinos.
When it aquires a VEV $v_\Phi$, it induces nonzero 
neutrino masses $m_\nu$ given by the mass matrix
\begin{equation}
(m_\nu)_{ij} = Y_{ij} v_\Phi \,.
\end{equation}
We assume that the smallness of neutrino masses is due to the smallness of 
$v_\Phi.$ 
In this work the tau Yukawa coupling is taken to be $Y_{\tau\tau}=0.01$, and
the rest of couplings are scaled accordingly. In particular, hierarchical
neutrino masses imply $Y_{ee},Y_{e\mu} \ll Y_{\tau \tau}$, consistent with
present experimental bounds.

In this framework there is a possibility to perform direct tests of the
neutrino mass generation mechanism at LHC, via pair production and
subsequent decays of scalar triplets. Here the Drell-Yan pair production
of the doubly charged
component
\begin{equation}
pp\to\Phi^{++}\Phi^{--}
\label{production}
\end{equation}
is studied, followed by leptonic decays~\cite{Gunion:1989in,Huitu:1996su,Muhlleitner:2003me,
Akeroyd:2005gt,Rommerskirchen:2007jv,Hektor:2007uu,Han:2007bk}.
Notice that in this process
(i) the production cross section only depends on $M_{\Phi^{++}}$ and known
SM parameters;
(ii) the smallness of $v_\Phi$ in this scenario, due to the smallness of
neutrino masses,  implies that decays $\Phi^{++}\to W^+W^+$ are negligible;
(iii) the $\Phi^{++}$ leptonic decay branching fractions do not depend on
the size of the Yukawa couplings but only on their ratios, which are known
from  neutrino oscillation experiments. For normal
hierarchy of neutrino masses and a very small value of the lightest neutrino
mass, the triplet seesaw model predicts
$\text{BR}(\Phi^{++} \to \mu^+ \mu^+) \simeq
\text{BR}(\Phi^{++} \to \tau^+ \tau^+) \simeq
\text{BR}(\Phi^{++} \to \mu^+ \tau^+) \simeq 1/3$. 
This scenario is testable at LHC experiments.

The production of the doubly-charged scalar has been implemented in the
{\tt PYTHIA}\ Monte Carlo generator~\cite{Sjostrand:2000wi}. Final and initial
state interactions and hadronisation have been taken into account. 
Four-lepton backgrounds with 
high $p_T$ leptons arise from three SM
processes: $t \bar t$, $t \bar t Z$ and $ZZ$ production.
{\tt PYTHIA}\ has been used to generate $t \bar t$ and $ZZ$ background, while
{\tt CompHEP}\ was used to generate the $t \bar t Z$ background via its {\tt PYTHIA}\
interface~\cite{Boos:2004kh,Belyaev:2000wn}. CTEQ5L parton distribution
functions have been used. 
Additional four-lepton backgrounds exist involving $b$-quarks in the final
state, for example, $b \bar b$ production. Charged leptons from 
such processes are very soft, and these backgrounds can be
eliminated~\cite{CMSTDR}. 
Possible background processes from physics beyond the SM
are not considered.

A clear experimental signature is obtained from the peak in the invariant mass 
of two like-sign muons and/or tau leptons:
\begin{equation}
(m_{\ell_1 \ell_2}^\pm)^2=(p_{\ell_1}^\pm + p_{\ell_2}^\pm)^2,
\label{NS:ST:eq1}
\end{equation}
where $p_{\ell_1,\ell_2}^\pm$ are the four-momenta of two like-sign leptons
$\ell_1^\pm$, $\ell_2^\pm$.
Since like-sign leptons originate from decay of a doubly charged Higgs boson,
their invariant mass peaks around $m_{\ell \ell'}^\pm = M_{\Phi^{\pm \pm}}$
in the case of the signal.
The four-muon final state allows to obtain invariant masses directly from
Eq.~(\ref{NS:ST:eq1}). In channels involving one or several tau leptons, which are
seen as $\tau$-jets or secondary muons (marked as $\mu'$ below), the momenta
of the latter has to be corrected according to the equation system
\begin{eqnarray}
{\vec p}_{\tau_i} &=& k_i \vec{p}_{\text{jet}_i / \mu'_i} \label{NS:ST:eq2} \,,\\
{\vec \ptmiss} &=& \sum_i ({\vec p}_{\nu_i})_T \label{NS:ST:eq3} \,,\\
m_{\ell_1 \ell_2}^+ &=& m_{\ell_1 \ell_2}^- \label{NS:ST:eq4} \,,
\end{eqnarray}
where $i$ counts $\tau$ leptons, $({\vec p}_{\nu_i})_T$ is the vector of
transverse momentum of the produced neutrinos, ${\vec \ptmiss}$ is the
vector of missing transverse momentum (measured by the detector) and
$k_i > 1$ are positive constants. Eq.~(\ref{NS:ST:eq2}) describes the standard approximation that the the decay products of a highly boosted $\tau$
are collinear. 
Eq.~(\ref{NS:ST:eq3}) assumes missing transverse energy only to be comprised of neutrinos from $\tau$ decays. In general, it is not a high-handed simplification, because the other neutrinos in the event are much less energetic and the detector error in missing energy is order of magnitude
smaller~\cite{DellaNegra:942733a}. Using the first two equations
it is possible to reconstruct up to two $\tau$ leptons per event. The additional requirement of Eq.~(\ref{NS:ST:eq4}) allows to reconstruct a third $\tau$,
although very small measurement errors are needed.

A clear signal extraction from the SM background can be achieved using a set of selection rules imposed on a reconstructed event in the following order:
\begin{itemize}
\item \emph{S1}: events with at least two positive and two negative muons or jets which have $|\eta| < 2.4$ and $p_T > 5$ GeV are selected.
\item \emph{S2}: The scalar sum of transverse momenta of the two most
energetic muons or $\tau$ jets has to be larger than a certain value
(depending on the $\Phi^{++}$ mass range studied).
\item \emph{S3}: If the invariant mass of a pair of opposite charge muons or
$\tau$ jets is close to the $Z$ boson mass ($85-95$ GeV), then the particles
are eliminated from the analysis.
\item \emph{S4}: as $\Phi^{\pm \pm}$ are produced in pairs, their reconstructed invariant masses  have to be equal (in each event). The condition
\begin{equation}
0.8 < m_{\ell_1 \ell_2}^{++} / m_{\ell_1 \ell_2}^{--} < 1.2.
\label{NS:SC:eq5}
\end{equation}
has been used. If the invariant masses are in this range then they are
included in the histograms, otherwise it is assumed that some muon may
originate from
$\tau$ decay, and it is attempted to 
find corrections to their momenta according to the method described above.
\end{itemize}

\begin{figure}[htb]
\begin{center}
\includegraphics[width=0.47\textwidth]{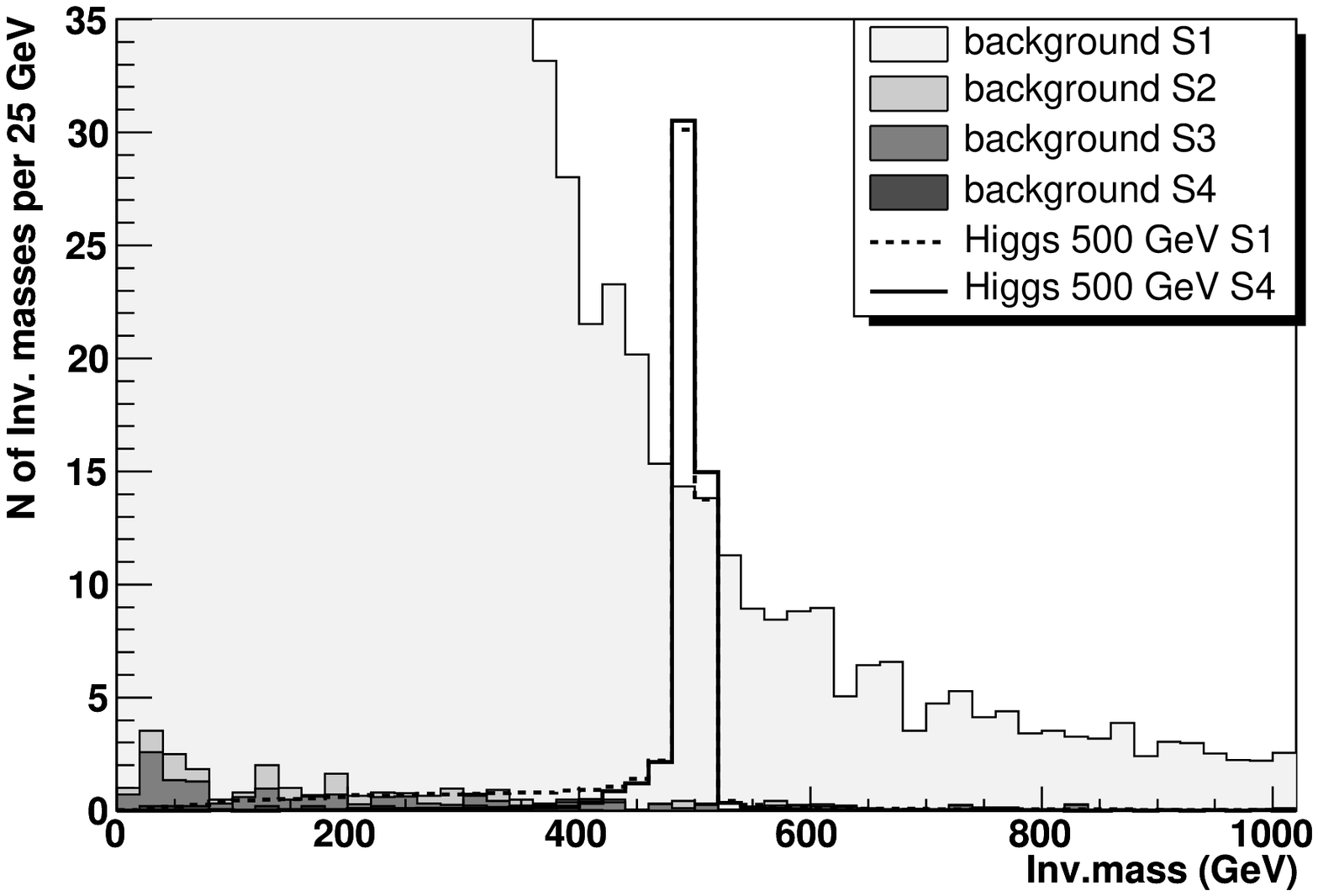}
\hfill
\includegraphics[width=0.47\textwidth]{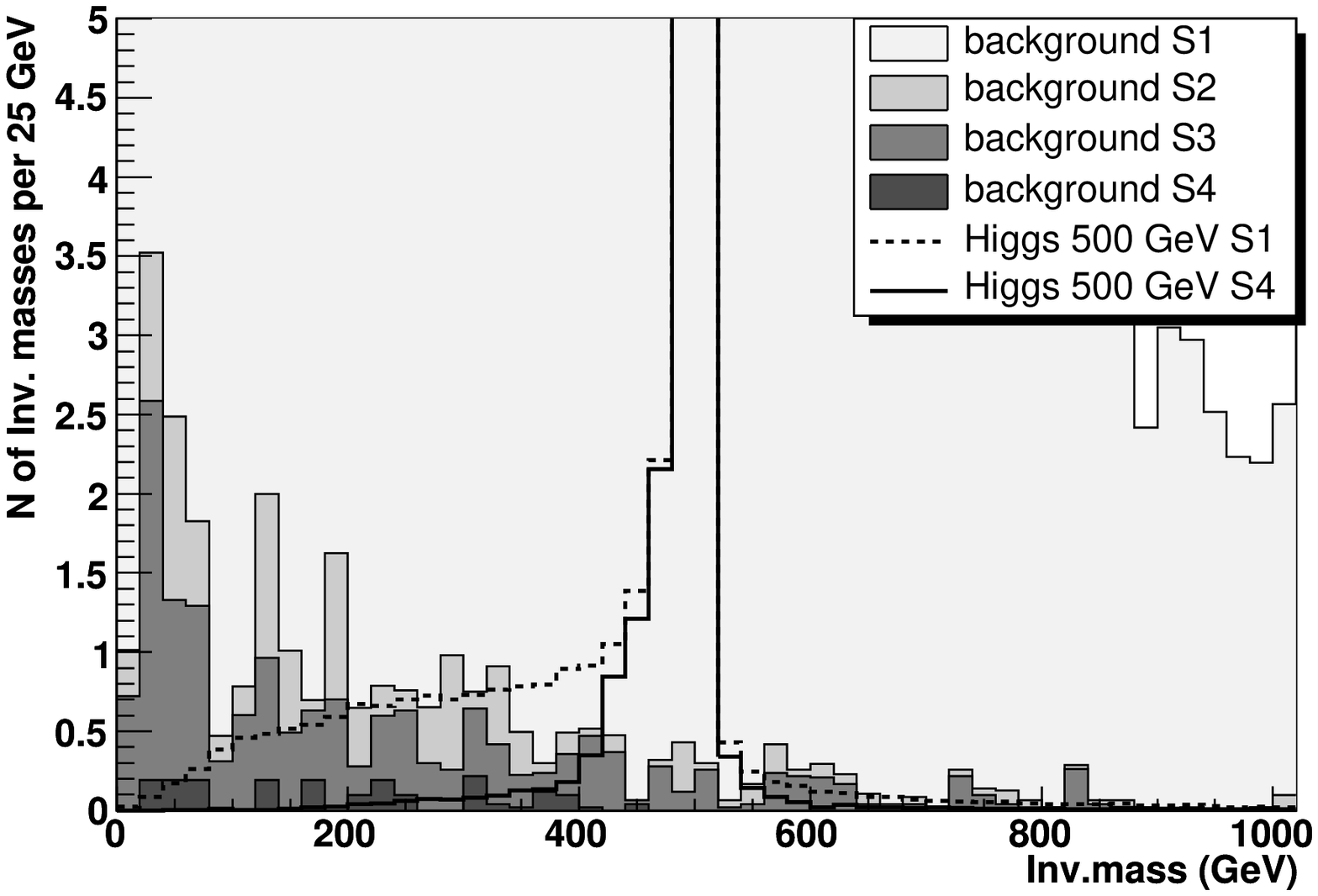}
\caption{Distribution of invariant masses of like-sign pairs
after applying selection rules (S1--S4) for scalar mass $M_{\Phi^{++}} = 500$ GeV and the SM background
(L=30 fb$^{-1}$). The histogram in the right panel is a zoom of the left
histogram to illustrate the effects of the selection rules S2--S4.}
\label{fig:NS:f3}
\end{center}
\end{figure}

An example of invariant mass distribution after applying selection rules is
shown in  Fig.~\ref{fig:NS:f3} for $M_{\Phi^{++}} = 500$ GeV.
A tabulated example is given for $M_{\Phi^{++}} = 200$, 500 and 800 GeV in
Table~\ref{tab:NS:t5}, corresponding to a 
luminosity $L=30$ fb$^{-1}$. The strength of the S2 cut is clearly visible:
almost no decrease in signal while the number of the background events
descends close to its final minimum value. A peculiar behavior of S4
 --- reducing the background, while also increasing the signal in its
peak --- is the effect of applying the $\tau \to \mu'$ correction method
described above. 

\begin{table}[htb]
\begin{center}
\caption{Effectiveness of the selection rules for the background and signal. All event 
numbers in the table are normalized for $L=30$ fb$^{-1}$. 
The numbers in brackets mark errors at 95\% confidence level for Poisson statistics. The signal increases after S4 due to the reconstructed $\tau \to \mu'$ decays.}
\begin{tabular}{|l|l|l|l|l|l|}
\hline
Process & \multicolumn{5}{c|}{N of like-sign pairs} \\\cline{2-6}
        & N of $\Phi$ & S1 & S2 & S3 & S4 \\
\hline
\multicolumn{6}{|c|}{Energy range $150-250$ GeV} \\
\hline
$M_\Phi$=200 GeV & 4670 & 1534 & 1488 & 1465 & 1539 \\
$t \bar t \to 4\ell$ & - & 1222 (168) & 172 (8.5) & 134 (6.9) & 17.6 (3.7) \\
$t \bar t Z$ & - & 21.3 (4.0) & 15.5 (1.0) & 6.3 (1.2) & 2.2 (1.1) \\
$ZZ$ & - & 95.0 (12.0) & 22.5 (0.7) & 9.8 (0.5) & 1.7 (0.2) \\
\hline
\multicolumn{6}{|c|}{Energy range $375-625$ GeV} \\
\hline
$M_\Phi$=500 GeV & 119.2 & 48.4 & 47.5 & 46.8 & 49.5 \\
$t \bar t\to 4\ell$ & - & 178 (28) & 2.1 (0.9) & 1.65 (0.87) & 0.10 (0.35) \\
$t \bar t Z$ & - & 6.6 (1.7) & 2.3 (1.0) & 1.0 (1.0) & 0.00 (0.1) \\
$ZZ$ & - & 9.4 (2.9) & 1.4 (0.2) & 0.68 (0.19) & 0.08 (0.09) \\
\hline
\multicolumn{6}{|c|}{Energy range $600-1000$ GeV} \\
\hline
$M_\Phi$=800 GeV & 11.67 & 5.05 & 5.00 & 4.92 & 5.21 \\
$t \bar t\to 4\ell$ & - & 77 (12) & 0.00 (0.22) & 0.00 (0.22) & 0.00 (0.07) \\
$t \bar t Z$ & - & 2.6 (1.2) & 0.39 (0.4) & 0.39 (0.4) & 0.00 (0.1) \\
$ZZ$ & - & 2.5 (0.8) & 0.34 (0.16) & 0.17 (0.09) & 0.00 (0.02) \\
\hline
\end{tabular}
\end{center}
\label{tab:NS:t5}
\end{table}

As it is seen in Table \ref{tab:NS:t5}, the SM background can be 
practically eliminated. In such an unusual situation
the log-likelihood ratio (LLR) statistical method
\cite{Read:2000ru,Barate:2003sz} has been used to
determine the $5\sigma$ discovery potential, demanding a significance
larger than $5\sigma$ in 95\% of ``hypothetical experiments'', generated
using a Poisson distribution. With this criterion,
$\Phi^{++}$ up to 300 GeV can be discovered in the first year of 
LHC ($L=1$ fb$^{-1}$) and $\Phi^{++}$ up to 800 GeV can be discovered for the integrated luminosity $L=30$ fb$^{-1}$.
Therefore the origin of neutrino mass can possibly be directly tested at
LHC.

\subsection{The discovery potential of the Babu-Zee model}
The new charged scalars of the model introduce new gauge
invariant Yukawa interactions, namely
\begin{equation}
  \label{eq:yuks}
  {\cal L} = f_{\alpha\beta} \epsilon_{ij} (L^{Ti}_{\alpha}CL^{j}_{\beta})h^+
  + h'_{\alpha\beta}(e^T_{\alpha}Ce_{\beta})k^{++} + {\rm h.c.}
\end{equation}
Here, $L$ are the standard model (left-handed) lepton doublets,
$e$ the charged lepton singlets, $\alpha ,\beta$ are
generation indices and $\epsilon_{ij}$ is the completely
antisymmetric tensor. Note that $f$ is antisymmetric while
$h'$ is symmetric. Assigning $L=2$ to $h^-$ and $k^{--}$,
eq.  (\ref{eq:yuks}) conserves lepton number. Lepton number
violation in the model resides only in the following term
in the scalar potential
\begin{equation}
\label{eq:scalar}
{\cal L} = - \mu h^+h^+k^{--} + {\rm h.c.}
\end{equation}
Vacuum stability arguments can be used to derive an upper bound
for the lepton number violating coupling $\mu$
\cite{Babu:2002uu}, namely, $\mu \le (6\pi^2)^{1/4}m_h$.
The structure of Eq. (\ref{eq:yuks}) and Eq. (\ref{eq:scalar})
generates Majorana neutrino masses at the two-loop level
(see ref. \cite{AristizabalSierra:2006gb} and \cite{Babu:2002uu}
for details).

Constraints on the parameter space of the model come from
neutrino physics experimental data and from the experimental
upper bounds on lepton flavour violation (LFV) processes.
Constraints on the antisymmetric couplings $f_{xy}$
are entirely determined by neutrino mixing angles and depend
on the hierarchy of the neutrino mass spectrum,
which in this model can be normal or inverse.
Analytical expressions, as well as numerical upper and
lower bounds, for the ratios $\epsilon=f_{13}/f_{23}$
and  $\epsilon'=f_{12}/f_{23}$ were calculated in
references~\cite{Babu:2002uu} and
\cite{AristizabalSierra:2006gb}.

The requirement of having a large atmospheric mixing angle
indicates that the symmetric Yukawa couplings
$h_{xy}$ ($x, y=\mu,\tau$) must follow the hierarchy
$h_{\tau\tau} \simeq (m_{\mu}/m_{\tau})h_{\mu\tau}
\simeq (m_{\mu}/m_{\tau})^2h_{\mu\mu}$. The couplings
$h_{ee}$, $h_{e\mu}$ and $h_{e\tau}$ are constrained by LFV
of the type $l_a\rightarrow l_bl_cl_d$
and have to be smaller than $0.4$, $4\cdot 10^{-3}$
and $7\cdot 10^{-2}$ \cite{Babu:2002uu}.
The most relevant constraint on $m_k$ come from
the LFV processes $\tau\to3\mu$ while for
$m_h$ is derived from $\mu\to e\gamma$. Lower bounds for both
scalar masses can be found (see ref. \cite{AristizabalSierra:2006gb}),
the results are $m_k\gtrsim 770$ GeV, $m_h\gtrsim 200$ GeV
(normal hierarchy case) and $m_h\gtrsim 900$ GeV (inverse hierarachy case).
\begin{figure}[t]
\centering
\includegraphics[width=6.6cm, height=6cm]{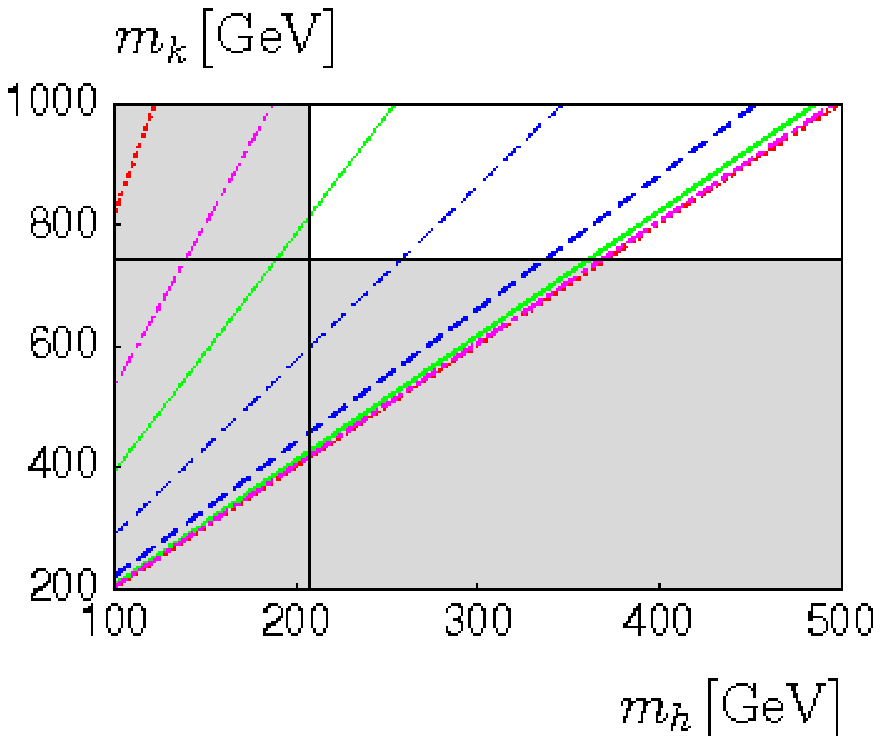}
\hspace{0.2cm}
\includegraphics[width=6.6cm, height=6cm]{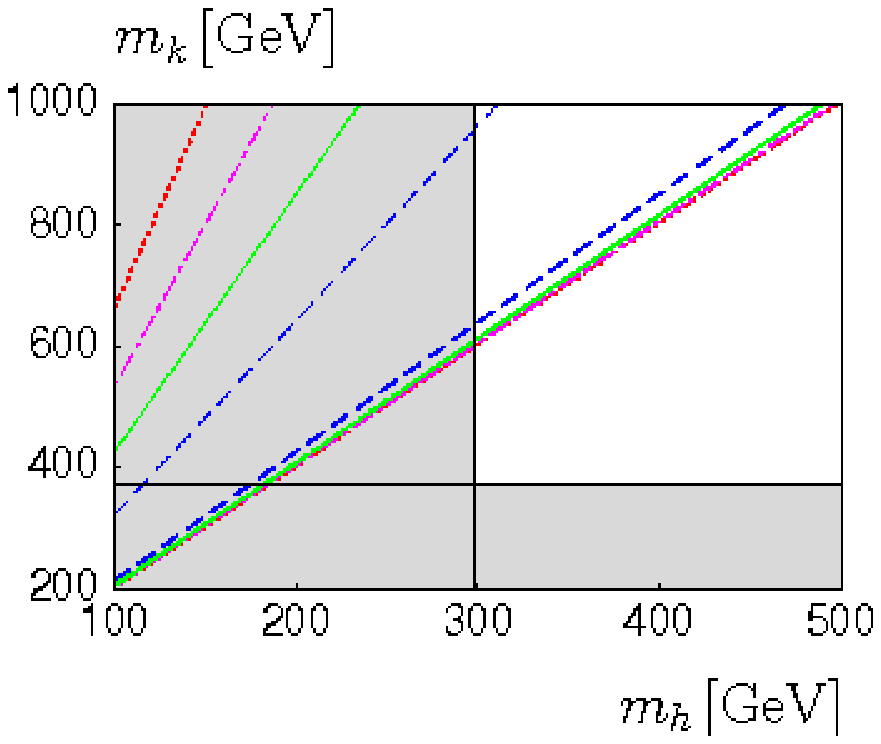}
\caption{Lines of constant BR$(k^{++} \rightarrow h^{+}h^{+})$,
assuming to the left $h_{\mu\mu}=1$: $\text{BR}^{hh}_k=0.1,0.2,0.3$ and $0.4$
for dotted, dash-dotted, full and dashed line. The vertical line
corresponds to $m_h=208\,\mbox{GeV}$ for which $\text{BR}(\mu\to e\gamma)=
1.2\times10^{-11}$ and horizontal line to $m_k=743\,\mbox{GeV}$ for
which $\text{BR}(\tau\to 3\mu)=1.9\times10^{-6}$, i.e. parameter combinations
to the left/below this line are forbidden. Plot on the right assumes
$h_{\mu\mu}=0.5$. Lines are for $\text{BR}^{hh}_k=0.4,0.5,0.6$ and $0.7$,
dotted, dash-dotted, full and dashed line. Again the shaded regions
are excluded by BR$(\mu\to e\gamma)$ and BR$(\tau\to 3\mu)$.}
\label{fig:brkhh}
\end{figure}
In \cite{Babu:2002uu} it has been estimated that at the LHC discovery of
$k^{++}$ might be possible up to masses of $m_{k}\le 1$ TeV approximately.
In the following it will therefore be assumed that $m_{k}\le 1$ TeV and,
in addition, $m_h \le 0.5$ TeV. The notation
$\text{BR}(h^{+} \rightarrow l_\alpha\sum_{\beta}{\nu_\beta})=
\text{BR}_h^{l_\alpha}$ and
$\text{BR}(k^{++} \rightarrow l_\alpha l_\beta)=\text{BR}_k^{l_\alpha l_\beta}$ will be used.
$h^{+}$ decays are governed by the parameters $\epsilon$ and
$\epsilon'$. Using the current 3$\sigma$ range for neutrino
mixing angles \cite{Maltoni:2004ei} it is possible to predict
\begin{eqnarray}
  \label{hbr}
  \text{BR}_h^e=[0.13,0.22],\quad
  \text{BR}_h^{\mu}=[0.31,0.50],\quad
  \text{BR}_h^\tau=[0.18,0.35],\\
  \label{hbr1}
  \text{BR}_h^e=[0.48,0.50],\quad
  \text{BR}_h^{\mu}=[0.17,0.34],\quad
  \text{BR}_h^\tau=[0.18,0.35].
\end{eqnarray}
For normal hierarchy (eq. (\ref{hbr})) or inverse hierarchy
(eq. (\ref{hbr1})).

The doubly charged scalar decay either to two same-sign
leptons or to two $h^+$ final states. Lepton pair final states decays
are controlled by the $h_{\alpha\beta}$ Yukawa couplings while
the lepton flavour violating decay $k^{++}\to h^+h^+$
is governed by the $\mu$ parameter (see Eq.~(\ref{eq:scalar})).
The hierarchy among the couplings $h_{\mu\mu}, h_{\mu\tau}$ and
$h_{\tau\tau}$ result in the prediction
\begin{equation}
  \label{eq:kratios}
  \text{BR}_k^{\mu\tau}/BR_k^{\mu\mu}\simeq(m_\mu/m_\tau)^2,\quad
  \text{BR}_k^{\tau\tau}/BR_k^{\mu\mu}\simeq(m_\mu/m_\tau)^4.
\end{equation}
Thus, the leptonic final states of $k^{++}$ decays are mainly like-sign
muon pairs.

Here it is important to remark that in general the decays
$k^{++}\to e^{+} l^+$ ($l=e, \mu, \tau$) are strongly
suppressed due to the LVF constraints on the $h_{e l}$ parameters.
However, if the Yukawa coupling $h_{ee}$ saturates its upper limit
then electron pair final states can be possibly observed.

The branching ratio for the process $k^{++}\to h^+h^+$
reads
\begin{equation}
  \label{eq:brkdech}
  \text{BR}(k^{++} \rightarrow h^{+}h^{+}) \simeq
  \frac{\mu^2\beta}{m_{k}^2h_{\mu\mu}^2+\mu^2\beta}.
\end{equation}
Here $\beta$ is the usual phase space suppression factor. From
eq.~(\ref{eq:brkdech}) it can be noted that if the process is
kinematically allowed the lepton violating coupling $\mu$ can
be measured by measuring this branching ratio. Here it should be
stressed that for $h_{\mu\mu}\lesssim 0.2$ the current limit on
BR$(\mu\to e\gamma)$ rules out all $m_h \lesssim 0.5$ TeV,
thus this measurement is possible only for
$h_{\mu\mu}\gtrsim 0.2$. Note that smaller values of $\mu$
lead to smaller neutrino masses, thus upper
bounds on the branching ratio for $\text{BR}^{hh}_k$ can be
interpreted as upper limit on the neutrino mass in this model.
Figure \ref{fig:brkhh} shows the resulting branching ratios
for 2 values of $h_{\mu\mu}$.

\chapter{Tools}
 {\small F.~Krauss, F.~Moortgat, G.~Polesello}
\label{chap:tools}

\section{Introduction}

\noindent
In the following, contributions highlighting the treatment of flavour aspects in 
publicly available calculational tools used in New Physics studies at colliders 
will be listed.  Such tools cover a wide range of applications; roughly speaking
there are a wide variety of computer programs discussed here:
\begin{itemize}
\item Analytical precision calculations:\\
	There, the results of analytical precision calculations for
	specific observables, often at loop level, are coded and thus
	made available for the public.  These observables usually are
	sets of single numbers, such as cross sections, decay widths,
	brancing ratios etc., calculated for a specific point in the
	respective models parameter space.  Examples for such tools
	covered here are {\tt HDECAY}, {\tt SDECAY},
	{\tt FchDecay} and {\tt FeynHiggs}.
\item Tools helping in or performing (mostly) analytical calculations:\\
	The best-known example for such a tool, the combinatin of
	{\tt FeynArts} and {\tt FormCalc} and its treatment of
	flavour aspects is discussed here.  In principle,
	{\tt FeynArts} allows for a automated construction of
	Feynman diagrams, including higher-order effects, and the
	corresponding amplitude.  {\tt FormCalc} can then be used
	to evaluate the loop integrals in a semi-automated fashion.
\item RGE codes:\\
	There, the renormalisation group equation is solved
	numerically in order to obtain from high-energy inputs the
	SUSY parameters at lower, physical scales.  These parameters
	usually are coupling constants, particle masses and widths and
	mixing matrices.  For this purpose, a number of codes exist,
	here {\tt SPheno} and {\tt SuSpect} are presented. It
	should be noted that many of these RGE codes also embed a
	number of relevant cross sections, branching ratios etc..
\item Matrix element generators/Parton level generators:\\
	These codes calculate, in a automated fashion, cross sections
	for multi-leg tree-level processes. Usually, they are capable
	of generating weighted or unweighted events at the parton
	level, i.e.\ without showering or hadronisation. This task is
	usually left for other programs, the neccessary information is
	passed by some standardised interface
	format~\cite{Boos:2001cv}. Examples for this type of code
	include {\tt CalcHep} and {\tt HvyN}.
\item Full-fledged event generators:\\
	These programs provide fully showered and hadronised
	events. Primary examples include {\tt PYTHIA},
	{\tt HERWIG}, and {\tt Sherpa}.
\end{itemize}
In addition interfaces are necessary to transfer data between the
various programs as will be discussed in the next section.

\section{A Brief Summary of The SUSY Les Houches Accord 2}
\label{sec:slha2}

The states and couplings appearing in the general minimal
supersymmetric standard model (MSSM) can be defined in a number of
ways. Indeed, it is often advantageous
to use different choices for different applications and hence 
no unique set of conventions prevails at present. In principle,
this is not a problem; translations between different conventions
can usually be carried out without ambiguity. From the point of
view of practical application, however, such translations
are, at best, tedious, and at worst they introduce an unnecessary
possibility for error. 

To deal with this problem, and to create a more transparent situation
for non-experts, the original SUSY Les Houches Accord (SLHA1) was proposed
\cite{Skands:2003cj}. This accord uniquely defines a 
set of conventions for supersymmetric models together with a common interface
between codes. However, SLHA1 was designed exclusively with the MSSM with real
parameters and R-parity conservation in mind. Some recent public
codes~\cite{Dreiner:1999qz,Skands:2001it,Allanach:2001kg,Sjostrand:2002ip,Porod:2003um,Ellwanger:2005dv,Frank:2006yh,Sjostrand:2007gs,FCHDECAY} 
are either implementing extensions to this base model or are anticipating such
extensions. We therefore here present extensions of the SLHA1 
relevant for R-parity violation (RPV), flavour violation, 
and CP-violation (CPV) in the minimal supersymmetric standard
model (MSSM).   
We also consider next-to-minimal models which we shall collectively
label by the acronym NMSSM. Full details of the SLHA2
agreement can be found in \cite{Allanach:2008qg}.

For simplicity, we still limit the scope of the SLHA2 in two regards: 
for the MSSM, we restrict our attention to \emph{either} CPV or
RPV, but not both. 
For the NMSSM, we define one catch-all model and extend the SLHA1 mixing only 
to include the new states, with CP, R-parity, and flavour still
assumed conserved.

The conventions described here are a superset of those
of the original SLHA1, unless explicitly stated otherwise. We  
use ASCII text for input and output, all dimensionful
parameters are taken to be in appropriate powers of GeV, and the 
output formats for SLHA2 data
\texttt{BLOCK}s follow those of SLHA1. All angles are in radians. In a few
cases it has been necessary to replace the original conventions. This
is clearly remarked upon in all places where it occurs, and the SLHA2
conventions then supersede the SLHA1 ones.

\subsection{The SLHA2 Conventions}

\subsubsection{Flavour Violation}
\label{flavchange}
The CKM basis is defined to be the one in which 
the quark mass matrix is diagonal. In the super-CKM basis 
\cite{Hall:1985dx} the squarks are rotated by exactly the same amount
as their respective quark superpartners, regardless of whether this makes
them (that is, the squarks) diagonal or not. Misalignment between the
quark and squark 
sectors thus results in flavour off-diagonal terms remaining in the squark
sector.

In this basis, the $6\times 6$ squark mass matrices are defined as 
\begin{equation}
{\cal L}^{\rm mass}_{\tilde q} ~=~ 
- \Phi_u^{\dagger}\,
{\cal M}_{\tilde u}^2\, 
\Phi_u
- \Phi_d^{\dagger}\,
{\cal M}_{\tilde d}^2\, 
\Phi_d~,
\end{equation}
where $\Phi_u = (\tilde u_L,\tilde c_L, \tilde t_L,
\tilde u_R,\tilde c_R, \tilde t_R)^T$ and 
$\Phi_d = (\tilde d_L,\tilde s_L, \tilde b_L,
\tilde d_R,\tilde s_R, \tilde b_R)^T$.
We diagonalise the squark mass matrices via $6\times6$ unitary
matrices $R_{u,d}$, such that $R_{u,d} \,{\cal M}_{{\tilde
    u},{\tilde d}}^2\,R_{u,d}^\dagger$ are diagonal matrices with increasing
mass squared values. We re-define the existing PDG codes for squarks
to enumerate the mass eigenstates in ascending order:

\noindent$(\tilde{d}_1,\tilde{d}_2,\tilde{d}_3,\tilde{d}_4,\tilde{d}_5,
\tilde{d}_6)$ =
\texttt{(1000001, 1000003, 1000005, 2000001, 2000003, 2000005)},

\noindent 
$(\tilde{u}_1,\tilde{u}_2,\tilde{u}_3,\tilde{u}_4,\tilde{u}_5,\tilde{u}_6)$ =
\texttt{(1000002, 1000004, 1000006, 2000002, 2000004, 2000006)}.

The flavour violating parameters of the model are specified in terms
of the CKM matrix together with five $3\times 3$ matrices of  
soft SUSY-breaking parameters given in the super-CKM basis
\begin{equation}
{\hat m_{\tilde Q}}^2 ~,~~~
{\hat m_{\tilde u}}^2 ~,~~~
{\hat m_{\tilde d}}^2 ~,~~~
{\hat T_{U}} ~,~~~ 
{\hat T_{D}} ~.
\label{eq:that}
\end{equation}

Analogous rotations and definitions are used for the lepton flavour
violating parameters, in this case using the super-PMNS basis. This
will be further elaborated on in the journal version of this report. 
Below, we refer to the combined basis as the super-CKM/PMNS basis. 

\subsubsection{R-parity Violation}
 \label{sec:rpv}

We write the R-parity violating superpotential as
\begin{eqnarray}
W_{RPV} &=& \epsilon_{ab} \left[
\frac{1}{2} {\hat \lambda}_{ijk} L_i^a L_j^b \bar{E}_k +
{\hat \lambda}'_{ijk} L_i^a Q_j^{bx} \bar{D}_{kx}
- {\hat \kappa}_i L_i^a H_2^b \right] \nonumber \\
&& + \frac{1}{2} {\hat \lambda}''_{ijk} \epsilon^{xyz} \bar{U}_{ix}
\bar{D}_{jy} \bar{D}_{kz} 
, \label{eq:Wrpv}
\end{eqnarray}
where $x,y,z=1,\ldots,3$ are fundamental SU(3)$_C$ indices and
$\epsilon^{xyz}$ is the totally antisymmetric tensor in 3 dimensions with
$\epsilon^{123}=+1$. 
In eq.~(\ref{eq:Wrpv}), ${\hat \lambda}_{ijk}, {\hat \lambda}'_{ijk}$ and 
${\hat \kappa}_i$ break
lepton number, whereas ${\hat \lambda}''_{ijk}$ violate baryon
number.  As in the previous section, all quantities are given in the
super-CKM/super-PMNS basis. Note, that in the R-parity violating case,
the PMNS is an output once lepton number is violated.

The trilinear R-parity violating terms in the soft SUSY-breaking potential
are 
\begin{eqnarray}
V_{3,RPV} &=& \epsilon_{ab} \left[
\frac12(\hat T)_{ijk} {\tilde L}_{iL}^a {\tilde L}_{jL}^b {\tilde e}^*_{kR} +
(\hat T')_{ijk} {\tilde L}_{iL}^a {\tilde Q}_{jL}^b {\tilde d}^*_{kR} \right]
 \nonumber
\\ 
&&+\frac12 (\hat T'')_{ijk}\epsilon_{xyz} {\tilde u}_{iR}^{x*} {\tilde d}_{jR}^{y*} {\tilde d}_{kR}^{z*}
+ \mathrm{h.c.} \label{eq:trilinear}~~~.
\end{eqnarray}
Note that we do not factor 
out the $\hat \lambda$ couplings (e.g.\ as in 
${\hat T_{ijk}}/{\hat \lambda_{ijk}}\equiv A_{\lambda,ijk}$).

When lepton number is broken, additional bilinear soft SUSY-breaking
potential terms can appear, 
\begin{eqnarray}
V_{2,RPV} = -\epsilon_{ab}{\hat D}_i {\tilde L}_{iL}^a H_2^b  + {\tilde
L}_{iaL}^\dag {\hat m}_{{\tilde L}_i H_1}^2 H_1^a + \mathrm{h.c.}~,
\label{eq:bilinear}
\end{eqnarray}
and the sneutrinos may acquire vacuum expectation values (VEVs)
$\langle {\tilde \nu}_{e,\mu,\tau} \rangle \equiv v_{e, \mu,\tau}/\sqrt{2}$.
The SLHA1 defined the tree-level VEV $v$ to be equal to $2 m_Z / \sqrt{g^2 +
  {g'}^2}\sim246$ GeV; this is now generalised to  
\begin{eqnarray}
v=\sqrt{v_1^2+v_2^2+v_{e}^2+v_{\mu}^2+v_{\tau}^2}~.
\end{eqnarray}
For $\tan\beta$ we maintain the SLHA1 definition, $\tan \beta=v_2/v_1$. 

The Lagrangian contains the (symmetric) neutrino/neutralino mass matrix as 
\begin{equation}
\mathcal{L}^{\mathrm{mass}}_{{\tilde \chi}^0} =
-\frac12{\tilde\psi^0}{}^T{\mathcal M}_{\tilde\psi^0}\tilde\psi^0 +
\mathrm{h.c.}~, 
\end{equation}
in the basis of 2--component spinors $\tilde\psi^0 =$
$( \nu_e, \nu_\mu, \nu_\tau, -i\tilde b, -i\tilde w^3,  
\tilde h_1, \tilde h_2)^T$. We define the
unitary $7 \times 7$ 
neutrino/neutralino mixing matrix $N$ (block {\tt RVNMIX}), such that:
\begin{equation}
-\frac12{\tilde\psi^0}{}^T{\mathcal M}_{\tilde\psi^0}\tilde\psi^0
= -\frac12\underbrace{{\tilde\psi^0}{}^TN^T}_{{{\tilde \chi}^0}{}^T}
\underbrace{N^*{\mathcal
    M}_{\tilde\psi^0}N^\dagger}_{\mathrm{diag}(m_{{\tilde \chi}^0})}
\underbrace{N\tilde\psi^0}_{{\tilde \chi}^0}~,  \label{eq:neutmass}
\end{equation}
where the 7 (2--component) neutral leptons ${\tilde \chi}^0$ 
are defined strictly mass-ordered, i.e.\ with the 1$^{st}$,2$^{nd}$,3$^{rd}$
lightest corresponding to the mass entries for the PDG codes 
\texttt{12}, \texttt{14}, and \texttt{16}, and the four heaviest to the
 PDG codes \texttt{1000022}, \texttt{1000023}, \texttt{1000025}, \texttt{1000035}.

Charginos and charged leptons may also mix in the case of $L$-violation. 
The Lagrangian contains 
\begin{equation}
\mathcal{L}^{\mathrm{mass}}_{{\tilde \chi}^+} =
-\frac12{\tilde\psi^-}{}^T{\mathcal M}_{\tilde\psi^+}\tilde\psi^+ +
\mathrm{h.c.}~, 
\end{equation}
in the basis of 2--component spinors $\tilde\psi^+ =$
$({e}^+,{\mu}^+,{\tau}^+,-i\tilde w^+, \tilde h_2^+)^T$,
$\tilde\psi^- =$ $({e}^-,{\mu}^-,{\tau}^-,-i\tilde w^-,\tilde h_1^-)^T$  
where $\tilde w^\pm = (\tilde w^1 \mp \tilde w^2) / \sqrt{2}$.
We define the
unitary $5 \times 5$
charged fermion mixing matrices $U,V$, blocks {\tt RVUMIX, RVVMIX}, 
such that:
\begin{equation}
-\frac12{\tilde\psi^-}{}^T{\mathcal M}_{\tilde\psi^+}\tilde\psi^+
= -\frac12\underbrace{{\tilde\psi^-}{}^TU^T}_{{{\tilde \chi}^-}{}^T}
\underbrace{U^*{\mathcal
    M}_{\tilde\psi^+}V^\dagger}_{\mathrm{diag}(m_{{\tilde \chi}^+})}
\underbrace{V\tilde\psi^+}_{{\tilde \chi}^+}~,  \label{eq:chargmass}
\end{equation}
where the generalised charged leptons $\tilde
\chi^+$ are defined as strictly mass ordered, i.e.\
with the 3 lightest states corresponding to the PDG codes \texttt{11},
\texttt{13}, and \texttt{15}, and the two heaviest to the codes
\texttt{1000024}, \texttt{1000037}.
For historical reasons, codes \texttt{11}, \texttt{13}, and
\texttt{15} pertain to the negatively charged field while codes
\texttt{1000024} and \texttt{1000037} pertain to the opposite charge. The
components of $\tilde{\chi}^+$ in ``PDG notation'' would thus be
\texttt{(-11,-13,-15,1000024,1000037)}. 
In the limit of CP
conservation, $U$ and $V$ are chosen to be real.

R-parity violation via lepton number violation implies that the
 sneutrinos can mix with the Higgs bosons. In the limit of 
 CP conservation the CP-even (-odd) Higgs bosons mix with real (imaginary)
 parts of the sneutrinos. We write the neutral scalars as $\phi^0
 \equiv \sqrt{2} \mathrm{Re}{(H_1^0, H_2^0, {\tilde \nu}_e, {\tilde \nu}_\mu, {\tilde
\nu}_\tau)^T}$, with the mass term
\begin{eqnarray}
{\mathcal L} = - \frac12 {\phi^0}^T {\mathcal M}_{\phi^0}^2 \phi^0~,
\end{eqnarray}
where ${\mathcal M}_{\phi^0}^2$ is a $5 \times 5$ symmetric mass matrix. 
We define the orthogonal $5 \times 5$ 
mixing matrix $\aleph$ (block {\tt RVHMIX}) by
\begin{equation}
-{\phi^0}{}^T{\mathcal M}_{\phi^0}^2
\phi^0
= -\underbrace{{\phi^0}{}^T {\mathbf \aleph}^T}_{{{
      \Phi}^0}{}^T} 
\underbrace{{\mathbf \aleph}{\mathcal
    M}_{\phi^0}^2\aleph^T}_{\mathrm{diag}(m_{{ \Phi}^0}^2)}
\underbrace{{\mathbf \aleph}\phi^0}_{{ \Phi}^0}~,  
\label{eq:sneutmass}
\end{equation}
where $\Phi^0$ are the
neutral scalar mass eigenstates in strictly increasing mass order
The states
are numbered sequentially by the PDG codes 
\texttt{(25,35,1000012,1000014,1000016)}, regardless of flavour
content. 

We write the neutral pseudoscalars 
as $\bar\phi^0 \equiv \sqrt{2} \mathrm{Im}{(H_1^0, H_2^0,
  {\tilde \nu}_e, {\tilde 
  \nu}_\mu, {\tilde \nu}_\tau)^T}$,
with the mass term
\begin{eqnarray}
{\mathcal L} = - \frac12 {{\bar \phi}^0}{}^T {\mathcal M}_{{\bar
  \phi}^0}^2 {\bar \phi}^0~,
\end{eqnarray}
where ${\mathcal M}_{\bar \phi^0}^2$ is a $5 \times 5$ symmetric mass
matrix. We define 
the $4 \times 5$ mixing matrix $\bar \aleph$ (block {\tt RVAMIX}) by
\begin{equation}
-{\bar \phi^0}{}^T{\mathcal M}_{\bar \phi^0}^2
\bar \phi^0
= -\underbrace{{\bar \phi^0}{}^T {\bar \aleph}^T}_{{{
      \bar \Phi}^0}{}^T} 
\underbrace{{\bar \aleph}{\mathcal
    M}_{\bar \phi^0}^2{\bar \aleph}^T}_{\mathrm{diag}(m_{{\bar \Phi}^0}^2)}
\underbrace{{\bar \aleph}\bar \phi^0}_{{\bar \Phi}^0}~,  
\label{eq:sneutmass2}
\end{equation}
where $\bar\Phi^0$ 
are the pseudoscalar mass eigenstates in increasing
mass order. The states
are numbered sequentially by the PDG codes
\texttt{(36,1000017, 1000018,1000019)}, regardless of 
flavour composition. The Goldstone boson $G^0$
 has been explicitly left out and the 
4 rows of $\bar\aleph$ form a set of orthonormal vectors. 

If the blocks {\tt RVHMIX, RVAMIX} are present, they supersede the
  SLHA1 {\tt ALPHA} variable/block.

The charged sleptons and charged Higgs bosons also mix in the $8 \times 8$ mass
squared matrix ${\mathcal M}^2_{\phi^\pm}$, which we diagonalise 
by a $7 \times 8$ matrix $C$ (block {\tt RVLMIX}):
\begin{eqnarray}
{\mathcal L}=- 
\underbrace{({H_1^-}^*, {H^+_2},\tilde{e}_{L_i}^*,\tilde{e}_{R_j}^*) C^\dagger}_{{\Phi^+}} 
\underbrace{C {\mathcal M}^2_{\phi^\pm} C^\dagger}_{ \mathrm{diag}({\mathcal M}^2_{\Phi^\pm})} 
C \left (
\begin{array}{c} {H_1^-}\\[1mm] {H_2^+}^* \\  \tilde{e}_{L_k} \\
\tilde{e}_{R_l} \end{array} \right )\, \label{eq:higgsslep}~,
\end{eqnarray}
where $i,j,k,l\in \{1,2,3\}$, $\alpha,\beta\in \{1,\ldots,6\}$ and
$\Phi^+=\Phi^-{}^\dagger$ are the charged scalar mass eigenstates
arranged in increasing mass order. 
These states are numbered sequentially by the PDG codes
\texttt{(37,1000011,1000013,1000015, 2000011,2000013,2000015)},
regardless of flavour composition.  The Goldstone boson $G^-$ has been explicitly left out and the 7 rows of $C$
form a set of orthonormal vectors.
  
\subsubsection{CP Violation}

When CP symmetry is broken, quantum corrections cause mixing between
the CP-even and CP-odd Higgs states. 
Writing the neutral scalar interaction eigenstates as 
$\phi^0 \equiv \sqrt{2} (\mathrm{Re}{H_1^0},$
$\mathrm{Re}{H_2^0},$ $\mathrm{Im}{H_1^0},$ $\mathrm{Im}{H_2^0})^T$ 
we define the $3 \times 4$  mixing matrix $S$ 
(block {\tt CVHMIX}) by 
\begin{equation}
-{\phi^0}{}^T{\mathcal M}_{\phi^0}^2
 \phi^0
= -\underbrace{{ \phi^0}{}^T {S}^T}_{{{
       \Phi}^0}{}^T} 
\underbrace{{S}^*{\mathcal
    M}_{\phi^0}^2{S}^\dagger}_{\mathrm{diag}(m_{{ \Phi}^0}^2)}
\underbrace{{S}\phi^0}_{{ \Phi}^0}~,  
\label{eq:cvhmass}
\end{equation}  
where $\Phi^0 \equiv (h_1^0,h_2^0,h_3^0)^T$ are the mass eigenstates
arranged in ascending mass order;
these states are numbered sequentially by the PDG codes 
\texttt{(25,35,36)}, regardless of flavour composition. 

For the neutralino and chargino mixing matrices, the default convention 
in SLHA1 
is that they be real matrices. One or more mass eigenvalues may then
have an apparent negative sign, which can be removed by a
phase transformation on $\tilde \chi_i$ as explained in
SLHA1~\cite{Skands:2003cj}. When going to CPV, the reason for
introducing the negative-mass convention in the first place, namely
maintaining the mixing matrices strictly real, disappears. 
We therefore here take
all masses real and positive, with $N$, $U$, and $V$ complex. This does
lead to a nominal dissimilarity with SLHA1 in the limit of vanishing CP
violation, but we note that the explicit CPV switch in \texttt{MODSEL}
can be used to decide unambiguously which convention to follow.

For the remaining MSSM parameters we use straightforward
generalisations to the complex case, see section \ref{sec:cpv-proposal}. 

\subsubsection{NMSSM}

We shall here define the next-to-minimal case as
having exactly the field content of the MSSM with the addition of one 
gauge singlet chiral superfield. As to couplings and parameterisations,
rather than adopting a particular choice, or treating each special
case separately, below we choose instead to work at the most general
level. Any particular special
case can then be obtained by setting different combinations of
couplings to zero. However, we do specialise to the
SLHA1-like case without CP violation, R-parity violation, or flavour
violation. Below, we shall use the acronym NMSSM for this class of
models, but we emphasise that we
understand it to relate to field content only, and not
to the presence or absence of specific couplings. 

In addition to the MSSM terms, the
most general CP conserving NMSSM superpotential is (extending the
notation of SLHA1):
\begin{eqnarray}\label{eq:nmssmsup}
W_{NMSSM} = W_{MSSM} - \epsilon_{ab}\lambda {S} {H}^a_1 {H}^b_2 + \frac{1}{3}
\kappa {S}^3 + \mu' S^2 +\xi_F S \ , \end{eqnarray} 
where $W_{MSSM}$ is the MSSM superpotential,
in the conventions of ref.~\cite[eq.~(3)]{Skands:2003cj}. 
A non-zero $\lambda$ in combination with a VEV $\left< S
\right>$ of the singlet generates a contribution to the effective 
$\mu$ term $\mu_\mathrm{eff}= \lambda \left< S
\right> + \mu$, where the MSSM $\mu$ term is normally assumed to be
zero, 
yielding $\mu_{\mathrm{eff}}=\lambda \left< S \right>$. 
The remaining terms represent a general cubic 
potential for the singlet; $\kappa$ is dimensionless, $\mu'$ has
dimension of mass, and $\xi_F$ has dimension of mass squared. 
The soft SUSY-breaking terms relevant to the NMSSM are
\begin{eqnarray}\label{eq:nmssmsoft}
V_\mathrm{soft} = V_{2,MSSM} + V_{3,MSSM} + m_\mathrm{S}^2 | S |^2 +
(-\epsilon_{ab}\lambda A_\lambda {S} {H}^a_1 {H}^b_2 + 
\frac{1}{3} \kappa A_\kappa {S}^3  
+ B'\mu' S^2 +\xi_S S
+ \mathrm{h.c.}) \ , \end{eqnarray}
where $V_{i,MSSM}$ are the MSSM soft terms, in the conventions of 
ref.~\cite[eqs.~(5) and (7)]{Skands:2003cj}. 

At tree level, there are thus 15 parameters (in addition to $m_Z$
which fixes the sum of the squared Higgs VEVs) 
that are 
relevant for the Higgs sector: 
\begin{eqnarray} 
\tan\!\beta,\ \mu,\ m_{H_1}^2,\ m_{H_2}^2,\ m_3^2,\
\lambda,\  \kappa,\ A_{\lambda},\  A_{\kappa},\ \mu',\ B',\ \xi_F,\
\xi_S,\ \lambda \left< S \right>,\ m_S^2~.\label{eq:nmssmpar}
\end{eqnarray}
The minimisation of the effective 
potential imposes 3 conditions on these
parameters, such that only 12 of them can be considered
independent. For the time being, we leave it up to each spectrum
calculator to decide on which combinations to accept. 
For the purpose of this accord, we note only that to specify a general
model exactly 12 parameters from eq.~(\ref{eq:nmssmpar}) should be
provided in the input, including explicit zeroes for parameters
desired ``switched off''. However, since
$\mu=m_3^2=\mu'=B'=\xi_F=\xi_S=0$ in the majority of phenomenological
constructions, for convenience we also allow for a six-parameter
specification in terms of the reduced parameter list:
\begin{eqnarray}
\tan\!\beta,\ m_{H_1}^2,\ m_{H_2}^2,\ 
\lambda,\  \kappa,\ A_{\lambda},\  A_{\kappa},\
 \lambda \left< S \right>,\ m_S^2~.\label{eq:nmssmpar-reduced}
\end{eqnarray}

To summarise, in addition to $m_Z$, the input to the accord should contain 
either 12 parameters
from the list given in eq.~(\ref{eq:nmssmpar}), including zeroes for parameters
not present in the desired model, or it should contain 6 parameters from
the list in eq.~(\ref{eq:nmssmpar-reduced}), in which case the
remaining 6 ``non-standard'' parameters, $\mu$, $m_3^2$, $\mu'$, $B'$,
$\xi_F$, and $\xi_F$, will be assumed to be zero;
in both cases 
the 3 unspecified parameters (as, e.g.,  
$m_{H_1}^2$, $m_{H_2}^2$, and $m_S^2$) are assumed to be determined by the
minimisation of the effective potential.

The CP-even neutral scalar interaction eigenstates are 
$\phi^0 \equiv\sqrt{2} \mathrm{Re}{(H_{1}^0, H_{2}^0, S)^T}$. 
We define the orthogonal $3 \times 3$ mixing matrix $S$ (block {\tt
  NMHMIX}) by 
\begin{equation}
-{\phi^0}{}^T{\mathcal M}_{\phi^0}^2 \phi^0
= -\underbrace{{\phi^0}{}^T {S}^T}_{{{
      \Phi}^0}{}^T} 
\underbrace{{S}{\mathcal
    M}_{\phi^0}^2{S}^T}_{\mathrm{diag}(m_{{\Phi}^0}^2)}
\underbrace{{S} \phi^0}_{{\Phi}^0}~,  
\end{equation}
where $\Phi^0 \equiv (h^0_1, h^0_2, h^0_3)$ are the mass eigenstates
ordered in mass. These states are
numbered sequentially by the PDG 
codes \texttt{(25,35,45)}. 
The format of {\tt BLOCK NMHMIX} is the same as for
the mixing matrices in SLHA1.

The CP-odd sector interaction eigenstates are 
$\bar\phi^0 \equiv\sqrt{2} \mathrm{Im}{(H_{1}^0, H_{2}^0, S)^T}$. 
We define the $2 \times 3$ mixing matrix $P$ (block {\tt NMAMIX}) by
\begin{equation}
-{\bar \phi^0}{}^T{\mathcal M}_{\bar \phi^0}^2
\bar \phi^0
= -\underbrace{{\bar \phi^0}{}^T {P}^T}_{{{
      \bar \Phi}^0}{}^T} 
\underbrace{{P}{\mathcal
    M}_{\bar \phi^0}^2{P}^T}_{\mathrm{diag}(m_{{\bar \Phi}^0}^2)}
\underbrace{{P}\bar \phi^0}_{{\bar \Phi}^0}~,  
\end{equation}
where $\bar\Phi^0 \equiv (A^0_1, A^0_2)$ are the mass eigenstates
ordered in mass. These states are numbered sequentially by the PDG
codes \texttt{(36,46)}. The Goldstone boson $G^0$ has
been explicitly left out and the 2 rows of $P$ form a set of
orthonormal vectors.  

If {\tt NMHMIX, NMAMIX} blocks are present, they {supersede} the
SLHA1 {\tt ALPHA} variable/block.

The Lagrangian contains the (symmetric) $5\times 5$ neutralino mass matrix as 
\begin{equation}
\mathcal{L}^{\mathrm{mass}}_{{\tilde \chi}^0} =
-\frac12{\tilde\psi^0}{}^T{\mathcal M}_{\tilde\psi^0}\tilde\psi^0 +
\mathrm{h.c.}~, 
\end{equation}
in the basis of 2--component spinors $\tilde\psi^0 =$ $(-i\tilde b,$
$-i\tilde w^3,$  $\tilde h_1,$ $\tilde h_2,$ $\tilde s)^T$. 
We define the unitary $5 \times 5$ neutralino mixing matrix $N$ (block {\tt
NMNMIX}), such that:
\begin{equation}\label{eq:nmssmneutmass}
-\frac12{\tilde\psi^0}{}^T{\mathcal M}_{\tilde\psi^0}\tilde\psi^0
= -\frac12\underbrace{{\tilde\psi^0}{}^TN^T}_{{{\tilde \chi}^0}{}^T}
\underbrace{N^*{\mathcal
    M}_{\tilde\psi^0}N^\dagger}_{\mathrm{diag}(m_{{\tilde \chi}^0})}
\underbrace{N\tilde\psi^0}_{{\tilde \chi}^0}~,  
\end{equation}
where the 5 (2--component) neutralinos ${\tilde \chi}_i$ are defined
such that the absolute value of their masses 
increase with $i$, cf.\ SLHA1 \cite{Skands:2003cj}. These states are
numbered sequentially by the PDG codes
\texttt{(1000022}, \texttt{1000023,1000025,1000035,1000045)}.  

\subsection{Explicit Proposals for SLHA2}

As in the SLHA1~\cite{Skands:2003cj}, for all running parameters in
the output of the spectrum file, we propose to use definitions in the
modified dimensional reduction ($\overline{\mrm{DR}}$) scheme. 

To define the general properties of the model, we propose to introduce
global switches in the SLHA1 model definition block \texttt{MODSEL}, as
follows.  Note that the switches defined here are in addition to the
ones in \cite{Skands:2003cj}.

\subsubsection{Model Selection}
\subsection*{\texttt{BLOCK MODSEL}\label{sec:modsel}}
Switches and options for model selection. The entries in this block
should consist of an index, identifying the particular switch in the
listing below, followed by another integer or real number, specifying
the option or value chosen: \\[2mm]
\numentry{3}{(Default=\texttt{0}) Choice of particle content.
Switches defined are:\\
\snumentry{0}{MSSM. This corresponds to SLHA1.}
\snumentry{1}{NMSSM. As defined here.}
}
\numentry{4}{(Default=\texttt{0}) R-parity violation. Switches defined are:\\
\snumentry{0}{R-parity conserved. This corresponds to the SLHA1.}
\snumentry{1}{R-parity violated. } 
} 
\numentry{5}{(Default=\texttt{0}) CP violation. Switches defined are:\\
\snumentry{0}{CP is conserved. No information even on the CKM phase
is used. This corresponds to the SLHA1.}
\snumentry{1}{CP is violated, but only by the standard CKM
phase. All other phases assumed zero.}
\snumentry{2}{CP is violated. Completely general CP phases
allowed.}
}
\numentry{6}{(Default=\texttt{0}) Flavour violation. Switches defined are:\\
\snumentry{0}{No (SUSY) flavour violation. This corresponds to the
SLHA1. 
}
\snumentry{1}{Quark flavour is violated. }
\snumentry{2}{Lepton flavour is violated.}
\snumentry{3}{Lepton and quark flavour is violated.}
}

\subsubsection{Flavour Violation}
 \label{sec:flvproposal}

\begin{itemize}
\item All input SUSY parameters are given at the scale $M_\mrm{input}$ as
  defined in the SLHA1 block \texttt{EXTPAR}, 
  except for \texttt{EXTPAR 26}, which, if present, is the {\em pole}
  pseudoscalar Higgs mass, and
  \texttt{EXTPAR 27}, which, if present, is the {\em pole}
  mass of the charged Higgs boson. If no $M_\mrm{input}$
  is  present, the GUT scale is used. 
\item For the SM input parameters, we take
the Particle Data Group (PDG) definition: 
lepton masses are all on-shell. 
The light quark masses $m_{u,d,s}$ are given at 2 GeV in the $\MSbar$
scheme, and the heavy quark masses are given as 
$m_c(m_c)^{\overline{\mrm{MS}}}$, $m_b(m_b)^{\overline{\mrm{MS}}}$ and
$m_{t}^{\mrm{on-shell}}$. 
The latter two quantities are already in the SLHA1. The others are
added to \texttt{SMINPUTS} in the following manner:\\[2mm]
\numentry{8}{$m_{\nu_3}$, pole mass.}
\numentry{11}{$m_e$, pole mass. }
\numentry{12}{$m_{\nu_1}$, pole mass.} 
\numentry{13}{$m_\mu$, pole mass. }
\numentry{14}{$m_{\nu_2}$, pole mass.} 
\numentry{21}{$m_d(2\ \GeV)^{\MSbar}$. $d$ quark running mass in the $\MSbar$
  scheme.}
\numentry{22}{$m_u(2\ \GeV)^{\MSbar}$. $u$ quark running mass in the $\MSbar$
  scheme.}
\numentry{23}{$m_s(2\ \GeV)^{\MSbar}$. $s$ quark running mass in the $\MSbar$
  scheme.}
\numentry{24}{$m_c(m_c)^{\MSbar}$. $c$ quark running mass in the $\MSbar$
  scheme.}
The FORTRAN 
format is the same as that of \texttt{SMINPUTS} in SLHA1 \cite{Skands:2003cj}.
\item $V_{\mrm{CKM}}$: the input CKM matrix, in the block
\texttt{VCKMIN} in terms of the Wolfenstein parameterisation:\\[2mm]
 \numentry{1}{$\lambda$}
 \numentry{2}{$A$}
 \numentry{3}{$\bar{\rho}$}
 \numentry{4}{$\bar{\eta}$}
The FORTRAN format is the same as that of \texttt{SMINPUTS} above. 
\item $U_{\mrm{PMNS}}$: the input PMNS matrix, in the block
  \texttt{UPMNSIN}. It should have the PDG parameterisation in terms of
  rotation angles~\cite{Yao:2006px}  (all in radians):\\[2mm]
 \numentry{1}{$\bar \theta_{12}$ (the solar angle)}
 \numentry{2}{$\bar \theta_{23}$ (the atmospheric mixing angle)}
 \numentry{3}{$\bar \theta_{13}$ (currently only has an upper bound)}
 \numentry{4}{$\bar \delta_{13}$ (the Dirac CP-violating phase)}
 \numentry{5}{$\alpha_1$ (the first Majorana CP-violating phase)}
\numentry{6}{$\alpha_2$ (the second CP-violating Majorana phase)}
The FORTRAN format is the same as that of \texttt{SMINPUTS} above. 
\item $(\hat{m}^2_{\tilde Q})_{ij}^{\overline{\mrm{DR}}}$,  
$(\hat{m}^2_{\tilde u})_{ij}^{\overline{\mrm{DR}}}$,
$(\hat{m}^2_{\tilde d})_{ij}^{\overline{\mrm{DR}}}$, 
$(\hat{m}^2_{\tilde L})_{ij}^{\overline{\mrm{DR}}}$,   
$(\hat{m}^2_{\tilde e})_{ij}^{\overline{\mrm{DR}}}$: the squark and
  slepton soft SUSY-breaking masses at the input scale 
in the super-CKM/PMNS basis, as defined above.
They will be given in the 
new blocks \texttt{MSQ2IN}, \texttt{MSU2IN},
\texttt{MSD2IN}, \texttt{MSL2IN}, \texttt{MSE2IN}, 
with the same format as matrices in SLHA1.  
Only the ``upper
triangle'' of these matrices should be given. 
If diagonal entries are present, these supersede the 
parameters in the SLHA1 block \texttt{EXTPAR}. 
\item $(\hat{T}_U)_{ij}^{\overline{\mrm{DR}}}$, 
 $(\hat{T}_D)_{ij}^{\overline{\mrm{DR}}}$, and 
$(\hat{T}_E)_{ij}^{\overline{\mrm{DR}}}$: 
the squark and slepton  
soft SUSY-breaking trilinear couplings at the input scale 
in the super-CKM/PMNS basis. They will be given in the 
new blocks \texttt{TUIN}, \texttt{TDIN},
\texttt{TEIN}, in the same format as matrices in SLHA1. 
If diagonal entries are present these supersede
 the $A$ parameters specified in the SLHA1 block \texttt{EXTPAR} 
\cite{Skands:2003cj}.
\end{itemize}
For the output, the pole masses are given in block \texttt{MASS} as in SLHA1, 
and the $\DRbar$ and mixing parameters as follows:
\begin{itemize}
\item $(\hat{m}^2_{\tilde Q})_{ij}^{\overline{\mrm{DR}}}$,  
$(\hat{m}^2_{\tilde u})_{ij}^{\overline{\mrm{DR}}}$,
$(\hat{m}^2_{\tilde d})_{ij}^{\overline{\mrm{DR}}}$,
  $(\hat{m}^2_{\tilde L})_{ij}^{\overline{\mrm{DR}}}$,   
$(\hat{m}^2_{\tilde e})_{ij}^{\overline{\mrm{DR}}}$: the squark and
  slepton soft SUSY-breaking masses at scale $Q$ in the super-CKM/PMNS basis. 
Will be given in the 
new blocks \texttt{MSQ2 Q=...}, \texttt{MSU2 Q=...}, \texttt{MSD2 Q=...},
\texttt{MSL2 Q=...}, \texttt{MSE2 Q=...}, with 
formats as the corresponding input blocks \texttt{MSX2IN}
above. 
\item $(\hat{T}_U)_{ij}^{\overline{\mrm{DR}}}$, 
 $(\hat{T}_D)_{ij}^{\overline{\mrm{DR}}}$, and 
$(\hat{T}_E)_{ij}^{\overline{\mrm{DR}}}$: 
The squark and slepton 
soft SUSY-breaking trilinear couplings in the super-CKM/PMNS
 basis. Given in the new blocks \texttt{TU Q=...},  \texttt{TD Q=...},  
\texttt{TE Q=...}, which supersede the SLHA1 blocks \texttt{AD}, \texttt{AU},
 and \texttt{AE}, see \cite{Skands:2003cj}.
\item $(\hat{Y}_U)_{ii}^{\overline{\mrm{DR}}}$,
$(\hat{Y}_D)_{ii}^{\overline{\mrm{DR}}}$,
$(\hat{Y}_E)_{ii}^{\overline{\mrm{DR}}}$: the diagonal 
$\overline{\mrm{DR}}$ Yukawas
in the super-CKM/PMNS basis, 
at the scale $Q$. Given in the SLHA1 blocks \texttt{YU Q=...},  \texttt{YD Q=...},  \texttt{YE Q=...},  see \cite{Skands:2003cj}. 
Note that although the SLHA1 blocks provide for off-diagonal
elements, only the 
diagonal ones will be relevant here, due to the CKM rotation.
\item The $\DRbar$ CKM matrix at the
scale $Q$. Will be given in the new block(s) \texttt{VCKM
Q=...}, with entries defined as for the input block \texttt{VCKMIN}
above. 
\item 
The new blocks $R_u=$\texttt{USQMIX} $R_d=$\texttt{DSQMIX}, $R_e=$\texttt{SELMIX}, and
$R_\nu=$\texttt{SNUMIX} connect the particle codes (=mass-ordered
basis) with the super-CKM/PMNS basis according to the following definitions:
\begin{equation}
\left(\begin{array}{c}
1000001 \\
1000003 \\
1000005 \\
2000001 \\
2000003 \\
2000005 \\
\end{array}\right) =
\left(\begin{array}{c}
\tilde{d}_1 \\
\tilde{d}_2 \\
\tilde{d}_3 \\
\tilde{d}_4 \\
\tilde{d}_5 \\
\tilde{d}_6 \\
\end{array}\right)_{\!\mathrm{mass-ordered}}\hspace*{-1cm} = 
\mbox{\texttt{DSQMIX}}_{ij} \left(\begin{array}{c}
\tilde{d}_L \\
\tilde{s}_L \\
\tilde{b}_L \\
\tilde{d}_R \\
\tilde{s}_R \\
\tilde{b}_R \\
\end{array}\right)_{\!\mathrm{super-CKM}}~,
\end{equation}
\begin{equation}
\left(\begin{array}{c}
1000002 \\
1000004 \\
1000006 \\
2000002 \\
2000004 \\
2000006 \\
\end{array}\right) =
\left(\begin{array}{c}
\tilde{u}_1 \\
\tilde{u}_2 \\
\tilde{u}_3 \\
\tilde{u}_4 \\
\tilde{u}_5 \\
\tilde{u}_6 \\
\end{array}\right)_{\!\mathrm{mass-ordered}}\hspace*{-1cm} = 
\mbox{\texttt{USQMIX}}_{ij} \left(\begin{array}{c}
\tilde{u}_L \\
\tilde{c}_L \\
\tilde{t}_L \\
\tilde{u}_R \\
\tilde{c}_R \\
\tilde{t}_R \\
\end{array}\right)_{\!\mathrm{super-CKM}}~.
\end{equation}
\begin{equation}
\left(\begin{array}{c}
1000011 \\
1000013 \\
1000015 \\
2000011 \\
2000013 \\
2000015 \\
\end{array}\right) =
\left(\begin{array}{c}
\tilde{e}_1 \\
\tilde{e}_2 \\
\tilde{e}_3 \\
\tilde{e}_4 \\
\tilde{e}_5 \\
\tilde{e}_6 \\
\end{array}\right)_{\!\mathrm{mass-ordered}}\hspace*{-1cm} = 
\mbox{\texttt{SELMIX}}_{ij} \left(\begin{array}{c}
\tilde{e}_L \\
\tilde{\mu}_L \\
\tilde{\tau}_L \\
\tilde{e}_R \\
\tilde{\mu}_R \\
\tilde{\tau}_R \\
\end{array}\right)_{\!\mathrm{super-PMNS}}~,
\end{equation}
\begin{equation}
\left(\begin{array}{c}
1000012 \\
1000014 \\
1000016 \\
\end{array}\right) =
\left(\begin{array}{c}
\tilde{\nu}_1 \\
\tilde{\nu}_2 \\
\tilde{\nu}_3 \\
\end{array}\right)_{\!\mathrm{mass-ordered}}\hspace*{-1cm} = 
\mbox{\texttt{SNUMIX}}_{ij} \left(\begin{array}{c}
\tilde{\nu}_{e} \\
\tilde{\nu}_{\mu} \\
\tilde{\nu}_{\tau} \\
\end{array}\right)_{\!\mathrm{super-PMNS}}~.
\end{equation}
{\bf Note!} A potential for inconsistency arises if the masses and
mixings are not calculated in the same way, e.g.\ if radiatively
corrected masses are used with tree-level mixing matrices. In this
case, it is possible that the radiative corrections to the masses
shift the mass ordering relative to the tree-level. This is especially
relevant when near-degenerate masses occur in the spectrum and/or when
the radiative corrections are large. In these cases, 
explicit care must be taken especially by the program writing the
spectrum, but also by the one reading it, to properly 
arrange the rows in
the order of the mass spectrum actually used.  

\end{itemize}

\subsubsection{R-Parity Violation}
 \label{sec:rpv-proposal}

The naming convention for input blocks is {\tt
BLOCK RV\#IN}, where the '{\tt \#}' character represents the name of
the relevant output block given below.
Default inputs for all R-parity violating couplings are zero.
The inputs are given at scale $M_{\mrm{input}}$, as described 
in SLHA1 (default is the GUT scale) and follow the output format given
below (with the omission of {\tt 
  Q= ...}). In addition, the known fermion masses should be given in
\texttt{SMINPUTS} as defined above. 

\begin{itemize}
\item The dimensionless couplings ${\hat \lambda}_{ijk}$, 
${\hat \lambda}_{ijk}'$, and 
${\hat \lambda}''_{ijk}$ are given in {\tt BLOCK RVLAMLLE, RVLAMLQD,
  RVLAMUDD Q= ...} respectively. The output standard should correspond to the
FORTRAN format 
\begin{verbatim}
(1x,I2,1x,I2,1x,I2,3x,1P,E16.8,0P,3x,'#',1x,A) .
\end{verbatim}
where the first three integers in the format correspond to $i$, $j$,
and $k$ and the double precision number is the coupling. 
\item The soft SUSY-breaking couplings ${\hat T}_{ijk}$,
${\hat T}'_{ijk}$, and ${\hat T}''_{ijk}$  should be 
given in {\tt BLOCK RVTLLE, RVTLQD, RVTUDD Q= ...}, in the same format
as the ${\hat \lambda}$ couplings above.

\item The bilinear superpotential and soft SUSY-breaking terms 
${\hat\kappa}_i$, 
${\hat D}_i$, and ${\hat m}_{{\tilde L}_iH_1}^2$ and the sneutrino VEVs 
are given in {\tt BLOCK RVKAPPA, RVD, RVM2LH1, RVSNVEV Q= ...} respectively, in
the same format as real-valued vectors in the SLHA1. 

\item The input/output blocks for R-parity violating couplings are summarised
in Tab.~\ref{tab:rpvSummary}.
\item The new mixing matrices that appear are described in section
  \ref{sec:rpv}. 
\end{itemize}
\begin{table}
\begin{center}\begin{tabular}{|l|l|l|}
\hline 
Input block & Output block  & data \\ \hline
 {\tt RVLAMLLEIN}& {\tt RVLAMLLE}&  $i$ $j$ $k$ ${\hat \lambda}_{ijk}$ \\
 {\tt RVLAMLQDIN}& {\tt RVLAMLQD}&  $i$ $j$ $k$ ${\hat \lambda}'_{ijk}$ \\
 {\tt RVLAMUDDIN}& {\tt RVLAMUDD}&  $i$ $j$ $k$ ${\hat \lambda}''_{ijk}$ \\
 {\tt RVTLLEIN}& {\tt RVTLLE}&  $i$ $j$ $k$ ${\hat T}_{ijk}$ \\
 {\tt RVTLQDIN}& {\tt RVTLQD}&  $i$ $j$ $k$ ${\hat T}'_{ijk}$ \\
 {\tt RVTUDDIN}& {\tt RVTUDD}&  $i$ $j$ $k$ ${\hat T}''_{ijk}$ \\
\hline
\multicolumn{3}{|c|}{NB: One of the following \texttt{RV...IN} blocks must be left out:}\\
\multicolumn{3}{|c|}{(which one up to user and RGE code)}\\
 {\tt RVKAPPAIN}& {\tt RVKAPPA}&  $i$ ${\hat \kappa}_{i}$ \\
 {\tt RVDIN}& {\tt RVD}&  $i$ ${\hat D}_{i}$ \\
 {\tt RVSNVEVIN}& {\tt RVSNVEV}&  $i$ $v_{i}$ \\
 {\tt RVM2LH1IN}& {\tt RVM2LH1}&  $i$ ${\hat m}_{{\tilde L}_{i}H_1}^2$ \\
\hline \end{tabular}
\caption{Summary of R-parity violating SLHA2 data blocks. 
Only 3 out of the last 4 blocks are independent. 
Which block to leave out of the input is in principle up to the
user, with the caveat that a given spectrum calculator may
not accept all combinations.}
\label{tab:rpvSummary}
\end{center}
\end{table}
As for the R-conserving MSSM, the bilinear terms (both SUSY-breaking
and SUSY-respect\-ing ones, including $\mu$) and the VEVs are not
independent parameters. They become related by the condition of
electroweak symmetry breaking.  This carries over to the RPV\ case,
where not all the parameters in the input blocks \texttt{RV...IN} in
Tab.~\ref{tab:rpvSummary} can be given simultaneously.  
Specifically, of the last 4
blocks only 3 are independent. One block is determined by minimising
the Higgs-sneutrino potential.  We do not here insist on a particular
choice for which of \texttt{RVKAPPAIN}, \texttt{RVDIN}, \texttt{RVSNVEVIN}, and
\texttt{RVM2LH1IN} to leave out, but leave it up to the spectrum
calculators to accept one or more combinations.

\subsubsection{CP Violation}
 \label{sec:cpv-proposal}

When adding CP violation to the MSSM model parameters and mixing
matrices,  
the SLHA1 blocks are understood to contain the real parts of the relevant
parameters. The imaginary parts should be provided with exactly the same
format, in a separate block of the same name but prefaced by {\tt IM}. 
The defaults for all imaginary parameters will be zero.

One special case is the $\mu$ parameter. When the real part of $\mu$
is given in \texttt{EXTPAR 23}, the imaginary part should be given in
\texttt{IMEXTPAR 23}, as above. However, when $|\mu|$ 
is determined by the conditions for 
electroweak symmetry breaking, only the phase $\varphi_\mu$ is taken 
as an input parameter. In this case, SLHA2 generalises the entry
\texttt{MINPAR 4} to contain the cosine of the phase (as opposed to just 
$\mathrm{sign}(\mu)$ in SLHA1), and we further introduce a new
block \texttt{IMMINPAR} whose entry \texttt{4} gives the sine of the
phase, that is:
\subsection*{\texttt{BLOCK MINPAR}}
\numentry{4}{CP conserved: $\mathrm{sign}(\mu)$.\\ CP violated:
  $\cos\varphi_\mu=\mathrm{Re}{\mu}/|\mu|$. 
}
\subsection*{\texttt{BLOCK IMMINPAR}}
\numentry{4}{CP conserved: n/a.\\ CP violated:
  $\sin\varphi_\mu=\mathrm{Im}{\mu}/|\mu|$. 
}
Note that $\cos\varphi_\mu$ coincides with  
  $\mathrm{sign}(\mu)$ in the CP-conserving cases. 

The new $3\times4$ block $S=$\texttt{CVHMIX} connects the particle codes
(=mass-ordered 
basis) with the interaction basis according to the following definition:
\begin{equation}
\left(\begin{array}{c}
25 \\
35 \\
36 \\
\end{array}\right) =
\left(\begin{array}{c}
{h}^0_1 \\
{h}^0_2 \\
{h}^0_3 \\
\end{array}\right)_{\!\mathrm{mass-ordered}}\hspace*{-1cm} = 
\mbox{\texttt{CVHMIX}}_{ij} \left(\begin{array}{c}
\sqrt2\mathrm{Re}{H_1^0} \\
\sqrt2\mathrm{Re}{H_2^0}\\
\sqrt2\mathrm{Im}{H_1^0}\\
\sqrt2\mathrm{Im}{H_2^0}\\
\end{array}\right)~.
\end{equation}

In order to translate between $S$ and other conventions, 
the tree-level angle $\alpha$ may be needed. This should be
given in the SLHA1 output {\tt BLOCK ALPHA}:
\subsection*{\texttt{BLOCK ALPHA}}
\entry{CP conserved: $\alpha$; precise definition up to spectrum
  calculator, see SLHA1.\\ CP violated:
  $\alpha_{\mathrm{tree}}$. Must be accompanied by the matrix $S$, as
  described above, in the block \texttt{CVHMIX}.
}

\subsubsection{NMSSM}

Firstly, as described above, 
{\tt BLOCK MODSEL} should contain the switch 3 with value 1, corresponding to
the choice of the NMSSM particle content. 

Secondly, for the parameters that are also present in the MSSM, we
re-use the corresponding SLHA1 entries. That is, $m_Z$ should be given
in \texttt{SMINPUTS} entry \texttt{4} and $m_{H_1}^2, m_{H_2}^2$ can be
given in the \texttt{EXTPAR} entries \texttt{21} and \texttt{22}. $\tan\beta$
should either be given in \texttt{MINPAR} entry 3 (default) or
\texttt{EXTPAR} entry 25 (user-defined input scale), as in SLHA1.  If
$\mu$ should be desired non-zero, it can be given in \texttt{EXTPAR}
entry \texttt{23}. The corresponding soft parameter $m_3^2$ can be given
in \texttt{EXTPAR} entry \texttt{24}, in the form
$m_3^2/(\cos\beta\sin\beta)$, see \cite{Skands:2003cj}.

Further, new entries in {\tt BLOCK EXTPAR} have been defined 
for the NMSSM specific input parameters, as follows. As in the SLHA1,
these parameters are all given at the common scale 
\mgut, which can either be left up to the spectrum calculator or given
explicitly using \texttt{EXTPAR 0} (see \cite{Skands:2003cj}):  

\subsection*{\texttt{BLOCK EXTPAR}}
\arrdes{Input parameters specific to the NMSSM (i.e., 
in addition to the entries defined in \cite{Skands:2003cj})}
\numentry{61}{$\lambda$. Superpotential trilinear Higgs $SH_2H_1$ coupling.}
\numentry{62}{$\kappa$. Superpotential cubic $S$ coupling.} 
\numentry{63}{$A_\lambda$. Soft trilinear Higgs $SH_2H_1$ coupling.}
\numentry{64}{$A_\kappa$. Soft cubic $S$ coupling.} 
\numentry{65}{$\lambda \left< S \right>$. Vacuum expectation value of
  the singlet (scaled by $\lambda$). } 
\numentry{66}{$\xi_F$. Superpotential linear $S$ coupling.} 
\numentry{67}{$\xi_S$. Soft linear $S$ coupling.}
\numentry{68}{$\mu'$. Superpotential quadratic $S$ coupling.}
\numentry{69}{$m_{S}'^2$.  Soft quadratic $S$ coupling (sometimes
  denoted $\mu'B'$).} 
\numentry{70}{$m_S^2$. Soft singlet mass squared.}

{\bf Important note:} only 12 of the parameters listed in
eq.~(\ref{eq:nmssmpar}) should be given as input at any one time
(including explicit zeroes for parameters desired ``switched off''),
the remaining ones being determined by the minimisation of the
effective potential. Which combinations to accept is left up to the
individual spectrum calculator programs. Alternatively, for minimal
models, 6 parameters of those listed in
eq.~(\ref{eq:nmssmpar-reduced}) should be given.

In the spectrum output, running NMSSM parameters corresponding to
the \texttt{EXTPAR} entries above can be given in the block \texttt{NMSSMRUN
Q=...}:

\subsection*{\texttt{BLOCK NMSSMRUN Q=...}}
Output parameters specific to the NMSSM, given in the \DRbar\ scheme, at the
scale $Q$. As in the SLHA1, several of these blocks may be given
simultaneously in the output, each then corresponding to a specific
scale, but at least one should always be present. See 
corresponding entries in \texttt{EXTPAR} above for definitions.\\[3mm]
\numentry{1}{$\lambda(Q)^\DRbar$.}
\numentry{2}{$\kappa(Q)^\DRbar$.}
\numentry{3}{$A_\lambda(Q)^\DRbar$.}
\numentry{4}{$A_\kappa(Q)^\DRbar$.}
\numentry{5}{$\lambda\left<S\right>(Q)^\DRbar$.}
\numentry{6}{$\xi_F(Q)^\DRbar$.}
\numentry{7}{$\xi_S(Q)^\DRbar$.}
\numentry{8}{$\mu'(Q)^\DRbar$.}
\numentry{9}{$m_{S}'^2(Q)^\DRbar$.}
\numentry{10}{$m_S^2(Q)^\DRbar$.}

The new $3\times3$ block $S=$\texttt{NMHMIX} connects the particle codes
(=mass-ordered 
basis) for the CP-even Higgs bosons 
with the interaction basis according to the following definition:
\begin{equation}
\left(\begin{array}{c}
25 \\
35 \\
45 \\
\end{array}\right) =
\left(\begin{array}{c}
{h}^0_1 \\
{h}^0_2 \\
{h}^0_3 \\
\end{array}\right)_{\!\mathrm{mass-ordered}}\hspace*{-1cm} = 
\mbox{\texttt{NMHMIX}}_{ij} \left(\begin{array}{c}
\sqrt2\mathrm{Re}{H_1^0} \\
\sqrt2\mathrm{Re}{H_2^0}\\
\sqrt2\mathrm{Re}{S}\\
\end{array}\right)~.
\end{equation}

The new $2\times3$ block $S=$\texttt{NMAMIX} connects the particle codes
(=mass-ordered 
basis) for the CP-odd Higgs bosons 
with the interaction basis according to the following definition:
\begin{equation}
\left(\begin{array}{c}
36 \\
46 \\
\end{array}\right) =
\left(\begin{array}{c}
{A}^0_1 \\
{A}^0_2 \\
\end{array}\right)_{\!\mathrm{mass-ordered}}\hspace*{-1cm} = 
\mbox{\texttt{NMAMIX}}_{ij} \left(\begin{array}{c}
\sqrt2\mathrm{Im}{H_1^0} \\
\sqrt2\mathrm{Im}{H_2^0}\\
\sqrt2\mathrm{Im}{S}\\
\end{array}\right)~.
\end{equation}

Finally, the new $5\times5$ block \texttt{NMNMIX} gives the neutralino
mixing matrix, with the fifth mass eigenstate labelled \texttt{1000045} and
the fifth interaction eigenstate being the singlino, $\tilde{s}$. 


\section{{\tt SuSpect, HDECAY, SDECAY} and {\tt SUSY-HIT}}

\subsection{{\tt SuSpect}}
The Fortran code {\tt SuSpect} calculates the supersymmetric and  
Higgs particle
spectrum in the MSSM. It deals with the ``phenomenological MSSM''  
with 22 free
parameters defined either at a low or high energy scale, with the  
possibility
of renormalization group evolution (RGE) to arbritary scales, and with
constrained models with universal boundary conditions at high scales.  
These
are the minimal supergravity (mSUGRA), the anomaly mediated SUSY  
breaking
(AMSB) and the gauge mediated SUSY breaking (GMSB) models. The basic
assumptions of the most general possible MSSM scenario are {\it (a)}  
minimal
gauge group, {\it (b)} minimal particle content, {\it (c)} minimal  
Yukawa
interactions and R-parity conservation, {\it (d)} minimal set of soft  
SUSY
breaking terms. Furthermore, {\it (i)} all soft SUSY breaking  
parameters are
real (no CP-violation); {\it (ii)} the matrices for sfermion masses and
trilinear couplings are diagonal; {\it (iii)} first and second sfermion
generation universality is assumed. Here and in the following we  
refer the
reader for more details to the user's manual~\cite{Djouadi:2002ze} . \smallskip

As for the calculation of the SUSY particle spectrum in constrained  
MSSMs,  in
addition to the choice of the input parameters, the general  
algorithm  contains
three main steps. These are {\it (i)} the RGE of parameters back and   
forth
between the low energy scales, such as $M_Z$ and the electroweak   
symmetry
breaking (EWSB) scale, and the high-energy scale characteristic for   
the various
models; {\it (ii)} the consistent implementation of (radiative)   
EWSB; {\it
(iii)} the calculation of the pole masses of the Higgs bosons and   
the SUSY
particles, including the mixing between the current eigenstates and  the
radiative corrections when they are important. Here the program  
mainly  follows
the content and notations of \cite{Pierce:1996zz}, and for the leading two-loop
corrections to the Higgs masses the results summrized in \cite{Allanach:2004rh}
 are taken.\smallskip

The necessary files for the use in {\tt SuSpect} are the input file   
{\tt
suspect2.in}, the main routine {\tt suspect2.f}, the routine  {\tt
twoloophiggs.f}, which calculates the Higgs masses, as well as  {\tt  
bsg.f} for
the calculation of the $b\to s\gamma$ branching ratio.  The latter is  
needed in
order to check if the results are in agreement with the experimental
measurments. In the input file one can  select the model to be  
investigated, the
accuracy of the algorithm and the  input data (Standard Model fermion  
masses and
gauge couplings). At each run  {\tt SuSpect} generates two output  
files: one
easy to read,  {\tt suspect2.out}, and the other in the SLHA format
\cite{Skands:2003cj}.

\subsection{{\tt HDECAY}}

The Fortran code {\tt HDECAY} \cite{Djouadi:1997yw} calculates the decay  
widths and branching ratios
of the Standard Model Higgs boson, and of the neutral and charged Higgs
particles of the MSSM according to the current theoretical knowledge
(for reviews see 
refs.~\cite{Djouadi:2005gi,Djouadi:2005gj,GomezBock:2005if,Spira:1997dg}). 
It includes:\vspace*{-0.2cm}
\begin{itemize}
\item[-] All kinematically allowed decay channels with branching ratios
larger than $10^ {-4}$; apart from the 2-body decays also the loop- 
mediated,
the most important 3-body decay modes, and in the MSSM the cascade and
SUSY decay channels. \vspace*{-0.1cm}
\item[-] All relevant higher-order QCD corrections to the decays into  
quark
pairs and to the quark loop mediated decays into gluons are  
incorporated.
\vspace*{-0.1cm}
\item[-] Double off-shell decays of the CP-even Higgs bosons into  
massive
gauge bosons, subsequently decaying into four massless fermions.  
\vspace*{-0.1cm}
\item[-] All important 3--body decays: with off-shell heavy
top quarks; with one off-shell gauge boson as well as heavy neutral  
Higgs
decays with one off-shell Higgs boson. \vspace*{-0.1cm}
\item[-] In the MSSM the complete radiative corrections in the effective
potential approach with full mixing in the stop and sbottom sectors; it
uses the RG improved values of the Higgs masses and couplings,
  the relevant NLO corrections are implemented 
\cite{Carena:1995bx,Carena:1995wu}.
\vspace*{-0.1cm}
\item[-] In the MSSM, all decays into SUSY particles when
kinematically allowed. \vspace*{-0.1cm}
\item[-] In the MSSM, all SUSY particles are included in the loop  
mediated
$\gamma\gamma$ and $gg$ decay channels. In the gluonic decay modes  
the large
QCD corrections for quark  and squark loops are also included.
\end{itemize}

{\tt HDECAY} has recently undergone a major upgrade. We have  
implemented the
SLHA format, so that the program can now read in any input
file in the SLHA format and also give out the Higgs decay widths and  
branching
ratios in this accord. So, the program can now be easily linked to any
spectrum or decay calculator. Two remarks are in order: \\[0.1cm]
\noindent 1)
{\tt HDECAY} calculates the higher order corrections to the Higgs boson
decays in the $\overline{\mbox{MS}}$ scheme whereas all scale dependent
parameters read in from an SLHA input file provided by a spectrum  
calculator
are given in the $\overline{\mbox{DR}}$ scheme. Therefore, {\tt HDECAY}
translates the input parameters from the SLHA file into the
$\overline{\mbox{MS}}$ scheme where needed. \\[0.1cm]
\noindent 2)
The SLHA parameter input file only includes the MSSM Higgs boson mass
values, but not the Higgs self-interactions, which are needed in {\tt  
HDECAY}.
For the time being, {\tt HDECAY} calculates the missing interactions
internally within the effective potential approach. This is not  
completely
consistent with the values for the Higgs masses, since the spectrum
calculator does not necessarily do it with the same method and level of
accuracy as {\tt HDECAY}. The difference is of higher order, though.

\subsection{{\tt SDECAY}}

The Fortran code {\tt SDECAY} ~\cite{Muhlleitner:2003vg}, which has implemented  
the MSSM in the same way
as it is done in {\tt SuSpect}, calculates the decay widths and  
branching
ratios of all SUSY particles in the MSSM, including the most  
important higher
order effects \cite{Boehm:1999tr,Djouadi:2000bx,Djouadi:2001fa}:
 \vspace*{-0.2cm}
\begin{itemize}
\item The usual 2-body decays for sfermions and gauginos are  
calculated at tree
level. \vspace*{-0.1cm}
\item A unique feature is the possibility of calculating the SUSY-QCD
corrections to the decays involving coloured particles. They
can amount up to several tens of per-cents in some cases. The bulk of  
the EW
corrections has been accounted for by taking running parameters where
appropriate. \vspace*{-0.1cm}
\item In GMSB models the 2-body decays into the lightest SUSY  
particle, the
gravitino, have been implemented. \vspace*{-0.1cm}
\item If the 2-body decays are closed, multibody decays will be  
dominant.
{\tt SDECAY} calculates the 3-body decays of the gauginos, the  
gluino, the
stops and sbottoms. \vspace*{-0.1cm}
\item Moreover, loop-induced decays of the lightest stop, the
next-to-lightest neutralino  and the gluino are
included. \vspace*{-0.1cm}
\item If the 3-body decays are kinematically forbidden, 4-body decays  
of the
lightest stop can compete with the loop-induced $\tilde{t}_1$
decay and have therefore been implemented.
\item Finally, the top decays within the MSSM have been programmed.
\end{itemize}

Recently, {\tt SDECAY} has been updated with some major changes being  
(other
changes related to {\tt SUSY-HIT} are listed below): $i$) For  
reasons  of
shortening the output file, only non-zero branching  ratios are  
written out  in
the new version. $ii)$ We have created common blocks for the  
branching ratios
and total widths  of the various SUSY particles.

\subsection{ {\tt SUSY-HIT}}

The previous three programs have been linked together in a program
called {\tt SUSY-HIT} \cite{Djouadi:2006bz}. Including higher order
effects in the calculations, the package allows the consistent
calculation of MSSM particle decays with the presently highest level
of precision. The following files are needed to run {\tt SUSY-HIT}:\\
\underline{Spectrum files:} The spectrum can either be taken from any  
input
file in the SLHA format or from {\tt SuSpect}. In the first case,
{\tt SUSY-HIT} needs an SLHA input file which has to be named
{\tt slhaspectrum.in}. In the latter case, we need the necessary {\tt  
SuSpect}
routines: {\tt suspect2.in}, {\tt suspect2.f}, {\tt twoloophiggs.f} and
{\tt bsg.f}. \\[0.1cm]
\underline{Decay files:} {\tt SDECAY} is the main program and now  
reads in
{\tt susyhit.in} and calls {\tt HDECAY} which is now a subroutine and,
in order to keep the package as small as possible, only one routine  
calculating the Higgs boson
masses and Higgs self-couplings has been retained in {\tt HDECAY} to  
extract
the Higgs self-interaction strengths not provided by the spectrum  
calculators;
also, {\tt HDECAY} does not create any output file within the package.
{\tt SDECAY} passes the necessary parameters from {\tt susyhit.in} to
{\tt HDECAY} via a newly created common block called {\tt SUSYHITIN}. As
before, it calls {\tt SuSpect} in case the spectrum is taken from  
there. The
SLHA parameter and spectrum input file {\tt slhaspectrum.in} is read  
in by
both {\tt HDECAY} and {\tt SDECAY}. The output file created by {\tt  
SDECAY} at
each run is called {\tt susyhit\_slha.out} if it is in the SLHA  
format or
simply {\tt susyhit.out} if it is in an output format easy to read. \\ 
[0.1cm]
\underline{Input file:}
The {\tt HDECAY} and {\tt SDECAY} input files have been merged into  
one input
file {\tt susyhit.in}. Here, first of all the user can choose among two
{\tt SUSY-HIT} related options:\\[0.1cm]
1. The three programs {\tt SuSpect}, {\tt HDECAY}, {\tt SDECAY} are
linked and hence {\tt SuSpect} provides the spectrum and the soft SUSY
breaking parameters at the EWSB scale. \\[0.1cm]
2. The two programs {\tt HDECAY} and {\tt SDECAY} are linked. The
necessary input parameters are taken from a file in the SLHA format
provided by any spectrum calculator.\\[0.1cm]
Furthermore, various options for running the {\tt SDECAY} program can be
chosen, such as whether or not to include QCD corrections to 2-body  
decays,
the multibody and/or loop decays, the GMSB decays and the top decays.  
The
scale and number of loops of the running couplings can be fixed.  
Finally,
some parameters related to {\tt HDECAY} can be set, like the charm and
strange quark masses, the $W,Z$ total widths, some CKM matrix  
elements etc.
All other necessary parameters are read in from the {\tt  
slhaspectrum.in}
input file.\\[0.1cm]
\noindent
\underline{Changes and how the package works:}
{\tt SuSpect}, {\tt HDECAY} and {\tt SDECAY} are lin\-ked via the  
SLHA format.
Therefore, the name of the output file provided by {\tt SuSpect} has  
to be the
same as the SLHA input file read in by {\tt HDECAY} and {\tt SDECAY}. We
called it {\tt slhaspectrum.in}.  This is one of the changes made in the
programs with respect to their original version. Further major
changes have been made. For the complete list of changes please refer  
to the
web page given below. \\[0.1cm]
\noindent \underline{Web page:} We have created a web page at the  
following
url address:\\
\centerline{{\tt http://lappweb.in2p3.fr/$\sim$muehlleitner/SUSY-HIT/}}
There the user can download all files necessary for the program package
as well as a {\tt makefile} for compiling the programs. We use the  
newest
versions of the various programs which will be updated regularly. Short
instructions are given how to use the programs. A file with updates and
changes is provided. Finally, some examples of output files are given.


\section{\tt FeynHiggs}

{\tt FeynHiggs}\ is a program for computing Higgs-boson masses
and related observables in the (NMFV) MSSM with real or complex
parameters. The observables comprise mixing angles, branching ratios, and
couplings, including state-of-the-art higher-order contributions.  
The centerpiece is a Fortran library for use with Fortran and C/C++.
Alternatively, {\tt FeynHiggs}\ has a command-line, Mathematica, and Web
interface. {\tt FeynHiggs}\ is available from {\tt www.feynhiggs.de}.

{\tt FeynHiggs}~\cite{Heinemeyer:1998yj,Heinemeyer:1998np,Degrassi:2002fi,Hahn:2005cu} is
a Fortran code for the evaluation of the masses, decays and production
processes of Higgs bosons in the (NMFV) MSSM with real or complex
parameters. 
The calculation of the higher-order corrections is based on the
Feynman-diagrammatic (FD) approach~\cite{Heinemeyer:1998jw,Heinemeyer:1998kz,Heinemeyer:1998np,Heinemeyer:2004gx}.
At the one-loop level, it consists of a complete evaluation, including the
full momentum and phase dependence, and as a further option the full 
$6 \times 6$ non-minimal flavor violation (NMFV) 
contributions~\cite{Heinemeyer:2004by,Hahn:2005qi}. At the two-loop
level all available corrections from the real MSSM have been included.
They are supplemented by the resummation of the leading effects from the
(scalar)~$b$ sector including the full complex phase dependence.

In addition to the Higgs-boson masses, the program also provides
results for the effective couplings
and the wave function normalization factors 
for external Higgs bosons~\cite{Hahn:2002gm}, taking into account NMFV
effects from the Higgs-boson self-energies.
Besides the computation of the Higgs-boson masses, effective couplings and 
wave function normalization factors,
the program also evaluates an estimate for 
the theory uncertainties of these quantities due to unknown
higher-order corrections. 

Furthermore {\tt FeynHiggs}\ contains the evaluation of all relevant Higgs-boson
decay widths%
\footnote{
The inclusion of flavor changing decays is work in progress.
}.
In particular, the following quantities are
calculated:
\begin{itemize}
\item
the total width for the neutral and charged Higgs bosons,

\item
the branching ratios and effective couplings of the three neutral Higgs
bosons to 
\begin{itemize}
\item SM fermions (see also Ref.~\cite{Heinemeyer:2000fa}),
      $h_i \to \bar f f$,
\item SM gauge bosons (possibly off-shell),
      $h_i \to \gamma\gamma, ZZ^*, WW^*, gg$,
\item gauge and Higgs bosons,
      $h_i \to Z h_j$,
      $h_i \to h_j h_k$,
\item scalar fermions,
      $h_i \to \tilde f^\dagger \tilde f$,
\item gauginos,
      $h_i \to \tilde\chi^\pm_k \tilde\chi^\mp_j$,
      $h_i \to \tilde\chi^0_l \tilde\chi^0_m$,
\end{itemize}

\item 
the branching ratios and effective couplings 
of the charged Higgs boson to 
\begin{itemize}
\item SM fermions,
      $H^- \to \bar f f'$,
\item a gauge and Higgs boson,
      $H^- \to h_i W^-$,
\item scalar fermions,
      $H^- \to \tilde f^\dagger \tilde f'$,
\item gauginos,
      $H^- \to \tilde\chi^-_k \tilde\chi^0_l$.
\end{itemize}

\item
the production cross sections 
of the neutral Higgs bosons at the Tevatron and the LHC in the
approximation where the corresponding SM cross section is rescaled 
by the ratios of the corresponding partial widths in the MSSM and the SM
or by the wave function normalization factors for external Higgs
bosons, see Ref.~\cite{Hahn:2006my} for further details.

\end{itemize}

\noindent
For comparisons with the SM, the following quantities are also evaluated
for SM Higgs bosons with the same mass as the three neutral MSSM Higgs
bosons:
\begin{itemize}
\item
the total decay width,

\item
the couplings and branching ratios of a SM Higgs boson to SM fermions,

\item
the couplings and branching ratios of a SM Higgs boson to 
SM gauge bosons (possibly off-shell).

\item
the production cross sections at the Tevatron and the
LHC~\cite{Hahn:2006my}.
\end{itemize}

\noindent
{\tt FeynHiggs}\ furthermore provides results for electroweak
precision observables that give rise to constraints on the 
SUSY parameter space (see Ref.~\cite{Heinemeyer:2004gx} and references
therein):
\begin{itemize}

\item
the quantity $\Delta\rho$ up to the two-loop level that
can be used to
indicate disfavored scalar top and bottom mass combinations, 

\item
an evaluation of $M_W$ and $\sin^2\theta_{\mathrm{eff}}$, 
where the SUSY contributions are 
treated in the $\Delta\rho$ approximation (see e.g.\
Ref.~\cite{Heinemeyer:2004gx}), 
taking into account at the one-loop level the effects of complex phases 
in the scalar top/bottom sector as well as NMFV
effects~\cite{Heinemeyer:2004by},

\item
the anomalous magnetic moment of the muon, including a full one-loop
calculation as well as leading and subleading two-loop
corrections,

\item
the evaluation of ${\rm BR}(b \to s \gamma)$ including NMFV
effects~\cite{Hahn:2005qi}.  
\end{itemize}

\noindent
Finally, {\tt FeynHiggs}\ possesses some further features:
\begin{itemize}
\item
Transformation of the input parameters from the $\overline{\rm{DR}}$ to
the on-shell scheme (for the scalar top and bottom parameters), including 
the full ${\cal O}{\alpha_s}$ and ${\cal O}{\alpha_{t,b}}$ corrections.

\item
Processing of SUSY Les Houches Accord (SLHA 2)
data~\cite{Skands:2003cj,Hahn:2004bc,Hahn:2006nq} including the full
NMFV structure.
{\tt FeynHiggs}\ reads the output of a spectrum generator file and evaluates the
Higgs boson masses, branching ratios etc.  The results are written in
the SLHA format to a new output file.

\item
Predefined input files for the SPS~benchmark
scenarios~\cite{Allanach:2002nj} and the Les Houches benchmarks for
Higgs boson searches at hadron colliders~\cite{Carena:2002qg} are included.

\item
Detailed information about all the features of {\tt FeynHiggs}\ are provided in
man pages.
\end{itemize}

\noindent
{\tt FeynHiggs}\ is available from {\tt www.feynhiggs.de}.

\section{\tt FchDecay}
\label{sec:tools:FchDecay}

\noindent
{\tt FchDecay} is a computer program to compute the Flavor Changing
Neutral Current (FCNC) decay branching ratios $BR(h\to bs)$ and
$BR(h\to tc)$ in the flavor violating Minimal Supersymmetric Standard
Model (MSSM). The input/output is performed in the SUSY Les Houches
Accord II (\texttt{SLHA})\cite{Skands:2003cj,Allanach:2006fy,SLHApage}
convention (using an extension of SLHALib~\cite{Hahn:2004bc}).  This
program is based on the work and results of
Refs.~\cite{Guasch:1999jp,Bejar:2001sj,Bejar:2004rz,Bejar:2005kv,Bejar:2006hd}.

\noindent
The approximations used in the computation are:
\begin{itemize}
\item The full one-loop SUSY-QCD contributions to the FCNC partial decay widths 
	$\Gamma(h\to bs,tc)$ is included;
\item The Higgs sector parameters (masses and CP-even mixing angle $\alpha$) have been treated using 
	the leading $m_t$ and $m_b\tan\beta$ approximation  to the one-loop result;
\item The Higgs bosons total decay widths $\Gamma(h\to X)$ are computed at leading order, including 
	all the relevant channels;
\item A Leading Order computation of $B(b\to s \gamma)$ (for checking the parameter space) is also
	included.
\end{itemize}

\noindent
The code implements the flavor violating MSSM, it allows complete intergenerational mixing in the 
Left-Left and Right-Right squark sector (but it does not allow for intergenerational mixing in the 
Left-Right sector).

\noindent
The program includes a (simplified) computation of the Higgs boson masses and total decay widths, and 
it will write them to the output file. However: 
\begin{itemize}
\item If the input file contains the Higgs sector parameters (masses and CP-even mixing angle $\alpha$) 
	it will use those values instead;
\item If the input file contains Higgs boson decay tables, it will just add the FCNC decays to that 
	table (instead of computing the full table).
\end{itemize}
This setup allows to use the computations of more sophisticated
programs for the Higgs boson parameters and/or total decay widths, and
then run the \texttt{FchDecay} program on the resulting output file to
obtain the FCNC partial decay widths.

\noindent
The program is available from the web page, \texttt{http://fchdecay.googlepages.com}, and comes with 
a complete manual (detailing the included physics models, and running instructions). The authors can 
be reached at \texttt{fchdecay@gmail.com}.

\section{MSSM NMFV in {\tt FeynArts} and {\tt FormCalc}} 
\label{sec:tool:feynarts}

\noindent
In the presence of non-minimal flavour violation (NMFV) the $2\times 2$ mixing of the squark within each 
family is enlarged to a full $6\times 6$ mixing among all three generations, such that the mixed states 
are
\begin{align}
\tilde u_i = (R_u)_{ij} \begin{pmatrix}
  \tilde u_L &
  \tilde c_L &
  \tilde t_L &
  \tilde u_R &
  \tilde c_R &
  \tilde t_R
\end{pmatrix}_j^T, \\
\tilde d_i = (R_d)_{ij} \begin{pmatrix}
  \tilde d_L &
  \tilde s_L &
  \tilde b_L &
  \tilde d_R &
  \tilde s_R &
  \tilde b_R
\end{pmatrix}_j^T.
\nonumber 
\end{align}
The matrices $R_q$ diagonalize the mass matrices
\begin{align}
M^2_q &= \begin{pmatrix}
M^2_{LL,q}     & M^2_{LR,q} \\
(M^2_{LR,q})^* & M^2_{RR,q}
\end{pmatrix} + \Delta_q\,, \\
\notag
M^2_{\begin{subarray}{l}
  AA,q \\
  A = L,R
\end{subarray}} &= 
\mathrm{diag}(M^2_{A,q_1},\,M^2_{A,q_2},\,M^2_{A,q_3}), \\ 
\notag
M^2_{LR,q} &= 
\mathrm{diag}(m_{q_1} X_{q_1},\,m_{q_2} X_{q_2},\,m_{q_3} X_{q_3})
\end{align}
where $q = \{u,d\}$, $\{q_1,q_2,q_3\} = u,c,t$ for the up- 
and $d,s,b$ for the down-squark mass matrix and
\begin{align}
\notag
M^2_{L,q_i}\!\! &= M_{\tilde Q,q_i}^2\!\! + 
  m_{q_i}^2 \!\!+ \cos 2\beta\,(T_3^q - Q_q s_W^2) m_Z^2\,, \\
\notag
M^2_{R,u_i}\!\! &= M_{\tilde U,u_i}^2\!\! +
  m_{u_i}^2 \!\!+ \cos 2\beta\,Q_u s_W^2 m_Z^2\,, \\
\notag
M^2_{R,d_i}\!\! &= M_{\tilde D,d_i}^2\!\! +
  m_{d_i}^2 \!\!+ \cos 2\beta\,Q_d s_W^2 m_Z^2\,, \\
X_{\{u,d\}_i}\!\! &= A_{\{u,d\}_i} - \mu\{\cot\beta,\tan\beta\}\,.
\end{align}
The actual dimensionless input quantities $\delta$ are
\begin{align}
\Delta_q &= \begin{pmatrix}
N^2_{LL,q} & N^2_{LR,q} \\
(N^2_{LR,q})^* & N^2_{RR,q}
\end{pmatrix} \delta_q\,, \\
\notag
N^2_{\begin{subarray}{l}
  AB,q \\
  A,B = L,R
\end{subarray}} &= \begin{pmatrix}
M_{A,q_1} \\
M_{A,q_2} \\
M_{A,q_3}
\end{pmatrix} \otimes \begin{pmatrix}
M_{B,q_1} \\
M_{B,q_2} \\
M_{B,q_3}
\end{pmatrix}.
\end{align}

\noindent
The new {\tt FeynArts} model file \texttt{FVMSSM.mod} generalizes the
squark couplings in \texttt{MSSM.mod} to the NMFV case.  It contains
the new objects
\begin{tabbing}
\texttt{UASf[$s$,$s'$,$t$]}~~\=
the squark mixing matrix $R_{u,d}$ \\[1ex]
\texttt{MASf[$s$,$t$]} \>
the squark masses, 
\end{tabbing}
with $s, s' = 1\dots 6,~~t = 3 (u), 4 (d).$

\noindent
The initialization of \texttt{MASf} and \texttt{UASf} is already built into {\tt FormCalc}'s 
\verb|model_mssm.F| but needs to be turned on by defining a preprocessor flag in \texttt{run.F}:
\begin{verbatim}
#define FLAVOUR_VIOLATION
\end{verbatim}
The NMFV parameters $(\delta_t)_{ss'}$ are represented by the \texttt{deltaSf} array:
\begin{tabbing}
\texttt{double complex deltaSf($s$,$s'$,$t$)}
\end{tabbing}
Since $\delta$ is a Hermitian matrix, only the entries on and above the
diagonal need to be filled.  For convenience, the following
abbreviations can be used for individual matrix elements:
\begin{align*}
\mathtt{deltaLLuc} &= (\delta_u)_{12} & \mathtt{deltaLRuc} &= (\delta_u)_{15}\\
\mathtt{deltaRLucC} &= (\delta_u)_{24} & \mathtt{deltaRRuc} &= (\delta_u)_{45}
\\[1ex]
\mathtt{deltaLLct} &= (\delta_u)_{23} & \mathtt{deltaLRct} &= (\delta_u)_{26}\\
\mathtt{deltaRLctC} &= (\delta_u)_{35} & \mathtt{deltaRRct} &= (\delta_u)_{56}
\\[1ex]
\mathtt{deltaLLut} &= (\delta_u)_{13} & \mathtt{deltaLRut} &= (\delta_u)_{16}\\
\mathtt{deltaRLutC} &= (\delta_u)_{34} & \mathtt{deltaRRut} &= (\delta_u)_{46}
\end{align*}
and analogous entries for the down sector. 

\noindent
Note the special treatment of the $RL$ elements: One has to provide the complex conjugate of the 
element. The original lies below the diagonal and would be ignored by the eigenvalue routine.

\noindent
The off-diagonal trilinear couplings $A$ acquire non-zero entries through the relations
\begin{equation}
m_{q,i} (A_q)_{ij} = (M^2_q)_{i,j+3}\,,
\quad
i, j = 1\dots 3\,.
\end{equation}

\noindent
In summary: NMFV effects (see~\cite{Hahn:2005qi}) can be computed with
{\tt FeynArts} \cite{Kublbeck:1990xc,Hahn:2000kx,Hahn:2001rv} and {\tt
FormCalc} \cite{Hahn:1998yk}. These packages provide a high level of
automation for perturbative calculations up to one loop. Compared to
calculations with the MFV MSSM, only three minor changes are required:
\begin{itemize}
\item
choosing \texttt{FVMSSM.mod} instead of \texttt{MSSM.mod},
\item
setting \verb|FLAVOUR_VIOLATION| in \texttt{run.F},
\item
providing values for the \texttt{deltaSf} matrix.
\end{itemize}
These changes are contained in {\tt FeynArts} and {\tt FormCalc},
available from www.feynarts.de.

\section{{\tt SPheno}}

\noindent
{\tt SPheno} is a program to calculate the spectrum of superymmetric
models, the decays of supersymmetric particles and Higgs bosons as
well as the production cross sections of these particles in $e^+ e^-$
annihilation.  Details of the algorithm used for the MSSM with real
parameters and neglecting mixing between the (s)fermion generations
can be found in~\cite{Porod:2003um}. This version can be found and
downloaded from
\begin{verbatim}
  http://theorie.physik.uni-wuerzburg.de/~porod/SPheno.html
\end{verbatim}
 
\noindent
In this contribution the model extensions regarding flavour aspects are
described. In the context of the MSSM the most general flavour structure
as well as all CP-phases are included in the RGE running and in the
computation of SUSY masses at tree-level as well as at the one-loop level.
In the Higgs sector, the complete flavour structure is included for the
calculation of the masses at the one-loop level. At the 2-loop level
there is still the approximation used that the 3rd generation does not
mix with the other ones. With respect to CP-phases, the induced mixing between
scalar and pseudoscalar Higgs bosons is not yet taken into account. For the
decays of supersymmetric particles and Higgs bosons, the complete flavour
structure is taken into account at tree-level using running 
$\overline{DR}$ couplings to
take into account the most important loop corrections. A few examples are
\begin{eqnarray}
\tilde \chi^0_i &\to& e^\pm \tilde \mu^\mp_R, \,\,
                      e^\pm  \mu^\mp \tilde \chi^0_j, \,\,
                      \bar{u} \tilde c_L, \,\,
                      \bar{u} c \tilde \chi^0_j, \,\,
                      \bar{u} b \tilde \chi^+_k; \,\,
\tilde g \to \bar{u} \tilde c_L, \,\,
                      \bar{u} c \tilde \chi^0_j, \,\,
                      \bar{u} b \tilde \chi^+_k; \\
H^+ &\to& \bar{b} c, \tilde{\bar{b}}_1 \tilde c_R; \,\,
H^0  \to \tilde e^\pm_R \tilde \tau^\mp_1 \,.
\end{eqnarray}
The complete list is given in the manual. Also in the case of production
in $e^+ e^-$ annihilation all flavour-off diagonal channels are available.
Flavour and $CP$ violating terms are already constrained by several
experimental data. For these reason, the following observables are
calculated taking into account all parameters: anomalous magnetic and
electric dipolements of leptons, the most important ones being
$a_\mu$ and $d_e$; the rare decays of leptons: $l \to l' \gamma$,
$l \to 3 l'$; rare decays of the $Z$-boson: $Z \to l l'$; 
$b\to s \gamma$, $b\to s  \mu^+ \mu^-$,
$B_{s,d} \to \mu^+ \mu^-$, $B^\pm_u \to \tau^\pm \nu$,
$\delta(M_{B_{s,d}})$ and $\Delta \rho$.

\noindent
This version of {\tt SPheno} also includes extended SUSY models: (a) the NMSSM
and (b) lepton number violation and thus R-parity violation. In both
model classes the masses are calculated at tree-level except for the
Higgs sector where radiative corrections are taken into account. In both cases
the complete flavour structure is taken into account in the calculation of the
masses, the decays of supersymmetric particles and Higgs bosons as well
as in the production of these particles in $e^+ e^-$ annihilation.
The low energy observables are not yet calculated in these models but
the extension of the corresponding routines to included these models
is foreseen for the near future.

\noindent
Concerning input and output the current version of the SLHA2 accord is
implemented as described in section~\ref{sec:slha2}
and in \cite{Allanach:2006fy}. The version described
here is currently under heavy testing and the write-up of the
corresponding manual has just started. As soon as the manual is in a useful
stage, the program can be found on the web page given above. In the meantime
a copy can be obtained be sending an email to 
\verb+porod@physik.uni-wuerzburg.de+.
\section{\tt SOFTSUSY}

\noindent
{\tt SOFTSUSY}~\cite{Allanach:2001kg} provides a SUSY spectrum in the
MSSM consistent with input low energy data, and a user supplied
high-energy constraint.  It is written in \textsc{C++} with an
emphasis on easy generalisability.  It can produce SUSY Les Houches
Accord compliant output~\cite{Skands:2003cj}, and therefore link to
Monte-Carlos (e.g. {\tt HERWIG}~\cite{Corcella:2002jc}) or programs
that calculate sparticle decays such as {\tt
SDecay}~\cite{Muhlleitner:2003vg}.  {\tt SOFTSUSY} can be obtained
from URL
\begin{verbatim}
  http://projects.hepforge.org/softsusy
\end{verbatim}

\noindent
{\tt SOFTSUSY} currently incorporates 3 family mixing in the limit of
CP conservation.  The high-energy constraint in {\tt SOFTSUSY} upon
the supersymmetry breaking terms may be completely non-universal,
i.e.\ can have 3 by three-family mixing incorporated within them. All
of the renormalisation group equations (RGEs) used to evolve the MSSM
between high-energy scales and the weak scale $M_Z$ have the full
three-family mixing effects incorporated at one loop in all MSSM
parameters. Two-loop terms in the RGEs are included in the dominant
third family approximation for speed of computation and so mixing is
neglected in the two-loop terms.  Currently, the smaller one-loop
weak-scale threshold corrections to sparticle masses are also
calculated in the dominant third-family Yukawa approximation, and so
family mixing is neglected within them.

\noindent
The user may request that, at the weak scale, all of the quark mixing
is incorporated within a symmetric up quark Yukawa matrix $(Y_U)'$, or
alternatively within a symmetric down quark Yukawa matrix $(Y_D)'$.
These are then related (via the {\tt SOFTSUSY}
conventions~\cite{Allanach:2001kg} for the Lagrangian) to the
mass-basis Yukawa matrices $Y_U, Y_D$ via
\begin{equation}
(Y_U)'=V_{CKM}^T (Y^U) V_{CKM} \mbox{~or~}
(Y_D)'=V_{CKM} (Y^D) V_{CKM}^T, \label{ckm}
\end{equation}
where by default $V_{CKM}$ contains the CKM matrix in the standard
parameterisation with central empirical values of the input angles
except for the complex phase, which is set to zero.  Even if one
starts at a high-energy scale with a completely family-universal model
(for example, mSUGRA), the off-diagonal quark Yukawa matrices induce
squark mixing through RGE effects.

\noindent
The second SUSY Les Houches Accord (SLHA2) has been  completed
recently, see section \ref{sec:slha2}.  The
flavour mixing aspects will be incorporated into {\tt SOFTSUSY}
as fast as possible,
allowing input and output of flavour mixing parameters in a common
format to other programs.

\section{{\tt CalcHep} for beyond Standard Model Physics}

\noindent
{\tt CalcHep} is a package for the computation of Feynman diagrams at tree-level, integration over 
multi-particle phase space, and partonic level event generation.  The main idea of {\tt CalcHep} is to
make publicly available the passing on from Lagrangians to final distributions. This is done effectively 
with a high level of automation. {\tt CalcHep} is a menu-driven system with help facilities, but it
also can be used in a non-interactive batch mode. 
 
\noindent
In principle, {\tt CalcHep} is restricted by tree level calculations
but there it can be applied to any model of particle interaction. {\tt
CalcHep} is based on the symbolic calculation of squared diagrams.  To
perform such a calculation it contains a built-in symbolic calculator.
Calculated diagrams are transformed into a C-code for further
numerical evaluations. Because of the factorial increase of the number
of diagrams with the number of external legs, {\tt CalcHep} is
restricted to $2->4$ processes.

\noindent
The Implementation of new models for {\tt CalcHep} is rather simple
and can be done with help of the {\tt LanHep} package. Currently,
there are publicly available realizations of the Standard Model, MSSM,
NMSSM, CPVMSSM, and Lepto-quark model.  Also there are {\it private}
realizations of models with extra dimensions and the Little Higgs
model. Models with flavour violation can also be implemented in {\tt
CalcHep}.

\noindent
WWW destination:
\begin{verbatim}
   http://theory.sinp.msu.ru/~pukhov/calchep.html
\end{verbatim}

\noindent
The basic references for {\tt CompHEP} can be found
in~\cite{Pukhov:1999gg,Pukhov:2004ca}.


\section{\tt HvyN}

The Monte Carlo program {\tt HvyN} allows to study heavy neutrino production
processes at hadron colliders.
It can be downloaded from
\begin{verbatim}
http://www.to.infn.it/~pittau/ALPGEN_BSM.tar.gz
\end{verbatim}
or
\begin{verbatim}
http://mlm.home.cern.ch/m/mlm/www/alpgen/
\end{verbatim}
and it is based on the {\tt Alpgen} package \cite{Mangano:2002ea}, 
from which inherits the main features and the interface facilities.

The code allows to study the following three processes, where a heavy Neutrino
$N$ (of Dirac or Majorana nature) is produced in association with a charged lepton
\begin{itemize}
\item[1)] $p p^{\!\!\!\!\!\textsuperscript{\tiny{(--)}}}\!\! \to \ell_1 N  
\to \ell_1\, \ell_2\, W  \to \ell_1\, \ell_2 \,f\, \bar f^\prime$; 
\item[2)] $p p^{\!\!\!\!\!\textsuperscript{\tiny{(--)}}}\!\! \to \ell_1 N  
\to \ell_1\,  \nu_{\ell_2}\, Z  \to  \ell_1\, \nu_{\ell_2}\,f\, \bar f$; 
\item[3)] $p p^{\!\!\!\!\!\textsuperscript{\tiny{(--)}}}\!\! \to \ell_1 N  
\to \ell_1\, \nu_{\ell_2}\,  H \, \to \ell_1\, \nu_{\ell_2}\,f\, \bar f$. 
\end{itemize}
The full $2 \to 4$ matrix element for the complete decay chain is implemented, 
so that spin correlations and finite width effects are correctly taken into account.
The only relevant subprocess is
\begin{equation}
q \bar q^\prime \to W^\star \to \ell_1 N\,,
\end{equation}
followed by the full decay chain. The appropriate 
Lagrangian can be found in \cite{delAguila:2006dx}.

  The above three processes are selected by setting an input  variable
({\tt indec}) to 1, 2 or 3, respectively.
The flavour of the outgoing leptons, not coming from the boson decay,
is controlled by 2 other variables
{\tt il1} and {\tt il2} (the values 1, 2, 3 correspond
to the fist, second and third lepton family).
In addition, the variable {\tt ilnv} should be set to 0 (1) if a
lepton number conserving (violating) process is considered.
Furthermore the variable {\tt ima} should be given the value 0 (1)
in case of Dirac (Majorana) heavy neutrinos.  

When {\tt indec= 1} and {\tt imode= 0,1} the $W$ decays into $e$ and $\nu_e$.
Other decay options can be implemented at the unweighting stage 
according to the following options
\begin{eqnarray}
{\tt 1} &=& e {\bar \nu_e}, \nonumber \\
{\tt 2} &=& \mu {\bar \nu_\mu}, \nonumber \\
{\tt 3} &=& \tau {\bar \nu_\tau}, \nonumber \\
{\tt 4} &=& l {\bar \nu}_l (l = e, \mu, \tau),  \nonumber  \\
{\tt 5} &=& q {\bar q}',  \nonumber  \\
{\tt 6} &=& {\rm fully} \,\, {\rm inclusive}. \nonumber 
\end{eqnarray}

When {\tt indec= 2} the decay mode of the $Z$ boson should be selected 
at the event generation level by setting the variable {\tt idf} to the following
values
\begin{eqnarray}
{\tt 0} &\Rightarrow& \sum_\ell \nu_\ell \bar \nu_\ell, \nonumber \\
{\tt 1} &\Rightarrow& \sum_\ell \ell^- \ell^+, \nonumber \\
{\tt 2} &\Rightarrow& u \bar u~{\rm and}~c \bar c, \nonumber \\
{\tt 3} &\Rightarrow& d \bar d~{\rm and}~s \bar s,  \nonumber  \\
{\tt 4} &\Rightarrow& b {\bar b},  \nonumber  \\
{\tt 11} &\Rightarrow&    e^-    e^+,           \nonumber \\
{\tt 13} &\Rightarrow&  \mu^-  \mu^+,    \nonumber \\
{\tt 15} &\Rightarrow& \tau^- \tau^+. \nonumber 
\end{eqnarray}

When {\tt indec= 3} the following decay modes of the $H$ boson 
can be selected, at the generation level, by setting the variable {\tt idf} 
according to the following scheme
\begin{eqnarray}
{\tt 1} &\Rightarrow& \tau^- \tau^+, \nonumber \\
{\tt 2} &\Rightarrow& c \bar c, \nonumber \\
{\tt 4} &\Rightarrow& b {\bar b}.  \nonumber  
\end{eqnarray}

\section{{\tt PYTHIA} for Flavour Physics at the LHC}

\noindent
{\tt PYTHIA}~\cite{Sjostrand:2006za} is a general-purpose event
generator for hadronic events in $e^+e^-$, $e h$, and $hh$ collisions
(where $h$ is any hadron or photon). The current version is always
available from the {\tt PYTHIA} web page, where also update notes
and a number of useful example main programs can be found. For recent
brief overviews relating to SM, BSM, and Higgs physics, see
\cite{Dobbs:2004qw},~\cite{Skands:2005vi}, and~\cite{Heinemeyer:2005gs}, respectively. For flavour 
physics at the LHC, the most relevant processes in {\tt PYTHIA} can
be categorised as follows:
\begin{itemize}
\item SUSY with trilinear $R$-parity violation~\cite{Skands:2001it,Sjostrand:2002ip}:\\ 
{\tt PYTHIA} includes all massive tree-level matrix
elements~\cite{Dreiner:1999qz} for 2-body sfermion decays and 3-body
gaugino/higgsino decays. (Note: RPV production cross sections are not
included.) Also, the Lund string fragmentation model has been extended
to handle antisymmetric colour topologies~\cite{Sjostrand:2002ip},
allowing a more correct treatment of baryon number flow when baryon
number is violated.
\item Other BSM:\\ 
	Production and decay/hadronization of {\sl 1)} Charged Higgs in 2HDM and SUSY models via 
	$\bar{q}g \to\bar{q}'H^+$, $gg/qq\to \bar{t}bH^+$, $q\bar{q}\to H^+ H^-$ (including the possibility
	of a $Z'$ contribution with full interference), $q\bar{q} \to H^\pm h^0/H^\pm H^0$, and 
	$t\to bH^+$, {\sl 2)} a $W'$ (without interference with the SM $W$), {\sl 3)} a horizontal 
	(FCNC) gauge boson $R^0$ coupling between generations, e.g.\ $s\bar{d} \to R^0 \to \mu^- e^+$, 
	{\sl 4)} Leptoquarks $L_Q$ via $qg \to \ell L_Q$ and $gg/q\bar{q}\to L_Q \bar{L}_Q$. 
  	{\sl 5)} compositeness (e.g.\ $u^*$), {\sl 6)} doubly charged Higgs bosons from L-R symmetry, 
	{\sl 7)} warped extra dimensions, and {\sl 8)} a strawman technicolor model. See
  	\cite{Sjostrand:2006za}, Sections 8.5-8.7 for details. 
\item Open heavy-flavour production ($c,b,t,b',t'$):\\
 	Massive matrix elements for QCD $2\to 2$ and resonant $Z/W$ (and $Z'/W'$) heavy flavour
 	production. Also includes flavour excitation and gluon splitting to massive quarks in the shower 
	evolution, see~\cite{Norrbin:2000zc}.
\item Closed heavy-flavour production ($J/\psi$, $\Upsilon$, $\chi_{c,b}$):\\
{\tt PYTHIA} includes a substantial number of colour singlet and (more
recently) NRQCD colour octet mechanisms. For details,
see~\cite{Sjostrand:2006za}, Section 8.2.3.
\item Hadron decays:\\ 
	A large number of $c$ and $b$ hadron (including -onia) decays are implemented. In both cases, 
	most channels for which exclusive branching fractions are known are explicitly listed. For the 
	remaining channels, either educated guesses or a fragmentation-like process determines the 
	flavour composition of the decay products. With few exceptions, hadronic decays are then 
	distributed according to phase space, while semileptonic ones incoporate a simple $V-A$
	structure in the limit of massless decay procucts. See~\cite{Sjostrand:2006za}, Section 13.3 for 
	more details. 
\end{itemize} 

\noindent
Additional user-defined production processes can be interfaced via the routines \texttt{UPINIT} and 
\texttt{UPEVNT} (see~\cite{Sjostrand:2006za}, Section 9.9), using the common Les Houches standard 
\cite{Boos:2001cv}.  Flavour violating resonance decays can also be introduced \emph{ad hoc} via the 
routine \texttt{PYSLHA}, using SUSY Les Houches Accord decay tables~\cite{Skands:2003cj}.

\section{{\tt Sherpa} for Flavour Physics}

\noindent
{\tt Sherpa}~\cite{Gleisberg:2003xi} is a multi-purpose Monte Carlo
event generator that can simulate high energetic collisions at lepton
and hadron colliders.  {\tt Sherpa} is publicly available and the
source code, potential bug-fixes, documentation material and also a
{\tt Sherpa} related WIKI can be found under:\\[-2ex]
\begin{verbatim}
  http://www.sherpa-mc.de
\end{verbatim}

\noindent
The ingredients of {\tt Sherpa} especially relevant for flavour physics at the LHC are the matrix 
elements for corresponding hard production processes and the hadronization and decay of flavours 
produced:
\begin{itemize}
\item The matrix elements for the hard production and decay processes within {\tt Sherpa} are delivered 
by its built-in matrix element generator \textsc{Amegic++}~\cite{Krauss:2001iv}.  At present, 
\textsc{Amegic++} provides tree-level matrix elements with up-to ten final state particles in the 
framework of the SM~\cite{Gleisberg:2003bi}, the THDM, the MSSM~\cite{Hagiwara:2005wg} and the ADD 
model~\cite{Gleisberg:2003ue}.  In general, the program allows all coupling constants to be complex.  

\noindent
The Standard Model interactions implemented allow for the full CKM mixing of quark generations 
including the complex phase. The implemented set of Feynman rules for the MSSM 
\cite{Rosiek:1989rs,Rosiek:1995kg} also considers CKM mixing in the supersymmetrized versions of the 
SM weak interactions, and the interactions with charged Higgs bosons.  A priori, \textsc{Amegic++} allows 
for a fully general inter-generational mixing of squarks, sleptons and sneutrinos, therefore allowing 
for various flavour changing interactions.  However, the MSSM input parameters being obtained from the
SLHA-conform files~\cite{Skands:2003cj}, only the mixing of the third generation scalar fermions is 
considered per default.  An extension of the SLHA inputs is straightforward and should also allow to 
consider complex mixing parameters. The implementation of bilinear R-parity violating supersymmetric 
interactions, triggering flavour violation effects as well, has currently being started. 

\noindent
Within Sherpa the multi-leg matrix elements of \textsc{Amegic++} are attached with the \textsc{Apacic++} 
initial- and final-state parton showers~\cite{Krauss:2005re} according to the merging algorithm of 
\cite{Catani:2001cc,Krauss:2002up,Krauss:2004bs,Schalicke:2005nv}.  This procedure allows for the 
incorporation of parton showering and, ultimately, hadronization and hadron decay models, independent 
of the energy scale of the hard process. 

\item Hadronization within {\tt Sherpa} is performed through an interface to {\tt PYTHIA}'s string 
fragmentation~\cite{Sjostrand:2001yu}, the emerging unstable hadrons
can then be treated by {\tt Sherpa}'s built-in hadron decay module
\textsc{Hadrons++}.  The current release, {\tt Sherpa}-1.0.9,
includes an early development stage, which already features complete
$\tau$-lepton decays, whereas the version currently under development
includes decay tables of approximately 100 particles.  Many of their
decay channels, especially in the flavour-relevant $K$, $D$ and $B$
decays, contain matrix elements and form factor models, while the rest
are decayed isotropically according to phase space. Throughout the
event chain of {\tt Sherpa} spin correlations between subsequent
decays are included.  A proper treatment of neutral meson mixing
phenomena is also being implemented.
  
\noindent
The structure of {\tt Sherpa} and its hadron decay module
\textsc{Hadrons++} allows for an easy incorporation of additional or
customized decay matrix elements.  In addition, parameters like
branching ratios or form factor parametrizations can be modified by
the user.
\end{itemize}

\section*{Acknowledgments}

We thank A.~Deandrea, J.~D'Hondt, A.~Gruzza,
I.~Hinchliffe, M.M.~Najafabadi, S.~Paktinat,
M.~Spiropulu and Z.~Was
for their presentatios in the WG1 sessions.

This work has been supported by the EU under the MRTN-CT-2004-503369
and MRTN-CT-2006-035505 network programmes.
A.~Bartl, K.~Hohenwarter-Sodek, T.~Kernreiter, and W.~Majerotto have been 
supported by the 'Fonds zur F\"orderung der
wissenschaftlichen Forschung' (FWF) of Austria, project. No. P18959-N16,
K.~Hidaka has been supported by RFBR grant No 07-02-00256.
The work of N.~Castro and F.~Veloso has supported by Funda\c{c}\~ao para
a Ci\^encia e Tecnologia (FCT) through the grants SFRH/BD/13936/2003 and
SFRH/BD/18762/2004.
The work of P.~M.~Fereira and R.~Santos is supported by FCT under
contract POCI/FIS/59741/2004. P.~M.~Fereira is supported by FCT under
contract SFRH/BPD/5575/2001. R.~Santos is supported by FCT under
contract SFRH/BPD/23427/2005.
J.~Guasch and J.~Sol\`a have been supported in part by MEC and
FEDER under project 2004-04582-C02-01 and by DURSI Generalitat de
Catalunya under project 2005SGR00564.
The work of S.\ Penaranda has been partially supported by the 
European Union under contract No.~MEIF-CT-2003-500030, and by the 
I3P Contract 2005 of IFIC, CSIC.  J.I.\ Illana acknowledges the
financial support by the EU\ (HPRN-CT-2000-149), the Spanish MCYT
(FPA2003-09298-C02-01) and Junta de Andaluc{\'\i}a (FQM-101).
The work of S.~B{\'e}jar has been supported by CICYT (FPA2002-00648), by
the EU (HPRN-CT-2000-00152), and by DURSI (2001SGR-00188).
S.\ Kraml is supported by an APART (Austrian Programme of Advanced Research
and Technology) grant of the Austrian Academy of Sciences. A.R.\ Raklev
acknowledges support from the European Community through a Marie Curie
Fellowship for Early Stage Researchers Training and the Norwegian
Research Council.
J. A. Aguilar-Saavedra acknowledges support by a MEC Ram\'on y Cajal contract.
P.~Skands  has been partially
supported by STFC and by Fermi Research Alliance, LLC, under Contract
No.\ DE-AC02-07CH11359 with the United States Department of
Energy.
M.~Misiak acknowledges support from the EU Contract MRTN-CT-2006-035482,
FLAVIAnet."

The material in \chapt{chap:susy}{WG1:sec:atlasslep} has been presented by
by I.~Borjanovi\'c for the ATLAS collaboration.
J.A.~Aguilar-Saavedra, J.~Carvalho, N.~Castro, A.~Onofre, F.~Veloso, 
T.~Lari and G.~Polesello
thank members of the ATLAS Collaboration for helpful
discussions. They have made use of ATLAS physics analysis and simulation
tools which are the result of collaboration-wide efforts

\providecommand{\href}[2]{#2}\begingroup\raggedright\endgroup

\end{document}